UNIVERSITÀ DEGLI STUDI DI PAVIA

DOTTORATO DI RICERCA IN FISICA – XXVI CICLO

# Classifying Unidentified Gamma-ray Sources

***David Salvetti***

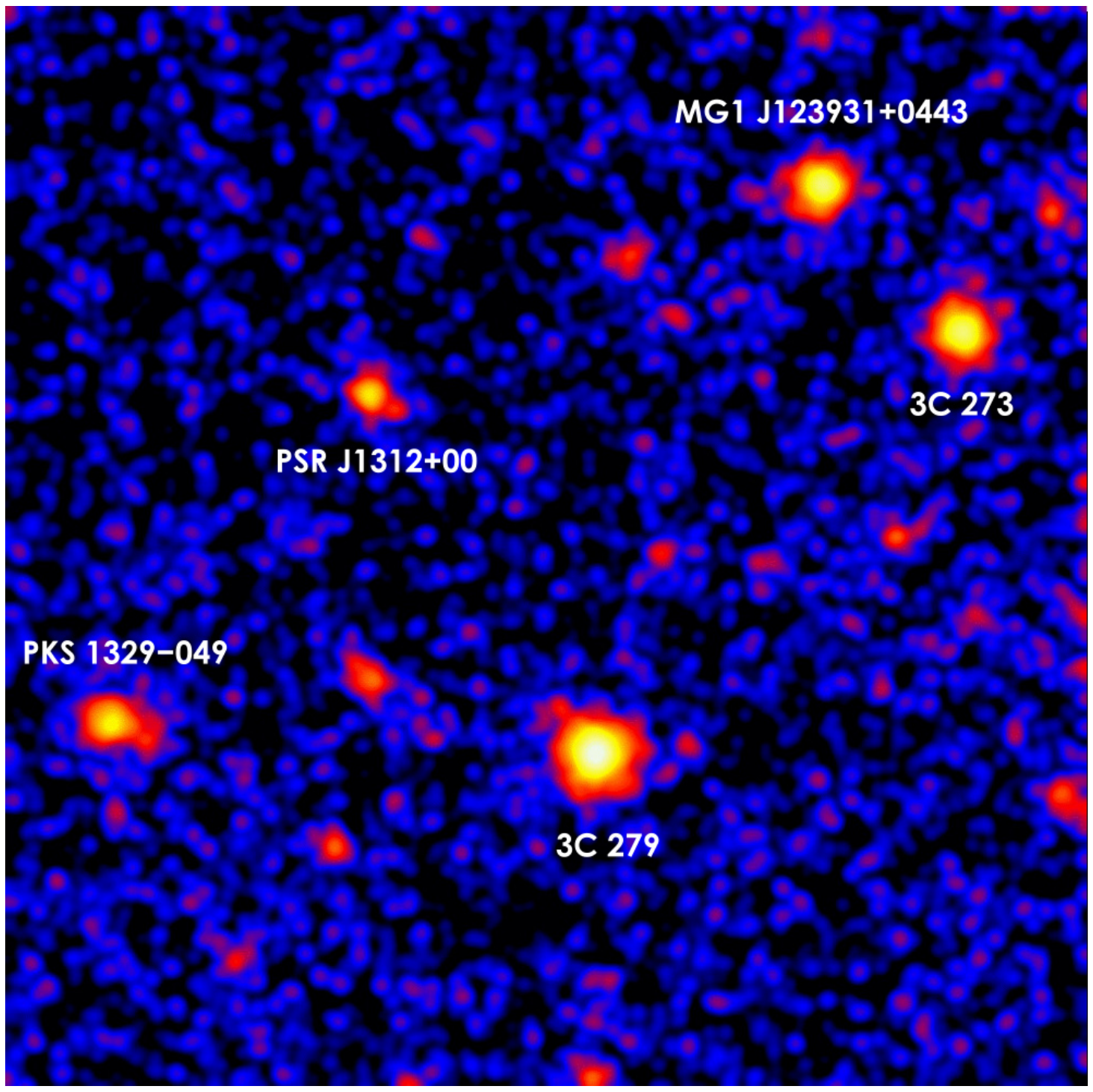

Supervisor: Dr. Patrizia A. Caraveo

Tesi per il conseguimento del titolo

Università degli Studi di Pavia

Istituto Nazionale di Astrofisica

Istituto Nazionale di Fisica Nucleare

DOTTORATO DI RICERCA IN FISICA – XXVI CICLO

# Classifying Unidentified Gamma-Ray Sources

By

***David Salvetti***

Submitted to the Graduate School in Physics
in Partial Fulfillment of the Requirements
for the Degree of

**DOTTORE DI RICERCA IN FISICA**

**DOCTOR OF PHILOSOPHY IN PHYSICS**

at the

University of Pavia

**Supervisor: Dr. Patrizia A. Caraveo**

Università degli Studi di Pavia

INAF/IASF, Istituto Nazionale di Astrofisica

**Cover:** *Fermi*-LAT image of five-year exposure of gamma rays with energies greater than 1 GeV. The image identifies several blazars and one pulsar (PSR J1312+00) in a 20-degree-wide patch in the constellation Virgo. Brighter colors indicate brighter gamma-ray sources. The view is centered at R.A. 13h 00m, Dec. -2d 00m.
Image credit: NASA/DOE/Fermi LAT Collaboration

*A Chiara e alla mia famiglia*

# Contents

# Introduction

The study of cosmic $\gamma$-ray sources is one of the most recent and exciting fields of research of modern astrophysics. The emission of $\gamma$ rays is related to the most violent and powerful phenomena in the Universe and provides a unique way to probe extreme physical environments.

Astrophysical sources of high-energy $\gamma$ rays (photons with energies above 10 MeV) have long been hard to identify. Only four of the 25 $\gamma$-ray sources in the second COS-B catalog (1981) had identifications, and over half the 271 sources in the third EGRET catalog (1999) had no associations with known objects at other wavelengths. The difficulty of finding counterparts of high-energy $\gamma$-ray sources was due the large positional errors in their measured locations, a result of the limited photon statistics and of poor angular resolution of the $\gamma$-ray instruments as well as the bright diffuse $\gamma$-ray emission from the Milky Way.

A major step forward for detection and identification of high-energy $\gamma$-ray sources came when the *Fermi Gamma-ray Space Telescope* was launched on 2008 June 11. *Fermi* is an international space mission that is studying the $\gamma$-ray sky covering the energy range from 10 keV to more than 300 GeV. It carries on board two instruments, the *Gamma-ray Burst Monitor*, that is mainly devoted to the study of the Gamma-Ray Bursts, powerful explosions in the cosmos that originate intense flashes of high-energy photons, and the *Large Area Telescope* (LAT), a pair conversion telescope based on advanced detectors for High Energy Physics that is the main *Fermi* instrument.

The optimal performances of the *Fermi*-LAT in terms of angular resolution and sensitivity allow studying the 100 MeV to $\sim$300 GeV $\gamma$-ray sky with unprecedented detail. During the first 2 years of mission it discovered a number of $\gamma$-ray sources one order of magnitude larger than those detected by EGRET during its 9 years of activity. Almost 70% of detected sources were associated with objects known in other wavelengths. By now, the *Fermi*-LAT instrument has discovered about one thousand of extragalactic objects (Active

Galactic Nuclei – AGN – of different classes), more then one hundred $\gamma$-ray pulsars (evenly divided between young radio-loud pulsars, young radio-quiet pulsars and radio-loud MSPs), and a handful of other Galactic sources, such as SNRs, HMXBs and globular clusters. In spite of the application of advanced techniques of identification and association of the LAT sources with counterparts at other wavelengths, the source uncertainty (ranging from 1 to 3 arcmin) prevents the association of $\sim$40% of the LAT sources. While greatly improved over the degree-scale uncertainties of previous instruments, these position uncertainties are still inadequate to allow identifications based solely on positional coincidence.
Thus, understanding the nature of the $\gamma$-ray unidentified sources is one of the most important open questions in high energy astrophysics.

The purpose of my Ph.D. work has been to pursue a statistical approach to identify unassociated $\gamma$-ray sources. To this aim, we implemented advanced machine learning techniques in order to classify sources, based on all the available $\gamma$-ray information about locations, energy spectra and time variability. The results from our analyses have been used for selecting targets for AGN and pulsar searches and planning multi-wavelength follow-up observations. In particular we have focused on the search of possible radio-quiet millisecond pulsar (MSP) candidates in the sample of the *Fermi*-LAT unidentified sources. These objects have not yet been detected but their discovery would have a formidable impact for our understanding of the MSP $\gamma$-ray emission mechanism.

The thesis is organized as follows.
In Chapter 1 we give a brief review of the main issues regarding $\gamma$-ray astrophysics, focusing on our knowledge on the $\gamma$-ray Universe before the launch of *Fermi*.
In Chapter 2 we give an overview of the procedure developed by the LAT collaboration to construct the *Fermi*-LAT source catalogs.
In Chapter 3 we discuss my first personal contribution within the *Fermi*-LAT collaboration to determine likely source classifications for the unidentified $\gamma$-ray sources in the first *Fermi*-LAT catalog using the Logistic Regression algorithm. This is a machine learning technique that uses identified objects to build up a classification analysis, yielding the probability for an unidentified source to belong to a given class based on its $\gamma$-ray properties.
In Chapter 4 we describe in detail our analysis devoted to the classification of sources in the second *Fermi*-LAT catalog. First, using the most recent $\gamma$-ray source catalog, we implemented a refined Logistic Regression analysis aimed at discriminating $\gamma$-ray pulsars and AGNs, the two most numerous classes of $\gamma$-ray sources. Second, we developed a more advanced machine learning technique, the Artificial Neural Networks (ANNs). These two classification techniques have never been applied for this purpose in astronomy. Finally, af-ter comparing the results obtained by our classification algorithms, we present the development of a complex neural network architecture in order to distinguish the pulsar and AGN subclasses.
In Chapter 5 we describe our multi-wavelength analyses of three radio-quiet MSP candidates selected by our classification algorithms. Multi-wavelength studies allow us to select the most probable counterparts of the putative radio-quiet MSP. Blind frequency searches for $\gamma$-ray pulsations from these unidentified sources can be efficiently run on much smaller sky area of the precise positions of their probable counterparts. Moreover, we describe a multi-wavelength study of an interesting fast-moving radio-quiet $\gamma$-ray pulsar, PSR J0357+3205, and its unusual "trail".
The Conclusions section summarizes the main results and provides perspectives future developments.
Appendix A is dedicated to an overview of the *Fermi* mission and to its main instruments, in particular to the *Large Area Telescope*, which is compared to its predecessor EGRET.
In Appendix B we describe in detail the theory of machine learning algorithms we have implemented and applied to classify $\gamma$-ray unidentified sources.
Appendix C contains likely source classification for the unidentified sources in the second *Fermi*-LAT source catalog on the basis of their $\gamma$-ray observables using the advanced artificial neural network developed and discussed in the Chapter 4.
Appendix D contains two papers we have published on astrophysical journals about the analyses of the X-ray emission from two isolated neutron stars, the *Central Compact Object* RX J0822-4300 and the *Magnetar* SGR 0418+5729.

# Chapter 1

# The gamma-ray Universe before Fermi

Gamma-ray astrophysics is presently one of the most interesting and exciting fields of research for several reason. The emission of $\gamma$ rays is related to the most violent and powerful phenomena in the Universe and provides a unique way to probe extreme physical environment characterized by the presence of intense magnetic fields and high energy particles. The $\gamma$-ray energy band extends from 100 keV up to multi TeV energies, making it the most energetic portion of the electromagnetic spectrum. Because Earth's atmosphere absorbs $\gamma$ rays it is necessary to put detectors at high altitude using balloons or satellites such as *Fermi Gamma-ray Space Telescope* (*Fermi*). At energies above 100 GeV it is possible to use the atmosphere itself as a detector to study the electromagnetic showers of the Very High Energy (VHE) $\gamma$ rays from the ground. This is the basic concept of the ground $\gamma$-ray *Čherenkov* telescopes like MAGIC, HESS, VERITAS or CANGAROO or large arrays like MILAGRO.

Gamma rays are extremely useful messengers since they are neutral, so that they are not deflected by cosmic magnetic fields (as happens e.g. for cosmic rays), and extremely energetic, so that they are unlikely absorbed by cosmic matter (as happens for photons of lower energy). The Universe is largely transparent to $\gamma$ rays and each $\gamma$-ray points directly back to its source. Thanks to these characteristic $\gamma$ rays permit to observe and study high energy cosmic sources extremely distant, up to $z \sim 5$, that act as natural engines accelerating particles up to extremely high energy.

The study of cosmic $\gamma$ rays is extremely important for different research fields, including Cosmology, Particle Physics and the search for Dark Matter. In this Chapter a review of the status of our knowledge about the $\gamma$-ray sky in the energy range of *Fermi* is presented, with an overview of the classes of $\gamma$-ray objects known before the launch of *Fermi* happened on June 11, 2008.

## 1.1 Gamma-ray astronomy

The development of the $\gamma$-ray astronomy have been carried mainly in the last decades, when the techniques to observe the high-energy Universe was developed. Three important facts explain why it was so hard develop specific techniques to observe the $\gamma$-ray sky:

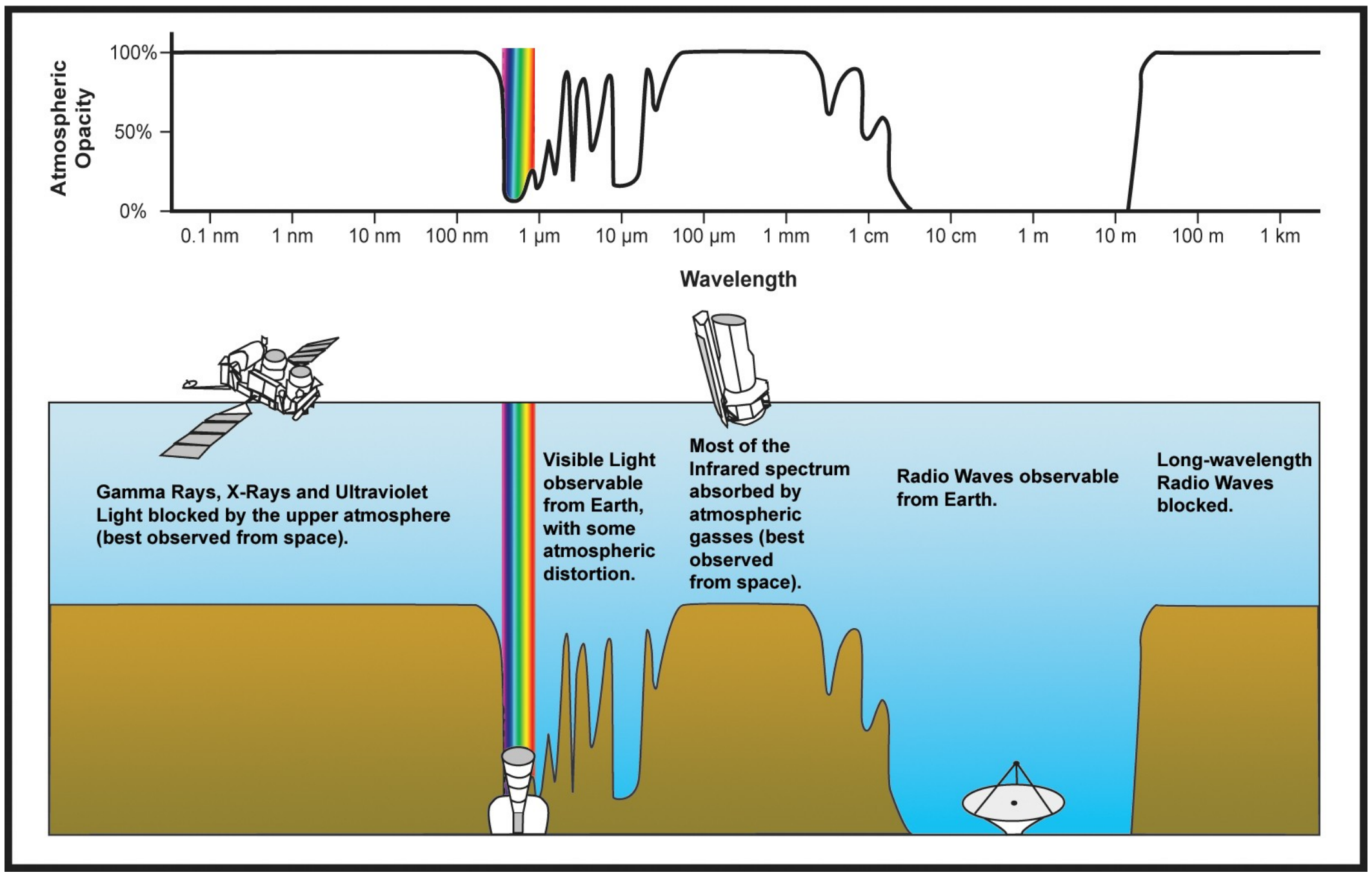

Figure 1.1: Representation of the atmosphere opacity for different wavelengths.

- As shown in Figure 1.1, the Earth's atmosphere is opaque to high-energy photons. At sea level, the atmosphere is 1033 g/cm$^2$ thick, this implies that an high-energy photon incident from the zenith can reach ground level without interacting electromagnetically with a probability of $\sim 10^{-10}$. Even at mountain altitudes, where the atmosphere is thinner, the probability that an high-energy photon can survive to ground is negligible. Only a detector above the atmosphere (satellite or balloon) can detect primary cosmic $\gamma$ rays.

- The flux of $\gamma$ rays from astrophysical sources is quite low and decrease rapidly with increasing energy. For example, the *Vela* pulsar, a very bright $\gamma$-ray object, has a flux equal to $\sim 10^{-5}$ photons cm$^{-2}$ s$^{-1}$ integrated above 100 MeV, i.e. about 1 photon cm$^2$ each day. This implies that a critical problem to detect a $\gamma$-ray source is the sensitivity, we need detectors with a large effective area. A satellite-based detector is too small to detect enough photons above about 100 GeV, this implies VHE $\gamma$-ray astronomy can be performed using only ground-based detectors, which reconstruct the energy and the direction of $\gamma$ rays from the study of their electromagnetic showers in the atmosphere.

- The flux of high-energy charged cosmic rays is much larger (a factor 3 higher in the energy range 100 GeV - 400 TeV) than the flux of $\gamma$ rays. This large cosmic-ray background has to be rejected in order to study the $\gamma$-ray sources. Therefore, detectors have to be able to distinguish efficiently cosmic rays from $\gamma$ rays.

There are two detector techniques, the *satellite-based techniques* and the *ground-based techniques*, this is the detection strategy of the Air *Čerenkov* Telescopes (ACTs) and of the Extended Arraw Shower Detectors (EASDs). Satellite experiments can operate until ∼300 GeV, otherwise ground-based experiments until ∼50 TeV.

### 1.1.1 Explorers of the gamma-ray sky

In the 1950s, works by Hayakawa [95], Hutchinson [106] and especially by Morrison [152] had led scientists to believe that a number of processes occurring in the Universe would result in $\gamma$-ray emission. These processes included cosmic ray interactions with interstellar medium, supernova explosions and interactions of energetic charged particles with intense magnetic fields. However, only in the 1960s the first cosmic $\gamma$ rays were detected.

Since the interaction of photons with an energy above 10 MeV is dominated by the pair production, all satellite-based telescope use a spark chamber or layers of tracker/converter made with high Z foil to estimate the incoming direction of the photons, a calorimeter to measure their energy and an anticoincidence shield to reduce the background due to charged particles. The results obtained with the first $\gamma$-ray space telescopes were affected by low statistics and large systematic errors.

The first $\gamma$-ray space mission was *Explorer XI* in 1961 [117] and it detected less than 100 photons uniformly distributed in the sky implying the presence of a sort of uniform "$\gamma$-ray background". Such a background would be expected from the interaction of cosmic rays with the interstellar medium. The next important step was the NASA *Orbiting Space Observatory III* (OSO III) mission in 1968 [118]. It detected about 600 photons concentrated on the Galactic plane attributable to the $\gamma$-ray production in the Milky Way. The main detector used scintillators and Čerenkov detectors and was able to detect photons above 50 MeV.

In 1969 and 1970 United States launched the Vela series spacecraft to detect X rays and $\gamma$ rays coming from the Earth or the Moon in order to determine if Soviets were complying with the nuclear test ban treaty. While no atmospheric $\gamma$ rays were detected, they serendipitously discovered transient flashes of radiation lasting in average of 10 ms to 10 s in random direction on the sky named Gamma-Ray Bursts (GRBs) [115].

**The SAS-2 observatory.** The first satellite exclusively designed for a $\gamma$-ray mission was the second *Small Astronomy Satellite* (SAS-2), launched in the 1972 [116]. SAS-2 carried a single telescope with a 32-level wire spark-chamber covering the energy range from 20 MeV to 1 GeV and with an effective area of 100 cm$^2$. SAS-2 provided the first detailed information about the $\gamma$-ray sky revealing a strong correlation between the diffuse radiation coming from the Galactic plane and and the Galactic structural features and it was the first satellite to detect the isotropic, apparently extragalactic, $\gamma$-ray emission. Moreover, SAS-2 resolved the first point sources detecting a pulsed $\gamma$-ray emission from 3 sources, the *Crab* and *Vela* pulsars and *Geminga*, identified as a pulsar more years later. A map profile along the Galactic plane obtained by SAS-2 is displayed in Figure 1.2.

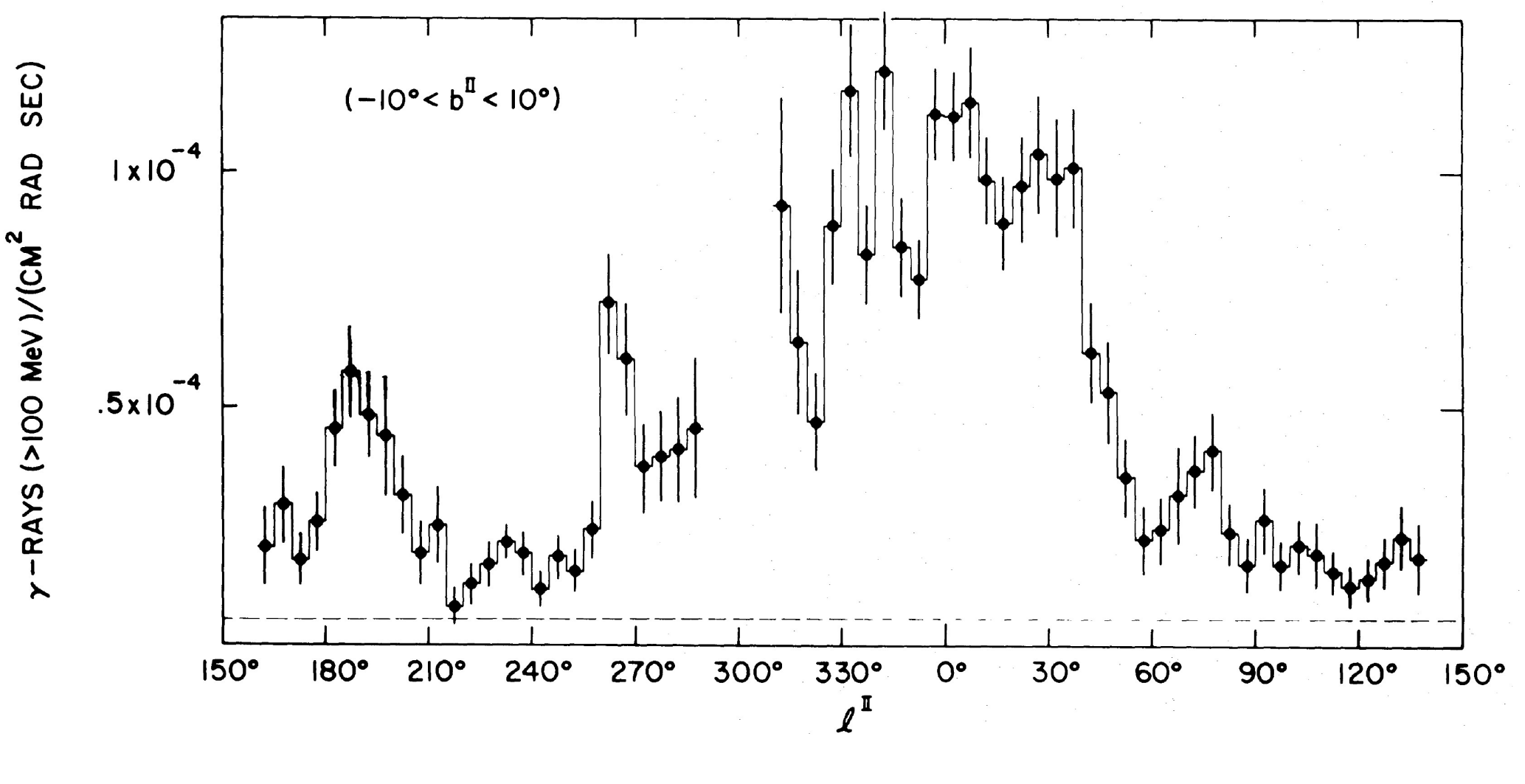

Figure 1.2: Distribution of high-energy (E>100 MeV) $\gamma$ rays along the Galactic plane observed by SAS-2. The SAS-2 data are summed from b=–10° to b=10°. The diffuse background level is shown by a dashed line [77].

**The COS-B observatory.** In 1975 the *European Space Agency* (ESA) launched COS-B [195]. As SAS-2, COS-B carried a wire spark-chamber that recorded the direction of the electron-positron pair created in thin Tungsten plates and a CsI calorimeter that measured the energy of the charged parti-cles. The upper part of the instrument was surrounded by an anticoincidence counter to assure that only neutral particles triggered the instrument. COS-B was able to detect photons in the energy range between 30 MeV and 5 GeV. The major result of the COS-B experiment was the creation of the first $\gamma$-ray catalog [35] that contained about 25 $\gamma$-ray sources from the observation made during the first three years of activity. The catalog included the *Crab* and *Vela* pulsars, the molecular cloud $\rho$-Oph and the first extragalactic $\gamma$-ray object, the Active Galactic Nucleus (AGN) 3C 273. The first full $\gamma$-ray map of the Galactic plane obtained by COS-B is shown in Figure 1.3.

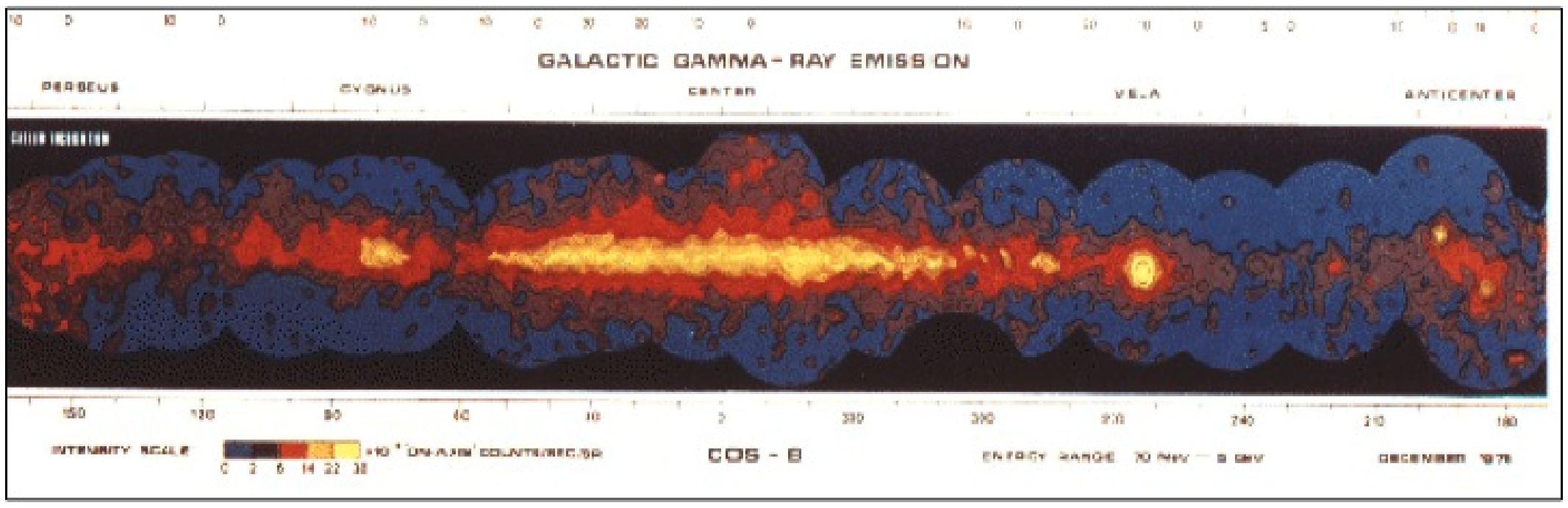

Figure 1.3: Gamma-ray map of the Galactic plane observed by COS-B in the energy range between 30 MeV and 5 GeV [35].

**The CGRO observatory.** After a stop of about 15 years because of the *Space Shuttle Challenger* disaster occurred in 1986, the NASA mission *Compton Gamma Ray Observatory* (CGRO) revolutionized our understanding of the $\gamma$-ray sky because of the improved sensitivity with respect to previous missions. The CGRO satellite was launched in 1991 and it carried four instruments, OSSE, COMPTEL, BATSE and EGRET, that covered an energy range between 20 MeV and 30 GeV. The *Oriented Scintillation Spectrometer Experiment* (OSSE) consisted of four NaI scintillation crystals and its main objective was spectroscopy of cosmic $\gamma$-ray sources and solar flares in the energy range from 50 keV to about 10 MeV. OSSE measured the distribution of the energy emitted from a number of $\gamma$-ray objects, studied nuclear lines in solar flares, radioactive decay of nuclei in supernova remnants and matter-antimatter annihilation taking place near the center of our Galaxy [109]. The *imaging COMpton TELescope* (COMPTEL) was an imaging detector, its main objective was to study active galaxies, supernova remnants and diffuse $\gamma$-ray emission from giant molecular clouds in the energy range between 1 and 30 MeV [109]. The *Burst and Transient Source Experiment* (BATSE) was the smallest instrument onboard CGRO. It was dedicated to monitor the full celestial sphere for transient $\gamma$-ray phenomena such as Gamma-Ray Bursts (GRBs) and bursts from other cosmic sources (including solar flares from the sun) all over the sky in the energy range from 20 keV to 1 GeV [109].

### The EGRET observatory

The *Energetic Gamma-Ray Experiment Telescope* (EGRET) was the main instrument onboard CGRO and it was devoted to the highest energy ever observed from the space, reaching the upper limit of E ∼30 GeV. EGRET was a pair-conversion telescope, the conversion of the high-energy photons into electron-positron pairs occurred in an upper stack of 28 Tantalum conversion foils of an average thickness of 90 $\mu$m interleaved with spark chamber modules. The spark chambers were filled with a gas mixture of neon, argon and ethane. The direction of the radiation was determined by a time of flight coincidence below the conversion chamber. The Total Absorption Spectrometer Calorimeter (TASC) was made of NaI, its thickness corresponded to about 8 radiation lengths and enabled to determine photon energies from 20 MeV up to about 30 GeV [109]. An anti-coincidence scintillation dome surrounded the instrument and it was able to distinguish a photon from a cosmic charged particle with an efficiency of about 0.9999. EGRET had a solid angle acceptance of $\sim 0.15\pi$ and an effective area of 0.12 $m^2$ at 1 GeV. The effective area decreased at higher energies, partly because backsplash from the calorimeter could trigger the anti-coincidence veto. The energy resolution was 9–12%, depending on energy. The angular resolution was 3.5° at 100 MeV and it improved to 0.8° at 1 GeV and 0.35° at 10 GeV. The instrument was extremely sensitive, about $5 \times 10^{-8}$ photons $cm^{-2}$ $s^{-1}$ with E > 100 MeV after $10^6$ s of exposure. The mission lasted for nine years and it revolutionized our understanding of the $\gamma$-ray sky because of its better performances with respect to the previous ones. During all the mission EGRET detected about 2 millions of photon with E > 100 MeV, allowing a detailed study of the Galactic and extragalactic diffuse emission and of the point-like $\gamma$-ray sources.

As shown in Figure 1.4, the Third EGRET Catalog (3EG) [93] consists of 271 sources (with E > 100 MeV): 6 pulsars (high-energy pulsed emission from *Geminga* was detected [44]), 93 blazars (a subclass of AGNs discovered to be a new class of $\gamma$-ray emitters [93]), one normal galaxy (the Large Magellanic Cloud) and a solar flare (the single 1991 solar flare). 170 $\gamma$-ray sources detected by EGRET, about 60%, had not a clear association with a class of objects known in a different wavelength and for this reason these “unknown” objects were called unidentified sources. Since the association of a $\gamma$-ray object with a source known in other wavelengths is primarily based on a positional coincidence, the large number of unidentified sources in the 3EG is related to the bad angular resolution of the $\gamma$-ray telescopes. Most of these objects may be members of existing $\gamma$-ray source classes, such as pulsars or blazars, but some of them may be represent discovery space for new source classes. A map of the $\gamma$-ray emission above 100 MeV obtained by EGRET is shown in Figure 1.5.

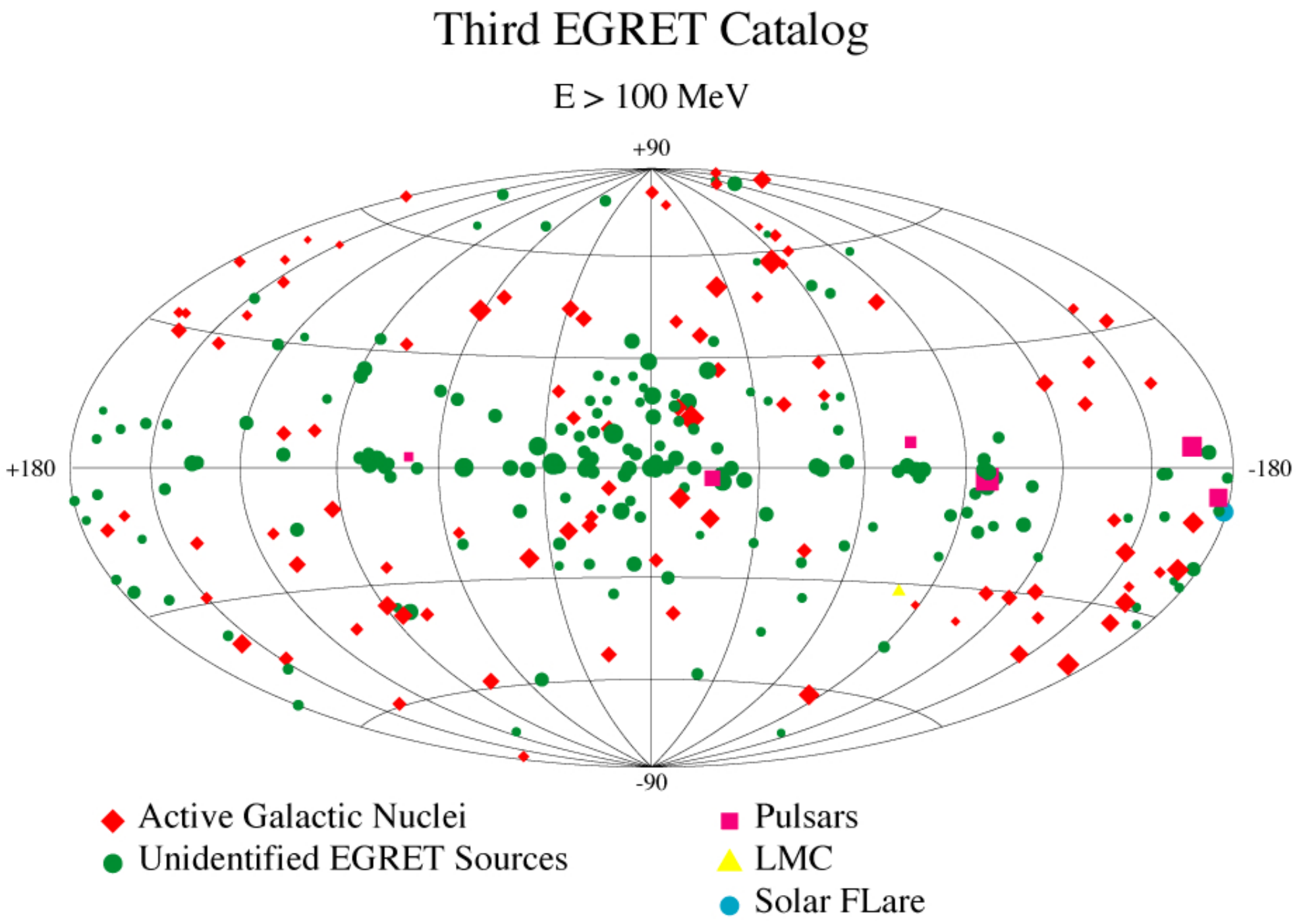

Figure 1.4: Third EGRET source catalog, shown in Galactic coordinates. The size of the symbol represents the highest intensity seen for this source by EGRET [93].

### The AGILE observatory

On April 23, 2007 it has been launched the AGILE mission by the Indian PSLV-C8 rocket from the Satish Dhawan Space Center SHAR, Sriharikota. AGILE (*Astrorivelatore Gamma ad Immagini LEggero*) is a completely italian high-energy astrophysics mission supported by the Italian Space Agency (ASI) with scientific and programmatic participation by INAF (Istituto Nazionale di AstroFisica), INFN (Istituto Nazionale di Fisica Nucleare) and several italian universities. The main scientific goal of AGILE is to explore the $\gamma$-ray Universe with a very innovative instrument combining a $\gamma$-ray imager (sensitive in the energy range of 30 MeV and 50 GeV) and a hard X-ray imager (sensitive in a energy range between 15 and 45 keV) [197]. The AGILE scientific payload

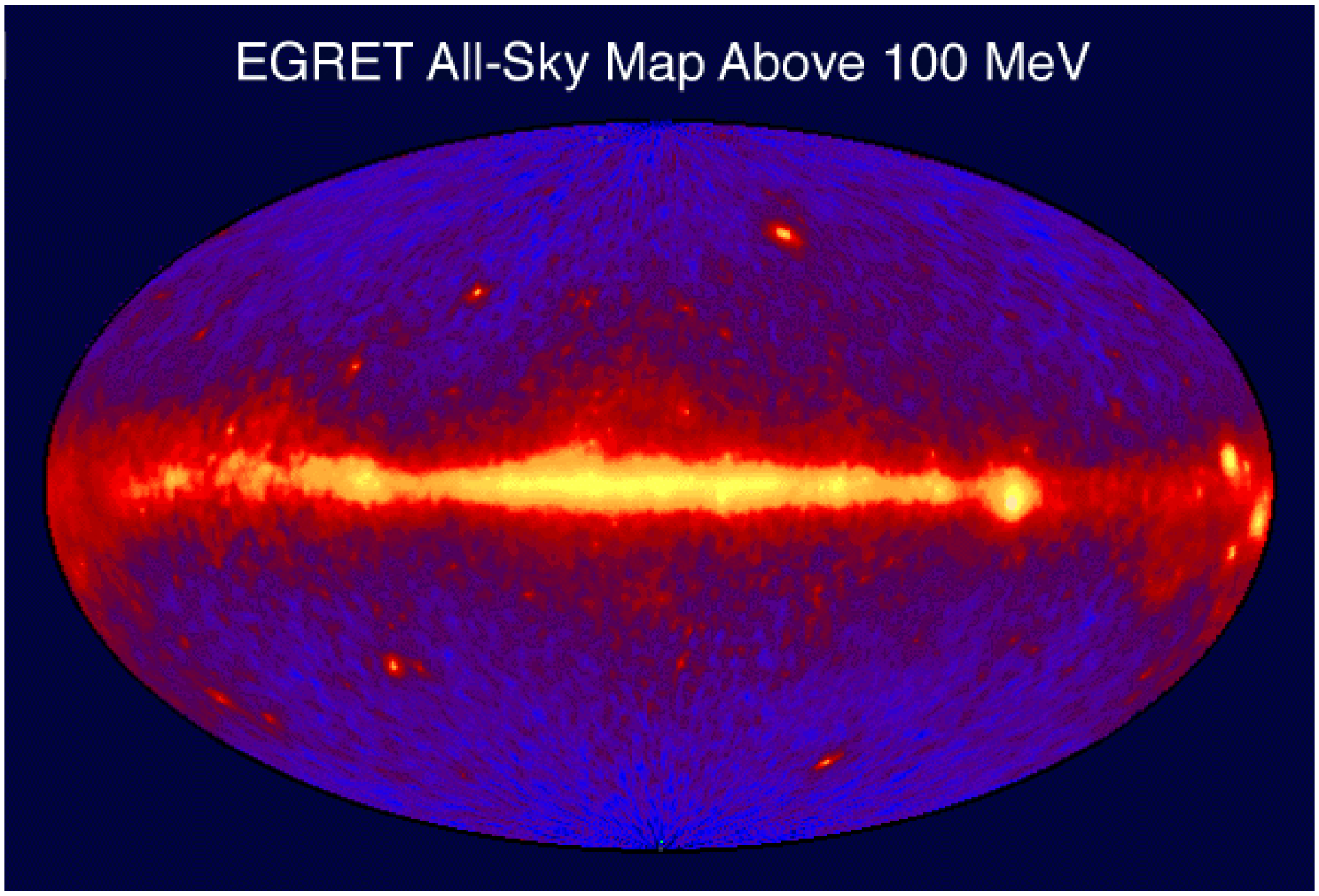

Figure 1.5: Distribution of $\gamma$-ray photons above 100 MeV obtained by EGRET telescope aboard the NASA *Compton Gamma Ray Observatory*. Credit: EGRET Team.

is a small satellite composed by three detectors combined into one integrated instrument, as shown in Figure 1.6. The first detector is the *Gamma-Ray Imaging Detector* (GRID) that is sensitive in the energy range between 30 MeV and 50 GeV and it has a sensitivity compared to EGRET. The detector consists of a Silicon Tracker (ST) providing the $\gamma$-ray imager that is based on photon conversion into electron-positron pair and is composed by a total of 12 trays. The first 10 trays are capable of converting $\gamma$ rays into charged particles by means of Tungsten converter layer positioned in the lower part of the tray. Tracking of charged particles is ensured by high resolution Silicon microstrip detectors positioned at the very top and bottom of the trays. Two more trays are inserted at the bottom of the Tracker without the Tungsten layers. The energy of the particles produced in the Silicon Tracker is deposited in the 30 CsI bars of the Cesium-Iodide Mini-Calorimeter (MCAL) and therefore it contributes to the determination of the total photon energy. All the AGILE detectors are completely surrounded by an AntiCoincidence (AC) System with the aim of rejecting charged particles. It is constitutes by three plastic scintillator layers and the signal is read by photomultiplier placed externally

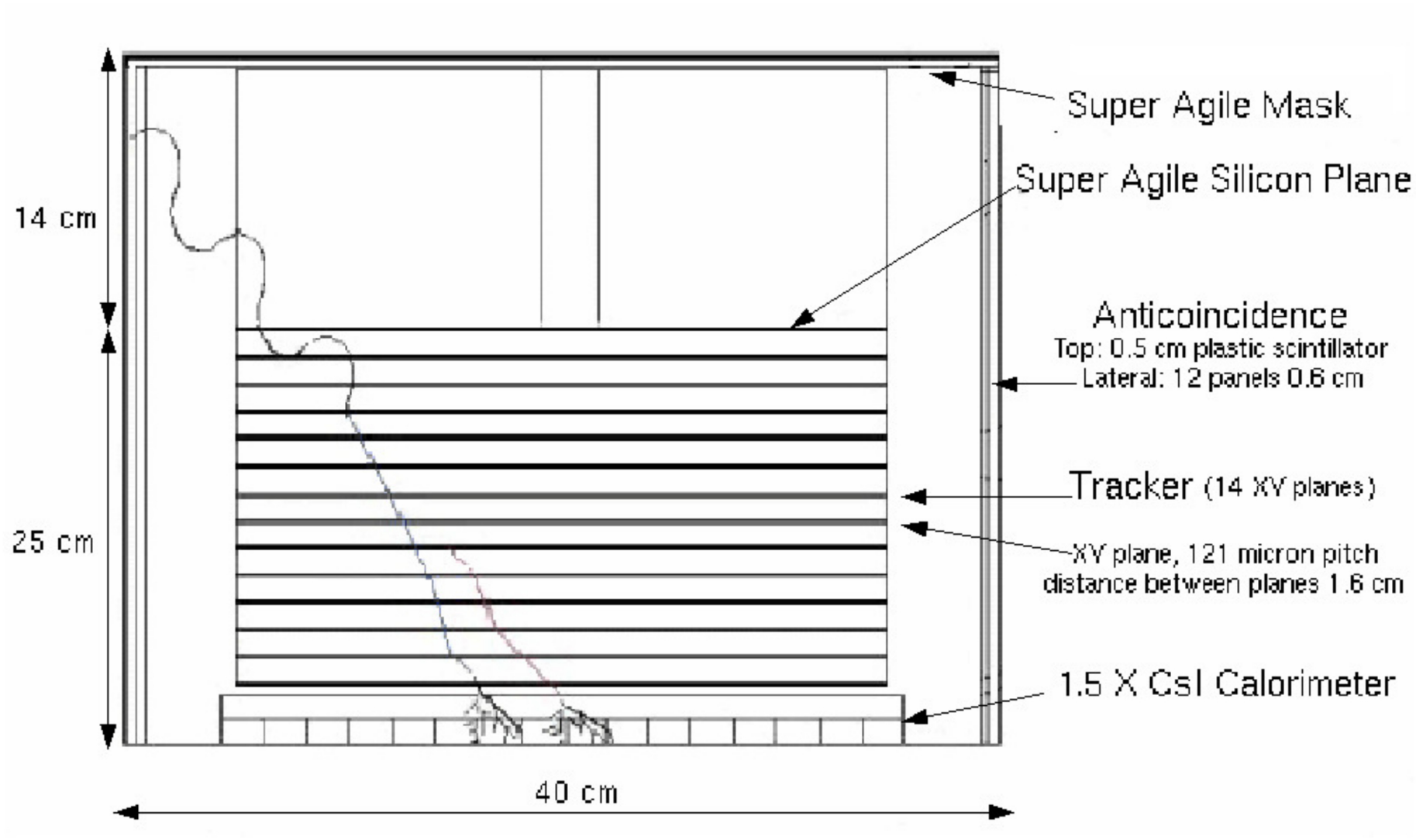

Figure 1.6: Scheme of AGILE detectors. Credit: AGILE Team
.

to the AC System.

The second detector of AGILE is the hard X-ray Imager (*Super-AGILE*) placed on the top of the GRID and it is sensitive in the energy range between 15 keV and 45 keV. The *Super-AGILE* (SA) detector is a coded-mask system made of a Silicon detector plane and a thin Tungsten mask positioned above it. This detector is aimed at simultaneous X-ray and $\gamma$-ray detection of high energy cosmic sources with excellent imaging capabilities.

The third detector is the Cesium-Iodide Mini-Calorimeter (MCAL). The MCAL is part of the GRID but it is also capable of independently detecting GRBs and other transient sources in the energy range between 300 keV and 100 MeV with optimal time capabilities [197].

The very innovative technology of the AGILE instrument allows it to have an excellent imaging capability in the energy range of 100 MeV and 50 GeV, a large Field-of-View (FoV) for both the GRID (FOV $\sim$2.5 sr) and the *Super AGILE* (FoV $\sim$1 sr), larger by a factor of $\sim$6 than the EGRET ones, a good sensitivity in the energy range of 30 MeV and 100 MeV, a very rapid response to $\gamma$-ray transients and a good localization accuracy ($\sim$15 arcmin at 400 MeV for GRID). To date, AGILE has been detecting more than ten million of photons giving the opportunity to study in detail a lot of Active Galactic Nuclei, Gamma-Ray Bursts and other transient objects on the Galactic plane, more than 20 $\gamma$-ray pulsars and some Pulsar Wind Nebulae. One of the main important results of the AGILE mission is the discovery of the $\gamma$-ray emission from accreting binary systems, which was already supposed but never observed. Finally, despite the better performance of AGILE instrument with respect to the previous ones, about 20% of the $\gamma$-ray sources in the first AGILE catalog [166] have not a clear association with a source known in other wavelengths. A map of the $\gamma$-ray emission above 100 MeV obtained by AGILE is shown in Figure 1.7.

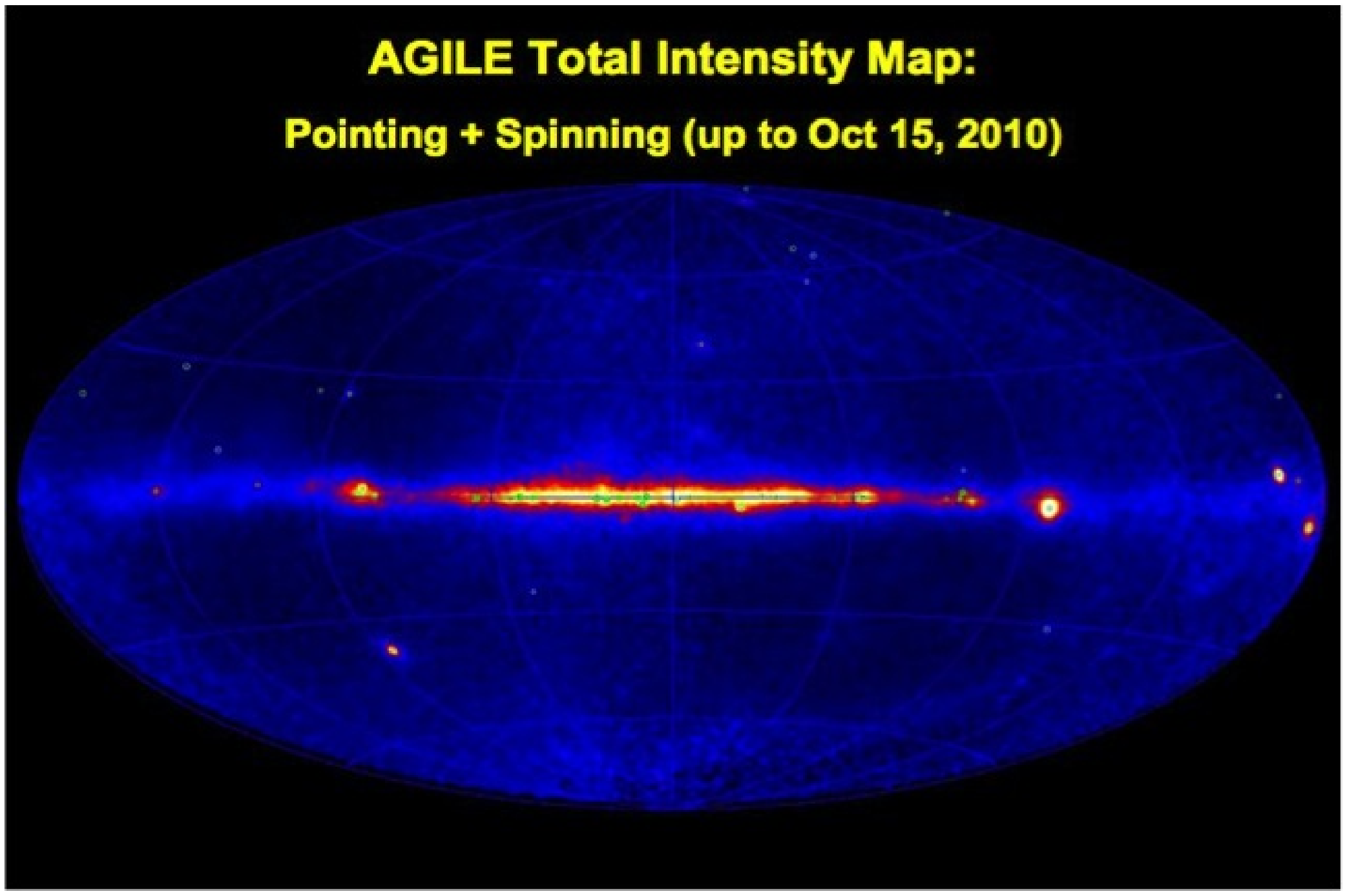

Figure 1.7: Total AGILE-GRID count map in Aitoff projection and Galactic coordinates, for energy above 100 MeV, accumulated during the period July 9, 2007-October 15, 2010. Credit: AGILE Team.

#### Ground-Based Telescopes

The upper part of the $\gamma$-ray spectrum is presently investigated from the ground by looking at the electromagnetic showers created when a high-energy $\gamma$ ray (E $>$ 100 GeV) enters the atmosphere. The predominant radiation-matter interaction at these energies is pair production, when a high-energy $\gamma$ ray enters the atmosphere it produces a pair formed by an electron and a positron, they propagate and produce photons via *Bremsstrahlung* initiating an electromagnetic shower. The resulting electromagnetic cascade grows nearly exponentially as it propagates through the atmosphere. The primary photon energy is distributed among more and more particles until electrons and positron approach to their critical energy (∼80 MeV in air). At this point the ionization energy-loss mechanism, that does not produce additional particles, becomes more important then *Bremsstrahlung*. As a consequence, energy is lost from the shower and the number of particles decreases as the shower continue to propagate.

The *Extensive Air Shower Detectors* (EADs) are large arrays that directly detect the secondary particles from the showers induced in the atmosphere by the interaction between high-energy $\gamma$ rays and air molecules. Examples of EADs are CYGNUS, CASA or MILAGRO.

The *Air Čerenkov Telescopes* (ACTs) detect the Čerenkov radiation produced by the secondary particles during their crossing of the atmosphere. Examples of ACTs are CANGAROO, HEGRA, STACEE, CELESTE, VERITAS, MAGIC or HESS. In order to improve imaging capability and background rejection more telescope are arranged in arrays working in *stereoscopic mode*. This is for example the case of HESS, CANGAROO, VERITAS and MAGIC. The future of the ATCs is the *Čerenkov Telescope Array* (CTA), its exact design is being studied and not yet precisely defined. The idea is to have a large of number of mid-size telescopes, in order to achieve full sky coverage it is planned to install two stations of several telescopes, one in the southern hemisphere and the other in the northern one.

Ground-Based Telescopes can detect only the upper part of the $\gamma$-ray spectrum (E > 100 GeV). To date, it has been discovered more than 100 high energy sources, belonging to the classes of pulsars and their nebulae, supernova remnants, $\gamma$-ray binary systems, star forming regions, starbursts and active galaxies. Moreover, about 18% of the detected high-energy sources have not a clear association with objects known in other wavelengths. A connection between the results obtained by satellite-based and ground-based telescopes is very important to study in detail the nature of the high-energy emission of the most energetic sources in the Universe.

## 1.2 Gamma rays from the sky

The knowledge of $\gamma$-ray sources before the launch of the *Fermi Gamma-Ray Space Telescope* (2008) came mainly from the experience of the CGRO experiment and in particular from the results of the Third EGRET Catalog [93]. It contained 271 sources (with E > 100 MeV) that can be divided into Galactic and extragalactic objects with a significant contribution due to the Galactic and extragalactic diffuse emission. The remaining 60% was not associated with an astrophysical object known at other wavelengths. Since 2000, several authors revisited the data of EGRET instrument because of the new discoveries in the field of $\gamma$-ray astronomy. One of the main results of this re-analysis was the production of a new catalog after a modified diffuse background subtrac-tion. The new 3EG (new Third EGRET Catalog) contained 23 new sources and 121 3EG objects was not detected during the new analysis [101].

Galactic $\gamma$-ray sources are mainly compact objects, such as neutron stars and their nebula or accreting black holes [184]. Supernova remnants (SNRs) are cosmic objects which may be related to the cosmic ray acceleration. Structures like shells that interact with Interstellar Medium (ISM) have been observed with high resolution telescopes in the X-ray band and these sites have been associated with shocks.

The main sources of extragalactic radiation are the Active Galactic Nuclei (AGN), and in particular blazars, a subclass of AGN whose jet is aligned with the line of sight. In the extragalactic Universe an important role is played by transient sources like GRBs, that are shining flashes of radiation.

In addition, a diffuse $\gamma$-ray emission given by the contribution of two different components, Galactic and extragalactic, dominates the entire high-energy Universe. The Galactic one is thought to be related to the interaction between cosmic rays and the matter in the Galactic disk; other side the extragalactic one is probably given by the contribution of thousands unresolved extragalactic point-like sources. Moreover, part of the extragalactic component may be related with the decay of exotic particles in the Primordial Universe.

In the following, a review of the celestial $\gamma$-ray sources known before the launch of the *Fermi Gamma-Ray Space Telescope* is given.

### 1.2.1 Diffuse gamma-ray emission

In Figure 1.5 is shown the map of the $\gamma$-ray sky (E > 100 MeV) based on data taken by the EGRET instrument. It is evident that the diffuse emission dominates the entire $\gamma$-ray sky with the highest intensity coming from the plane of our Galaxy. The diffuse $\gamma$-ray emission can be divided in two components, the Galactic diffuse emission, placed along the Galactic plane and coming from our Galaxy, and the extragalactic diffuse emission, characterized by an isotropic distribution in the sky.

#### Galactic diffuse emission

In 1968 was detected for the first time an emission of high-energy photons coming from our Galaxy and not attributable to point-like sources. A detailed map of the Galactic diffuse $\gamma$-ray emission was accomplished after the SAS-2 and COS-B observations between 1972 and 1982. About 90% of $\gamma$ rays with energies above 100 MeV detected by EGRET came from the Milky Way galaxy. The Galactic diffuse $\gamma$-ray emission can be explained through the energetic interactions between cosmic rays and the interstellar medium. Cosmic Rays (CRs) are very high-energy particles, composed primarily of protons, atomic nuclei and leptons, and their origin is still a mystery. Once they are accelerated up to relativistic velocities through not well known mechanisms, they move through the interstellar medium where are trapped by the Galactic magnetic field. The Galactic diffuse $\gamma$-ray emission is then produced via *Bremsstrahlung*, if a high-energy electron is deflected by nuclei of the ISM, Inverse Compton (IC), if a high-energy electron transfers part of its energy to a soft photon coming from the stellar radiation, and $\pi^0$ decay, if a high-energy proton or atomic nucleus collides an interstellar proton creating a neutral pion [153]. All these three type of interactions product $\gamma$ rays and their spectra are very different.

Detection of this diffuse $\gamma$-ray emission should give information about the production and propagation of cosmic rays in the Milky Way galaxy. After removal of identified galactic point-like sources, the Galactic diffuse emission shows a structure that reflects the main features of the mass distribution in the Galaxy known in other wavelengths. The study of the diffuse emission gives information about spectra and intensities of CR species at distant locations and allows to study CR acceleration in Supernova Remnants (SNRs) and other sources and their propagation in the ISM. On the other hand, the Galactic diffuse emission is a structured background source for point-like sources and its accurate determination is essential in order to understand if an excess of photon emission in a specific region of the sky is related to the existence of a $\gamma$-ray source or to statistical fluctuation of the background and it is also important for accurate localization of such source and its spectrum [102].

### Extragalactic diffuse emission

An apparently isotropic, presumably extragalactic, component of the diffuse $\gamma$-ray flux above 30 MeV was discovered by the SAS-2 satellite and confirmed with EGRET. The low sensitivity and the poor angular resolution of EGRET did not allow an identification of this emission as the contribution of many point-like sources.

The hypothesis on the origin of the extragalactic $\gamma$-ray background emission were various, from the most conservative, such as the summed contribution of thousands of unresolved AGNs, to more exotic, such as the contribution of the annihilation from exciting particles which came from some unknown processes that took place in the primordial Universe. Other explanations involve particles deriving from extension of the Standard Model to supersymmetric particles (SUSY), which can contribute substantially to the Dark Matter content of the Universe and that can be found in the Galactic halos.

The extragalactic diffuse $\gamma$-ray emission is well described by a power law with index $2.1 \pm 0.3$ over EGRET energies and it is consistent with the average index for blazars detected by EGRET, which lends some support to the hypothesis that the isotropic flux is from unresolved AGN sources [190].

Diffuse emission has also been observed from the Large Magellanic Cloud (LMC), characterized by a flux consistent with production by cosmic rays.

### 1.2.2 Active Galactic Nuclei and Blazars

In the Universe there are billions of galaxies, which differ basically from their morphology in the Hubble diagram. Galaxies are composed of stars and ISM (gas and dust), which contribute in the total luminosity of the galaxy (for Milky Way L $\sim 10^{11}$ $L_{\odot}$).

In the Forties, the American astronomer C. Seyfert discovered a new class of galaxies, with intense emission from a star-like nuclei and with broad emission lines, that are more than $10^3$ km/s wide. This kind of objects are now called *Seyfert galaxies* and later, together with other extragalactic sources such as *Quasars* and *BL Lac objects*, they were associated with a new class of galaxies, the *Active Galactic Nuclei* (AGNs) [185]. The observed luminosity of these objects is extremely high (L $\sim 2 \times 10^{46}$ erg/s), which corresponds to more than about twenty Milky Way galaxies. Before CGRO telescope had been launched in 1991, we observed the $\gamma$-ray emission of only one extragalactic objects, the quasar *3C 273*, detected by COS-B about thirty years ago [196]. Among the high-energy objects detected by EGRET, a large fraction was associated with AGNs and in particular with *blazars* [126]. Blazars are very powerful objects characterized by variability, non-thermal emission at $\gamma$ rays and a broad multiwavelength emission that extend from radio to TeV.

According to current models, blazars are a class of AGN where the collimated jet points directly toward the observer. The radiation is boosted by the bulk Lorentz Factor and photons have been observed up to TeV energies. Blazars are also characterized by high degree of polarization and by a variability up to a factor of 100% on the order of a day. The emission above 100 MeV is a significant fraction of the total luminosity and in flaring state the $\gamma$-ray luminosity can exceed the luminosity in all other the energy bands by a factor of $\sim$10 or more. The Spectral Energy Distribution (SED) is characterized by a two peak component, one between radio and X rays, probably caused by Synchrotron emission in the jet, and the other one in the $\gamma$-ray band, peaked in the GeV energies. In Figure 1.8 is shown the SED of the blazars *3C 273* and *3C 279* containing also the EGRET observation [94].

In order to explain the different phenomenology among AGNs, we resort to the current *Unified Model* of AGNs. According to the model, the engine that powers the AGN emission is a supermassive black hole (M $\sim 10^8$ $M_{\odot}$) surrounded by an accreting disk that extends up to about 100 A.U. from the central black hole [204]. The size of the accretion disk can also be inferred from causality arguments by observing the typical variability timescale of an AGN.

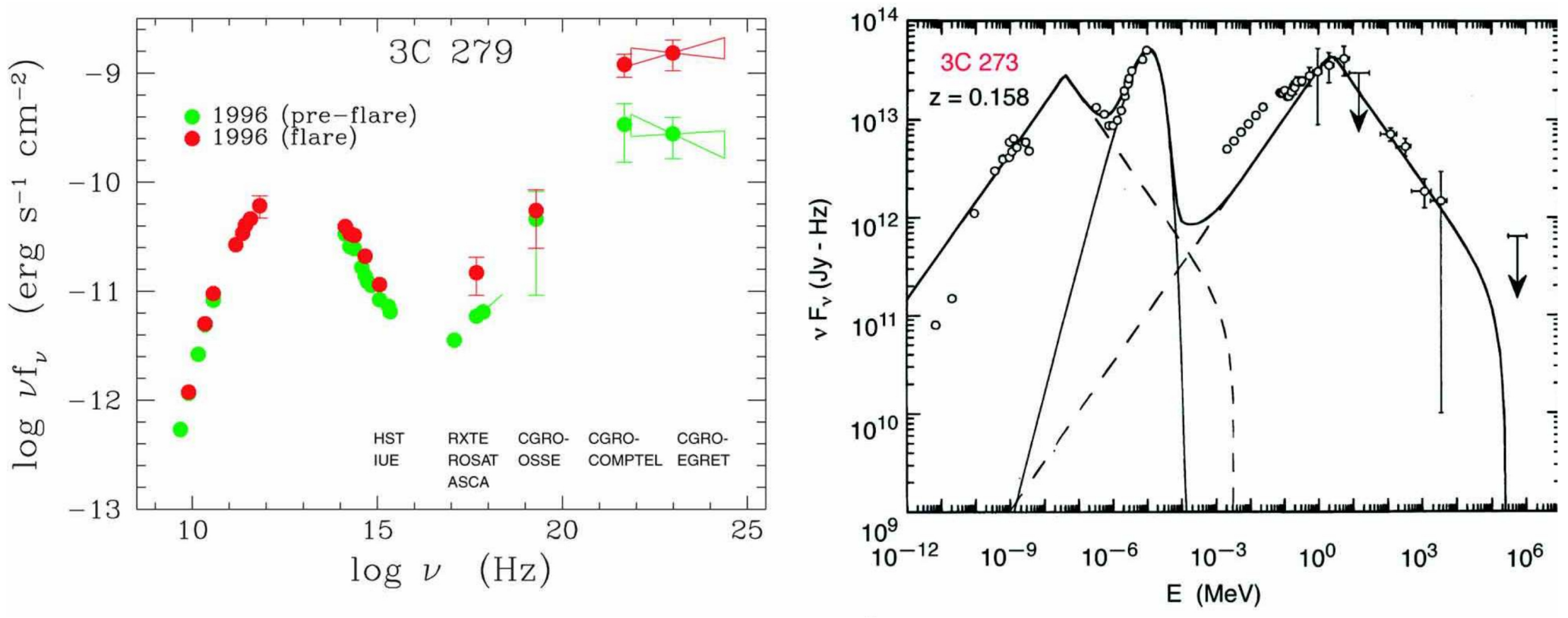

Figure 1.8: Spectral Energy Distribution of the quasar *3C 279* (left) and *3C 273* (right) [74].

Assuming a variability timescale of about 1 day and a redshift of about 0.1 this leads to an estimate of R $\sim 10^{10}$ km, i.e. about 100 A.U.. The accretion is also connected to the presence of large jets where particles are accelerated up to high energies. According to this model at greater distance from the center there is a torus of matter extending from about 1.5 pc to 30 pc.

Between the accretion disk and the torus there are some fast-moving clouds (v $>$ 2000 km/s), that are illuminated by the central engine and originate the emission lines observed in the AGNs spectra, that are widened because of Doppler effect (*Broad Line Regions*) due to their motion around the central supermassive black hole.

At greater distance from the supermassive black hole the model includes some slow moving clouds which emit emission lines with narrower widening due to lower speed (v $<$ 2000 km/s). These clouds are the origin of the narrower emission lines (*Narrow Line Regions*) observed in some class of AGN, e.g. the Seyfert of type I.

According to the *Unified Model* we observe various classes of AGNs because of the different angle under the AGN is seen as shown in Figure 1.9. Current models for the $\gamma$-ray emission in blazars can be divided in two main classes, the *leptonic models* and the *hadronic models*.

According to the leptonic models $\gamma$-ray emission is produced from the interaction of the accelerated electrons and positrons with the environmental soft photons. The origin of these photons depends on the adopted scenario. The *Synchrotron Self Compton* (SSC) scenario assumes the soft photons are emitted directly by the same electrons and positrons in the jet via Synchrotron radiation [133] . The *External Compton Scattering* (ECS) assumes the soft photons

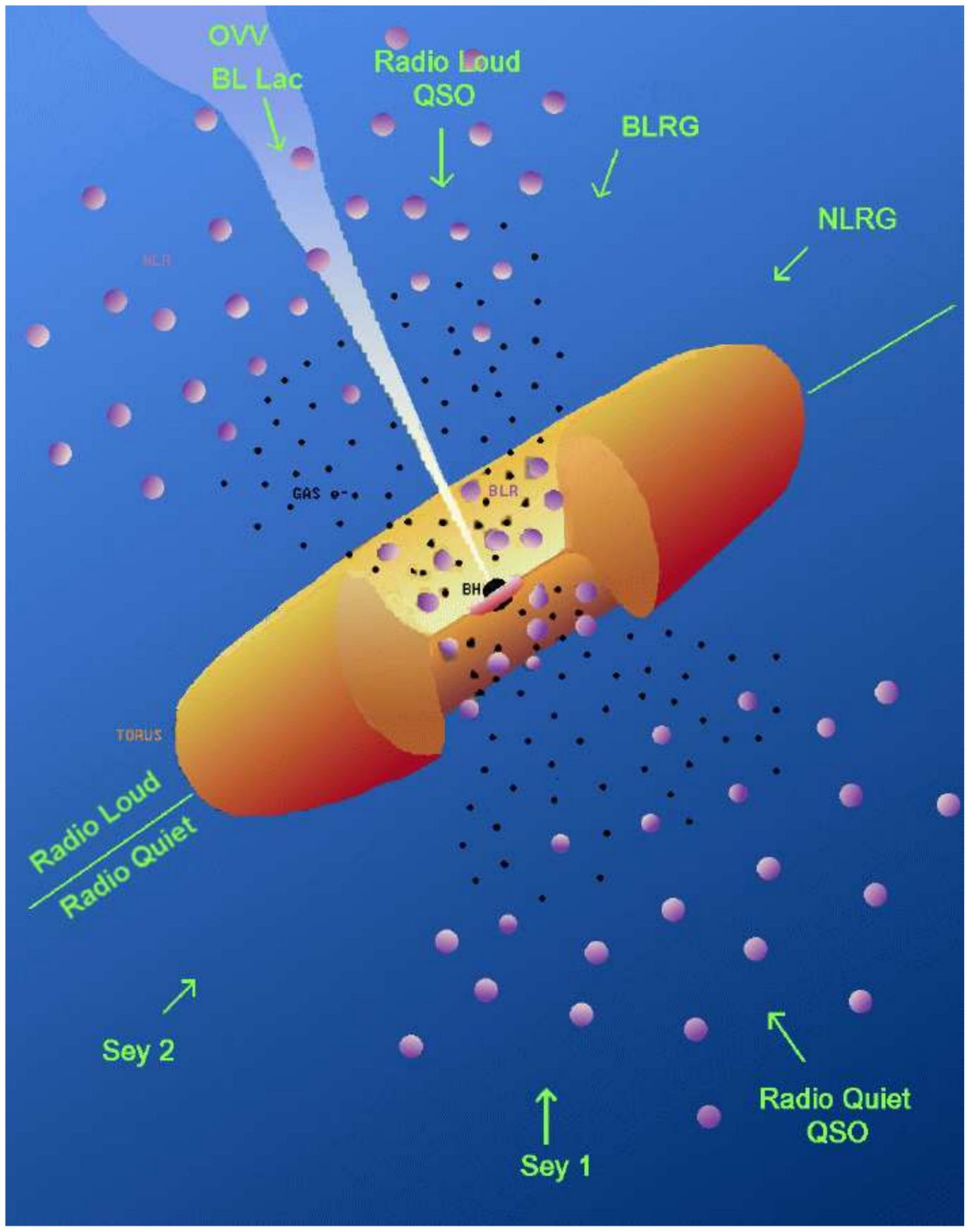

Figure 1.9: Scheme of the Unified Model for AGNs [204]. The observational difference between AGN classes is due to a geometric effect.

originate in the thermal emission of the accretion disk and are injected into the jet [72].

The hadronic models assume $\gamma$ rays are created as a consequence of the presence of accelerated protons, which do not radiate Synchrotron and can be accelerated up to E = $10^{20}$ eV. At these extremely high energies become possible processes of photoproduction of pions and electron-positron pairs. The pions can decay into $\gamma$ rays and leptons can cool via Inverse Compton and produce $\gamma$ rays [132].

Probably a realistic scenario contains both sets of processes and observing $\gamma$-ray emission during flares can help to determine which contribution dominates in each type of source.

Blazars compose the largest fraction of identified $\gamma$-ray sources detected by EGRET, with 66 high-confidence and 27 lower confidence identifications according to the criteria adopted using the maximum likelihood method in the Third EGRET Catalog [93]. For this reason we expect a large fraction of EGRET unidentified sources, especially at high latitude ($|b| > 30°$), may be associated with blazars.

### 1.2.3 Gamma-ray Pulsars and Pulsar Wind Nebulae

A pulsar (*PULSating StAR*) is a rapidly-rotating neutron star, i.e. the stellar remnant of a massive star ($M > 8M_\odot$) after its gravitational collapse, with a very intense dipole magnetic field, which emits a beam of detectable electromagnetic radiation with observed periods ranging from about 1 ms to 10 s. The period is observed increase in time. The radiation can only be observed when the beam of emission is pointing toward the Earth. From timing measurements, it is possible to estimate the strength of the magnetic field on the surface of the star, the age of the pulsar and other physical parameters [198].

Pulsars were generally discovered in radio wavelengths, but they can emit in all wavelengths of the electromagnetic spectrum, in particular the first two $\gamma$-ray emissions from point-like sources were observed from the *Crab* and *Vela* pulsars by SAS-2 [77]. The first radio pulsar was discovered in August 1967 by A. Hewish and J. Bell during a radio astronomy project [99]. By the end of 2004 there were about 1500 radio pulsars detected, but only a few pulsars have been observed to emit in optical wavelength essentially because of the interstellar medium absorption, some tens of pulsars are known to emit in X-ray band and seven pulsars were observed to emit in $\gamma$-ray energy band by EGRET [93]. In Figure 1.10 the light curves of the seven $\gamma$-ray pulsars are shown in different energy bands of the electromagnetic spectrum.

The $\gamma$-ray spectrum of pulsars are extremely flat showing a peak of emission in the GeV (*cut-off* energy), above this energy the spectrum decreases very quickly. Light curves of $\gamma$-ray pulsars are extremely regular showing one or two peaks not always in phase among energy bands.

A neutron star has a radius of about 15 km, a mass of about 1.4 $M_\odot$ (this means an average density of about $10^{15}$ g/cm$^3$) and a very intense magnetic field with a strength at poles of order of $10^{12}$ – $10^{13}$ Gauss. There are other classes of isolated neutron stars characterized by the lack of a steady radio band emission and a very different strength of the magnetic field, with a value of about $10^{10}$ Gauss for the CCO (*Central Compact Object*) [67] up to $10^{15}$ Gauss for the Magnetars [144]. Because of these objects do not emit $\gamma$ rays they will not be a matter of this thesis but, since we worked on these objects during the Ph.D., the results of our analyses are explained in the Appendix D.

In describing the emission from pulsars, an analogy has been drawn with a *lighthouse*. The light in the lighthouse is always on, but we can only see it when the beam points in our direction. In this model of neutron stars, there is some emission point on the surface of the star, producing a beam of radiation, like a lighthouse beam. We can only see the radiation when the beam is pointed in our direction. If we stay in one place, we will see the radiation appear to flash on briefly once per cycle. The lighthouse nature of the pulsar mechanism has an important consequence. The beam only traces out a cone on the sky. If the observer is not on that cone, the pulsar will never be visible. Any given pulsar will only be seen by about some per cent of the potential observers, depending on the size of the light cone. This means that our galaxy contains many more pulsars than we actually observe.

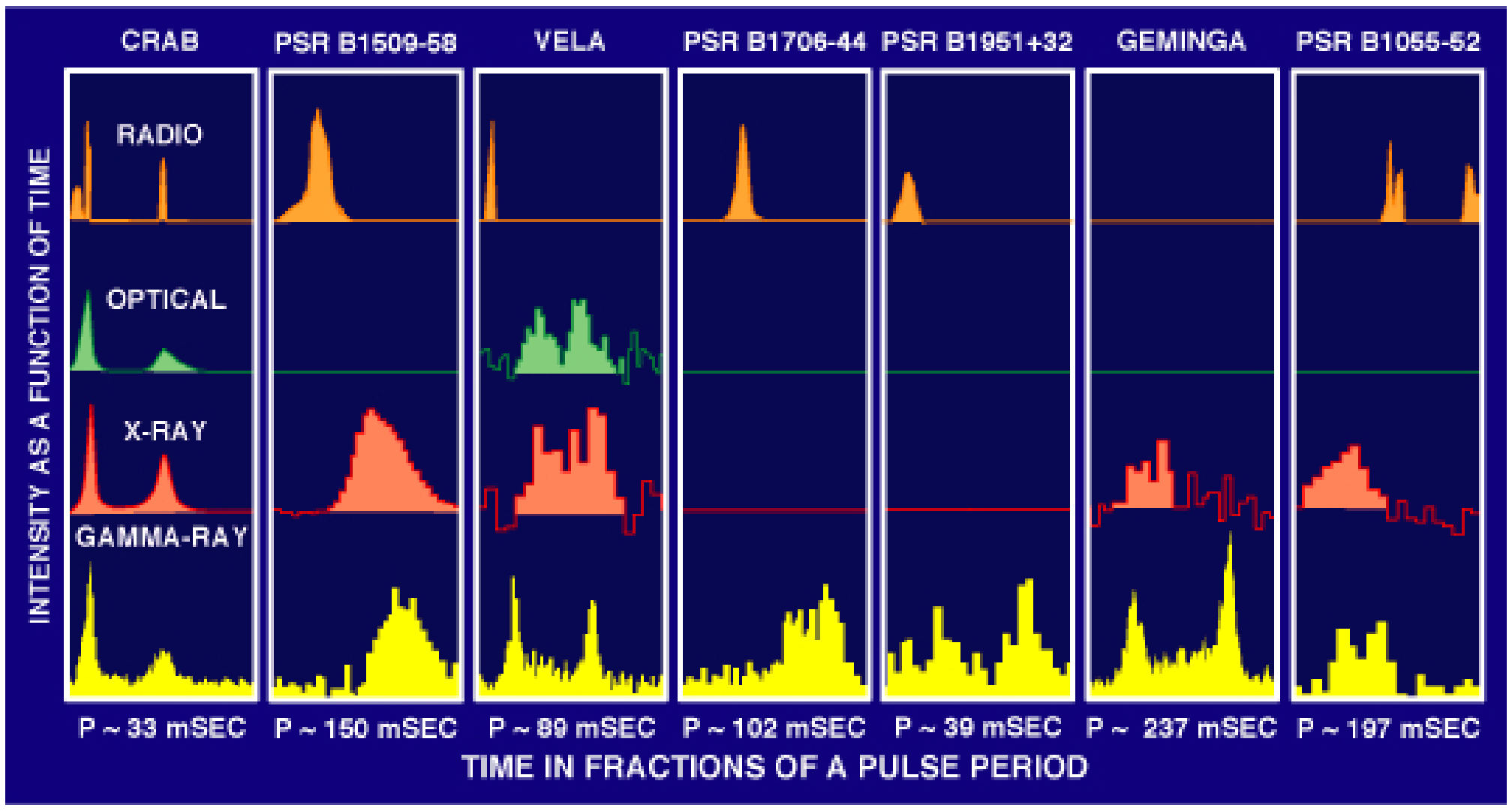

Figure 1.10: Light curves in different energy bands of the seven $\gamma$-ray pulsars detected by EGRET [93].

The emission mechanism may be related to the strong magnetic field that neutron stars must have. Current models assume the magnetic axis of the star is probably not aligned with the rotation axis. If the beam of radiation is somehow collimated along the magnetic axis, we only see it when the beam points in our direction. We may even see two pulsed as opposite magnetic poles pass by.

The details of the emission mechanism are not clear. One possibility is that the spinning magnetic field generate electric fields strong enough to remove electrons and positrons from the surface of the neutron star. These charged particles are anchored to the magnetic field lines into a region called *magnetosphere* where the plasma is forced to spin with the star inside a region called *light cylinder* arranging themselves in order to have $\vec{E} \cdot \vec{B} = 0$. The previous condition prohibits particle acceleration, acceleration can occur only in region near the magnetic poles, called *vacuum gaps*, in which $\vec{E} \cdot \vec{B} \neq 0$ and the field lines are opened [102].

There are currently two types of theoretical models proposed to explain the $\gamma$-ray emission of pulsars. The two models, *Polar Cap* and *Outer Gap*, differ essentially for the different production regions of $\gamma$ rays, as shown in Figure 1.11.

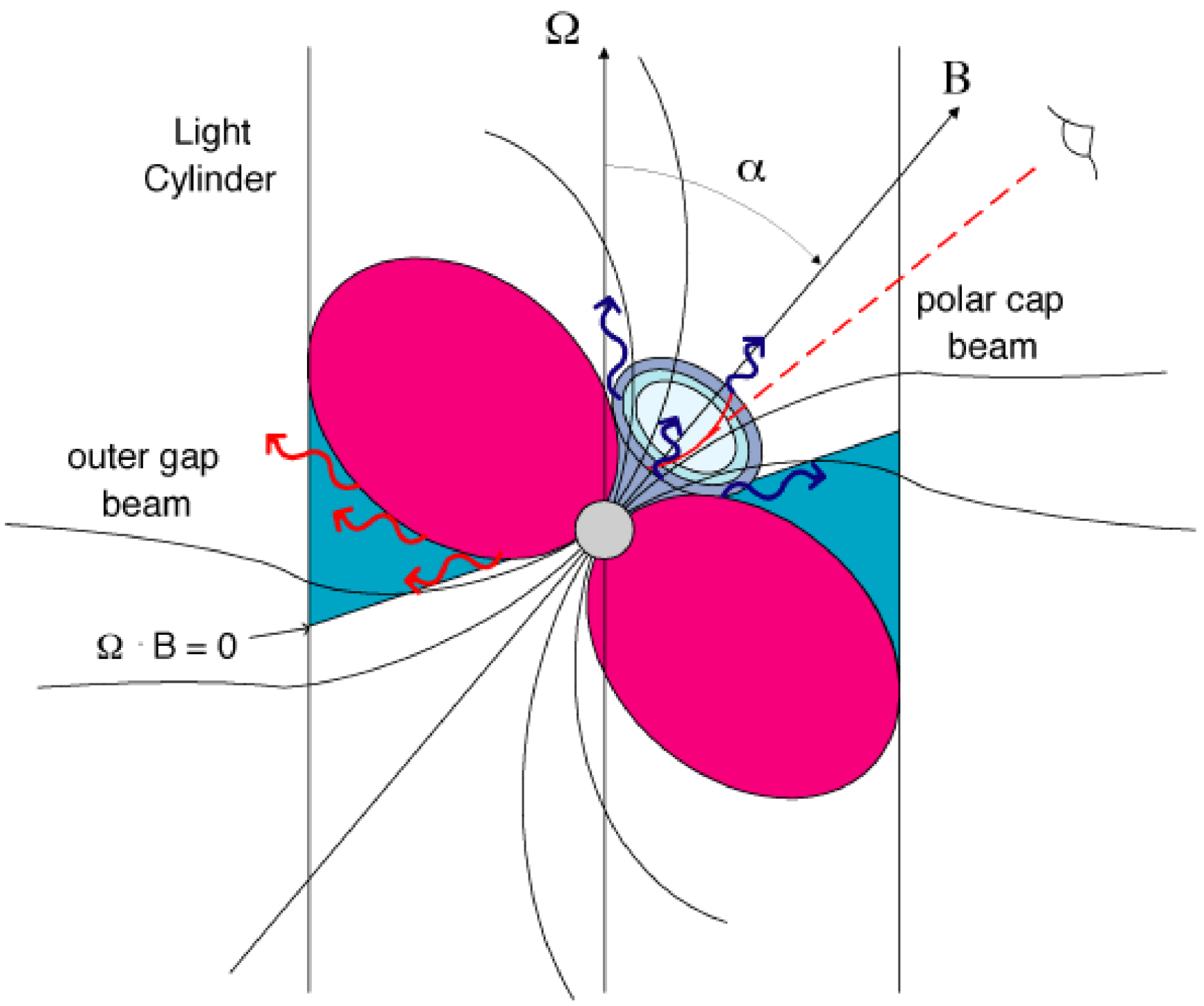

Figure 1.11: Schematic view of a pulsar with the *polar cap* and *outer gap* beams.

According to the *Polar Cap* model, particles are accelerated at the magnetic poles along open magnetic field lines near the neutron star surface by parallel electric fields and induce pair cascades by either curvature radiation or Inverse Compton (IC) radiation [64].

According to the *Outer Gap* model, primary particles are accelerated in vacuum gaps in the outer magnetosphere and induce pair cascades through $\gamma$–$\gamma$ pair production [177].

The $\gamma$-ray observations of pulsars shown they can be divided in two categories, *radio-loud* pulsars, such as the *Crab* and the *Vela* pulsar, if also a radio emission is detected, and the *radio-quiet* pulsars, such as *Geminga*, if the radio emission is not detected. The radio emission mechanism of pulsars is still poorly understood, this is probably related to some coherent emission processes. The detection of $\gamma$-ray pulsars those do not emit in the radio band tells us that the $\gamma$-ray light cone is probably wider than the radio one. Because of the identification of a $\gamma$-ray pulsar is primarily based on the periodicity and because of the little number of detected radio pulsar, we expect some unidentified EGRET sources, especially those along the Galactic plane, may be associated with pulsars and particularly to *radio-quiet* pulsars.

We know a different type of pulsar, the *Millisecond Pulsars* (MSPs). A MSP is a pulsar with a rotational period in the range of about 1 and 10 milliseconds (this is the origin of their name), a weaker magnetic field (less than $10^{10}$ Gauss) and their frequency decreases much slower in time. The origin of millisecond pulsars is still unknown. The leading theory is that they begin life as longer period pulsars but are spun up through accretion. For this reason, millisecond pulsars are sometimes called "recycled" pulsars. MSPs are very old pulsars (about $10^9$ years) and are thought to be related to low-mass X-ray binary systems. In these systems, X-rays are emitted by the accretion disk of a neutron star produced by the outer layers of a companion star that has overflowed its Roche lobe. The transfer of angular momentum from this accretion event can increase the rotation rate of the pulsar to hundreds of times a second, as is observed in MSPs.

MSPs are found near from us, because they are fainter than normal pulsars, or in globular clusters, in according with the spin-up theory of their formation, as the extremely high stellar density of these clusters implies a much higher likelihood of a pulsar having a giant companion star. MSPs have been detected in the radio and X-ray portions of the electromagnetic spectrum but one of the $\gamma$-ray sources detected by EGRET has an high probability to be associated with a MSP, this object would be the first MSP detected in the $\gamma$-ray band [119].

The properties of MSPs are very different from those of normal pulsars, and high-energy emission processes of MSPs are still unclear, but similar *Outer Gap* [219] and *Polar Cap* [192] models have been proposed also for MSPs to explain the non-thermal $\gamma$-ray emission. Both models predict that a large number of MSPs may emit $\gamma$ rays, this opens an interesting perspective for future $\gamma$-ray missions, discover a new class of $\gamma$-ray emitters.

Pulsed emission represents only a little fraction of the total energy emitted by a pulsar. The rapidly rotating, superstrong magnetic field of the spinning pulsars accelerates charged particles of the magnetosphere to ultrarelativistic speed, creating the pulsar wind. About 90% of the total energy emitted by a pulsar is through the wind of high energy charged particles. The pulsar wind streams into the interstellar medium, creating a standing shock wave, where it is decelerated to sub-relativistic speed. In this region, ultrarelativistic charged particles interact and are confined by the ram pressure of the ambient medium and by the magnetic field of the pulsar, setting up the *Pulsar Wind Nebula* (PWN). Beyond this radius Synchrotron emission increases in the magnetized flow.

The only PWN detected by EGRET was the *Crab* nebula [66] as an unpulsed component up to 10-20 GeV. The spectrum showed a Synchrotron emission with cut-off at around 100 MeV and an Inverse Compton emission above 100 MeV, its emission has been detected also at TeV energies. According to the current models, lower energy emission in $\gamma$ rays is due mainly to Synchrotron emission from leptons from the pulsar relativistic wind, and the higher emission is due to Inverse Compton scattering produced by leptons and lower energy Synchrotron photons, Cosmic Microwave Background (CMB) or infrared photons coming from the pulsar. Leptons are accelerated as results from interaction between the pulsar wind and the nebula. It is also possible that $\gamma$ rays can be produced by interaction of hadrons in the pulsar wind with the ambient matter can contribute to the observed spectrum, probably for younger nebulae.

### 1.2.4 Supernova Remnants

Supernova remnants (SNRs) represent the relics of a supernova explosion, that cause a burst of radiation that often briefly outshines an entire galaxy. They are important because connected to the study of the late stages of stellar evolution of high-mass stars ($M > 8\ M_{\odot}$), of the properties of the explosive nucleosynthesis and because of their interaction with the surrounding space, that is contaminated and energized by the products of the supernova explosion.

The importance of SNRs in astroparticles physics is related to the origin of the Cosmic Rays (CRs). Cosmic Rays, relativistic cosmic particle from the space, have been studied since early in the twentieth century. To date, the question about the origin of cosmic rays nuclei remains only partially answered, with widely accepted theoretical expectations but incomplete observational confirmation.

Theoretical models and indirect observational evidence support the idea that CRs with energy below $\sim 10^{15}$ eV are produced and accelerated in the Galaxy by SNRs. The main mechanism which is believed to be responsible for the CR production is the shock acceleration, happening when the Supernovae shell shocks interact with the interstellar medium. The shock mechanism is an efficient particle accelerator up to very high energy (TeV energies) and in the case of SNe on time scales of $10^3 - 10^4$ years. The accelerated CR escape from the SNR remaining trapped in the Galactic magnetic field. It has been calculated that roughly 10% of kinetic energy of a SNR must be transferred to CRs [101]. Observing charged particles, there is no possibility to directly observe the sites of their production, due to chaotic magnetic deviation. Cosmic rays interact with the interstellar gas and dust and photons, producing $\gamma$ rays. Photons are not deviated by the Galactic magnetic field and a direct observation of the accelerator sites is then possible.

The low angular resolution of EGRET did not permit to resolve the SNR structure and the study of different acceleration sites, for this reason many questions are still opened.

During their explosion supernovae can accelerate protons and heavy ions to high energies. The protons can interact in the interstellar medium, producing cascades of secondary particles, such as neutral pions ($\pi^0$) that decay quickly into $\gamma$ rays. The heavy ions are radioactive and they decay, emitting $\gamma$ rays. Moreover, the electrons accelerated by the shock wave during the supernova explosion can interact with the photons of the Galactic background radiation and the radiation fields of the SNR (this process is responsible also for the X-ray emission from the SNR) by the *Bremsstrahlung* and the Inverse Compton. To date, it is not clear the leptonic and hadronic contribution by the various spectral components observed for the $\gamma$-ray SNRs because of the not precise measure of their spectra by EGRET.

### 1.2.5 Gamma-Ray Bursts

Gamma-Ray Bursts (GRBs) are the most powerful and most distant known sources of $\gamma$ rays. The brightest GRB at GeV energy is about $10^4$ times brighter than the brightest AGN. GRBs are intense flashes of $\gamma$ rays, lasting from some milliseconds up to hundreds of seconds and they are detected with a frequency of about one GRB per day.

GRBs were discovered serendipitously in the late 1960s by VELA satellites, a series of U.S. military satellites designed to detect covert nuclear weapons tests, but their discovery was first reported in 1973. An advance in understanding these mysterious objects occurred in 1991 with the launch of the *Compton Gamma-Ray Observatory* (CGRO). The all-sky survey from BATSE onboard CGRO, which measured about 3000 bursts, showed that they are isotropically distributed, as shown in Figure 1.12, suggesting a cosmological origin for these objects [146]. Analysis of the distribution of the observed duration for a large number of GRBs showed a clear bimodality, suggesting the existence of two separate populations: a "short" population with an average duration of about 0.3 seconds and a "long" population with an average duration of about 30 seconds [161]. Both distributions are very broad with a significant overlap region in which the identity of a given event is not clear from duration alone. Their spectra are very hard, with a peak of emission at some hundreds of keV, they follow a power law characterized by a spectral index that vary during the explosion.

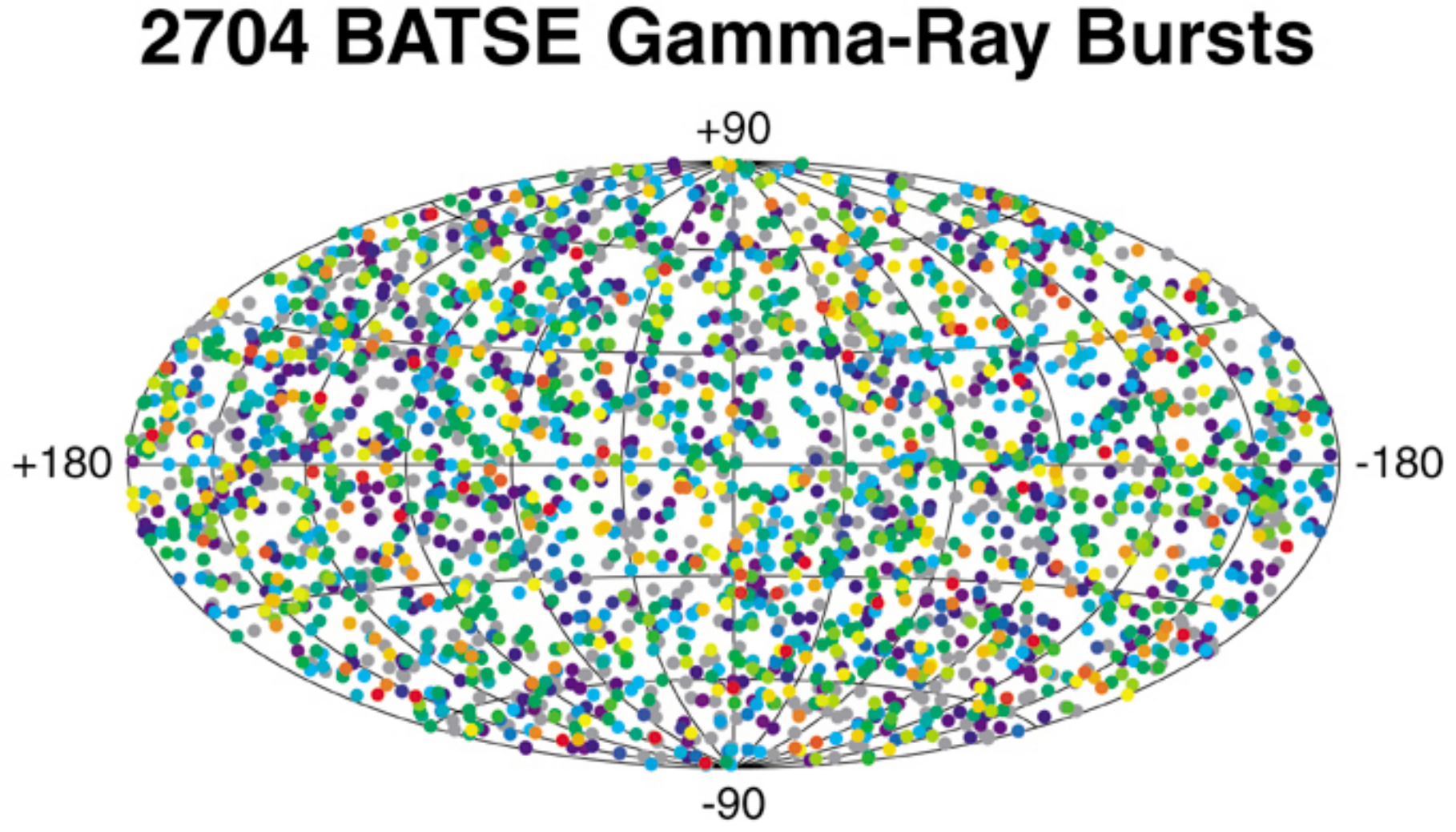

Figure 1.12: Angular distribution of 2704 GRBs detected from 1991 to 2000 by BATSE. The colors differentiate the brightness of the bursts. Credit: NASA

The prompt $\gamma$-ray emission from a GRB is sometimes followed by a second transient event called *afterglow* at energies with longer lasting emission in the X rays, optical and radio. The first X-ray GRB afterglow was measured by the *BeppoSax* mission (1996 - 2002). *BeppoSax* was a program of the Italian Space Agency (ASI) with participation of the Netherlands Agency for Aerospace programs (NIVR). On 28 February 1997 *BeppoSax* detected the X-ray afterglow from GRB 970228 [209]. This was the first accurate determination of the distance to a GRB and, together with the discovery of the host galaxy of 970228, it proved that GRBs occur in extremely distant galaxies.

The questions about which type of celestial object can emit GRBs is still unsolved. GRBs show an extraordinary degree of diversity. The near complete lack of observational constraint led to a profusion of theories, including evaporating black holes, magnetic flares on white dwarfs, accretion of matter onto neutron stars, hypernovae, rapid extraction of rotational energy from supermassive black holes and fusion of two neutron stars or one neutron star and one black hole of a binary system. There are at least two different types of progenitors of GRBs: one responsible for the long-duration, soft-spectrum bursts and one responsible for short duration, hard-spectrum bursts. The progenitors of long GRBs are believed to be massive, low-metallicity stars exploding due to the collapse of their cores (*collapsar model*) [142]. The progenitors of short GRBs are still unknown but mergers of neutron stars is probably the most plausible model (*merger model*). Both models suggest the final creation of a black hole, surrounded by an accretion torus which realizes gravitational energy that feeds the explosion.

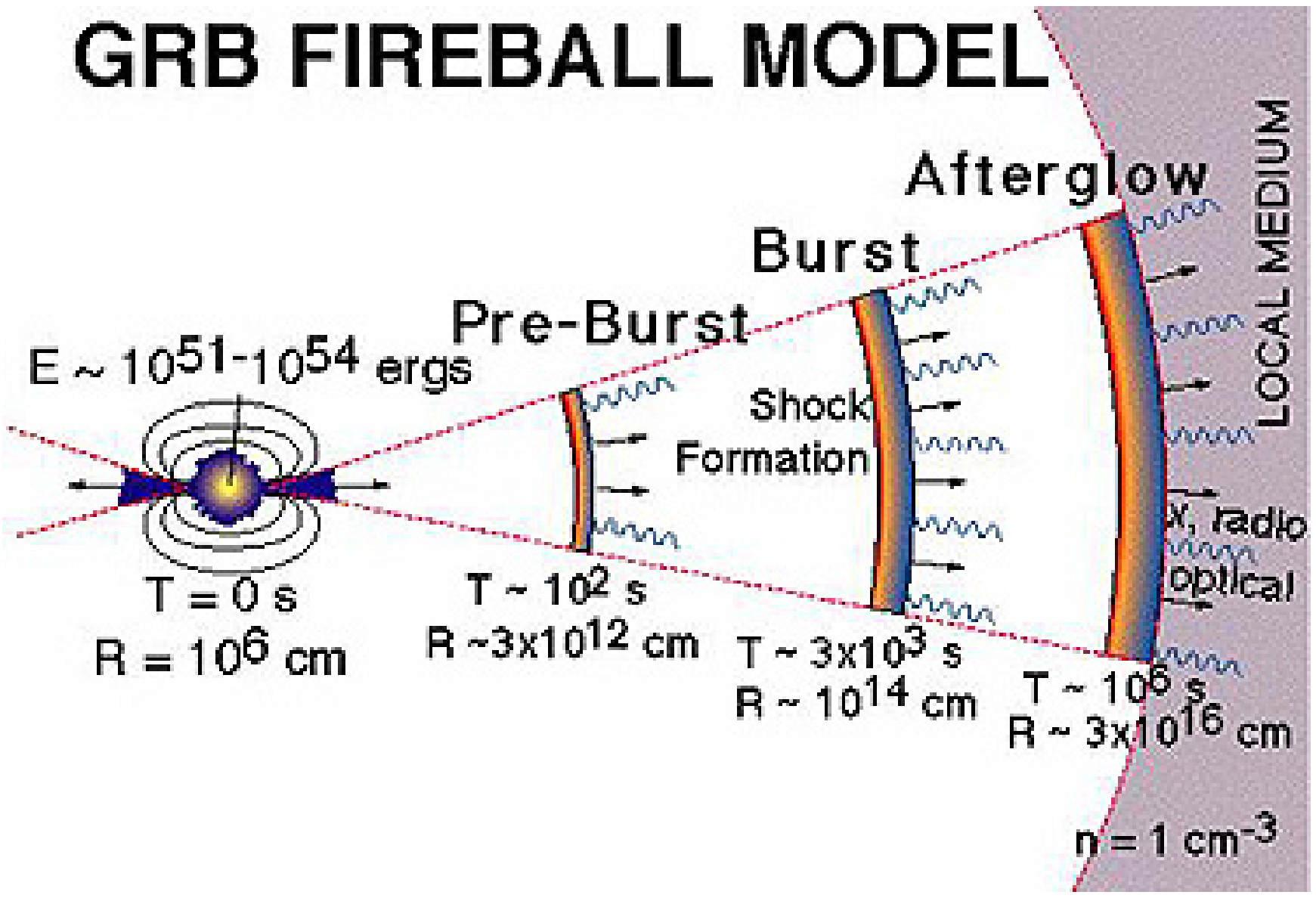

Figure 1.13: Schematic view of the fireball model for GRBs.

Several models have been developed in order to explain the $\gamma$-ray emission from GRBs. Probably the *Fireball model* (FBM), introduced by Piran in 1999 [165], is the most plausible one. The term "fireball" refers to an opaque radiation-plasma ball (composed by electrons, positrons and $\gamma$ rays) which would expand relativistically. Two different types of shocks may arise in this scenario. In the first case, the expanding fireball runs into external medium, the ISM or a pre-ejected stellar wind. The second possibility is that even before external shocks occur, internal shocks develop in the relativistic wind itself. The model hypothesis is that $\gamma$-ray burst is due to internal shocks, while afterglow is due to the relativistic expanding wind, which decelerates producing radiation of lower energy, going from X rays to optical and radio, as time goes on [145]. A schematic view of the fireball model for GRBs is shown in Figure 1.13.

### 1.2.6 Solar flares

The Sun, the star of our planetary system, has been known to produce $\gamma$ rays during its flaring period with energies greater than several MeV. This emission was detected for the first time by the American satellite OSO-VII between August 4 and 7, 1972. Subsequently, other missions were dedicated to the study of the $\gamma$-ray emission by the solar flares but important results came from EGRET and COMPTEL telescopes onboard of the CGRO observatory, which discovered that the Sun is a source of GeV $\gamma$ rays.

Accelerated charged particles interact with the ambient solar atmosphere, radiating high-energy $\gamma$ rays via *Bremsstrahlung* (see. e.g., [170], [156]). Secondary $\pi^{\pm}$ are produced by nuclear interaction and yield to $\gamma$ rays with a spectrum that extends to the energies of the primary particles. Protons and heavy ions interactions also produce $\gamma$ rays through $\pi^0$ decay, resulting in a spectrum that has a maximum at 68 MeV and is distinctly different from the *Bremsstrahlung* spectrum. The processes that accelerate the primary particles are not well known, but stochastic acceleration through MagnetoHydroDynamics (MHD) turbulence or shocks ([79], [179]) are though to be the most creditable mechanisms. Particles are accelerated in large magnetic loops that are energized by flares, and they get trapped due to magnetic field, generating $\gamma$ rays ([130], [131]). However, it is not clear where the acceleration takes place and whether protons are accelerated along with the electrons.

Figure 1.14 shows the extraordinary flare of June 11, 1991, detected by the EGRET telescope. The contribution from electron *Bremsstrahlung* and $\pi^0$ decay are separately shown.

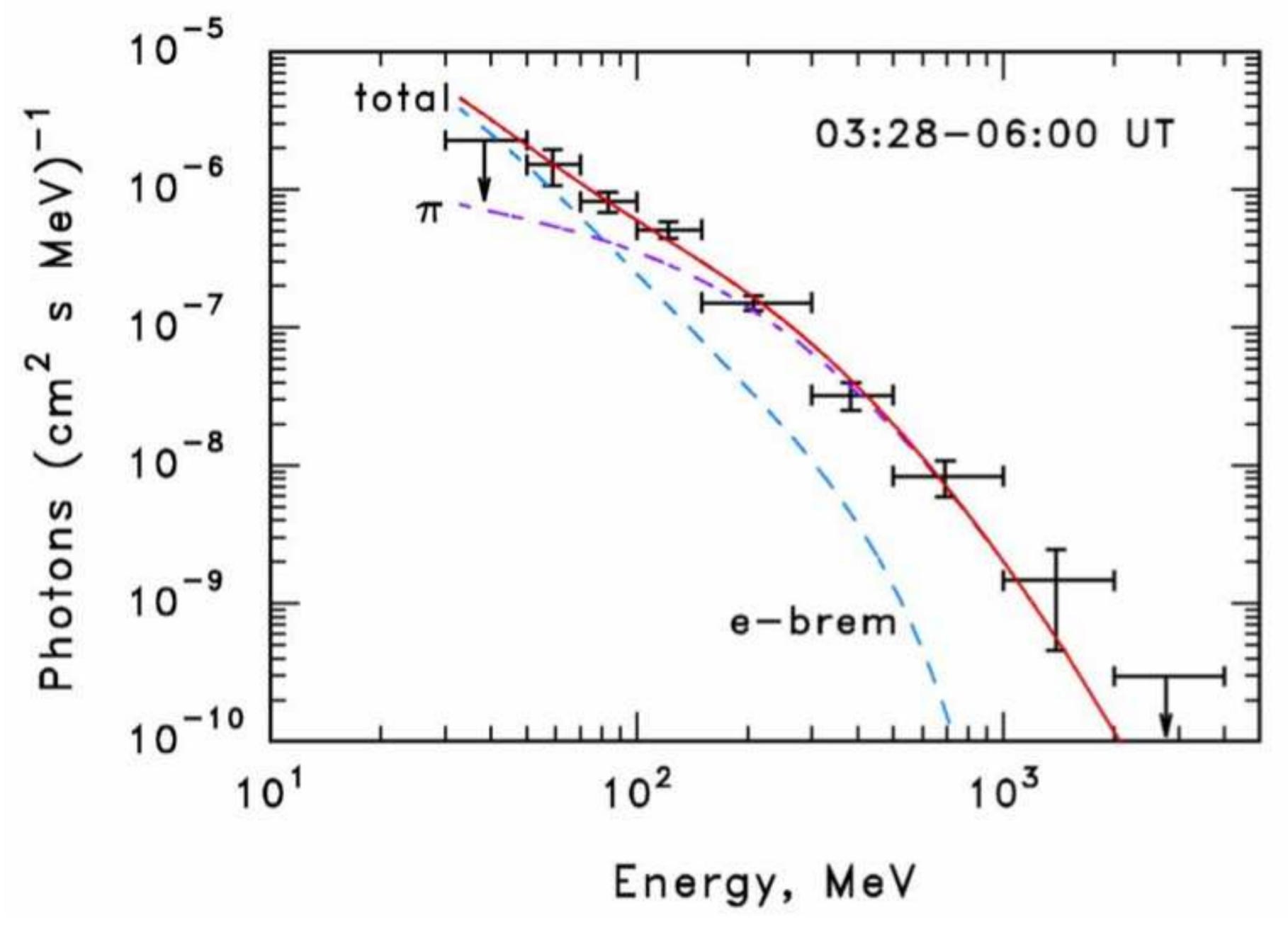

Figure 1.14: The extraordinary flare of June 11, 1991 detected by EGRET telescope, which produces $\gamma$ rays up to GeV energies. The contribution from electron bremsstrahlung and from pion decay are displayed separately [74].

Some models are proposed for production of $\gamma$ rays from the Sun also in quiescent state, e.g. from nuclear gamma decay of nuclei like $^{58}$Co, or from *microflares*, already observed in UV and X rays but never seen in $\gamma$ rays.

### 1.2.7 Dark Matter

To date, it is not known if dark matter can be a source of $\gamma$ rays. As discussed in Section 1.2.1, EGRET identified about 70 AGNs but probably an important fraction of the extragalactic diffuse $\gamma$-ray emission is related to the emission of unresolved AGNs. However, any remaining extragalactic diffuse emission would be of great interest. It is thought that a fraction of the extragalactic $\gamma$-ray diffuse emission could originate from the decay of exotic particles in the primordial Universe. The energy spectrum of this component should be different from the AGNs contribution. The difficulty in detecting $\gamma$ rays from dark matter is distinguishing which are produced by dark matter annihilations from those generated by numerous other sources in the Universe.

This contribution could be related to the possible decay of supersymmetric particles. Current models assumes the existence of the Dark Matter (DM) in the halo of our Galaxy, this hypothesis is also sustained by the comparison between the rotational curves of the spiral galaxies and the baryonic visible matter, which tell us that the visible mass is not sufficient to explain the rotational velocities of the stars in the spiral galaxies. Current theory suggests that DM is composed by WIMPs (*Weakly Interacting Massive Particles*) which are massive particles that do not emit or absorb light. Such particles are predicted by supersymmetry, a theory that extends the Standard Model of particle physics. According to supersymmetry, WIMPs act as their own antimatter particles. When two WIMPs interact, they annihilate each other and release secondary particles such as $\gamma$ rays. Dark matter interacts much more weakly than ordinary matter, but it is not spread out evenly through space and should form clumps in and around galaxies.

The lightest supersymmetric particle is the *Neutralino* ($\chi$) and it is perhaps the most promising candidate for the WIMPs ([214], [87]). Although the highest energy accelerators have begun to probe regions of supersymmetric parameter space, the limits set at this time are not very restrictive. The mass of the Neutralino particle can be constrained, in order to make up the overall dark matter in the Universe. The required mass is in the range 30 GeV $<$ $M_\chi < 10$ TeV, depending on the model chosen. If Neutralinos make up the dark matter of the Milky Way, they can annihilate into the $\gamma\gamma$ final state giving rise to photons with unique energies, which are $\gamma$-ray lines depending on the preferred channel. The signature would be spatially diffuse, narrow line emission peaked toward the Galactic center. Figure 1.15 shows the predicted signal from Neutralino annihilation into $\gamma\gamma$, with an assumed mass of about 47

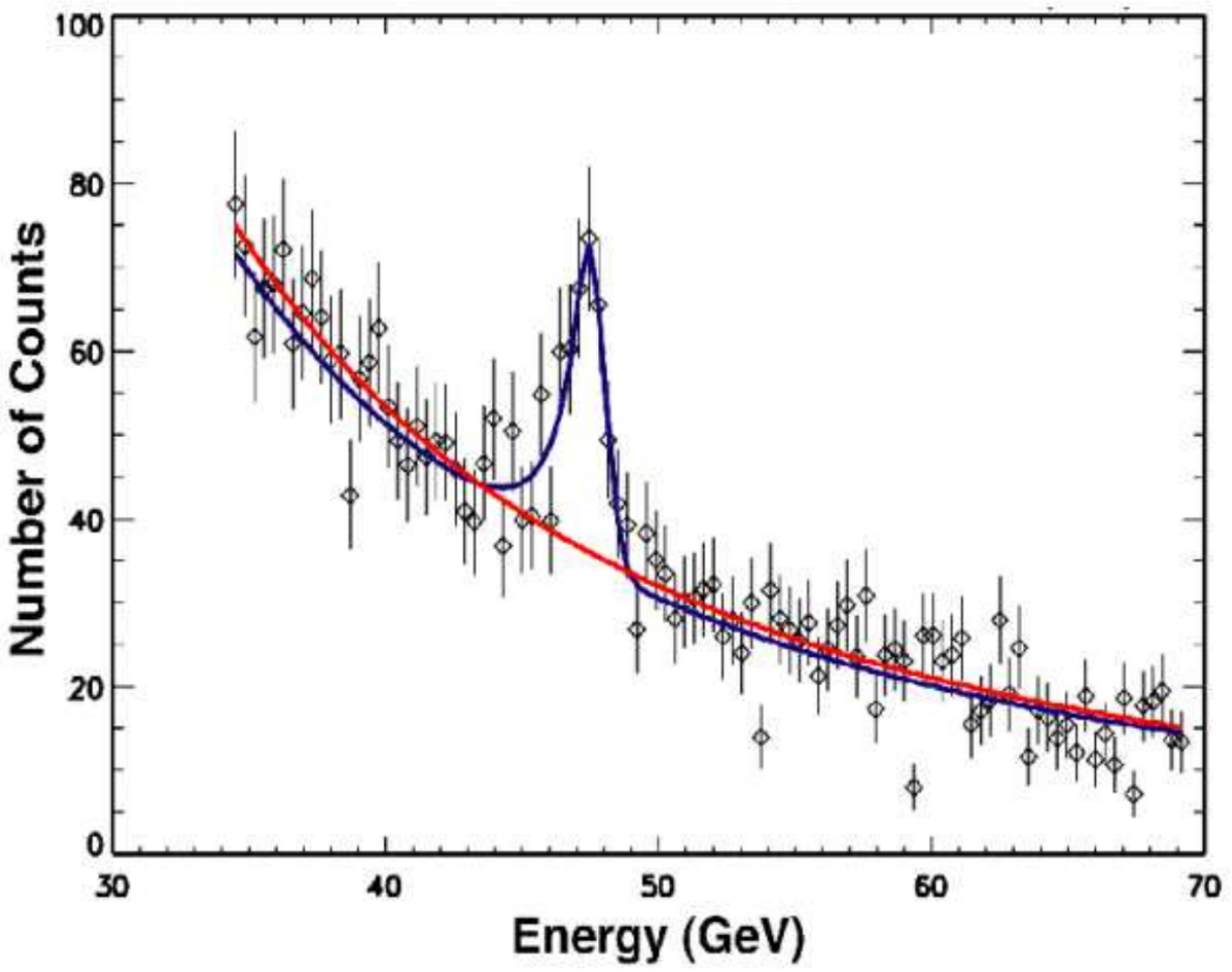

Figure 1.15: Simulation of the signature of a galactic Neutralino annihilation into $\gamma\gamma$. The red line is the contribution from an AGN while the blue line is the contribution of the particle relic [74].

GeV. While the signal would be the most spectacular of all possible indirect signals, its rates are suppressed with respect to other Neutralino annihilation channels. On the other hand, photons may also be produced in the cascade decays of other primary annihilation products. In contrast to the line signal, cascade decays produce a large flux of photons with a continuum of energies detectable as an excess in the $\gamma$-ray flux.

### 1.2.8 Unidentified Sources

Only four of the 25 sources in the second COS-B catalog had identifications [195], and over half the 271 sources in the third EGRET catalog had no associations with known objects in other wavelengths [93]. Since association is primarily based on positional coincidence of possible counterparts known at other wavelengths with a $\gamma$-ray source, a principal reason for the difficulty of finding counterparts to high-energy $\gamma$-ray sources has been the large positional errors in their measured locations, which are related to the limited photon statistics, associated with a relatively small effective area, and poor angular resolution of the $\gamma$-ray observations, e.g. EGRET detector had an angular resolution of $\sim 5.8°$. Also the bright diffuse $\gamma$-ray emission from the Milky Way is a limit in the procedure of association of $\gamma$-ray sources, only a very detail knowledge of the model of the $\gamma$-ray emission of our Galaxy can help in distinguishing if an excess of $\gamma$-ray photons in a specific region in the Galactic plane is related to a statistical fluctuation of the background or to the presence of a $\gamma$-ray source.

Gamma-ray sources are tracers of the most energetic processes in the Universe, they are very exotic objects, characterized by very intense magnetic fields and the presence of very high energy particles. For this reason, understanding the nature of the $\gamma$-ray unidentified sources is one of the most important open questions in high energy astrophysics. Gamma-ray unidentified sources represent discovery space for new members of existing $\gamma$-ray source classes, or new source classes. These sources should have an high value of the ratio $L_\gamma/L_\lambda$, where $L_\gamma$ is the $\gamma$-ray luminosity of the source and $L_\lambda$ is the luminosity of the source at lower energy. This makes them possible powerful accelerator of particles.

Since pulsars and blazars represent the two most numerous $\gamma$-ray source classes, the first suggested hypothesis was to associate the EGRET unidentified sources with these two classes [140]. In particular, less than one third of these are far from the Galactic plane, probably associated with blazars because extragalactic, with the remaining most likely within the Milky Way. Further works suggest that many of these unidentified sources are associated with nearby Gould Belt of star-forming regions that surrounds the solar neighborhood [85], while apparently-steady sources are likely to be *radio-quiet* pulsars [92].

Among new $\gamma$-ray source classes, some of the unidentified EGRET sources may be associated with Galactic *microquasars*. Microquasars are a subclass of X-ray Binaries (XRBs) that show a jet of mild relativistic accelerated particles. They are believed to be a binary system made up of a compact object, perhaps a neutron star or a black hole, orbiting around a massive star. The jet of particles should be the basic of $\gamma$-ray emission both leptonic or hadronic scenarios. The name *microquasar* have been proposed since they mimic on smaller scales the phenomenology of the quasars, their investigation is believed to be much important for the understanding of the AGN physics. Recently two microquasars have been observed by the AGILE detector and some ground-based satellites, the LS I +61 303 by MAGIC [18] and LS 5039 by HESS [17] both shown in Figure 1.16. In both cases the orbital modulation of the $\gamma$-ray flux have been observed.

To identify $\gamma$-ray sources, the study of the emission at other wavelengths is crucial. One of the most significant example was the *Geminga* pulsar. In that case, a search for pulsed emission using X-ray data led to find the characteristic spin period of a neutron star. Gamma-ray observations confirmed this pulsation and so *Geminga* was found to be a $\gamma$-ray pulsar.

The progress in finding the identity of all $\gamma$-ray sources are correlated to the ability of new generation $\gamma$-ray experiments to localize $\gamma$-ray objects with higher precision in order to find exactly the other wavelength counterparts.

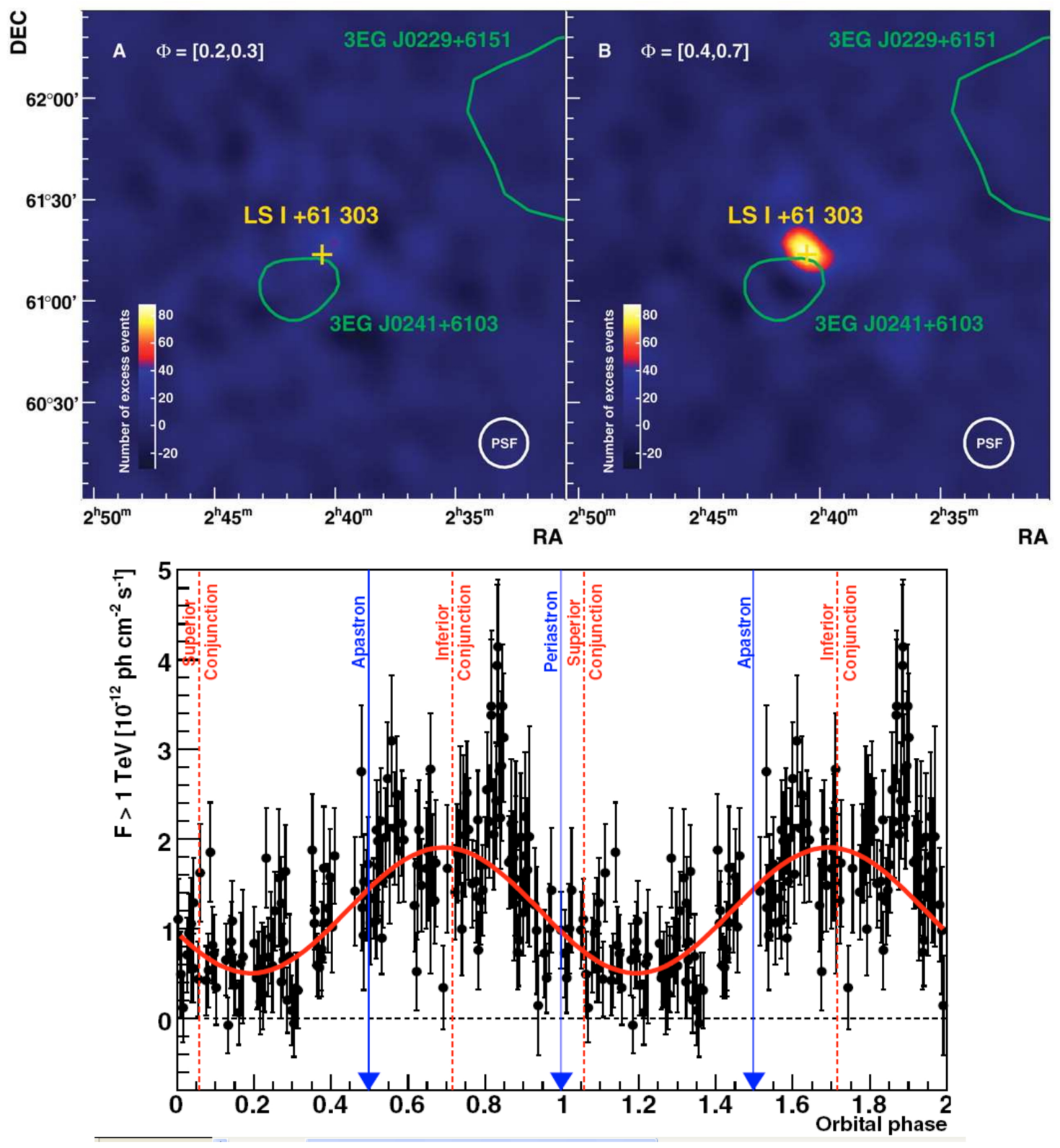

Figure 1.16: Top: the microquasar LS I +61 303 observed by MAGIC [18] in two states. Bottom: the orbital modulation of the LS 5039 microquasar observed by HESS [17].

In the end, the spatial, spectral and variability properties may provide a framework that could allow to predict the expected source classes for the sources that remain unassociated. Because of the limits of the EGRET detector, these information were not enough accurate to compare intrinsic properties of the EGRET sources for both associated and unassociated populations, potentially providing insight into the likely classes of the unidentified sources. An increase of the performances of new generation $\gamma$-ray experiments will allow to do it by using the properties of the associated sources to define a model that describes the distributions and correlations between measured properties of the $\gamma$-ray behavior of each source class. This model will be able to be then compared to the $\gamma$-ray properties of each unidentified source in order to classify some of them as likely members of one of these source types on the basic of their $\gamma$-ray properties.

## 1.3 Summary

In this Chapter the main issues regarding $\gamma$-ray astrophysics have been reviewed. The development of $\gamma$-ray astronomy have been carried mainly from space and the last important mission was the CGRO observatory. It has revolutionized and improved our knowledge about $\gamma$-ray Universe. A discussion about the results found by EGRET and a review of the discovered $\gamma$-ray sources have been presented.

Although the EGRET mission has allowed to improve our knowledge of the $\gamma$-ray sky, many questions are still open. The *Fermi Gamma-Ray Space Telescope* is a member of the new generation of $\gamma$-ray telescopes together with *AGILE*, it was launched on 2008 and it is contributing in a decisive way in several topics of modern understanding of the $\gamma$-ray Universe, from the study of Galactic and extragalactic cosmic accelerators to the detailed investigation of the diffuse emission and the nature of the unidentified sources.

The energy range of *Fermi* is guaranteeing the exploration of the energy range between the EGRET upper limit and the lower limit of the ground-based telescopes, a spectral window that would contain a lot of new high-energy sources. *Fermi* is also complementary to the ground-based VHE $\gamma$-ray instruments, so that inter calibration is possible and a more complete multi-wavelength investigation is accomplished.

# Chapter 2

# The Fermi-LAT Source Catalogs

The *Fermi Gamma-ray Space Telescope* (a detailed description of the telescope is reported in Appendix A) has been routinely surveying the sky with the *Large Area Telescope* (LAT) since the mission began in 2008 August. The excellent performances of the LAT in terms of deep and fairly uniform exposure, good per-photon angular resolution and stable response, provide the best resolved survey of the sky in the 100 MeV to 100 GeV energy range. For this reason the LAT survey data provide a detailed characterizations for each $\gamma$-ray source detected as for as localization, time variability and spectral shape are concerned.

In this Chapter an overview of the procedure to construct the *Fermi*-LAT catalogs will be given, we describe how it is established if an excess of $\gamma$-ray photons in a region of the sky is related to a point-like source emission and how this $\gamma$-ray source can be associated with an object known in other wavelengths. The subject of this thesis is based on the study of the $\gamma$-ray point-like sources presented in the *Fermi-LAT Source Catalogs*.

We will present the list of $\gamma$-ray sources detected by the LAT instrument with a threshold likelihood *Test Statistic* (TS) greater than 25, corresponding to a significance of just over $4\sigma$, in the energy range between 100 MeV and 100 GeV, and their detailed characterization. This means that we will not discuss the Bright Source List (BLS, [7]), that listed sources detected by the LAT during the first 3 months of mission and characterized by a TS $> 100$, nor the First *Fermi*-LAT Catalog of Sources above 10 GeV (1FHL, [199]), that listed $\gamma$-ray objects with an energy greater than 10 GeV detected by the LAT.

## 2.1 The First Fermi-LAT Source Catalog (1FGL)

The data analyzed for the First *Fermi*-LAT Source Catalog (1FGL) were obtained during the first 11 months of mission, between 2008 August 4 and 2009 July 4 [3]. *Fermi* is currently in an almost circular orbit at an altitude of 565 km, an inclination of 25.6° and an orbital period of 96 minutes. After an initial period of engineering data taking and on-orbit calibration [5], the observatory was put into sky-scanning survey mode in which the normal to the front of the instrument ($z$-axis) is ±35° above and below the rocking orbital plane on alternate orbits. In this way, after 2 orbits, corresponding about 3 hours, the sky exposure is almost uniform. For particularly interesting targets of opportunity, the observatory can be inertially pointed.

In order to limit the contamination from albedo $\gamma$ rays from interactions of cosmic rays with the upper atmosphere of the Earth, which is a very bright $\gamma$-ray source [202], a cut on zenith angle (angle between the foresight of the LAT and the local zenith) at 105° was applied for the construction of the catalog.

During the survey, observations are nearly continuous, although a few data gaps are present due to operational issues, special calibration runs, or in rare cases, data loss in transmission. This results in a total live time of satellite during this period of about 245.6 days, which corresponds to an absolute efficiency of 73.5%. Most of the inefficiency is due to readout dead time (9.2%) and to time lost during passages through the South Atlantic Anomaly (∼13%), which is responsible for the nonuniformity of the exposure (about 30% difference between minimum and maximum).

Onboard the satellite there is a first analysis of the data. Immediately the data are filtered and analyzed reconstructioning and classifying each detected event [22]. In reconstructing the events from the hits in the LAT, various cuts are made classifying the events on the basis of probability that they result from photons and the quality of the reconstruction. The events are then separated into various event classes, each class is characterized by its own set of instrument response functions[1]. For the construction of the catalog, the class with the least residual contamination from charged particle background events is chosen.

The instrument response functions (IRFs) – effective area, energy redistribution, and point-spread function (PSF) – use the analyses derived from Monte Carlo simulations of the LAT, which were calibrated before [22] and after [171] the launch using the event-selection criteria corresponding to the chosen event classes. After these analyses, only the events with energy above

[1]see `http://fermi.gsfc.nasa.gov/ssc/data/analysis/documentation/Cicerone/Cicerone_Data/LAT_DP.html`

100 MeV were selected for the construction of the catalog because below 100 MeV the effective area is relatively small and strongly dependent on energy and because at low energy the width of the PSF increases (scaling approximately as $0.8^{\circ}(\mathrm{E}/1\mathrm{GeV})^{-0.8}$).

After these cuts, the data set contains $1.1 \times 10^{7}$ events with energies above 100 MeV. The Figure 2.1 summarizes the data set used for the construction of the catalog: it shows the $\gamma$-ray intensity map for energies above 300 MeV. The map is dominated by a dramatic increase of the brightness of the $\gamma$-ray sky at low Galactic latitudes.

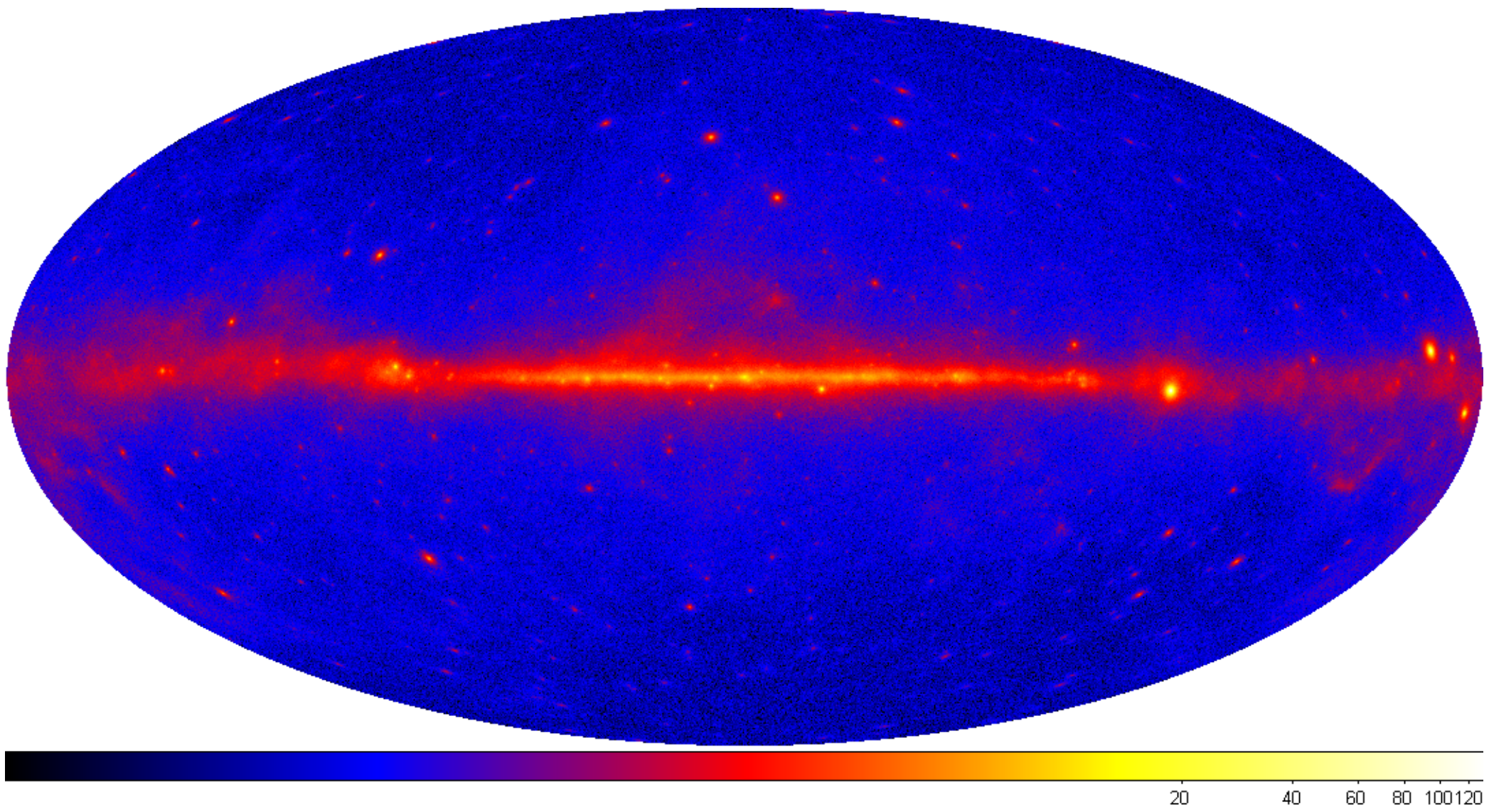

Figure 2.1: Aitoff projection in Galactic coordinates of the sky map of the LAT data detected during the first 11 months of mission. The image shows $\gamma$-ray intensity for energies above 300 MeV, in units of photons $\mathrm{m}^{-2}\ \mathrm{s}^{-1}\ \mathrm{sr}^{-1}$ [3].

### 2.1.1 Construction of the catalog

A fundamental input to the construction of the catalog is a detailed model for the diffuse $\gamma$-ray emission. As explained in the Section 1.2.1, the diffuse emission can be divided in two components: the Galactic diffuse emission, related to the interaction between cosmic rays and interstellar gas and photons, and the extragalactic diffuse emission, related to the isotropic unresolved emission of extragalactic sources. In addition, also residual charged particle background, i.e. cosmic rays, misclassified as $\gamma$ rays from the LAT, provides another approximately isotropic background. The models for the Galactic diffuse emission and isotropic backgrounds were developed by the LAT team and are available, along with descriptions of their derivation, from the *Fermi* Science Support Center[2]. A detailed model of the $\gamma$-ray emission is essential for assessing if an excess of photons is related to the presence of a $\gamma$-ray source and for characterizing it. Unfortunately it is extremely difficult to develop a very accurate model, especially for the Galactic diffuse emission, because of a lot of parameters, such as the distribution of interstellar gas in Galactocentric rings and the propagation of the cosmic rays in our Galaxy [193], are not completely known. New information about these parameters can be used to update the models. This can change significantly all the results.

In constructing the catalog, the source detection step was applied only to the data from the full 11 month period as a whole. This means that a $\gamma$-ray source can be included in the catalog only if it was detected on the basis of its average flux. In this way, bright flaring sources detected only on shorter timescales, as GRBs, are not included in the catalog. The procedure used to build the 1FGL catalog follows three important steps applied in sequence: the detection, the localization and the significance estimation. In this scheme the threshold for inclusion in 1FGL is defined at the last step, but the completeness is controlled by the first one. After the list is defined, the source characteristics are determined (flux in 5 energy bands, spectral shape and time variability). Hereafter flux $F$ means photon flux and spectral index $\Gamma$ is for power law photon spectra (i.e. $F \propto E^{-\Gamma}$ )

### Detection

The detection is based on three energy bands, combining *Front* and *Back* events to preserve spatial resolution [3]. The detection does not use events below 200 MeV, which have poor angular resolution, and it uses events up to 100 GeV. The soft band starts at 200 MeV for *Front* and 400 MeV for *Back* events. The medium band starts at 1 GeV for *Front* and 2 GeV for *Back* events. The hard band starts at 5 GeV for *Front* and 10 GeV for *Back* events. The sky is then partitioned into 24 planar projections and the methods used to look for sources on top of the diffuse emission model are *Point find* [89], the *Minimum Spanning Tree* algorithm [43] and wavelet-based algorithms, as *mr_filter* [191] and *PGWave* [63] [54]. The "seed" positions from those four methods were then combined in order to minimize the number of missed sources. After this source detection analysis the total number of seeds was 2433.

[2] `http://fermi.gsfc.nasa.gov/ssc/data/access/lat/BackgroundModels.html`

### Localization

The image-based detection algorithms provide estimates of the source positions, but the positions are not optimal because no information about the energy is taken into account. These methods also do not provide error estimates on the positions. At this point an advanced statistical method is introduced to localize each detected source.

The method used to localize the sources is an iterative binned likelihood technique [3]. Each source is treated independently in descending order such that brighter sources are included in the background model for fainter sources. The photons are assigned to 12 energy bands (four per decade from 100 MeV to 100 GeV) and HEALpix-based spatial bins for which the size is selected to be small compared with the scale set by the PSF. For each band, the likelihood is defined as a function of the position and flux of the assumed point source, while the background is the sum of Galactic diffuse, isotropic diffuse and any nearby, i.e. within 5°, other point sources in the catalog. The function of the position ($\mathbf{p}$) is defined as $2(\log(\mathrm{L}_{max})\text{-}\log(\mathrm{L}(\mathbf{p}))$, where $L$ is the product of the band likelihoods. According to Wilks' theorem [216], this is the probability distribution for the coordinates of the point source consistent with the observed data. The width of this distribution is a measure of the uncertainty, and it scales directly with the width of the PSF. The distribution is fitted by a two-dimensional quadratic form with five parameters describing the expected elliptical shape: the coordinates (R.A. and dec.) of the center of the ellipse, semimajor and semiminor axis extents ($\alpha$ and $\beta$), and the position angle $\phi$ of the ellipse.

### Significance and Thresholding

Although the detection and localization steps provide estimates of source significances, these estimates are not sufficiently accurate since the detection step does not use the energy information and the localization step fits only one source at a time. To better estimate the source significances the LAT team used a 3-dimensional maximum likelihood algorithm in unbinned mode (i.e., the position and energy of each event is considered individually) applied on the full energy range from 100 MeV to 100 GeV [3]. This is part of the standard *Science Tools* software package. The tool does not vary the source position, but does adjust the source spectrum. The tool provides the best-fit parameters for each source and the *Test Statistic* TS $= 2\Delta\log(\text{likelihood})$ between models with and without the source. The TS associated to each source is a measure of the source significance. For this stage the sources are modelled with simple power-law spectra.

The sky is split into 445 overlapping circular Regions-of-Interest (RoIs)

with radii ranging between 9° and 15° to cover the 2433 seed positions. The RoI sizes are set so that not more than 8 sources are free at a time. The parameters for the sources in the central part of each RoI (RoI radius minus 7°) are free and the value of the best-fit parameters for each one are found iteratively. Considering a threshold of TS = 25, out of 2433 starting "seeds" the procedure selects 1451 sources.

The TS of each source can be related to the probability that such an excess can be obtained from background fluctuations alone. The probability distribution in such a situation (source over background) is not known precisely [169]. However, since only positive fluctuations are considered, and each fit involves four degrees of freedom (two for position, plus flux and spectral index), the probability to get at least TS at a given position in the sky is close to half of the $\chi^2$ distribution with four degrees of freedom [139], so that TS = 25 corresponds to a false detection probability of about $4\sigma$.

**Flux determination, spectral shape and time variability**

Since the spectra of most sources do not follow a single power law over the considered energy range, the maximum likelihood method described earlier does not provide very accurate estimates of the fluxes of the sources detected with TS > 25. Within the two most populous categories, the AGNs often have broken power-law spectra and the pulsars have power-law spectra with an exponential cutoff. In both cases fitting a single power law over the entire range overpredicts the flux in the low-energy region of the spectrum, which contains the majority of the photons from the source, biasing the fluxes to higher values.

To provide better estimates of the source fluxes, the LAT team split the range into five energy bands from 100 to 300 MeV, 300 MeV to 1 GeV, 1 to 3 GeV, 3 to 10 GeV, and 10 to 100 GeV [3]. Since it is generally not possible to fit the spectral index in each of those energy bands (and the flux estimate does not depend very much on the index), the spectral index of each source was frozen to the best fit over the full interval.

The five fluxes provide a rough spectrum, allowing departures from a power law to be quantified. Examples of those rough spectra are given in Figure 2.2, on the left is shown a bright pulsar (*Vela*) and on the right a bright blazar (*3C 454.3*). In order to quantify departures from a power-law shape, it was introduced a *Curvature Index*:

$$C = \sum_i \frac{\left(F_i - F_i^{PL}\right)^2}{\sigma_i^2 + \left(f_i^{rel} F_i\right)^2} \tag{2.1}$$

where $i$ runs over all bands and $F_i^{PL}$ is the flux predicted in that band from the global power-law fit. $f_i^{rel}$ reflects the relative systematic uncertainty on effective area and its values are 0.1, 0.05, 0.1, 0.15 and 0.2 in each energy band [3]. $C$ follows a $\chi^2$ distribution with 5- - 2 = 3 degrees of freedom if the power-law hypothesis is true because the fit involves two parameters, the normalization and the spectral index. At the 99% confidence level, the spectral shape is significantly different from a power law if $C > 11.34$. Only 225 1FGL sources met that condition. The curvature index is not an estimate of curvature itself, just a statistical indicator. A faint source with a strongly curved spectrum can have the same curvature index as a bright source with a slightly curved spectrum. Moreover, any kind of deviation from the best-fit power law can trigger that index, thus the curvature index is not exclusively an indicator of curvature.

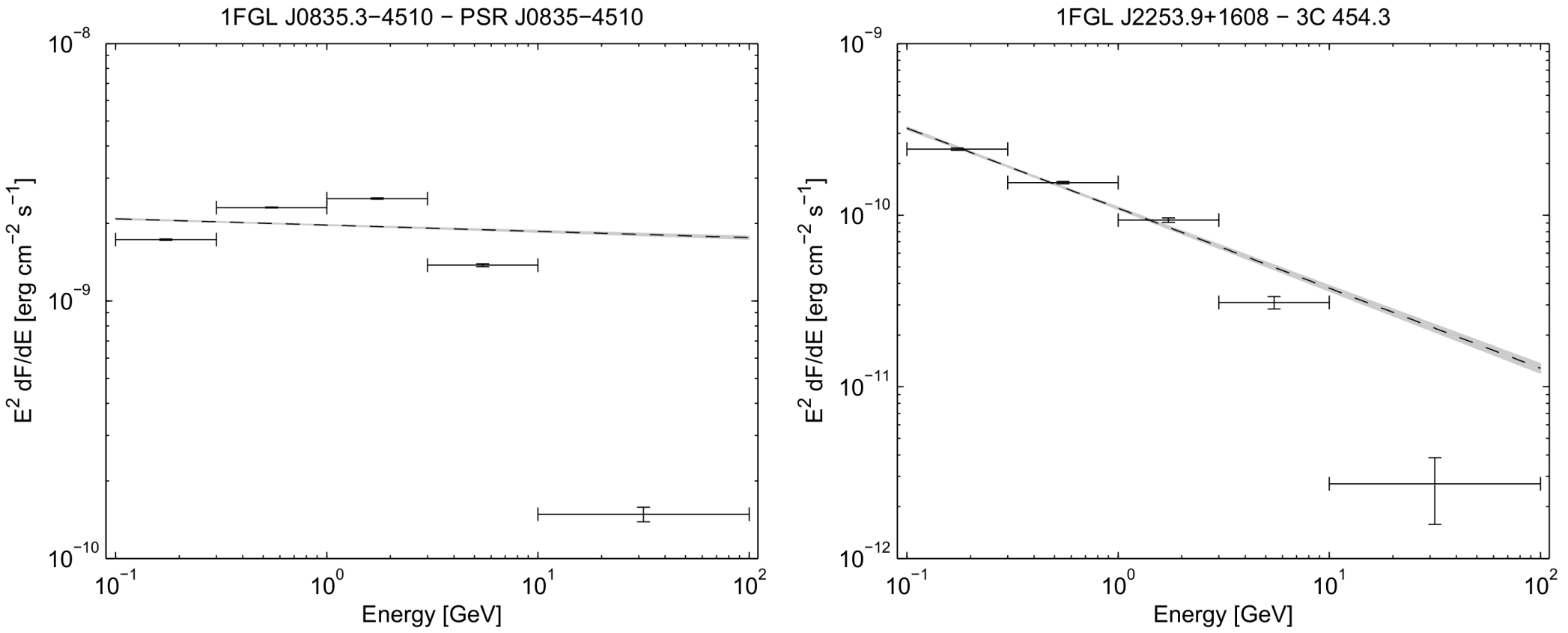

Figure 2.2: On the left is shown the spectrum of the *Vela* pulsar while on the right the spectrum of the bright blazar *3C 454.3*. The dashed lines quantify the uncertainties on index of the power-law fit to the full energy range [3].

Moreover, in order to estimate the time variability for each source it was also introduced a *Variability Index* computed by splitting the full 11 month interval into $N_{int} = 11$ intervals of about 1 month each freezing the spectral index of each source to the best fit over the full interval. In this way, it is possible to detect if a source varies above a specific threshold, but not to characterize such variation [3]. The variability index is defined as a $\chi^2$ criterion:

$$w_i = \frac{1}{\sigma_i^2 + (f_{rel}F_i)^2} \tag{2.2}$$

$$F_{wt} = \frac{\sum_i w_i F_i}{\sum_i w_i} \tag{2.3}$$

$$V = \sum_i w_i (F_i - F_{wt})^2 \tag{2.4}$$

where $i$ runs over the 11 intervals and $\sigma_i$ is the statistical uncertainty in $F_i$. $f_{rel}$ = 3% of the flux for each interval $F_i$, related to the time scale where the instrument and the event classification are stable. Since the weighted average flux $F_{wt}$ is not known a priori, V is expected, in the absence of variability, to follow a $\chi^2$ distribution with $N_{int} - 1 = 10$ degrees of freedom. At the 99% confidence level, the light curve is significantly different from a flat one if V $> 23.21$. That condition is met by 241 sources. Examples of light curves are given in Figure 2.3 for a bright constant source (the *Vela* pulsar) and a bright variable source (the blazar *3C 454.3*). With a 3% systematic uncertainty no pulsar is found to be variable.

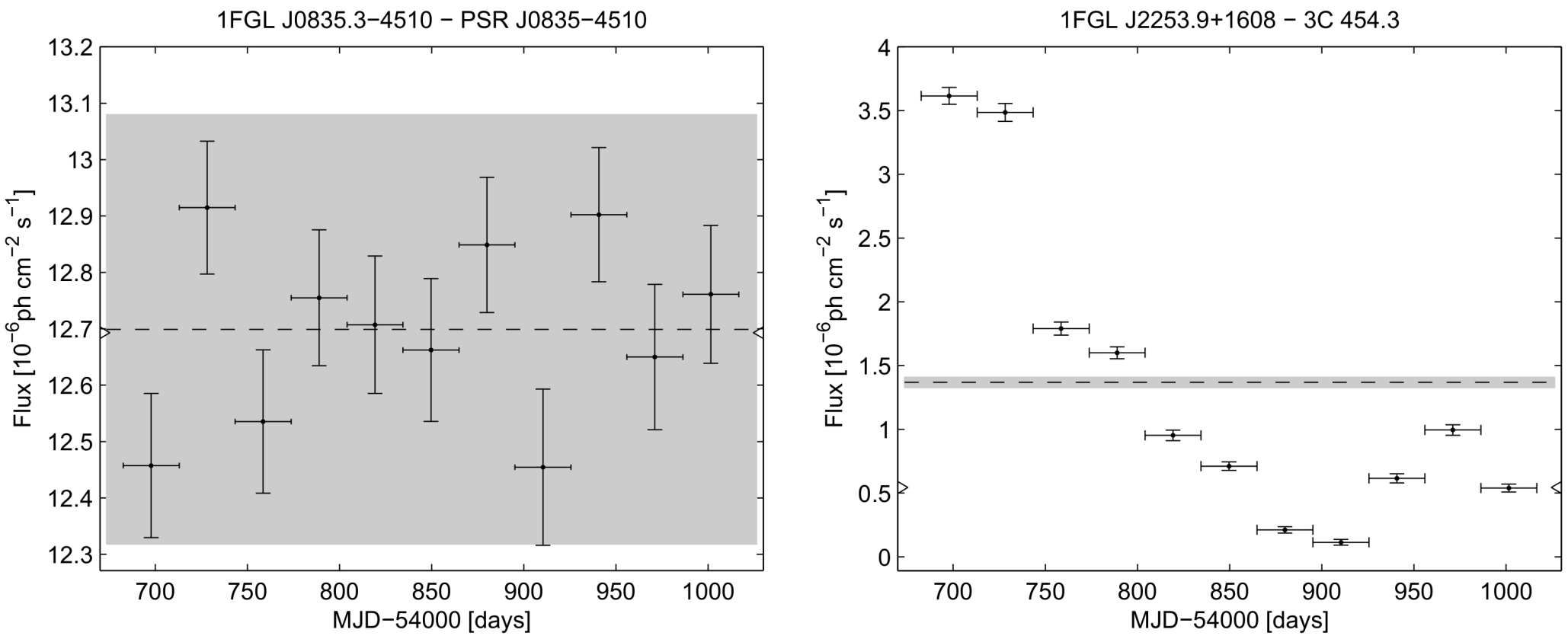

Figure 2.3: On the left is shown the light curve of the *Vela* pulsar while on the right the light curve of the bright blazar *3C 454.3*. The gray band shows the time-averaged flux with the 3% systematic error that was adopted for evaluating the variability index. [3].

### 2.1.2 Source identification and association

Even with the good angular resolution of LAT, source location accuracy is typically not precise enough to make a firm identification based on positional coincidence alone. A typical LAT error region contains numerous stars, galaxies, X-ray sources, infrared sources and radio sources. Determining the nature of a given LAT source must therefore rely on more information than simply location, including time variability, spectral information and availability of sufficient energy at the source and a plausible physical process to produce $\gamma$ rays.

The LAT team introduced a distinction between a source *identification* and an *association* with an object known at other wavelengths. A firm identifica-tion of a source is based on a periodicity for a pulsar or a binary system or on a variability correlated with observations at another wavelength, in the case of a blazar, or on measurement of finite angular extent, which is the case for some Galactic sources, e.g., SNRs. Otherwise, an association is defined as a positional coincidence that is statistically unlikely to have occurred by chance between a plausible $\gamma$-ray-producing object and a LAT source [3].

In order to associate the LAT sources with a plausible $\gamma$-ray emitter an automated source association algorithm is used. The approach to automated source association of the LAT team is based on a list of 32 catalogs containing potential counterparts of LAT sources on the basis either of a priori knowledge about classes of high-energy $\gamma$-ray emitters or on theoretical expectations. The selected catalogs contain AGNs, nearby and starburst galaxies, pulsars and their nebulae, massive stars and star clusters, X-ray binaries and MSPs. The complete list of catalogs, the numbers of objects they contain, and the references are presented in [3]. This approach follows the ideas developed by Mattox [141] for the identification of EGRET sources with flat-spectrum radio sources. It is easy to show that for each catalog in the list, the a posteriori probabilities $P_{ik}$ that an object $i$ from the catalog is the correct association of the LAT source $k$ can be computed using the Bayes' theorem:

$$P_{ik} = \left(1 + \frac{1 - P_{prior}}{P_{prior}} 2\pi \rho_k a_k b_k e^{\Delta_k}\right)^{-1} \tag{2.5}$$

where $P_{prior}$ is the prior probability that a counterpart $i$ is detectable by the LAT and it is determined through Monte Carlo simulations, $a_k$ and $b_k$ are the axes of the ellipse at $1\sigma$, $\rho_k$ is the local counterpart density around source $k$ and

$$\Delta_k = \frac{r^2}{2}\left(\frac{\cos^2(\phi - \phi_k)}{a_k^2} + \frac{\sin^2(\phi - \phi_k)}{b_k^2}\right) \tag{2.6}$$

for a given position angle $\phi$ between LAT source $k$ and the counterpart $i$, $\phi_k$ being the position angle of the error ellipse, and $r$ being the angular separation between LAT source $k$ and counterpart $i$. For the automated association of the 1FGL catalog a LAT source is associated with an objects in the selected catalogs if the a posteriori probability is greater than a threshold set as $P_{thr} = 0.8$, which means that each individual association has $\leq 20\%$ chance of being spurious.

### 2.1.3 Results

Out of 1451 sources in the 1FGL catalog 821 (56%) were associated with a least one non-$\gamma$-ray counterpart by the automated procedure at the 80% confidence level. Sources without firm identifications that are in regions of enhanced diffuse $\gamma$-ray emission along the Galactic plane or are near local interstellar cloud complexes (like Orion), sources that lie along the Galactic ridge ($300° < l < 60°$, $|b| < 1°$), and sources that are in regions with source densities great enough that their position error estimates overlap in the $\gamma$-ray data are called c-sources to indicate "caution" or "confused region". In Figure 2.4 the locations on the sky of the 1451 1FGL catalog sources is shown. Each $\gamma$-ray source class is coded with a different symbol.

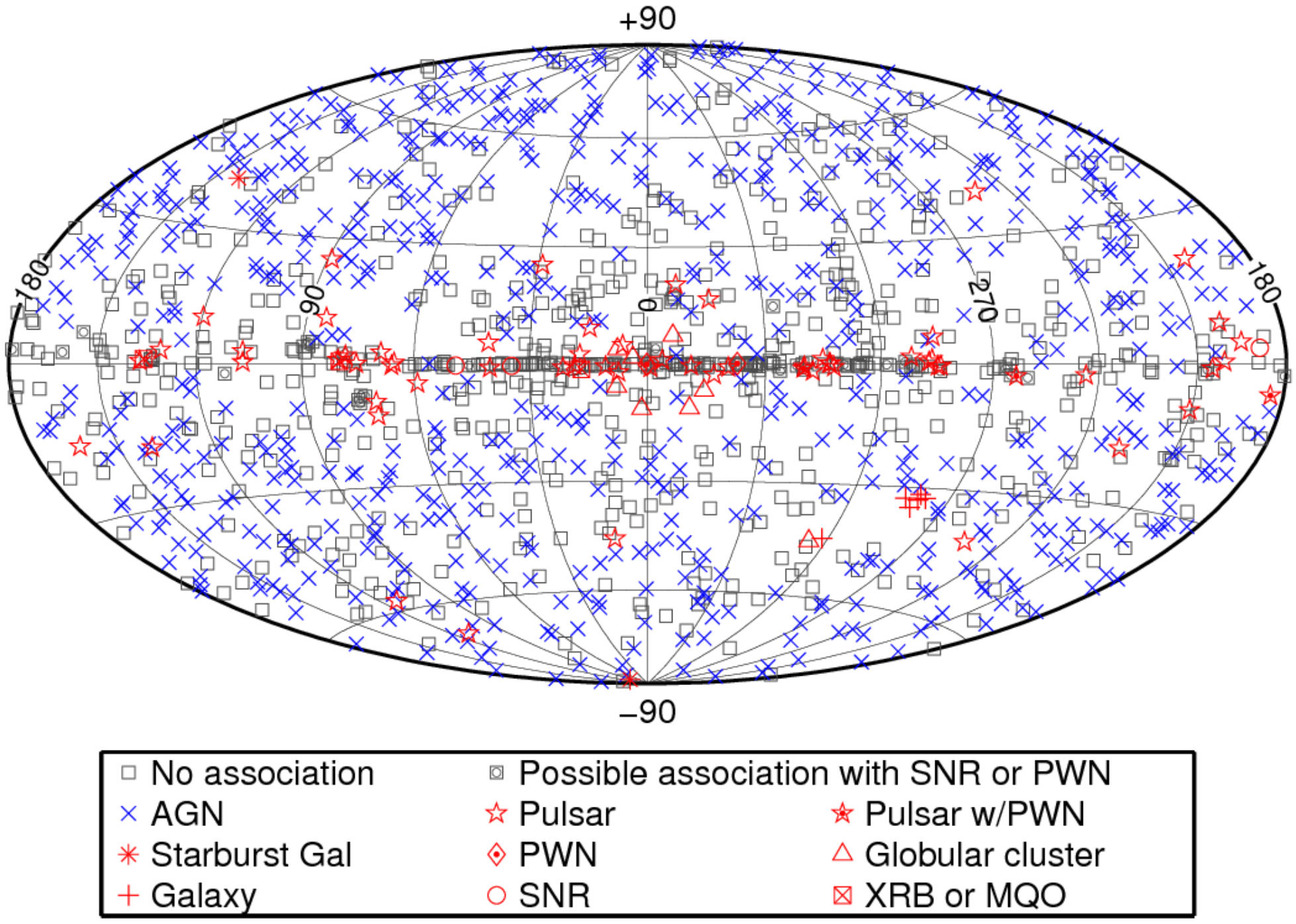

Figure 2.4: Aitoff projection in Galactic coordinates of the distribution on the sky of the 1451 1FGL catalog sources. Each associated source class is coded according to the legend [3].

The automated association results of the LAT for each source class are discussed in some detail in the following:

**AGNs:** 689 1FGL sources were associated with blazars, they represent the most numerous $\gamma$-ray source class, as already observed by the EGRET instrument. Out of them, 282 have also been observed to be radio sources and only 4 were identified through a correlation with variability seen at another wavelength because the procedure of firm identification was not yet carried out systematically for the LAT data. The blazar catalogs selected for the procedure of association are typically incomplete at low Galactic latitudes because of the presence of the Milky Way galaxy that absorbs efficiently a large fraction of low energy photons emitted by extraGalactic objects. Since the $\gamma$-ray photons are not particularly absorbed by our Galaxy we expect that a large fraction of 1FGL unidentified sources situated close to the Galactic plane may be associated with blazars. Moreover, we do not expect an observed isotropical distribution of $\gamma$-ray blazars but a depression of their distribution close to the Galactic plane owing to the higher detection threshold.
Most of the 1FGL non-blazar AGN associations seem to be associated only with non-blazar Seyfert galaxies.

**Normal galaxies:** 2 1FGL sources were associated to nearby starburst galaxies (NGC 253 and M 82) and seven coincide with the Large Magellanic Cloud (LMC) and the Small Magellanic Cloud (SMC) probably corresponding to local maxima of extended emission features.

**Pulsars, PWNe and globular clusters:** 56 1FGL sources have been identified as pulsars through their high-confidence (statistical probability of chance occurrence less than $10^{-6}$) $\gamma$-ray pulsations caused by the rotation of the neutron star. Regarding the PWNe is important to understand whether the LAT indeed detects these objects, or whether the $\gamma$-ray emission arises from the yet unknown pulsars that power the nebulae, or potentially from an associated SNR. Finally, 8 1FGL sources were associated with globular clusters. None of those have alternative associations different from MSPs or low-mass X-ray binaries (LMXBs; which both are known source populations residing in globular clusters).

**Supernova Remnants:** 41 1FGL sources were associated with SNRs through the automated association procedure. Of those, five are associated with small angular size (diameter less than 20') SNRs. Except for two of 1FGL SNRs, the presence of alternative associations with PWNe or an LMXB inside the error radius, makes the physical association of these sources with SNRs questionable. Only 3 1FGL sources were identified as SNRs on the basis of morphology analyses.

**X-ray binaries:** Three 1FGL sources were identified by their orbital modulations as high-mass X-ray binaries (HMXB), they are LS I61 303, LS 5039 and CygX-3. None of the LMXB associations gives strong evidence that this class can emit $\gamma$ rays. All of them are situated in globular clusters, where a combined emission from MSPs appears to be the more plausible counterpart of the 1FGL sources detected.

Despite the application of advanced techniques of identification and association of the LAT sources with counterparts at other wavelengths, 630 ($\sim$40%) 1FGL objects remain without a clear association. These sources are defined as "unassociated" $\gamma$-ray sources by the *Fermi*-LAT team [3] but in this Ph.D. thesis we will mainly refer to these astrophysical objects as "unidentified" $\gamma$-ray sources because their nature is unknown. In order to determine the plausible counterparts of each 1FGL unidentified source multi-wavelength campaigns, mainly radio and X-rays, are in process. Moreover, the excellent performances of the LAT with respect to the previous $\gamma$-ray missions allowed to characterized in great detail the 1FGL sources in terms of location, spectral shape and variability. All the primary $\gamma$-ray information of the unidentified sources can be correlated with the $\gamma$-ray properties of known source classes to try to classify them by specific statistical analyses as explained in Section 3. The results of the source classification can then be used for planning multi-wavelength follow-up observations.

## 2.2 The Second Fermi-LAT Source Catalog (2FGL)

The Second *Fermi*-LAT source Catalog (2FGL) [158] is the most recent high-energy $\gamma$-ray (energy range 100 MeV - 100 GeV) source catalog, it is the successor to the 1FGL catalog [3] and it is based on 2 years of flight data. Since the work explained in the next chapters is based on the data coming from the 2FGL catalog, in this section the procedure to produce the catalog will be described. This procedure follows the same principles used to produce the 1FGL catalog (see the Section 2.1), for this reason only the important improvements compared to the 1FGL catalog will be explained. A first important improvement is that the 2FGL catalog is based on data from 24 months of observations, this means an increase of $\gamma$-ray photons detected allowing a better characterization of the 2FGL sources. Moreover, the data and instrument response functions (IRFs) use the newer and updated event selection, this increases the performances of the analyses. The catalog employs a new, higher-resolution model of the diffuse Galactic and isotropic emissions allowing to distinguish more exactly if an excess of photon in a particular region of the sky is related to the presence of a $\gamma$-ray emitter or to a statistical fluctuation of the background. Moreover, a specific analysis for the spatially extended sources is introduced. The sources detected are analyzed using spectra other than simple power laws (PL) and spectral shapes and variability are characterized in a more refined manner. In the end, the source association process described in the Section 2.1 has been refined and expanded.

### 2.2.1 Improvements with respect to the 1FGL and results

The data analyzed for the 2FGL catalog were taken during the period 2008 August 4 and 2010 August 1. During this time the observing efficiency was very high and limited primarily by interruptions of data taking during the passage of *Fermi* through the South Atlantic Anomaly (SAA; $\sim 13\%$) and trigger dead time fraction ($\sim 9\%$). Thanks to the experience with the data the process of selection of the probable $\gamma$-ray events to use for the construction of the catalog was improved decreasing the instrumental background at energies above 10 GeV and increasing the effective area at energies below 200 MeV. The new class event selections[3] are accompanied by a corresponding revised set of IRFs, including an energy-dependent PSF calibrated using known celestial point sources. These improvements mean the 2FGL catalog is not simply derived from an extension of the 1FGL data set but from a new data set [158]. Other changes were applied for the selection of the events in order to increase the performances of the analysis, e.g. it was applied a more conservative cut on the zenith angles of the $\gamma$-rays, $100°$ instead of the $105°$ used for the 1FGL catalog, because of a larger rocking angle for survey-mode observations ($50°$ instead of $35°$).

#### Diffuse $\gamma$-ray background model

The first step to construct a $\gamma$-ray catalog is modelling in detail the diffuse $\gamma$-ray emission. An accurate model is essential to understand if an excess of photons in a specific region on the sky is related to the presence of a $\gamma$-ray source. Since our Galaxy is the brightest $\gamma$-ray emitter and the emission is particularly structured, the detection of high-energy sources is extremely difficult if they are close to the Galactic plane. For these reasons a lot of efforts went into improving the modeling of the diffuse emission [158]. Recent studies helped to model in great detail the Galactic diffuse emission, a template for the emission from the Earth limb not completely removed was created and also the isotropic background emission was modelled in a more sophisticated manner. The models for the Galactic diffuse emission and the isotropic background spectrum, along with more detailed descriptions of their derivation, are available from the *Fermi* Science Support Center[4].

[3] see `http://www.slac.stanford.edu/exp/glast/groups/canda/archive/pass7v6/lat_Performance.htm`

[4] `http://fermi.gsfc.nasa.gov/ssc/data/access/lat/BackgroundModels.html`

### Construction of the catalog

The procedure used to construct the 2FGL catalog has a number of improvements with respect to what was done for the 1FGL catalog. As for the 1FGL catalog, the basic analysis steps are source detection, localization (position refinement), and significance estimate. Once the final source list was determined, by applying a significance threshold, the flux in five bands, the spectral shape and the variability are evaluated for each source. As for the 1FGL analysis, the source detection step was applied only to the data from the full 24 month time interval of the data set. No transient sources that may have been bright for only a small fraction of the 24 month interval are included.

The procedure to built the 2FGL catalog does have a number of important differences with respect to 1FGL. The analysis is based on a binned likelihood algorithm for different reasons, such as the computing time, which increases linearly with observing time in unbinned likelihood, and then because the scale factors for the diffuse emission models terms returned by binned likelihood are not biased as for unbinned one.

Moreover, many bright sources are fitted with curved spectra instead of simple PL. This provides more detailed descriptions of bright sources and improves the reliability of the procedure for neighboring sources, since it greatly reduces the spectral residuals, which otherwise might have been picked up by neighboring sources. A good representation of pulsar spectra is discovered to be an exponentially cutoff PL [9], i.e. the combination of a PL and an exponential:

$$\frac{dN}{dE} = K\left(\frac{E}{E_0}\right)^{-\Gamma}\exp\left(-\frac{E-E_0}{E_c}\right) \tag{2.7}$$

where $K$ and $\Gamma$ are the same parameters used for a simple PL, $E_c$ is the cutoff energy and $E_0$ is the reference energy arbitrarily chosen [158]. All the detected $\gamma$-ray pulsars with significant LAT pulsations were fitted using this model. Moreover, analysis of the bright blazars [12] indicated that a broken PL was the best spectral representation. Since this model would add two free parameters making it not stable enough for moderately bright sources, the LAT team chosen to use a log-normal representation, called LogParabola, which adds only one parameter while decreasing more smoothly at high energy than the PLExpCutoff form:

$$\frac{dN}{dE} = K\left(\frac{E}{E_0}\right)^{-\alpha-\beta\log(E/E_0)} \tag{2.8}$$

where $K$ is the normalization, $\alpha$ the spectral slope at $E_0$ (the reference energy arbitrarily chosen) and $\beta$ is the curvature. Other bright sources are also well represented by LogParabola spectra, but to limit the number of parameters only those in which the curvature was significant are fitted with this model [158].
Another important difference from the construction of the 1FGL catalog is to include as extended sources the 12 objects that have been shown to be extended in the LAT data. The extended sources include seven SNRs, two PWNe, the LMC and the SMC, and the radio galaxy Centaurus A. For these extended sources a dedicated analysis was developed and applied to characterize their spatial template and spectral form [122]. Since this thesis is based on the study of the point-like LAT sources, these extended objects will never be taken into account in the next chapters.
For a more detailed description about the improvements with respect to the 1FGL catalog refer to [158].

**Flux determination, spectral shape and variability**

As in 1FGL, the source photon fluxes reported in the 2FGL catalog are split into five energy bands: 100–300 MeV, 300 MeV to 1 GeV, 1–3 GeV, 3–10 GeV, 10–100 GeV. The fluxes were obtained by freezing the spectral index to that obtained in the fit over the full range and adjusting the normalization in each spectral band. The spectral index in a band of the curved spectra was set to the local spectral slope at the logarithmic mid-point of the band $\sqrt{E_n E_{n+1}}$ [158].
For the 1FGL sources the spectral departure from a power-law shape was quantified introducing the curvature index (Equation 2.1). In the 2FGL catalog this parameter was refined. An improved estimator of how much the spectrum deviates from a PL based on a likelihood ratio test was introduced as follow:

$$TS_{curve} = 2(\log \mathcal{L}(\text{SpectralModel}) - \log \mathcal{L}(\text{PL})) \tag{2.9}$$

where $\mathcal{L}$ represents the likelihood function, changing only the spectral representation of that source, this means that $TS_{curve}$ is a measure of the significance of the spectral model selected. In order to take into account how much the systematic uncertainties influence the spectrum, in the 2FGL the departure of the spectrum from a PL is quantified by the curvature significance (*Signif_Curve*) defined as follow:

$$\text{Signif_Curve} = \sqrt{TS_{curve}\left(\frac{C^{PL}_{syst}}{C^{PL}_{nosyst}}\right)} \tag{2.10}$$

where $C^{PL}_{syst}$ is the curvature index defined in the Equation 2.1 and $C^{PL}_{nosyst}$ is the same with no $f^{rel}_i$ term. The *Signif_Curve* is defined in $\sigma$ units. The spectrum of a LAT source is significantly curved if *Signif_Curve* $> 4$ [158].
As for the 1FGL sources, for each 2FGL source a light curve was produced dividing the data into 24 time bins of about 1 month each and applying the likelihood procedure to each. As in 1FGL, the spectral shape of each source is frozen to the best fit over the full interval in the light curve analysis. To test for variability in each source a refined variability index was constructed, combining the value of the likelihood in the null hypothesis, that the source flux is constant across the full two-year period, and the value under the alternate hypothesis where the flux in each bin is optimized. Moreover, in 1FGL the brightest pulsars detected by the LAT were flagged as being variable because of systematic errors in the calculation of the source exposure. In order to take into account these effects a correction factor which combines the error on the flux in each time bin in quadrature with a fixed fraction of the overall flux was introduced, so that the bright pulsars are steady. The refined *Variability Index* ($TS_{var}$) was introduced as follow:

$$TS_{var} = 2\sum_i \frac{\Delta F_i^2}{\Delta F_i^2 + f^2 F_{Const}^2} V_i^2 \tag{2.11}$$

where $f = 0.02$, i.e. a 2% systematic correction factor, smaller than the 3% correction required in 1FGL and:

$$V_i^2 = \log \mathcal{L}_i(F_i) - \log \mathcal{L}_i(F_{Const}) \tag{2.12}$$

where $\mathcal{L}_i$ is the likelihood for the individual time bands.

### Results

After the procedure of detection, localization and significance evaluations, the 2FGL catalog lists 1873 $\gamma$-ray sources. In order to associate the LAT sources with a plausible $\gamma$-ray emitter, an improved automated source association is used (see [158]). The improved automated source association of the LAT team is based on a list of 35 catalogs that contain potential counterparts of LAT sources. The designations of the classes that were used to categorize the 2FGL sources are listed in Table 2.1 along with the numbers of sources assigned to each class. As for the 1FGL objects, distinction between associated and identified sources is used, with associations depending primarily on close positional correspondence and identifications requiring measurement of correlated variability at other wavelengths or characterization of each 2FGL source by its angular extent. Sources associated with SNRs are often also associated with PWNe and pulsars, and the SNRs themselves are often not pointlike. Figure 2.5 illustrates where the different classes of sources are located in the sky. At this point it is interesting analyzing how many sources of the 1451 listed in the 1FGL catalog can be associated with a 2FGL object. Associations between

| **Description** | **Associated** | **Identified** |
|---|---|---|
| Pulsar, identified by pulsation | 83 | ... |
| Pulsar, no pulsations seen in LAT yet | ... | 25 |
| Pulsar wind nebula | 0 | 3 |
| Supernova remnant | 4 | 6 |
| Supernova remnant/ pulsar wind nebula | 58 | ... |
| Globular cluster | 11 | 0 |
| Nova | 0 | 1 |
| BL Lac object type of blazar | 429 | 7 |
| FSRQ type of blazar | 353 | 17 |
| Non-blazar active galaxy | 10 | 1 |
| Radio galaxy | 10 | 2 |
| Seyfert galaxy | 5 | 1 |
| Active galaxy of uncertain type | 257 | 0 |
| Normal galaxy | 4 | 2 |
| Starburst galaxy | 4 | 0 |
| Class uncertain | 1 | ... |
| Unassociated | 575 | .. |
| Total | 1746 | 127 |

Table 2.1: LAT 2FGL Source Classes [158].

2FGL and 1FGL sources can based on the following relation:

$$\Delta \leq d_x = \sqrt{\theta^2_{x_{1FGL}} + \theta^2_{x_{2FGL}}} \tag{2.13}$$

where $\Delta$ is the angular distance between the sources and $d_x$ is defined in terms of the semimajor axis of the $x\%$ confidence error ellipse for the position of each source, e.g., the 95% confidence error for the automatic source association procedure. In total, 1099 2FGL sources were automatically associated with entries in the 1FGL catalog. Using 95% source location confidence contours the 2FGL catalog contains 774 (out of 1873) new $\gamma$-ray sources, while 352 sources previously listed in 1FGL do not have a counterpart in the 2FGL catalog [158]. These results show clearly how the 2FGL catalog is different from the 1FGL one, this difference is not simply related to the number of $\gamma$-ray source detected but to the improvements developed by the LAT team for each step in the catalog construction.

The automated association results of the LAT for each source class are discussed in some detail in the following:

**Active Galactic Nuclei:** AGNs, and in particular blazars, are the most prominent class of associated sources in 2FGL. In total, the automatic association procedure finds 917 2FGL sources that are associated with AGNs, of which 894 are blazars, 9 are radio galaxies, 5 are Seyfert galaxies and 9 are other AGNs. One of the most interesting results is that among the five Seyfert galaxies, four are narrow-line Seyfert 1 galaxies that have been established as a new class of $\gamma$-ray active AGNs [15]. AGNs observed by the LAT are also sources of radio (and X-ray) emission, and a clear trend that AGNs associated with 2FGL sources have larger radio fluxes than the average object in the counterpart catalogs was observed. For the AGN associations presented in the 2FGL catalog, the results of the sophisticated 2LAC procedure [83] combined with the results of the automatic association pipeline were adopted in order to increase the number of AGNs associated reducing considerably the chance coincidence probabilities.

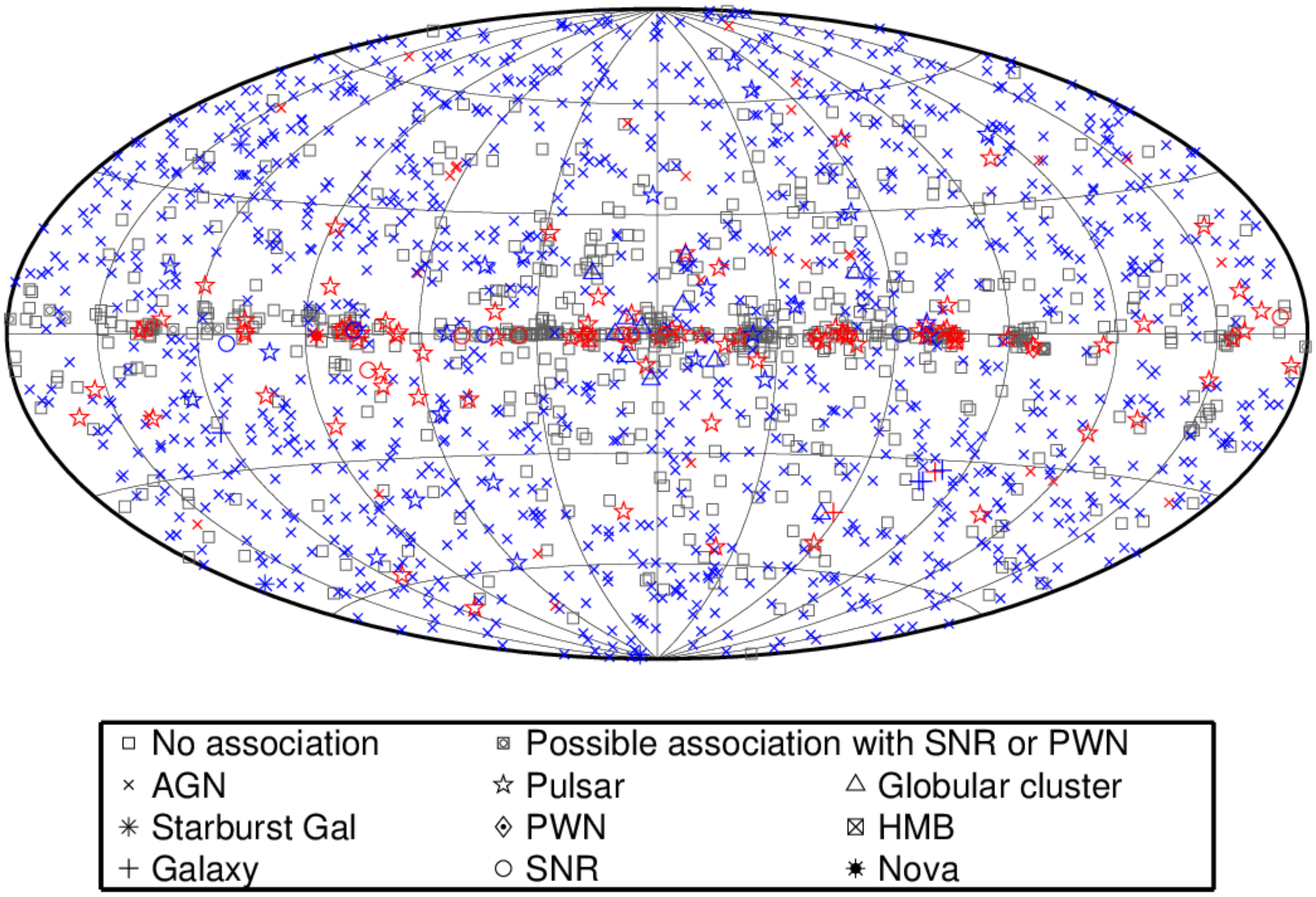

Figure 2.5: Full sky map showing sources by source class (see Table 2.1). Identified sources are shown with a red symbol, associated sources in blue. [158].

**Normal Galaxies:** Normal galaxies are now established as a class of high-energy $\gamma$-ray emitters [4] and seven 2FGL sources were associated with such objects. Of those, the SMC and the LMC are treated as extended sources and analyzed with a dedicated analysis. From the remaining five, four are classified as starburst galaxies (M82, NGC 253, NGC 4945 and NGC 1068) and the fifth is the Andromeda galaxy M31. An interesting result is that the $\gamma$-ray fluxes of Local Group and starburst galaxies were found correlate well with star formation rates [4], which in turn correlate with infrared luminosity.

**Pulsars:** 4 of the 87 pulsars firmly identified by the LAT through the detection of $\gamma$-ray pulsations did not pass TS $> 25$ in the catalog analysis and therefore they were excluded from the 2FGL catalog. Moreover, 3 of the remaining 83 were found to be close to 2FGL sources, but their angular separation from these sources exceeds their effective 99.9% location error radius. In addition to the identified pulsars, four 2FGL sources were associated with radio pulsars but no $\gamma$-ray pulsation has been detected. The automatic association procedure also found 21 2FGL sources to be associated with MSPs. Nineteen of those have unassociated counterparts in the 1FGL catalog and have been discovered in radio pulsar searches of unassociated 1FGL sources ([172], [55], [114], [98], the other two MSPs have no 1FGL counterparts.

**Pulsar Wind Nebulae:** 69 2FGL sources were associated with PWNe, but except for three, all of them are also associated with known pulsars. Among those are three sources for which a dedicated analysis allowed to identify both the pulsar and the PWN [158] and the 2FGL catalog contains both the pulsar and the PWN as separate associated.

**Globular Clusters:** Eleven 2FGL sources are associated with globular clusters. Among those, nine have been published previously: 47 Tuc, NGC 6266, NGC 6388, Terzan 5, NGC 6440, NGC 6626, NGC 6652, Omega Cen and M 80. The two new associations are IC 1257 and 2MS-GC01.

**Supernova Remnants:** 6 2FGL sources were associated with point-like SNR, of which two are also associated with firmly identified pulsars. Other six of the 2FGL sources correspond to SNRs that were firmly identified as $\gamma$-ray sources based on their spatial extensions (IC 443, W28, W30, W44, W51C, and the Cygnus Loop). Moreover, 59 2FGL sources may be associated with an extended SNR on the basis of the overlapping of the 95% confidence error radius of the 2FGL source with the circular extension of the SNR. Since the high chance coincidence rate of these objects, they are considered potential SNRs.

**Binaries:** The 2FGL catalog includes four HMB systems, all of which have been firmly identified by their orbital modulation. They are LSI +61 303 [10], LS5039 [11], Cygnus X-3 [76] and 2FGL J1019.0-5856 [58], recently discovered in a blind search using the LAT data.

Despite the application of advanced techniques of identification and association of the LAT sources with counterparts at other wavelengths, among the 1873 sources in the 2FGL catalog, 575 (31%) have not a clear association, this fraction is diminished with respect to the 1FGL one but is still considerable. In order to determine the plausible counterparts of each 2FGL unidentified source multi-wavelength campaigns, mainly radio and X-rays, are in process. Moreover, the improvements in the 2FGL catalog with respect to the 1FGL one allow to characterized in major detail the 2FGL sources in terms of location, spectral shape and time variability. All the primary $\gamma$-ray information of the unidentified sources can be correlated with the $\gamma$-ray properties of known source classes to try to classify them by advanced statistical analyses as explained in the next Chapter. The results of the source classification can then be used for planning multi-wavelength follow-up observations.

## 2.3 Summary

In this Chapter an overview of the procedure to construct the *Fermi*-LAT catalogs have been given. The high sensitivity, improved angular resolution, and nearly uniform sky coverage of the LAT made it a powerful tool for detecting and characterizing large numbers of $\gamma$-ray sources. A plausible counterpart is a member of a known or likely $\gamma$-ray emitting class located close to the 95% uncertainty radius of a given 1FGL source, with an association confidence of 80% or higher. A different technique of association was introduced by the LAT team, the firm identification, based on a periodicity or on a variability correlated with observations at another wavelength or on measurement of finite angular extent.

Moreover, an overview of the construction of the 2FGL catalog, the most recent $\gamma$-ray source catalog, including the differences and the improvements with respect to the previous 1FGL catalog has been presented. 2FGL catalog is not simply derived from an extension of the 1FGL data set but from a new data set on the basis of the improvements developed. The 2FGL catalog lists 1873 high-confidence sources and the information about location, spectral shape and variability have been refined. Despite the improvements developed for the construction of the 2FGL catalog, approximately 30% of these sources remain unidentified.

Chapter 3

# Classifying unidentified gamma-ray sources: the case of the 1FGL Source Catalog

In this Chapter we will present a brief report of the ongoing efforts to observe the unidentified $\gamma$-ray source fields in other wavebands and to analyze in more detail LAT data in order to find the most plausible counterpart of each unidentified source. A particular attention will be given to my personal contribution within the *Fermi*-LAT collaboration to determine likely source classifications for the unidentified $\gamma$-ray sources using the Logistic Regression algorithm. This is a machine learning technique that uses identified objects to build up a classification analysis, yielding the probability for an unidentified source to belong to a given class based on its $\gamma$-ray properties. We will describe the automated method we have developed: this was the starting point of my Ph.D. work. Classification methods can provide important guidance on what types of follow-up observations are likely to be fruitful for many of these unidentified sources.

The high sensitivity, improved angular resolution, and nearly uniform sky coverage of the LAT made it a powerful tool for detecting and characterizing large numbers of $\gamma$-ray sources. 630 of the 1451 1FGL sources remain unassociated with plausible counterparts at other wavelengths. A plausible counterpart is a member of a known or likely $\gamma$-ray emitting class located close to the 95% uncertainty radius of a given 1FGL source, with an association confidence of 80% or higher as explained in Section 2.1.2. The 95% uncertainty radii for 1FGL source locations are typically 10'. While greatly improved over the degree-scale uncertainties of previous instruments, these position measurements are still inadequate to make firm identifications based solely on loca-tion. The first step of the thesis was to take a statistical approach toward understanding these unassociated 1FGL sources, using all the available information about the $\gamma$-ray sources. Information about locations, spectra, and time variability has been combined with properties of the established $\gamma$-ray source classes and multi-wavelength counterpart searches.

## 3.1 Properties of the 1FGL unidentified sources

The positions, time variability, and spectral information given in the 1FGL catalog provide an important starting point for the characterization of LAT unidentified sources. Intrinsic properties of the 1FGL sources such as spectral index, curvature index, and flux in different energy bands can be easily compared for both associated and unassociated populations, potentially providing insight into the likely classes of the unidentified sources. For the 1FGL catalog, the limiting flux for detecting a source with photon spectral index $\Gamma = 2.2$ and Test Statistic of 25 varied across the sky by about a factor of five [3]. This non-uniform flux limit is due to the non-uniform Galactic diffuse background and non-uniform exposure (mostly arising from the passage of the Fermi observatory through the South Atlantic Anomaly).

The spatial distribution of the 1FGL unidentified sources (see the Figure 2.4) is characterized by a significant excess at low Galactic latitudes ($|b| < 10°$) where 64% of the $\gamma$-ray detected sources have no formal counterparts, compared with only 36% unassociated at $|b| > 10°$ [16]. As for the EGRET (3EG) catalog sources [93], the distribution of the unidentified sources is clearly not isotropic. One consideration when interpreting the distribution of unassociated 1FGL sources is that a number of the remaining unidentified sources are in low Galactic latitude regions where catalogs of AGNs have limited or no coverage, reducing the fraction of AGN associations. If we bin the different source types by Galactic latitude (Figure 3.1), we see a clear absence of AGN associations in the central 10° of the Galaxy ($|b| < 5°$), while in the same region there is a spike in the number of unidentified sources. The absence of AGNs associations is also related to the fact that in the Galactic plane a $\gamma$-ray source must be brighter than at high latitudes in order to be detected above the strong Galactic diffuse emission. Moreover, as was the case for COS-B and EGRET, we expect that probably a subset of the unidentified sources situated close to the Galactic plane may be spurious, resulting from an imperfect Galactic diffuse model.

During the 1FGL analysis all sources were fit with a power-law spectral form and the spectral indices were included in the catalog. In addition, the catalog includes a curvature index, which for each source measures the deviation of the spectrum from the simple power law. Indeed, the curvature index is more a measure of the quality of the power-law spectral fit than of the intrinsic spectral shape. Moreover, in the 1FGL catalog a variability index was introduced for each source to measure the $\chi^2$ of the deviations of eleven monthly (30 day) source flux measurements from the average source flux. Now we look in detail the distribution of these parameters as a function of the flux to try to understand which are the best parameters that distinguish pulsars from AGNs.

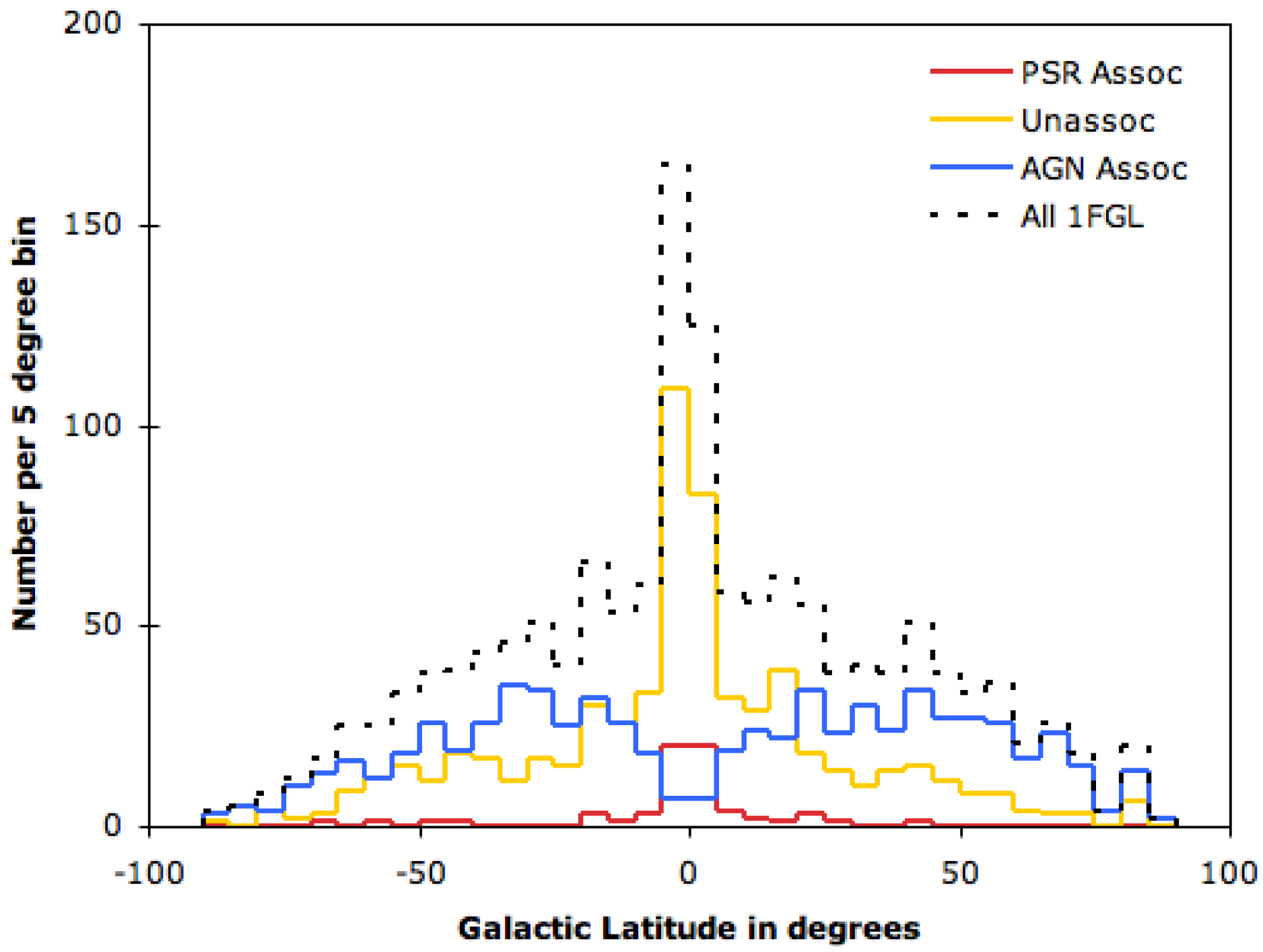

Figure 3.1: Distribution of 1FGL source types by Galactic latitude. The sources associated with AGNs (blue line) show a clear deficit at low latitudes, while the same region hosts a large number of unidentified sources (yellow line) and identified pulsars (red line) [16].

Analyzing the top and the middle panels of the Figure 3.2 we can assert the spectral index and the curvature index do not appear to discriminate well the AGNs from the pulsars. In addition, the relationship is nearly linear for the curvature index, indicating that this parameter is strongly correlated with flux. That is, fainter sources have relatively poorly measured spectra that cannot be measured to be significantly different from power laws. This means that faint $\gamma$-ray sources provide less discriminating information than bright sources. AGN spectra are well described by a broken power law, while pulsar spectra are not well described by a simple power law, this means the spectral index of a power-law fit do not describe the intrinsic spectral properties of AGNs and pulsar and it cannot be a good discriminator between these two source classes [16]. Examining the bottom panel of the Figure 3.2 it is clear that while the variability index increases with flux for AGNs, it does not do so for the pulsars, making variability a much better discriminator between the two major classes. This trend reflects what we know about these source classes (see Chapter 1), AGNs are frequently significantly variable in $\gamma$-rays, their fluxes can vary up to a factor of five on timescales of a few hours and by a factor of 50 or more over several months. On the other hand pulsars are generally steady sources making pulsars extremely different from AGNs in the $\gamma$-ray regime.

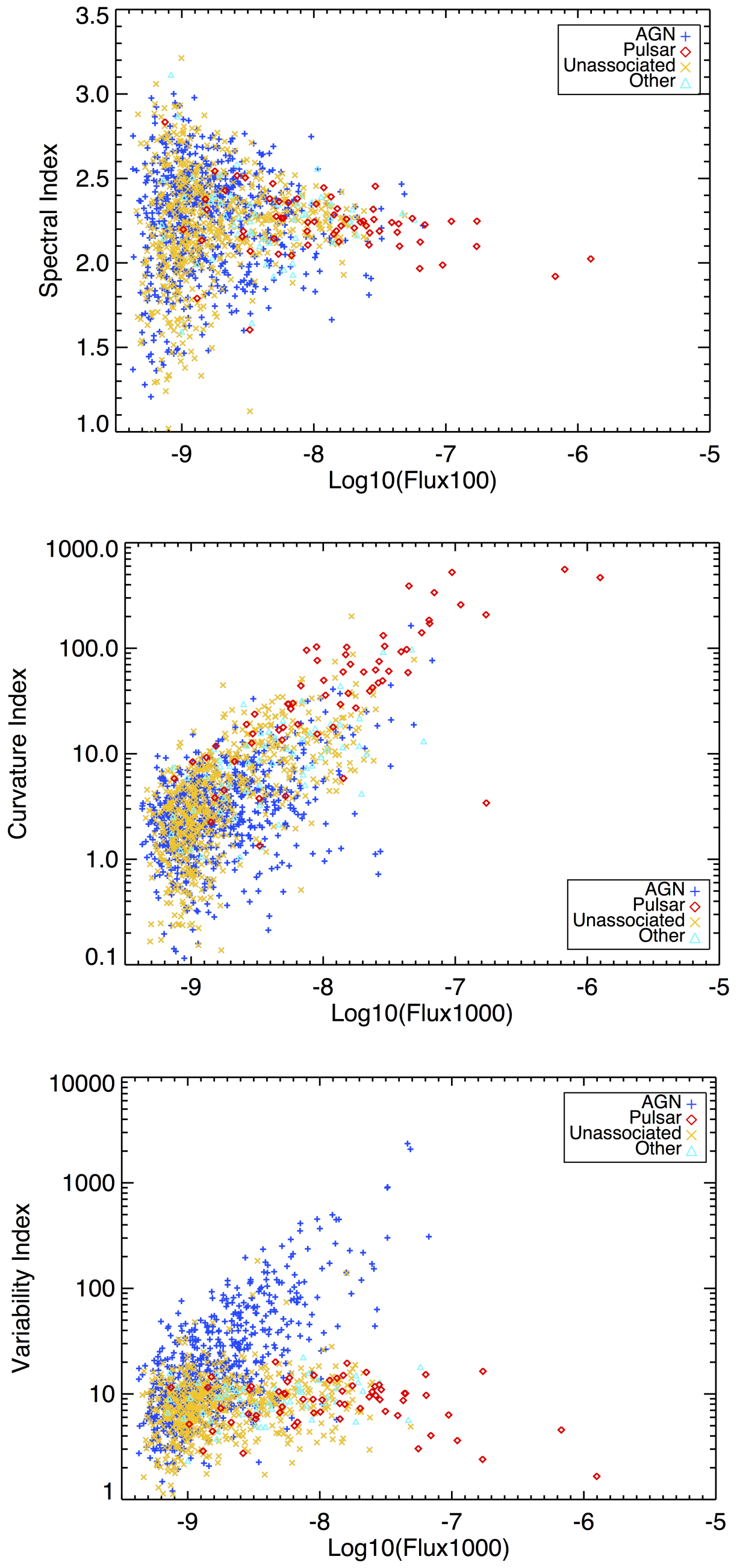

Figure 3.2: Distributions with respect to flux of the spectral index (top), curvature index (middle), and variability index (bottom) for the 1FGL sources. It is clear that the curvature index is dependent on source flux for both AGN (blue crosses) and pulsar (red diamonds) populations.

As shown in Figure 3.3, when the variability and spectral curvature properties of 1FGL sources are compared against each other, a clear separation is visible between bright sources with AGN associations and those with pulsar associations. We are analyzing the distribution of only these two classes because we expect that a large fraction of 1FGL unidentified sources may be associated with AGNs and pulsars since they represent the two most numerous object classes detected in $\gamma$ rays. Pulsars lie in the lower right-hand quadrant and AGNs lie in the upper half. However, the two classes mix in the lower left-hand quadrant, making it difficult to distinguish between them. This region of parameter space is home to much of the unidentified source population. These and other properties of the known sources may give clues to statistical methods of separating the two major types, allowing to classify some of the unidentified sources as likely members of one of these two source types on the basis of their $\gamma$-ray observables (Section 3.3).

This preliminary rough analysis gives us some important information about which are the best discriminator between AGNs and pulsars, these results can be used to predict the most probable classification of each 1FGL unidentified object on the basis of its $\gamma$-ray characteristics. The introduction of new $\gamma$-ray parameters which do not depend too much on the flux, and the use of an accurate statistical approach based on a multivariate analysis would improve the performances of this rough analysis.

## 3.2 Multi-wavelength observations and blind searches

Even with the improved source locations provided by the Fermi-LAT with respect to the previous $\gamma$-ray detectors, positional coincidence with a particular object is not enough to claim an association. If potential candidates can be found, then additional tests, based on spatial morphology, correlated variability, or physical modeling of multi-wavelength properties offer the opportunity to expand the list of associations. X-ray, optical, or radio candidate counter-parts all have better localizations than the $\gamma$-ray sources, so that a candidate in one of these wavelength bands can be matched with those in others. Moreover, most of the catalogs and observations used to find new associations are not complete surveys of the sky. Therefore the lack of an association for a 1FGL source does not mean that the source cannot be associated.

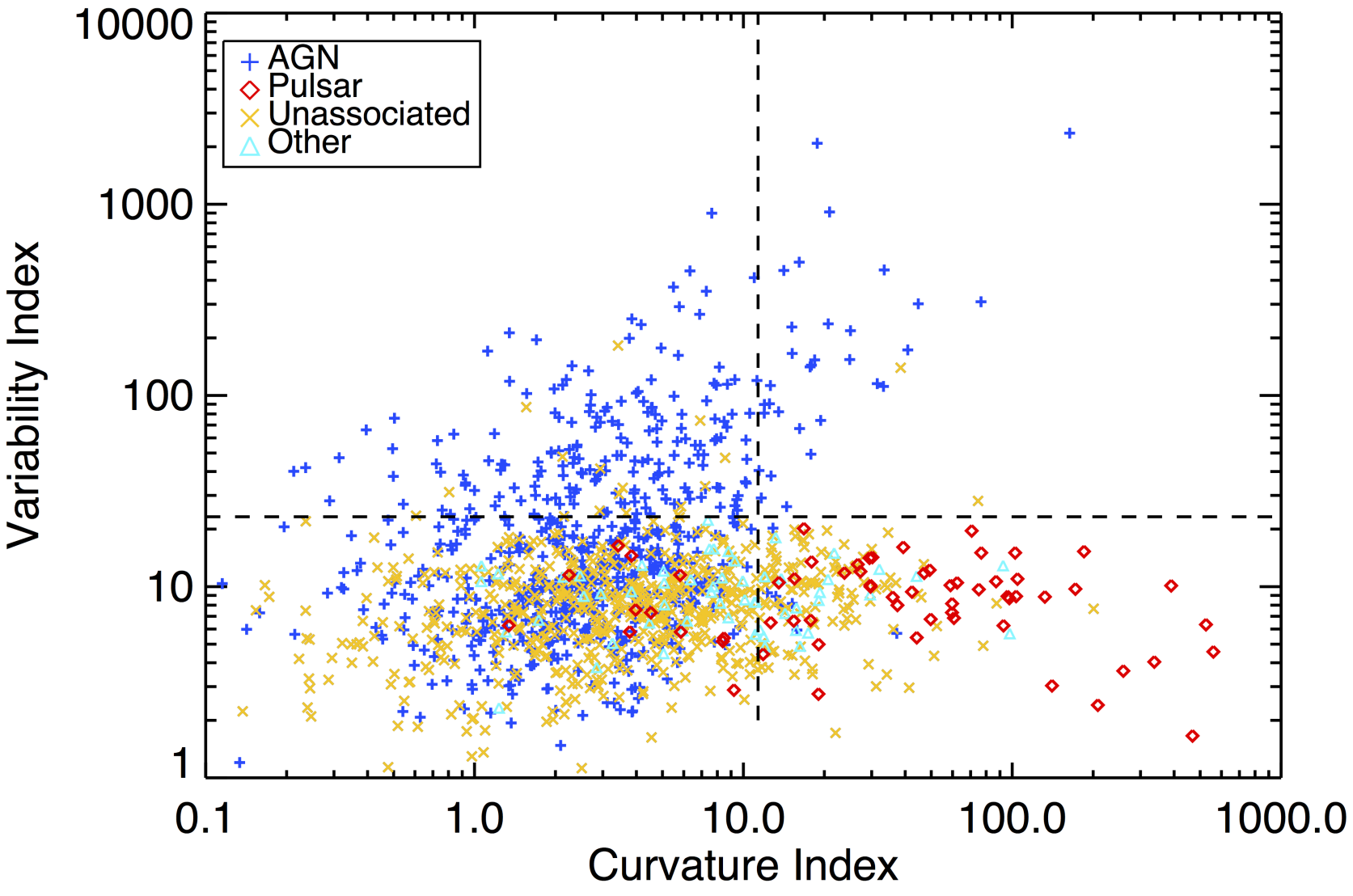

Figure 3.3: Variability index plotted as a function of curvature index. The horizontal dashed line shows where we set the variable source limit, at $V > 23.21$. The vertical dashed line shows where the spectra start deviating from a power law, at $C > 11.34$.

### 3.2.1 Radio and IR searches for pulsars and AGNs

Of the 56 $\gamma$-ray emitting pulsars identified in 1FGL, 32 were detected by folding the $\gamma$-ray data using timing solutions from observations of known radio pulsars. The pulsars' ephemerides were collected by a global consortium of radio astronomers who devote a portion of their observing time to this task [189]. These 32 pulsars (23 young $\gamma$-ray pulsars and nine MSPs) had all been discovered in the radio band prior to their detection by the LAT. In addition to folding data using the properties of known radio pulsars, a promising technique for identifying unidentified sources is searching for previously unknown radio pulsars that might be powering the $\gamma$-ray emission. This technique was used on many of the EGRET unidentified sources ([48], [60], [113], for example) with modest success, because the error boxes were many times larger than a typical radio telescope beam. With the LAT, the unidentified source localizations are a much better match to radio telescope beam widths and, generally, each LAT source can be covered with a single pointing. So far, over 450 unassociated LAT sources, mostly at high Galactic latitudes, have been searched by the *Fermi Pulsar Search Consortium* (PSC; [172]) at 350, 820, or 1400 MHz with preference for those that displayed low variability and a spectrum consistent with an exponential cutoff in the few GeV range, as seen in the identified $\gamma$-ray pulsar population [9]. Members of the PSC have used the following radio telescopes for these observations: Green Bank Telescope (GBT), Parkes, Effelsberg, Nancay Radio Telescope (NRT), Arecibo Telescope, the Lovell Telescope at Jodrell Bank and the Giant Metrewave Radio Telescope (GMRT). Searches by the PSC are ongoing. In the Galactic plane, high dispersion measures and sky temperatures demand higher frequency observations with smaller beam sizes. Young, energetic pulsars can be very faint in the radio ([40], [42]). Nevertheless, we expect that deep observations will continue to turn up more discoveries of radio pulsars in unassociated 1FGL sources in the near future.

The first step in searching for (or excluding) AGN counterparts of *Fermi*-LAT unidentified sources is to consult catalogs of radio sources because they are also radio emitters with compact, flat-spectrum cores [3]. Almost all radio AGN candidates of possible interest are detected either in the NRAO VLA Sky Survey (NVSS; [57]) or the Sydney University Molonglo Sky Survey (SUMSS; [33]). NVSS covers the entire $\delta > -40°$ sky and provides interferometric flux density measurements at 1.4 GHz. SUMSS covers the remainder of the sky and offers interferometric measurements at 0.843 GHz. Different radio follow-up programs are ongoing in order to discover radio counterpart of the blazar candidates, e.g. VLA programs [96]. Moreover, serendipitous radio identification surveys of 1FGL sources have been independently carried out using the recently released Australia Telescope 20 GHz radio source catalog [157], which contains entries for 5890 sources observed at $\delta < 0°$. Recently, using the Wide-field Infrared Survey Explorer (WISE) survey, it was discovered that blazars can be recognized also on the basis of their infrared (IR) colors, the analysis of these data are very useful to discover if the counterpart of 1FGL unidentified sources is a blazar [138].

### 3.2.2 X-ray observations of unidentified source fields

To look for additional possible counterparts the list of unassociated 1FGL sources was cross-correlated with existing X-ray source catalogs. The resulting compilation has no claim of completeness since the match with cataloged X-ray sources depends on the serendipitous sky coverage provided by the X-ray observations and the integration time of the observation. While it is possible that candidate X-ray counterparts to the LAT unidentified sources may be singled out on the basis of, e.g., their brightness and/or spectral properties, most are recognized only through a coordinated multi-wavelength identification approach. In order to find X-ray counterparts of 1FGL unidentified sources different X-ray catalogs were considered cross-correlating the 95% confidence ellipse of the LAT source with the positional uncertainty for the X-ray source in the specific catalog. The catalogs considered are the 2XMM source catalog derived from pointed XMM-*Newton* observations [213], in particular the fourth incremental release of the catalog (2XMMi), the ROSAT *ALL Sky Survey* catalogs ([211], [212]), the *Swift*-BAT source catalog [61] and the 4th IBIS/SGRI Soft $\gamma$-ray Survey Catalog [30]. Recently, a *Swift* program to perform follow-up observation of *Fermi* unidentified sources was performed in an attempt to find X-ray counterparts to the unidentified $\gamma$-ray sources[1].

### 3.2.3 TeV observations of unidentified sources

*Fermi*-LAT spectra have been shown to be good predictors of TeV emission, with 55 1FGL sources having very high energy (VHE) counterparts ([8], [14]). The energy range from ∼50 GeV to ∼300 GeV is the only range where the LAT data directly overlap with other instruments. The 1FGL unidentified sources were cross-checked with the list of TeV sources from TeVCat[2] and current publications. A TeV source is considered coincident with a LAT source if its extension overlaps with the 95% confidence ellipse of the LAT source. Note that the 1FGL association procedure did not assign an association to a coincident TeV source if that TeV source had no identification in another waveband [16].

### 3.2.4 The Blind Search technique

A small number of sources have been associated or identified since the release of the 1FGL catalog by using LAT data alone. The improved sensitivity of the LAT has resulted in the detection of an order of magnitude more $\gamma$-ray pulsars than were previously known. In addition to detecting $\gamma$-ray pulsations from known radio pulsars, the LAT is the first $\gamma$-ray telescope to independently discover pulsars through blind searches in $\gamma$-ray data. Searching for pulsars in $\gamma$-ray data poses significant challenges, the main one being the scarcity of photons. Despite its huge improvement in sensitivity, the LAT still only detects a relatively small number of $\gamma$-ray photons from a given source. Typical $\gamma$-ray pulsars result in tens or at most hundreds of photons per day. The detection of $\gamma$-ray pulsations therefore requires observations spanning long periods of time

[1] *Swift*-XRT Survey of *Fermi* unidentified sources:
`http://www.swift.psu.edu/unassociated/`

[2] `http://tevcat.uchicago.edu`

(up to years), during which the pulsars not only slow down, but often also experience significant timing irregularities, such as timing noise or glitches.

In order to lessen the impact of the long integrations required for blind searches of $\gamma$-ray pulsars, a new technique, known as "time-differencing", was developed, in which FFTs are computed on the time differences of events, rather than the times themselves. By limiting the maximum time window up to which differences are computed to $\sim$ days, rather than months or years, the required number of FFT bins is greatly reduced. The reduced frequency resolution results in a larger step size required in frequency derivative, $\dot{f}$, thus greatly reducing the number of $\dot{f}$ trials needed to cover the requisite parameter space, with the added bonus of making such searches less sensitive to timing irregularities than a traditional coherent search. The net result is a significant reduction in the computational and memory costs, relative to the standard FFT methods, with only a modest effect on the overall sensitivity [183].

Recently, a new blind search technique was developed and applied to discovery pulsars using only the $\gamma$-ray data coming from regions of the 1FGL unidentified sources. The new method, designed to find isolated pulsars spinning at up to kHz frequencies, is computationally efficient and incorporates several advances, including a metric-based gridding of the search parameter space (frequency, frequency derivative, and sky location) and the use of photon probability weights [167]. From the summer 2011, *Einstein@Home*, an on-going distributed computing project[3], is used to search for isolated $\gamma$-ray pulsars in data from *Fermi* satellite's LAT using the previous algorithm increasing of some order of magnitude the number of CPUs and so the computing power.

In the end, regarding the AGNs, the first LAT catalog of AGNs (1LAC; [2]) listed high-confidence AGN associations for 671 high Galactic latitude 1FGL sources, with an additional 155 LAT sources included in the low-latitude and lower confidence association lists. The 1LAC association method was the same as for 1FGL, but the acceptance threshold for association was lower than for 1FGL, thus it included more AGNs than the 1FGL catalog.

## 3.3 Classification of 1FGL unidentified sources

The spatial, spectral, and time variability properties are a framework that allows us to try to predict the expected source classes for the sources that remain unassociated. This is done by using the properties of the associated sources to define a model that describes the distributions and correlations between measured properties of the $\gamma$-ray behavior of each source class. This model is then compared to the $\gamma$-ray properties of each unidentified source. Generating the model requires an associated source parent population with enough members to describe its behavior well. For this reason, we have focused only on AGNs and pulsars as the input source populations.

[3]See `http://einstein.phys.uwm.edu`

To create a model, it is necessary to use $\gamma$-ray properties that are clearly different between the parent populations. In addition, it is important that the properties used to generate the model not be related to source significance, as this will bias the results. To generate valid classifications, we must first define new parameters that allow intrinsic properties to be compared rather than relative fluxes. With the new parameters in hand, we can generate classification predictions using multiple methods and compare these predictions to each other.

To mitigate the effect of low fluxes on the determination of the band spectra, it is necessary to define additional comparative parameters that remove the significance dependency. In this case, the 1FGL catalog provides a set of fluxes in five bands for each source from which we can find hardness ratios [16]. To get a normalized quantity the hardness ratios are constructed as:

$$HR_{ij} = \frac{EnergyFlux_j - EnergyFlux_i}{EnergyFlux_j + EnergyFlux_i} \tag{3.1}$$

This quantity is always between –1 and 1, –1 for a very soft source and +1 for a very hard source. Here, energy flux in $\log(E)$ units (i.e., $\nu \mathrm{F}\nu$ ) is used instead of photon flux because the definition works well only when the quantities are of the same order. This is true for the energy fluxes (because the spectra are not too far from an $\mathrm{E}^{-2}$ power law) but not for photon fluxes [16].

To remove the source significance dependency for variability, we use the *Fractional Variability* (as defined in [3]) instead of the variability index. The fractional variability is:

$$FracVar = \sqrt{\frac{\sum_i (Flux_i - Flux_{av})^2}{(N_{int}-1)Flux_{av}^2} - \frac{\sum_i \sigma_i^2}{N_{int} Flux_{av}^2} - f_{rel}} \tag{3.2}$$

where $N_{int}$ is the number of time intervals (11 in 1FGL), $\sigma_i$ is the statistical uncertainty in $F_i$, and $f_{rel}$ is an estimate of the systematic uncertainty on the flux for each interval. For some 1FGL sources the quantity inside the square root is negative. Those sources are assigned a fractional variability of 2% [3].

### 3.3.1 Machine Learning methods

Machine learning concerns the construction and study of systems that can learn from data [128]. For example, a machine learning system could be trained on associated $\gamma$-ray sources to learn to distinguish between the different source classes, e.g. AGNs and pulsars, on the basis of some parameters. After learning, it can then be used to classify unknown objects, such as unidentified sources, into AGN or pulsar candidates. The core of machine learning deals with representation and generalization. Representation of data instances and functions evaluated on these instances are part of all machine learning systems. Generalization is the property that the system will perform accurately on new, unseen data instances after having experienced a learning data set. The training examples come from some generally unknown probability distribution (considered representative of the space of occurrences) and the learner has to build a general model about this space that enables it to produce sufficiently accurate predictions in new cases. There is a wide variety of machine learning tasks and successful applications.

Machine learning is commonly confused with data mining, as they often employ the same methods and overlap significantly [128]. Machine learning focuses on prediction, based on known properties learned from the training data, while data mining focuses on the discovery of (previously) unknown properties in the data. The two areas overlap in many ways, data mining uses many machine learning methods, but often with a slightly different goal in mind. On the other hand, machine learning also employs data mining methods as "unsupervised learning" or as a preprocessing step to improve learner accuracy.

Machine learning algorithms can be organized into a taxonomy based on the desired outcome of the algorithm or the type of input available during training the machine [128]. Supervised learning algorithms are trained on labelled examples, i.e. input where the desired output is known. The supervised learning algorithm attempts to generalize a function or mapping from inputs to outputs which can then be used to speculatively generate an output for previously unseen inputs. Unsupervised learning algorithms operate on unlabeled examples, i.e. input where the desired output is unknown. Here the objective is to discover structure in the data (e.g. through a cluster analysis), not to generalize a mapping from inputs to outputs.

In the followings we show the list of the most common machine learning methods [128]:

**Artificial neural networks:** an artificial neural network algorithm, is a learning algorithm that is inspired by the structure and functional aspects of biological neural networks. Computations are structured in terms of an interconnected group of artificial neurons, processing information using a connectionist approach to computation. Modern neural networks are non-linear statistical data modeling tools. They are usually used to model complex relationships between inputs and outputs, to find patterns in data, or to capture the statistical structure in an unknown joint probability distribution between observed variables (see Section 4.3).

**Bayesian networks:** a Bayesian network is a probabilistic graphical model that represents a set of random variables and their conditional independencies via a directed acyclic graph. Formally, Bayesian networks are directed acyclic graphs whose nodes represent random variables in the Bayesian sense: they may be observable quantities, latent variables, unknown parameters or hypotheses. Edges represent conditional dependencies; nodes that are not connected represent variables that are conditionally independent of each other. Each node is associated with a probability function that takes as input a particular set of values for the node's parent variables and gives the probability of the variable represented by the node. Efficient algorithms exist that perform inference and learning in Bayesian networks.

**Classification trees:** classification trees are classification tools that have a tree structure in which each internal (non-leaf) node is labeled with an input feature. The arcs coming from a node labeled with a feature are labeled with each of the possible values of the feature. Each leaf of the tree is labeled with a class or a probability distribution over the classes. A tree can be "learned" by splitting the source set into subsets based on an attribute value test. This process is repeated on each derived subset in a recursive manner called recursive partitioning. The recursion is completed when the subset at a node has all the same value of the target variable, or when splitting no longer adds value to the predictions.

**Clustering:** cluster analysis is the assignment of a set of observations into subsets (called clusters) so that observations within the same cluster are similar according to some predesignated criterion or criteria, while observations drawn from different clusters are dissimilar. Different clustering techniques make different assumptions on the structure of the data, often defined by some similarity metric and evaluated for example by internal compactness (similarity between members of the same cluster) and separation between different clusters. Clustering is a method of unsupervised learning, and a common technique for statistical data analysis.

**Logistic regression:** logistic regression is part of a class of generalized linear models. Binary logistic regression forms a multivariate relation between a dependent variable that can only take values from "0" to "1" and several independent variables. Logistic regression is used to predict the probability of being a specific source class based on the values of the independent variables (see Section 3.3.2).

**Support vector machines:** a support vector machine constructs a hyperplane or set of hyperplanes in a high- or infinite-dimensional space, which can be used for classification, regression, or other tasks. Intuitively, a good separation is achieved by the hyperplane that has the largest distance to the nearest training data point of any class (so-called functional margin), since in general the larger the margin the lower the generalization error of the classifier.

**Random forest:** random forest is an ensemble classifier that grows a large forest of classification trees. To classify a new object, each tree in the forest votes on the class. The proportion of votes that agree on a decision provides a measure of the accuracy of the classification.

In this thesis we will use the logistic regression algorithm because it is one of simplest machine learning method, it discriminates each object class using a linear separation, and the artificial neural networks because are the natural extension of the logistic regression as we will explain in the next Chapter, they discriminate each object class using a more complex non-linear separation.

### 3.3.2 Classification using Logistic Regression

Two different machine learning techniques were implemented to determine likely source classifications for the 1FGL unidentified sources: Logistic Regression (LR) and Classification Trees (CT). Both techniques use identified objects to build up a classification analysis which provides the probability for an unidentified source to belong to a given class. We applied these techniques to the sources in 1FGL to provide a set of classification probabilities for each unidentified source.

The approach to assign likely classifications for the 1FGL unidentified sources we have developed is the Logistic Regression (LR) analysis method [103]. LR allows us to quantify the probability of correct classification based on fitting a model form to the data (a more detailed description of this method is explained in the next Chapter and in Appendix B.1). We do not describe the CT technique that was developed by the LAT team, the application of this method is very similar to the LR, for a detailed description of this technique see [16]. In this Section we will explain in detail only the LR technique because this was the first step of my thesis. I started to develop and apply this advanced statistical method to classify 1FGL unidentified sources on the basis of their $\gamma$-ray observables. At the end of this Section the LR and CT results will be compared in order to obtain more robust results.

LR is part of a class of generalized linear models and is one of the simplest machine learning techniques. LR forms a multivariate relation between a dependent variable that can only take values from 0 to 1 and several independent variables. When the dependent variable has only two possible assignment cate-gories, the simplicity of the LR method can be a benefit over other discriminant analyses.

In our case, the dependent variable is a binary variable that represents the classification of a given 1FGL unidentified source. Quantitatively, the relationship between the classification and its dependence on several variables can be expressed as:

$$P = \frac{1}{(1 + e^{-z})} \tag{3.3}$$

where $P$ is the probability of the classification, and $z$ can be defined as a linear combination:

$$z = b_0 + b_1 x_1 + b_2 x_2 + ... + b_n x_n \tag{3.4}$$

where $b_0$ is the intercept of the model, the $b_i$ (i = 0, 1, 2, ..., n) are the slope coefficients of the LR model and the $x_i$ (i = 0, 1, 2, ..., n) are the independent variables. Therefore, LR evaluates the probability of association with a particular class of sources as a function of the independent variables (e.g. spectral shape or time variability).

Similarly to linear regression, LR finds a "best fitting" equation. However, the principles on which the two methods are based are rather different. Instead of using a least-squared deviations criterion for the best fit, it uses a maximum-likelihood method, which maximizes the probability of matching the associations in the training sample by optimizing the regression coefficients. As a result, the goodness of fit and overall significance statistics used in LR are different from those used in linear regression.

### Selection of the training sample and the predictor variables

Since LR is a supervised machine learning technique, it must be trained on known objects in order to predict the membership of a new object to a given class on the basis of its observables. We trained the predictor using the pulsar and AGN associated sources in the 1FGL catalog [3] because, as we have seen, they represent the two most numerous $\gamma$-ray classes and because they can be well distinguished by their $\gamma$-ray observables. The output of this training process is the probability that an unidentified source has characteristics more similar to an AGN than to a pulsar.

To evaluate the best predictor variables for the LR analysis, we used the likelihood ratio test, comparing the likelihoods of the models not including (null hypothesis) and including (alternative hypothesis) the predictor variable under examination. We started by using the fractional variability, the spectral index, the hardness ratios for the 5 energy bands in the catalog and the sky position (i.e. the Galactic latitude and longitude). The value of the likelihood ratio test is the p-value, and is useful in determining if a predictor variable is significant in distinguishing an AGN from a pulsar. If the p-value for a given predictor variable is smaller than the significance threshold $\alpha$ (0.05) then the predictor variable is included in the multivariate LR model. We do not include the curvature value ($HR_{23} - HR_{34}$) in this evaluation because the LR analysis does not work well with predictor variables that are linearly dependent on other predictor values.

We then calculated the significance of each predictor variable to find the resulting LR coefficients. The list of the LR predictor variables with the relative values of the maximum likelihood ratios can be found in Table 3.1.

| **Variable** | **Coefficient** | **Standard Error** | **p-value** |
|---|---|---|---|
| Intercept | -22.17 | 4.97 | <0.001 |
| Fractional Variability | 10.61 | 1.49 | <0.001 |
| Spectral Index | 11.30 | 2.47 | <0.001 |
| $\text{Hardness}_{23}$ | -3.84 | 1.27 | 0.002 |
| $\text{Hardness}_{34}$ | 8.14 | 1.53 | <0.001 |
| $\text{Hardness}_{45}$ | 3.72 | 0.76 | <0.001 |
| $\text{Hardness}_{12}$ | ... | ... | 0.242 |
| glat | ... | ... | 0.333 |
| glon | ... | ... | 0.144 |

Table 3.1: List of the Predictor Variables for the LR Model. Variables selected for the Logistic Regression analysis are listed at top. Those rejected are listed below the line [16].

While AGNs are isotropically distributed and pulsars are concentrated along the Galactic plane, we wanted to see whether our multivariate LR model was able to recognize this effect. The results indicate that Galactic latitude and longitude are not significant at the $\alpha = 0.05$ (5% significance) level. Moreover we find that also $HR_{12}$ is not highly significant in the LR analysis. It is interesting to note that $HR_{12}$, in the univariate LR model, is quite significant (p-value=0.02) to distinguish between AGNs and pulsars but in a multivariate LR analysis it loses its significance. In Table 3.1 those predictor variables selected for the LR model are above the line and those we did not select lie below the line.

### Defining thresholds

Next we derive the predictor variable for 1FGL unidentified sources by applying the trained classification analysis to those sources. Since the LR analysis uses AGNs as primary source type, the output parameter ($A$) listed in Table 4 of [16] describes the probability that an unidentified source is an AGN. The probability that an unidentified source is a pulsar is $P = 1 - A$ (because we are modeling the behavior of AGNs as "opposite" of the behavior of the pulsars based on the predictor variables).

In principle, the dependent variable is a binary variable that represents the presence or absence of a particular class of objects. We could have selected "pulsars" and "non-pulsars" (e.g. all other 1FGL associated sources) to teach the model to recognize the new pulsars, and done similarly for the AGNs. We did not follow this approach because there are no source populations in 1FGL other than AGNs and pulsars with sufficient numbers to significantly affect the results. By focusing on "opposing" characteristics, we improve the efficiency of classifying new AGN or pulsar candidates.

Later, we defined two threshold values, one to classify an AGN candidate ($C_A$) and one to classify a pulsar candidate ($C_P$). We chose these two thresholds by analyzing the *Receiver Operating Characteristic* (ROC) curves by plotting all combinations of true positives and the proportion of false negatives generated by varying the decision threshold so that 80% of the AGN associations in 1FGL would have a predictor value greater than $C_A$ and 80% of the pulsars would have a predictor value smaller than $C_P$, and to result in very low contamination. Using this principle we set $C_A$ to 0.98 and $C_P$ to 0.62. With these thresholds, only 1% of AGNs are misclassified as pulsars, while 3% of pulsars are classified as AGN.

To estimate how accurately our predictive model performs in practice, we cross-validated using only the 756 pulsars and AGNs in the 1FGL catalog. We held out 75 sources to be the testing data set, and we used the remaining 681 for training. We repeated this procedure 10 times, using a different set of 75 test sources in each data set. At the end, this 10-fold cross-validation showed that the average testing efficiency rates for these threshold values are 75% for pulsars and 80% for AGNs, and that the average testing error rates (false positives) are very low, 0.05% for pulsars and 0.02% for AGNs. The 5% lower success rate for the pulsars is likely due to low statistics in the test samples.

It must also be noted that the sources associated with a different class than AGN or pulsar have been excluded from this training procedure, for a total of 24 sources. We cannot treat these 24 sources uniformly as "background", because of the smallness of their sample and the diversity of their spectral properties. However, it is possible to estimate the contamination to the candidate AGN and pulsar samples deriving from the likely presence of these "other" sources in the unassociated sample.

### Results

If we apply the model to the 1FGL unidentified sources we find that 368 are classified as AGN candidates ($P > 0.98$), 122 are classified as pulsar candidates ($P < 0.62$) and 140 remain unclassified after the LR analysis. The distributions of 1FGL associated and unidentified sources as a function of the probability of being AGNs are shown in the Figure 4.5. The thresholds for assigning pulsar candidates and AGN candidates are indicated in the figure. It is important to note that in order to meet the acceptance threshold of 80% of the known pulsars, we are including a large range of predictor values with very few pulsars. This may result in over-predicting the number of pulsars in unidentified sources.

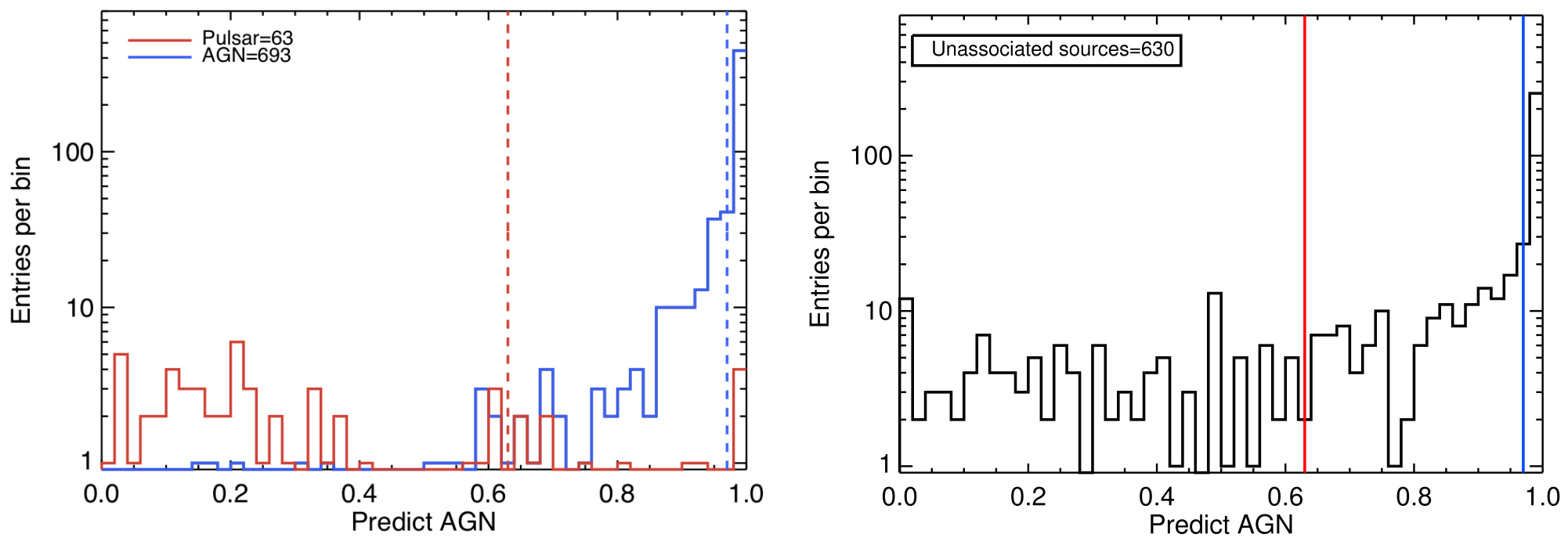

Figure 3.4: Distribution of the Logistic Regression predictor. Vertical lines indicate the value of the thresholds we set to identify pulsar candidates (Predictor < 0.62) and AGN candidates (Predictor > 0.98). Left: sources of the 1FGL catalog identified as pulsars (red) and AGNs (blue). Right: for 1FGL unidentified sources [16].

The predictor distribution for the 24 sources that were not used during the training procedure can be used to estimate the contamination from these sources to the AGN and pulsar candidate distributions. According to the LR analysis, those 24 sources are equally distributed between AGN-like objects, pulsar-like objects, and still unclassified objects. Moreover, it is reasonable to assume that those sources will not be overrepresented in the unassociated population compared with the associated one. Therefore, we expect that up to 2% of the newly classified AGN candidates and up to 5% of the newly classified pulsar candidates will indeed belong to one of the “other” classes (galaxies, globular clusters, supernova remnants, etc.).

### Validating the results

To test the capability of the LR algorithm for identifying the different source classes, we applied the LR analysis to the new sources identified after the release of the 1FLG catalog [3]. Of the 177 newly associated AGNs, 142 were correctly classified as AGN candidates by the LR analysis (efficiency: 80%), only 7 were classified as pulsar candidates (4%), while the other 28 sources remained unclassified (16%). For the 37 newly pulsars, we noticed a different performance between "new pulsar detections", for which pulsations detected in the LAT data, and "new pulsar candidates", i.e. pulsars found at another wavelength in the unidentified source field for which pulsations have been detected only in the radio. For the 20 objects detected as pulsars by the LAT, we correctly classify 11 pulsars (efficiency: 55%), we misclassify only one source (5%) and we leave the remaining sources as unclassified (efficiency: 40%). On the other hand, for the 17 sources that are considered pulsar candidates (as opposed to detections), the classification rate was much worse. We correctly classified only 4 objects as pulsars (efficiency: 23.5%), we misclassified 4 objects as AGN (23.5% of the new pulsar candidates) and left the 9 remaining objects as still unassociated (53%). These results are interesting, as the definition of the pulsar fiducial threshold in the LR analysis appeared that it might overestimate the pulsar candidates. However, the LR actually has a somewhat poorer success rate for finding new pulsars and pulsar candidates than for finding new AGN candidates. Looking more closely at the 1FGL properties of the misclassified pulsar candidates, we find that 12 of the 17 new pulsar candidates have only upper limits for the 0.3 – 1 GeV band, a prime portion of the typical spectrum in the LAT. In contrast, 80% of the new pulsar detections were significantly detected in this portion of the LAT spectrum. This difference in characteristics for the two pulsar groups may indicate the need for additional criteria when selecting sources for follow-up observations.

The sky distribution (left) and the latitude (right) distribution of the newly classified sources are shown in Figure 4.24. Note that both the AGN and pulsar distributions are as expected, even though we have not used the Galactic latitude as an input to either classification method. The pulsar candidates are mainly distributed along the Galactic plane, with a few high-latitude exceptions that suggest additional nearby MSPs, while the AGN candidates are nearly isotropically distributed over the sky. From this we can conclude that, using only the $\gamma$-ray properties of the *Fermi*-LAT sources, and the firm associations of the 1FGL, we were able to develop a predictive method for AGN and pulsar classification that nearly matches our expectations (i.e., pulsar candidates are not variable, have a curved spectrum, and are mainly distributed along the Galactic plane, while AGN candidates are mostly extraGalactic, variable sources).

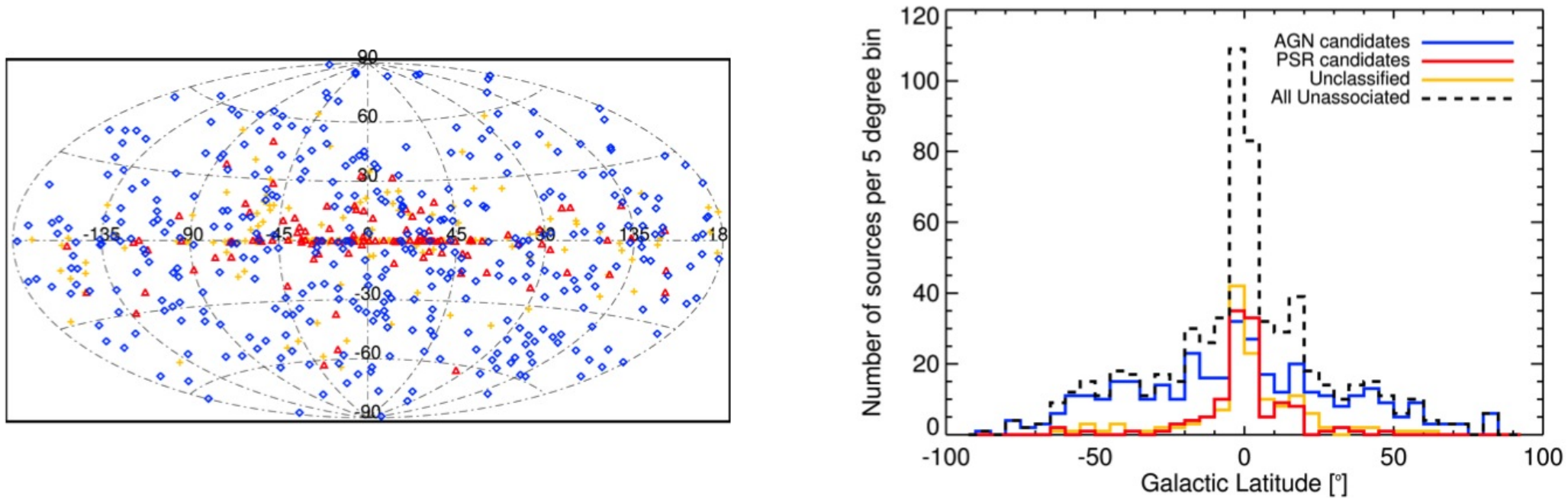

Figure 3.5: Distribution of the newly classified sources. Left: spatial distribution in galactic coordinates. Right: latitude distribution. In red are shown the pulsar candidates, in blue the AGN candidates and in green the unclassified sources.

### 3.3.3 Combining Logistic Regression with Classification Trees

The LAT team implemented another machine learning technique to determine likely source classifications for the 1FGL unidentified sources: the Classification Trees (CT), which iteratively split the original sample trying to maximize the separation at each cut[4]. The LR results were then combined with those found using the CT. In this way the combined classifier would have a higher performance in the classification of 1FGL unidentified sources. For this reason, since the purpose of this analysis is to provide candidate sources for follow-up multi-wavelength studies, the positive results from both techniques are used to generate our candidate lists. It was noticed the two classification techniques gave somewhat different results. Of the 630 unidentified sources in 1FGL, both techniques agreed on the appropriate classification for 57.6% of the sources (363), while they gave conflicting classifications for 5.4% (34 sources). The remaining 253 sources were left unclassified by one or both techniques (see [16]). We can now synthesize a final set of classifications of 1FGL unidentified sources as follow: AGN candidates must be classified by at least one method, and the other method must not disagree (that is, not classify it as a pulsar); pulsar candidates must be classified by at least one method, and the other method must not disagree (that is, not classify it as an AGN); unclassified sources are not classified by either method; "Conflicting" sources are those that have been assigned opposite classifications (one AGN and one pulsar) by the two different methods. Based on these definitions, in the 1FGL unidentified source list there are 396 AGN candidates (269 are classified as AGNs by both methods), 159 pulsar candidates (72 classified as pulsars by both methods), 41 unclassified sources, and 34 conflicting sources. Figure 3.6 shows on the left the curvature-variability distribution of the newly classified AGN and pulsar candidates based on this synthesis of the two methods and on the right their spatial distribution [16].

[4] Details on the analysis and the results using Classification Trees can be found in [16]

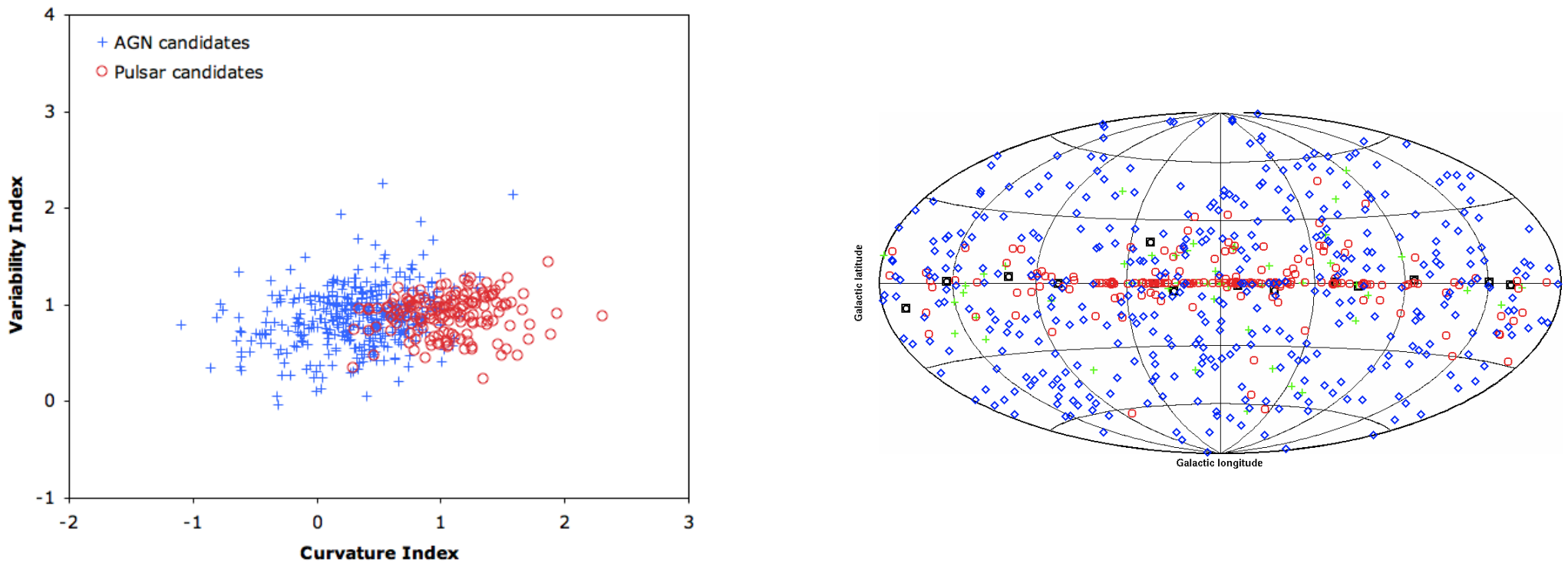

Figure 3.6: (left) Variability index vs. curvature index for 1FGL unidentified sources classified as AGN (blue crosses) and pulsar candidates (red circles). (right) Spatial distribution of the combined classification sample, in Galactic coordinates. Sources are classified as AGN candidates (blue diamond), pulsar candidates (red circles), unclassified (green crosses), or in conflict (black squares) [16].

Now we can compare the new associations to the combined classifications. Of the 214 newly associated AGNs and pulsars, 171 sources (151 new AGNs, 16 new pulsar detections, and 4 new pulsar candidates) match the classification given by the combined analysis, and 26 sources (15 new AGNs, only one new pulsar detection, and 10 new pulsar candidates) are in direct conflict with the classification source type. This gives an efficiency of 85% for AGN classification and 80% for classification of new pulsar detections, but only 59% for new pulsar candidates. Seventeen of the newly associated sources are unclassified by either method, and only one source has conflicting source classification. The one conflicting source turns out to be a new pulsar candidate that also has an AGN association, suggesting the LAT source could be the sum of these two objects. The overall efficiency for this combined sample is $\sim$ 80%, comparable to the value we were seeking when we set the fiducial values for the two methods. The combined sample has a false negative rate of $\sim$ 12%.

### Discussion

The results of the classification analyses demonstrate that source properties measured with the *Fermi*-LAT can provide important guidance on what types of follow-up observations are likely to be fruitful for many of these unidentified sources. The emphasis in follow-up observations of LAT sources has been on radio imaging and timing observations for a large number of sources, as well as targets X-ray observations for sources of interest (e.g., flaring sources or new radio pulsar candidates). In addition, there is an on-going program to observe all the bright, well-localized *Fermi*-LAT unidentified sources with *Swift* with the aim to add important new insights into these sources as a group. The list containing what follow-up observations are recommended in several wavebands on the basis of these analyses is shown in [16].

The efficiency of the LR method at classifying new AGN is high, with a low rate of false negatives, while the efficiency for new pulsar candidates is much lower than expected. The main reason is related to the number of objects in the two training classes, AGN class includes an order of magnitude of objects larger than pulsar class, for this reason our classification method tends to classify better AGNs than pulsars. Moreover, its performances may improve with different criteria selection of input parameters and classification thresholds. In the first section of the next Chapter we will describe the results of a refined LR analysis, in which we will use different criteria selection. In the end, the efficiency of the combined classification methods at classifying new AGN and pulsar candidates is higher as we expect. These results suggest us to combine more classification methods in order to obtain a stronger classifier.

## 3.4 Summary

The *Fermi*-LAT source catalogs list a number of $\gamma$-ray sources one order of magnitude greater with respect to the previous $\gamma$-ray catalogs. The fraction of sources without a firm association is decreasing but is still significant. The association of $\gamma$-ray sources is primarily based on positional coincidence, correlated variability, and pulsation. The low statistics and poor localization (>1') of $\gamma$-ray instruments hamper the association process. Other $\gamma$-ray information can be used to investigate the nature of unidentified sources. We have reported the study and the classification of the 1FGL unidentified sources: this was the starting point of my Ph.D. work. The predictions from these statistical analyses are useful for planning multi-wavelength follow-up observations and selecting targets for pulsar and blazars searches.

630 of the 1451 1FGL sources remain unassociated with plausible counterparts at other wavelengths. The 95% uncertainty radii for 1FGL source locations are typically 10'. While greatly improved over the degree-scale uncertainties of previous instruments, these position measurements are still inadequate to make firm identification based solely on location. The first step of the thesis was to develop and apply an advanced statistical approach based on the Logistic Regression toward understanding the 1FGL unidentified sources, using all the available information about the $\gamma$-ray sources. Information about locations, spectra, and time variability has been combined with properties of the established $\gamma$-ray source classes. These results, combined with those found using other machine learning techniques (e.g. Classification Trees), are useful for planning multi-wavelength follow-up observations and selecting targets for pulsar and AGN searches.

Chapter 4

# Classification of the 2FGL unidentified sources

The Second *Fermi*-LAT Source Catalog (2FGL) is the most recent $\gamma$-ray source catalog. It lists 1873 sources detected during the first 24 months of operation by the LAT in the 100 MeV to 300 GeV energy range. For each LAT source, the proposed associations with sources in other astronomical catalogs is based primarily on positional coincidence, correlated variability and pulsation. In particular, a plausible counterpart is a member of a known or likely $\gamma$-ray emitting class located close to the 95% uncertainty radius of given 2FGL source, with an association confidence of 80% or higher. The procedure used to construct the 2FGL catalog was improved with respect to that used for the 1FGL catalog. Moreover, since the 2FGL catalog is based on data from 2 years of observations, the number of detected $\gamma$-ray photons is considerably increased, allowing for a better characterization of each 2FGL source. All this allowed to improve the localization of each $\gamma$-ray source and to refine the parameters that describe its spectral shape and time variability. Despite all these improvements, 575 ($\sim$ 30%) 2FGL sources remain unassociated. As shown in Figure 4.1, the fraction of unidentified sources is lower than of the previous $\gamma$-ray source catalogs but, because of the low statistics and poor localization ($>$ 1') of $\gamma$-ray instruments, is still significant. Other $\gamma$-ray information can be used to investigate the nature of unidentified sources. Understanding the nature of these sources is very important for different reasons, these objects represent discovery space for new source classes, or new members of existing source classes and their study can help us to better characterize the $\gamma$-ray Universe.

Advanced statistical techniques will be applied to assign the probability of a $\gamma$-ray source to belong to a specific source class on the basis of its $\gamma$-ray observables. Two machine learning techniques will be used for this purpose, a Logistic Regression (LR) refined with respect to that used for the study of 1FGL unidentified sources and an Artificial Neural Network (ANN), which represents the natural extension of the LR method. The results of these two techniques will be compared and from the outcomes multi-wavelength follow-up observations will be planned.

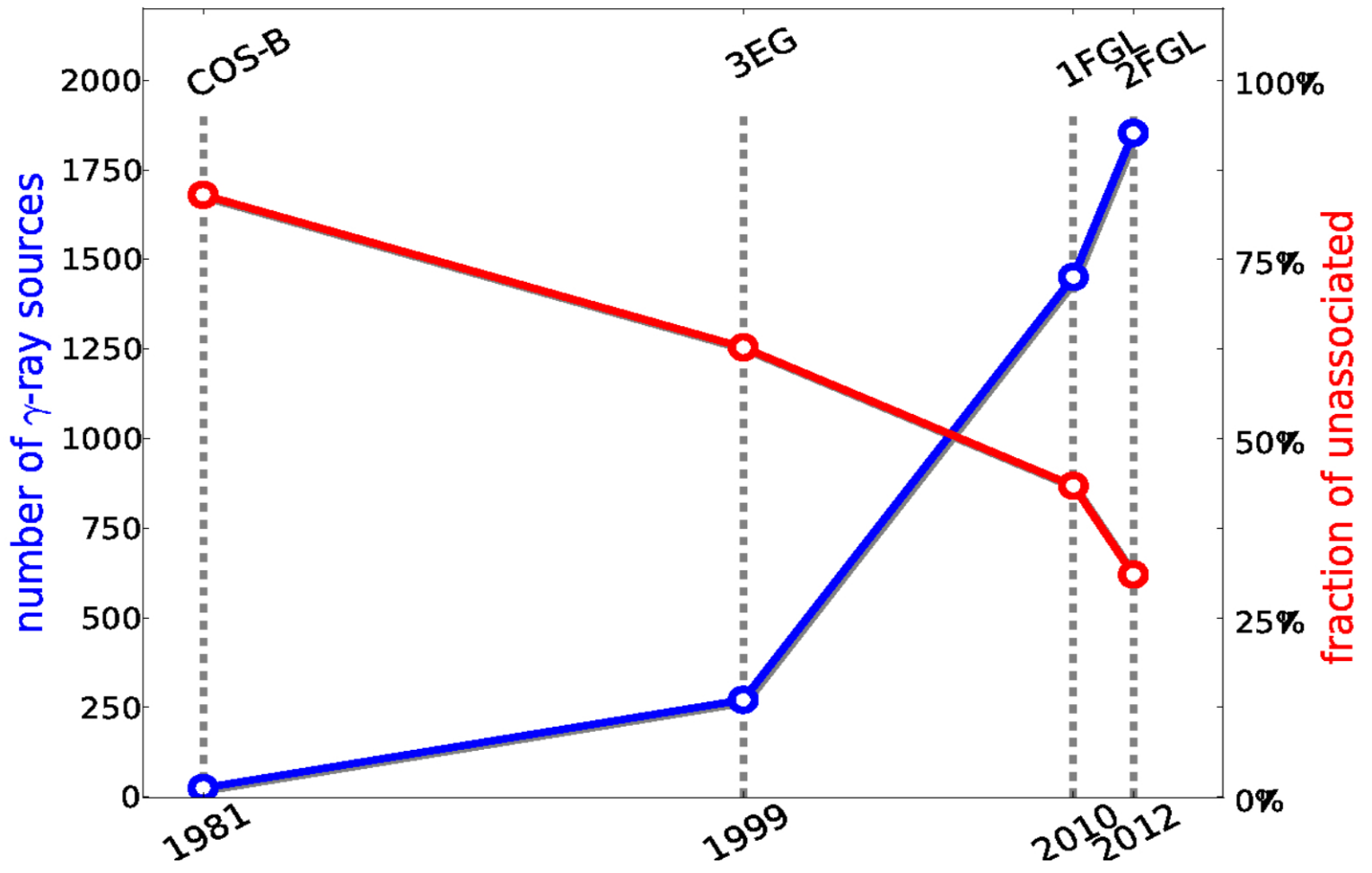

Figure 4.1: Number of detected $\gamma$-ray sources (blue) and fraction of unidentified sources (red).

## 4.1 Properties of the 2FGL unidentified sources

As for the 1FGL unidentified sources, the position, time variability and spectral information given in the 2FGL catalog provide an important starting point for the characterization of the LAT unidentified sources. Since the procedure to build the 2FGL catalog is based on a refined definition of the parameters which are used to describe the spectral shape and the time variability for each source, it is necessary to analyze the "refined" intrinsic properties distribution of the associated sources in order to compare them with the distribution of the unidentified sources.

Figure 4.2 shows on the left the distribution on the sky of the 2FGL sources color coded on the basis of their classifications while on the right the distribution of the different classes as a function of the Galactic latitude. From these figures it is clear that AGNs and pulsars represent the two most numerous source classes in the $\gamma$-ray catalogs and they are characterized by a different distribution. AGNs are characterized by a nearly isotropical distribution that reflects the extragalactic nature of these objects. Inspecting their distribution near Galactic latitude, the absence of AGN associations close to the Galactic plane is striking. As explained in the previous Chapter this absence is related to two facts, first the AGN catalogs used for the procedure of association are not complete at low Galactic latitude, moreover in the Galactic plane, a $\gamma$-ray source must be brighter than at high latitudes in order to be detected above the Galactic diffuse emission. On the contrary, pulsars are primarily situated along the Galactic plane to witness the Galactic nature of these objects. However, a few number of $\gamma$-ray sources associated with pulsars are located at higher latitudes. These objects are primary MSPs, they are fainter than young $\gamma$-ray pulsars and only the nearer ones can be detected by the LAT. They are in our Galaxy but they are located at high latitudes only for a projection effect. The distribution of the 2FGL unidentified sources is clearly not isotropic, as was observed for the 1FGL unidentified sources. There is a significant excess of unidentified sources at low Galactic latitudes ($|b| < 10°$) where 55% of the detected sources have no formal counterparts, compared with only 21% unassociated at $|b| > 10°$. This distribution is probably the sum of a Galactic and extragalactic component. Almost all the objects at high latitude ($|b| > 10°$) may be associated with extragalactic objects, primarily with AGNs and in particular blazars, the most abundant $\gamma$-ray source class. Likely a small fraction of these unidentified sources may be associated with Galactic objects such as MSP. Probably, as was the case for COS-B and EGRET, a subset of the unidentified sources situated close to the Galactic plane may be spurious, resulting from an imperfect Galactic diffuse model, this fact may explain a fraction of the spike in their Galactic latitude distribution. Almost all the "real" unidentified sources located close to the Galactic plane may be associated with Galactic objects, primarily pulsars, because they are the most numerous $\gamma$-ray Galactic source class, the others may be associated to SNRs or other Galactic sources. Owing to the limitation of the AGNs catalogs used during the pro-cedure of association we expect that a small fraction of unidentified sources located along the Galactic plane may be associated with AGNs. In Figure 4.2 the objects called "*psr*" represent the LAT sources associated with a pulsar by the automated source association procedure although no significant pulsation has been found analyzing LAT data. This means that the association of some of these objects may be spurious and we decide not to consider these objects as pulsars but as "potential" pulsars. Otherwise, the objects called "*spp*" are LAT sources for which the 95% confidence error radius overlaps with an extended SNR but their association is not clear, they may be associated with a pulsar or a PWN. Since the high-chance coincidence rate of these objects, we do not consider them as SNRs but as "potential" SNRs.

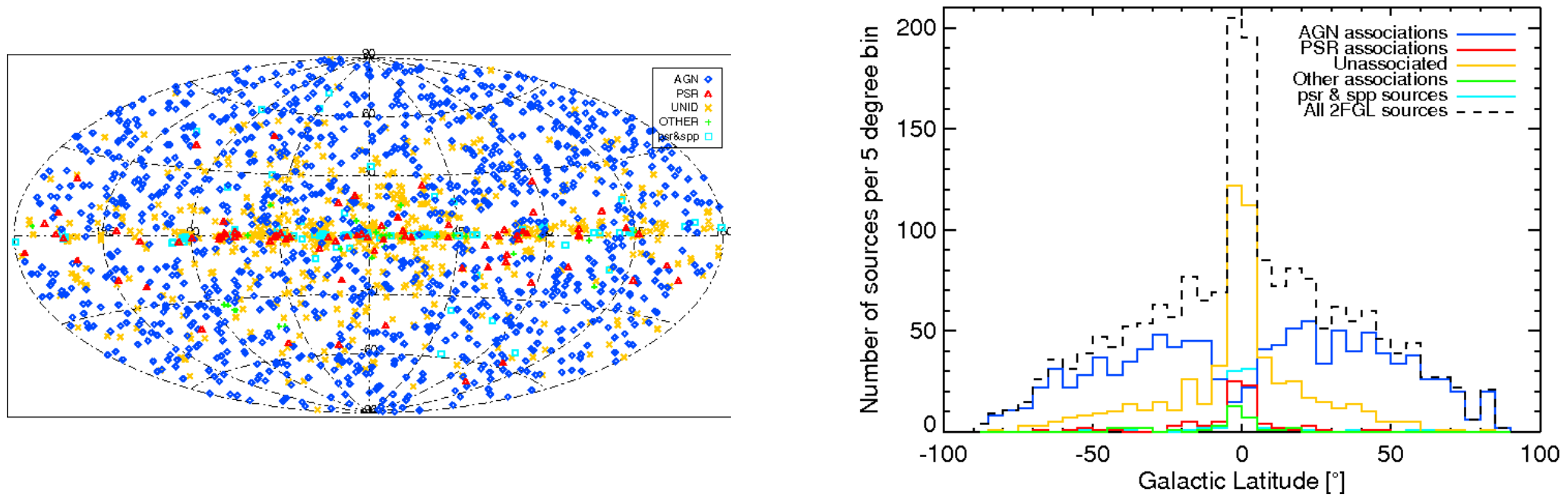

Figure 4.2: Distribution of the 2FGL sources color coded on the basis of their classifications. Left: spatial distribution in Galactic coordinates. Right: latitude distribution.

During the construction of the 2FGL catalog, the brighter sources were fitted with curved spectra instead with a simple power law. For example, all the $\gamma$-ray pulsars with significant LAT pulsations were fitted with an exponential cutoff power law (PLExpCutoff, see Equation 2.7), while bright blazars and other bright LAT sources were fitted with a log-normal model (LogParabola, see equation 2.8). In order to compare the properties of the 2FGL sources with those of unidentified sources we need an homogeneous parameters sample. For this reason all sources were also fitted with a simple power-law spectral form and the spectral indices were included in the 2FGL catalog. This index is called *PowerLaw Index* and it is different from the spectral index that in the 2FGL catalog represents the best-fit photon number power-law index for power law spectra, while for LogParabola spectra is the index at pivot energy and for PLExpCutoff spectra is the low-energy index [158]. In addition, the 2FGL catalog includes an estimate of the curvature significance based on a likelihood ratio test, which measures the significance of the spectral model selected for a specific LAT source with respect to a simple power-law form. This is an improvement of the 1FGL curvature index definition. In the 2FGL catalog a refined variability index was introduced for each source. It derives from the value of the likelihood in the null hypothesis, that the source flux is constant across the full 24-month period, and the value under the alternative hypothesis where the flux in each bin is optimized. Now we inspect the distribution of these parameters as a function of the flux detected with energy above 100 MeV (*F100*) with the aim to understand which are the best parameters in distinguishing pulsars from AGNs.

Analyzing the top panel of the Figure 4.3 we can see that, as done for the 1FGL sources, the PowerLaw index does not appear to discriminate well the AGNs from the pulsars. While AGN spectra are well described by a broken power law, which is very similar to a power law, pulsar spectra are not well described by a simple power law, implying that the spectral index of a power-law fit cannot be a good discriminator between pulsar and AGN classes. Moreover, an instrumental bias is limiting the value of the PowerLaw index since a very faint $\gamma$-ray object cannot be fitted by an hard power-law model because of the low statistics which masks any spectral peculiarity.

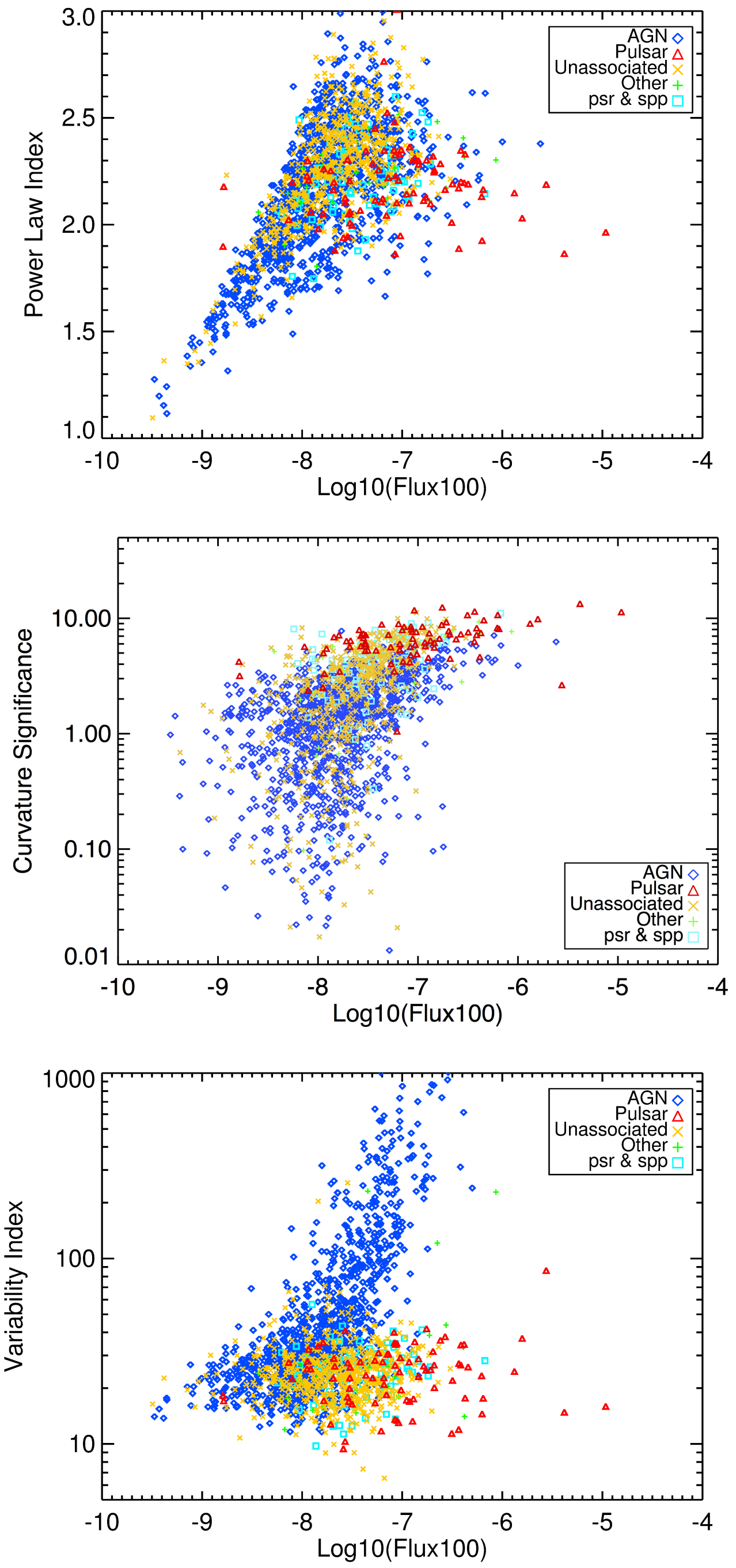

Figure 4.3: Distributions as a function of the photon flux (E > 100 MeV) in units of $cm^{-2}$ $sec^{-1}$ of the PowerLaw index (top), curvature significance (middle), and variability index (bottom) for the 2FGL sources.

Observing the middle panel of the Figure 4.3 it is clear that, differently from the 1FGL curvature index, the relationship between the curvature significance and the flux is not linear, this indicates that the refined curvature parameter is not tightly correlated with flux. Moreover, the curvature significance is a good discriminator to distinguish pulsars and AGNs. This is what we expect on the basis of the typical spectral shape of these two source classes. The broken power-law model describes very well the spectral shape of the AGNs, this model is not very different from a simple power law and thus the value of the curvature significance for these objects is quite low. Conversely the pulsar spectra follow an exponential cutoff power-law model that is very different from a simple power law, especially at energies greater then the cutoff energy, thus their curvature significance is rather high.

An inspection of the bottom panel of the Figure 4.3 shows clearly that, while for AGNs the refined variability index increases with flux, this does not happen for pulsars, making variability index a very good discriminator between the two major classes of 2FGL sources.

If we plot for all source classes the refined variability index and the curvature significance, a clear separation is visible between bright sources with AGNs and those with pulsars (see Figure 4.4). Pulsars lie in the lower right-hand quadrant while AGNs lie in the upper half. These two $\gamma$-ray parameters seem to split the two source classes more efficiently than in the 1FGL catalog. Indeed, the two classes do not mix too much in the lower left-hand quadrant, making it easier to distinguish between them.

This simple analysis allows us to understand which $\gamma$-ray observables separate more efficiently the two major $\gamma$-ray source classes, AGNs and pulsars. In this way we can classify some of the unidentified sources as likely members of one of these two source types on the basis of these $\gamma$-ray observables. However, the analysis performed is characterized by two important limits. First, we can compare by eye against each at most 3 parameters at a time. This is an human limit, our eyes cannot assess if adding new parameters result in a more efficient splitting of AGNs and pulsars. Second, in order to establish if a specific parameter is significant above a certain threshold to distinguish between pulsars and AGNs and in order to assign a probability to each 2FGL unidentified source to be more similar to one of these $\gamma$-ray objects, we need to introduce some statistical rules. As a result we need to develop a statistical method that combines the input parameters, processes the result and assigns the probability of the LAT source to belong to a specific source class, here pulsar or AGN class. The main input parameters will be the PowerLaw index, the curvature significance and the refined variability index but other observables must be tested in order to understand if they can be used to distinguish the two main $\gamma$-ray source classes. To do so we will test all the $\gamma$-ray parameters included in the 2FGL catalog that do not depend too much on the source significance.

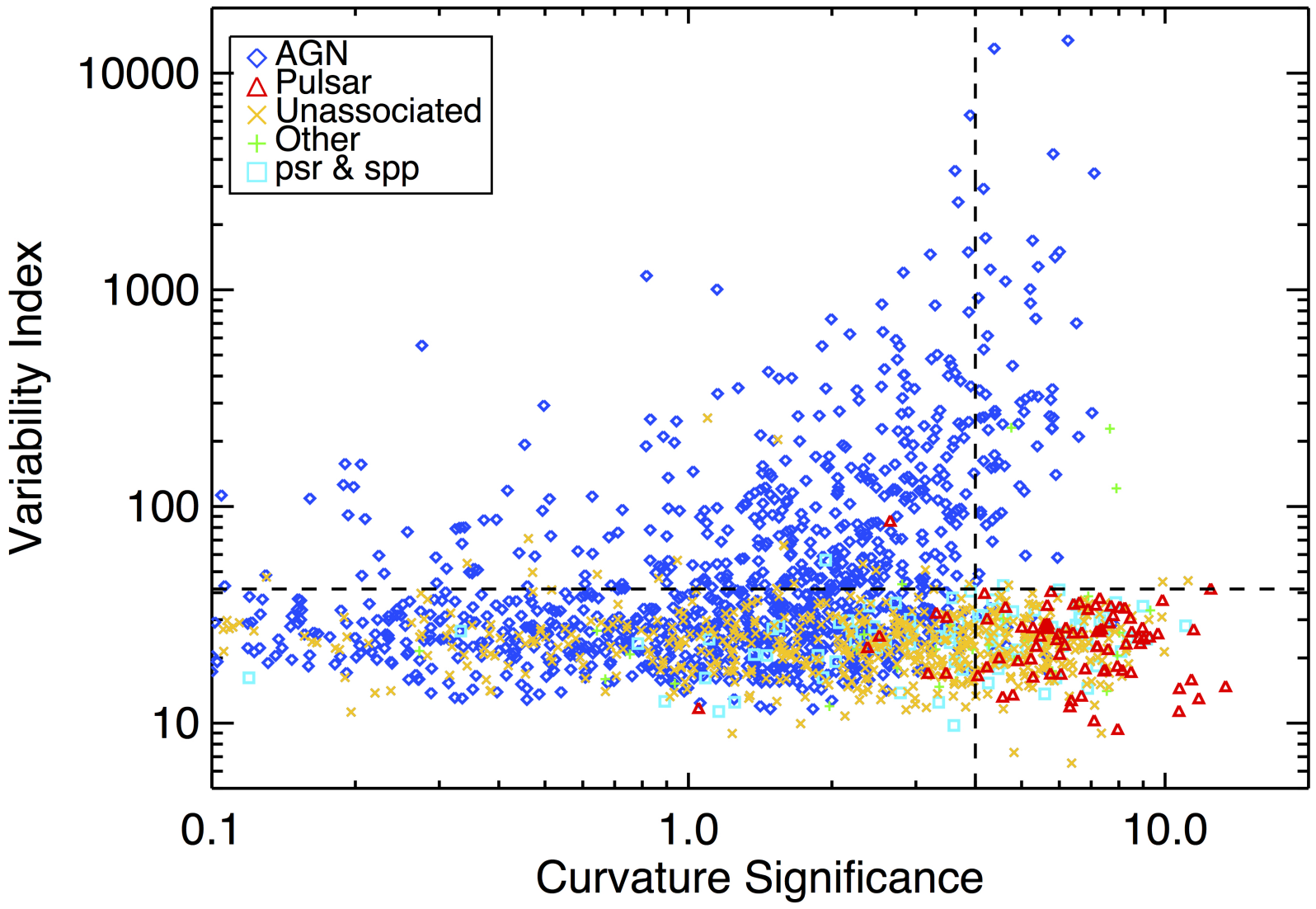

Figure 4.4: Variability index plotted as a function of the curvature significance for different sources classes. The horizontal dashed line mask a variability index of 41.6, above which sources are likely to be variable. The vertical dashed line is set to a curvature significance of 4.0, above which curved spectra are needed.

## 4.2 Classification using Logistic Regression

We have implemented two different machine learning techniques to determine likely source classifications for the 2FGL unidentified sources: Logistic Regression (LR) and Artificial Neural Network (ANN). The advantage of these statistical methods is, in addition to evaluating the probability that an unidentified source is more similar to an AGN or a pulsar on the basis of its $\gamma$-ray observables, to be able to establish which are the parameters useful to distinguish the two major source classes testing for their significance. Following up on the previous section we expect the curvature significance and the variability index to be good discriminators between pulsars and AGNs.

The Logistic Regression method analysis has already been used to classify the 1FGL unidentified sources but the method we will describe here has a number of improvements with respect to the previous LR analysis and they will highlight during the description of the construction and the relative application of the model. The theoretical foundations about the Logistic Regression are reported in Appendix B.1.

### 4.2.1 Construction of the Logistic Regression model

The Logistic Regression model is part of a class of generalized linear models and it is based on the logistic function defined in Equation 3.3. The construction of the LR model to classify LAT objects is based on a number of steps:

1. Selection of the training sample. This step must be based on specific considerations, e.g. the number of objects in each known LAT class or which are the classes characterized by a marked different phenomenology.

2. Selection of the predictor variables. This step consists in determining which are the observables that significantly distinguish the classes selected during the previous step. During this step the regression coefficients are estimated by the maximum likelihood method.

3. Defining thresholds. This step sets a number of classification thresholds to single out each LAT class selected during the first step. The classification thresholds must be set in order to optimize accuracy minimizing contamination and misclassification.

4. Validating. This step consists in evaluating the performances of the classification rules set during the previous step on the basis of a 10-fold cross-validation test.

Once the LR model is built we can apply it to determine likely source classifications for the 2FGL unidentified sources. At this point an additional validation test can be implemented analyzing the performances of the Logistic Regression algorithm in classifying LAT objects associated after the publication of the 2FGL catalog through multi-wavelength studies or blind searches.

#### Selection of the training sample

As LR is a supervised machine learning technique, it must be trained on known objects in order to predict the membership of a new object to a given class on the basis of its observables. We decide to train the predictor using the 83 firmly identified pulsars and all the 1096 AGNs (blazars, non-blazar active galaxies, Seyfert galaxies and active galaxies of uncertain type) listed in the 2FGL catalog [158] because, as we have seen in the previous section, they represent the two most abundant $\gamma$-ray source classes and because they can be well distinguished using their $\gamma$-ray observables. We set the predictor value $P$, defined in the Equation 3.3, equal to 1 for the identified pulsars and equal to 0 for all the AGNs. In such a way the output of this training process is the probability that an unidentified source has characteristics more similar to a pulsar than to an AGN. Moreover, we do not include in the training sample categories other than pulsars and AGNs because of the smallness of their samples. Since we are primarily interested in searching for pulsar candidates in the unidentified sample, we do not include the 25 “potential” pulsars in the training sample.

**Selection of the predictor variables**

The criteria for including a variable in a model is a key point for statistical model building. The approach we use for building the LR model involves seeking the most parsimonious model that explains the data. The rationale for minimizing the number of variables is that the resultant model is more likely to be numerically stable, and is more easy generalized. The more variables included in a model, the greater the estimated standard errors become, and the more dependent the model becomes on the observed data [103].

The several steps we follow to select the variables for the Logistic Regression model are presented in the following:

1. The selection process begins with a careful univariate analysis of each $\gamma$-ray observables. This involves fitting an univariate logistic regression model to obtain the estimated coefficient, the estimated standard error and the likelihood ratio test for the significance of each coefficient (see Appendix B.1).

2. Upon completion of the univariate analyses, we select variables for the multivariable analysis. Any variable whose univariate test obtained during the first step has a *p-value* $< 0.25$ is a candidate for the multivariate model. Once the variable has been identified, we begin with a model containing all of the selected parameters. Our choice to use 0.25 level as a screening criterion for variable selection is based on the detailed studies on Logistic Regression [103]. These studies show that use of a more traditional level of 0.05 often fails to identify variables known to be important. This is related to the fact that any univariate approach has the problem that it ignores the possibility that a collection of variables, each weakly associated with the outcome, can become an important predictor when taken together. Use of such higher level has the disadvantage of including variables that are of questionable importance at the model building stage. For this reason, we will critically review all variables added to a model before deciding on the final model.

3. At this point, following the fit of the multivariate model, the importance of each variable included in the model is verified. This includes an examination of the likelihood ratio for each variable and a comparison of each coefficient estimated from the model containing only that variable. Variables that do not contribute to the model based on these criteria are eliminated and a new model is built. The new model is then compared to the full one by the maximum likelihood test. This process of deleting, refitting and verifying continues until it appears that all of the important variables are included in the model.

4. At this point, any variable not selected in the step 2 for the original multivariable model are added back into the model. This step is helpful in identifying variables that, by themselves, are not significantly related to the outcome but make an important contribution in the presence of other variables. We refer to the model at the conclusion of this step as *final Logistic Regression model*.

To evaluate the best predictor variables for the LR analysis, we follow the procedure previously explained. We start by using the curvature significance, the variability index, the PowerLaw index, the fluxes and the hardness ratios for the 5 energy bands in the catalog and the position on the sky (i.e. the Galactic latitude and longitude). As for the LR model applied to the 1FGL sources, we do not include the curvature value ($HR_{23} - HR_{34}$) in this evaluation. We decide to include the refined variability index instead of the fractional variability because the first one does not depend too much on significance of 2FGL sources as for the variability index defined in the 1FGL catalog [3]. During the procedure of construction of the LR model we use linearly normalized variables in the range [0, 1] in order to stabilize and improve the efficiency of the procedure.

The Table 4.1 shows the results of the univariate analysis applied to pulsars and AGNs of the training sample. In this table, for each observable listed in the first column, the following are shown. (1) The estimated coefficient obtained by our univariate logistic regression fitting. (2) The relative standard error. (3) The p-value as the value of the likelihood ratio test for the parameters significance.

At this point, following step 2 of the selection procedure, only predictor variables with a p-value smaller than the significance threshold $\alpha = 0.25$ are included in the multivariate LR model. The excluded predictor variables are the PowerLaw index and the Galactic latitude and longitude. They will be added back into the multivariate model after the importance of each variable is verified.

We then calculate the significance of each predictor variable to find the resulting LR coefficients. The list of the LR predictor variables with the relative values of the maximum likelihood ratios can be found in Table 4.2. If the p-value for a given predictor variable is smaller than the significance threshold $\alpha = 0.05$ then the predictor variable is included in the final multivariate LR model, the others are rejected. In the table predictor variables selected for the LR model are above the line and those we did not select lie below the line.

| Variable | Coefficient | Standard Error | p-value |
|---|---|---|---|
| Curvature Significance | 17.08 | 1.48 | <0.001 |
| Variability Index | -11.52 | 2.31 | <0.001 |
| PowerLaw Index | 0.74 | 1.40 | 0.60 |
| $Flux_{0.1-0.3GeV}$ | -35.28 | 7.48 | <0.001 |
| $Flux_{0.3-1GeV}$ | -109.73 | 14.60 | <0.001 |
| $Flux_{1-3GeV}$ | 191.82 | 23.25 | <0.001 |
| $Flux_{3-10GeV}$ | -126.68 | 16.54 | <0.001 |
| $Flux_{10-100GeV}$ | -6.91 | 2.34 | 0.003 |
| $Hardness_{12}$ | -1.65 | 0.45 | <0.01 |
| $Hardness_{23}$ | 1.56 | 0.62 | 0.013 |
| $Hardness_{34}$ | 1.62 | 0.68 | 0.019 |
| $Hardness_{45}$ | 7.85 | 0.97 | <0.001 |
| glat | -0.73 | 0.48 | 0.331 |
| glon | 0.32 | 0.41 | 0.427 |

Table 4.1: List of the Predictor Variables for the univariate LR Model. Variables selected for the multivariate Logistic Regression model has a p-value $<$ 0.25.

While AGNs are isotropically distributed and pulsars are concentrated along the Galactic plane, we want to verify whether our final multivariate LR model is able to recognize this effect. The results indicate that Galactic latitude and longitude are not significant at the $\alpha = 0.05$ (5% significance) level as we expect. Moreover, we find that also $HR_{12}$, $HR_{23}$, $HR_{45}$, $Flux_{0.1-0.3}$ and $Flux_{10-100}$ are not significant in the multivariate LR analysis. It is interesting to notice that all these predictor variables, in the univariate LR model, are quite significant (p-value $< 0.02$) to distinguish between AGNs and pulsars but in a multivariate LR analysis they loose their significance. PowerLaw index is not significant to distinguish between AGNs and pulsars in a univariate LR analysis, it becomes significant in the multivariate LR model making an important contribution in the presence of other variables. The positive sign of a predictor coefficient reveals that the source is likely to be pulsar-like. The signs of each predictor coefficient shown in Table 4.2 reveal that a pulsar-like object must be characterized by low variability and by a very curved spectral shape as we expect on the basis of the previous section. The different spectral shapes of pulsars and AGNs are clearly seen in the signs of the fluxes and the $HR_{34}$. The pulsar fluxes are lower than those of AGNs in the energy bands 0.3 – 1 GeV and 3 – 10 GeV, while are higher in the energy band 1 – 3 GeV, where typical cutoff energies of pulsar spectra are found [200]. Above the cutoff energy the pulsar spectrum decreases exponentially becoming very different from a typical AGN spectrum. This may explain the significance of the $HR_{34}$, because in such energy bands the pulsar spectrum change significantly its shape.

| **Variable** | **Coefficient** | **Standard Error** | **p-value** |
|---|---|---|---|
| Intercept | -13.40 | 2.68 | <0.001 |
| Curvature Significance | 20.97 | 2.68 | <0.001 |
| Variability Index | -42.64 | 11.13 | <0.001 |
| PowerLaw Index | 13.88 | 5.74 | 0.016 |
| $Flux_{0.3-1GeV}$ | -743.01 | 200.06 | <0.001 |
| $Flux_{1-3GeV}$ | 2005.48 | 567.03 | <0.001 |
| $Flux_{3-10GeV}$ | -738.68 | 281.07 | 0.008 |
| $Hardness_{34}$ | 5.60 | 2.17 | 0.010 |
| $Flux_{0.1-0.3GeV}$ | ... | ... | 0.122 |
| $Flux_{10-100GeV}$ | ... | ... | 0.454 |
| $Hardness_{12}$ | ... | ... | 0.971 |
| $Hardness_{23}$ | ... | ... | 0.456 |
| $Hardness_{45}$ | ... | ... | 0.850 |
| glat | ... | ... | 0.977 |
| glon | ... | ... | 0.473 |

Table 4.2: List of the Predictor Variables for the LR Model. Variables selected for the final Logistic Regression model are listed at top. Those rejected are listed below the line.

### Defining thresholds

We define two threshold values, one to classify an AGN candidate ($C_A$) and one to classify a pulsar candidate ($C_P$). We chose these two thresholds in order to optimize the accuracy, that measures the proportion of actual positives which are correctly identified as such (it is also called *sensitivity*), and minimize the misclassification and the contamination by plotting these values by varying the decision threshold as shown in Figure 4.5. On the left are shown the distribution of the parameters we analyzed to set the threshold (indicated as a red vertical line) to single out pulsars, conversely on the right those we analyzed to set the threshold (indicated as a blue line) to single out AGNs. Using this principle we set $C_A$ to 0.05 and $C_P$ to 0.65. With these thresholds, 82% of pulsars are correctly classified as a pulsar while only 3.6% are misclassified as an AGN, conversely 95% of AGNs are correctly classified as an AGN while only 0.3% are misclassified as a pulsar. Regarding the contamination, only 4.2% of the objects classified as a pulsar are indeed AGNs and only 0.3% classified as an AGN are indeed pulsars. These results are shown in the table of classification in Figure 4.6. If we analyze in detail the pulsars misclassified we find 3 objects: PSR J1357–6429, PSR J1823-3021A and PSR J2240+5832. The first one is a young and energetic pulsar ($\tau_c$=7.3 ks and $\dot{E} \sim 10^{36}$ erg s$^{-1}$) [41] surrounded by a bright PWN observed in X-ray and at TeV energies [50] [97]. Thus, if the PWN is contributing significantly, this may explain the misclassification. The second one is an energetic MSP located 8.4 kpc away in the core of the cluster NGC 6624 [80]. This object is very peculiar, characterized by the highest $\gamma$-ray luminosity observed for any MSP with an unusually large rate of change of its period. Their peculiarities may be the reason of the misclassification of these objects. The last one seems to be a typical young radio-loud pulsar [188] with no peculiar behaviour. The 3 AGNs misclassified are an associated FSRQ blazar (2FGL J2206.6+6500) and two active galaxies of uncertain type, associated through statistical methods different from those in the 2FGL catalog [83]. Recently, 1 of the active galaxies of uncertain type (2FGL J0621.9+3750) has been associated with the pulsar J0622+3749 found in a blind search analyzing LAT data [167]. Probably also the misclassification of the other AGNs is related to an incorrect association.

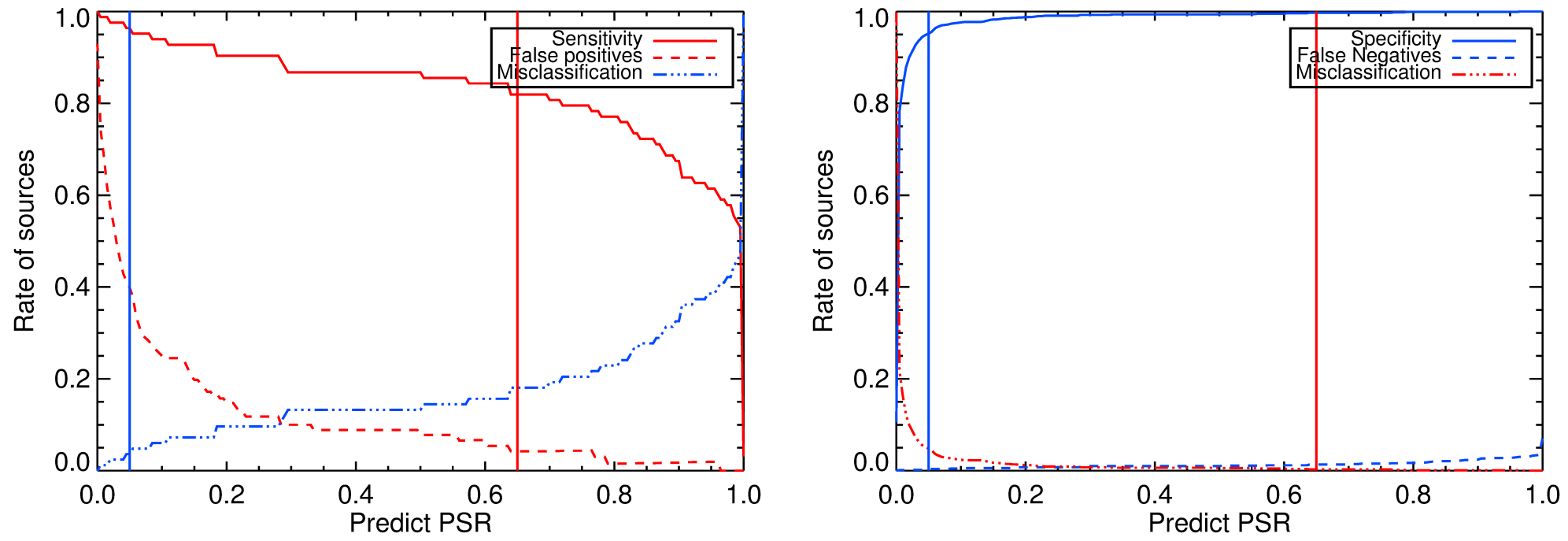

Figure 4.5: Distribution of the accuracy (here called sensitivity), misclassification and contamination by varying the decision threshold. (left) Distribution related to pulsars: false positives represent the contamination given by AGNs and misclassification is related to AGNs classified as pulsars. (right ) Distribution related to AGNs: false negatives represent the contamination given by pulsars and misclassification is related to pulsars classified as AGNs. Vertical lines indicate the value of the thresholds we set to single out pulsars (red) and AGNs (blue).

| | PSR observed | AGN observed | Efficiency/ Contam. |
|---|---|---|---|
| PSR predicted | 68 | 3 | 95.8% 4.2% |
| AGN predicted | 3 | 1043 | 99.7% 0.3% |
| Unclassified | 12 | 50 | |
| Accuracy/ Misclass. | 82% 3.6% | 95% 0.3% | 94% 0.5% |

Figure 4.6: Table of classification based on the classification rules described in the text. The objects called observed represent the sources in the training sample, otherwise the objects called predicted represent the results of the LR algorithm applied to the observed sources using our classification rules. In green are represented the correct classifications and in red the incorrect ones.

The distribution of the LR predictor for the sources of the 2FGL catalog identified as pulsars and AGNs is shown in Figure 4.7.

It must be noted that the sources associated with a different class than AGN or pulsar, for a total of 35 sources, have been excluded from this training procedure. They are composed by Galactic objects, such as SNRs, HMBs, globular clusters and a nova, and extragalactic objects, such as normal galaxies, SMC and LMC. Here we have not considered the "potential" pulsars nor the "potential" SNRs because their association is uncertain. We cannot treat the "other" 35 sources uniformly as "background", because of the smallness of their sample and the diversity of their spectral properties. However, it is possible to estimate the contamination to the candidate AGN and pulsar samples due to the likely presence of these "other" sources in the unassociated sample. In Figure 4.8 is shown how we can use the predictor distribution for the 35 "other" sources to estimate the further contamination from these sources to the AGN and pulsar candidate distributions.

According to the LR analysis, the 35 sources are nearly equally distributed between AGN-like objects, pulsar-like objects, and still unclassified objects. In particular, 46% (6) are classified as pulsar candidates, while 23% (8) as

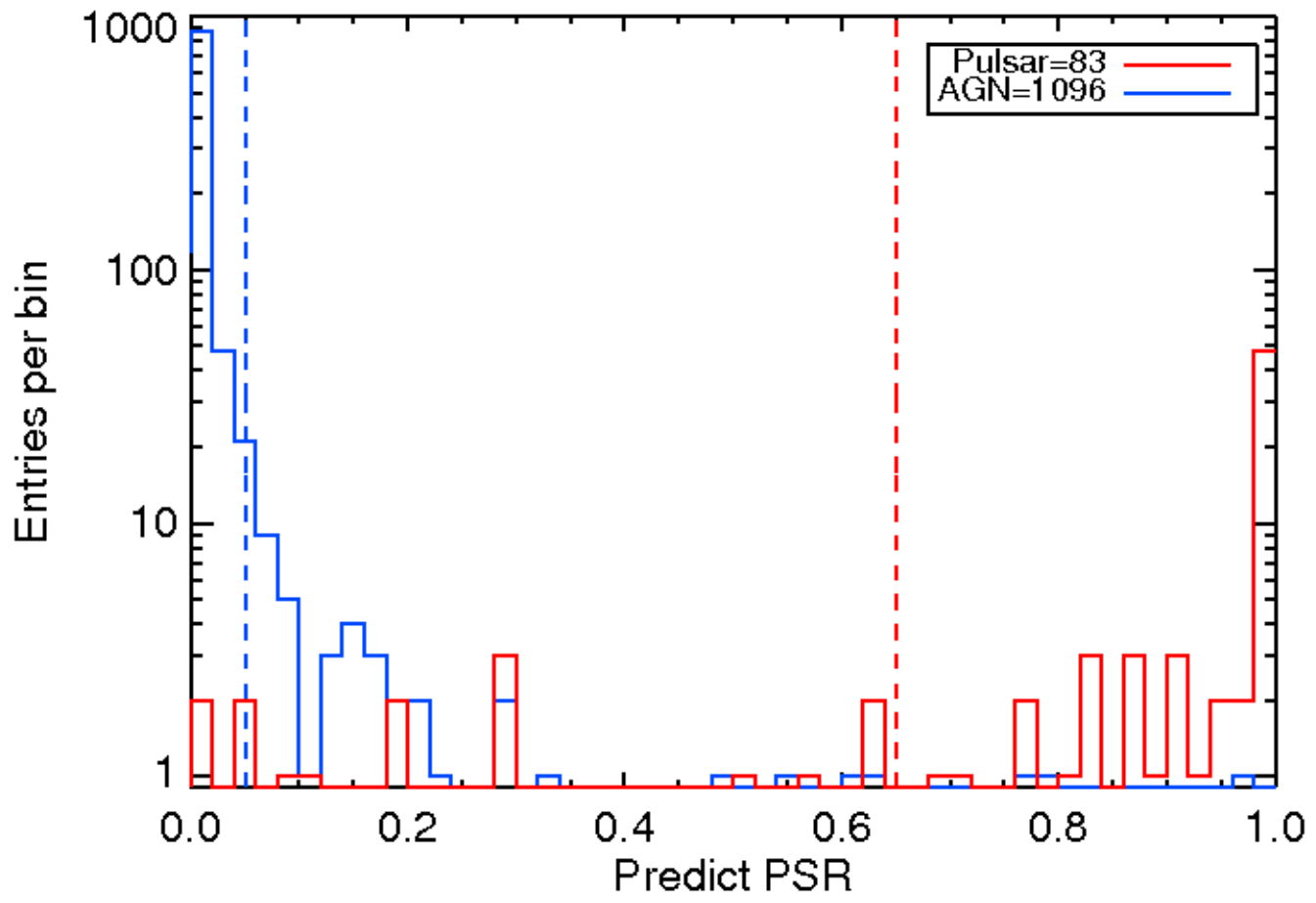

Figure 4.7: Distribution of the LR predictor for sources of the 2FGL catalog identified as pulsars (red) and AGNs (blue). Vertical lines indicate the value of the thresholds we set to identify pulsar candidates (Predictor > 0.65) and AGN candidates (Predictor < 0.05).

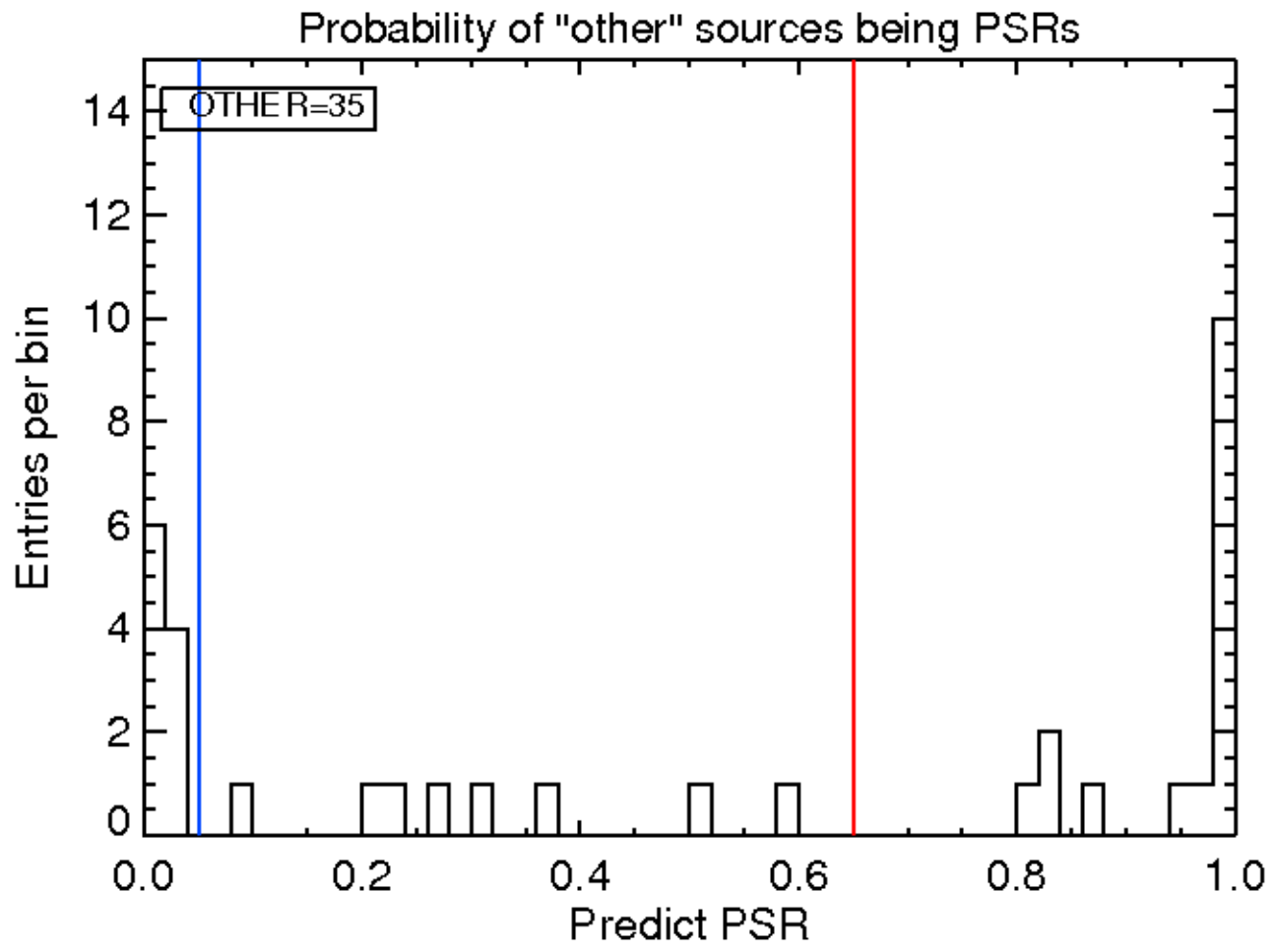

Figure 4.8: Distribution of the LR predictor for the "other" 2FGL sources defined in the text. Vertical lines indicate the value of the thresholds we set to identify pulsar candidates (Predictor > 0.65) and AGN candidates (Predictor < 0.05)

AGN candidates. Note that the contamination due to these sources does not change the accuracy and the misclassification of the LR model but contributes only to the contamination. The pulsar candidates are all the 6 SNRs firmly identified as $\gamma$-ray sources on the basis of their spatial extensions (IC 443, W28, W30, W44, W51C ad Cygnus Loop), 2 associated point-like SNRs (with angular diameters $< 20'$), 3 HMBs (LS I+61 303, 1FGL J1018.6–5856 and LS 5039) whose $\gamma$-ray observables seem very similar to those of a young pulsar [90] [19], and 5 globular clusters, whose $\gamma$-ray emission may be related to the contribution of a number of MSPs in the cluster as explained in Chapter 2. The AGN candidates are associated with the source Small Magellanic Cloud (SMC), 2 objects in the field of the LMC, the Andromeda galaxy M31, 3 PWNe, 1 HMB (Cygnus X-3, which is different from the other 3 $\gamma$-ray HMBs detected by the LAT because this is a microquasar, 2 globular clusters (M80 and NGC 6440) and the nova (V407 Cyg), whose $\gamma$-ray emission lasted only 2 weeks [13]. From this analysis it is clear that only Galactic pulsar-like objects are classified as a pulsar, otherwise only extragalactic or AGN-like objects are classified as an AGN. This means that our algorithm is able to distinguish pulsar-like from AGN-like objects and not only pulsars from AGNs. Moreover, it is reasonable to assume that those sources will not be overrepresented in the unassociated population compared with the associated one. We expect that up to 0.7% of the newly classified AGN candidates and up to 18% of the newly classified pulsar candidates will indeed belong to one of the "other" classes (galaxies, globular clusters, supernova remnants, etc.). The higher contamination rate for the pulsars is likely due to low statistics in the test samples and is primarily associated with a Galactic contamination. These results are shown in the table of classification in Figure 4.9.

| | "Other" observed | Total Efficiency/ Contamination |
|---|---|---|
| **PSR predicted** | **16 (45.5%)** | **79%**<br>**3% AGN**<br>**18% "other"** |
| **AGN predicted** | **11 (31.5%)** | **99%**<br>**0.3% PSR**<br>**0.7% "other"** |
| **Unclassified** | **8** | |

Figure 4.9: Table of classification based on the classification rules described in the text where the contamination given by the "other" 2FGL sources is included. In green are represented the correct classifications and in red the incorrect ones. Note that the inclusion of the "other" 2FGL sources does not change the accuracy and the misclassification of the LR model.

To estimate how accurately our predictive model performs, we cross-validate it using all the 1179 pulsars and AGNs in the 2FGL catalog. We randomly selected 118 sources containing both pulsars and AGNs to be the testing data set, and we used the remaining 1061 sources for training. We repeated this procedure 10 times, using different sets of 118 test sources. At each step the "other" sources are used to evaluate the total contamination. At the end, this 10-fold cross-validation shows that the average testing accuracy rates for these threshold values are 81% for pulsars and 95.5% for AGNs, that the average misclassification rates are 6% for pulsars and 1% for AGNs and that the average total contamination rates are 24% for pulsars (7.5% AGNs and 16.5% "other" sources) and 1.5% for AGNs (1% pulsars and 0.5% "other" sources). These results are consistent with those found including all pulsars and AGNs in the training sample.

### 4.2.2 Results and their validation

Applying the model to the 2FGL unidentified sources we find that 325 are classified as AGN candidates ($P < 0.05$), 108 are classified as pulsar candidates ($P > 0.65$) and 143 remain unclassified after the LR analysis. The distribution of 2FGL unidentified sources as a function of the probability of being pulsars is shown in the Figure 4.10. The thresholds for assigning pulsar candidates and AGN candidates are indicated in the figure. It is important to notice that in order to meet the threshold of $\sim$80% of the known pulsars, we are including a large range of predictor values with very few pulsars. This may result in over-predicting the number of pulsars in unidentified sources. As a result, if we take into account the estimated contamination and misclassification rates related to our classification rules we expect up to 100 $\gamma$-ray pulsars and up to 350 AGNs to be discovered in 2FGL catalog .

The spatial (top) and the latitude (bottom) distribution of the newly classified sources are shown in Figure 4.11. Their distributions give us the opportunity to cross check our results. Note that both the AGN and pulsar distributions are as expected, even though we have not used the Galactic latitude as an input to either classification method. The pulsar candidates are mainly distributed along the Galactic plane (85% are situated in $|b| < 10°$), with a few high-latitude exceptions that suggest additional nearby MSPs, while the AGN candidates are nearly isotropically distributed on the sky.

If we combine the AGN candidate population with the 2FGL sources that already have AGN associations (Figure 4.12, left panel), we find that the shape of the AGN distribution matches reasonably well with a nearly isotropical distribution that we expect, at low Galactic latitudes there is still a lack of AGNs primarily because of the bright Galactic diffuse $\gamma$-ray emission. On the contrary, if we combine the pulsar candidate population with the 2FGL sources that already have pulsar associations (Figure 4.12, right panel), we find that their distribution is characterized by an evident peak at low Galactic latitudes. Since our studies about the contamination in the pulsar candidate sample, we expect that a fraction of these objects may be associated with another Galactic source classes, e.g. SNRs or HMBs, but probably a certain fraction of these sources can be spurious.

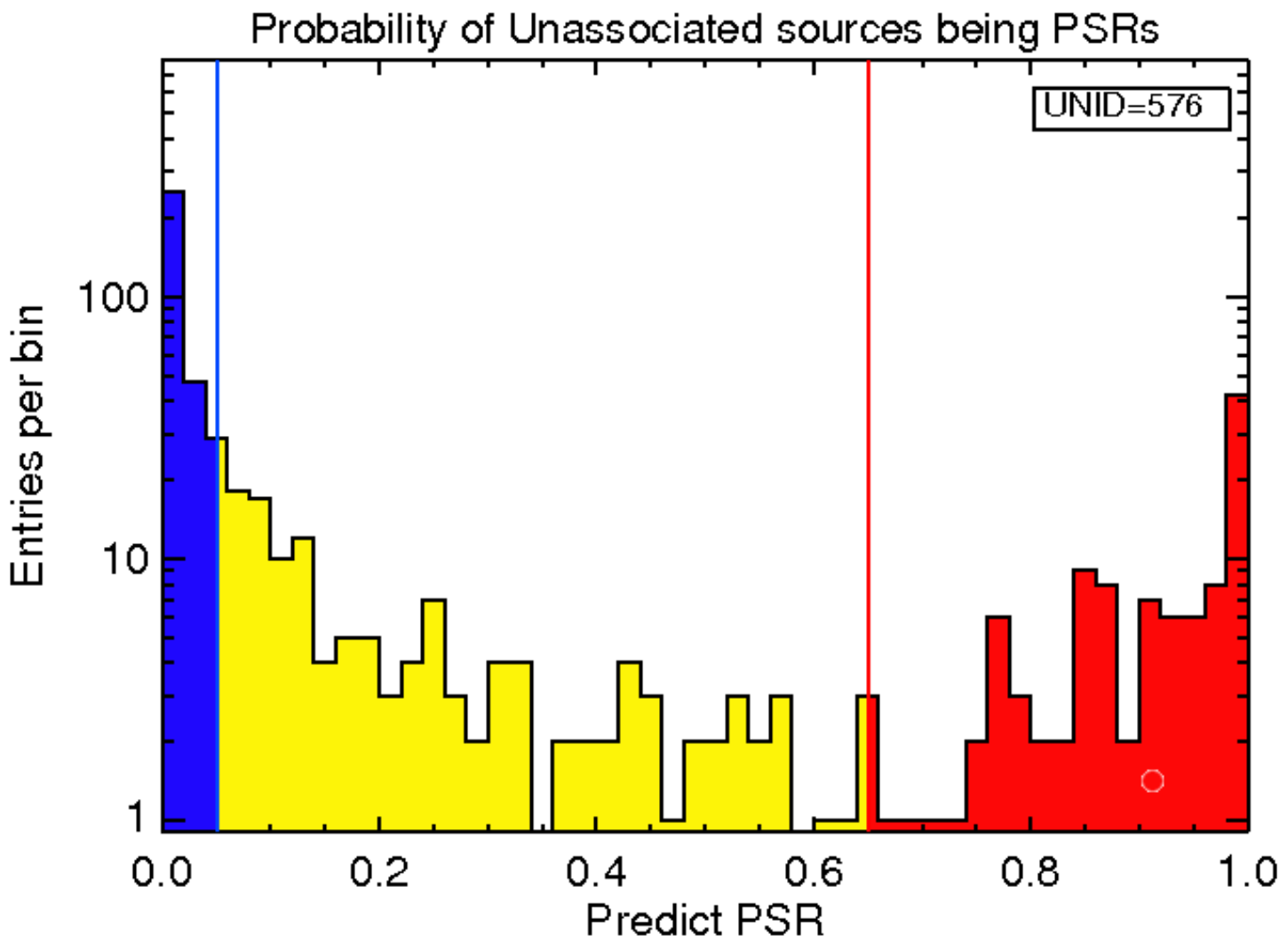

Figure 4.10: Distribution of the Logistic Regression predictor for 2FGL unidentified sources. Vertical lines indicate the value of the thresholds we set to identify pulsar candidates (Predictor > 0.65) and AGN candidates (Predictor < 0.05).

In Figure 4.13 is shown curvature-variability distribution of the newly classified AGN and pulsar candidates on the basis of the LR method. We see that the unidentified sources have been separated into two populations with some overlap between them. Comparing this distribution to Figure 4.4, we see it follows the separation between the associated AGNs and pulsars.

From these examples we can conclude that using only the $\gamma$-ray properties of the *Fermi*-LAT sources, and the firm associations of the 2FGL, we are able to develop a predictive model for AGN and pulsar classification that nearly matches our expectations (i.e., pulsar candidates are not variable, have a curved spectrum, and are mainly distributed along the Galactic plane, while AGN candidates are mostly extragalactic, variable sources).

In order to test the capability of the LR algorithm for identifying the different source classes, we must apply the LR analysis to an independent set of associations from those used to train the LR model. For this, we look to association efforts that have taken place since the release of the 2FGL cata-log. Regarding the 2FGL unidentified sources, the follow-up multi-wavelength association efforts and specific analyses of the LAT data discussed in Section 3.2 have resulted in 46 new Galactic source associations (18 young pulsars and 28 MSPs), and 47 new AGN associations (2 were unidentified in 2FGL, 1 was a "potential" SNR and 41 were active galaxies of uncertain type). We cannot consider the 41 objects classified as active galaxies of uncertain type because they were a component of the training sample. Of the 3 newly associated AGNs, 2 are correctly classified as AGN candidates by the LR analysis, no one is classified as pulsar candidates, while the other sources remain unclassified.

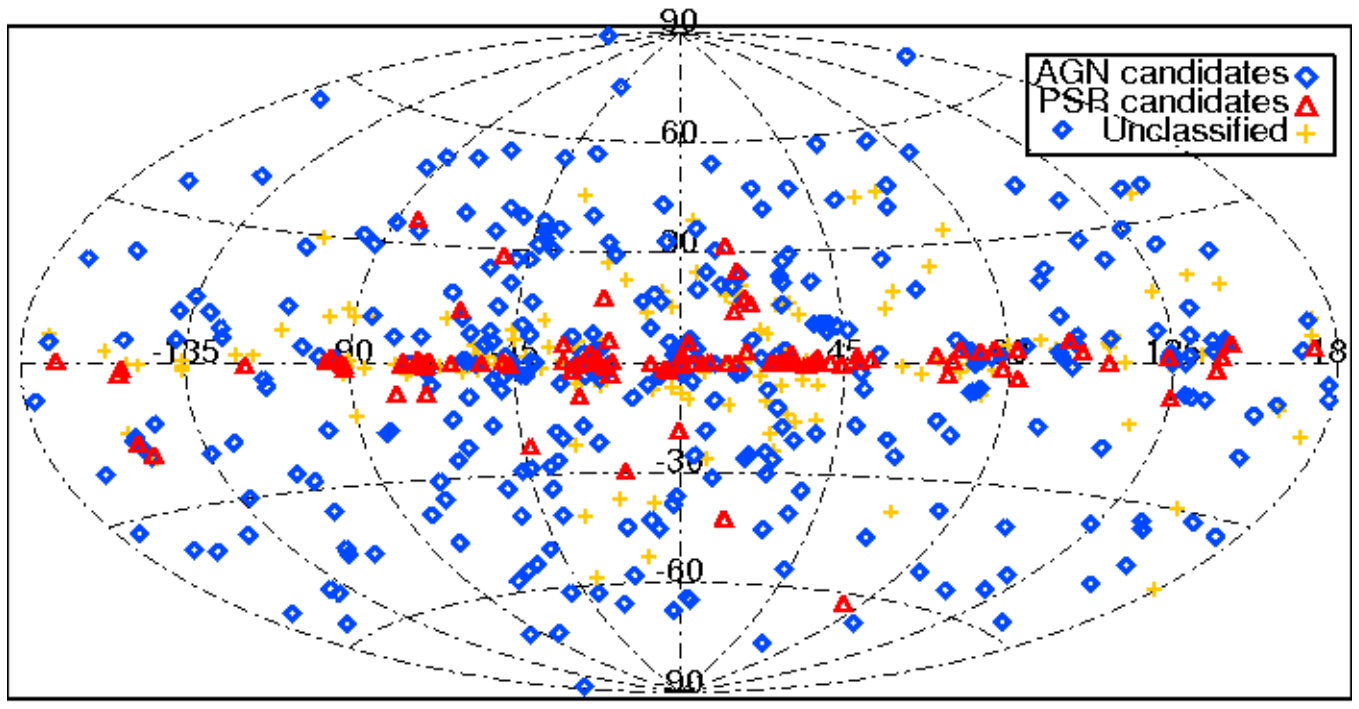

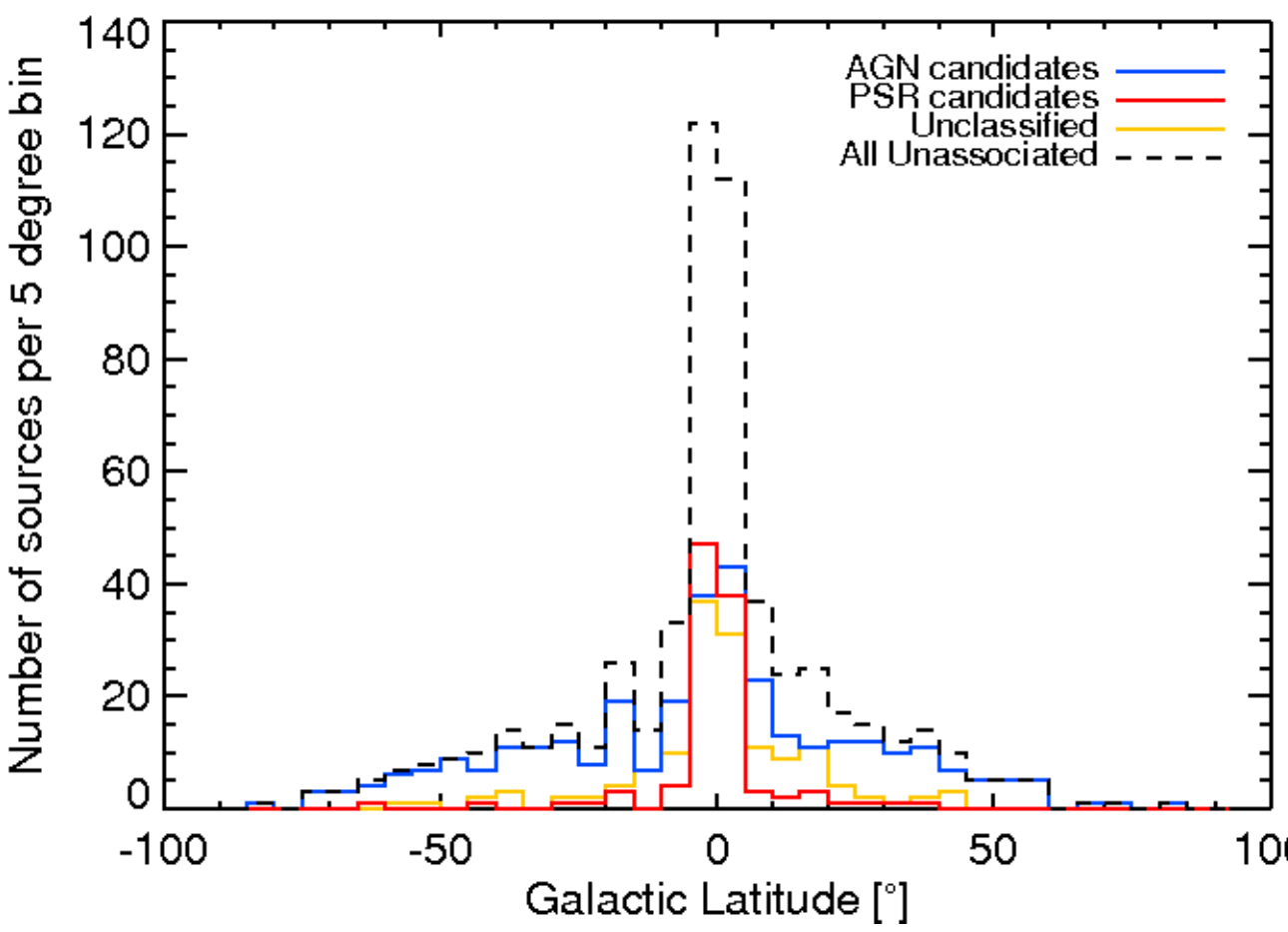

Figure 4.11: Distribution of the newly classified sources. Top: spatial distribution in galactic coordinates. Bottom: latitude distribution. In red are shown the pulsar candidates, in blue the AGN candidates and in yellow the unclassified sources.

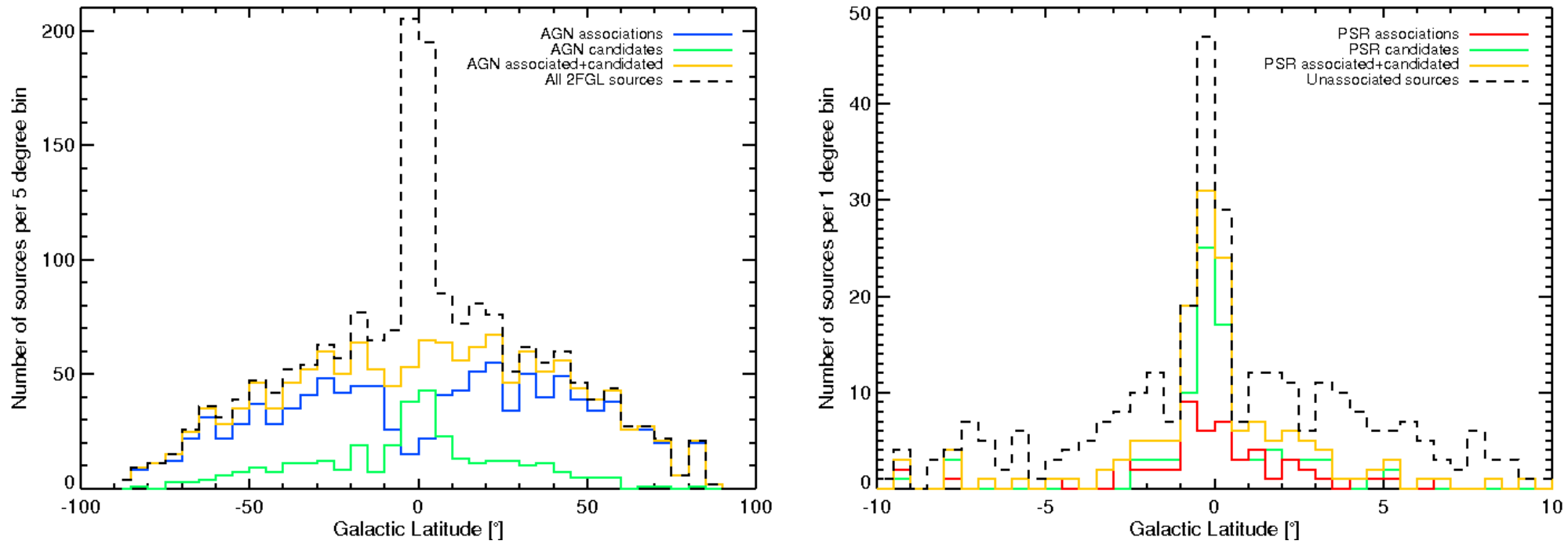

Figure 4.12: Left: distribution of AGN candidates binned by Galactic latitude. The yellow line is the sum of the 2FGL AGN associations (blue line) plus the sources classified as AGN candidates (green line). The dashed line is the distribution for all 2FGL sources. Right: distribution of pulsar candidates binned by Galactic latitude. The yellow line is the sum of the 2FGL pulsar associations (red line) plus the sources classified as pulsar candidates (green line). The dashed line is the distribution for all 2FGL sources. Note that the latitude bins of the two figures are different, AGN distribution covers the entire sky while the pulsar distribution is just ±10° around the plane of our Galaxy.

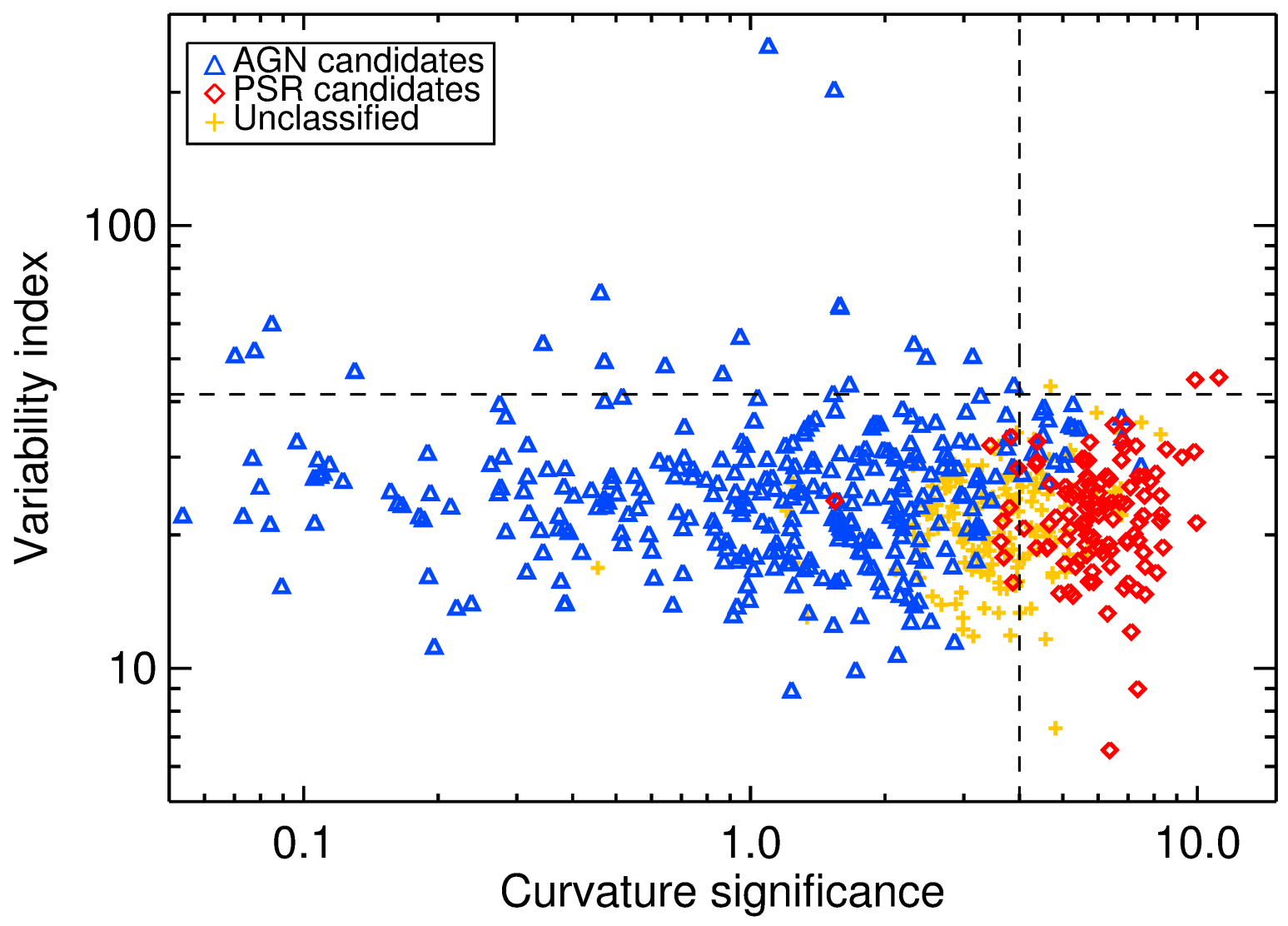

Figure 4.13: Variability index vs. curvature significance for 2FGL unidentified sources classified as AGN (blue triangles) and pulsar candidates (red diamonds) and unclassified (yellow crosses).

For pulsars, 21 are correctly classified as pulsar candidates by the LR analysis (efficiency: 46%), only 6 are classified as AGN candidates (false negative: 13%), while the other 19 sources remain unclassified (41%).
We cannot assess the efficiency of the algorithm at classifying new AGNs owing to the smallness of the sample. We notice that source significance of all the unclassified new AGNs is very low (TS $<$ 100), this means that their $\gamma$-ray observables cannot be well characterized hampering the source classification. For new pulsars, we noticed a different performance between new young pulsars and new MSPs. For new objects identified as young pulsars, we correctly classify 14 pulsars (efficiency: 79%), we misclassify only 1 source (false negative: 5%) and we leave the remaining sources as unclassified (efficiency: 16%). The misclassified new young pulsar is 2FGL J1112–6103, for which an extended emission in the off-peak phase was detected, possibly pointing to an associated PWN [200]. On the other hand, for new sources identified as MSPs, the classification rate is much worse. We correctly classify only 7 objects as pulsars (efficiency: 25%), we misclassify 5 objects as AGN (false negative: 18%) and left the 16 remaining objects as still unassociated (57% of the new MSPs). These misclassifications are mainly due to low source significance (TS $<$ 100), so that only upper limits are available in many energy bands making a spectral characterization very difficult.

In conclusion, the efficiency of the algorithm at classifying new relatively bright young pulsars is very high (almost 80%), which is consistent with the efficiency we obtained for the pulsars in the training sample. Also, the low rate of false negative is a very encouraging result, that makes us confident in using the results from the LR analysis while planning multi-wavelength observation of unidentified sources. We cannot conclude anything about the efficiency of the algorithm at classifying new AGNs or new MSPs because these results are affected by low statistics and bad characterization related to their low source significance.

## 4.3 Classification using Artificial Neural Networks (ANNs)

In this section we describe the use of Artificial Neural Networks (ANNs) as a very promising method for understanding the nature of *Fermi*-LAT unidentified sources, this is the first time this machine learning technique is used for this purpose. This technique uses identified objects as a training sample, learning to distinguish each source class on the basis of parameters that describe its $\gamma$-ray properties. By applying the algorithm to unknown objects, such as the unidentified sources, it is possible to quantify their probability of belonging to a specific class. There are different packages available to perform an ANN analysis (e.g. *MATLAB Neural Network Toolbox*[1] or *PyBrain*[2]) but we decided to develop our own algorithms to address our specific problem because, although ANNs exist in many different models and architectures, the relatively low complexity of astronomical data does not pose special constrains in all the steps of the method which will be discussed below. We used a very simple neural model known as MultiLayer Perceptron (MLP; see Appendix B.2), which is probably the most widely used architecture for practical applications of neural networks. We wrote our algorithms in *Python* programming language[3], our choice gives us a number of advantages, first of all our ANN does not work as a "*black box*", this is a typical problem of any available ANN package for which the learning process is always unknown. Since we have implemented our algorithms, we can examine step by step how our network is learning to distinguish 2FGL source classes. Another important advantage is that our ANN is easy-to-use because we have implemented the algorithms and we have a clear idea on how they operate. Moreover, our package is very flexible and, if we need a new specific analysis, we can easily implement it. Before using our algorithms to understand the nature of 2FGL unidentified sources we carefully tested them in very simple situations.

Neural networks were originally conceived as simple models for the behavior of the brain, but have since found many real world applications in fields as diverse as medicine, linguistics, or high-energy physics. Multilayer perceptrons [127] are well suited to many situations where one is searching for a functional relationship between a set of input variables and response variables. Artificial neural networks can be considered as an extension of generalized linear models (e.g. logistic regression), and are applied to approximate complicated functional relationships. At variance with generalized linear models, it is not necessary to specify a priori the type of relationship between the input and output variables. This often leads to superior results, compared to simple logistic regression analyses [206] (see Figure 4.14). The use of ANNs in astronomy goes back over 20 years (e.g. Odewahn et al. 1992 [160] applied it to the problem of star/galaxy discrimination or [24], [210], [23], [45], [75]). In $\gamma$-ray astronomy, ANNs are often used for such applications as background rejection, though other techniques (e.g. classification trees) are also used for such purposes. In [16] and in the previous section we have explored the application of logistic regression to source classification, based on some $\gamma$-ray observables, and shown that it is very efficient at sorting sources into pulsar-like and AGN-like classes showing that there is much to be gained in developing an automated system of sorting (and ranking) sources according to their probability of being a pulsar or an AGN. Our goal is to extend our studies of source classification using ANNs, aiming at improving on the results already obtained.

[1] `http://www.mathworks.com/products/neural-network/`
[2] `http://pybrain.org`
[3] `http://www.python.org`

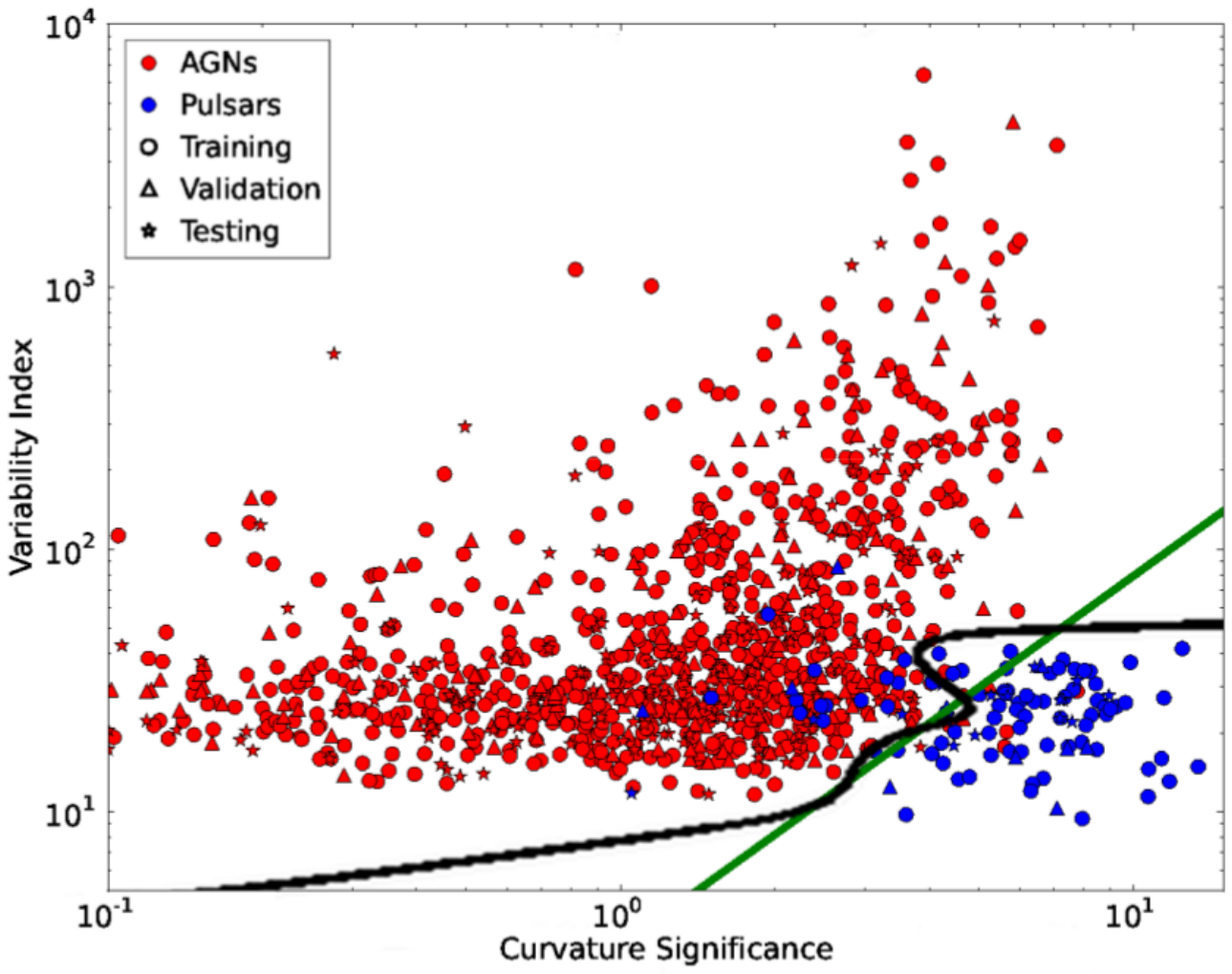

Figure 4.14: Simple classification test comparing logistic regression (LR) and a multilayer perceptron (MLP) with only 2 input parameters. The green and black lines represent the thresholds to single out AGNs and pulsars using the LR and the MLP respectively. The full sample containing all AGNs and pulsars has been split in "Training", "Validation" and "Testing" subsamples as described in Appendix B.2 in order to train and optimize the ANN. LR has been trained using all sources in the Training and Validation subsamples. Applying the optimized models to the Testing subsample we estimate an higher efficiency of MLP in classifying new sources.

### 4.3.1 Selection of the training sample and the predictor variables

The first step of the ANN analysis is to select a sample of data to build the predictor variable. We decide to focus on the 83 firmly identified pulsars and the 834 associated AGNs. This choice derives from what we have learned through the Logistic Regression analysis discussed in the previous section. We do not include in the training sample the 25 "potential" pulsars for the same reason explained for the LR analysis. Moreover, we decide not to include the active galaxies of uncertain type for which, as we have seen in the previous section, the association is not certain. Indeed, some of them may be spurious while others may be associated with pulsars or other Galactic objects. We intend to apply the optimized ANN to these uncertain associations in order to classify them. We exclude from the training procedure also the sources associated with a class different than AGNs or pulsars because of the smallness of their sample. However, we use them to estimate the contamination to the candidate AGN and pulsar samples due to the likely presence of these "other" source classes in the unassociated sample.

The second important step is to select the most appropriate set of variables for training the ANN. This choice must follow some criteria, the selected variables must not depend too much on flux and significance of the source otherwise the ANN would tend to classify the 2FGL objects in the training sample on the basis of these parameters and not on the basis of their intrinsic features losing efficiency in the classification. Moreover, physical considerations about the $\gamma$-ray properties of each source class can guide us in the choice of the most effective variables for discriminating AGNs from pulsars. After exploring most of the parameters in the 2FGL catalog, we select a set of variables that includes the curvature significance, the variability index, the PowerLaw index, the fluxes and the hardness ratios for the 5 energy bands in the 2FGL catalog. We decide not to use the Galactic latitude and longitude as input to the ANN in order to avoid biasing our selection against AGNs situated close to the Galactic plane and pulsars (especially MSPs) situated at high Galactic latitudes. Furthermore, this choice gives us the opportunity to use the position on the sky of the different populations as a cross check of our result. The new pulsar candidates should be mainly distributed along the Galactic plane, while the AGN candidates should be isotropically distributed.

### 4.3.2 Architecture of the ANN

The basic building block of an ANN is the *neuron*. Information is passed as inputs to the neuron, which processes them and produces an output. The output is typically a simple mathematical function of the inputs. The power of the ANN comes from assembling many neurons into a network. The network is able to model very complex behavior from input to output. Since the relatively low complexity of our data, we decide to use a very simple neural model known as *Feed Forward MultiLayer Perceptron* (MLP; see Appendix B.2) and in particular a two-layer feed-forward network (2LP). It consists of a layer of input neurons, a layer of "hidden" neurons and a layer of output neurons. In such an arrangement each neuron is referred to as a *node*. Figure B.2 shows a schematic design of such a network. We use this simple architecture because the Weierstrass' theorem ascertains that for a multilayer perceptron a single hidden layer is sufficient to approximate a given continuous correlation function to any precision, provided that a sufficiently large numbers of nodes is used in the hidden layer. An important corollary of this result is that, in the context of a classification problem, networks with two layers can approximate any decision boundary to arbitrary accuracy. Thus, such networks also provide universal nonlinear discriminant functions. More generally, the capability of such networks of approximating general smooth functions allows them to model a posteriori probabilities of class membership.

The output of the hidden layer and the output layer are related to their inputs as follows:

$$\text{hidden layer:} \qquad h_j = g^{(1)}\left(f_j^{(1)}\right); \qquad f_j^{(1)} = \sum_i w_{ji}^{(1)} x_i + b_j^{(1)}, \tag{4.1}$$

$$\text{output layer:} \qquad y_k = g^{(2)}\left(f_k^{(2)}\right); \qquad f_k^{(2)} = \sum_j w_{kj}^{(2)} h_j + b_k^{(2)}, \tag{4.2}$$

where the output of the hidden layer $h$ and output layer $y$ are given for each hidden node $j$ and each output node $k$. The index $i$ runs over all input nodes. The functions $g^{(1)}$ and $g^{(2)}$ are called activation functions. The non-linear nature of $g^{(1)}$ and $g^{(2)}$ is a key ingredient in constructing a viable and practically useful network. This non-linear function must be bounded, smooth and monotonic; we use $g^{(1)}(x) = g^{(2)}(x) = tanh(x)$ where $tanh(x)$ is the hyperbolic tangent. The layout and number of nodes are collectively termed the architecture of the network (see Appendix B.2). For a given architecture, the weights $w$ and biases $b$ define the operation of the network and are the quantities we will determine by a specific training algorithm explained in Section 4.3.3. Following the usual approach for starting the learning algorithm we initialize the weights randomly in the range $[-1, 1]$ and moreover we decide to not include biases in our network ($\mathbf{b} = 0$). As weights vary during training, a very wide range of non-linear mappings between inputs and outputs is possible.

In our application, we will construct a classification network. The aim of any classification method is to place members of a set into groups based on inherent properties or features of the individuals, given some pre-classified training data. Formally, classification can be summarized as finding a classifier $\mathcal{C} : x \to C$ which maps an object from some (typically multi-dimensional) feature space $x$ to its classification label $C$, which is typically taken as one of $1, ..., N$ where $N$ is the number of distinct classes, in our case AGNs and pulsars. Thus the problem of classification is to partition feature space into regions, assigning each region a label corresponding to the appropriate classification.

In building a classifier using a neural network, it is convenient to view the problem *probabilistically.* To this end we consider a 2LP consisting of an input layer ($x_i$), a hidden layer ($h_j$), and an output layer ($y_i$). In classification networks, however, the outputs must be post-processed in order to have a probability that tells us the input feature vector $x$ belongs to the $k$th class. We decide to transform the outputs with the following procedure:

$$p_k = \frac{y_m + 1}{\sum_k (y_k + 1)} \tag{4.3}$$

such that they are all non-negative and sum to unity. In this way the network produces an output vector for each 2FGL sources. The $k$th component of this vector can be viewed as the probability for $k$ given the the input parameters $P(C_k|\mathbf{x})$. In fact, it can be proved theoretically ([86], [176]) that the output of an ANN is indeed a Bayesian a posteriori probability. The advantage of this *probabilistic* approach is that we gain the ability to make statistical decisions on the appropriate classification in very large feature spaces where a direct linear partition would not be feasible.

In order to define the architecture of our network we must determine the number of nodes in each layer. The number of input and output nodes is strictly related to the problem we are analyzing. The number of input nodes is given by the number of selected predictor variables, in our case 12. During the ANN analysis a pre-processing is performed normalizing each variable to the range [0, 1]. All the variables are normalized logarithmically except for the hardness ratios, the curvature significance and the PowerLaw index which are linearly normalized. The number of output nodes is given by the number of source classes we want to classify: in our case 2 (pulsars and AGNs). To the $l$th source in the training sample are associated two vectors ($\mathbf{x}^{(l)}$, $\mathbf{t}^{(l)}$) in which the target vector $\mathbf{t}^{(l)}$ for the network outputs has unity in the element corresponding to the true class of the $l$th feature vector $\mathbf{x}^{(l)}$ and minus ones elsewhere. Thus for pulsars $\mathbf{t} = [1, -1]$ while for AGNs $\mathbf{t} = [-1, 1]$. Determining the number of hidden nodes is crucial, especially for its practical implications in learning and generalization. Theory and experience ([71], [125], [52]) show that networks with few hidden nodes exhibit a better generalization performance. Moreover, knowledge embedded in small trained networks is presumably easier to interpret and thus the extraction of simple rules can hopefully be facilitated. Lastly, from an implementation stand point, small networks only require limited resources in any physical computational environment. To solve the problem of choosing the right number of hidden nodes we pursue an incremental approach known as *pruning* [175]. Pruning starts with a relatively large number of hidden nodes and excise those unnecessary so as to arrive at an optimal network architecture. We use the following procedure, first we train the network with a large number of hidden nodes, we start with 20, using a standard training algorithms (see Section 4.3.3 for details), then we examine the value of the mean squared error (mse, defined as in the Equation 4.4) and we remove one of the nodes in the hidden layer. At this point we retrain the pruned network and we re-examine the value of the mse, then we remove another hidden nodes and we continue until an additional pruning process results in increasing significantly the mse decreasing the performances of our network. Using these procedure combined to a detailed analysis we decide to use 12 hidden nodes. In Figure 4.15 in blue is shown the value of the mean squared error defined in Equation 4.4 as a function of the number of hidden nodes while in red the performance of our network in terms of the fraction of correct classifications of the validation sample (defined below) as a function of the number of hidden nodes.

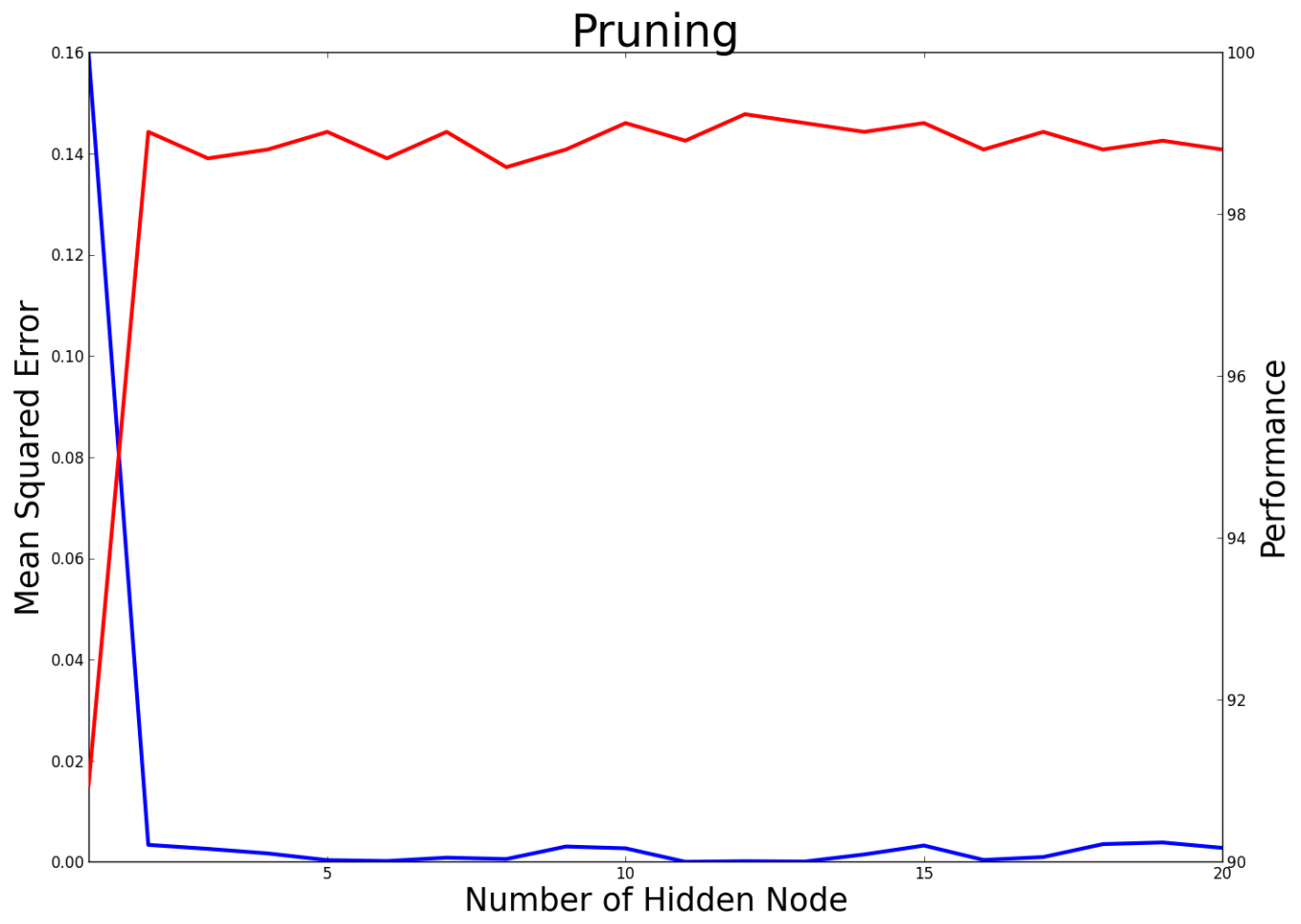

Figure 4.15: (blue) Value of the mean squared error (mse) defined in Equation 4.4 as a function of the number of hidden nodes. (red) Performance of the network in terms of fraction of correct classifications of the validation sample (defined in the text) as a function of the number of hidden nodes.

As a result our feed-forward 2LP is built up of 12 input nodes, 12 hidden nodes and 2 output nodes, each node in a layer is linked to all the nodes of the next layer and to each link is associated a weight randomly initialized in the range $[-1, 1]$.

### 4.3.3 Training session

For a given network architecture the first step is the "training" of the ANN. In this step the weights $\mathbf{w}$ (the "free parameters") are determined by a specific learning algorithm. The basic learning algorithm for 2LP is the so-called back-propagation (see Appendix B.2.3) which is based on the error-correction learning rule. In essence, back-propagation consists of two passes through the different layers of the network: a forward pass and a backward pass. In the forward pass an input vector is applied to the input nodes of the network, and its effect propagates through the network layer by layer. Finally, a set of outputs is produced as the actual response of the network. During the backward pass, on the other hand, the weights are all adjusted in accordance with the error correction rule. Specifically, the actual response of the network is subtracted from a desired (target) response which we denote as a vector $\mathbf{t} = (t_1, t_2, ..., t_c)$ to produce an error signal. This error signal is then propagated backward through the network. There are several choices for the form of the error signal and this choice still depends on the nature of the problem, we choose the sum-squared error ($E$) defined in the Equation B.14. Moreover, we define the mean square error (mse) as the sum of the errors given by each object in the training sample:

$$mse = \frac{1}{N} \sum_n E^{(n)} \tag{4.4}$$

where $E^{(n)}$ is the error related to the $n$th source in the training sample and $N$ is the number of sources in the training sample. The weights are adjusted to make the actual response of the network move closer to the desired response in a statistical sense. The set of weights that minimizes the error function are found using the method of gradient descendent. Starting from a random set of weights $\mathbf{w}$ the weights are updated by moving a small distance in $w$-space into the direction $-\nabla_{\mathbf{w}} E$ where $E$ decreases most rapidly:

$$\mathbf{w}_{new} = \mathbf{w}_{old} - \eta \frac{\partial E}{\partial \mathbf{w}} \tag{4.5}$$

where the positive number $\eta$ is the learning rate and we set it to 0.2. The algorithm is stopped when the value of the error function has become sufficiently small. We analyzed also the possibility to use the "heavy ball method" defined in the Equation B.17 adding a momentum $\alpha = 0.9$ in the Equation 4.5 but without a significant improvement in the performance of the network thus we decided to not use this learning algorithm.

We use the learning algorithm in the *online* version, in which the weights of the connections are updated after each example are processed by the network. One epoch corresponds to the processing of all examples one time and the value of the mse can be obtained.

In order to avoid under- and overfitting, and obtain a good generalization of our network we split the training sample in 3 subsamples, the training sample, the validation sample and the testing sample as explained in Appendix B.2. The first one is used to optimize the weights and classify correctly the 2FGL sources; the second one is used to avoid over-fitting during the training, this is not used for updating the weights but during the training it monitors the generalization error. The best epoch corresponds to the lowest validation error, and the training is stopped when the validation error rate "starts to go up". The last one is independent both of the training and validation subsample and is used to monitor the accuracy of the 2LP, after each training the network is applied to this subsample and the error on the testing subsample provides an unbiased estimate of the generalization error. We choose a training sample as large as possible ($\sim 70\%$ of the full sample) while keeping the other independent samples homogeneous ($\sim 15\%$ each one). Since we use an "online" version of the learning algorithm, we decide to shuffle the training sample after each epoch, this choice allows us to maintain a good generalization of our network.

Moreover, in order to rank the relative importance of the different variables at distinguishing AGNs from pulsars we implemented a variable ranking that uses the sum of the weights-squared of the connection between the variable's nodes in the input layer and the hidden layer. The importance $I_i$ of the input variable $i$ is given by:

$$I_i = \bar{x}_i^2 \sum_{j=1}^{N_h} \left( w_{ij}^{(1)} \right)^2, \qquad i = 1, ..., N_{var} \tag{4.6}$$

where $\bar{x}_i$ is the sample mean of input variable $i$.

Table 2.3 ranks the relative importance of the different variables at distinguishing AGNs from pulsars according to the Equation 4.6. As we expect variability index is the most important variable, here $HR_{12}$ seems very important, a result very different from what we obtained from the LR analysis. The others hardness ratios as well as the curvature significance and the PowerLaw index are relatively important, indicating that the spectral shapes associated to pulsars and AGNs are rather different, for the first ones the spectrum is described by a power law with exponential cutoff, the others by a broken power law. The fluxes in each energy bands are the least important parameters at distinguishing AGNs from pulsars.

In Figure 4.16 the correlation matrix of the variables selected for our analysis is shown. The $i, j$ element of the matrix is the correlation coefficient $\rho_{ij}$ defined as:

$$\rho_{ij} = \frac{C_{ij}}{\sqrt{C_{ii} \times C_{jj}}} \tag{4.7}$$

where $C_{ij}$ is the covariance of the variables $x_i$ and $x_j$ while $C_{ii}$ is the variance of the variable $x_i$. The values of $\rho$ are between –1 and 1, inclusive, and in Figure 4.16 to each correlation coefficient is associated a color on the basis of its value.

The performance of our neural network is shown in Figure 4.17. At the top panel the mean squared error for training, validation and testing samples during the learning process is shown as a function of the epoch. We have stopped the learning process and consider that the ANN is optimized when the mean squared error of the validation sample "starts to go up" and in the figure is specified by a dotted vertical line. Our ANN algorithm is not very fast because the neural network is optimized after ~150 epochs but it works well because it is able to reach the minimum of the error function for the validation sample. After the epoch we have stopped the learning process, the error function of the training sample continues to decrease as we expect, while the error function of the testing sample increases similarly to the error of the validation sample. The sawtooth trend of the error function is related to the learning algorithm we use, that is an "online" version of the learning algorithm where after each epoch the training sample is shuffled, this causes a statistical fluctuation of the mean squared error during the learning process. In the middle panel is shown the classification table, pulsars are indicated as class 1 while AGN as class 2. Since the output of our neural network is a vector with two components, the first one defines the probability that the object is a pulsar and the other one the probability that it is an AGN, at this step an object is classified as a pulsar if the value of the first component of the output vector is greater of those of the second one. This table tells us that our trained neural networks works very well at distinguish between pulsars and AGNs because the accuracy and the precision are very high, the total performance of our ANN is given by the total accuracy in the block blue and it is $> 99\%$. In this case we are considering that all the 2FGL sources are pulsars and AGNs but there are other source classes, for this reason we must define some classification thresholds in order to classify pulsar and AGN candidates more efficiently (see Section 4.3.4). On the bottom panel the distribution of the ANN predictor for the 2FGL pulsars (Class 1) and AGNs (Class 2) is shown, this corresponds to the value of the first element of the post-processed output vector shown in the Equation 4.3. Such predictor (P) describes the probability that a 2FGL source is a pulsar, the probability that the same object is an AGN is $A = 1 - P$ because $\sum_k p_k = 1$ ($k$ is the source class) since we are considering only two source classes, thus in this case we are modelling the behavior of pulsars as "opposite" of the behavior of AGNs. Also this distribution tells us the optimized neural network works very well because almost all the pulsars are characterized by an high predictor value, while almost all AGNs by a low predictor value.

| Variable | Importance |
|---|---|
| (0) Curvature Significance | 2.89 (3.81%) |
| (1) Variability Index | 26.05 (34.29%) |
| (2) PowerLaw Index | 4.88 (6.43%) |
| (3) $\text{Flux}_{0.1-0.3GeV}$ | 2.53 (3.34%) |
| (4) $\text{Flux}_{0.3-1GeV}$ | 1.34 (1.77%) |
| (5) $\text{Flux}_{1-3GeV}$ | 1.62 (2.14%) |
| (6) $\text{Flux}_{3-10GeV}$ | 1.04 (1.37%) |
| (7) $\text{Flux}_{10-100GeV}$ | 1.50 (1.98%) |
| (8) $\text{Hardness}_{12}$ | 17.38 (22.89%) |
| (9) $\text{Hardness}_{23}$ | 3.44 (4.54%) |
| (10) $\text{Hardness}_{34}$ | 6.75 (8.88%) |
| (11) $\text{Hardness}_{45}$ | 6.51 (8.57%) |

Table 4.3: List of the training variables for the ANN: each variable is ranked according to its relevance in the discrimination process (see Equation 4.6), as computed by the ANN algorithm.

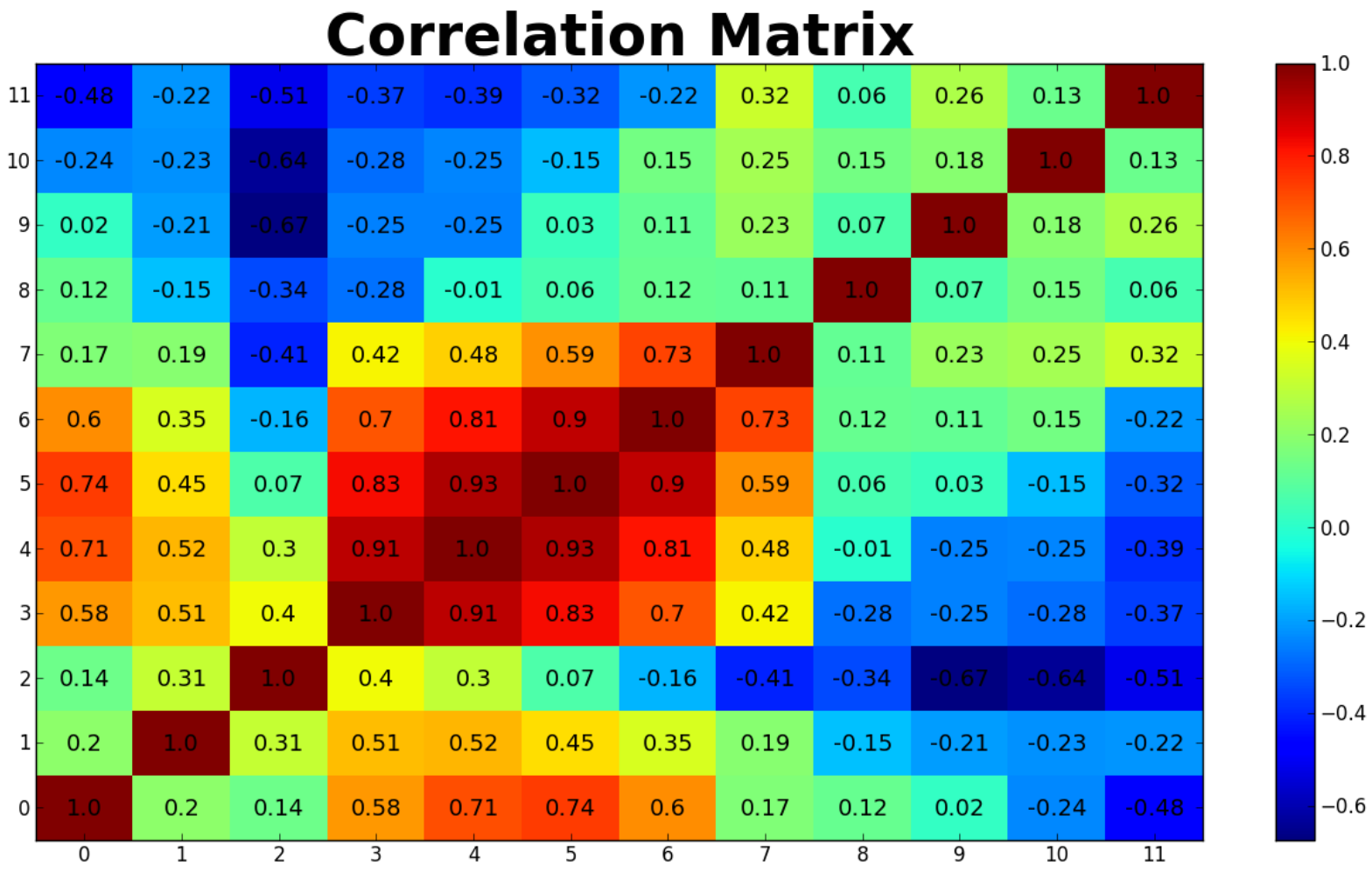

Figure 4.16: Correlation matrix of the variables selected for our analysis. The number of each parameter is expressed in Table 4.3 and to each correlation coefficient is associated a color on the basis of its value.

### 4.3.4 Defining thresholds

The distribution of the predictor value of the associated sources (at the bottom of the Figure 4.17) clearly shows that we can select a set of AGN and pulsar candidates with high confidence, when choosing the appropriate fiducial regions. We define two threshold values, one to classify a pulsar candidate ($C_P$) and one to classify an AGN candidate ($C_A$) so that all the sources with a predictor greater than 0.9962 are classified as pulsar candidates while all the sources with a predictor smaller than $3.4 \times 10^{-7}$ are classified as AGN candida-tes. All the sources with an intermediate value of the predictor remain unclassified after the ANN analysis. The choice of these boundaries is optimized for an accuracy of 80% for the two source classes in order to keep the misclassification fraction under 2%. Here, 80% of pulsar associations in 2FGL have a predictor greater than $C_P$ and 80% of AGNs have a predictor smaller than $C_A$. We use these thresholds in order to maintain the same level of accuracy for AGNs and pulsars. The results are shown in the table of classification in Figure 4.18. The only pulsar misclassified is PSR J1823–3021A, this is an energetic MSP located in the core of the cluster NGC 6624, 8.4 kpc away. This object was one of the pulsars misclassified after the application of the LR algorithm and this may be related to the peculiarities of this source. Otherwise the only AGN misclassified (2FGL J2206.6+6500) is associated with a FSRQ blazar. Also this object was misclassified by the LR algorithm and the reason may be related to an incorrect association.

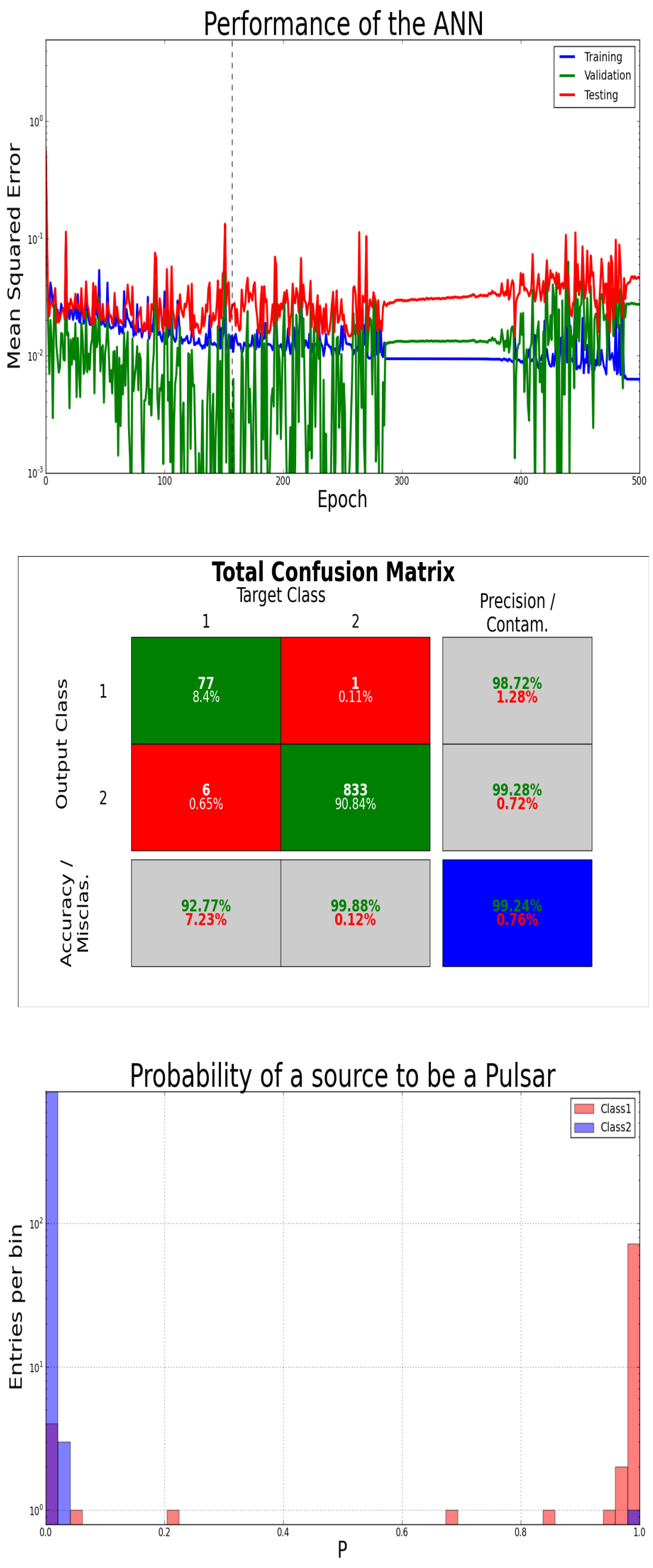

Figure 4.17: Performance of our ANN. Top: mean squared error for training, validation and testing samples during the learning process as a function of the epoch. The vertical dotted line specifies the epoch the learning process is stopped (optimized network). Middle: classification table. Bottom: distribution of the ANN predictor for the 2FGL pulsars (indicated as "Class 1") and AGNs (indicated as "Class 2"). Violet indicates pulsars which are put on top of AGNs.

| | PSR observed | AGN observed | Efficiency/ Contam. |
|---|---|---|---|
| PSR predicted | 67 | 1 | 98.7%<br>1.3% |
| AGN predicted | 1 | 669 | 99.9%<br>0.1% |
| Unclassified | 15 | 164 | |
| Accuracy/ Misclass. | 80%<br>1.2% | 80%<br>0.1% | 80%<br>0.2% |

Figure 4.18: Table of classification based on the classification rules described in the text. In green are represented the correct classifications and in red the incorrect ones.

As for the LR analysis, the sources associated with a different class than AGN or pulsar have been excluded from this training procedure. They are composed by Galactic objects, such as SNRs, HMBs, globular clusters and a nova, and extragalactic objects, such as normal galaxies, SMC and LMC. We do not include in this sample the 6 SNRs identified on the basis of their spatial extensions and the 3 PWNe because they are extended sources and different specific analyses were applied to their in the 2FGL catalog [158]. We decide to use the "other" sources to estimate the contamination to the candidate AGN and pulsar samples from the likely presence of these objects in the unassociated sample. In Figure 4.19 is shown how we can use the predictor distribution for the 28 "other" sources to estimate the further contamination from these sources to the AGN and pulsar candidate distributions.

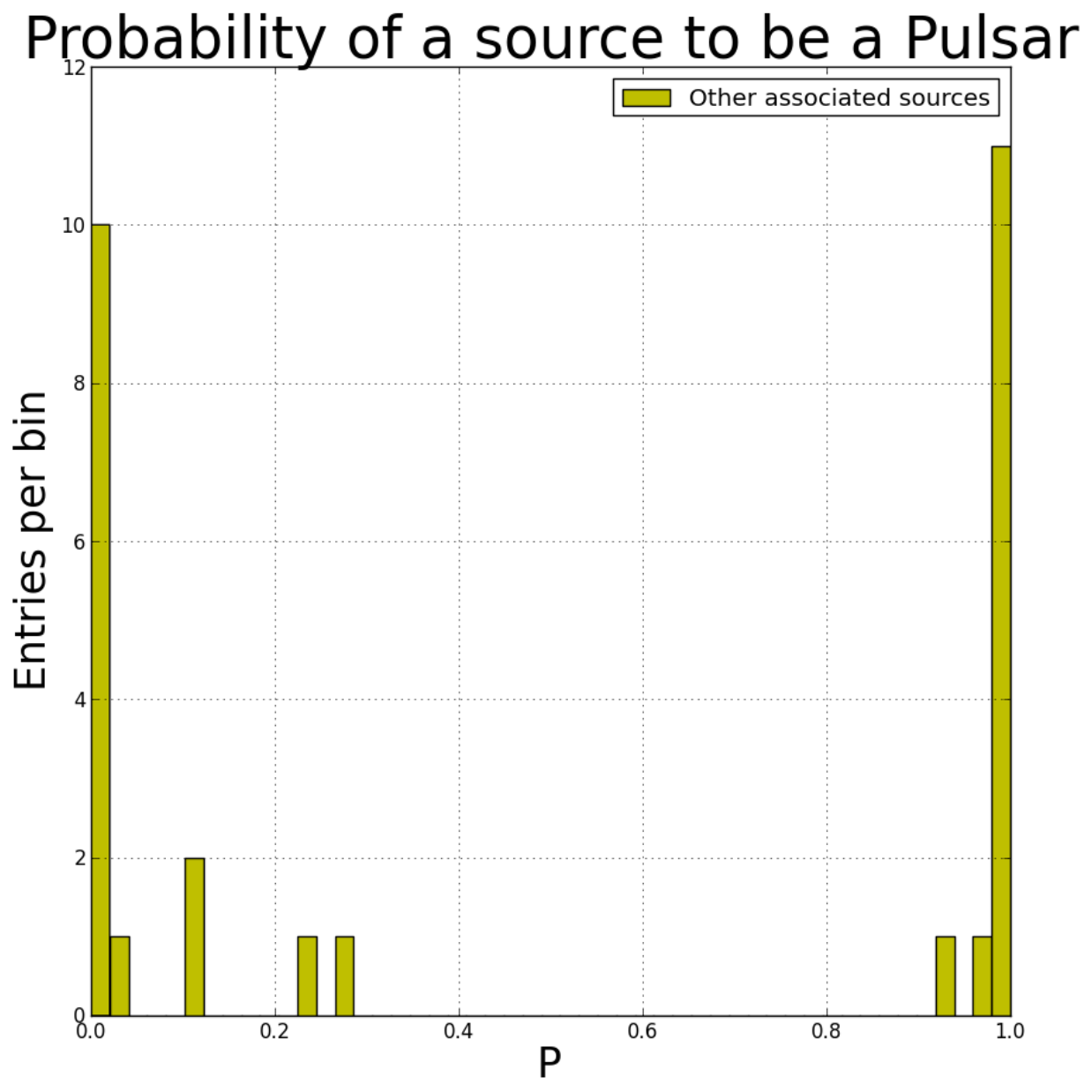

Figure 4.19: Distribution of the predictor value for the "other" 28 2FGL sources defined in the text.

According to the ANN analysis, those 28 sources are nearly equally distributed between AGN-like objects, pulsar-like objects, and still unclassified objects. In particular, 32% (9) are classify as a pulsar candidate, while 25% (7) as an AGN candidate. Notice that the contamination due to these sources does not change the accuracy and the misclassification of the ANN model but only the contamination. The pulsar candidates are 2 associated point-like SNRs (with angular diameters $< 20$'), 2 HMBs (1FGL J1018.6–5856 and LS 5039) whose $\gamma$-ray observables seem very similar to those of a young pulsar [19], and 5 globular clusters, whose $\gamma$-ray emission is probably due to the contribution of a number of MSPs. The AGN candidates are 2 objects in the field of the LMC, the Andromeda galaxy M31, 3 starburst galaxies and the nova (V407 Cyg). From this analysis it is clear that only Galactic pulsar-like objects are classified as a pulsar candidate, otherwise only extragalactic or AGN-like objects are classified as an AGN candidate. We expect that up to 0.1% of the newly classified AGN candidates and up to 11.5% of the newly classified pulsar candidates will indeed belong to one of the "other" classes (galaxies, globular clusters, supernova remnants, etc.). The higher contamination rate for the pulsars is likely due to low statistics in the 2FGL source sample but it is primarily associated with a Galactic contamination. These results are shown in the table of classification in Figure 4.20.

| | "Other" observed | Total Efficiency/ Contamination |
|---|---|---|
| PSR predicted | 9 (32%) | 87.5% 1% AGN 11.5% "other" |
| AGN predicted | 7 (25%) | 98.9% 0.1% PSR 1% "other" |
| Unclassified | 12 | |

Figure 4.20: Table of classification based on the classification rules described in the text where the contamination given by the "other" 2FGL sources is assessed. In green are represented the correct classifications and in red the incorrect ones. Note that the inclusion of the "other" 2FGL sources does not change the accuracy and the misclassification of the ANN model.

Figure 4.21 shows on the left the distribution of the accuracy, misclassification and contamination by varying the decision threshold related to pulsars while on the right the same distribution related to AGNs.

### 4.3.5 Results and their validation

Applying the trained neural network to the 2FGL unidentified sources we find that 131 are classified as pulsar candidates ($P > C_P$), 176 are classified as AGN candidates ($P < C_A$) and 269 remain unclassified. The distribution of 2FGL unidentified sources as a function of the probability of being pulsars is shown in the Figure 4.22. It is important to note that we defined more restrictive thresholds with respect to LR analysis, this produces a decrease of the contamination and an increase of the efficiency of the classification, only the objects characterized by a very high value of ANN predictor are classified as pulsar candidates, conversely for AGN candidates. In this way we avoid to overpredict the number of pulsars and AGNs in unidentified sources. As a result, if we take into account the estimated contamination and misclassification rates related to our classification rules we expect about 100 $\gamma$-ray pulsars and up to 250 AGNs to be discovered in 2FGL catalog .

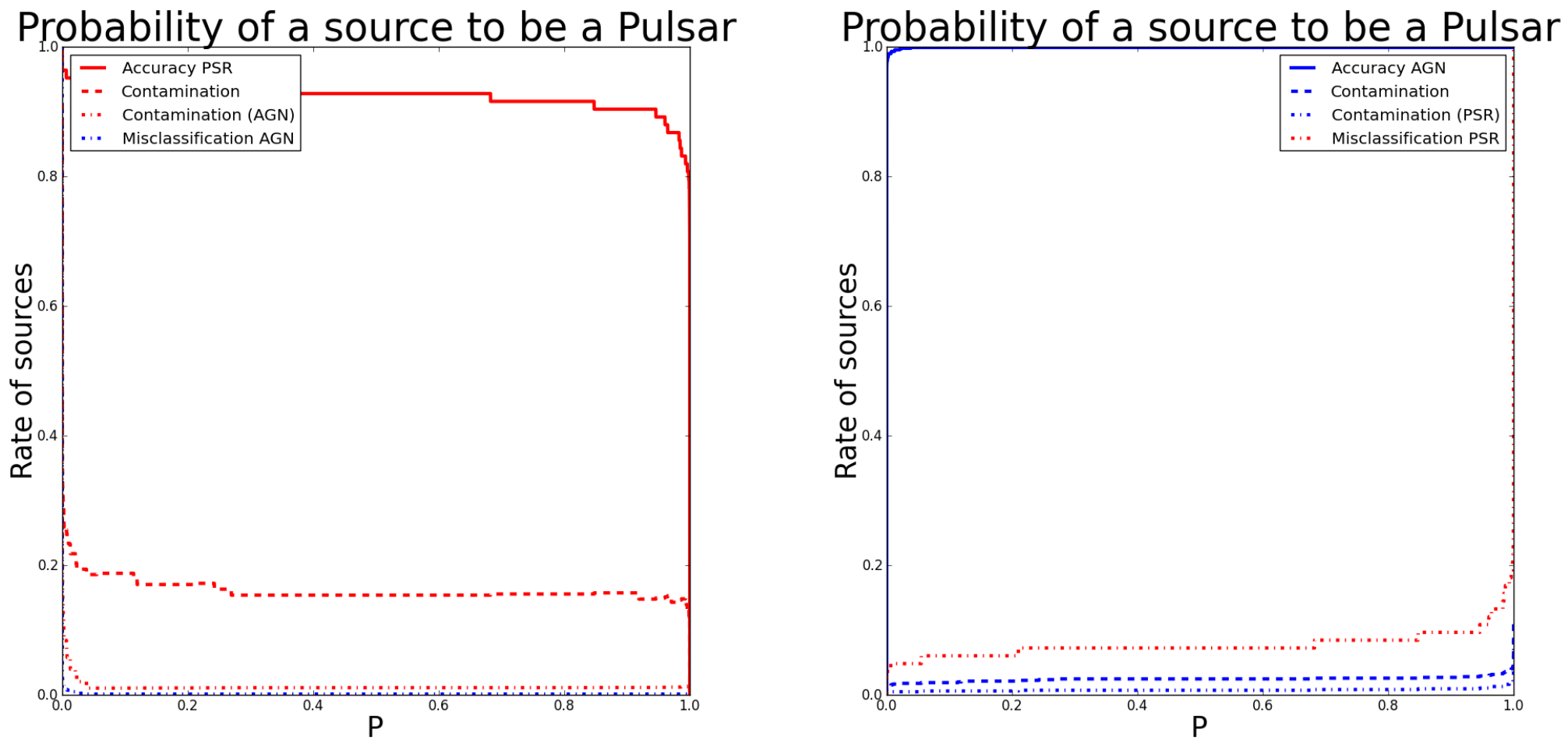

Figure 4.21: Distribution of the accuracy, misclassification and contamination as a function of the varying decision threshold. (Left) Distribution related to pulsars, the contamination is given by AGNs and "other" sources and misclassification is related to AGNs classified as a pulsar. (Right ) Distribution related to AGNs, the contamination given by pulsars and "other" sources and misclassification is related to pulsars classified as an AGN.

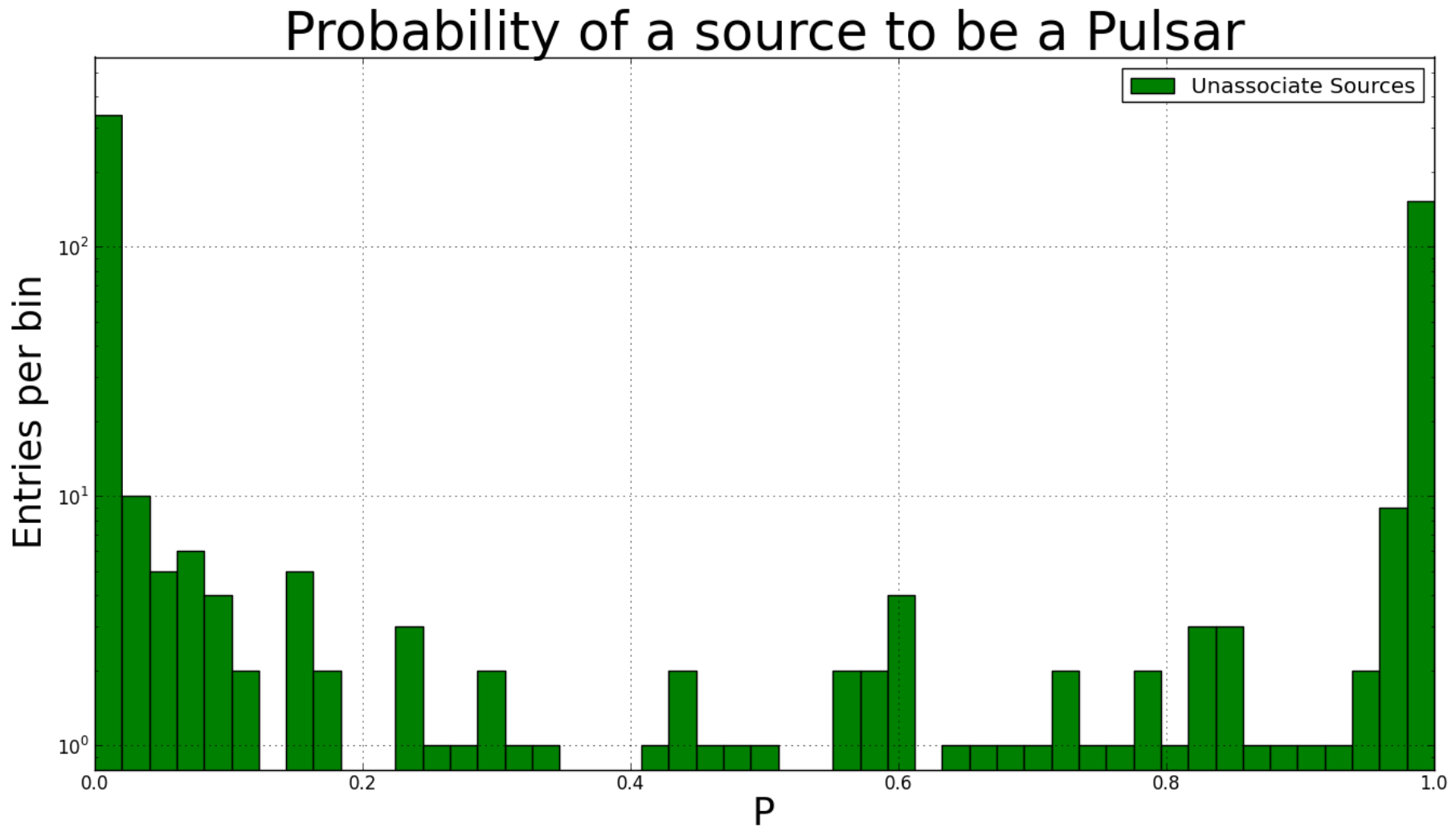

Figure 4.22: Distribution of the ANN predictor for the 576 2FGL unidentified sources.

The spatial distribution of the newly classified sources, shown in Figure 4.23, gives us the opportunity to cross check our results. Notice that both the AGN and pulsar distributions are as expected, even though we have not used the Galactic latitude as an input to either classification method. The pulsar candidates are mainly distributed along the Galactic plane, with a few high-latitude exceptions that suggest additional nearby MSPs, while the AGN candidates are nearly isotropically distributed on the sky.

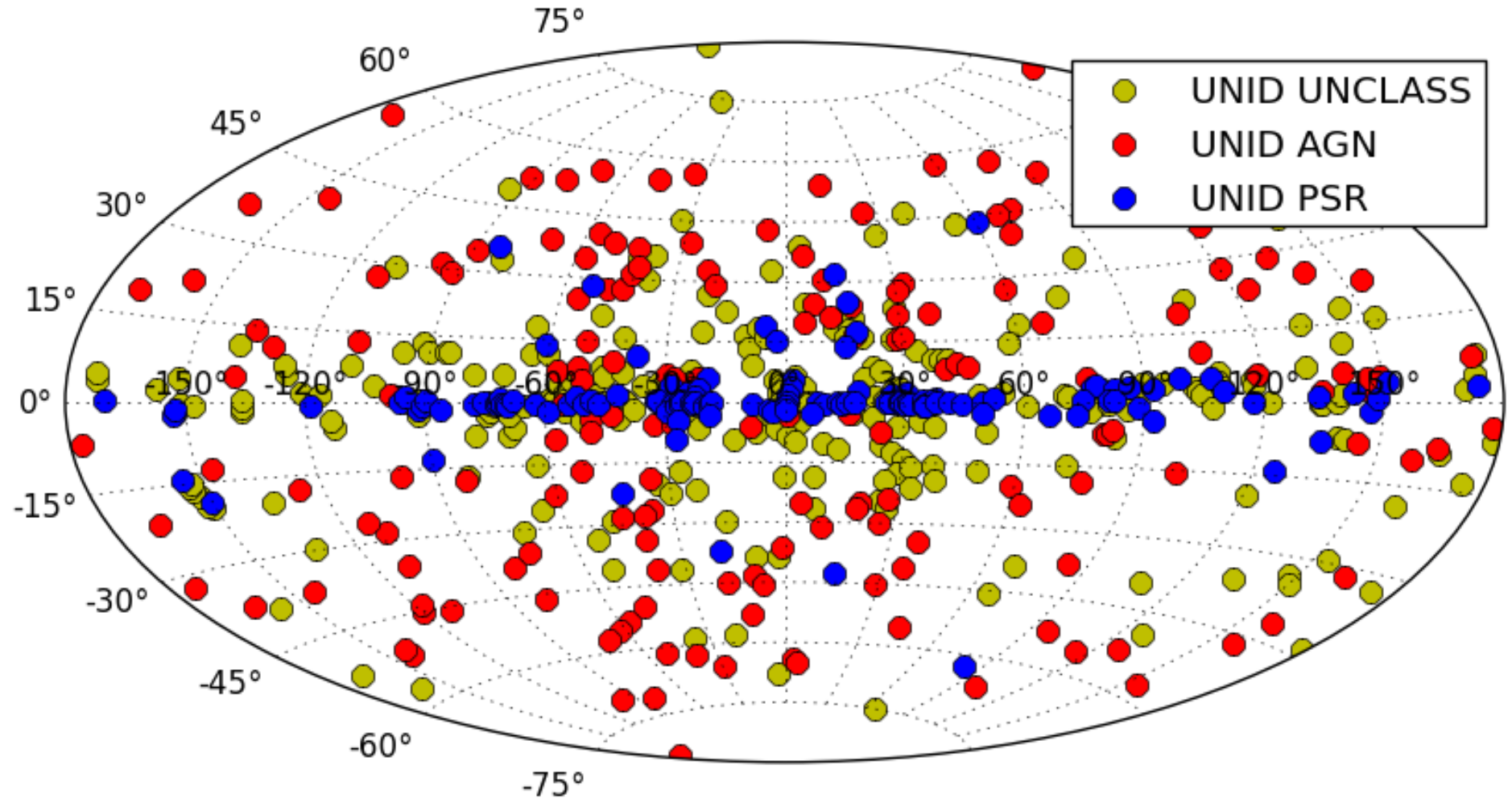

Figure 4.23: Spatial distribution in galactic coordinates of the newly classified sources. In blue are shown the pulsar candidates, in red the AGN candidates and in yellow the unclassified sources

In the top left panel of Figure 4.24 the distribution in Galactic latitude of the associated pulsars and AGNs is shown, while on the top right panel the distribution of the pulsar and AGN candidates derived by the ANN is presented. It is evident that these two distributions are very similar when comparing pulsars with pulsar candidates while they are quite different when comparing AGNs and AGN candidates. This difference may be related to the selected thresholds, our choice seems to be too conservative to single out AGNs resulting in under-predicting the number of AGN candidates in the unidentified sources. If we combine the AGN candidate population with the 2FGL sources that already have AGN associations (Figure 4.24, bottom left panel), we find that the shape of the AGN distribution matches reasonably well the nearly isotropical distribution that we expect, at low Galactic latitudes there is still an important lack of AGNs because of the bright Galactic diffuse $\gamma$-ray emission and our conservative choice regarding $C_A$. Otherwise, if we combine

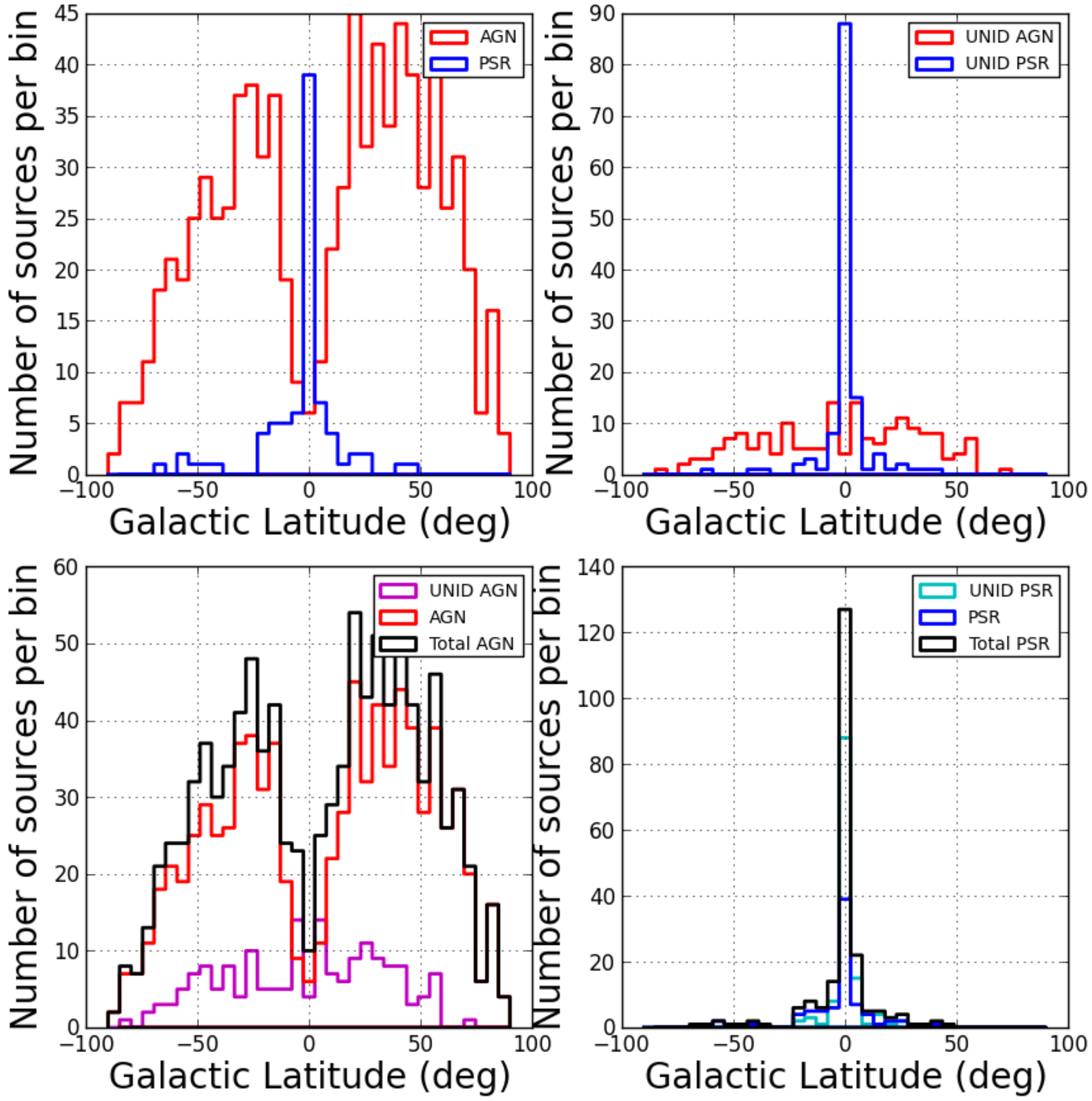

Figure 4.24: (Top) Latitude distribution of associated pulsars and AGNs (left) and of pulsar and AGN candidates (right). (Bottom) Left: distribution of AGN candidates binned by Galactic latitude. Black line is the sum of the 2FGL AGN associations (red line) plus the sources classified as AGN candidates (magenta line). Right: distribution of pulsar candidates binned by Galactic latitude. Black line is the sum of the 2FGL pulsar associations (blue line) plus the sources classified as pulsar candidates (cyan line).

the pulsar candidate population with the 2FGL sources that already have pul-sar associations (Figure 4.24, bottom right panel), we find their distribution is characterized by an evident peak at low Galactic latitudes. In viewing of our studies on the contamination in the pulsar candidate sample, we expect that a fraction of these objects may be associated with other Galactic source classes, e.g. SNRs or HMBs, but probably a certain fraction of these sources may also be spurious related to a bad model describing the Galactic diffuse $\gamma$-ray emission

In Figure 4.25 curvature-variability distribution of the newly classified AGN and pulsar candidates on the basis of the ANN analysis is shown. We see that the unidentified sources have been separated into two populations with few overlap between them because of our conservative choice on thresholds. Comparing this distribution to Figure 4.4, we see that this separation follows the separation seen between the associated AGNs and pulsars.

These results allow us to assert that our ANN algorithms work as well as we expect, i.e., pulsar candidates are not variable, have a curved spectrum, and are mainly distributed along the Galactic plane, while AGN candidates are mostly extragalactic, variable sources.

In order to test the capability of the ANN algorithms for identifying the different source classes, we must apply the neural network analysis to an independent set of associations from those used to train the ANN. To this purpose, as for the logistic regression analysis, we consider 46 new Galactic source associations (18 young pulsars and 28 MSPs), and 41 new AGN associations yielded after the publication of the 2FGL catalog [158]. Of the 41 newly associated AGNs, 28 are correctly classified as AGN candidates by the ANN analysis (efficiency: 68%), no one is classified as pulsar candidates (false negative: 0%), while the other 17 sources remain unclassified (32%). For pulsars, 24 are correctly classified as pulsar candidates by the ANN analysis (efficiency: 52%), only 3 are classified as AGN candidates (false negative: 7%), while the other 19 sources remain unclassified (41%). The efficiency of the algorithm at classifying new AGNs is good (efficiency: $\sim$70%) and no newly AGN is classified as a pulsar candidate. We notice that 12 of the 17 unclassified newly AGNs have a predictor value $< 5 \times 10^{-5}$, this means that a little change of the classification threshold to single out AGN would lead to a significant increase of the efficiency at classifying new AGNs.
For newly pulsars, we notice the same difference in performance between young pulsars and MSPs observed using the LR analysis. For newly objects identified as young pulsars, we correctly classify 15 pulsars (efficiency: 84%), no source is misclassified (false negative: 0%) and we leave the remaining sources as unclassified (efficiency: 16%). On the other hand, for newly sources identified as MSPs, the classification rate is much worse. We correctly classify only 9 objects as pulsars (efficiency: 32%), we misclassify 3 objects as AGN (false negative: 17%) and left the remaining objects as still unassociated (51% of the new MSPs). As was noticed after the logistic regression analysis, these misclassifications are mainly due to the fact that, in contrast to the new young pulsars, almost all the new MSPs have a very low source significance (TS $<$ 100), thus they have only upper limits in many energy bands making very difficult an accurate spectral characterization.

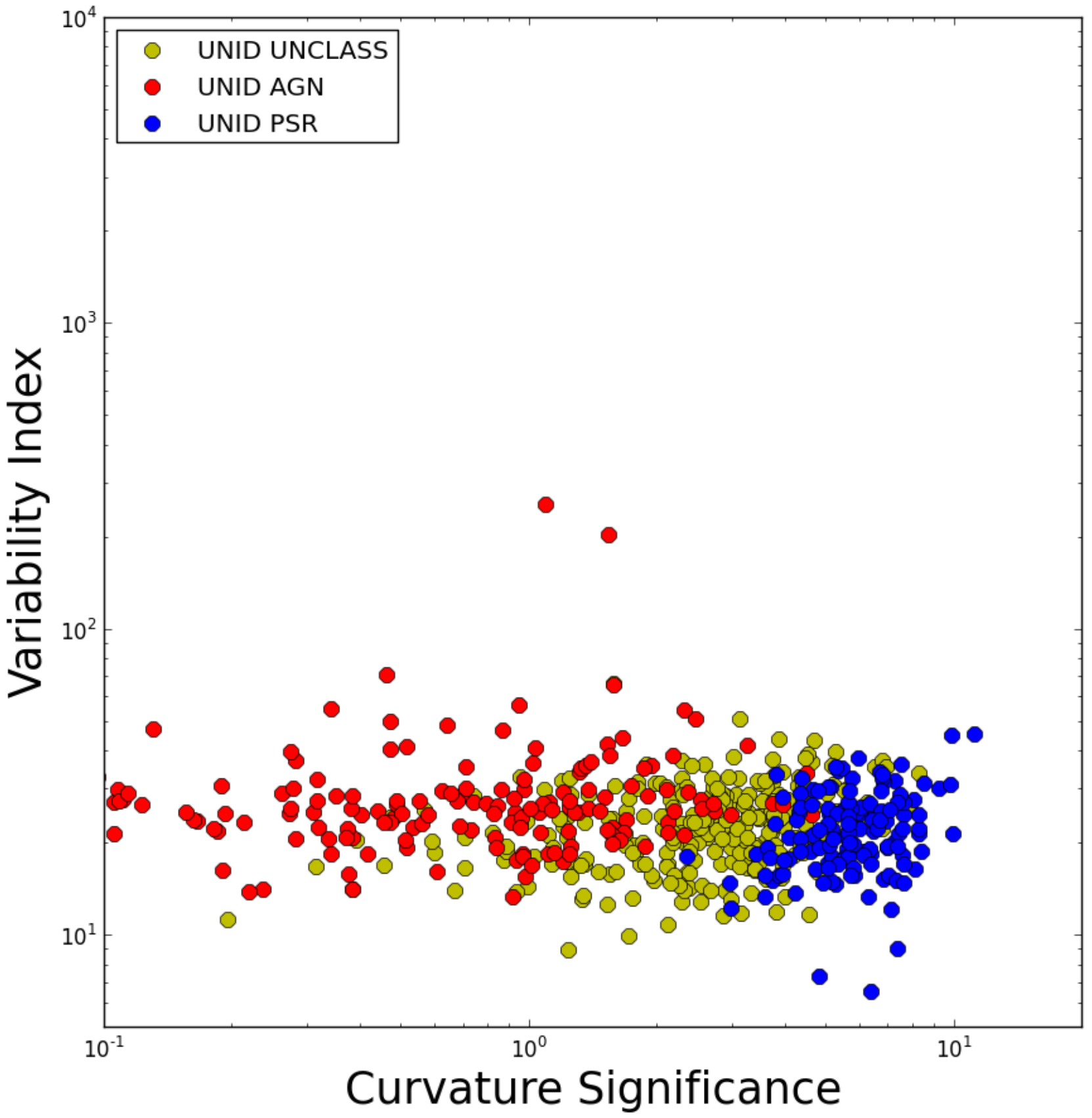

Figure 4.25: Variability index vs. curvature significance for 2FGL unidentified sources classified as AGN (red) and pulsar candidates (blue) and unclassified (yellow).

As a result, the efficiency of the algorithm at classifying new young pulsar detections is very high (more than 80%), which is consistent with the efficiency we obtained for the pulsars in the training sample. Also the efficiency of our neural network at classifying new AGNs is rather high ($\sim$ 70%). Moreover, the absence of false negative is a very encouraging result, that makes us confident in using the results from the ANN analysis while planning multi-wavelength observation of unidentified sources. Otherwise, we cannot assert anything about the efficiency of the algorithm at classifying new MSPs because these results are affected by low statistics and bad characterization related to the low source significance.

Differently from the LR analysis, we did not include the active galaxies of uncertain type in the training sample because their association may be uncertain and it is based on different techniques with respect to those applied in the construction of the 2FGL catalog ([158], [83]). Applying the trained ANN to these sources we find that 162 (63%) are classified as AGN candidates ($P < C_A$), 3 (1%) are classified as pulsar candidates ($P > C_P$) and 92 (37%) remain unclassified. Among the misclassified active galaxies of uncertain type, 1 is now associated with a young $\gamma$-ray pulsar (PSR J0622+3749 [167]) and the other two have a "c" appended to indicate "caution" because these objects has an high probability to be spurious [158]. Moreover, in the preliminary construction of the new *Fermi*-LAT catalog developed by the LAT team, where the most recent diffuse $\gamma$-ray Galactic emission model and 5 years of data are used, these two objects disappear. The distribution of 2FGL active galaxies of uncertain type as a function of the probability of being pulsars is shown in the Figure 4.26. If we compare this distribution with those of associated AGNs (see the bottom panel of Figure 4.17) we notice that active galaxies of uncertain type are characterized by a broader distribution with some of them with an high probability to be a pulsar, indicating that probably some of these objects are spurious or associated with another source class. Active galaxies of uncertain type are typically characterized by a low source significance (TS $< 100$) thus their features are not well detected, this is particularly true for time variability, making these objects a bit different from a "standard" $\gamma$-ray AGN.

## 4.4 Comparing the two classification methods

We have implemented two different machine learning techniques to determine likely source classifications for the 2FGL unidentified sources: Logistic Regression and Artificial Neural Networks. Both techniques use identified objects to build up a classification analysis which compute the probability for an unidentified source to belong to a given class. We applied these techniques to the sources in 2FGL and obtained a set of classification probabilities. LR is a generalized linear model and it classifies the 2FGL sources through a linear distinction in the predictor variables space, while ANNs can be considered an extension of this simple linear model classifying the 2FGL sources through a more complex non-linear separation, this is evident in Figure 4.14. We cannot directly compare the results obtained by the logistic regression and the neural networks because we used two different training samples (for the ANN analysis we do not include active galaxies of uncertain type), we selected classification thresholds based on different assumptions and we used a different set of predictor variables. However, we can compare their performance at classifying pulsars from AGNs.

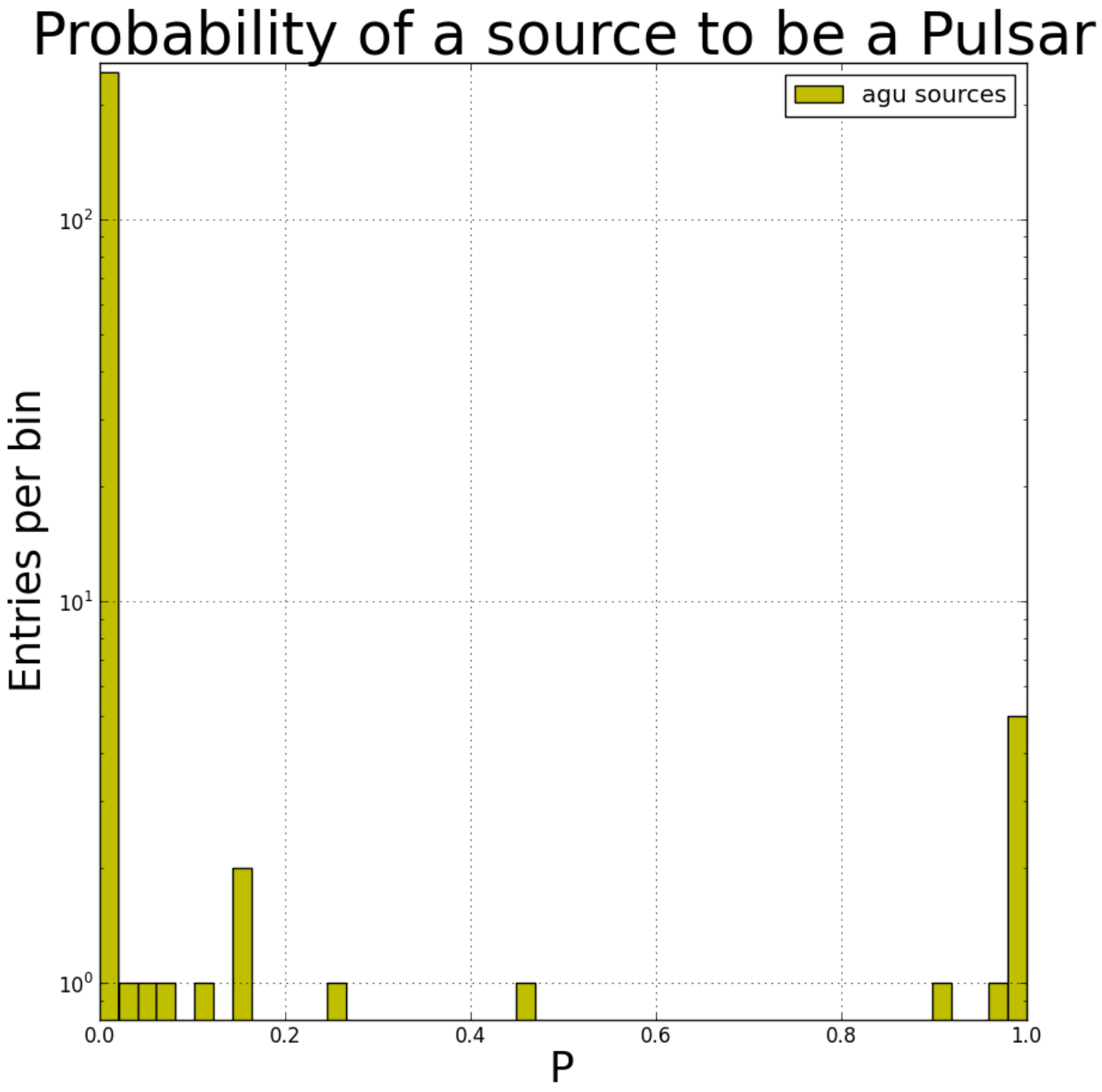

Figure 4.26: Distribution of the ANN predictor for the 257 2FGL active galaxies of uncertain type.

Comparing the LR and ANN predictor distributions shown in Figure 4.7 and at the bottom of Figure 4.17 it is clear they are very different. ANN seems to separate pulsars from AGNs better than LR, in fact these objects are primarily situated near to the extreme values of the predictor (P=1 and P=0). A significant comparison between these two classification techniques can be done analyzing the relative efficiency at classifying pulsars and AGNs associated after the publication of the 2FGL catalog. The more conservative classification thresholds selected for the ANN analysis misclassify very few newly associated sources (only a total of 3 objects) with respect to the LR analysis but the significant improvement given by the ANN is related to the fraction of correct classifications. Since the conservative classification thresholds selected, we would expect the rate of correct classifications is higher using the LR analysis but ANN algorithm correctly classify a larger fraction of young pulsars, MSPs and AGNs. This means that the efficiency to determine likely source classification for the 2FGL unidentified sources is considerably higher if it is performed by ANN analysis independently from our assumptions. This is a general result, we must apply a non-linear analysis to efficiently classify the 2FGL sources.

In the end, the number of pulsar candidates in 2FGL unassociated sample is similar for the LR and ANN analysis because their accuracy at classifying pulsar is comparable (see Figures 4.6 and 4.18)), while the number of AGN candidates is lower using an ANN analysis because using an acceptance threshold of 80% of the known AGNs, we are not including a large number of AGN candidates.

## 4.5 Applying an ANN analysis to distinguish young pulsars from MSPs

One of the early surprising and mostly unexpected results from the LAT, was the fact that MSPs are strong $\gamma$-ray emitters. A MSP is a very old pulsar (about $10^9$ years) with a rotational period in the range of about 1 and 10 ms, a magnetic field $\lesssim 10^{10}$ Gauss and a spindown rate $\lesssim 10^{-17}$ s/s. These characteristics are very different from those of LAT young pulsars, which have a rotational period in the range of about 0.03 and 1 s a very intense magnetic field of order of $10^{12} - 10^{13}$ Gauss and a spindown rate in the range of about $10^{-14}$ and $10^{-12}$ s/s. Pulsar period-spindown rate distribution is shown in Figure 4.27.

Out of 83 2FGL identified pulsars, 57 are young pulsars and 26 are MSPs and their distribution on the sky is shown in Figure 4.28. Young pulsars are primarily located along the Galactic plane which reflects the Galactic nature of these objects, while MSPs are seen at higher latitudes. MSPs are typically fainter in $\gamma$ rays than young pulsars and only the near ones can be detected by the LAT, they are in our Galaxy but they are very local, thus distributed all over the sky.

Despite their intrinsic differences, which are evident in Figure 4.27, after detailed inspections, the $\gamma$-ray properties (e.g. light curves and spectra) of MSPs appear to be very similar to those of young $\gamma$-ray pulsars [200]. Their $\gamma$-ray similarity is evident observing the curvature-variability distribution of the young pulsar and the MSPs shown in Figure 4.29. We see that these two populations overlap in the lower right-hand quadrant making impossible their distinction on the basis of only these $\gamma$-ray observables.

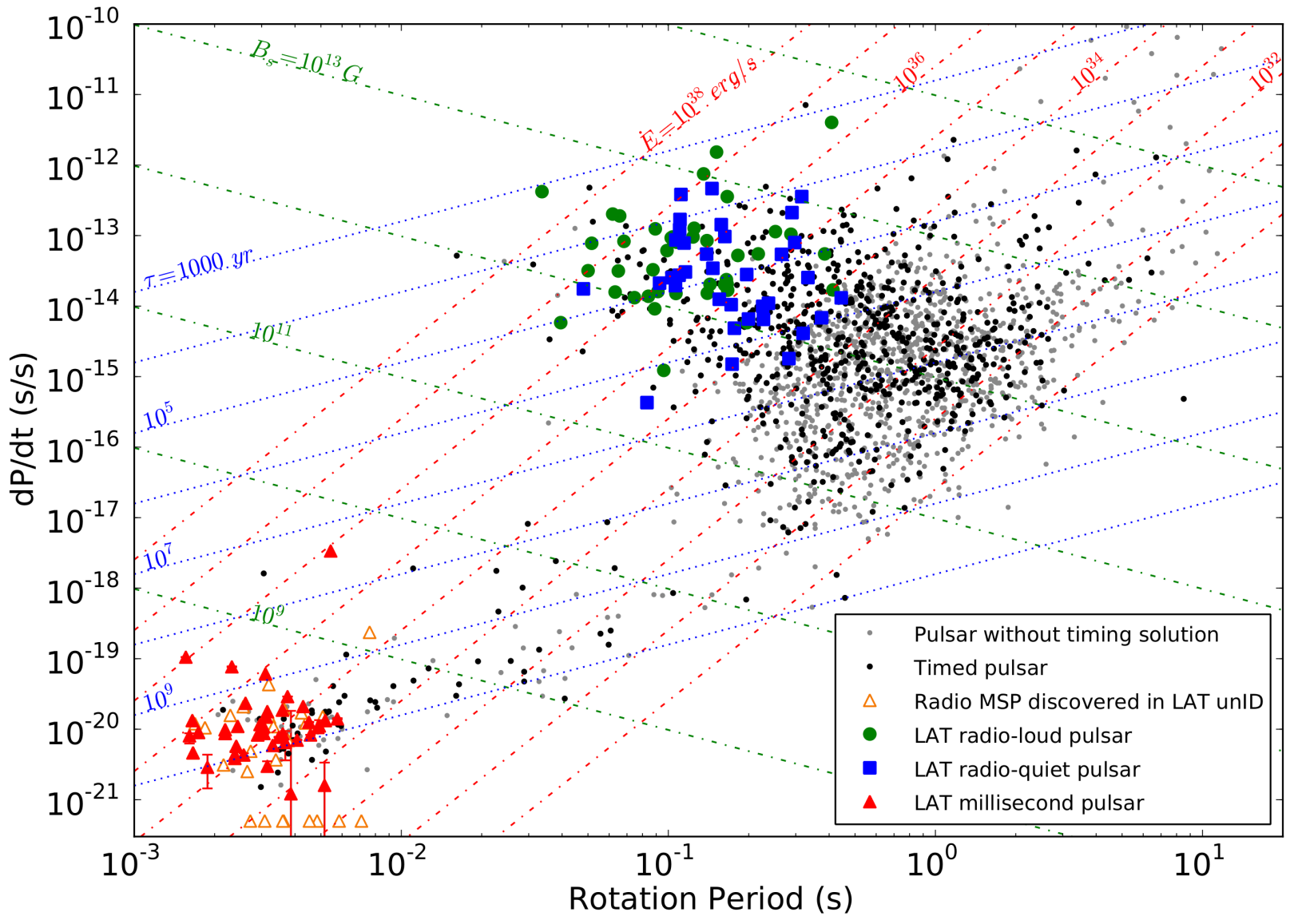

Figure 4.27: Pulsar spindown rate versus rotation period [200].

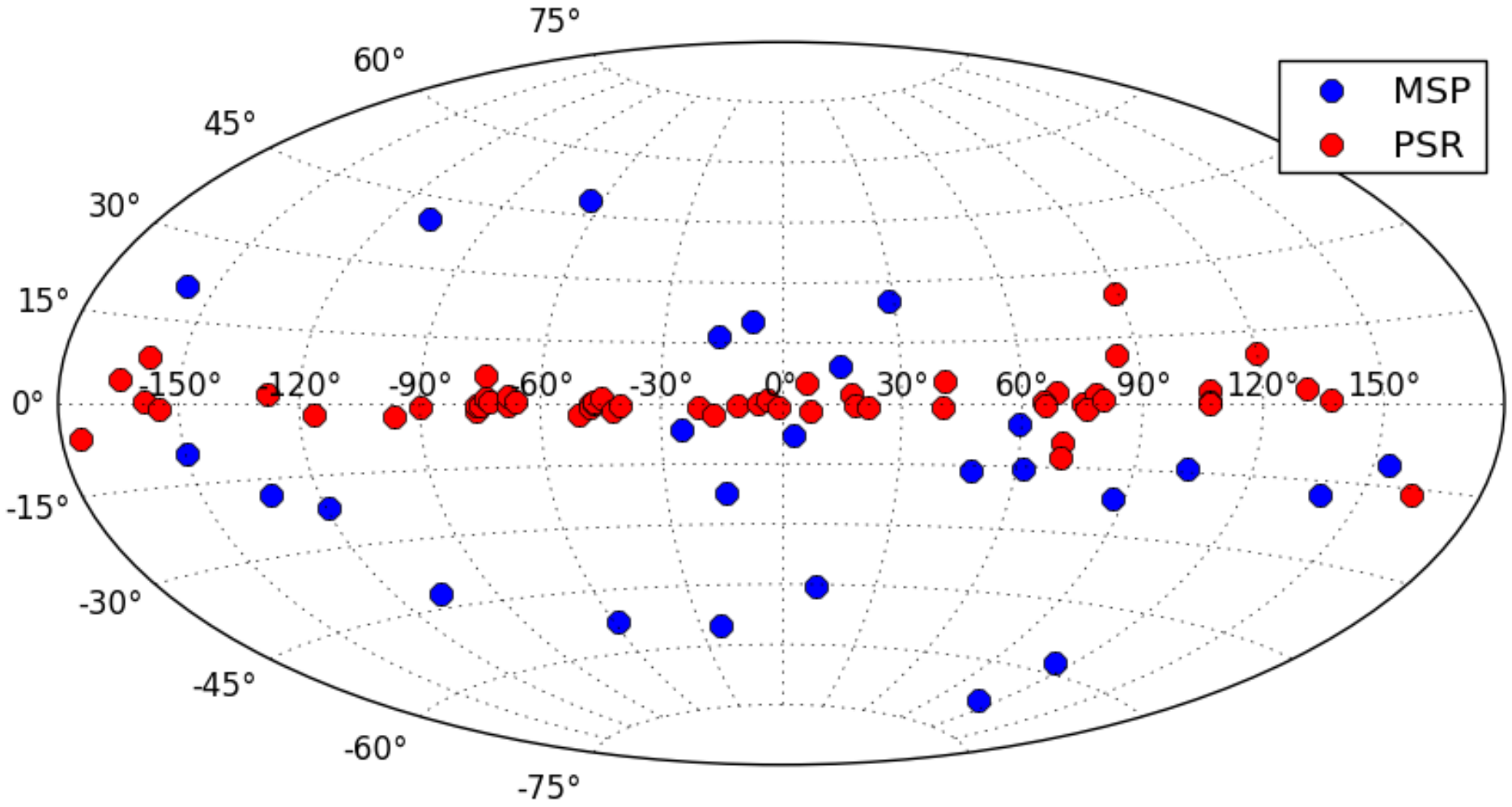

Figure 4.28: Spatial distribution in Galactic coordinates of 2FGL young pulsars and MSPs.

The current sample of 83 2FGL pulsars and a refined definition of $\gamma$-ray parameters in 2FGL catalog allow us to perform detailed statistical comparisons between the properties of these two types of pulsars. The excellent results obtained by ANNs to classify 2FGL unidentified sources as pulsar or AGN candidates on the basis of their $\gamma$-ray observables prompted us to build a different neural network to try to distinguish young pulsars from MSPs. Our goal is to explore all the parameter space of these $\gamma$-ray sources and search for a set of input variables that might lead to some discriminating power between these sources.

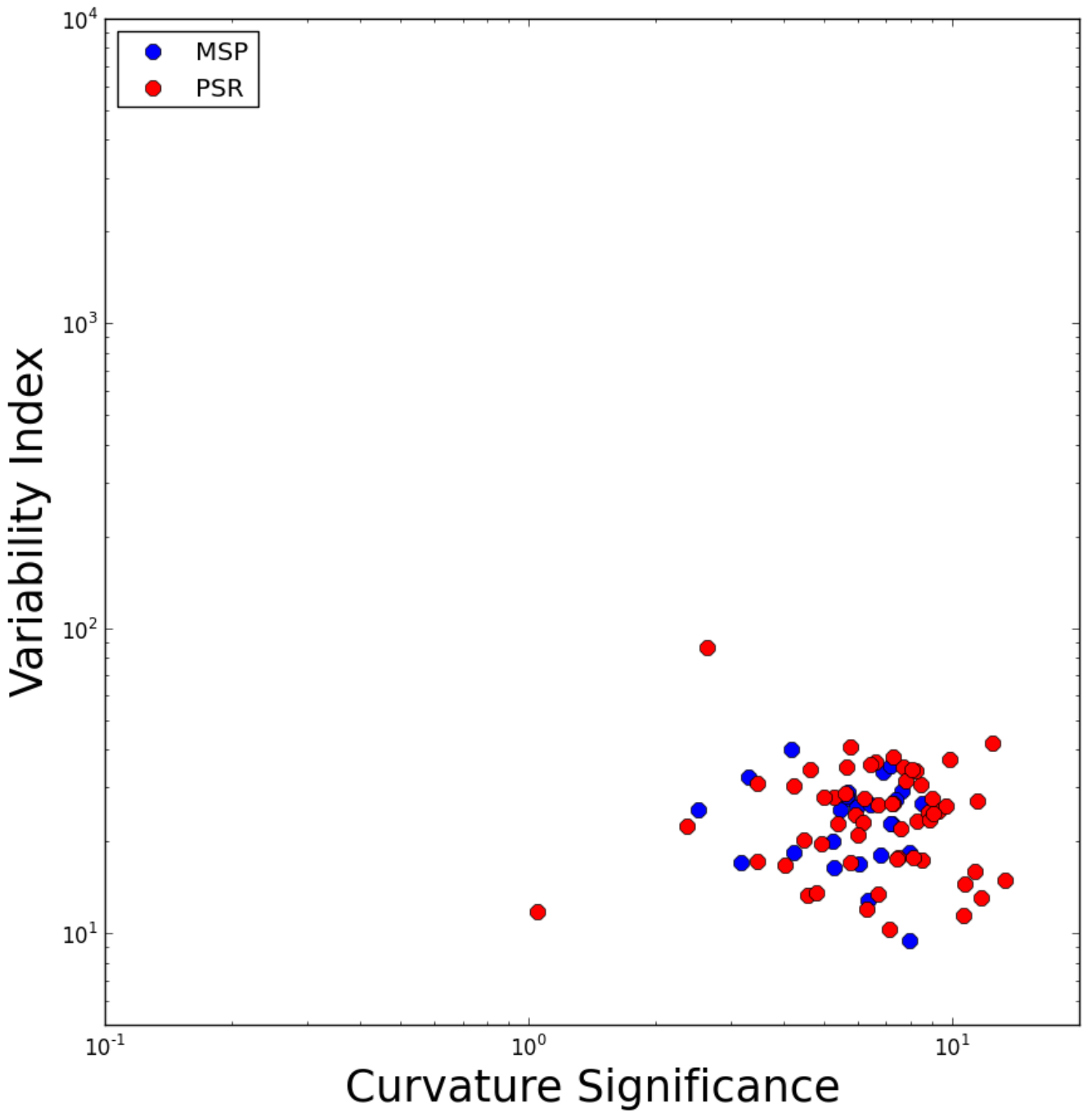

Figure 4.29: Variability index plotted as a function of curvature significance.

It is important to assert that if we want to classify 2FGL unidentified sources as young pulsar and MSP candidates we cannot directly apply a neural network trained with these two source classes because in this way almost all non-pulsar objects (e.g. AGNs) in the unassociated sample will be erroneously classified as MSP or young pulsar candidates. For this reason we must use a more complex architecture and we decide to use a hierarchical neural network model composed by two simple neural networks. The first one (defined as *ANN1*) is constructed to distinguish pulsars from AGNs (we use the ANN described in Section 4.3) and the second one (defined as *ANN2*) to distinguish young pulsars from MSPs. In this way each 2FGL source “enters” the *ANN1* and if it is classified as a pulsar candidate can “enter” the *ANN2*, which decides if such object is a young pulsar or a MSP candidate.

In order to perform an ANN analysis to try to discriminate young pulsars from MSPs in the $\gamma$-ray observables space we follow the procedure explained in Section 4.3. The first step consists of selecting a sample of data to build the predictor value P, to this aim we focus on the 57 young pulsars an the 26 MSPs. Then, we must select the most appropriate set of variables for training the *ANN2*, variability index and curvature significance are not useful at distinguish young pulsars from MSPs as shown in Figure 4.29. One of the advantages of ANN is that we can use all $\gamma$-ray parameters included in 2FGL catalog which do not depend too much on flux and significance of the source. The trained ANN will decide which parameters are not important on the basis of the weights associated to them (see Equation 4.6). In particular, weights associated to an unimportant parameters will tend to zero making irrelevant its presence in the network. According to the previous consideration, we select a set of variables that includes the curvature significance, the variability index, the PowerLaw index, the fluxes and the hardness ratios for the 5 energy bands in the catalog. Also for the *ANN2* we do not choose to use the Galactic latitude and longitude as input to the neural network because we want to use the position on the sky of the different populations as a cross check of our result. The new young pulsar candidates should be mainly distributed along the Galactic plane, while the MSP candidates should be located at higher latitudes.

### 4.5.1 Architecture of the ANN

In order to define the architecture of *ANN2* we must determine the number of nodes in each layer. The number of input nodes is given by the number of selected predictor variables, in our case 12. During the ANN analysis a pre-processing is performed normalizing each variable to the range [0, 1]. All the variables are normalized as explained in Section 4.3. The number of output nodes is given by the number of source classes we want to classify, in our case 2 (young pulsars and MSPs). Since for *ANN2* we use an hyperbolic tangent activation function both for hidden and output nodes, the target vector for the network outputs associated to the $l$th source in the training sample ($\mathbf{t}^{(l)}$) has 1 in the element corresponding to the true class and minus ones elsewhere. Thus for young pulsars $t = [1, -1]$ while for MSPs $\mathbf{t} = [-1, 1]$. The number of hidden nodes is determined by the pruning procedure as explained is Section 4.3.2. In Figure 4.30 in blue is shown the value of the mean square error defined in Equation 4.4 as a function of the number of hidden nodes while in red the performance of our network in terms of the fraction of correct classifications of the validation as a function of the number of hidden nodes. Using these procedure we decide to use 8 hidden nodes.

As a result our feed-forward 2LP is built up of 12 input nodes, 8 hidden nodes and 2 output nodes, each node in a layer is linked to all the nodes of the next layer and to each link is associated a weight randomly initialized in the range $[-1, 1]$.

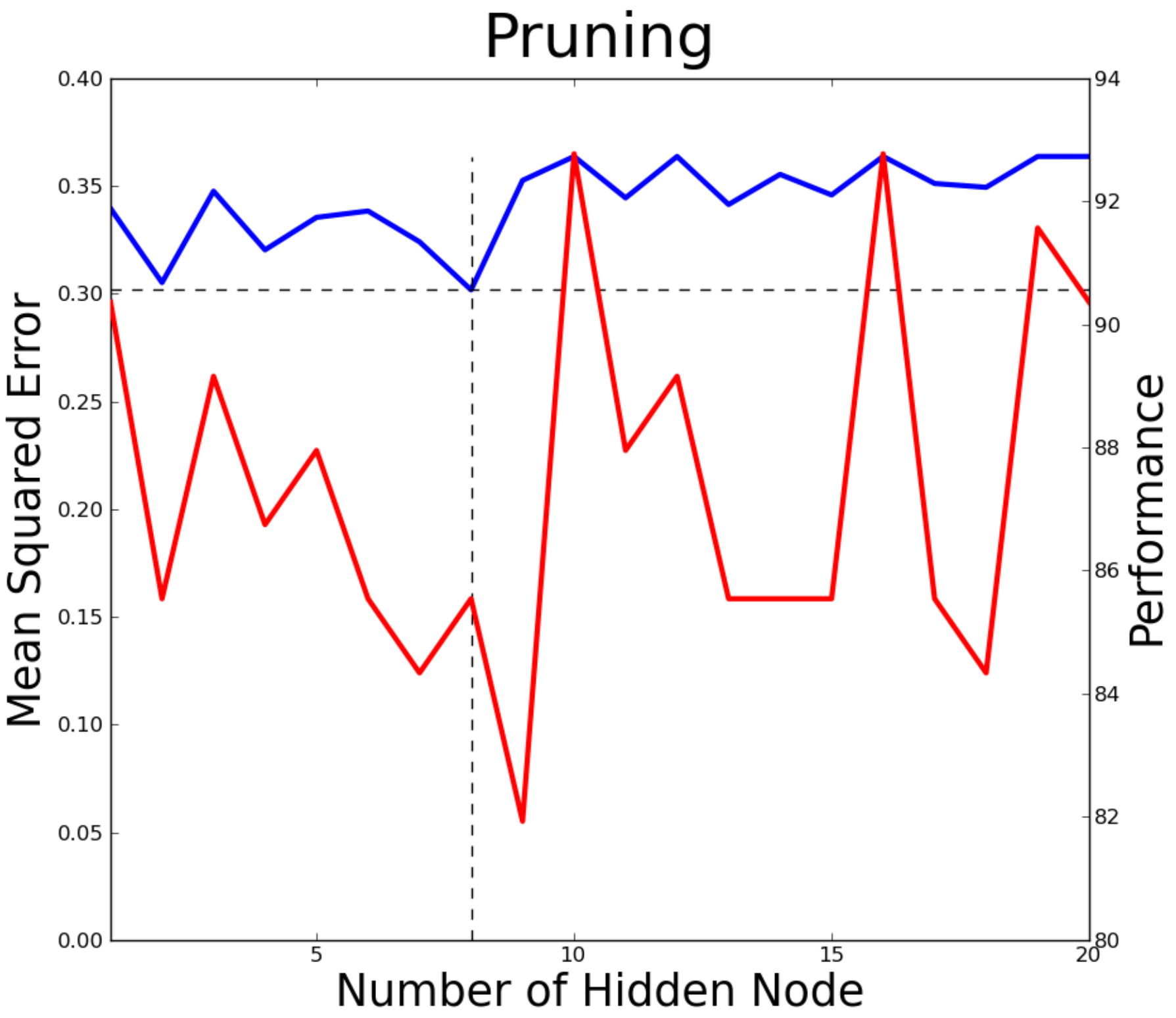

Figure 4.30: (blue) Value of the mean square error (mse) defined in Equation 4.4 as a function of the number of hidden nodes. (red) Performance of the *ANN2* in terms of fraction of correct classifications of the objects in the validation sample as a function of the number of hidden nodes.

## 4.5.2 Training session

Given the network architecture, we must train the *ANN2* to distinguish young pulsars from MSPs. We follow the procedure described in the Section 4.3.3 and we use the back-propagation learning algorithm, which is based on the error-correction learning rule. We use this learning algorithm in the "online" version, in which the weights of the connections are updated after each example are processed by the network (see Equation 4.5). We set the learning rate $\eta$ to 0.2.

Table 4.4 ranks the relative importance of the different variables at distin-guishing MSPs from young pulsars according to the Equation 4.6. As expected, variability index and curvature significance are not very important at distinguishing the two source classes, the same for the fluxes at energies greater than 300 MeV. $HR_{45}$ and PowerLaw index are the two less important $\gamma$-ray parameters. The most relevant result is that the flux in the lower energy band and the hardness ratios (except for $HR_{45}$) are the most important $\gamma$-ray observables to distinguish MSPs from young pulsars. Young pulsars are typically brighter than MSPs in the energy range between 100 and 300 MeV, this may explain the importance of the flux in this energy range. Although curvature significance and PowerLaw index, which are related to spectral shape, are not very significant parameters at discriminating young pulsars from MSPs, hardness ratios (except for $HR_{45}$) seem very important. This means that spectra of these two source classes are very different from a simple pawer law but, because curvature significance does not say anything about spectral shape, probably these two pulsar types are characterized by a different spectral shape, this result is very interesting and it must be analyzed in more detail.

| **Variable** | **Importance** |
|---|---|
| (0) curvature significance | 1.63 (2.37%) |
| (1) Variability Index | 1.73 (2.51%) |
| (2) PowerLaw Index | 0.61 (0.89%) |
| (3) $\text{Flux}_{0.1-0.3GeV}$ | 20.47 (29.79%) |
| (4) $\text{Flux}_{0.3-1GeV}$ | 2.39 (3.48%) |
| (5) $\text{Flux}_{1-3GeV}$ | 1.61 (2.35%) |
| (6) $\text{Flux}_{3-10GeV}$ | 1.51 (2.2%) |
| (7) $\text{Flux}_{10-100GeV}$ | 0.64 (0.93%) |
| (8) $\text{Hardness}_{12}$ | 11.27 (16.39%) |
| (9) $\text{Hardness}_{23}$ | 15.30 (22.27%) |
| (10) $\text{Hardness}_{34}$ | 11.51 (16.75%) |
| (11) $\text{Hardness}_{45}$ | 0.04 (0.06%) |

Table 4.4: List of the training variables for the ANN: each variable is ranked according to its relevance in the discrimination process (see Equation 4.6), as computed by the ANN algorithm.

Figure 4.31 shows the correlation matrix of the variables selected for our analysis as described by the Equation 4.7.

Figure 4.32 shows the results of the learning process and the performance of our trained neural network. The top panel shows the mean squared error for training, validation and testing samples during the learning process as a function of the epoch. In the middle panel is shown the classification table.

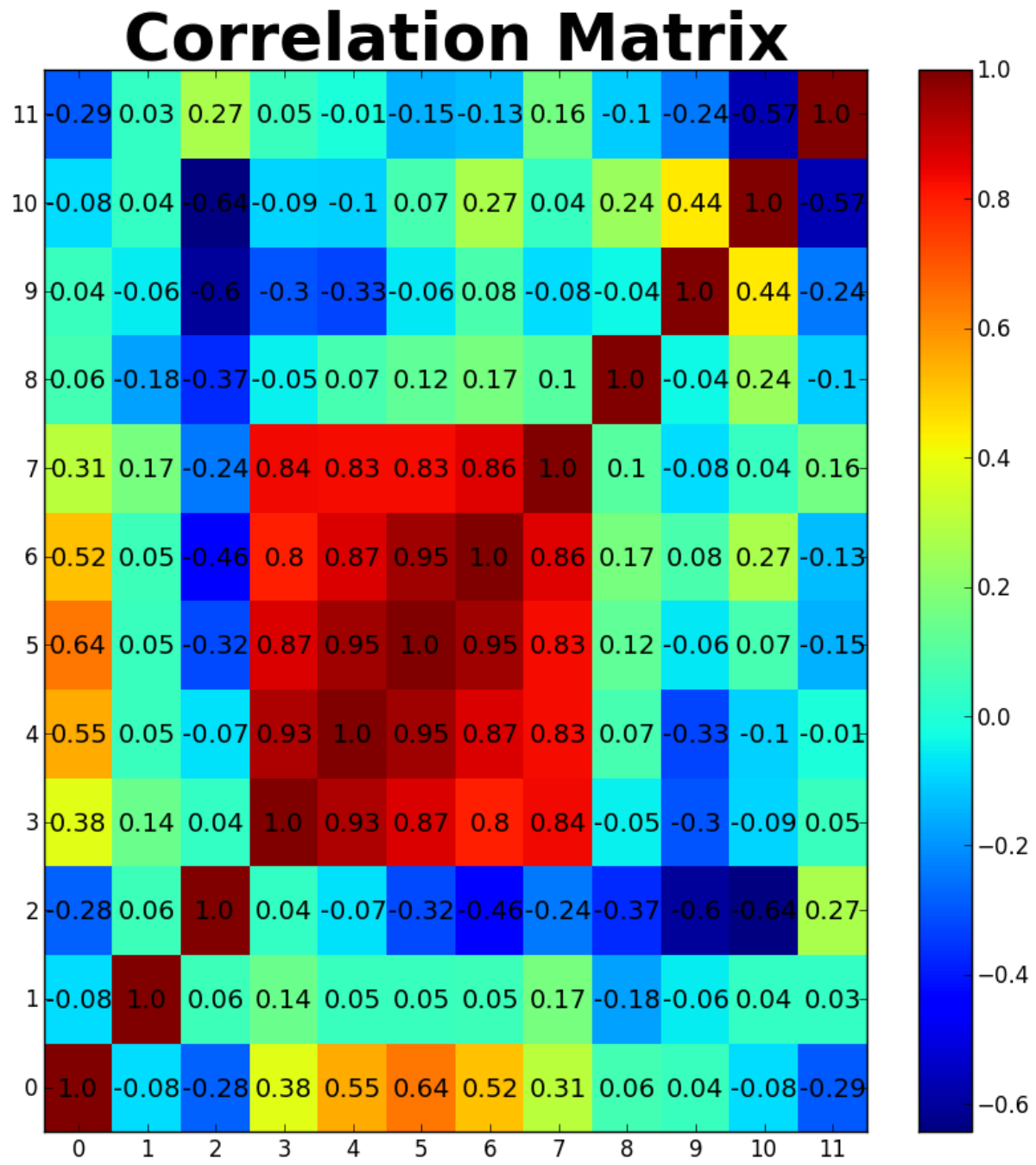

Figure 4.31: Correlation matrix of the variables selected for our analysis. The number of each parameter is expressed in Table 4.4 and to each correlation coefficient is associated a color on the basis of its value.

This table suggests that our trained neural network is able to distinguish young pulsars from MSPs on the basis of their $\gamma$-ray observables because the accuracy and the precision are very high. The total performance of our ANN is given by the total accuracy in the block blue and it is $> 93\%$. In order to efficiently classify 2FGL unidentified sources as young pulsar or MSP candidates we must select some classification thresholds as we will explain in the next section. In the bottom panel the distribution of the *ANN2* predictor for the 2FGL young pulsars (Class 1) and MSPs (Class 2) is shown. This distribution shows our algorithm is able to discriminate very well young pulsars from MSPs because almost all the MSPs are characterized by a low predictor value, while almost all young pulsars by an high predictor value, with few misclassifications with very low predictor value.

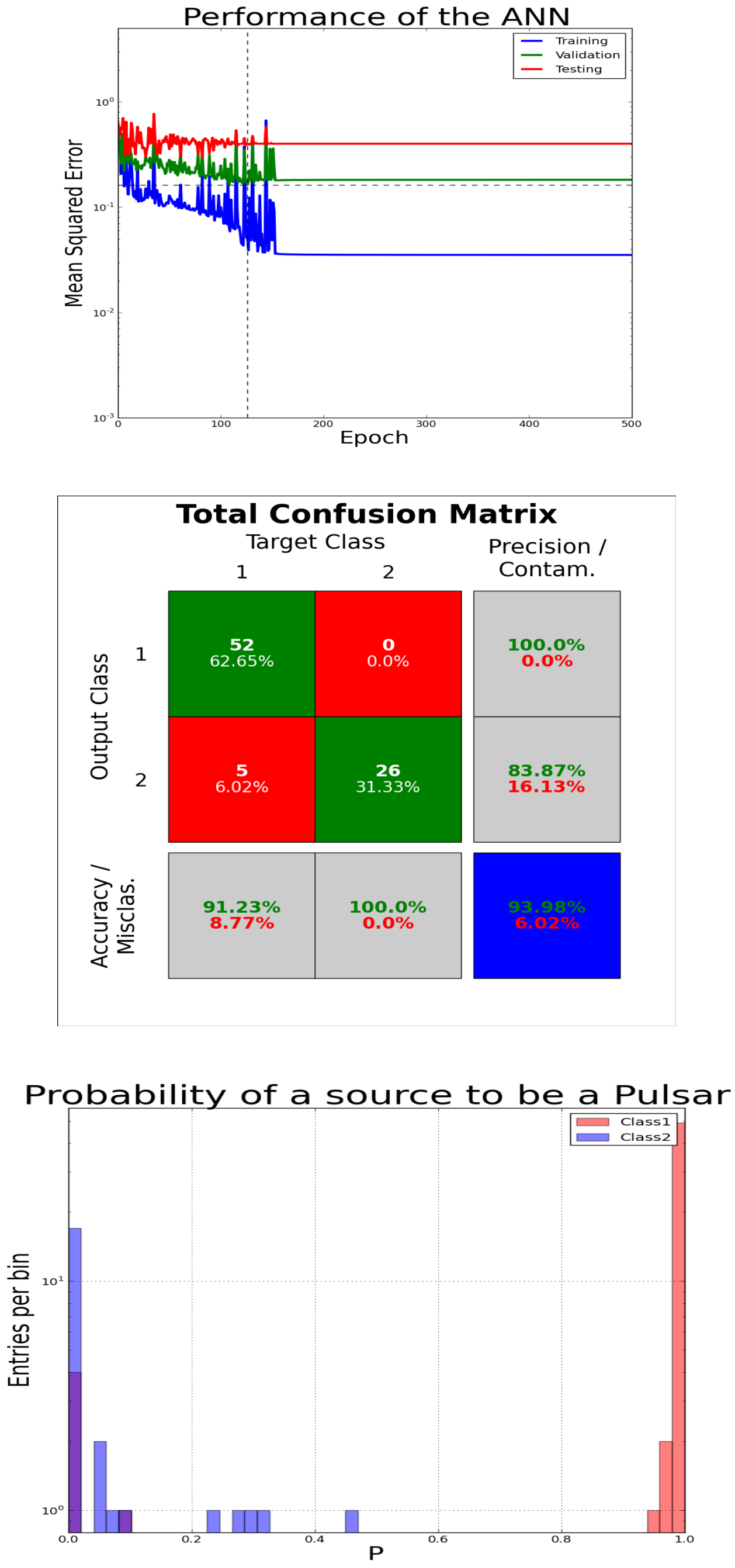

Figure 4.32: Performance of *ANN2*. Top: mean squared error for training, validation and testing samples during the learning process as a function of the epoch. The vertical dotted line specifies the epoch the learning process is stopped (optimized network). Middle: classification table. Bottom: distribution of the *ANN2* predictor for the 2FGL young pulsars (indicated as "Class 1") and MSPs (indicated as "Class 2"). Violet indicates young pulsars which are put on top of MSPs.

### 4.5.3 Defining thresholds

Now we set two classification thresholds, one to classify a young pulsar candidate ($C_{YP}$) and one to classify a MSP candidate ($C_{MSP}$) so that all the sources with a predictor greater than 0.984 are classified as young pulsar candidates while all the sources with a predictor smaller than 0.085 are classified as MSPs candidates. All the sources with an intermediate value of the predictor remain unclassified after the ANN analysis. As a result of this choice, 80% of young pulsars in 2FGL have a predictor greater than $C_{YP}$ while 80% of MSPs have a predictor smaller than $C_{MSP}$. The results are shown in the table of classification in Figure 4.33. The misclassified young pulsars are 4: PSR J0631+1036, PSR J1846+0919, PSR J2030+3641 and PSR J2043+2740, while no MSP was misclassified as a young pulsar by our trained neural network. PSR J0631 is a bright young $\gamma$-ray pulsars and it was one of the first pulsars detected by the LAT [215]. This young pulsar, located in the direction of the Galactic anticenter, could be inside the dark cloud LDN 1605, which is part of the active star-forming region 3 Mon, and could be interacting with it [9]. These peculiarities make this pulsar very different from a "standard" young pulsar and this may be the reason of its misclassification. PSR J1846+0919 is located off the plane and is not as energetic as other young pulsars, moreover it has a low magnetic field and a rather large characteristic age [183]. Thus, also this pulsar is different from a "standard" young pulsars and this may be the reason of its misclassification. PSR J2030+3641 is a middle-aged pulsar located in the Cygnus region [39]. The reason of its misclassification is not clear, it may be related to a bad model of the Galactic diffuse emission in that difficult region. PSR J2043+2740 is one of the oldest non-recycled $\gamma$-ray pulsar (characteristic age of 1.2 Myr) with a very short period but with a relative high spin-down luminosity [159]. These anomalous features may be the reason of its misclassification.

Since we use a hierarchical neural network, we cannot estimate the contamination to the candidate young pulsar and MSP samples from the likely presence in the unassociated sample of the other 2FGL associated objects excluded from the training procedure (e.g. AGNs, SNRs, HMBs, etc.). In order to estimate their contamination, we must follow the hierarchical structure of our ANN. First, each source associated with a different class than MSP or young pulsar "enters" the *ANN1* optimized at distinguishing pulsars from AGNs (see Section 4.3) and if an object is classified as a pulsar-like source can "enter" the *ANN2* just constructed and trained to distinguish young pulsars and MSPs. At this point the contamination of the "other" 2FGL associated sources can be estimated.

| | YP observed | MSP observed | Efficiency/ Contam. |
|---|---|---|---|
| YP predicted | 45 | 0 | 100%<br>0% |
| MSP predicted | 4 | 21 | 84%<br>16% |
| Unclassified | 8 | 5 | |
| Accuracy/ Misclass. | 80%<br>7% | 80%<br>0% | 94%<br>6% |

Figure 4.33: Table of classification based on the rules described in the text. In green are represented the correct classifications and in red the incorrect ones.

### 4.5.4 Results and their validation

First of all we apply the trained hierarchical neural network to 83 2FGL pulsars in order to determine the total efficiency and accuracy at distinguishing young pulsars from MSPs. Of 46 young pulsars classified as pulsar candidates by the *ANN1* (see Section 4.3), 41 are correctly classified as young pulsar candidates by the *ANN2* (efficiency: 89%) and only 1 is classified as MSP candidates (false negative: 2%), while the other source remains unclassified (9%). For 21 MSPs classified as pulsar candidates, 18 are correctly classified as MSP candidates by the *ANN2* (efficiency: 86%), no one is classified as young pulsar candidates (false negative: 0%), while the other 3 sources remain unclassified (14%). These results are encouraging because applying a hierarchical neural network the efficiency and the accuracy at distinguishing young pulsars from MSPs is higher than those found using a single neural network.

Applying the optimized hierarchical neural network to the "other" 2FGL associated sources we can estimate their contamination to the candidate MSP and young pulsar samples from the likely presence of these objects in the unassociated sample. The *ANN1* has classified 13 non-pulsar objects as pulsar candidates, they are 1 AGN, 3 active galaxies of uncertain type, 2 HMBs, 2 point-like SNRs and 5 globular clusters. Of these, 6 are classified as young pulsar candidates by the *ANN2* (2 SNRs, 2 HMBs and the globular clusters Terzan 5 and 2MS-GC01), 3 are classified as MSP candidates (3 globular clusters) while the other 4 sources remain unclassified (1 AGN and 3 active galaxies of uncertain type). This result is very encouraging because all AGN-like objects remain unclassified, almost all globular clusters are classified as MSP candidates, a reasonable choice since their $\gamma$-ray emission is related to the contribution of a number of MSPs, and all the objects whose emission may be related to a young pulsar are classified as young pulsar candidates. Here we observe that the contamination to the candidate MSP and young pulsar samples is very low, thus irrelevant.

Applying the trained hierarchical neural network to the 2FGL unidentified sources we find that, out of 131 classified as pulsar candidates by the *ANN1*, 75 are classified as young pulsar candidates ($P > C_{YP}$), 34 are classified as MSP candidates ($P < C_{MSP}$) and 26 are pulsars of uncertain type. In Figure 4.34 is shown the distribution of the 131 unidentified sources classified as pulsar candidates by the *ANN1* as a function of the probability of being young pulsars. The *ANN2* predictor distribution of 131 pulsar-like unidentified sources is characterized by two noticeable peaks close to $P = 1$ and $P = 0$, these objects are very similar to 2FGL young pulsars and MSPs respectively, between these two values the pulsar-like unidentified sources are uniformly distributed.

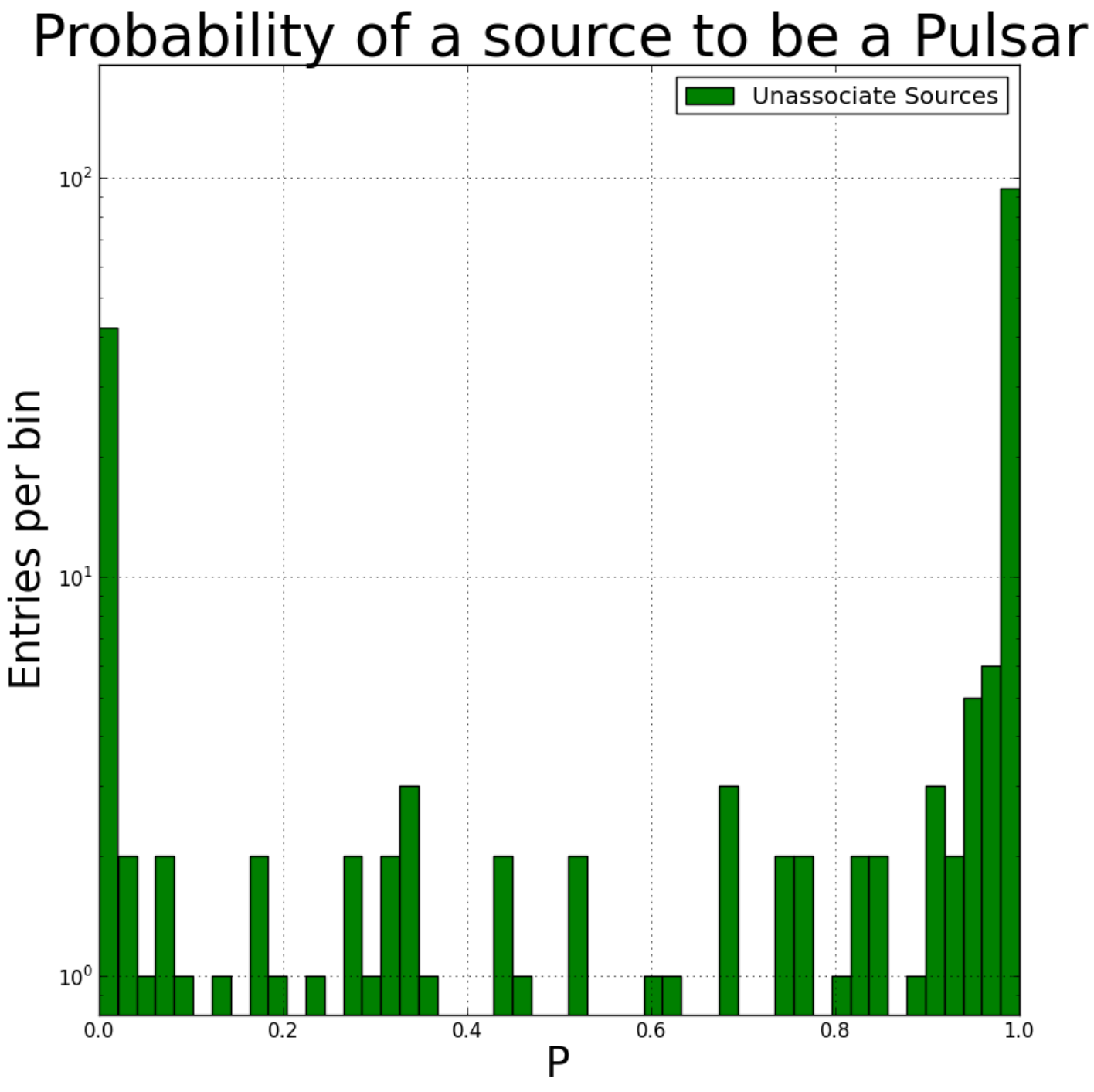

Figure 4.34: Distribution of young pulsar candidates vs. MSP candidates on the basis of the hierarchical ANN predictor for the 131 unidentified sources classified as pulsar candidates by the *ANN1*.

The spatial distribution of the sources newly classified by our hierarchical neural network model is shown in Figure 4.23. Comparing this distribution with those in Figure 4.28 we have the opportunity to cross check our results. Note that both the MSP and young pulsar distributions are as expected, even though we have not used the Galactic latitude as an input to either classification method. The young pulsar candidates are primarily distributed along the Galactic plane, while MSP candidates are mainly distributed at higher latitudes.

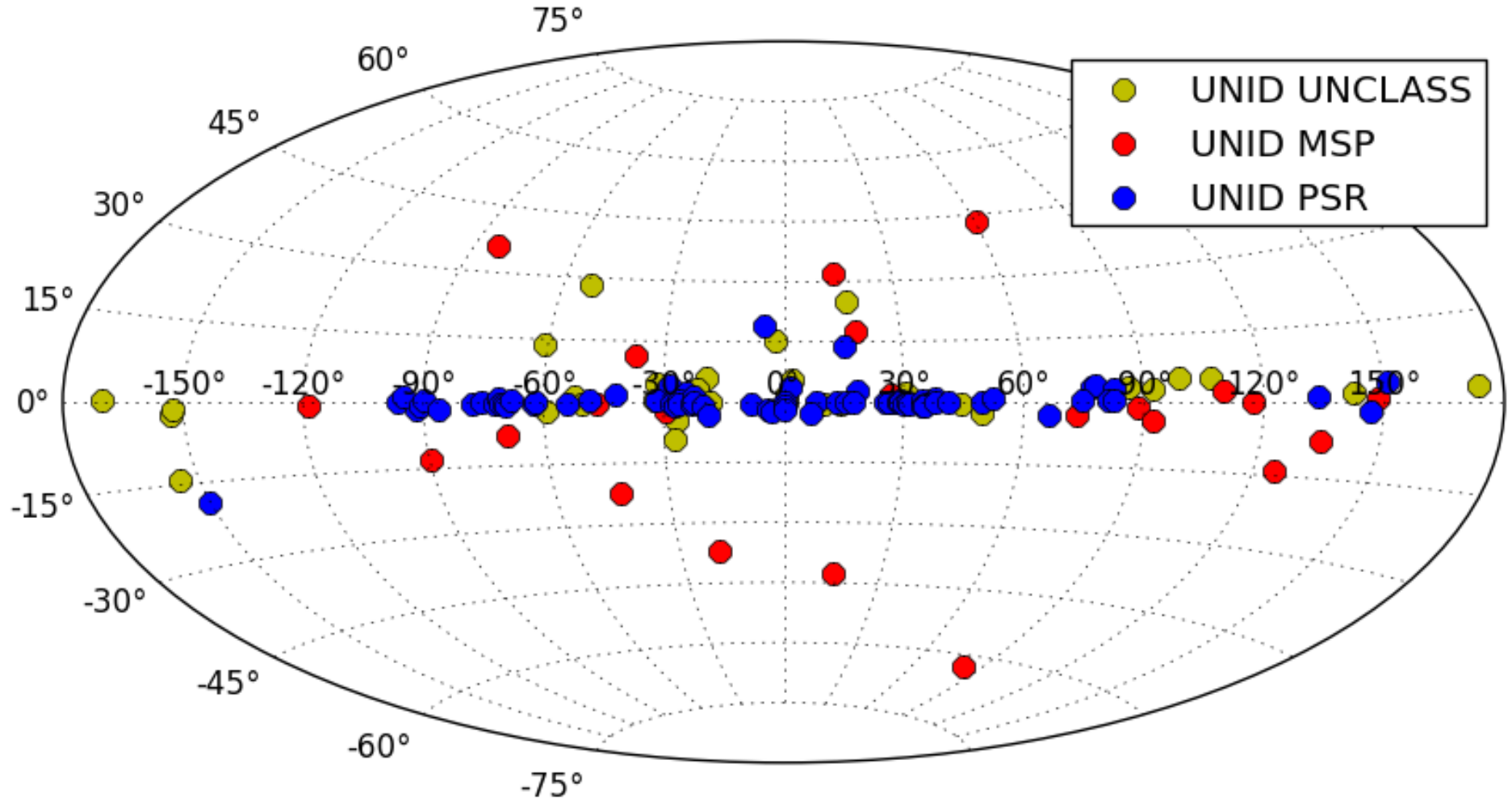

Figure 4.35: Spatial distribution in galactic coordinates of the newly classified sources by our hierarchical neural network model. In blue are shown the young pulsar candidates, in red the MSP candidates and in yellow the pulsars of uncertain type

In Figure 4.36 we show the curvature-variability distribution of the newly classified MSP and young pulsar candidates on the basis of the hierarchical ANN analysis. Comparing this distribution to Figure 4.29, we see that these $\gamma$-ray parameters are not able to separate the associated young pulsars and MSPs (see Table 4.31).

In order to test the capability of the hierarchical neural network model for identifying young pulsar and MSP candidates, we apply the trained ANN model to 24 new pulsars detected (15 young pulsars and 9 MSPs) after the publication of the 2FGL catalog which were classified as pulsar candidates by the *ANN1* (see Section 4.3). Of the 15 newly associated young pulsars, 9 are correctly classified as young pulsar candidates by the ANN analysis (efficiency: 60%) and 4 are classified as MSP candidates (false negative: 27%), while the other source remains unclassified (13%). For MSPs, 7 are correctly classified as MSP candidates by the ANN analysis (efficiency: 78%), no one is classified as young pulsar candidates (false negative: 0%), while the other 2 sources remain unclassified (22%). The young pulsars classified as MSP candidates are PSR J0106+4855, PSR J1422-6138, PSR J2111+4606 and PSR J2139+4716. The association of the second object with that young pulsar has not already been confirmed, further specific analysis will determine its correctness. The other 3 objects were associated with a young pulsar by blind searches [167], PSR J2111+4606 is a young and energetic Galactic-plane pulsar, the other 2 are older and less energetic. PSR J2139+4716 has one of the lowest spin down power among non-recycled pulsars and PSR J0106+4855 has a large characteristic age (3 Myr) and a small surface magnetic field. All these peculiarities may be the origin of their misclassification which keeps occurring in the ANN classification.

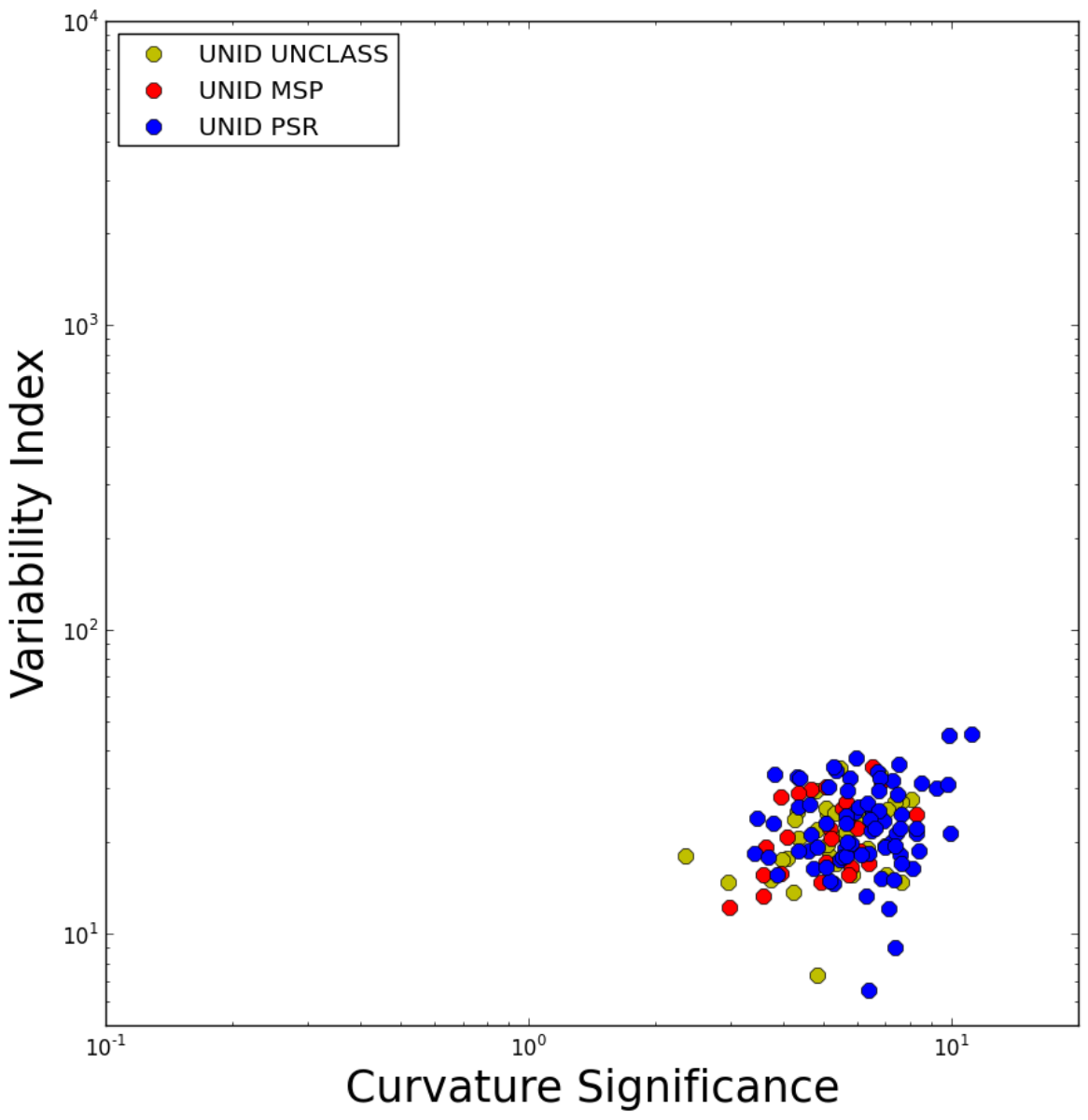

Figure 4.36: Variability index vs. curvature significance for 131 pulsar-like 2FGL unidentified sources classified as MSP (red) and young pulsar candidates (blue) and unclassified (yellow).

As a result, the efficiency of the hierarchical algorithm at classifying new MSP detections is very high (nearby 80%), which is consistent with the efficiency we obtained for the MSPs in the training sample, in spite of the smallness of the sample. Moreover, the absence of false negative is a very encouraging result. Conversely, the efficiency of our hierarchical neural network at classifying new young pulsars is not extremely high (60%) and misclassified objects may be related to some peculiarities which make them "non-standard" young pulsars.

## 4.6 Applying an ANN analysis to distinguish AGN subclasses

The purpose of my Ph.D. thesis is to develop an advanced classification technique to find good radio-quiet MSP candidates in the sample of 2FGL unidentified sources but our hierarchical neural networks can be used to discriminate and classify AGN subclasses. Since non-blazar AGN sources represent only $\sim$ 3% of the entire 2FGL AGN sample, we assume that all AGN-like candidates classified by *ANN1* defined in the Section 4.3 are blazars. For this reason we will develop an ANN to distinguish the two blazar subclasses, which are BL Lacertae (BL Lacs) and flat-spectrum radio quasars (FSRQs), on the basis of their $\gamma$-ray observables.

Blazars are jet-dominated extragalactic objects characterized by the emission of strongly variable and polarized non-thermal radiation across the entire electromagnetic spectrum, from radio waves to $\gamma$-rays [204]. As the extreme properties of these sources are due to the relativistic amplification of radiation emitted along a jet pointing very close to the line of sight [32, 205]. Blazars can be categorized by their optical properties and the shape of their broad-band spectral energy distributions (SEDs). Blazar SEDs always show two broad bumps in the $\log \nu - \log \nu F_\nu$ space; the lower energy one is usually attributed to synchrotron radiation, while the more energetic one is attributed to inverse Compton scattering. Blazars displaying strong and broad optical emission lines are usually called flat-spectrum radio quasars (FSRQs), while objects with no broad emission lines (i.e., rest-frame equivalent width, EW, $<$ 5) are called BL Lac objects. In [163] the terms LBL and HBL was introduced to distinguish between BL Lacs with low and high values of the peak frequency of the synchrotron bump ($\nu^s_{peak}$). Recently this definition was extended to all types of blazars and defined the terms LSP, ISP, and HSP (corresponding to low, intermediate, and high synchrotron peaked blazars) for the cases where $\nu^s_{peak} < 10^{14}$ Hz, $10^{14}$ Hz $< \nu^s_{peak} < 10^{15}$ Hz, and $\nu^s_{peak} > 10^{15}$ Hz, respectively [1]. In the same paper authors showed that the synchrotron peak frequency is positioned between $10^{12.5}$ and $10^{14.5}$ Hz in FSRQs and between $10^{13}$ and $10^{17}$ Hz in featureless BL Lacertae objects. In the rest of this section, we use only the BL Lac and FSRQ nomenclature.

Out of 806 2FGL associated blazars, 436 are BL Lacs and 370 are FSRQs and their distribution on the sky is shown in Figure 4.37. The distribution reflects the extragalactic nature of these objects, while the lack of blazars at low latitude is the effect of diffuse radio emission, Galactic point sources and heavy optical extinction which make the low-latitude sky a difficult region for blazar studies, and catalogs of blazars and blazar candidates often avoid it partially or entirely.

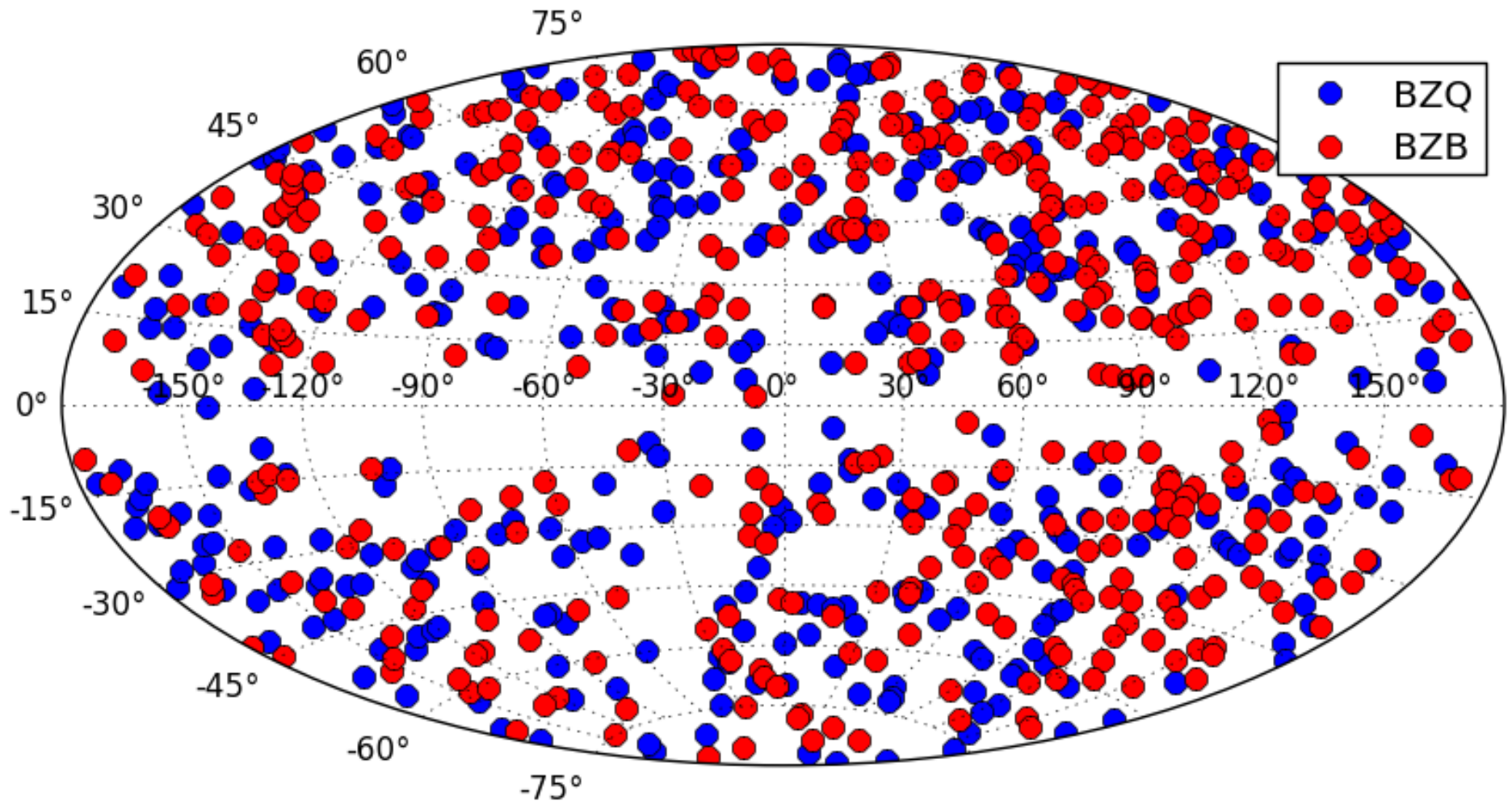

Figure 4.37: Distribution on the sky of 2FGL BL Lacs (indicated as “BZB”) and FSRQs (indicated as “BZQ”).

Figure 4.38 shows the curvature-variability distribution of BL Lacs FSRQs. Curvature significance does not seem significant to distinguish the two blazar subclasses while FSRQs are typically more variable than BL Lacs as discussed in [83]. Differences between variability properties between BL Lac objects and FSRQs at GeV energies are important for understanding the jet location and jet radiation mechanisms, considering that rapid variability is more likely to be related to emission sites near the central nucleus, whereas extended ($\gtrsim$ kpc) jets can only make weakly variable or quiescent emission.

Figure 4.39 shows the PowerLaw index distribution for the two different classes of blazars, BL Lacs and FSRQs. In the LAT energy range FSRQ spectra are softer than BL Lac ones with a moderate overlap as discussed in [83]. Fitting a Gaussian model we obtain that the resulting PowerLaw index distribution mean values and rms are $1.99 \pm 0.25$ and $2.4 \pm 0.19$ for FSRQs and BL Lac objects making this $\gamma$-ray observable a good parameter to distinguish the two blazar subclasses. This feature shows a strong correlation between the $\gamma$-ray spectral slope and the synchrotron peak energy as expected in synchrotron-inverse Compton scenarios [83].

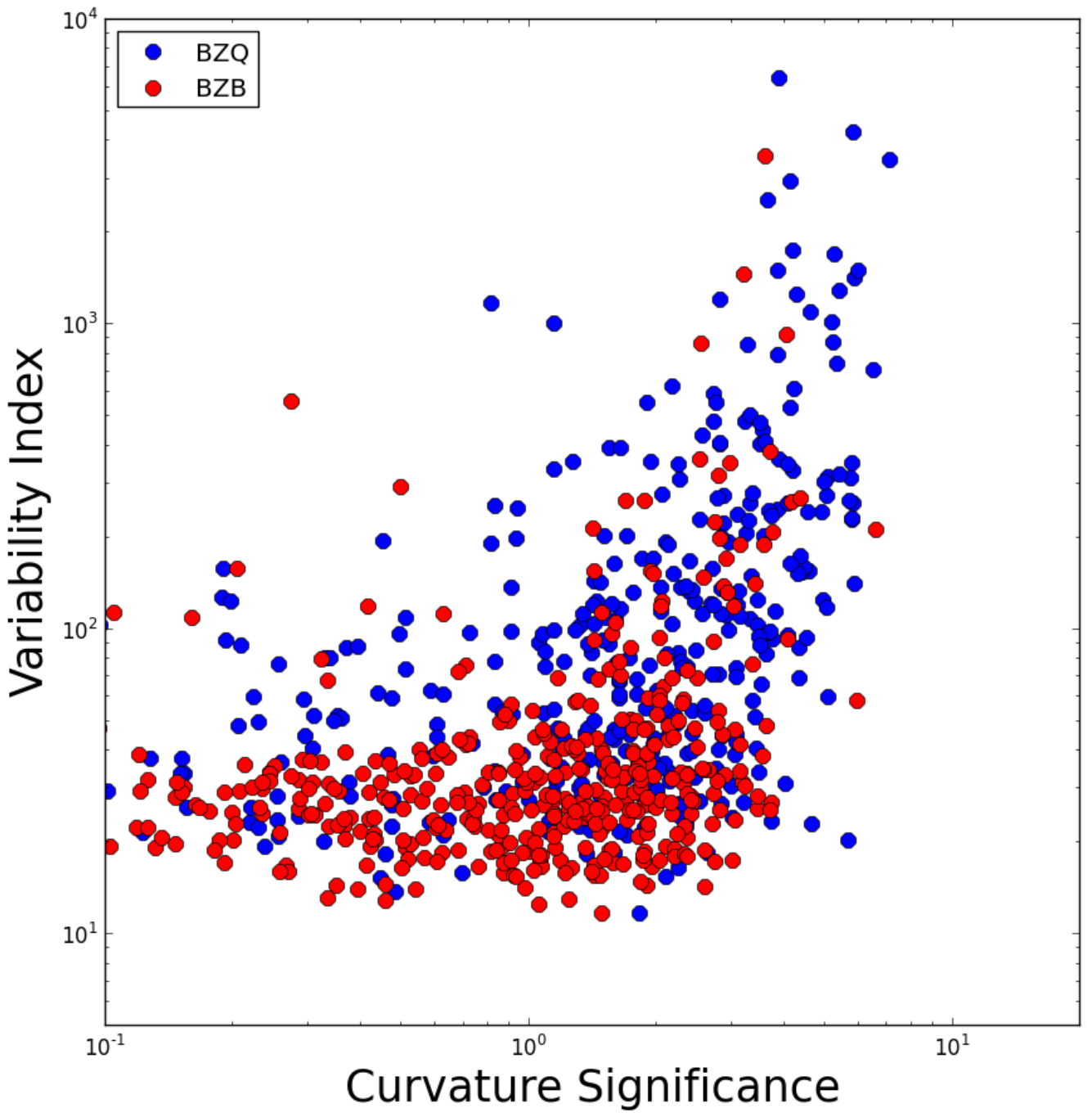

Figure 4.38: Variability index plotted as a function of curvature significance. BZB indicates BL Lac objects while BZQ FSRQs.

In order to perform an ANN analysis to discriminate 2FGL BL Lacs from FSRQs in the $\gamma$-ray observables space we follow the procedure explained in Section 4.3. The first step consists of selecting a sample of data to build the predictor value ($P$), to this aim we focus on the 436 BL Lacs an the 370 FSRQs. The full sample is split in three subsamples used for training, validation and testing. Then we must select the most appropriate set of variables for training the *ANN3*. We select a set of variables that includes the curvature significance, the variability index, the PowerLaw index, the fluxes and the hardness ratios for the 5 energy bands in the catalog. We do not use the Galactic latitude and longitude as input because blazars are extragalactic objects and their distribution on the sky is rather isotropical.

## 4.6.1 Architecture of the ANN

The architecture of *ANN3* is defined by the number of nodes in each layer. The number of input nodes is given by the number of selected predictor variables, in our case 12. During the ANN analysis a pre-processing is performed normalizing each variable to the range [0, 1] as explained in Section 4.3. The number of output nodes is given by the number of source classes we want to classify, in our case 2 (BL Lacs and FSRQs). For *ANN3* we use an hyperbolic tangent activation function both for hidden and output nodes and the target vector for BL Lacs is $t = [1, -1]$ while for FSRQs is $\mathbf{t} = [-1, 1]$. The number of hidden nodes is determined by the pruning procedure as explained is Section 4.3.2. In Figure 4.40 in blue is shown the value of the mean square error defined in Equation 4.4 as a function of the number of hidden nodes while in red the performance of our network in terms of the fraction of correct classifications of the validation as a function of the number of hidden nodes. Using this procedure we decide to use 8 hidden nodes.

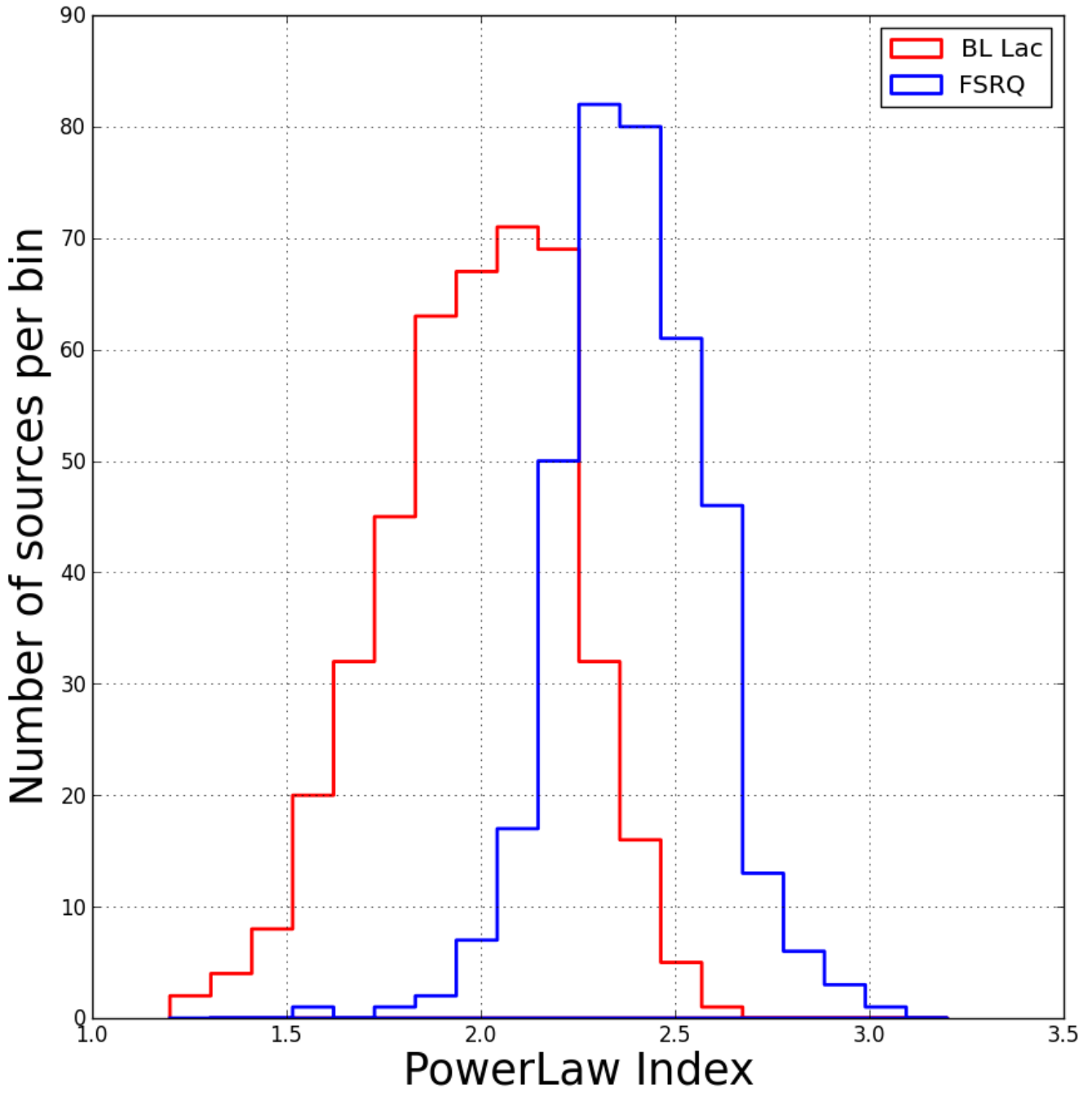

Figure 4.39: PowerLaw Index distribution for 2FGL blazar subclasses.

As a result the *ANN3* is built up of 12 input nodes, 8 hidden nodes and 2 output nodes, each node in a layer is linked to all the nodes of the next layer and to each link is associated a weight randomly initialized in the range $[-1, 1]$.

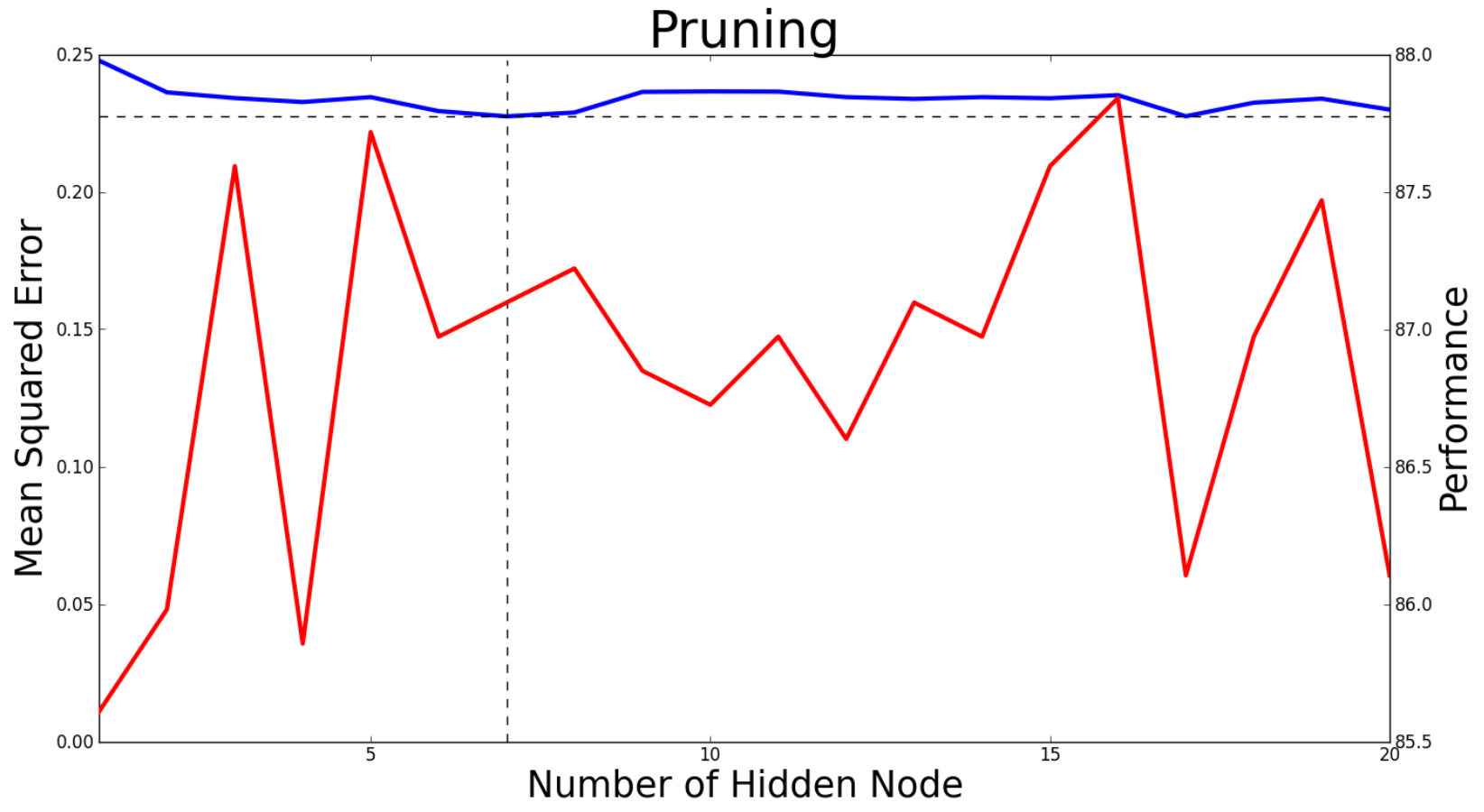

Figure 4.40: (blue) Value of the mean square error (mse) defined in Equation 4.4 as a function of the number of hidden nodes. (red) Performance of the network in terms of fraction of correct classifications of the objects in the validation sample as a function of the number of hidden nodes.

### 4.6.2 Training session and classification thresholds

Given the network architecture, we can train the *ANN3* to distinguish BL Lacs from FSRQs. We follow the procedure described in the Section 4.3.3 and we use the back-propagation learning algorithm, which is applied in the "online" version (see Equation 4.5).

Table 4.5 ranks the relative importance of the different variables at distinguishing BL Lacs from FSRQs according to the Equation 4.6. As expected, PowerLaw index is the most significant $\gamma$-ray parameter at distinguishing FSRQs from BL Lacs, also variability index is relatively important, while curvature significance is the less important. An interesting result is that also $HR_{23}$ and $HR_{34}$ are rather significance at distinguishing the two blazar subclasses, this means that, despite their spectral shape can be described very well from a broken power law, their spectra are significantly different in the energy range between 0.3 and 10 GeV, i.e. where there is the "break" energy. This result is very important because can be useful to constrain any $\gamma$-ray emission model for FSRQs and BL Lacs, they are typically characterized by the same spectral model (i.e. broken power law), but the break energy is different for the two blazar subclasses. All other $\gamma$-ray parameters, fluxes in the five energy range, $HR_{12}$ and $HR_{45}$, are not important at distinguishing FSRQs from BL Lacs. These results are very encouraging because confirm what we expect in synchrotron-inverse Compton scenarios. Moreover, the result suggesting a different brake energy for FSRQs and BL Lac is very interesting and it must be analyzed in more detail to improve the relative $\gamma$-ray emission models.

| **Variable** | **Importance** |
|---|---|
| (0) curvature significance | 0.16 (0.37%) |
| (1) Variability Index | 6.90 (16.05%) |
| (2) PowerLaw Index | 17.36 (40.39%) |
| (3) $\text{Flux}_{0.1-0.3GeV}$ | 0.60 (1.39%) |
| (4) $\text{Flux}_{0.3-1GeV}$ | 0.25 (0.59%) |
| (5) $\text{Flux}_{1-3GeV}$ | 1.04 (2.42%) |
| (6) $\text{Flux}_{3-10GeV}$ | 0.26 (0.6%) |
| (7) $\text{Flux}_{10-100GeV}$ | 0.44 (1.03%) |
| (8) $\text{Hardness}_{12}$ | 1.58 (3.68%) |
| (9) $\text{Hardness}_{23}$ | 6.52 (15.17%) |
| (10) $\text{Hardness}_{34}$ | 6.57 (15.3%) |
| (11) $\text{Hardness}_{45}$ | 1.29 (3.0%) |

Table 4.5: List of the training variables for the *ANN3*: each variable is ranked according to its relevance in the discrimination process (see Equation 4.6), as computed by the *ANN3* algorithm.

Figure 4.41 shows the correlation matrix of the variables selected for our analysis as desribed by the Equation 4.7.

Figure 4.42 shows the results of the learning process and the performance of *ANN3*. The top panel shows the mean squared error for training, validation and testing samples during the learning process as a function of the epoch. In the middle panel is shown the classification table. This table suggests that the optimized *ANN3* is able to distinguish BL Lacs from FSRQs on the basis of their $\gamma$-ray observables because the accuracy and the precision are rather high. The total performance of *ANN3* is given by the total accuracy in the block blue and it is $> 86\%$. In the bottom panel the distribution of the ANN predictor for the 2FGL BL Lacs (Class 1) and FSRQs (Class 2) is shown. This distribution shows that BL Lacs are typically characterized by an high predictor value, while FSRQs by a lower one.

In order to efficiently classify 2FGL unidentified sources as BL Lac or FSRQ candidates we select two classification thresholds, one to classify a BL Lac candidate ($C_{BL}$) and one to classify a FSRQ candidate ($C_{FSRQ}$) so that 2FGL sources with a predictor greater than 0.758 are classified as BL Lac candidates while the sources with a predictor smaller than 0.429 are classified as FSRQ candidates. Sources with an intermediate value of the predictor remain blazar of uncertain type. As a result of this choice, 80% of BL Lacs in 2FGL have a predictor greater than $C_{BL}$ while 80% of FSRQs have a predictor smaller than $C_{FSRQ}$. The results are shown in the table of classification in Figure 4.43.

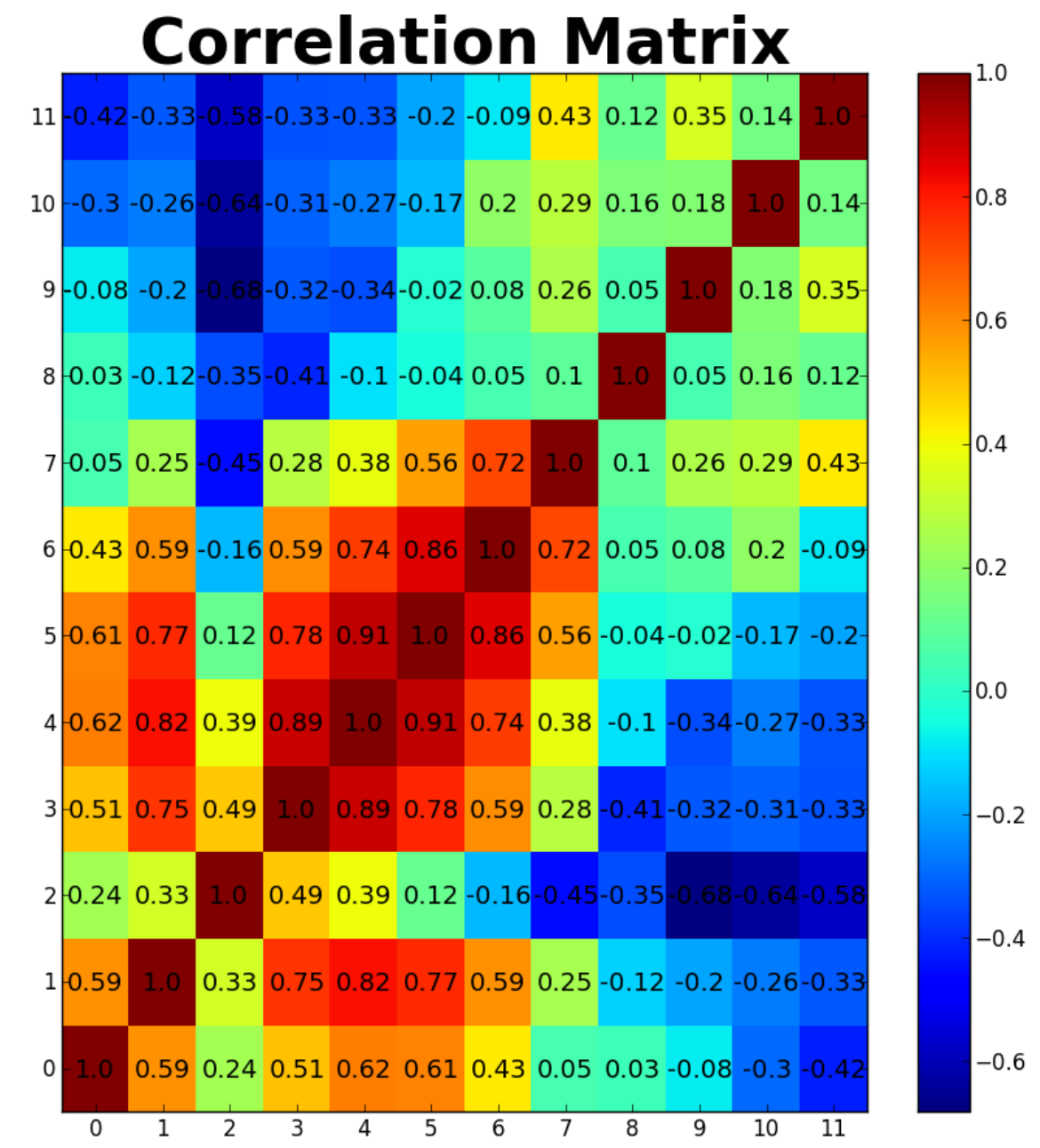

Figure 4.41: Correlation matrix of the variables selected for our analysis. The number of each parameter is expressed in Table 4.5 and to each correlation coefficient is associated a color on the basis of its value.

### 4.6.3 Results and their validation

At this point we apply the trained hierarchical neural network to 806 2FGL blazars in order to determine the total efficiency and accuracy at distinguishing BL Lacs from FSRQs. Of 327 BL Lacs classified as AGN candidates by the *ANN1* (see Section 4.3), 251 are correctly classified as BL Lac candidates by the *ANN3* analysis (efficiency: 77%) and 29 are classified as FSRQ candidates (false negative: 9%), while the other sources remain unclassified (14%). For 321 FSRQs classified as AGN candidates, 263 are correctly classified as FSRQ candidates by the *ANN3* analysis (efficiency: 82%), 19 are classified as BL Lac candidates (false negative: 6%), while the other sources remain unclassified (12%). These results are encouraging because the efficiency and accuracy of the hierarchical neural network at distinguishing BL Lacs from FSRQs are compatible with those found using a single neural network (*ANN3*).

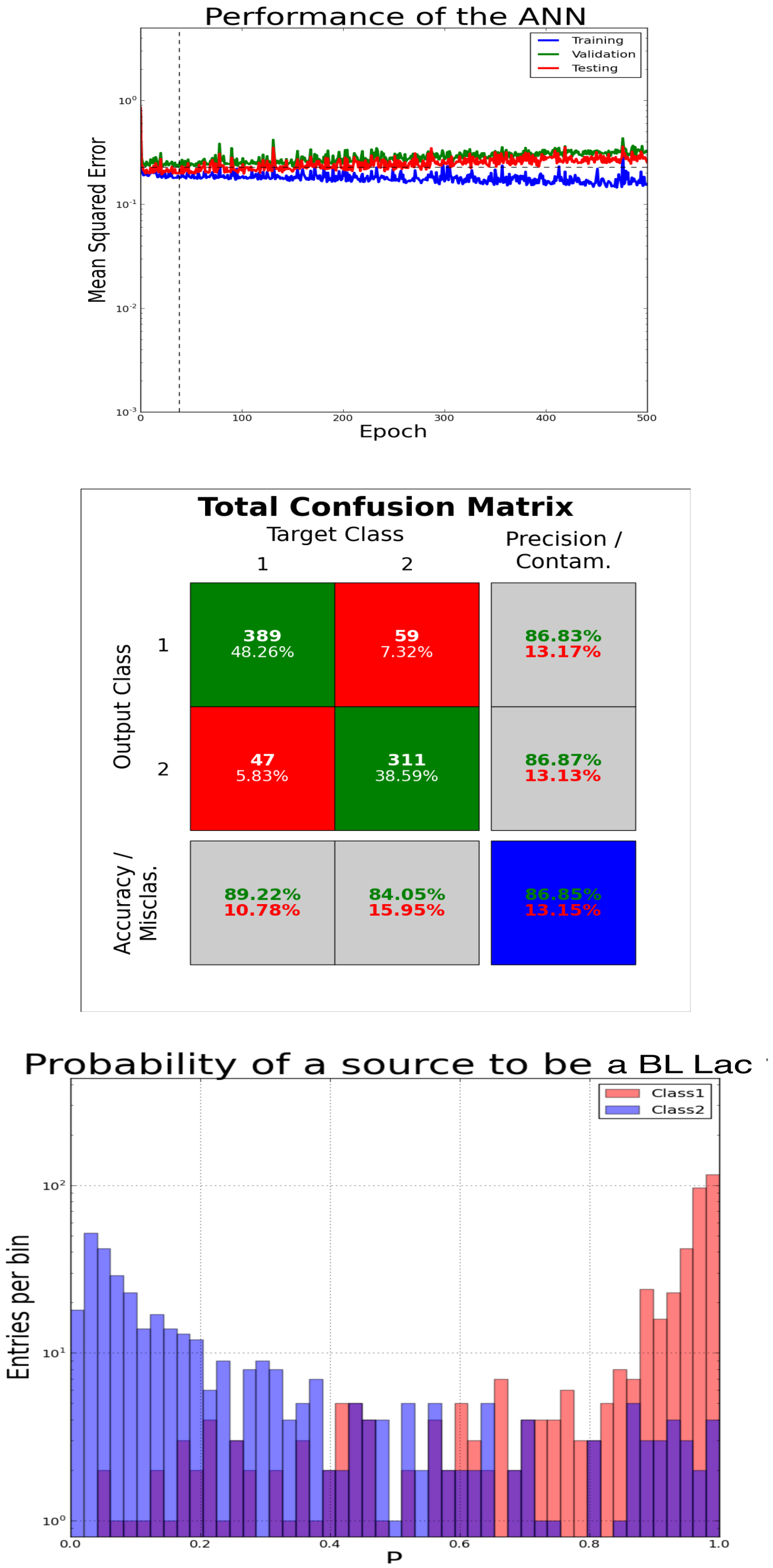

Figure 4.42: Performance of *ANN3*. Top: mean squared error for training, validation and testing samples during the learning process as a function of the epoch. The vertical dotted line specifies the epoch the learning process is stopped (optimized network). Middle: classification table. Bottom: distribution of the *ANN2* predictor for the 2FGL BL Lacs (indicated as “Class 1”) and FSRQs (indicated as “Class 2”). Violet indicates a superimposition of the distribution of BL Lacs and FSRQs.

| | BL Lac observed | FSRQ observed | Efficiency/ Contam. |
|---|---|---|---|
| BL Lac predicted | 348 | 28 | 93%<br>7% |
| FSRQ predicted | 37 | 297 | 89%<br>11% |
| Unclassified | 51 | 45 | |
| Accuracy/ Misclass. | 80%<br>8% | 80%<br>8% | 91%<br>9% |

Figure 4.43: Table of classification based on the rules described in the text. In green are represented the correct classifications and in red the incorrect ones.

Applying the optimized hierarchical neural network to the 2FGL sources associated with a class different from blazar we can estimate their contamination to the candidate BL Lac and FSRQ samples from the likely presence of these objects in the unassociated sample. The *ANN1* has classified 23 non-blazar objects as AGN candidates, they are 1 pulsar (2FGL J1823.4-3014), 3 starburst galaxies, 7 radio galaxies, 8 AGNs, 2 objects in the field of the LMC, the Andromeda galaxy M31 and the nova. Of these, 13 are classified as BL Lac candidates by the *ANN3* analysis (the pulsar, 2 starburst galaxies, 4 radio galaxies, 4 AGNs and the 2 objects in the field of the LMC), 6 are classified as FSRQ candidate (the nova, 3 AGNs and 2 radio galaxies) while 4 sources remain unclassified (1 starburst galaxy, 1 radio galaxy, 1 AGN and the galaxy M31). We observe that the contamination to the candidate BL Lac and FSRQ samples is very low, thus irrelevant.

Applying the trained hierarchical neural network to the 2FGL unidentified sources we find that, out of 176 classified as AGN candidates by the *ANN1*, 84 are classified as BL Lac candidates ($P > C_{BL}$), 66 are classified as FSRQ candidates ($P < C_{FSRQ}$) and 26 remain unclassified. In Figure 4.44 is shown the distribution of the 176 2FGL unidentified sources classified as AGN candidates by the *ANN1* as a function of the probability of being BL Lacs.

The PowerLaw index distribution of the sources newly classified by our hierarchical neural network model is shown in Figure 4.45. Comparing this distribution with those in Figure 4.39 we have the opportunity to cross check our results. Fitting a Gaussian model we obtain that the resulting PowerLaw index distribution mean values and rms are $2.02 \pm 0.21$, $2.56 \pm 0.16$ and $2.33 \pm 0.2$ for FSRQs and BL Lac candidates and blazars of uncertain type. These results are very encouraging because are compatible with those found for associated blazars, FSRQ candidates are characterized by softer specra than BL Lac candidates ones, while blazars of uncertain type have an intermediate value as we expect.

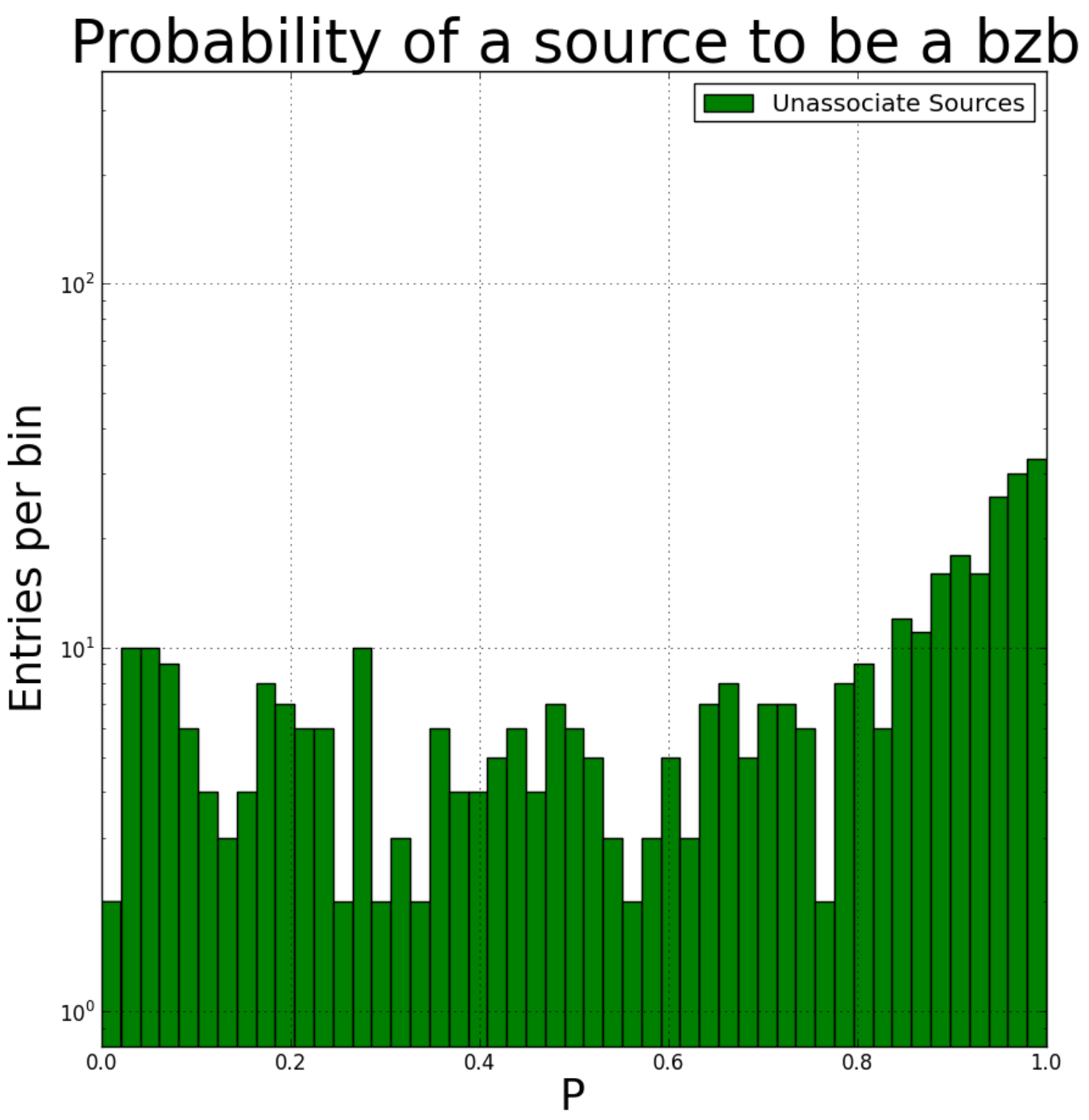

Figure 4.44: Distribution of BL Lac candidates vs. FSRQ candidates on the basis of the hierarchical ANN predictor for the 176 2FGL unidentified sources classified as AGN candidates by the *ANN1*. Sources with $P > 0.758$ are classified as BL Lac candidates, while sources with $P < 0.429$ are classified as FSRQ candidates, the others are AGNs of uncertain type.

In order to test the capability of the hierarchical neural network model for identifying BL Lac and FSRQ candidates, we apply the trained hierarchical ANN model to 28 new blazars associated (23 BL Lacs and 5 FSRQs) after the publication of the 2FGL catalog which were classified as AGN candidates by the *ANN1* (see Section 4.3). Of the 23 newly associated BL Lacs, 17 are correctly classified as BL Lac candidates by the ANN analysis (efficiency: 74%) and 3 are classified as FSRQ candidates (false negative: 13%), while the other sources remain unclassified (13%). For FSRQs, 4 are correctly classified as FSRQ candidates by our analysis (efficiency: 80%), one is classified as BL Lac candidate (false negative: 20%), while no one remains unclassified (0%). The efficiency of the hierarchical algorithm at classifying new BL Lacs and FSRQs is very high (nearby 80%), which is consistent with the efficiency we obtained for them in the training sample, in spite of the smallness of the sample.

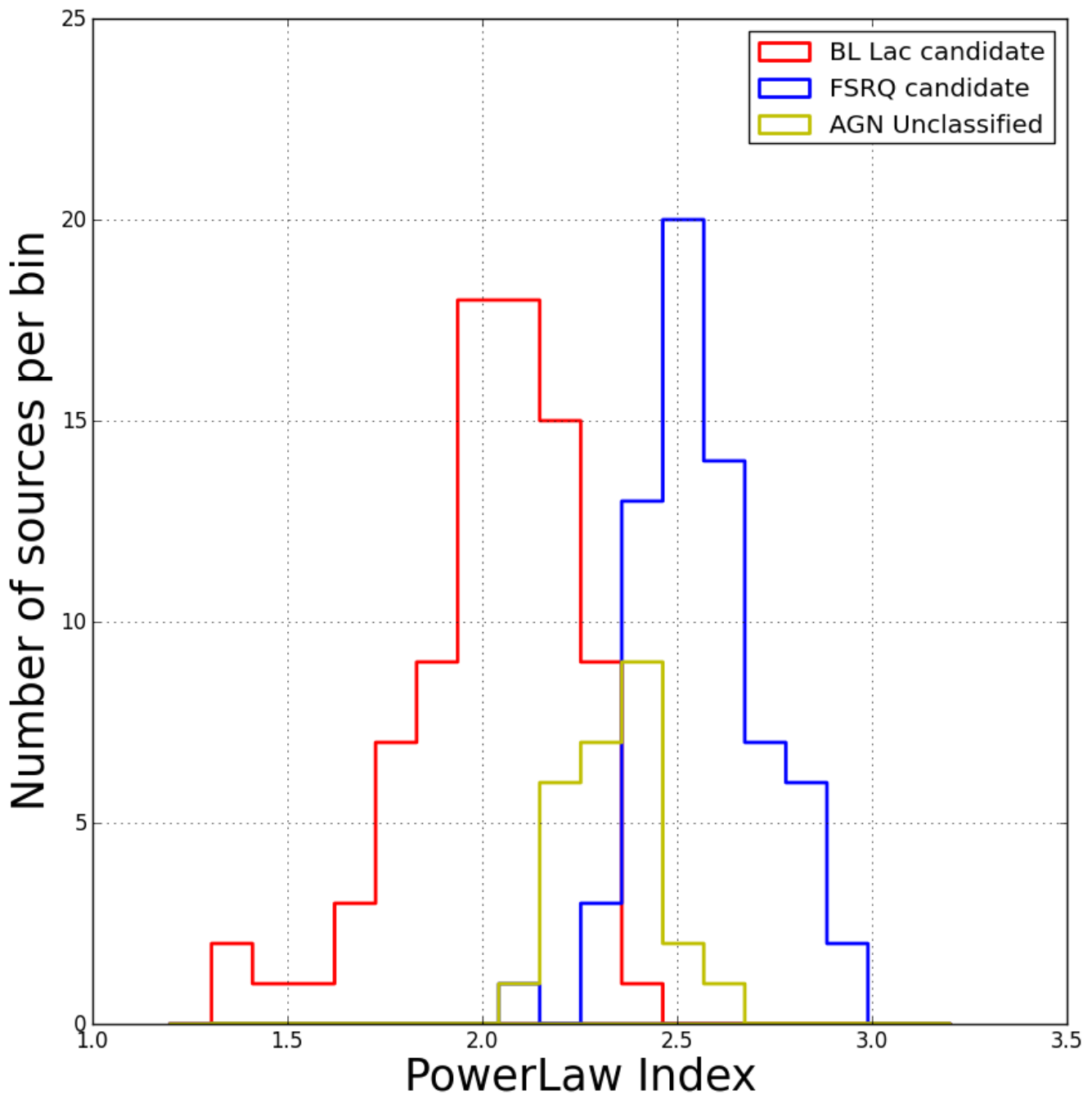

Figure 4.45: Distribution of the PowerLaw Index of the newly classified sources by our hierarchical neural network model. In red are shown BL Lac candidates, in blue FSRQ candidates and in yellow the blazars of uncertain type.

We have not included active galaxies of uncertain type in the "contamination" sample because they are associated with AGNs on the basis of advanced techniques described in [83] which are different from those applied in the 2FGL catalog ([158]), their counterparts are without a good optical spectrum or an optical spectrum at all making impossible their classification as blazar. We can apply our hierarchical neural network to classify them as BL Lac or FSRQ candidates. Applying the trained *ANN3* to these sources we find that, out of 162 classified as AGN-like sources by the *ANN1*, 84 (52%) are classified as BL Lac candidates ($P < C_BL$), 59 (36%) as FSRQ candidates ($P > C_FSRQ$) and 19 (12%) remain blazar of uncertain type. The distribution of 2FGL active galaxies of uncertain type as a function of the probability of being BL Lacs is shown in the Figure 4.46.

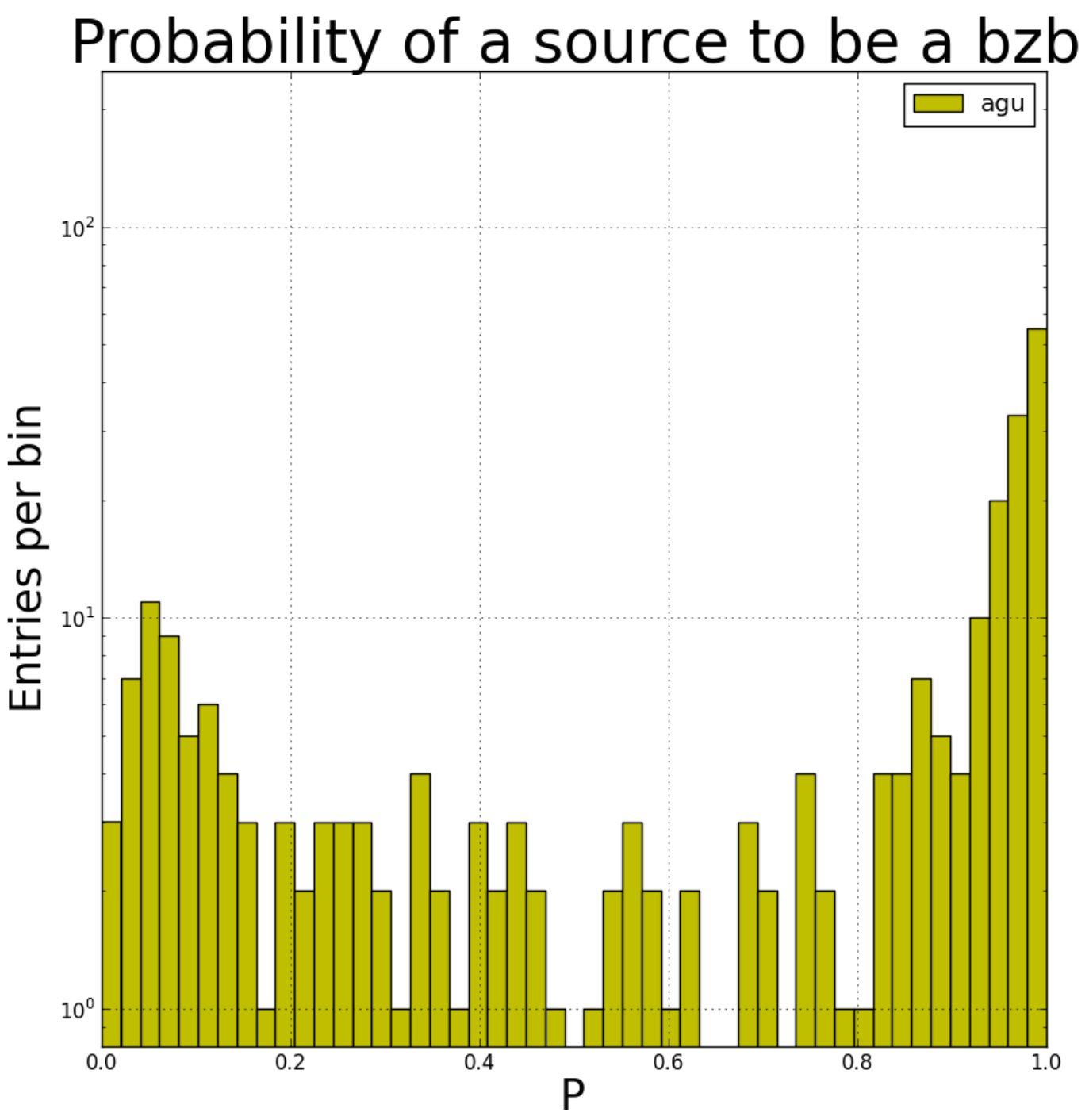

Figure 4.46: Distribution of the *ANN3* predictor for the 162 2FGL active galaxies of uncertain type classified as AGN-like sources using *ANN1*.

## 4.7 Summary

Gamma-ray sources are tracers of the most energetic processes in the Universe, they are exotic objects, characterized by very intense magnetic and electric fields and very high-energy particles. For this reason, understanding the nature of the $\gamma$-ray unidentified sources is one of the most important open question in high-energy astrophysics. These sources represent an important discovery space for new members of existing $\gamma$-ray source classes (e.g. AGNs or pulsars), or new source classes. In this Chapter the development and the application of two different machine learning techniques aimed at determining likely source classifications for the 2FGL unidentified sources on the basis of their $\gamma$-ray observables has been presented. The algorithms presented are the starting point for selecting targets for multi-wavelengths pulsar searches as shown in the next Chapter.

Encouraged by the results obtained through the application of the logistic regression at classifying 1FGL unidentified sources, we have started to imple-ment a refined logistic regression to discriminate 2FGL pulsars and AGNs, the two most numerous source classes in the 2FGL source catalog. The larger sample and the refinement of the $\gamma$-ray parameters in the 2FGL catalog have allowed us to significantly increase the performance of our classification technique at classifying 2FGL unidentified sources. Logistic regression is a generalized linear classifier and it classifies 2FGL sources through a linear distinction in the $\gamma$-ray variable space, this is a limit of this technique. Performance in classifying pulsar and AGN candidates in 2FGL source sample increases considerably using a neural network techniques, which classifies 2FGL sources through a more complex non-linear separation. In order to control each step of the procedure, we have developed our own code. This has been the first time that ANN technique is applied for this purpose.

Encouraged by ANN results we have implemented a more complex neural network architecture to distinguish the two pulsar subsamples, young pulsars and MSPs. For this purpose we have used a hierarchical neural network model. Although $\gamma$-ray properties of young pulsars seems very similar to MSPs, our model is able to efficiently discriminate these two source classes showing that source properties measured with the *Fermi*-LAT can provide important guidance on target selections as well as the types of follow-up observations. Our results has shown some differences in the $\gamma$-ray emission between young pulsars and MSPs, they should be used to develop a theory that explains these differences in order to better understand the emission mechanisms of these exotic objects.

In the end, we have modified the hierarchical neural network previously developed in order to discriminate also the two blazar subclasses, BL Lacs and FSRQs. Our results are very stimulating because confirm what we expect from synchrotron-inverse Compton scenarios validating the accuracy of our approach. Moreover, our results provide some indications to improve and better understand the $\gamma$-ray emission models of blazar subclasses.

The complete list of our hierarchical neural network predictions unassociated *Fermi*-LAT sources in the 2FGL catalog is shown in the Table C.1 in Appendix C.1.

After the publication of the 2FGL catalog different association methods were applied in order to classify the 2FGL unidentified sources. Some of them use only $\gamma$-ray information (e.g. in [148] the authors assign a probability to each 2FGL unidentified source to be more similar to a pulsar or an AGN on the basis of a Random Forest algorithm, an extension of the Classification Tree described in [16]), while others include also multi-wavelength information (e.g. in [138] the authors use IR information to recognize $\gamma$-ray blazar candidates in the sample of 2FGL unidentified sources). Combining our results with those found by other association techniques it is possible to build a classifier char-acterized by a higher performance in the classification of 2FGL unidentified sources. The positive results from all techniques could be used to generate a new candidate list aimed at providing "stronger" candidate sources for followup multi-wavelength studies.

Chapter 5

# Multi-wavelength analysis of some interesting 2FGL sources

After only 24 months of mission the *Fermi*-LAT has detected a number of $\gamma$-ray sources that is about an order of magnitude larger than those detected by its predecessor EGRET during its entire mission. About 70% of the sources in the 2FGL catalog have been associated with an object known at other wavelengths through an objective procedure. Some of the counterparts was known before detecting their high-energy emission by the LAT while others were discovered by multi-wavelength observations in the field of the 2FGL sources or by using LAT data alone. Many pulsars were discovered using blind frequency searches for $\gamma$-ray pulsations from the bright unidentified sources. These are typically young pulsars, for which the solid angle of the radio beam is likely to be much smaller than the $\gamma$-ray one [9]. By now, no radio-quiet millisecond $\gamma$-ray pulsar was discovered using blind search algorithms. A limit of these algorithms is that they cannot be performed scanning the entire 2FGL error circle to search for a possible periodicity of some *millisecond* directly in $\gamma$ rays because of the prohibitive request for computing power. To increase the efficiency of the algorithm it is possible to perform deep X-ray observations of the most probable candidates MSPs so that a blind search can be run on much smaller sky area covered by the error regions of the detected X-ray sources, with a large gain in computer time. In Section 5.1 we will describe multi-wavelength analysis of the counterparts of 3 MSP candidates. We have selected these targets on the basis of the results of the hierarchical neural network analysis described in the previous chapter, their position in the sky and the unsuccessful radio searches within their error circle.

*Fermi*-LAT is changing our knowledge of $\gamma$-ray Universe, it is giving us the opportunity to understand which astrophysical objects are able to emit $\gamma$ rays and which models are able to explain their $\gamma$-ray emission mechanisms.

We are understanding that not all members of the same source class are able to generate and emit $\gamma$ rays, multi-wavelength studies are necessary to discover which are the causes and the differences among these sources. Moreover, since $\gamma$-ray sources are tracers of the most energetic processes in the Universe, multi-wavelength studies are very important to understand the physics of these exotic objects characterizing their and exploring their environment. Regarding pulsars, the most important objects to constrain their high-energy emission models are the "extreme" ones, accounting for the trails of the population distribution in energy, age and magnetic field. In this respect, in Section 5.2 we will describe X-ray studies on pulsar J0357+3205. Its X-ray counterpart [69] has been identified after that the *Fermi*-LAT detected its $\gamma$-ray emission. In the same X-ray investigation, it has also been identified a very peculiar elongated structure of diffuse X-ray emission, originating at the pulsar position and extending for 9 arcminutes in the sky. PSR J0357+3205 is a middle-aged radio-quiet $\gamma$-ray pulsar and it was discovered with a blind search algorithm in the first three months of *Fermi*-LAT scanning [7]. PSR J0357+3205 is not the oldest rotation-powered pulsar, it is the slowest rotator among Fermi pulsars, with a spin-down luminosity very low, which makes it one of the not-recycled $\gamma$-ray pulsar with the smallest rotational energy loss detected so far. This suggests that PSR J0357+3205 is quite close to us, making PSR J0357+3205 a natural target for X-ray observations.

## 5.1 Searching for radio-quiet MSPs among the 2FGL unidentified sources

The *Fermi* Large Area Telescope (LAT) revolutionized pulsar astronomy by discovering pulsed $\gamma$-ray emission from more than 100 rotation-powered pulsars [200]. Approximately a third are radio-loud young pulsars, a third are radio-quiet young pulsars (discovered in blind searches of LAT data) and a third are radio-loud, MilliSecond PSRs (MSP). Such samples opened the possibility to understand the physics of pulsar magnetospheres (the "pulsar engine") and the evolution of the pulsar population with age. It is now apparent that the $\gamma$ rays in young pulsars come from the outer magnetosphere, leading to a much broader beam than what is observed in radio.

MSPs are - so far - the least understood objects. The emitting regions and beaming pattern of such extremely fast-rotating sources seem rather different with respect to the ones of normal young pulsars, possibly due to the compactness of their magnetospheres. All $\gamma$-ray MSPs seen by the LAT were detected by folding the $\gamma$-ray event times using a radio ephemeredid. One key question remains to be answered, and that is whether, like the case of young pulsars, there exist a population of radio-quiet millisecond $\gamma$-ray pulsars. Discovering a radio-quiet MSP would have a formidable impact for our understanding of the configuration of MSP magnetospheres. However, this is a very challenging task, and requires the help of X-ray astronomy, as explained in the followings.

While for radio-loud pulsars an accurate, contemporaneous radio ephemeris eases the $\gamma$-ray pulsation search, since $\gamma$-ray photon time of arrivals can be directly folded according to the known radio timing solution, for a radio-quiet pulsar the periodicity has to be searched directly in $\gamma$-ray data, using a "blind search" algorithm (see Chapter 3). LAT blind frequency searches have been very successful in discovering normal young pulsars. However, blind search algorithms are very sensitive to positional offsets, and the higher the frequency of the pulsar, the more sensitive the search is to incorrect starting positions. For a MSP, a positional error as small as 0.5 arcsec can completely wash out the pulsed signal. Thus, a very fine step have to be used to scan the error region of the $\gamma$-ray source, repeating the search for each position of the grid. Using a $\sim 0.2''$ step to scan a LAT source error region with a radius of (typically) $\sim 5$ arcmin results in a prohibitive request for computing power, which has hampered any search for a radio-quiet MSP so far. A possible solution is to explore the error region with a deep X-ray observation that could detect the X-ray counterpart of the putative pulsar. Any blind search could be run on the much smaller sky area covered by the error regions of the detected X-ray sources, with a large gain in computer time, assuming an X-ray error circle of $3''$, the gain in computer time would be of order $10^3$, making the detection of the first radio-quiet MSP a distinct possibility.

We have based the selection of our targets on both their probability of being a MSP (as measured by P) based on the hierarchical neural network analysis described in the Section 4.5, and their Galactic latitude, in order to avoid the confused region of the Galactic plane, and because we expect nearby MSPs to be preferentially located off the plane. Our targets must have already been the object of deep unsuccessful radio searches, so that the discovery of a MSP would ensure that it is at least extremely radio faint, if not radio-quiet. The blind frequency search for pulsars with LAT data has been very successful during the first years. However, most of the pulsars discovered so far have been very bright, and fairly high spin-down luminosity pulsars. Pushing down the sensitivity of the blind searches to lower fluxes and weaker pulsars will require improvements in the current search techniques. As longer integration times become necessary, further complications arise. We have selected our targets so that their source significance is high (TS>100). Also positional uncertainties are important. While the LAT represents a significant improvement over EGRET, the typical 95% error radii of LAT sources are still on the order of several arcminutes, in the best of cases. An error in the position of $\epsilon$ (in radians), leads to a Doppler shift form the Earth's orbital motion as well as to a shift in the frequency derivative [49]. Since this shift is proportional to frequency, they are particularly important for MSP searches. Indeed, given a position accurate to one arcsecond, blind search codes are able to detect known isolated $\gamma$-ray MSPs and young pulsars in blind searches of LAT data. However, such task becomes daunting when the entire LAT error region of several square arcminutes must be searched on this fine step grid, considering that it currently takes us ∼10–15 minutes to search each position. Even X-ray observations result in ∼10 possible counterparts, the gain in computer time make possible the detection of the first radio-quiet $\gamma$-ray MSP.

| Target 2FGL J | b (°) | r95% (′) | Flux erg cm$^{-2}$ s$^{-1}$ | TS | P (%) | X-ray observation | time ks |
|---|---|---|---|---|---|---|---|
| 1036.1–6722 | -7.84 | 3.7 | $1.99 \times 10^{-11}$ | 330 | >99.9 | XMM | 25 |
| 1539.2–3325 | 17.53 | 4.3 | $1.05 \times 10^{-11}$ | 130 | >99.9 | *Swift* | 85 |
| 1744.1–7620 | -22.48 | 3.3 | $2 \times 10^{-11}$ | 450 | >99.9 | XMM | 25 |

Table 5.1: List of requested and analyzed targets.

On the basis of criteria explained above we have selected and observed in X-ray the 3 most promising MSP radio-quiet candidates. These 2FGL unidentified sources and some relative information are shown in Table 5.1. Column (1) LAT 2FGL catalog source name; (2) Galactic latitude; (3) positional uncertainty (in arcmin); (4) energy flux at energies above 100 MeV; (5) source significance (Test Statistic); (6) probability that the source is a MSP, according to the hierarchical neural network analysis; (4) X-ray satellite requested for the observations; (5) X-ray observation exposure time.

### 5.1.1 Multi-wavelength analysis procedure

In this section we describe the procedure to select the most probable multi-wavelength counterparts for each selected 2FGL unidentified sources. X-ray observations are used to detect X-ray sources within the 99% error circle of the *Fermi*-LAT source, for these objects we produce X-ray spectra and light curve and we search for optical/IR counterparts in order to classify the X-ray sources.

#### XMM-Newton data analysis

For each *XMM-Newton* observation we have analyzed data collected by the EPIC instruments, the PN camera [194] and the two MOS detectors [203]. All the data reduction are performed using the most recent release of the *XMM-Newton Science Analysis Software* (SAS) v13.0. First of all we perform standard data processing, using the `epproc` and `emproc` tools, and screening for high particle background time intervals following [70].

Source detection using maximum likelihood fitting is done simultaneously on each of the EPIC-PN, MOS1, and MOS2 using the SAS tool `edetect_chain`. This tool runs on the event lists and invokes several other SAS tools to produce background, sensitivity, and vignetting-corrected exposure maps. A likelihood threshold of 10 is used for source detection, corresponding to a significance level of $3.6\sigma$. We focus on the X-ray analysis of the sources within the 99% error circle of the *Fermi*-LAT source in order to produce X-ray spectra, light curves and find optical/IR counterparts.

We accumulate the source spectra by selecting only events within a circle of $20''$ radius containing the source detected with PATTERN = 0–4 for the PN and PATTERN = 0–12 for the two MOS detectors and the standard selection filter FLAG = 0 and we generate ad hoc response matrices and ancillary files using the SAS tasks `rmfgen` and `arfgen`. Background photons are extracted within a circle of $45''$ radius from suitable regions on the same CCD chip containing the source counts. Before spectral fitting, all spectra are binned with `grppha` with a minimum of 25 counts per bin in order to be able to apply the $\chi^2$ minimization technique. In this process, the background count rate is rescaled using the ratio of the source and background areas. Then we fit the source spectra with three spectral models: *power-law*, well-suited for AGNs as well as pulsars, *apec*, well-suited for stellar coronae, and *black-body*, well-suited for pulsars[1]. In all cases we also include the absorption by the interstellar medium, we try to both leave it as a free parameter and fix it to the value of the hydrogen column density along the line of sight [73]. For each emission model, we calculate the 90% confidence level error on both the hydrogen column density and the temperature/photon-index.

The X-ray spectral parameters are used to compute the sources' X-ray flux values, to be compared to the $\gamma$-ray one in the framework of the $f_\gamma/f_X$ identification tool in order to identify the probable pulsar subclass for the $\gamma$-ray unidentified source [135].

Finally, we produce the light curves for each X-ray source in the 99% radius error-circle of the $\gamma$-ray source using the standard SAS procedure in order to detect any time variability. We extract a source+background light curve, using the selection region containing the source (see above) detected with PATTERN = 0–4 for the PN and PATTERN = 0–12 for the two MOS detectors and bin size = 100 s. We then extract a background light curve, using all the selection defined for the source+background light curve. At this point we subtract the background light curve to the source+background light curve with the FTOOL task `lcmath`. In this process, the background count rate is rescaled using the ratio of the source and background areas. We test the significant of the null hypothesis (non variability - flat light curve) on a binned light curve generated by the FTOOL task `lcurve`. Light curves are generated using bin times of 1000 s, 2500 s and 5000 s in order to have at least 25 counts per bin to apply the $\chi^2$ test. All error bars are reported at the $1\sigma$.

[1] *Power-law*, *apec* and *black-body* are respectively `pow`, `apec` and `bbody` in XSPEC

### Swift data analysis

For each *Swift* [84] observation we analyze data collected by the *X-Ray Telescope* (XRT) [38]. XRT uses a CCD detector sensitive to photons with energies between 0.2 and 10 keV. Two types of XRT archival data can be obtained from the *Swift* Data Center, Level 1 and Level 2. Level 1 data can be calibrated by yourselves in a way recommended by *Swift* team[2], while Level 2 cleaned data have gone through the standard pipeline process. We decide to use Level 2 cleaned data. All of the observations were obtained in Photon Counting (PC) mode [100]. The data are processed with standard procedures using FTOOLS[3] tasks under the `Heasoft` package v6.13.

In the XRT image analysis, we try to detect the X-ray counterparts of the 2FGL unidentified source, and localize each source. First, we extract X-ray images in the energy range of 0.3–10 keV, where the PC response matrices is well calibrated, using `xselect`. Next by using `ximage`, we searched for "possible" X-ray sources which are $> 3\sigma$ confidence level in photon statistics against background. The position of these sources are determined with a typical accuracy of $\sim 5''$ using `xrtcentroid`.

We analyze the spectra of the most likely X-ray counterparts, which are situated within the 99% error circle of the *Fermi*-LAT source and have XRT data with a signal-to-noise ratio above $3\sigma$, in order to perform a reliable spectral analysis. Events for spectral analysis are extracted within a circular region of radius $20''$, centered on the source position; this region encloses about 90% of the PSF at 1.5 keV [151]. The background is taken from a source-free region, using a circular region of radius $150''$. Since the data show a maximum count rate <0.3 counts $s^{-1}$, no pile-up correction is necessary.

A total *Swift* observation of a specific target is always performed splitting it in a large number of short observations. Our individual data sets have too few counts for a meaningful spectral analysis. Therefore, we extract a cumulative spectrum using the task `mathpha` rescaling the total count rate with those of a single observation. However, the total count statistics of the detected sources is too low, some tens of counts, to bin the spectra with `grppha` with a minimum of 25 counts per bin in order to be able to apply the $\chi^2$ minimization technique. We decide to perform an unbinned analysis adopting the *Cash statistics* (c-stat) [47]. The goodness-of-fit is calculated using the task `goodness`[4] implemented in XSPEC. The source spectrum of a single observation is extracted from the corresponding event file using the task `xselect`. The relative ancillary response file `arf` is generated with `xrtmkarf` and it accounts for different extraction regions, vignetting and point-spread function corrections, while we used the current spectral redistribution matrix files (RMFs) in CALDB (`swxpc0to12s6_20010101v013.rmf`). The cumulative ancillary response file is extracted by the task `addarf`. The energy band used for the spectral analysis, performed with XSPEC v.12.8.0 [21], depends on the statistical quality of the data and typically ranges from 0.3 to 10 keV. As for EPIC, we fit the source spectra with three spectral models: *power-law*, *apec*, and *black-body*. In all cases we include the absorption by the interstellar medium, we try to to both leave it as a free parameter and fix it to the value of the hydrogen column density along the line of sight [73]. All quoted errors correspond to 90% confidence level for a single parameter of interest.

[2]The SWIFT XRT Data Reduction Guide:`http://heasarc.nasa.gov/docs/swift/analysis/xrt swguide_v1_2.pdf`

[3]See `http://heasarc.gsfc.nasa.gov/docs/software/`

We identify the probable pulsar subclass for the $\gamma$-ray unidentified source [135] comparing the the sources' X-ray flux values with the $\gamma$-ray one in the framework of the $f_\gamma/f_X$ identification tool

Finally, we produce the light curves for each X-ray source in the 99% radius error-circle of the $\gamma$-ray source in order to detect a timing variability. We cannot produce light curves to each observation and not even full light curves with a bin size of one observation per bin because the statistics is too poor. We decide to produce full light curves using bin size minimum of 20 counts bin$^{-1}$ in order to apply the $\chi^2$ test (we use only 5 and 10 bins). From the observations in each bin time we extract a cumulative spectrum for the source and background and we generate the relative ancillary response file as explained above. At this point we fit the source spectra with three spectral models: *power-law*, *apec* and *black-body*. The flux we obtain from the best fit is so the flux in each bin time. All error bars are reported at the $1\sigma$.

### X-ray source hardness ratios analysis

Since the count statistics (usually a few tens of photons) of the detected sources is too low to produce significant spectra, we perform a qualitative spectral analysis using the count rate (CR) measured in the three energy ranges (*soft*: 0.3–1 keV; *medium*: 1–2 keV; *hard*: 2–10 keV) to compute two different *Hardness Ratios* (HRs):

$$HR12 = [CR(1-2) - CR(0.3-1)]/[CR(1-2) + CR(0.3-1)]$$
$$HR23 = [CR(2-10) - CR(1-2)]/[CR(2-10) + CR(1-2)]$$

[4] `http://heasarc.gsfc.nasa.gov/xanadu/xspec/manual/XSgoodness.html`

Adopting the above definition, sources with a small/large HR12 value are little/very absorbed, while sources with a small/large HR23 value are characterized by a soft/hard spectrum. All error bars of the distribution of the HRs of the X-ray sources are reported at the $1\sigma$.

To obtain a further indication on the sources spectra, we compare the measured HRs with the expected ones computed for three different template spectral models, namely: a *black-body*, with temperatures kT increasing from 0.1 to 1.1 keV, a *power-law*, with photon indexes $\Gamma$ increasing from 1.5 to 2.5 and an *apec*, with temperatures kT increasing from 0.5 to 5.5 keV. Each spectral model is computed using the interstellar medium absorption given by [73] and one third of that value. In such a way we can identify the spectral model more appropriate for a given source and tentatively assign its most likely spectral parameters.

### Optical/IR candidate counterparts analysis

In order to identify the plausible X-ray counterparts, we cross-correlate their positions with two optical catalogues, the *United States Naval Observatory* (USNO B1.0) catalogue [149], with limiting magnitudes V $\sim$ 21, $0.2''$ astrometric accuracy and $\sim$0.3 mag photometric accuracy, and the last version of the *Guide Star Catalogue* (GSC) version 2.3 [53]; and two infrared (IR) catalogues, *Two Micron All Sky Survey* (2MASS) [187] and *Wide-field Infrared Survey Explorer* (WISE) [217] from the preliminary data release (WISEP) [62]. Moreover, we have also analyzed the NASA/IPAC extragalactic Database (NED[5]) and the *NRAO VLA Sky Survey* (NVSS) [57] radio catalog but we have not found any plausible counterparts in the radius error-circle of each X-ray source. The search for optical or IR counterparts is performed by selecting candidates at $< 5''$ from the X-ray position.

The possibility to find more than one optical or IR source within the rather conservative $5''$ radius error-circle suggests that we cannot ignore the possible foreground contamination, which could affect our cross-correlation. The probability of chance coincidence between a X-ray and an optical source is calculated by $P = 1 - e^{-\pi \mu r^2}$, where $r$ is the X-ray error-circle radius and $\mu$ the surface density of the optical sources [186].

Each measured magnitude values are de-absorbed as follows: defining $m_i$ as the measured magnitude in the $i$ filter and $m'_i$ as the absorbed one then $m_i = m'_i - A_i$, where $A_i$ is the extinction given by the interstellar medium in the $i$ filter. From [168], the visual extinction can be derived by the X-ray Galactic hydrogen column density as $N_H = A_V(1.87 \times 10^{21})$ cm$^{-2}$. $i$ extinction can be obtained by the simple relation $A_i = f(\lambda) A_V$ [46], where $f(\lambda)$ is a function depending on the wavelength. We then determine the stellar spectral classes which are compatible with the optical/IR candidate counterpart on the basis of its optical colors [220] (B-V, V-R, V-I, V-J, V-H and V-K), and infrared colors [78]( J-H, H-K, R-K and J-K). If the visual magnitude of a candidate counterpart is not reported we extrapolate its value as the average between absorbed B and R magnitudes.

[5] `http://ned.ipac.caltech.edu/`

The X-ray spectral parameters are used to compute the sources' X-ray flux values, to be compared to the optical ones in the framework of the $f_X/f_{opt}$ identification tool [121]. The optical flux of the candidate counterpart is based on B magnitude or V magnitude if reported. B and V magnitudes are converted to the flux in the B or V filter assuming the standard values $f_B(m_B = 0) = 6.39 \times 10^{-6}$ erg s$^{-1}$ cm$^{-2}$ and $f_V(m_V = 0) = 4.27 \times 10^{-6}$ erg s$^{-1}$ cm$^{-2}$.

We use the V magnitude to calculate the source flux and if the source in the optical catalogue has not the value for V we use the B magnitude, while for the X-ray sources with no counterpart, we use V=21 as the optical upper limit.

### 5.1.2 2FGL J1036.1–6722

The first object we analyze is the unidentified source 2FGL J1036.1–6722 (see Table 5.1). This object was discovered for the first time by the *Fermi*-LAT and it was included in the 1FGL source catalog as 1FGL J1036.2–6719 [3]. 2FGL J1036.1–6722 is a rather bright $\gamma$-ray source, situated far from the Galactic plane (b=-7°.84, l=290°.45) and it is characterized by low variability (variability index=35.50) and high curvature significance (sign_curv = 6.51). The variability and spectral curvature properties of this source are very similar to those of a typical $\gamma$-ray pulsars as shown in the bottom of Figure 5.1. 2FGL J1036.1–6722 $\gamma$-ray spectrum and light curve are shown at the top of Figure 5.1.

In this Section we report on recent monitoring data taken with *XMM-Newton* with the aim of detecting the X-ray counterpart of the putative $\gamma$-ray millisecond pulsar radio-quiet.

#### Observation and X-ray analysis

Our deep *XMM-Newton* observation 2FGL J1036.1–6722 started on 2012 June 16 at 12:17:48 UT and lasted 24.9 ks. The PN camera [194] of the EPIC instrument was operating in *Extended Full Frame* mode (time resolution of 200 ms over a $26' \times 27'$ field of view (FoV)), while the MOS detectors [203] were set in *Full Frame* mode (2.6 s time resolution on a $15'$ radius FoV). The thin optical filter was used for the PN camera while we chose to use the medium filter for the MOS detectors. After standard data processing explained in the previous section, the good, dead-time corrected exposure time is 14.4 ks for the PN, 20.6 ks for the MOS1 and 21.4 ks for the MOS2 detector.

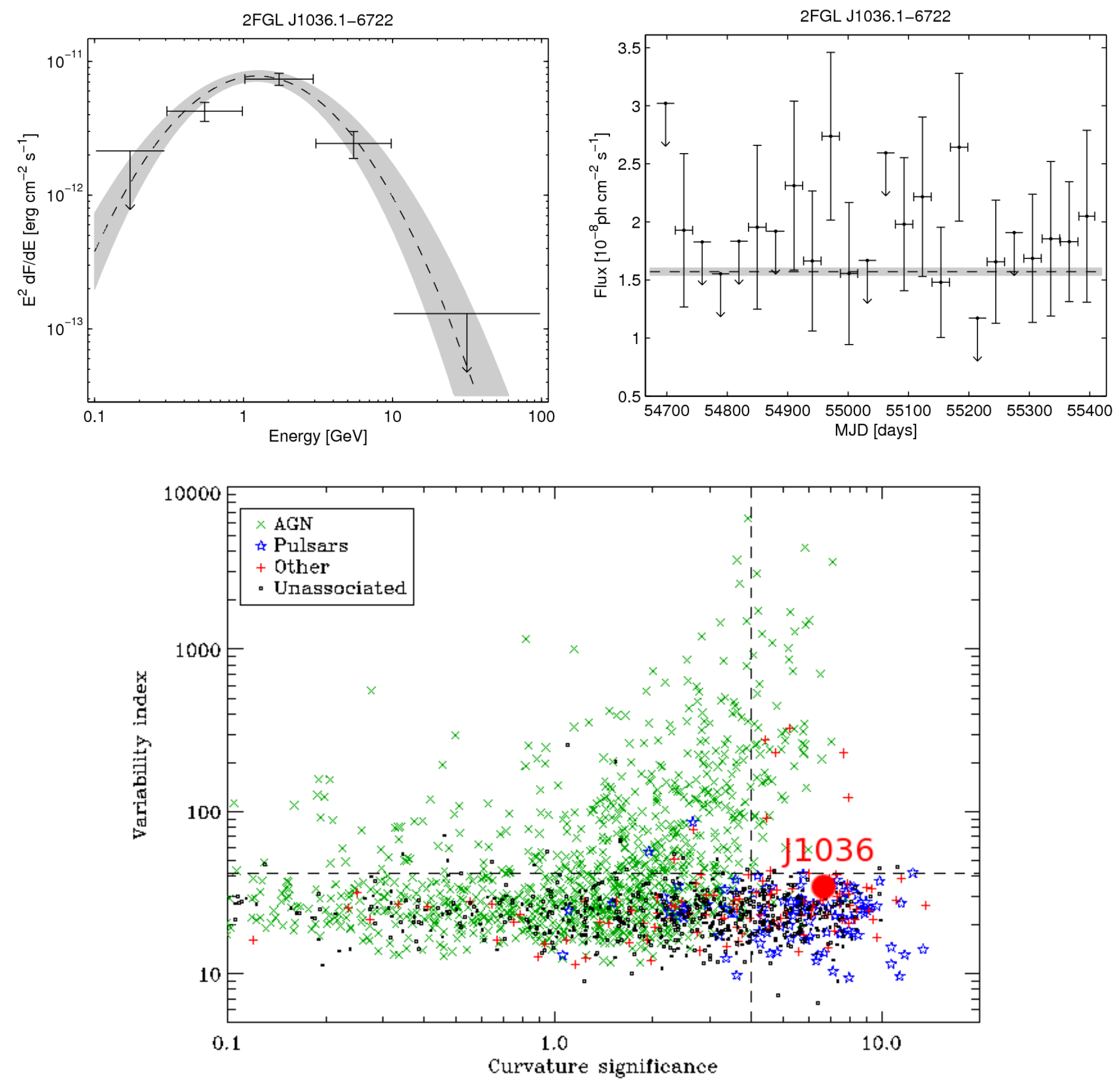

Figure 5.1: Top: 2FGL J1036.1–6722 $\gamma$-ray spectrum (left) and light curve (right). Bottom: variability index plotted as a function of the curvature significance for 2FGL source classes. Red dot represents the position of 2FGL J1036.1–6722 in this parameters space. This object is not variable and the spectrum is curved.

We detect 26 X-ray sources in the PN FoV with a source significance greater than 3.6$\sigma$. Of these, only 3 are located within the 99% error circle of the *Fermi*-LAT source, for these we produce X-ray spectra, light curves and find optical counterparts. Figure 5.2 shows the 0.3–10 keV exposure-corrected *XMM-Newton* PN FoV image. We applied a Gaussian filter with a kernel radius of 5”.

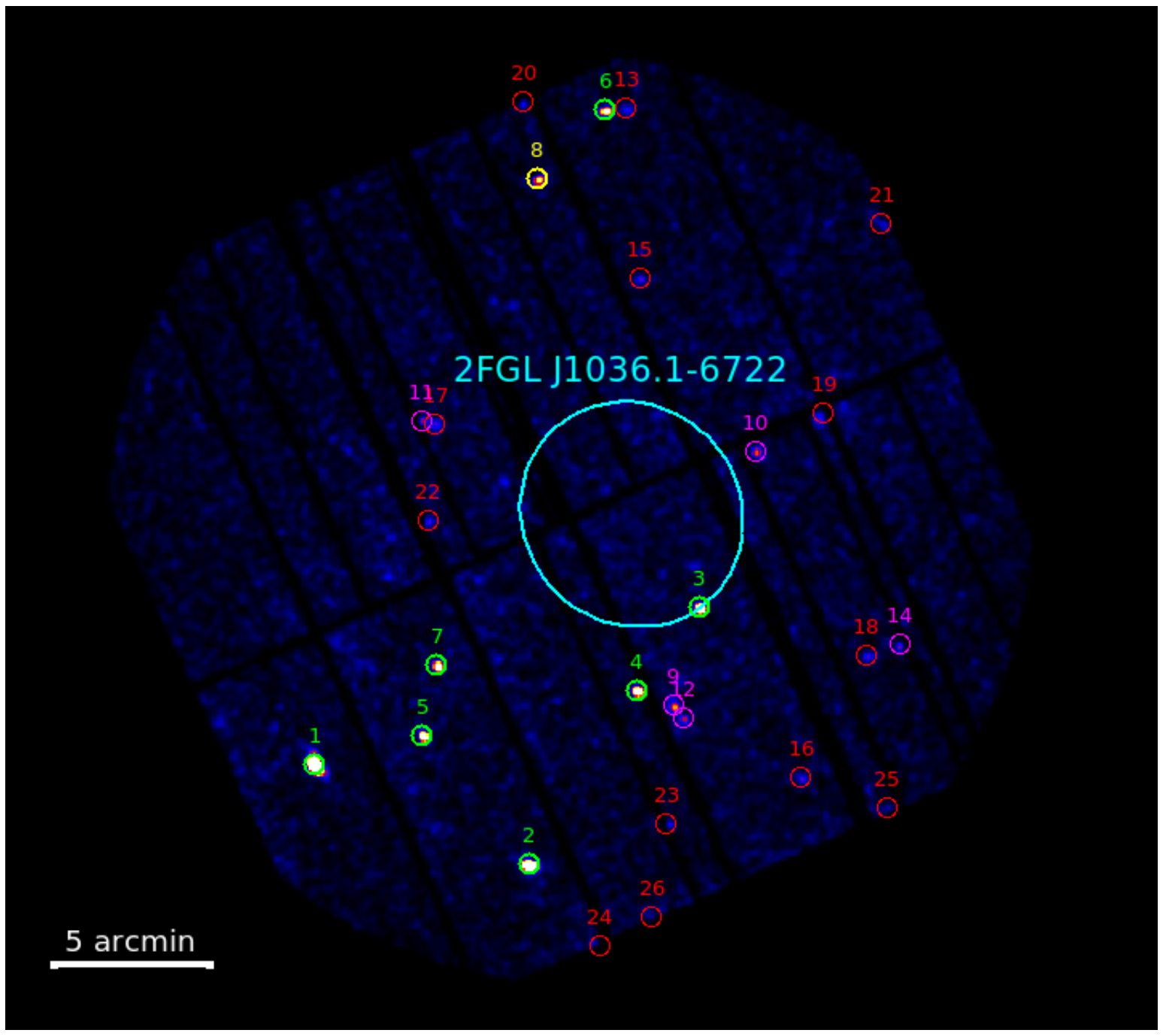

Figure 5.2: 0.3–10 keV exposure-corrected *XMM-Newton* PN FoV image . A gaussian filter with a kernel radius of 3″ is applied. 2FGL J1036.1–6722 with 95% confidence error ellipse is plotted in cyan. Each X-ray source detected by the PN is plotted with a circle of 10″ radius. Colors represent the source significance: in red sources with TS<25, in magenta sources with 25<TS<50, in yellow sources with 50<TS<100 and in green sources with TS>100.

In Figure 5.3 is shown the *Digital Sky Survey* image in the field of 2FGL J1036.1–6722 with the distribution of the X-ray sources detected by *XMM-Newton*. Within the 15 arcmin radius image area, the USNO B1.0 catalogue provides a total of ∼16000 sources, corresponding to a surface density $\mu \sim 6.4 \times 10^{-3}$ sources arcsec$^{-2}$. Since the X-ray error-circle is 5 arcsec, we estimate that the probability of chance coincidence between a X-ray and an optical source is ∼ 0.4. Therefore up to 40% of the selected counterparts could be spurious candidates.

### Notes on individual X-ray sources

In the following, we present results on the most likely candidate X-ray counterparts to the unidentified $\gamma$-ray source.

**Source #3:** The X-ray analysis yields 110 counts in the energy band 0.3–10 keV, with a source significance TS≃140 in the PN data and TS≃85 in the two MOS data; its count rate in the total energy band is $1.2 \times 10^{-2}$ cts $s^{-1}$. This object is situated within the 2FGL 95% confidence error ellipse at (RA, Dec)=(158°.94, -67°.42). In Figure 5.4 is shown the spectrum of this X-ray sources.

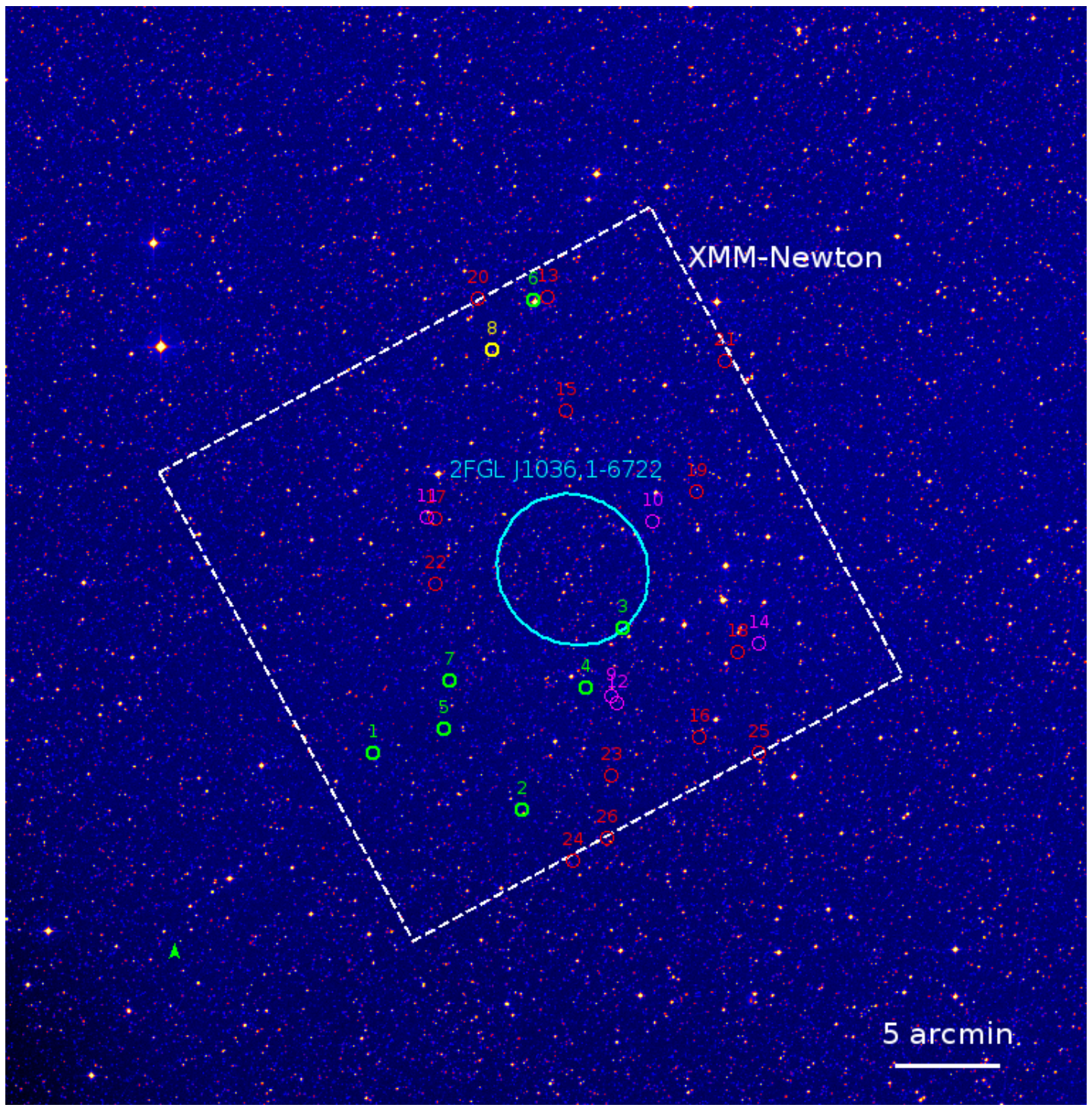

Figure 5.3: *Digital Sky Survey 2* image with red filter of the field of unidentified source 2FGL J1036.1–6722, for which 95% confidence error ellipse is plotted in cyan. Each X-ray source detected by the PN is overplotted with a circle of 10″ radius colored as described in Figure 5.2. White dashed line represents the FoV of the *XMM-Newton* observation.

The X-ray spectral fitting results are summarized in Table 5.2. We notice that we find physical plausible best-fit spectral parameter values only if we fix the interstellar medium absorption to $N_H = 0.2 \times 10^{22}$ cm$^{-2}$ [73].

In Figure 5.5 is shown the light curved with a bin time of 2500 s.

Within 5″ radius X-ray error circle there is only an optical candidate counterpart of the source #3, its properties are shown in Table 5.3.
If the optical counterpart of the X-ray source is not detected because too faint, above the limiting magnitude of USNO B1.0 (V > 21), then $\log(f_X/f_V) > (-0.27 \div 0.03)$, using an unabsorbed visual flux $f_V = 5.47\times$

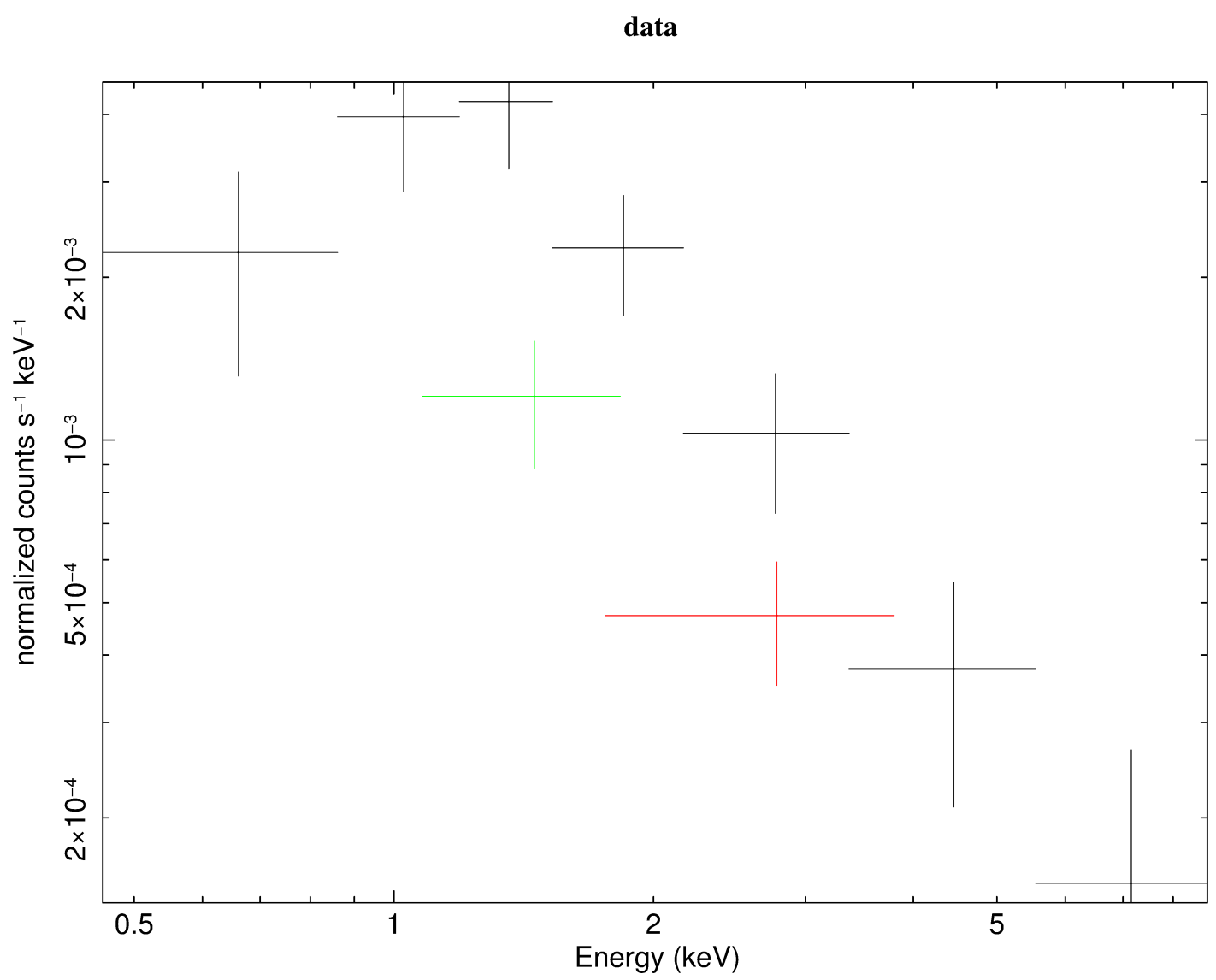

Figure 5.4: Spectrum of the X-ray source #3 binned with a minimum of 25 counts per bin. The spectrum shows that the X-ray object is rather absorbed at low energies and that the PN detected photons until ≃7 keV.

| Parameter | Power-law model | Apec model | Black body model |
|---|---|---|---|
| $N_H{}^a$ | 0.2 (fixed) | 0.2 (fixed) | 0.2 (fixed) |
| $\Gamma$ | $1.58^{+0.31}_{-0.30}$ | - | - |
| kT (keV) | - | $6.50^{+26.58}_{-2.96}$ | $0.59^{+0.17}_{-0.13}$ |
| d.o.f. | 9 | 9 | 9 |
| $\chi^2_\nu$ | 0.602 | 0.177 | 1.04 |
| $\text{Flux}_{\text{unabs}}{}^b$ | 5.80 | 5.27 | 2.91 |

$^a$in units of $10^{22}$ cm$^{-2}$

$^b$in units of $10^{-14}$ erg cm$^{-2}$ s$^{-1}$ and in the energy range of 0.3–10 keV

Table 5.2: Spectral parameters for the X-ray source #3 within the 95% error ellipse of 2FGL J1036.1–6722. $Flux_{unabs}$ corresponds to the unabsorbed flux. We cannot assert if this X-ray source is a pulsar, a star or an AGN because no spectral model can be rejected by the $\chi^2$ test. The $\gamma$-ray (E>100 MeV) to soft X-ray (0.3–10 keV) flux ratio is $F_\gamma/F_X \simeq (300 \div 700)$, this flux ratio is compatible with the value of a typical $\gamma$-ray MSP and radio-loud pulsar [135].

$10^{-14}$. This value does not exclude any source class as counterpart of the X-ray source, notice that a pulsar is characterized by $\log(f_X/f_{opt}) > 2$ [121].

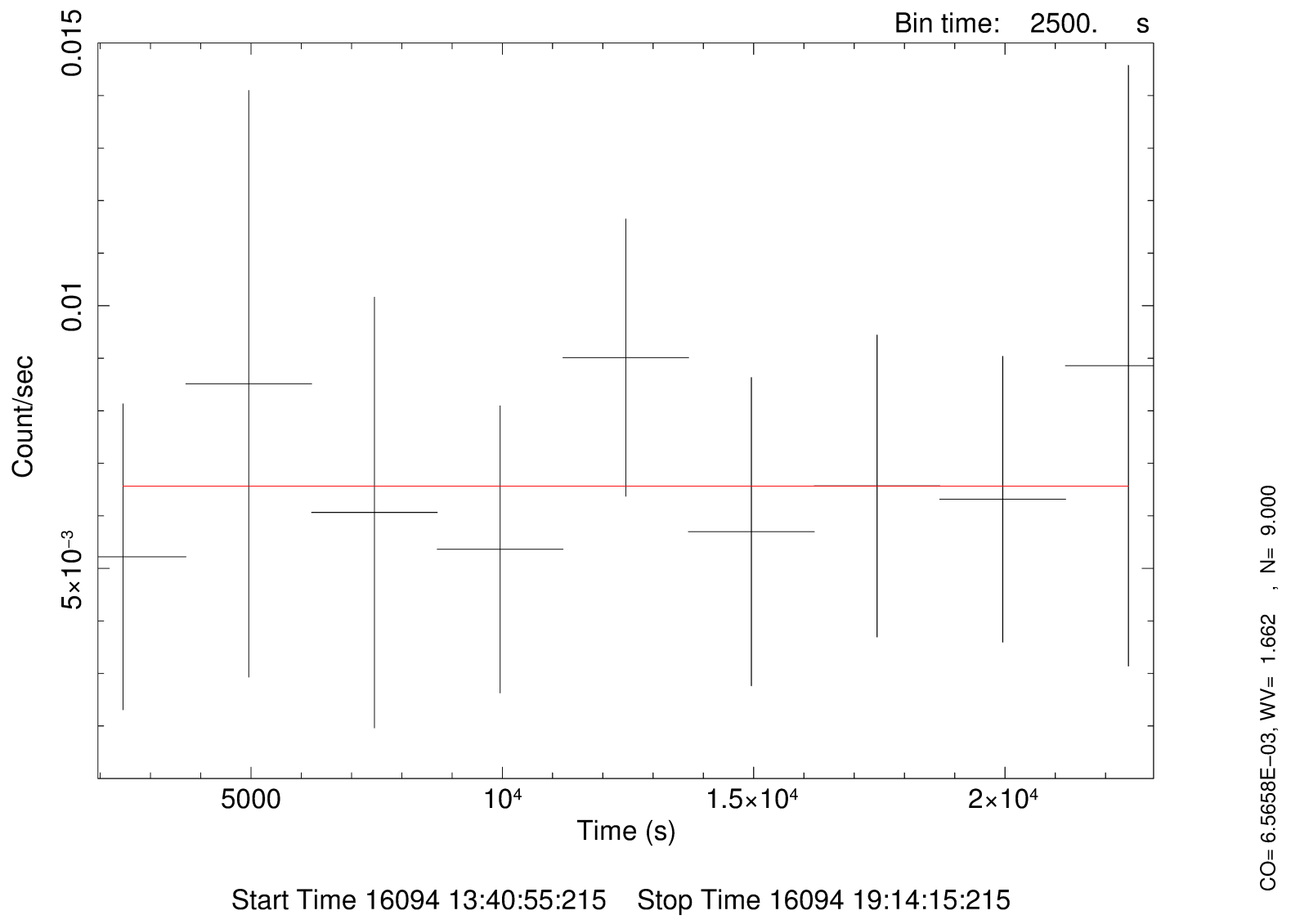

Figure 5.5: Light curve for the X-ray source #3 in the energy range between 0.3 keV and 10 keV within the $\sim$ 15 ks observation span. The red line depicts the best fit assuming a flat light curve. No significant X-ray variability is detected on the basis of the $\chi^2$ test.

**Source #4:** The X-ray analysis yields 129 counts in the energy band 0.3—10 keV, with a source significance TS$\simeq$150 in the PN data and TS$\simeq$125 in the two MOS data; its count rate in the total energy band is $1.5 \times 10^{-2}$ cts s$^{-1}$. This object is situated in the limit of the 2FGL 99% confidence error ellipse at (RA, Dec)=(159°.03, -67°.47). In Figure 5.6 is shown the spectrum of this X-ray sources.

The X-ray spectral fitting results are summarized in Table 5.4. We notice that we find physical plausible best fit spectral parameter values only if we fix the interstellar medium absorption to $N_H = 0.2 \times 10^{22}$ cm$^{-2}$ [73]. Only the *black-body* model is significant and has plausible best-fit parameters. However, we cannot assert if this X-ray source is a pulsar or an AGN because the *black-body* model is well-suited for both source classes.

In Figure 5.7 is shown the light curved with a bin time of 2500 s.

Within 5″ radius X-ray error circle there are three optical/IR candidate counterparts of the source #4, their properties are shown in Table 5.5.
If the optical counterpart of the X-ray source is not detected because too faint, above the limiting magnitude of USNO B1.0 (V > 21), then $\log(f_X/f_V) > (0.24 \div 0.42)$, using an unabsorbed visual flux $f_V = 5.47 \times$ $10^{-14}$. This value does not exclude any source class as counterpart of the X-ray source, remember that a pulsar is characterized by $\log(f_X/f_{opt}) > 2$ [121].

| Source #3 – optical/IR analysis | |
|---|---|
| Optical counterpart | GSC |
| IR counterpart | - |
| Detection year | 1978 |
| Distance | 2.2” |
| De-absorbed magnitude (mag) | B=19.98 |
| Stellar spectral class | - |
| $f_{opt}$ (erg cm$^{-2}$ s$^{-1}$) | $6.51 \times 10^{-14}$ |
| $f_X/f_{opt}$ ($\log_{10}$) | -0.37÷-0.12 |
| Suggested class | star, CV or galaxy |

Table 5.3: Properties of the optical candidate counterpart of the X-ray source #3. Rows (1) and (2): optical and IR catalogs where the candidate counterpart of the X-ray source is listed. Row (3): Detection year of the candidate counterpart. Row (4): distance between the X-ray source and its candidate counterpart. Row (5): de-absorbed magnitudes of the candidate counterpart. Row (6): Stellar spectral class of the candidate counterpart based on [220] and [78]; optical colors are B-V, V-R, V-I , V-J, V-H and V-K while IR colors are J-H, H-K, R-K and J-K; the symbol “?” indicates that colors are not compatible with any stellar class. Row (7): optical flux of the candidate counterpart based on the B magnitude or V magnitude if measured. Row (8): logarithmic values of the X-ray-to-optical flux ratio; the X-ray flux is based on all acceptable models. Row (9): proposed X-ray source classification based on [121]; “CV” indicates cataclysmic variable.

**Source #10:** The X-ray analysis yields 78 counts in the energy band 0.3—10 keV, with a low source significance TS$simeq$40 in the PN data and TS≃20 in the two MOS data; its count rate in the total energy band is $0.6 \times 10^{-2}$ cts s$^{-1}$. This object is situated in the limit of the 2FGL 99% confidence error ellipse at (RA, Dec)=(158°.87, -67°.34). In Figure 5.8 is shown the spectrum of this X-ray sources.

The X-ray spectral fitting results are summarized in Table 5.6. We notice that we find physical plausible best-fit spectral parameter values only if we fix the interstellar medium absorption to $N_H = 0.2 \times 10^{22}$ cm$^{-2}$ [73].

In Figure 5.9 is shown the light curved with a bin time of 2500 s.

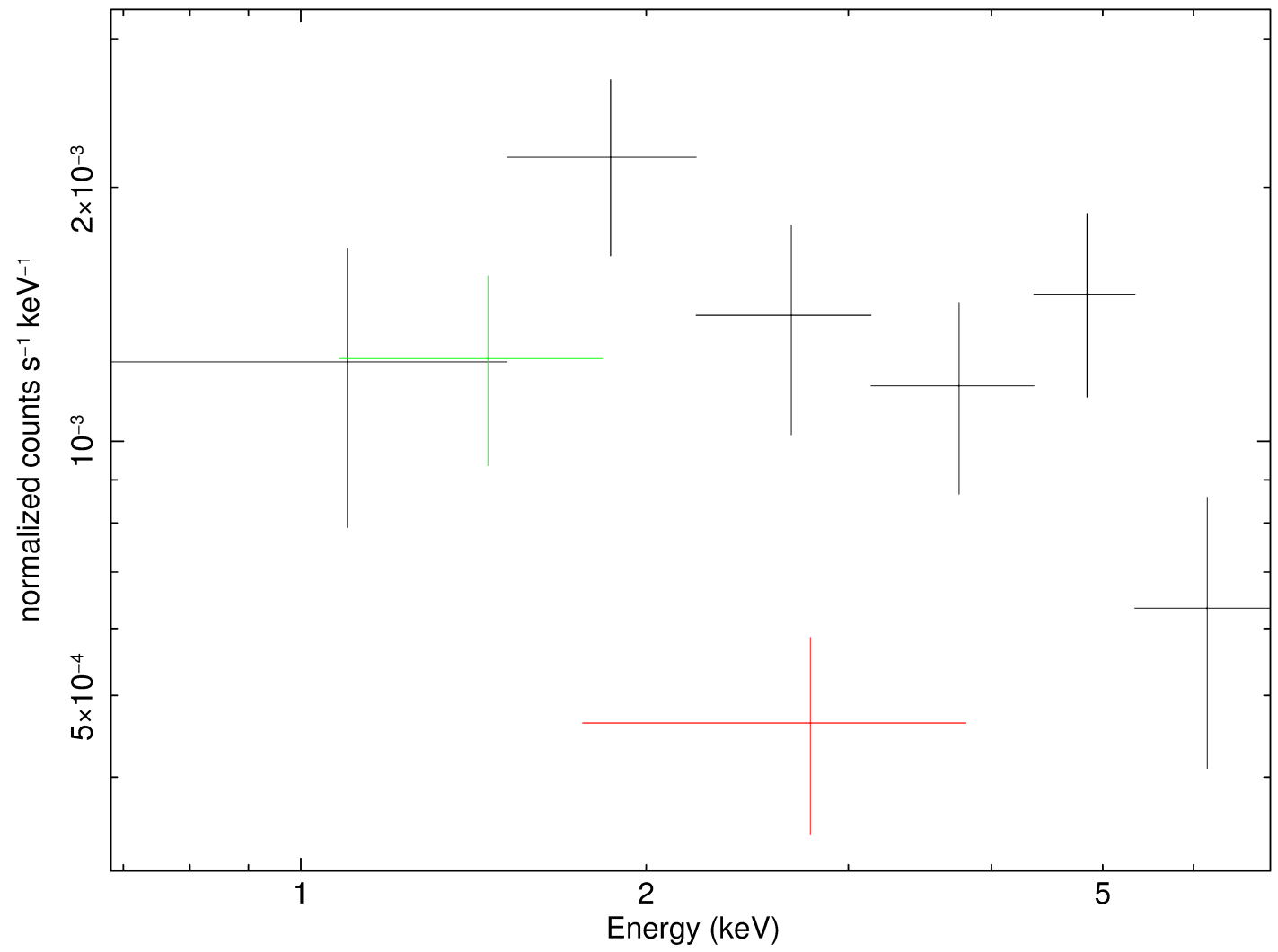

Figure 5.6: Spectrum of the X-ray source #4 binned with a minimum of 25 counts per bin. The spectrum shows that the X-ray object is rather absorbed at low energies and that the PN detected photons until ≃6 keV.

| Parameter | Power-law model | Apec model | Black body model |
|---|---|---|---|
| $N_H{}^a$ | 0.2 (fixed) | 0.2 (fixed) | 0.2 (fixed) |
| $\Gamma$ | $0.40^{+0.31}_{-0.32}$ | - | - |
| kT (keV) | - | $64^{+46.58}_{-37.07}$ | $1.51^{+0.52}_{-0.35}$ |
| d.o.f. | 8 | 8 | 8 |
| $\chi^2_\nu$ | 0.998 | 3.835 | 1.30 |
| $\mathrm{Flux_{unabs}}^b$ | 14.3 | 6.94 | 9.57 |

$^a$in units of $10^{22}$ cm$^{-2}$

$^b$in units of $10^{-14}$ erg cm$^{-2}$ s$^{-1}$ and in the energy range of 0.3–10 keV

Table 5.4: Spectral parameters for the X-ray source #4 within the 2FGL 95% error ellipse of 2FGL J1036.1–6722. $Flux_{unabs}$ corresponds to the unabsorbed flux. Only the *black-body* model is significant and has plausible best-fit parameters. The $\gamma$-ray (E>100 MeV) to soft X-ray (0.3–10 keV) flux ratio is $F_\gamma/F_X \simeq 200$ considering a *black-body* spectral model, which is compatible with the value of a typical $\gamma$-ray MSP [135]

Within 5″ radius X-ray error circle there is no optical or IR candidate counterpart for the source #10. If the optical counterpart for the X-ray source has not detected because too faint, above the limiting magnitude of USNO B1.0 (V>21), then $\log(f_X/f_V) > (-0.57 \div -0.25)$, adopting the unabsorbed visual flux $f_V = 5.47 \times 10^{-14}$. This value does not exclude any source class as

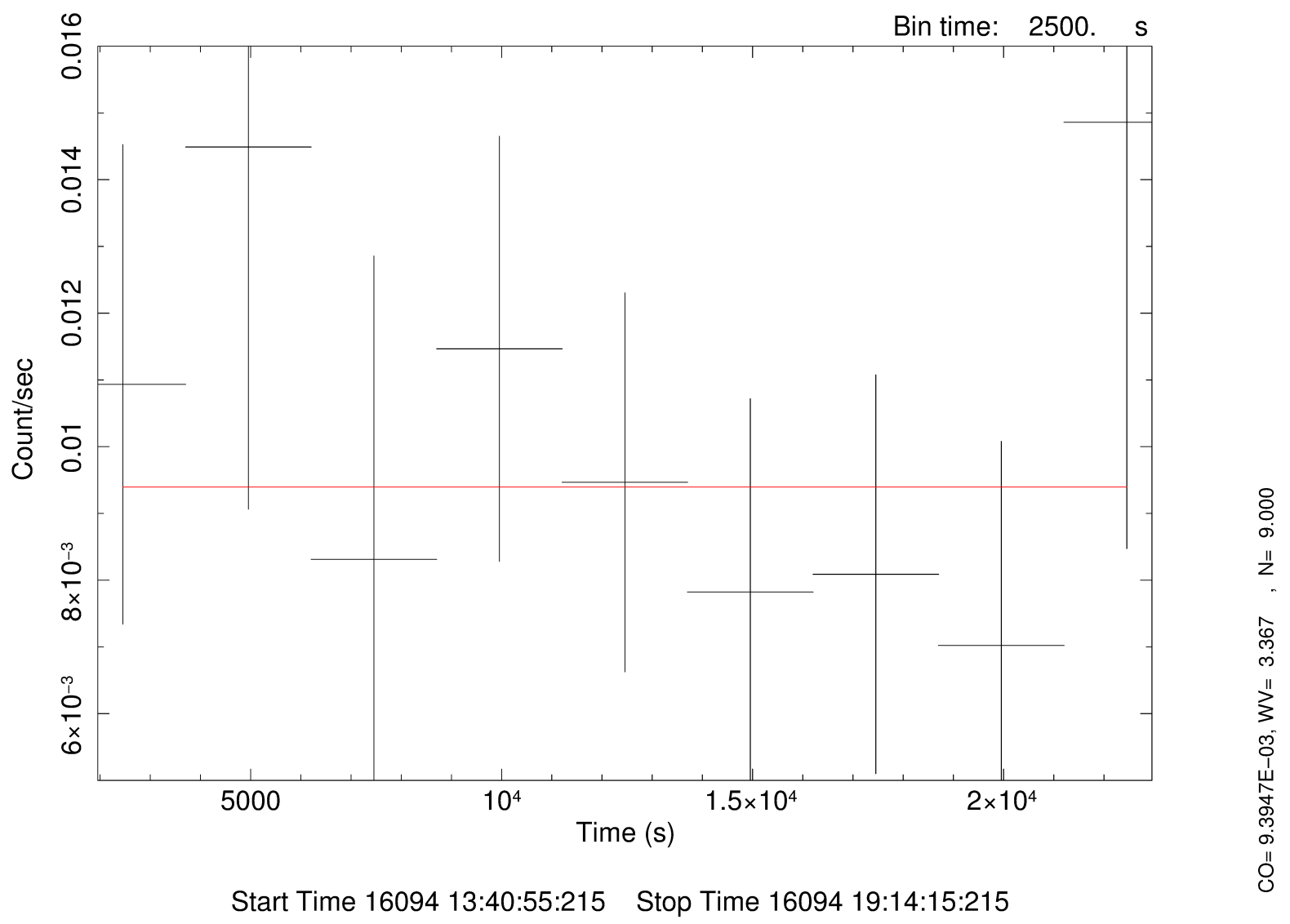

Figure 5.7: Light curve for the X-ray source #4 in the energy range between 0.3 keV and 10 keV within the ∼ 15 ks observation span. The red line depicts the best fit assuming a flat light curve. No significant X-ray variability is detected on the basis of the $\chi^2$ test.

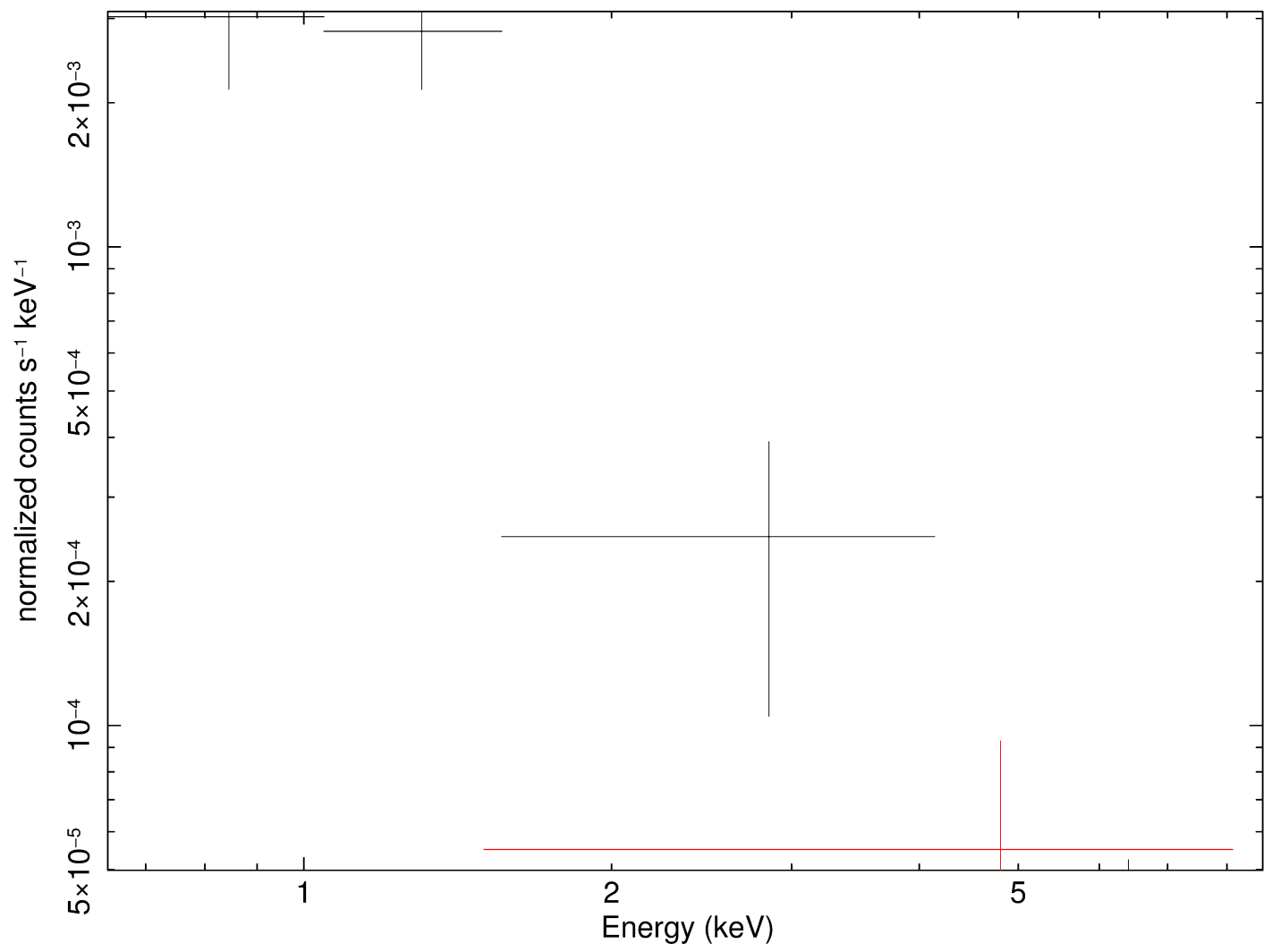

Figure 5.8: Spectrum of the X-ray source #10 binned with a minimum of 25 counts per bin. The spectrum shows that the X-ray object is not significantly absorbed at low energies and that the PN detected photons only until ≃3 keV.

| Source #4 – optical/IR analysis | | | |
|---|---|---|---|
| Optical counterpart | USNO B1.0 | USNO B1.0 | USNO B1.0 |
| IR counterpart | 2MASS | - | - |
| Detection year | 1985 | 1985 | 1989 |
| Distance | $1''$ | $4''$ | $4.8''$ |
| | B=19.99 | B=16.46 | B=17.14 |
| De-absorbed | V=18.44[a] | V=17.57[a] | V=17.94[a] |
| magnitudes | R=17.77 | R =18.47 | R=18.63 |
| (mag) | J=16.27 | - | - |
| | H=15.44 | - | - |
| | K=15.1 | - | - |
| Stellar spectral class | F, G, M or K | ? | ? |
| $f_{opt}$ (erg cm$^{-2}$ s$^{-1}$) | $1.62 \times 10^{-13}$ | $1.66 \times 10^{-12}$ | $8.9 \times 10^{-13}$ |
| $f_X/f_{opt}$ ($\log_{10}$) | -023÷-0.05 | -1.24÷-1.06 | -1.24÷-0.97 |
| Suggested class | star, CV or galaxy | star | star |

[a]value extrapolated as the average between absorbed B and R magnitudes

Table 5.5: Properties of the optical/IR candidate counterparts of the X-ray source #4. For a description of each row see Table 5.5. Last two optical candidate counterparts are classified as stars but these results are not compatible with their colors, this may be related to the simple extrapolation of the visual magnitude or to the $N_H$ value we have assumed, or maybe because they are not the optical counterpart of the X-ray source.

| Parameter | Power-law model | Apec model | Black body model |
|---|---|---|---|
| $N_H{}^a$ | 0.2 (fixed) | 0.2 (fixed) | 0.2 (fixed) |
| $\Gamma$ | $2.91^{+0.82}_{-0.65}$ | - | - |
| kT (keV) | - | $1.44^{+0.62}_{-0.30}$ | $0.28^{+0.08}_{-0.07}$ |
| d.o.f. | 5 | 5 | 5 |
| $\chi^2_\nu$ | 0.726 | 0.12 | 0.092 |
| $\mathrm{Flux_{unabs}}{}^b$ | 3.06 | 1.5 | 1.47 |

[a]in units of $10^{22}$ cm$^{-2}$

[b]in units of $10^{-14}$ erg cm$^{-2}$ s$^{-1}$ and in the energy range of 0.3–10 keV

Table 5.6: Spectral parameters for the X-ray source #10 within the 2FGL 95% error ellipse of 2FGL J1036.1–6722. $Flux_{unabs}$ corresponds to the unabsorbed flux. We cannot assert if this X-ray source is a pulsar a star or an AGN because no spectral model can be rejected by the $\chi^2$ test. The $\gamma$-ray (E>100 MeV) to soft X-ray (0.3–10 keV) flux ratio is $F_\gamma/F_X \simeq (650 \div 1330)$, which is compatible with the value of those a typical $\gamma$-ray MSP and radio-loud pulsar [135].

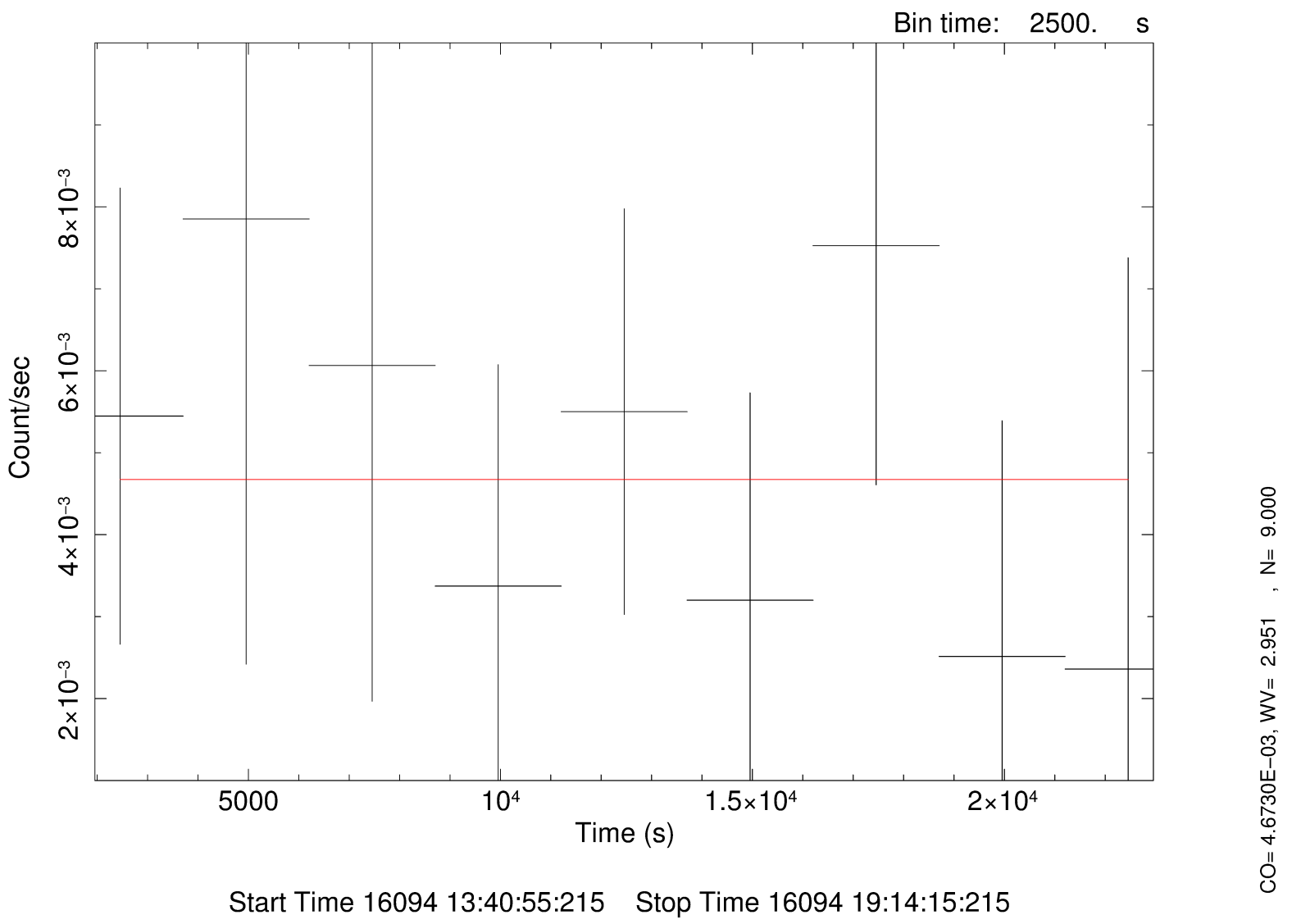

Figure 5.9: Light curve for the X-ray source #10 in the energy range between 0.3 keV and 10 keV within the ~ 15 ks observation span. The red line depicts the best fit assuming a flat light curve. No significant X-ray variability within the ~ 15 ks observation span on the basis of the $\chi^2$ test.

counterpart of the X-ray source [121].

**Source hardness ratios distribution**

In Figure 5.10 is shown the distribution of the HRs of the X-ray sources detected by the *XMM-Newton* within the 99% error circle of the *Fermi*-LAT source. All error bars are reported at the $1\sigma$. Source #10 is little absorbed and is characterized by a soft spectrum ($HR12 \simeq 0$ and $HR23 \simeq -1$), probably it is a nearby star. The other two sources are very absorbed and characterized by a rather hard spectrum ($HR12 > 0$ and $HR23 \geq 0$), one of these two objects may be the counterpart of the 2FGL unidentified source. The values of the expected HRs are overplotted in Figure 5.10. As can be seen, the distributions are compatible with a rather wide range of temperatures and photon indexes, thus suggesting that we are probably sampling different types of sources. This conclusion is not surprising, since the area is far from the Galactic plane and therefore are expected to contain both Galactic and extragalactic X-ray sources.

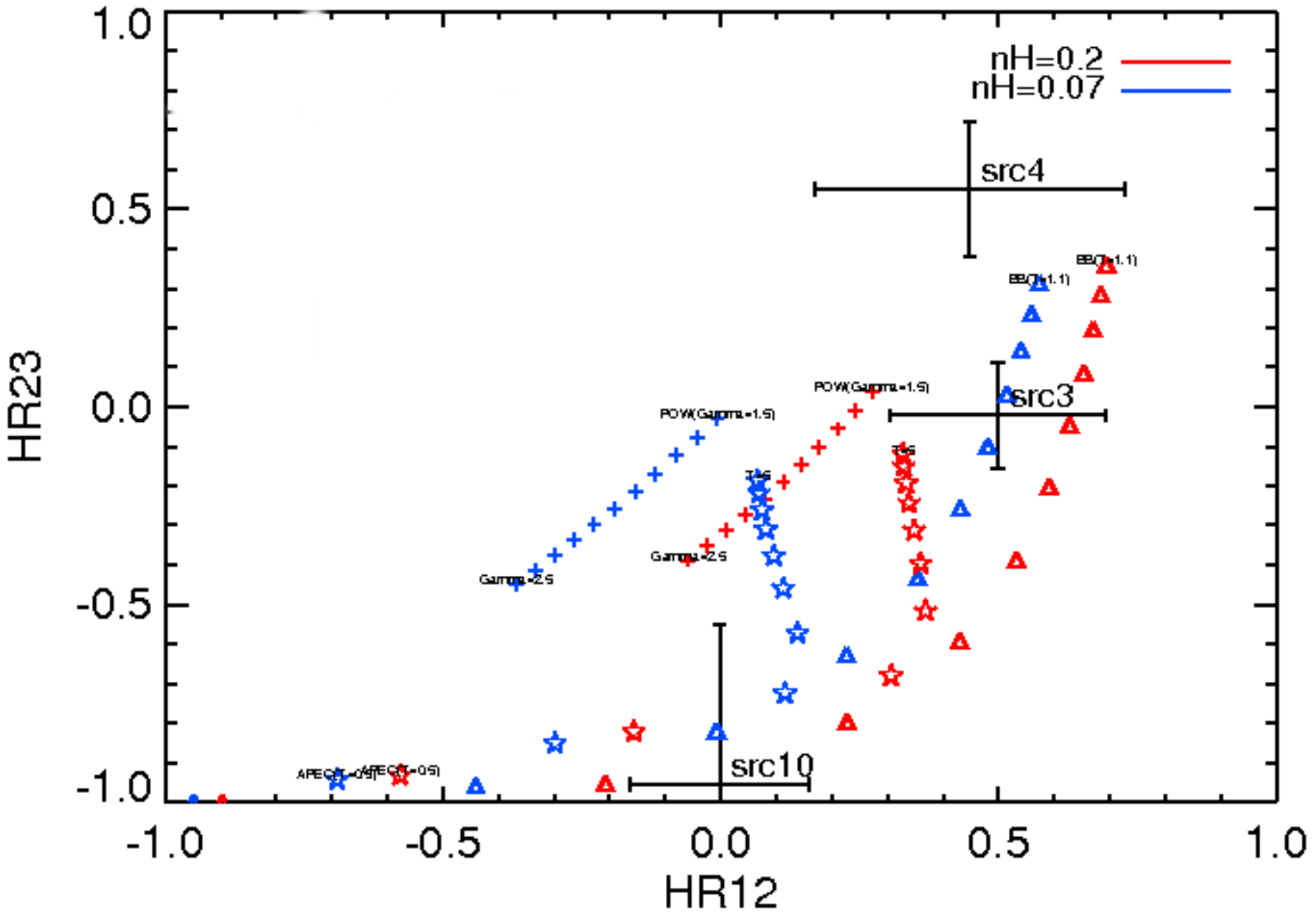

Figure 5.10: Distribution of HR12 vs. HR23 of the X-ray sources detected by the *XMM-Newton* within the 99% error circle of the *Fermi*-LAT source. Error bars are reported at $1\sigma$. Crosses indicate the expected HR12 vs. HR23 computed for power law spectra with $\Gamma$ from 1.5 and 2.5. Stars indicate the expected HR12 vs. HR23 computed for apec spectra with kT from 0.5 to 5.5 keV. Triangles indicate the expected HR12 vs. HR23 computed for black body spectra with kT from 0.1 to 1.1 keV. Each spectral model is computed using the interstellar medium absorption given by [73] (red) and one third of this value (blue).

### Discussion

In the *XMM-Newton* FoV of the $\gamma$-ray unidentified source 2FGL J1036.1–6722 we have detected 26 X-ray sources between 0.3 and 10 keV, only 3 of them are situated within the 99% error circle of the *Fermi*-LAT source, we can consider them as plausible counterparts of the putative MSPs. For each plausible X-ray counterpart we have not detected any significant X-ray variability and by an analysis of $f_\gamma/f_X$ each one seems to be a plausible counterpart of the 2FGL unidentified source. No optical or IR candidate counterpart was found for the source #10 but analyzing its X-ray spectrum it seems a nearby star. On the basis of the characteristics of 3 optical/IR candidate counterparts and the spectrum of the source #4, we may identify this X-ray object as a likely AGN.

Source #3 is the most interesting plausible counterpart of the putative MSP. The optical object detected within 5″ radius X-ray error circle does not seem the counterpart of source #3 and from its HRs we can classify it as a pulsar or at least an AGN.

If we hypothesize that the "real" counterpart of the unidentified source has not been detected above $3\sigma$ inside the 2FGL 99% error ellipse, we can derive the upper limit assuming an absorbed power law spectral model for the putative counterpart with $N_H = 0.2 \times 10^{22}$ cm$^2$ and $\Gamma = 2$. This means the putative counterpart should have an unabsorbed flux of $f_X \sim 9 \times 10^{-15}$ erg cm$^{-2}$ s$^{-1}$ in the energy range between 0.3 and 10 keV to be detected above $3\sigma$ with our observation. In this way we obtain the inferior limit of the $\gamma$-ray flux to X-ray flux ratio as $f_\gamma / f_X \gtrsim 2200$.

### 5.1.3 2FGL J1539.2–3325

Now we analyze the unidentified source 2FGL J1539.2–3325 (see Table 5.1). This object was discovered for the first time by the *Fermi*-LAT and included in the 1FGL source catalog as 1FGL J1539.0-3328 [3]. 2FGL J1539.2–3325 is a rather bright $\gamma$-ray source, situated at high Galactic latitude (b=17°.53, l=338°.74) and it is characterized by low variability (variability index=29.53) and high curvature significance (sign_curv = 5.54). The variability and spectral curvature properties of this source are very similar to those of a typical $\gamma$-ray pulsar as shown in the bottom of Figure 5.11. The $\gamma$-ray spectrum and the light curve of 2FGL J1539.2–3325 are shown at the top of Figure 5.11.

In this Section we report on recent monitoring data taken with *Swift X-ray Telescope* (XRT) with the aim of detecting the X-ray counterpart of the putative $\gamma$-ray millisecond pulsar radio-quiet.

#### Observations and X-ray analysis

The set of XRT observations are 27 and each one was typically taken as 0.5–15 ks exposures with various sampling intervals. The total *Swift-XRT* data set includes 84 ks of observations extending from 2010 October 1 to 2012 September 29.

We detect 31 X-ray sources which are $> 3\sigma$ confidence level in photon statistics against background in the XRT FoV. 5 of these X-ray detected sources are situated within the 99% error circle of the *Fermi*-LAT source, for these we produce X-ray spectra, light curves and find optical counterparts. Figure 5.12 shows the 0.3–10 keV exposure-corrected *Swift-XRT* FoV image.

We have noticed that in the Field-of-View of 2FGL J1539.2–3325 there is a dark cloud named *Lupus1*, a component of the Lupus dark-cloud complex [20, 91, 56]. Its mass is estimated to be ∼150 $M_\odot$, this is the most massive region in the Lupus complex, and the distance is determinate to be between 140 pc and 240 pc. This is a star forming site in which a large number of T-Tauri stars were discovered. Lupus1 is a very dense dark cloud, characterized by an average visual extinction of $A_V = 4.5$, which can be converted to column density using the relation suggested in [168]: $N_H = A_V(1.87 \times 10^{21}) = 0.85 \times 10^{22}$ cm$^{-2}$. In Figure 5.13 is shown the *Digital Sky Survey* image in the field of 2FGL J1539.2–3325 with the distribution of the X-ray sources detected by *Swift-XRT*. Within the 15 arcmin radius image area, the USNO B1.0 catalogue provides a total of ∼3800 sources, corresponding to a surface density $\mu \sim 1.5 \times 10^{-3}$sources arcsec$^{-2}$. Since the X-ray error-circle is 5 arcsec, we estimate that the probability of chance coincidence between a X-ray and an optical source is $\sim 0.1$. Therefore up to 10% of the selected counterparts could be spurious candidates, in the dark cloud region this value decreases.

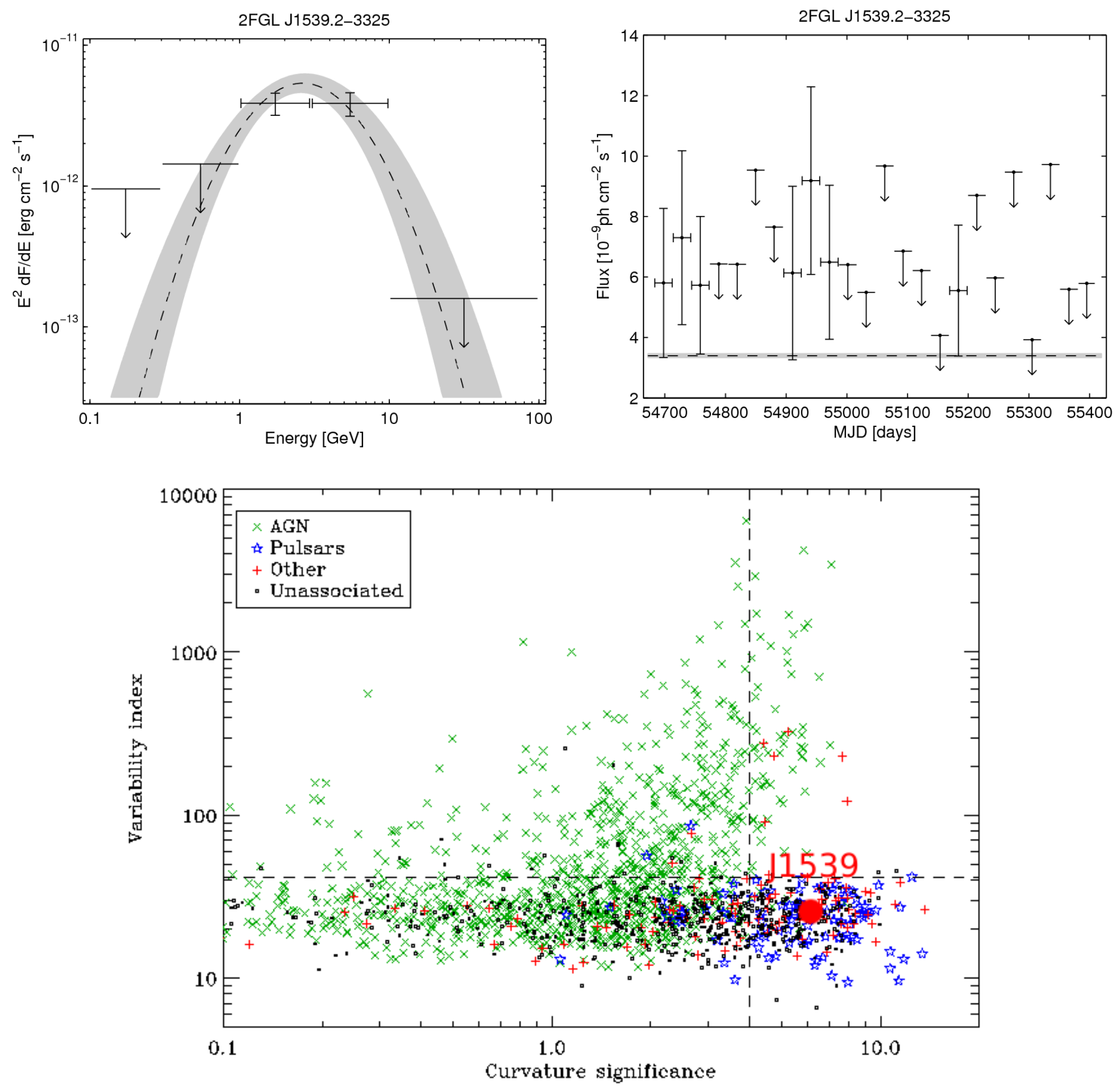

Figure 5.11: Top: $\gamma$-ray spectrum (left) and light curve (right) of the unidentified source 2FGL J1539.2–3325. Bottom: variability index plotted as a function of the curvature significance for different broad classes of sources. Red dot represents the position of 2FGL J1539.2–3325 in this space. This object is not variable and the spectrum is curved.

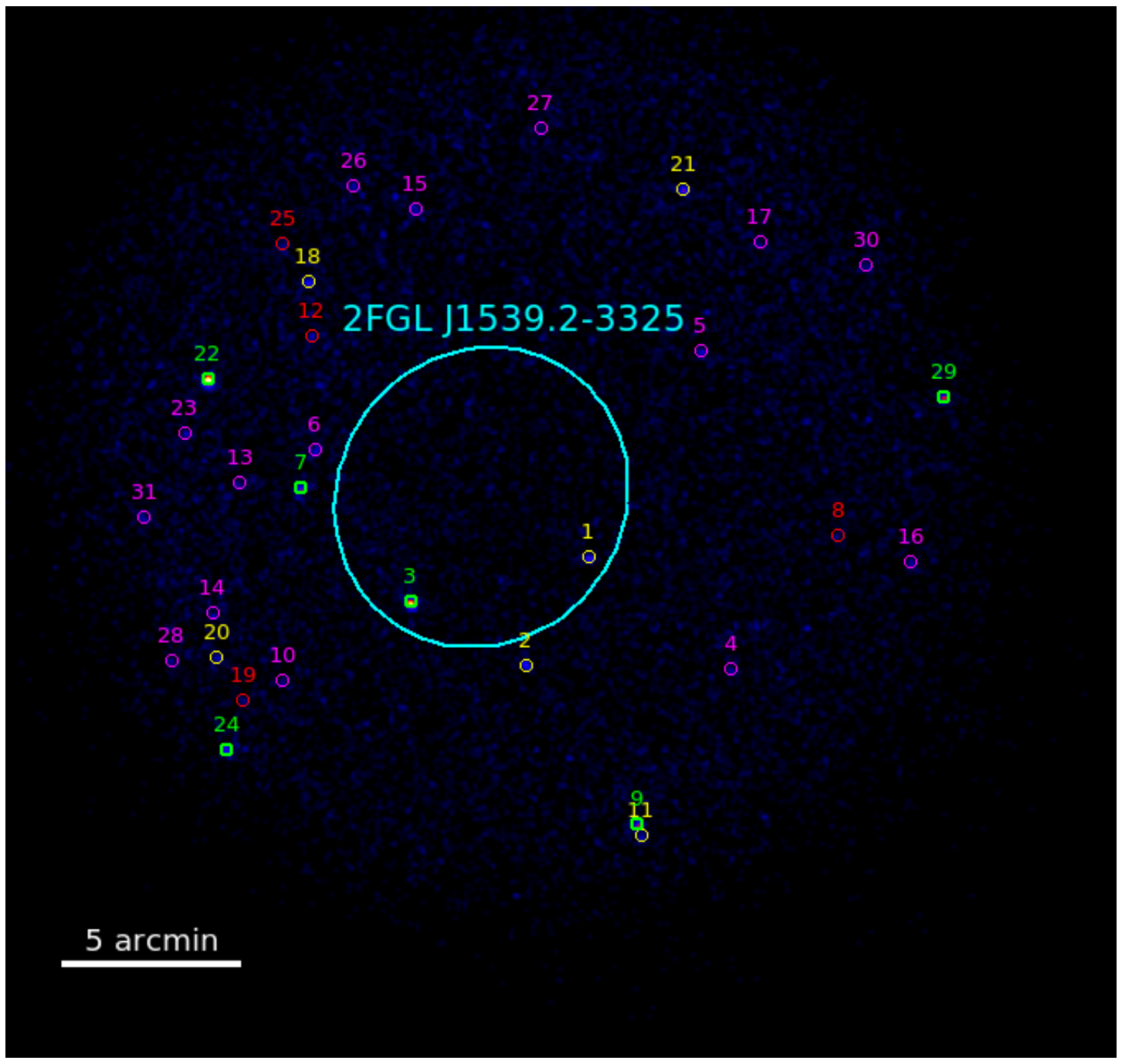

Figure 5.12: 0.3–10 keV exposure-corrected *Swift-XRT* FoV image. A gaussian filter with a kernel radius of 3″ is applied. 2FGL J1539.2–3325 with 95% confidence error ellipse is plotted in cyan. Each X-ray source detected by the XRT is plotted with a circle of 10″ radius. Colors represent the signal-to-noise (SN) ratio of each source: in red sources with SN<3, in magenta sources with 3<TS<4, in yellow sources with 4<TS<5 and in green sources with TS>5.

### Notes on individual X-ray sources

In the following, we present results on the most likely candidate X-ray counterparts to the unidentified $\gamma$-ray source.

**Source #1:** The X-ray analysis yields only 42 counts in the energy band 0.3–10 keV, with a signal-to-noise ratio of 4.7; its count rate in the total energy band is $5.02 \times 10^{-4}$ cts s$^{-1}$. This object is situated within the 2FGL 95% confidence error ellipse at (RA, Dec)=(234°.75, −33°.46). In Figure 5.14 is shown the spectrum of this X-ray sources.

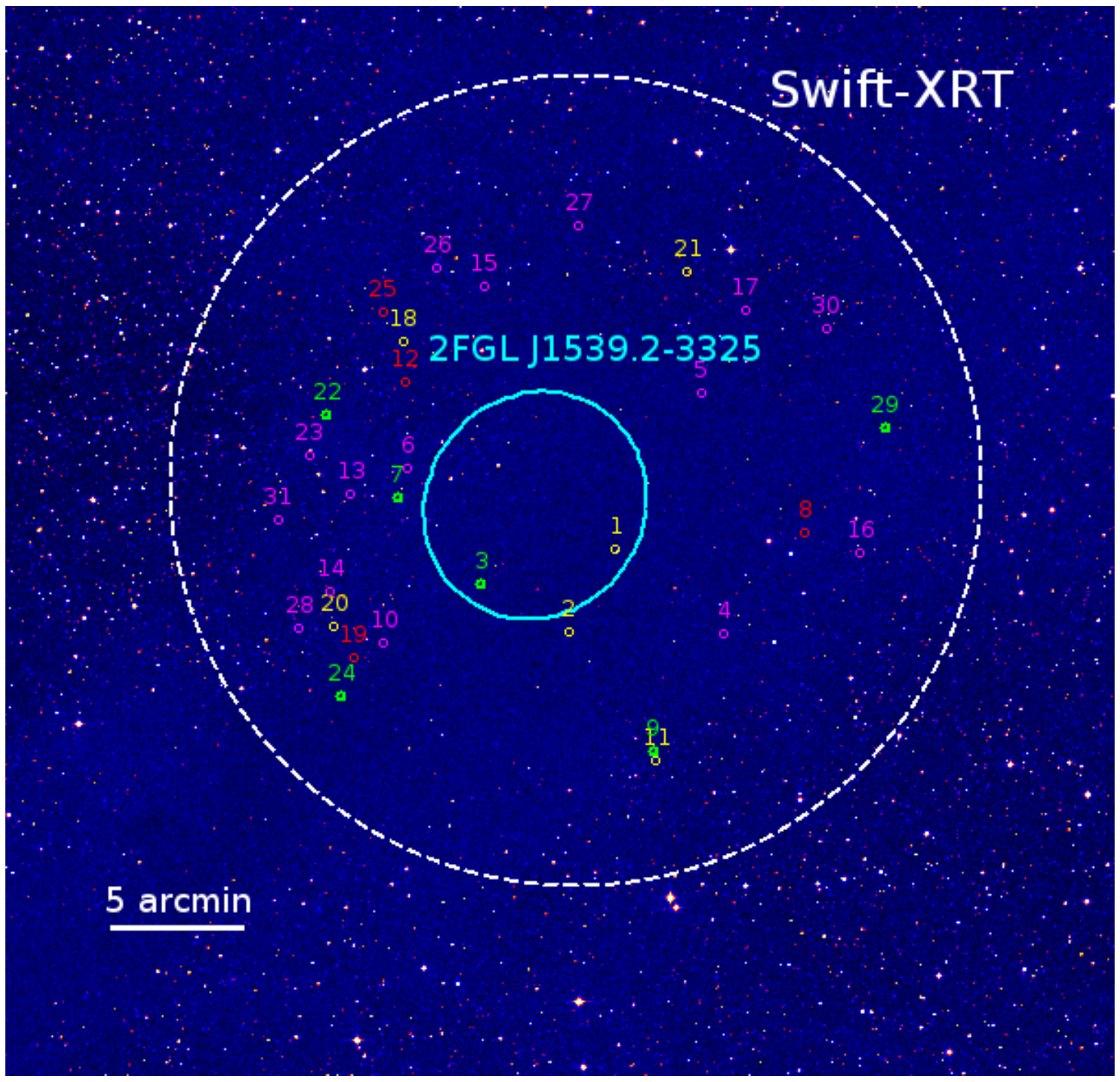

Figure 5.13: *Digital Sky Survey 2* image with red filter of the field of unidentified source 2FGL J1539.2–3325, for which 95% confidence error ellipse is plotted in cyan. Each X-ray source detected by the XRT is overplotted with a circle of 10″ radius colored as described in Figure 5.12. White dashed line represents the FoV of the *Swift*-XRT observation. It is evident the presence of the dense dark cloud Lupus1 in the central region of the image.

The X-ray spectral fitting results are summarized in Table 5.7. We notice that only if we fix the interstellar medium absorption to $N_H = 0.09 \times 10^{22}$ cm$^{-2}$ we find physical plausible best-fit spectral parameter values, this means the X-ray source is closer to us than the dark cloud, thus it cannot be an extragalactic object.

In Figure 5.15 is shown the light curve with 5 bins.

Within 5″ radius X-ray error circle there are three optical/IR candidate counterparts of the source #1, their properties are shown in Table 5.8.
If the optical counterpart of the X-ray source is not detected because too faint, above the limiting magnitude of USNO B1.0 (V>21), then $\log(f_X/f_V) > (0.02 \div 0.32)$, using an unabsorbed visual flux $f_V = 1.07 \times 10^{-14}$. This value does not exclude any source class as counterpart of the X-ray source.

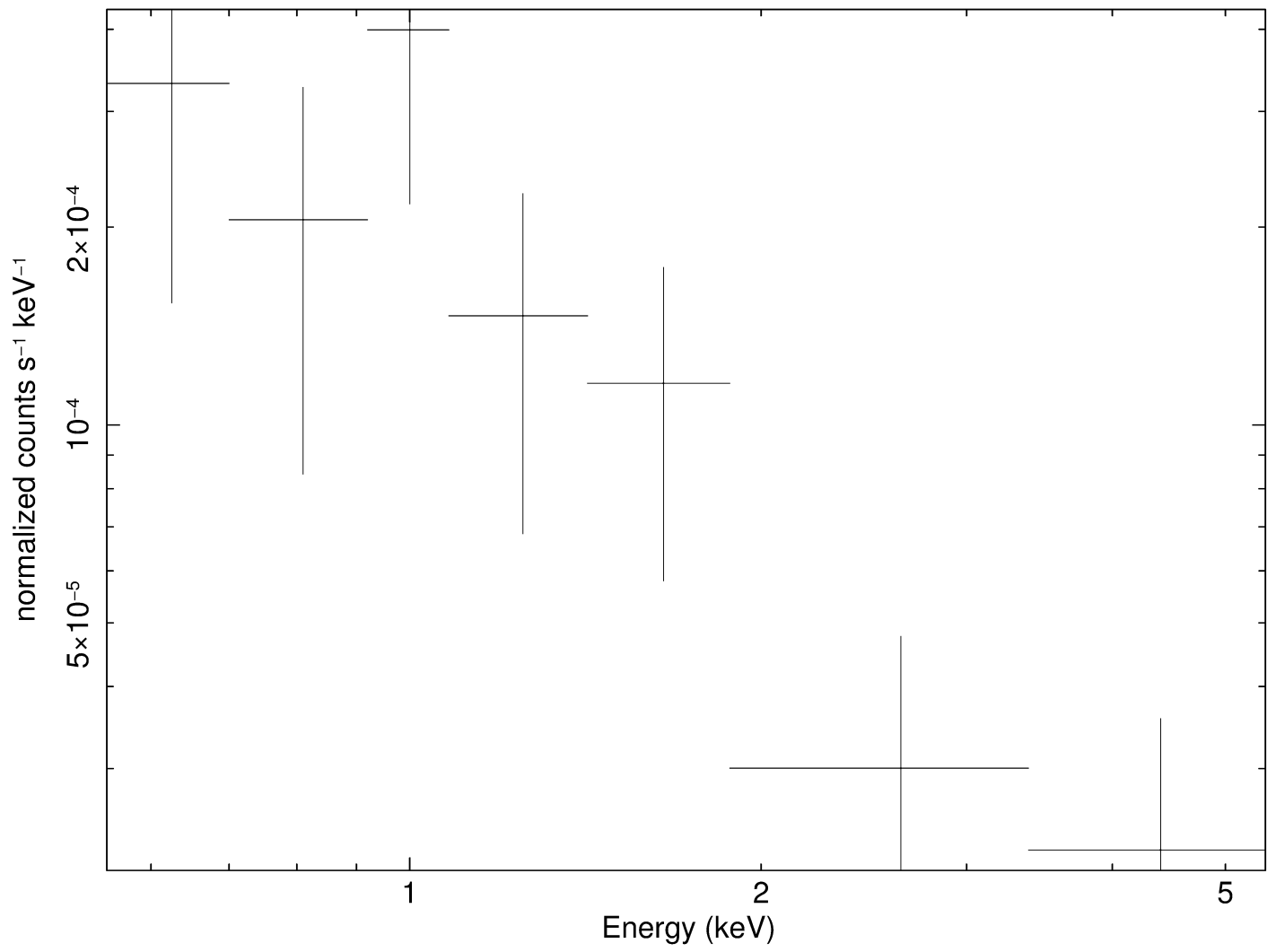

Figure 5.14: Spectrum of the X-ray source #1 binned with a minimum of 5 counts per bin. The binned spectrum shows that the X-ray object is rather absorbed at low energies and that the XRT detected photons until ≃5 keV.

| Parameter | Power-law model | Apec model | Black body model |
|---|---|---|---|
| $N_H{}^a$ | 0.09 (fixed) | 0.09 (fixed) | 0.09 (fixed) |
| $\Gamma$ | $2.05^{+0.65}_{-0.55}$ | - | - |
| kT (keV) | - | $1.59^{+3.48}_{-0.48}$ | $0.28^{+0.12}_{-0.07}$ |
| d.o.f. | 967 | 967 | 967 |
| c-stat | 184.71 | 183.29 | 183.87 |
| goodness$^b$ | 40.9% | 88.7% | 95.3% |
| $\mathrm{Flux_{unabs}}^c$ | 2.16 | 1.26 | 1.12 |

$^a$in units of $10^{22}$ cm$^{-2}$

$^b$fraction of $10^4$ Monte Carlo simulations with fit statistic less then c-stat

$^c$in units of $10^{-14}$ erg cm$^{-2}$ s$^{-1}$ and in the energy range of 0.3–10 keV

Table 5.7: Spectral parameters for the X-ray source #1 within the 95% error ellipse of 2FGL J1539.2–3325. $Flux_{unabs}$ corresponds to the unabsorbed flux. We cannot assert if this X-ray source is a pulsar or star because no spectral model can be rejected by the C-statistic value. The $\gamma$-ray (E>100 MeV) to soft X-ray (0.3–10 keV) flux ratio is $F_\gamma/F_X \simeq (500 \div 950)$, which is compatible with the value of a typical $\gamma$-ray MSP and radio-loud pulsar [135].

**Source #2:** The X-ray analysis yields only 36 counts in the energy band 0.3–10 keV, with a signal-to-noise ratio of 4.3; its count rate in the total energy band is $4.33 \times 10^{-4}$ cts s$^{-1}$. This object is situated within the 2FGL 99% confidence error ellipse at (RA, Dec)=(234°.79, −33°.51). In Figure 5.16 is shown the spectrum of this X-ray sources.

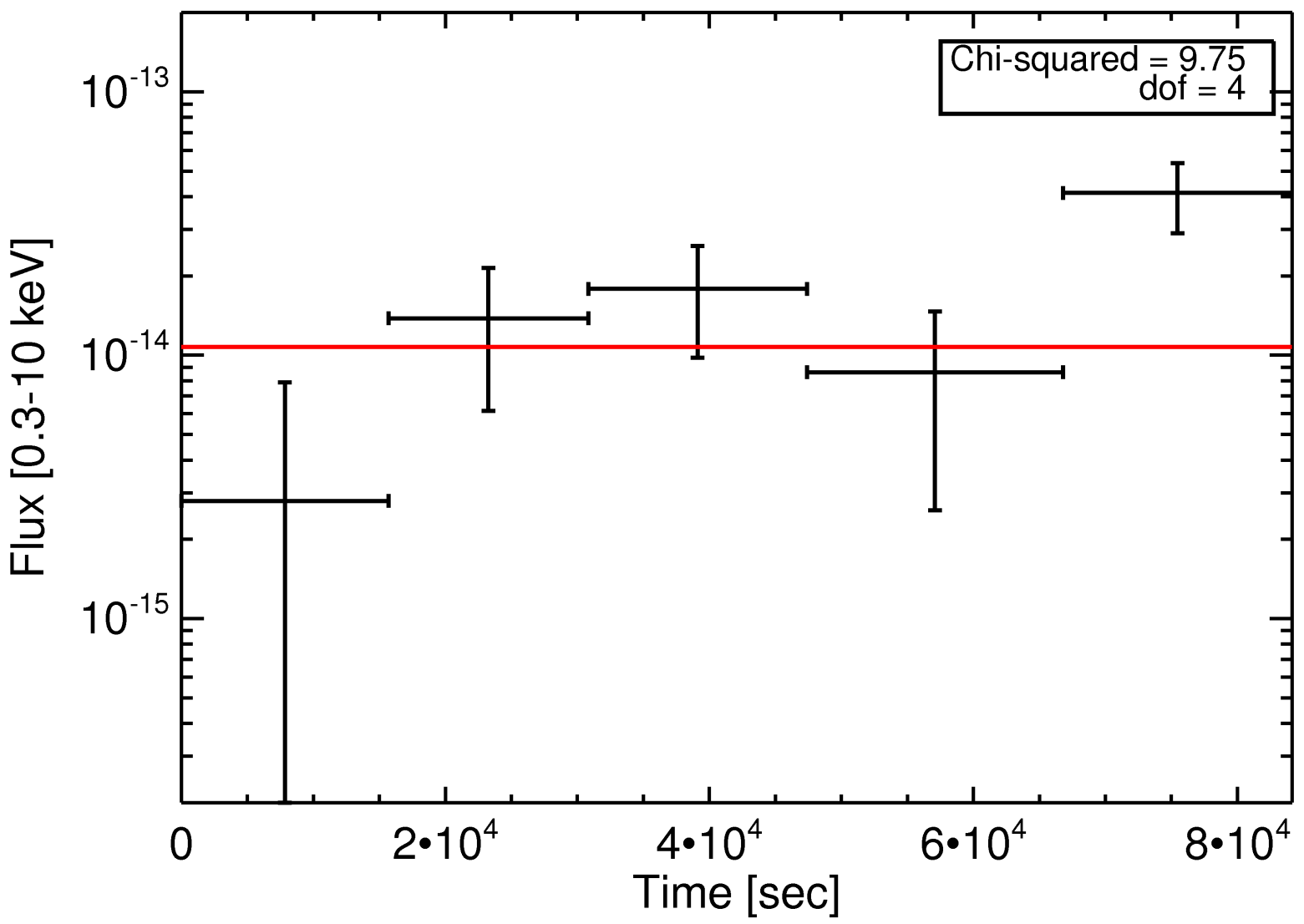

Figure 5.15: Light curve with 5 bins for the X-ray source #1 in the energy range between 0.3 keV and 10 keV within the $\sim$ 84 ks observation span assuming a spectral power law model. The red line depicts the best fit assuming a flat light curve. Error bars for the flux are reported at $1\sigma$. No significant X-ray variability within the $\sim$ 84 ks observation span is detected on the basis of the $\chi^2$ test

The X-ray spectral fitting results are summarized in Table 5.9. We notice that only if we fix the interstellar medium absorption to $N_H = 0.85 \times 10^{22}$ $cm^{-2}$ we find physical plausible best-fit spectral parameter values, this means the X-ray source is in the dark cloud or farther than it.

In Figure 5.17 is shown the light curve with 5 bins.

Within 5″ radius X-ray error circle there is no optical or IR candidate counterpart of the source #2. If the optical counterpart of the X-ray source is not detected because too faint, above the limiting magnitude of USNO B1.0 (V>21), then $\log(f_X/f_V) > (-1.78 \div -1.49)$, using an unabsorbed visual flux $f_V = 1.17 \times 10^{-12}$. This value does not exclude any source class as counterpart of the X-ray source.

**Source #3:** The X-ray analysis yields only 135 counts in the energy band 0.3–10 keV, with a signal-to-noise ratio of 8.9; its count rate in the total energy band is $1.6 \times 10^{-3}$ cts $s^{-1}$. This object is situated within the

| Source #1 – optical/IR analysis | | | |
|---|---|---|---|
| Optical counterpart | GSC | USNO B1.0 | USNO B1.0 |
| IR counterpart | - | 2MASS | - |
| Detection year | 1990 | 1990 | 1968 |
| Distance | 2.4″ | 2.7″ | 3.2″ |
| | B=16.51 | - | B=16.32 |
| De-absorbed | V=16.01[a] | - | V=15.37[a] |
| magnitudes | R=14.59 | R =18.47 | R=13.99 |
| (mag) | I=12.94 | - | - |
| | J=16.27 | J=11.52 | - |
| | H=15.44 | H=11.01 | - |
| | K=15.1 | K=10.72 | - |
| Stellar spectral class | G, K or M | K or M | M |
| $f_{opt}$ (erg cm$^{-2}$ s$^{-1}$) | $1.59 \times 10^{-12}$ | - | $1.89 \times 10^{-12}$ |
| $f_X/f_{opt}$ ($\log_{10}$) | -2.15÷-1.86 | - | -2.23÷-1.94 |
| Suggested class | star | - | star |

[a]value extrapolated as the average between absorbed B and R magnitudes

Table 5.8: Properties of the optical/IR candidate counterparts of the X-ray source #1. For a description of each row see Table 5.5. We assert that the three candidate counterparts are probably the same object because of the value of their proper motion and their magnitudes.

2FGL 95% confidence error ellipse at (RA, Dec)=(234°.85, −33°.48). In Figure 5.18 is shown the spectrum of this X-ray sources.

The X-ray spectral fitting results are summarized in Table 5.10. We notice that only if we fix the interstellar medium absorption to $N_H = 0.85 \times 10^{22}$ cm$^{-2}$ we find physical plausible best-fit spectral parameter values, this means the X-ray source is in the dark cloud or farther than it.

In Figure 5.19 is shown the light curve with 5 bins.

Within 5″ radius X-ray error circle there are two optical/IR candidate counterparts of the source #3, their properties are shown in Table 5.11.

If the optical counterpart of the X-ray source is not detected because too faint, above the limiting magnitude of USNO B1.0 (V>21), then $\log(f_X/f_V) > (-1.11 \div -0.75)$, using an unabsorbed visual flux $f_V = 1.17 \times 10^{-12}$. This value does not exclude any source class as counterpart of the X-ray source.

**Source #6:** The X-ray analysis yields only 35 counts in the energy band

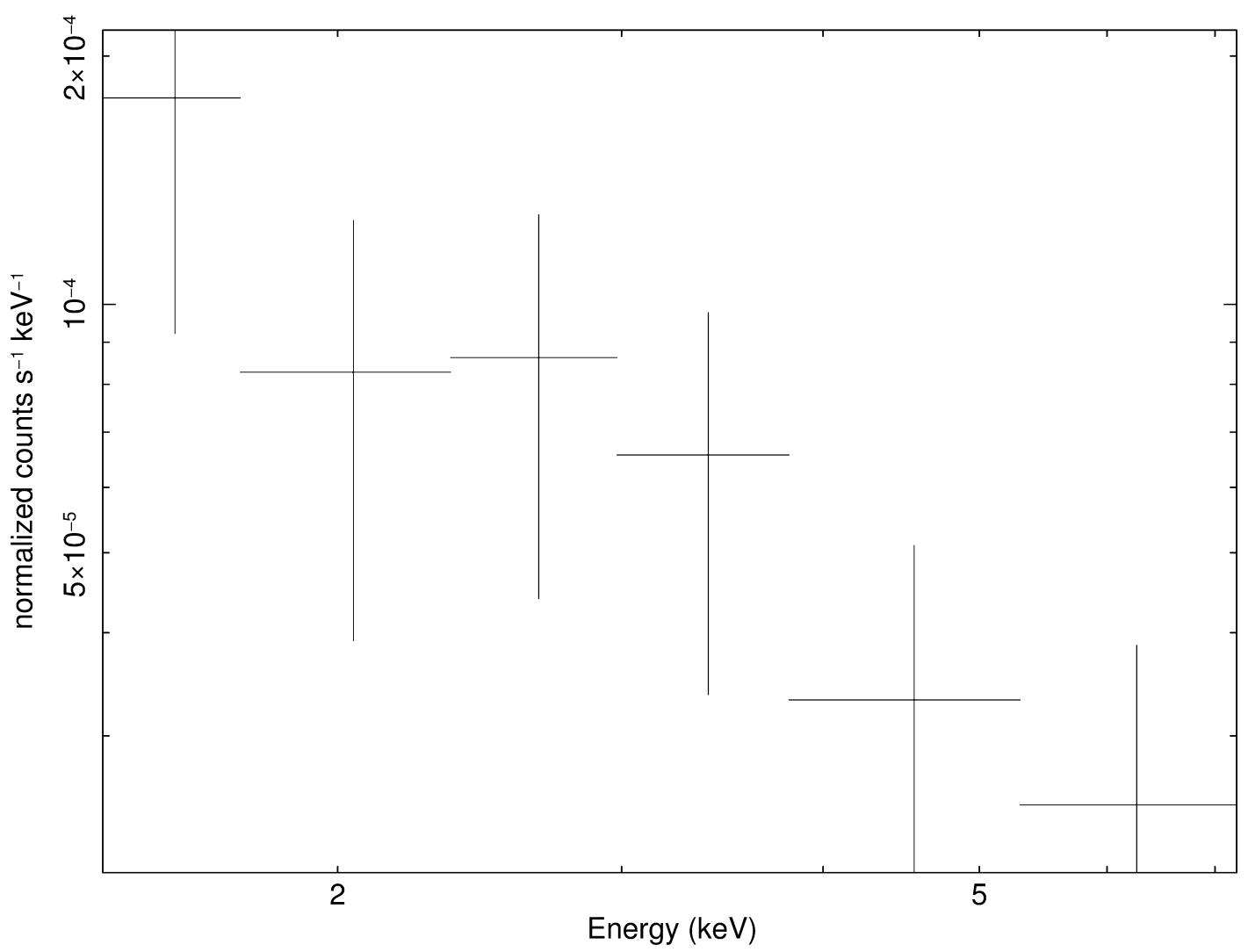

Figure 5.16: Spectrum of the X-ray source #2 binned with a minimum of 5 counts per bin. The binned spectrum shows that the X-ray object is very absorbed at low energies and that the XRT detected photons until ≃7 keV.

| Parameter | Power-law model | Apec model | Black body model |
|---|---|---|---|
| $N_H{}^a$ | 0.85 (fixed) | 0.85 (fixed) | 0.85 (fixed) |
| $\Gamma$ | $1.66^{+1.01}_{-0.86}$ | - | - |
| kT (keV) | - | $5.45^{+23.12}_{-3.33}$ | $0.65^{+0.39}_{-0.20}$ |
| d.o.f. | 967 | 967 | 967 |
| c-stat | 207.45 | 206.38 | 205.06 |
| goodness$^b$ | 68.1% | 70.3% | 92.3% |
| $\mathrm{Flux_{unabs}}^c$ | 3.76 | 3.40 | 1.93 |

$^a$in units of $10^{22}$ cm$^{-2}$

$^b$fraction of $10^4$ Monte Carlo simulations with fit statistic less then c-stat

$^c$in units of $10^{-14}$ erg cm$^{-2}$ s$^{-1}$ and in the energy range of 0.3–10 keV

Table 5.9: Spectral parameters for the X-ray source #2 within the 2FGL 99% error ellipse of 2FGL J1539.2–3325. $Flux_{unabs}$ corresponds to the unabsorbed flux. We cannot assert if this X-ray source is a pulsar a star or an AGN because no spectral model can be rejected by the C-statistic value. The $\gamma$-ray (E>100 MeV) to soft X-ray (0.3–10 keV) flux ratio is $F_\gamma/F_X \simeq (280 \div 550)$, which is compatible with the value of a typical $\gamma$-ray MSP and radio-loud pulsar [135].

0.3–10 keV, with a signal-to-noise ratio of 3.8; its count rate in the total energy band is $4.1 \times 10^{-4}$ cts s$^{-1}$. This object is situated within the 2FGL 99% confidence error ellipse at (RA, Dec)=(234°.91, −33°.41). In

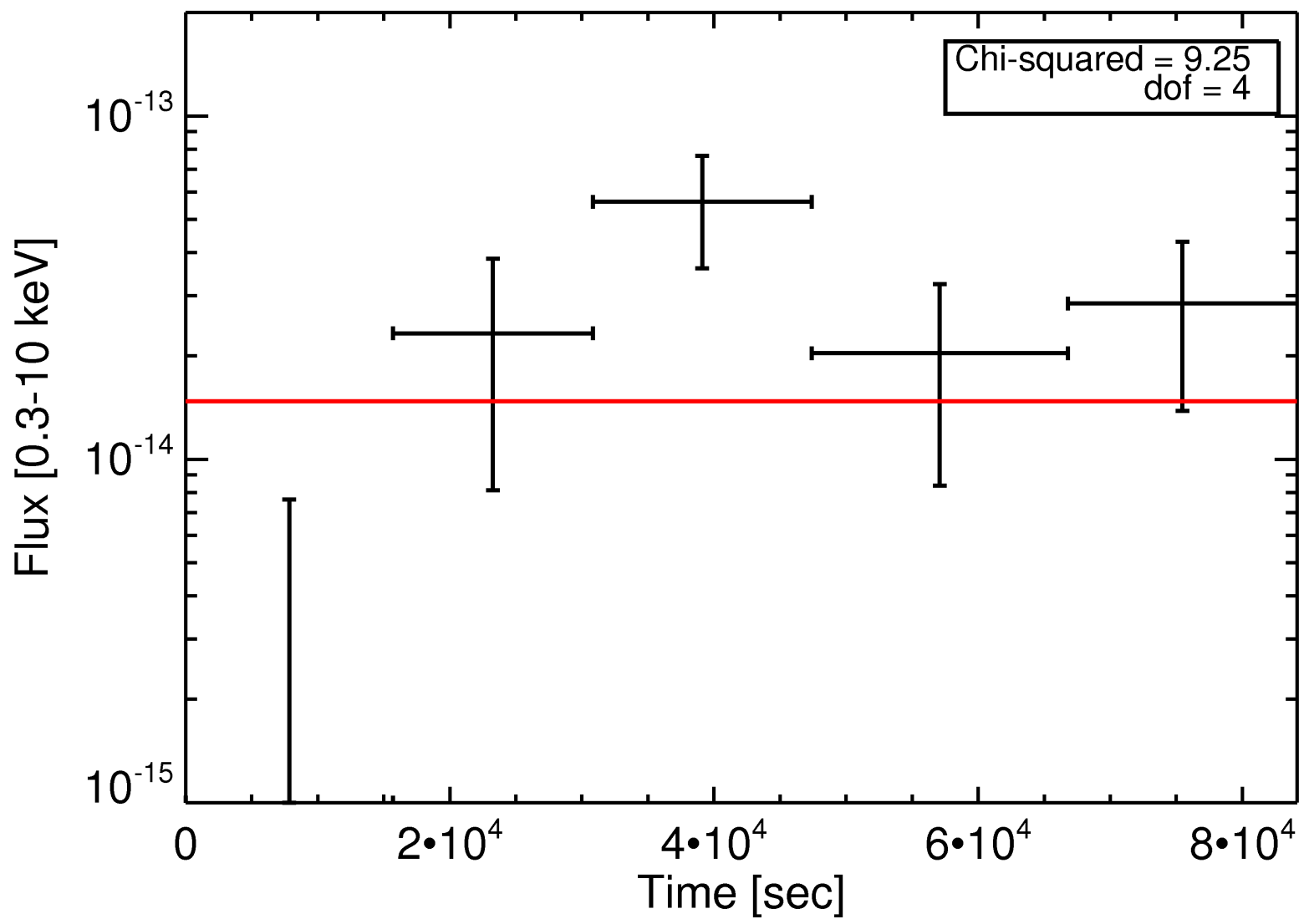

Figure 5.17: Light curve with 5 bins for the X-ray source #2 in the energy range between 0.3 keV and 10 keV within the $\sim$ 84 ks observation span assuming a spectral power law model. The red line depicts the best fit assuming a flat light curve. Error bars for the flux are reported at $1\sigma$. No significant X-ray variability within the observation span is detected on the basis of the $\chi^2$ test

| Parameter | Power-law model | Apec model | Black body model |
|---|---|---|---|
| $N_H{}^a$ | 0.85 (fixed) | 0.85 (fixed) | 0.85 (fixed) |
| $\Gamma$ | $2.45^{+0.38}_{-0.36}$ | - | - |
| kT (keV) | - | $2.64^{+1.02}_{-0.59}$ | $0.54^{+0.08}_{-0.07}$ |
| d.o.f. | 967 | 967 | 967 |
| c-stat | 346.44 | 348.19 | 344.85 |
| goodness$^b$ | 48.2% | 56.9% | 93.7% |
| $\text{Flux}_{\text{unabs}}{}^c$ | 20.5 | 13.2 | 9.13 |

$^a$in units of $10^{22}$ cm$^{-2}$

$^b$fraction of $10^4$ Monte Carlo simulations with fit statistic less then c-stat

$^c$in units of $10^{-14}$ erg cm$^{-2}$ s$^{-1}$ and in the energy range of 0.3–10 keV

Table 5.10: Spectral parameters for the X-ray source #3 within the 95% error ellipse of 2FGL J1539.2–3325. $Flux_{unabs}$ corresponds to the unabsorbed flux. We cannot assert if this X-ray source is a pulsar, a star or an AGN because no spectral model can be rejected by the C-statistic value. The $\gamma$-ray (E>100 MeV) to soft X-ray (0.3–10 keV) flux ratio is $F_\gamma/F_X \simeq (50 \div 100)$, which is compatible only with the value of a typical $\gamma$-ray MSP [135].

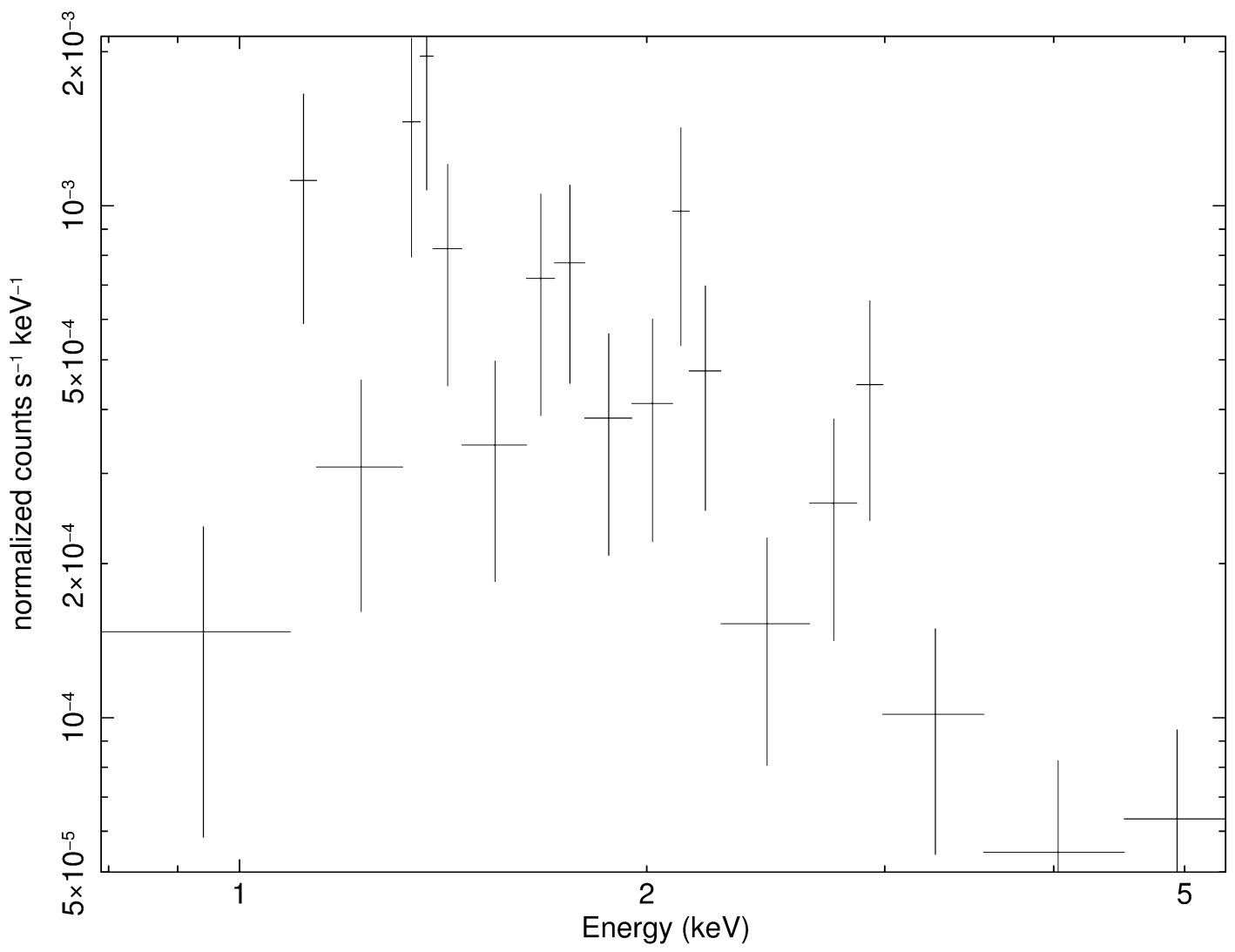

Figure 5.18: Spectrum of the X-ray source #3 binned with a minimum of 5 counts per bin. The binned spectrum shows that the X-ray object is very absorbed at low energies and that the XRT detected photons until ≃8 keV.

Figure 5.20 is shown the spectrum of this X-ray sources.

The X-ray spectral fitting results are summarized in Table 5.12. We notice that only if we fix the interstellar medium absorption to $N_H = 0.85 \times 10^{22}$ cm$^{-2}$ we find physical plausible best-fit spectral parameter values, this means the X-ray source is in the dark cloud or farther than it.

In Figure 5.21 is shown the light curve with 5 bins.

Within 5″ radius X-ray error circle there is only an IR candidate counterpart of the source #6 detected by WISE but with no magnitude values. If the optical counterpart of the X-ray source is not detected because too faint, above the limiting magnitude of USNO B1.0 (V>21), then $\log(f_X/f_V) > (-1.79 \div -1.43)$, using an unabsorbed visual flux $f_V = 1.17 \times 10^{-12}$. This value does not exclude any source class as counterpart of the X-ray source.

**Source #7:** The X-ray analysis yields only 54 counts in the energy band 0.3–10 keV, with a signal-to-noise ratio of 5.1; its count rate in the total energy band is $6.4 \times 10^{-4}$ cts s$^{-1}$. This object is situated within the 2FGL 99% confidence error ellipse at (RA, Dec)=(234°.91, −33°.42). In Figure 5.22 is shown the spectrum of this X-ray sources. The X-ray spectral fitting results are summarized in Table 5.13. We notice that only if we fix the interstellar medium absorption to $N_H = 0.09 \times 10^{22}$ cm$^{-2}$ we find physical plausible best-fit spectral parameter values, this means the X-ray source is closer to us than the dark cloud, it cannot be an extragalactic object.

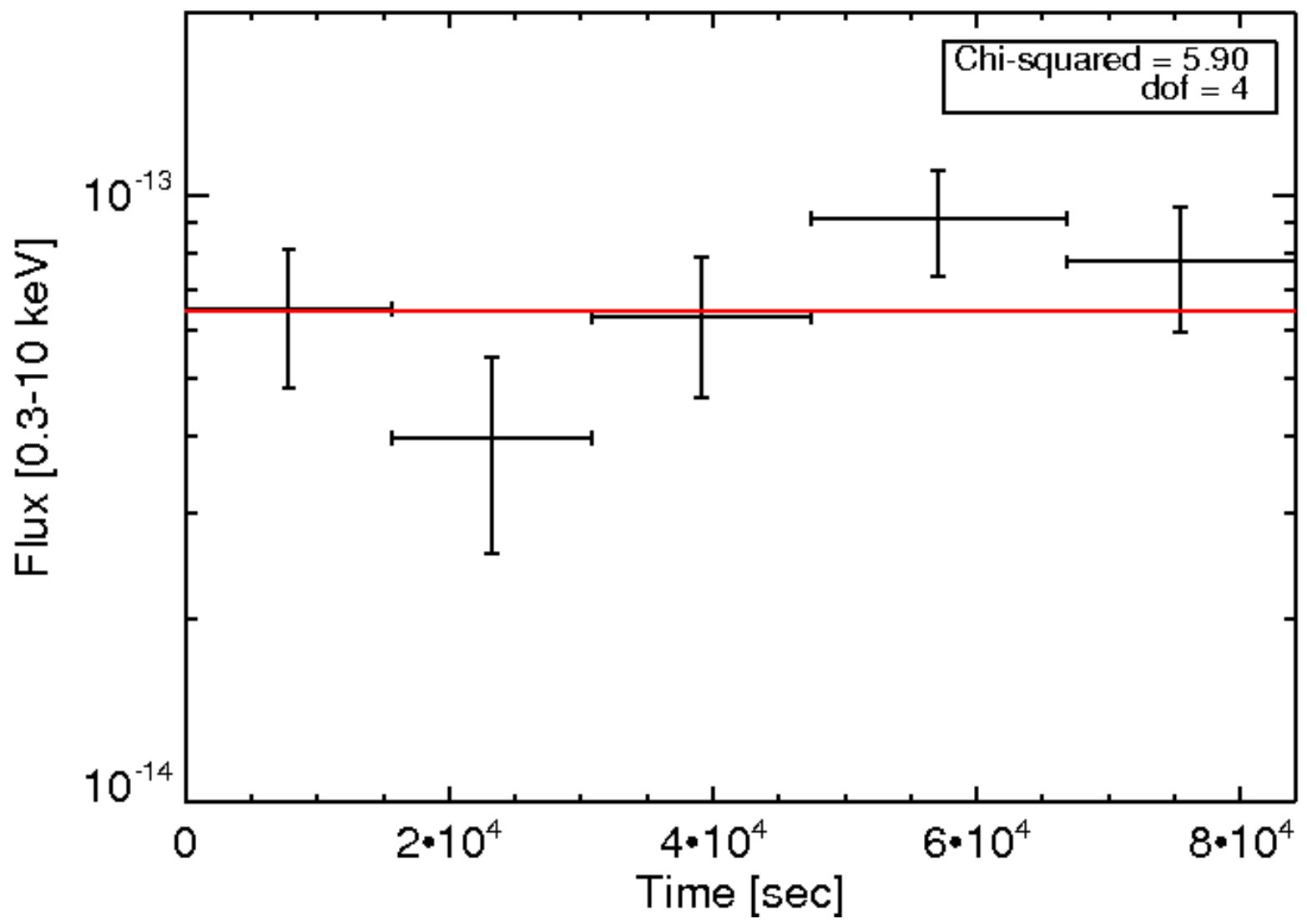

Figure 5.19: Light curve with 5 bins for the X-ray source #3 in the energy range between 0.3 keV and 10 keV within the $\sim$ 84 ks observation span assuming a spectral power law model. The red line depicts the best fit assuming a flat light curve. Error bars for the flux are reported at $1\sigma$. No significant X-ray variability within the observation span is detected on the basis of the $\chi^2$ test

In Figure 5.23 is shown the light curve with 5 bins.

Within 5″ radius X-ray error circle there are two optical/IR candidate counterparts of the source #7, their properties are shown in Table 5.14.

If the optical counterpart of the X-ray source is not detected because too faint, above the limiting magnitude of USNO B1.0 (V>21), then $\log(f_X/f_V) > (-0.14 \div 0.08)$, using an unabsorbed visual flux $f_V = 2.69 \times 10^{-14}$. This value does not exclude any source class as counterpart of the X-ray source.

### Source hardness ratios distribution

In Figure 5.24 is shown the distribution of the HRs of the X-ray sources detected by the *Swift-XRT* within the 99% error circle of the *Fermi*-LAT source.

| Source #3 – optical/IR analysis | | |
|---|---|---|
| Optical counterpart | USNO B1.0 | GSC |
| IR counterpart | 2MASS | - |
| Detection year | 1986 | 1990 |
| Distance | 2.7″ | 3.6″ |
| | B=13.6 | B=15.57 |
| De-absorbed | V=14.64[a] | V=16.25[a] |
| magnitudes | R=15.3 | R=16.54 |
| (mag) | - | I=16.68 |
| | J=15.05 | - |
| | H=14.42 | - |
| | K=14.56 | - |
| Stellar Stellar spectral class | ? | ? |
| $f_{opt}$ (erg cm$^{-2}$ s$^{-1}$) | $2.3 \times 10^{-11}$ | $3.78 \times 10^{-12}$ |
| $f_X/f_{opt}$ ($\log_{10}$) | -2.4÷-2.05 | -1.62÷-1.27 |
| Suggested class | star | star |

[a]value extrapolated as the average between absorbed B and R magnitudes

Table 5.11: Properties of the optical/IR candidate counterparts of the X-ray source #3. For a description of each row see Table 5.5. The candidate counterparts are classified as stars but these results are not compatible with their colors, this means either that these objects may be not the candidate counterpart of the X-ray source or that they may be two interacting objects in a binary system (e.g. a star and a pulsar) making this X-ray source very interesting. Since the X-ray source is probably located in the dark cloud, the identification is uncertain because of the $N_H$ value we have assumed, thus this object may also be a star in formation.

Sources #1 and#7 are little absorbed and are characterized by a rather soft spectrum ($HR12 \simeq 0$ and $HR23 < 0$), probably they are nearby stars. Source #6 is probably located at the limit of the dark cloud because is rather absorbed. The other two sources are very absorbed, they are probably situated in the dark cloud or farther, and source #2 is characterized by a very hard spectrum ($HR12 \simeq 1$ and $HR23 \simeq 0.5$). Source #2 and #3 are very interesting and probably one of these two objects may be the counterpart of the 2FGL unidentified source.

The values of the expected HRs are overplotted in Figure 5.24. As can be seen, the distributions are compatible with a rather wide range of temperatures and photon indexes.

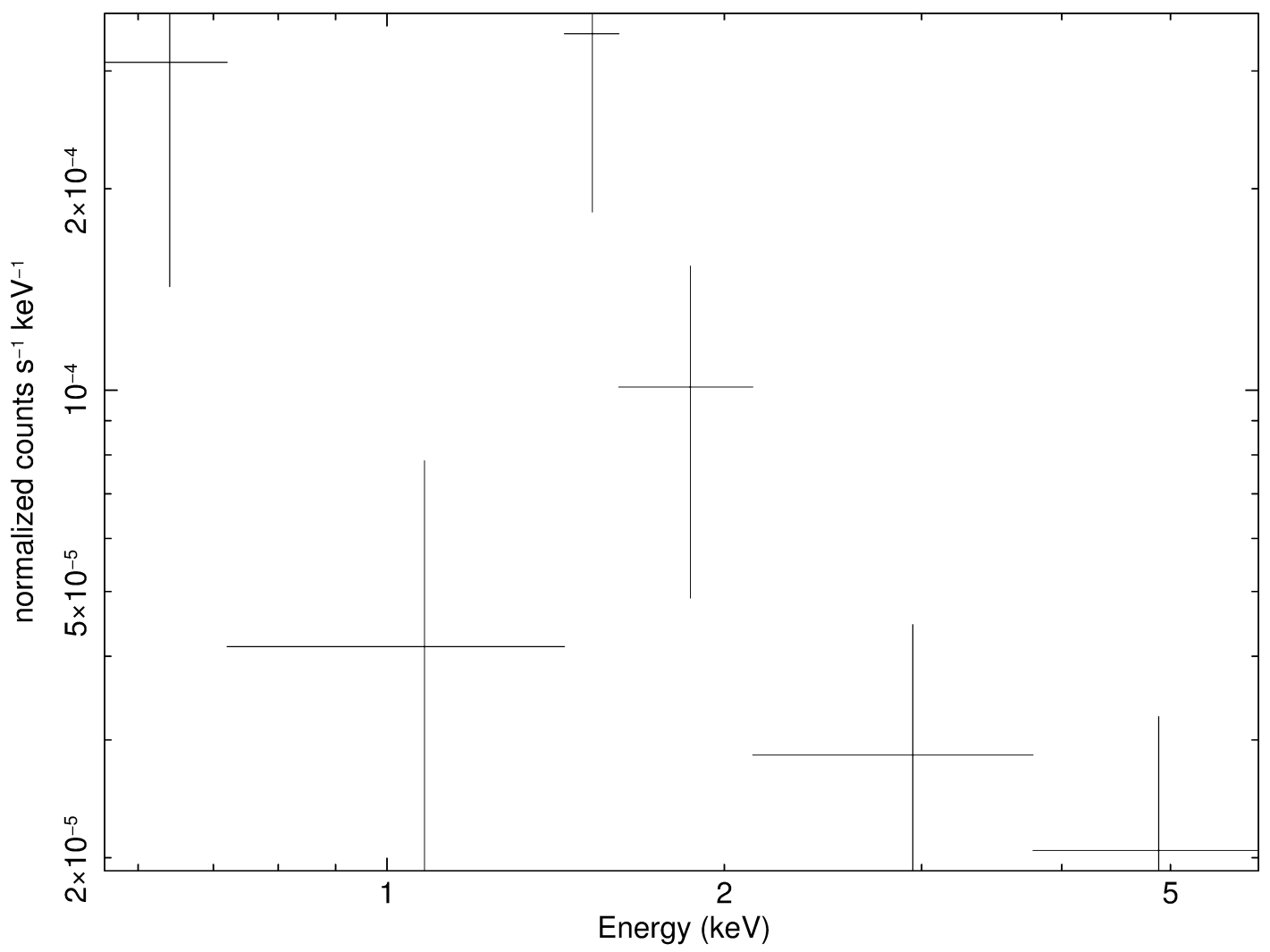

Figure 5.20: Spectrum of the X-ray source #6 binned with a minimum of 5 counts per bin. The binned spectrum shows that the X-ray object is not very absorbed at low energies and that the XRT detected photons until ≃5 keV.

| Parameter | Power-law model | Apec model | Black body model |
|---|---|---|---|
| $N_H{}^a$ | 0.85 (fixed) | 0.85 (fixed) | 0.85 (fixed) |
| $\Gamma$ | $2.43^{+1.56}_{-1.09}$ | - | - |
| kT (keV) | - | $9.71^{+28.47}_{-7.80}$ | $0.36^{+0.24}_{-0.11}$ |
| d.o.f. | 967 | 967 | 967 |
| c-stat | 195.64 | 197.91 | 193.73 |
| goodness$^b$ | 85.2% | 64.4% | 99.4% |
| $\mathrm{Flux_{unabs}}{}^c$ | 4.34 | 3.54 | 1.89 |

$^a$in units of $10^{22}$ cm$^{-2}$

$^b$fraction of $10^4$ Monte Carlo simulations with fit statistic less then c-stat

$^c$in units of $10^{-14}$ erg cm$^{-2}$ s$^{-1}$ and in the energy range of 0.3–10 keV

Table 5.12: Spectral parameters for the X-ray source #6 within the 2FGL 99% error ellipse of 2FGL J1539.2–3325. $Flux_{unabs}$ corresponds to the unabsorbed flux. We cannot assert if this X-ray source is a pulsar, a star or an AGN because no spectral model can be rejected by the C-statistic value. The $\gamma$-ray (E>100 MeV) to soft X-ray (0.3–10 keV) flux ratio is $F_\gamma/F_X \simeq (240 \div 560)$, which is compatible with the value of a typical $\gamma$-ray MSP and radio-loud pulsar [135].

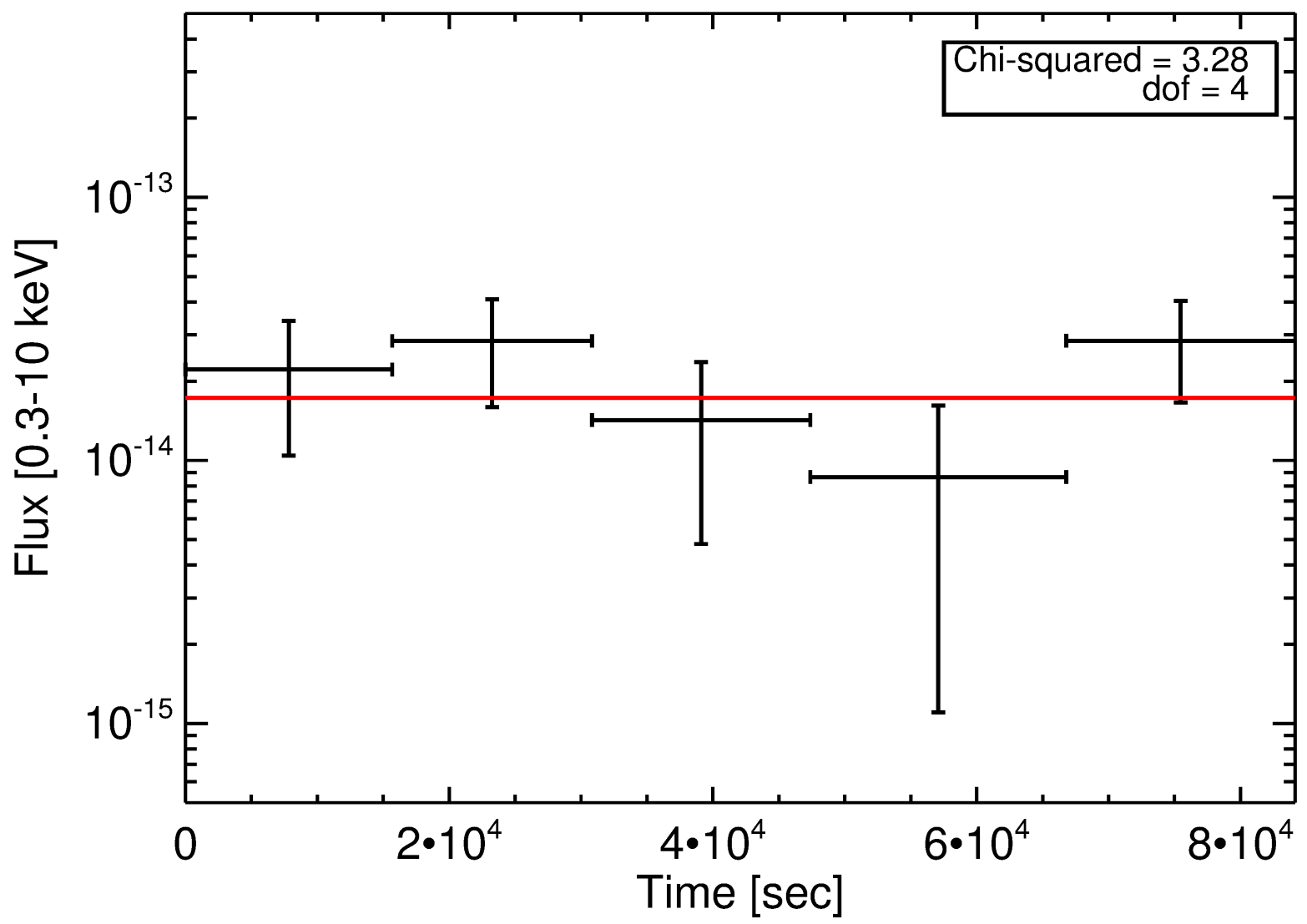

Figure 5.21: Light curve with 5 bins for the X-ray source #6 in the energy range between 0.3 keV and 10 keV within the $\sim$ 84 ks observation span assuming a spectral power law model. The red line depicts the best fit assuming a flat light curve. Error bars for the flux are reported at $1\sigma$. No significant X-ray variability within the observation span is detected on the basis of the $\chi^2$ test

| Parameter | Power-law model | Apec model | Black body model |
|---|---|---|---|
| $N_H{}^a$ | 0.09 (fixed) | 0.09 (fixed) | 0.09 (fixed) |
| $\Gamma$ | $2.47^{+0.46}_{-0.43}$ | - | - |
| kT (keV) | - | $1.33^{+1.36}_{-1.04}$ | $0.26^{+0.06}_{-0.05}$ |
| d.o.f. | 967 | 967 | 967 |
| c-stat | 197.52 | 204.27 | 194.44 |
| goodness$^b$ | 24% | 90% | 84.3% |
| $\text{Flux}_{\text{unabs}}{}^c$ | 3.29 | 1.93 | 1.99 |

$^a$in units of $10^{22}$ cm$^{-2}$

$^b$fraction of $10^4$ Monte Carlo simulations with fit statistic less then c-stat

$^c$in units of $10^{-14}$ erg cm$^{-2}$ s$^{-1}$ and in the energy range of 0.3–10 keV

Table 5.13: Spectral parameters for the X-ray source #7 within the 2FGL 99% error ellipse of 2FGL J1539.2–3325. $Flux_{unabs}$ corresponds to the unabsorbed flux. We cannot assert if this X-ray source is a pulsar or a star because no spectral model can be rejected by the C-statistic value. The $\gamma$-ray (E>100 MeV) to soft X-ray (0.3–10 keV) flux ratio is $F_\gamma/F_X \simeq (320 \div 530)$, which is compatible with the value of a typical $\gamma$-ray MSP and radio-loud pulsar [135].

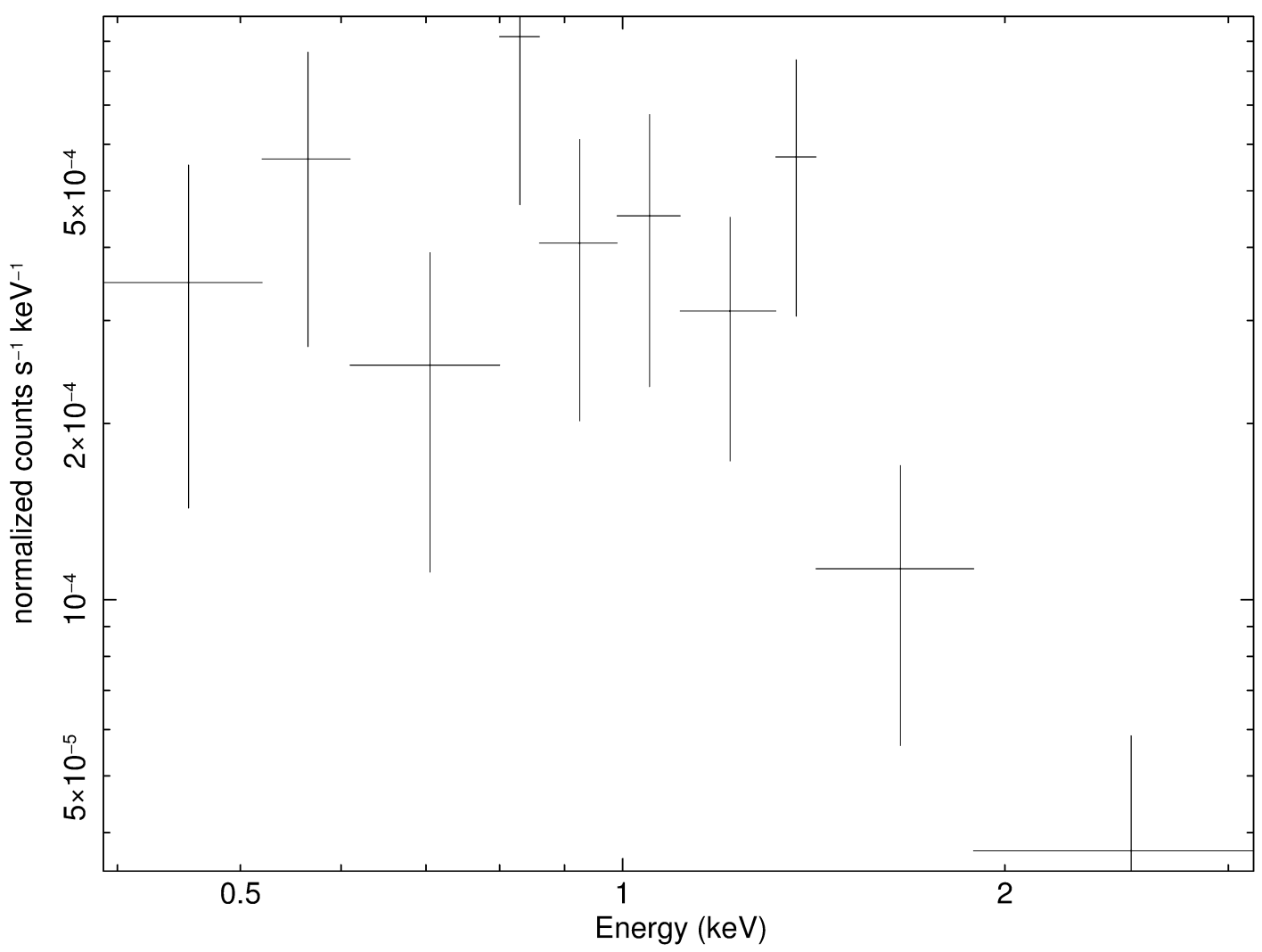

Figure 5.22: Spectrum of the X-ray source #7 binned with a minimum of 5 counts per bin. The binned spectrum shows that the X-ray object is not very absorbed at low energies and that the XRT detected photons only until $\simeq$2.5 keV.

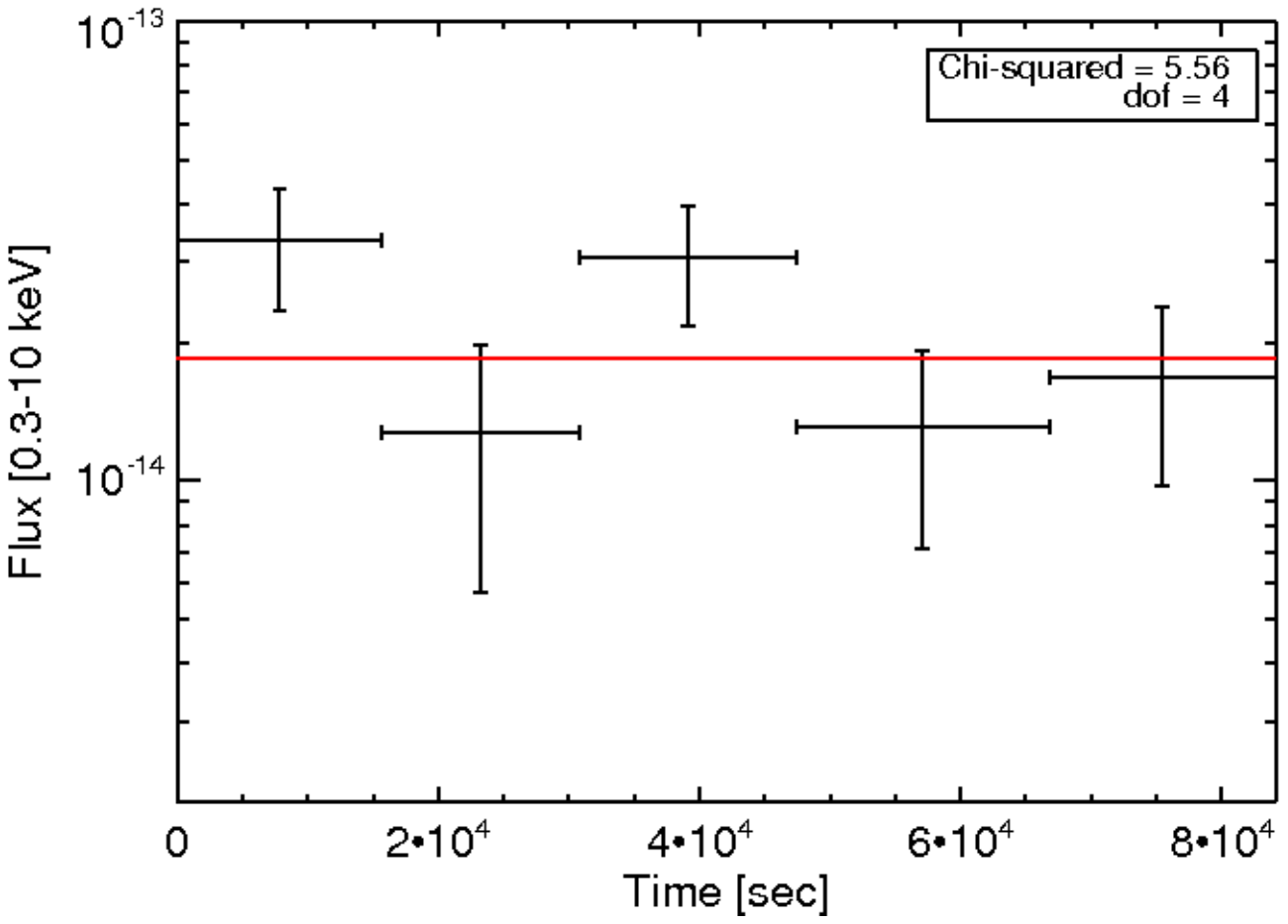

Figure 5.23: Light curve with 5 bins for the X-ray source #7 in the energy range between 0.3 keV and 10 keV within the $\sim$ 84 ks observation span assuming a spectral power law model. The red line depicts the best fit assuming a flat light curve. Error bars for the flux are reported at $1\sigma$. No significant X-ray variability within the observation span is detected on the basis of the $\chi^2$ test

| Source #7 – optical/IR analysis | | |
|---|---|---|
| Optical counterpart | USNO B1.0 | GSC |
| IR counterpart | 2MASS | - |
| Detection year | 1975 | 1990 |
| Distance | 0.8″ | 1.4″ |
| | B=17.74 | B=17.92 |
| De-absorbed | V=17 | V=17.13[a] |
| magnitudes | R=16.36 | R=16.23 |
| (mag) | - | I=14.15 |
| | J=12.85 | - |
| | H=12.28 | - |
| | K=12.01 | - |
| Stellar spectral class | M or K | G, K or M |
| $f_{opt}$ (erg cm$^{-2}$ s$^{-1}$) | $6.77 \times 10^{-13}$ | $4.34 \times 10^{-13}$ |
| $f_X/f_{opt}$ ($\log_{10}$) | -1.53÷-1.31 | -1.34÷-1.12 |
| Suggested class | star or CV | star |

[a]value extrapolated as the average between absorbed B and R magnitudes

Table 5.14: Properties of the optical/IR candidate counterparts of the X-ray source #7. For a description of each row see Table 5.5. We assert that the two candidate counterparts may be the same object on the basis of their proper motion and their magnitudes.

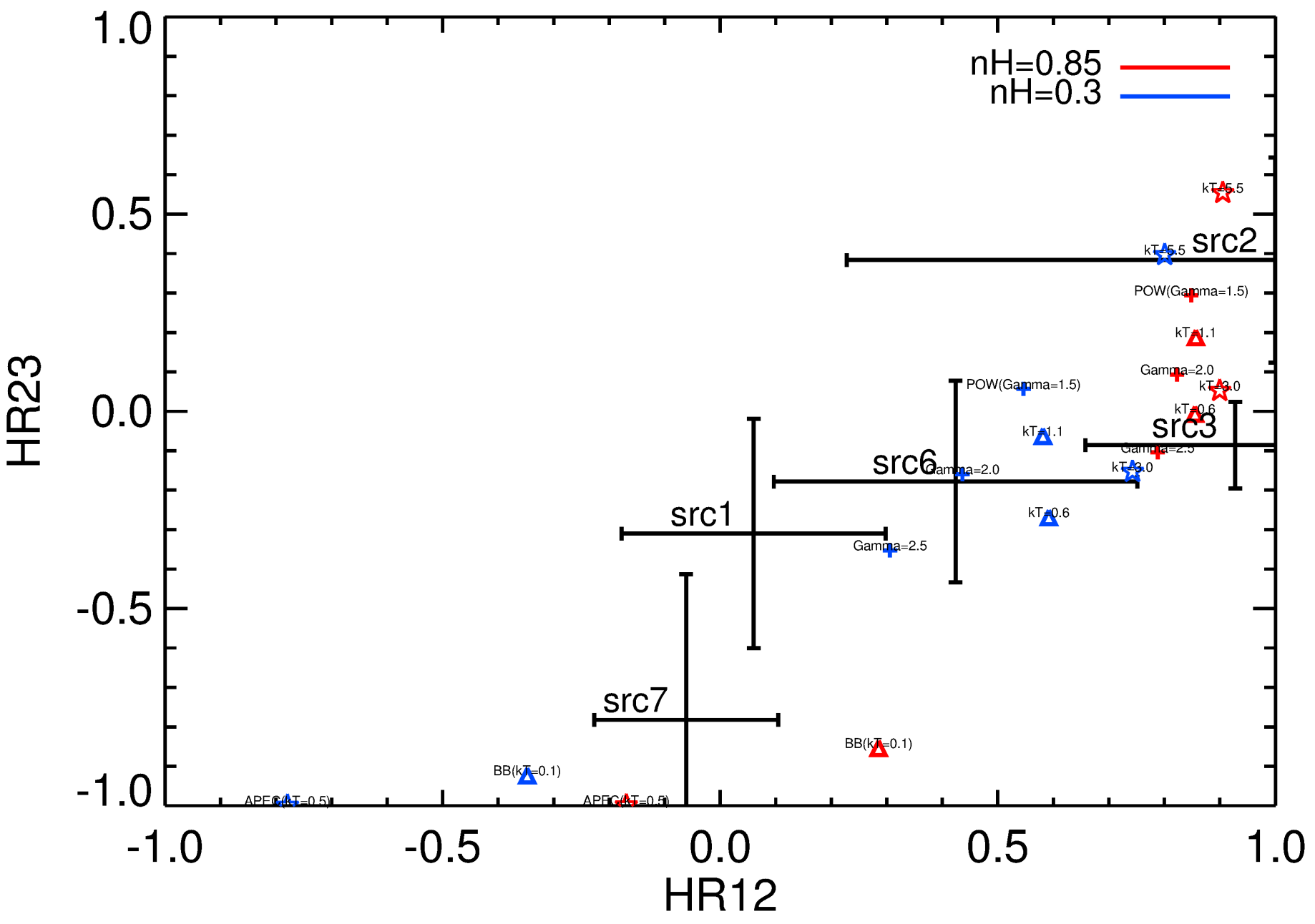

Figure 5.24: Distribution of HR12 vs. HR23 of the X-ray sources detected by the *Swift-XRT* within the 99% error circle of the *Fermi*-LAT source. Error bars are reported at $1\sigma$. Crosses indicate the expected HR12 vs. HR23 computed for power law spectra with $\Gamma$ from 1.5 and 2.5. Stars indicate the expected HR12 vs. HR23 computed for apec spectra with kT from 0.5 to 5.5 keV. Triangles indicate the expected HR12 vs. HR23 computed for black body spectra with kT from 0.1 to 1.1 keV. Each spectral model is computed using the interstellar medium absorption given by its value in the dark cloud (red) and one third of this value (blue).

### Discussion

In the *Swift-XRT* FoV of 2FGL J1539.2–3325 we have detected 31 X-ray sources between 0.3 and 10 keV, only 5 of them are situated within the 99% error circle of the *Fermi*-LAT source, we can consider them as plausible counterparts of the putative MSPs. For each candidate X-ray counterpart we have not detected any significant X-ray variability and by an analysis of $f_\gamma/f_X$ each one seems to be a plausible counterpart of the 2FGL unidentified source. Sources #1 and #7 can be associated with nearby stars, located closer to us than the Lupus1 dark cloud, for each one we have identified the optical/IR counterpart. Source #6 has an IR candidate counterpart within 5″ radius X-ray error circle detected by WISE but without any magnitude value. It is likely situated at the edge of the dark cloud, analyzing its spectrum we observe that this X-ray object may be the counterpart of the putative MSP. In the end, sources #2 and #3 are situated inside or farther than the dark cloud because their spectra are very absorbed. Probably the source #3 is situated inside the dark cloud and it may be a star in formation (e.g. a T-Tauri star). Source #2 has not any optical/IR candidate counterpart within 5″ radius X-ray error circle and probably it is located farther than the dark cloud and analyzing its spectrum the X-ray source may be a pulsar or an AGN.

If we hypothesize that the "real" counterpart of the unidentified source has not been detected above $3\sigma$ inside the 2FGL 99% error ellipse, we can derive the upper limit assuming an absorbed power law spectral model for the putative counterpart with $N_H = 0.09 \times 10^22$ cm$^2$ and $\Gamma = 2$. This means the putative counterpart should have an unabsorbed flux of $f_X \sim 8.2 \times 10^{-15}$ erg cm$^{-2}$ s$^{-1}$ in the energy range between 0.3 and 10 keV to be detected above $3\sigma$ with our observation. In this way we obtain the inferior limit of the $\gamma$-ray flux to X-ray flux ratio as $f_\gamma/f_X \gtrsim 1300$.

## 5.1.4 2FGL J1744.1–7620

Finally, we have analyzed the unidentified source 2FGL J1744.1–7620 (see Table 5.1). This object was discovered for the first time by the *Fermi*-LAT and included in the 1FGL source catalog as 1FGL J1743.8-7620 [3]. 2FGL J1744.1–7620 is a rather bright $\gamma$-ray source, situated off the Galactic plane (b=-22°.48, l=317°.09) and it is characterized by low variability (variability index=27.14) and high curvature significance (sign_curv=5.66). The variability and spectral curvature properties of this source are very similar to those of a typical $\gamma$-ray pulsars as shown in the bottom of Figure 5.25. $\gamma$-ray spectrum and light curve of 2FGL J1744.1–7620 are shown on the top of Figure 5.25.

In this Section we report on recent monitoring data taken with XMM*Newton* with the aim of detecting the X-ray counterpart of the putative $\gamma$-ray millisec-ond pulsar radio-quiet.

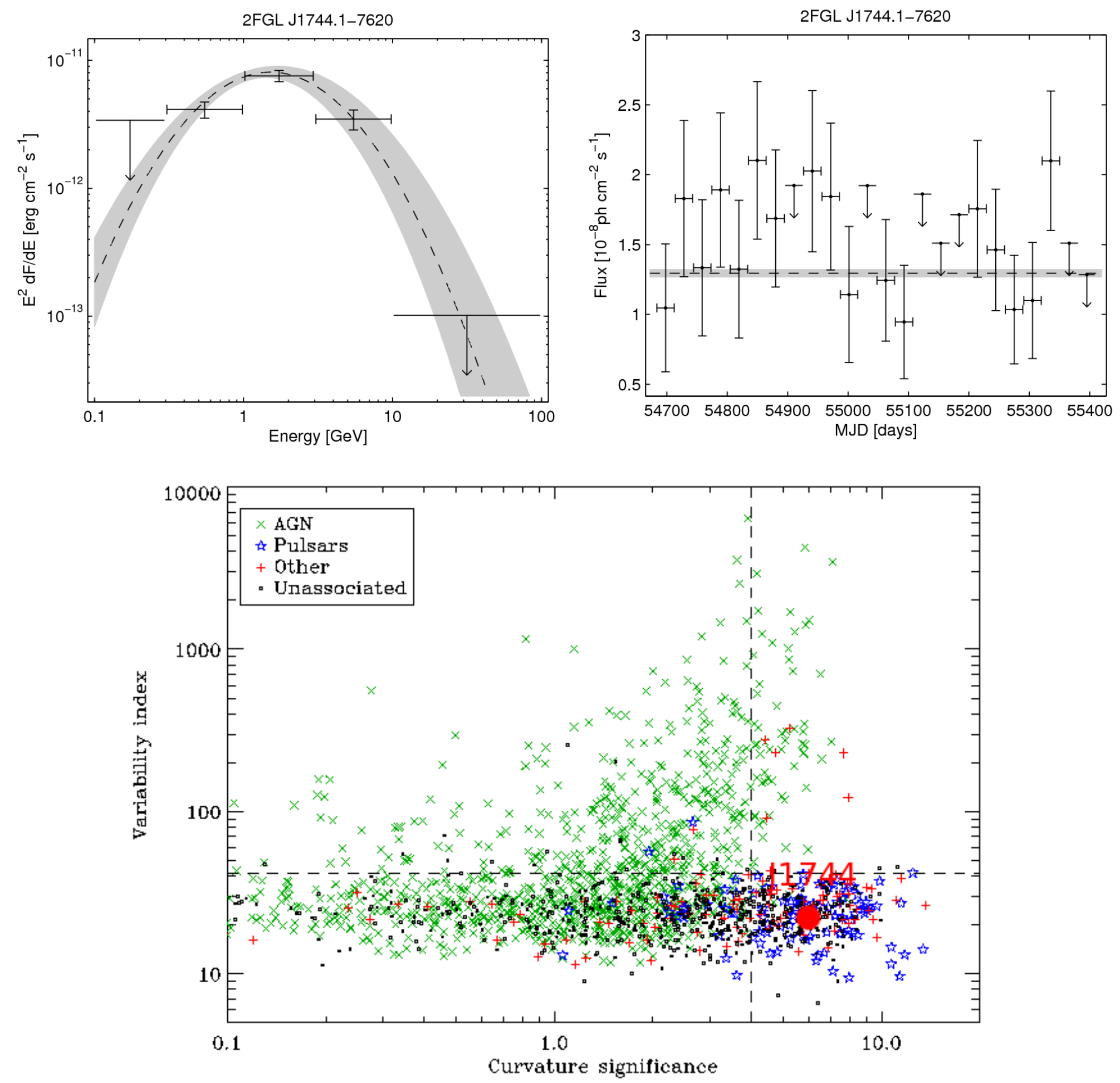

Figure 5.25: Top: $\gamma$-ray spectrum (left) and light curve (right) of the unidentified source 2FGL J1744.1–7620. Bottom: variability index plotted as a function of the curvature significance for different broad classes of sources. Red dot represents the position of 2FGL J1744.1–7620 in this space. This object is not variable and the spectrum is curved.

### Observation and X-ray analysis

Our deep *XMM-Newton* observation 2FGL J1744.1–7620 started on 2012 August 21 at 03:13:20 UT and lasted 25.9 ks. The PN camera [194] of the EPIC instrument was operating in *Extended Full Frame* mode (time resolution of 200 ms over a $26' \times 27'$ Field-of-View), while the MOS detectors [203] were set in *Full Frame* mode (2.6 s time resolution on a 15′ radius FoV). The thin optical filter was used for the PN camera while we chose to use the medium filter for the MOS detectors. After standard data processing , the good, dead-time corrected exposure time is 18.7 ks for the PN, 25.4 ks for the MOS1 and the MOS2 detectors.

We detect 31 sources with a source significance greater than 3.6$\sigma$ in the PN FoV. Of these sources, only 3 are situated within the 99% error circle of the *Fermi*-LAT source, for these we produce X-ray spectra, light curves and find optical counterparts. We decide to exclude *source #21* from our analysis because X-ray spectral and time variability study can be performed only for sources with a source significance level TS > 25.

Figure 5.26 shows the 0.3–10 keV exposure-corrected *XMM-Newton* PN FoV image.

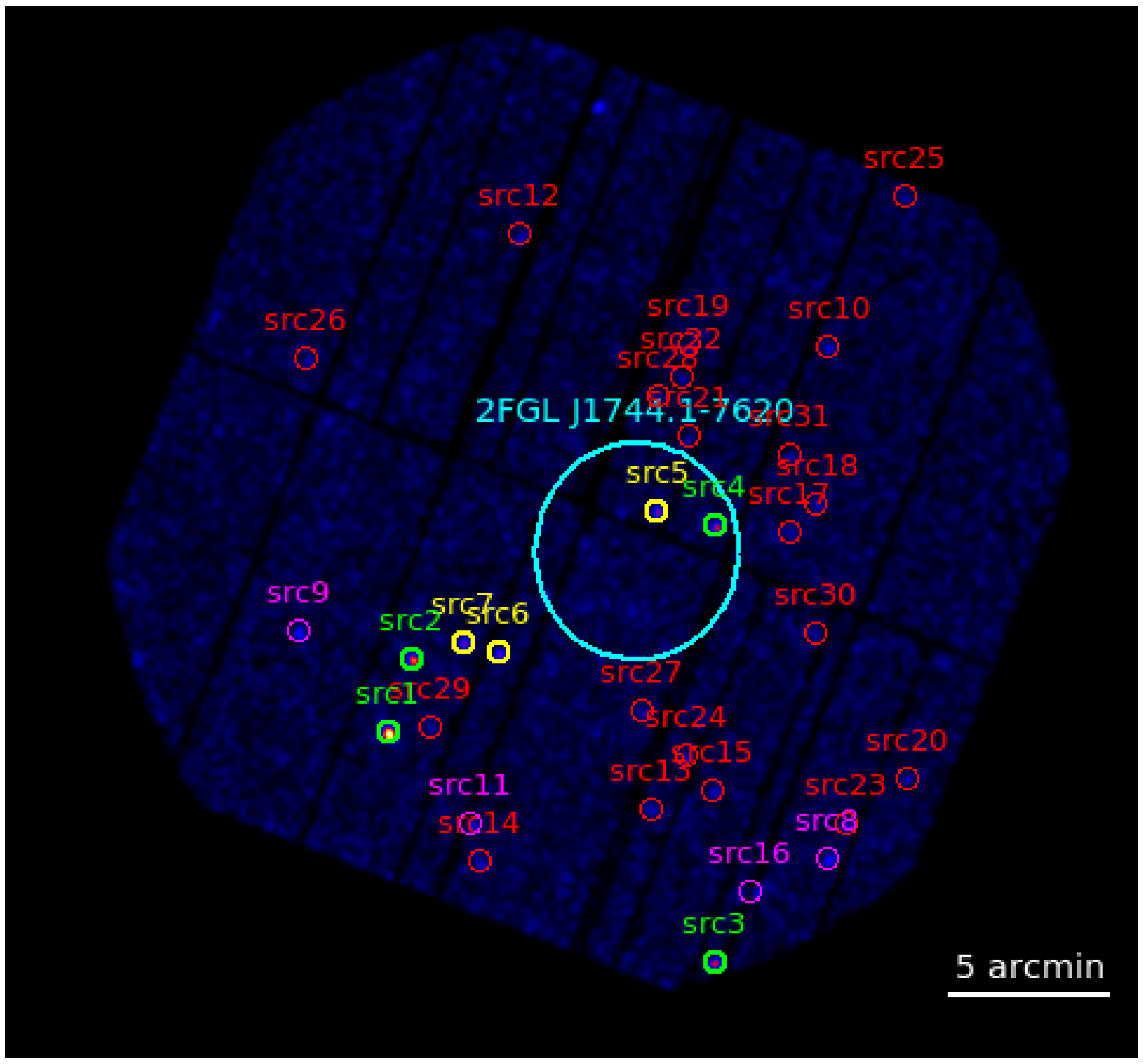

Figure 5.26: 0.3–10 keV exposure-corrected *XMM-Newton* PN FoV image . A gaussian filter with a kernel radius of 3″ is applied. 2FGL J744.1-7620 with 95% confidence error ellipse is plotted in cyan. Each X-ray source detected by the PN is plotted with a circle of 10″ radius. Colors represent the source significance: in red sources with TS<25, in magenta sources with 25<TS<50, in yellow sources with 50<TS<100 and in green sources with TS>100.

In Figure 5.27 is shown the *Digital Sky Survey* image in the field of 2FGL

J1744.1–7620 with the distribution of the X-ray sources detected by *XMM-Newton*. Within the 15 arcmin radius imaged area, the USNO B1.0 catalogue provides a total of ∼3500 sources, corresponding to a surface density $\mu \sim 1.4 \times 10^{-3}$ sources arcsec$^{-2}$. Since the X-ray error-circle is 5 arcsec, we estimate that the probability of chance coincidence between a X-ray and an optical source is ∼ 0.1. Therefore up to 10% of the selected counterparts could be spurious candidates.

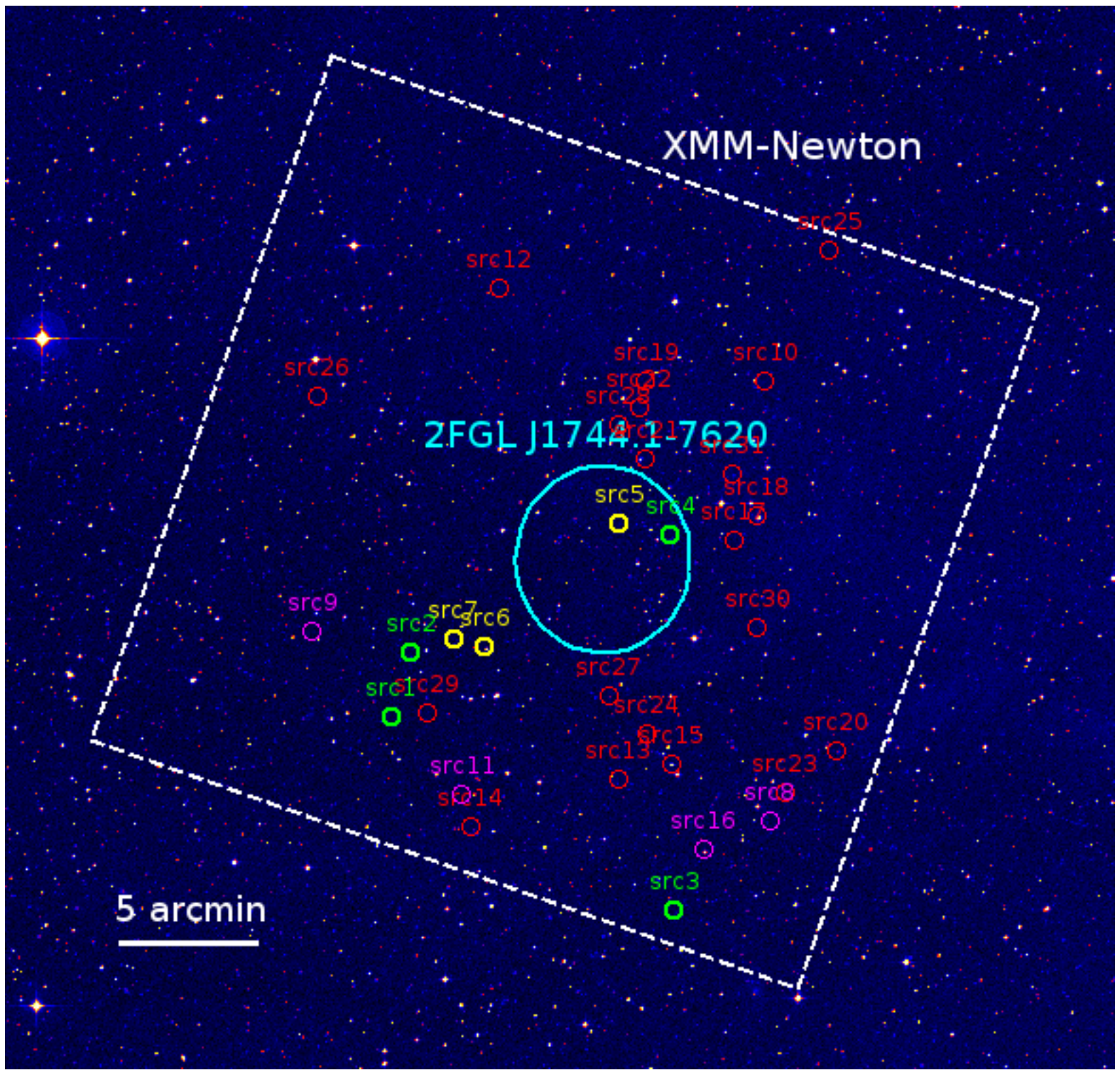

Figure 5.27: *Digital Sky Survey 2* image with red filter of the field of unidentified source 2FGL J1744.1–7620, for which 95% confidence error ellipse is plotted in cyan. Each X-ray source detected by the PN is overplotted with a circle of 10″ radius colored as described in Figure 5.26. White dashed line represents the FoV of the PN.

### Notes on individual X-ray sources

In the following, we present results on the most likely candidate X-ray counterparts to the unidentified $\gamma$-ray source.

**Source #4:** The X-ray analysis yields 74 counts in the energy band 0.3–10 keV, with a source significance TS=105 in the PN and TS≃60 in the two

MOS; its count rate in the total energy band is $0.6 \times 10^{-2}$ cts s$^{-1}$. This object is situated within the 2FGL 95% confidence error ellipse at (RA, Dec)=(265°.87, -76°.33). In Figure 5.28 is shown the spectrum of this X-ray sources.

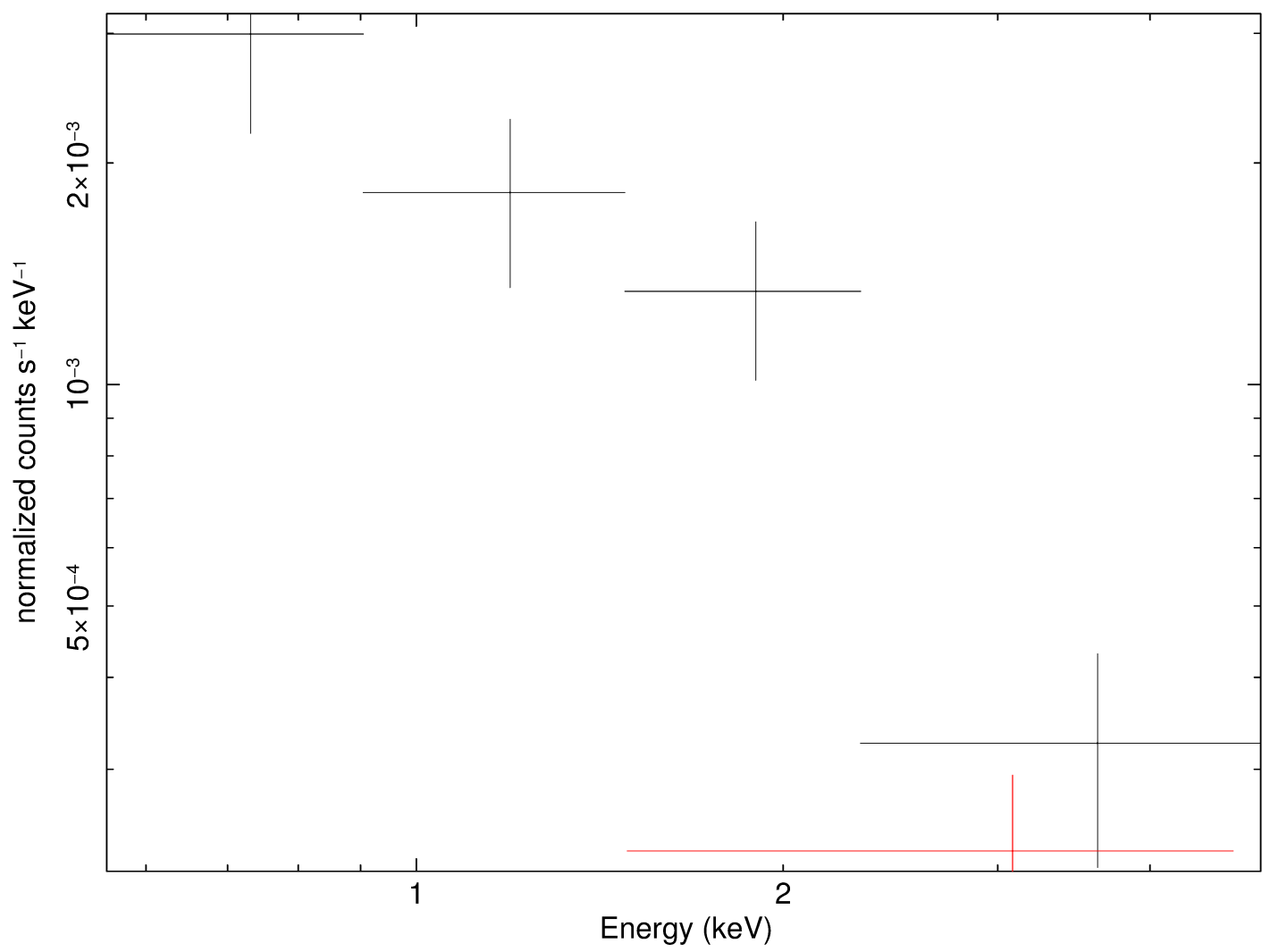

Figure 5.28: Spectrum of the X-ray source #4 binned with a minimum of 15 counts per bin. The spectrum shows that the X-ray object is rather absorbed at low energies and that the PN detected photons until ≃5 keV.

The X-ray spectral fitting results are summarized in Table 5.17. We notice that only if we fix the interstellar medium absorption to $N_H = 0.08 \times 10^{22}$ cm$^{-2}$ [73] we find physical plausible best-fit spectral parameter values. If we consider that this X-ray source is the counterpart of the 2FGL unidentified sources then the $\gamma$-ray (E>100 MeV) to soft X-ray (0.3–10 keV) flux ratio is $F_\gamma/F_X \simeq (800 \div 1500)$ on the basis of the X-ray spectral model we assume. In addition, if 2FGL J1744.1–7620 have been a pulsar then the $\gamma$-ray to X-ray flux ratio would have been compatible with the value of a typical $\gamma$-ray MSP and radio-loud pulsar [135].

In Figure 5.29 is shown the light curved with a bin time of 2500 s.

Within 5″ radius X-ray error circle there is only an optical candidate counterpart of the source #4, its properties are shown in Table 5.16.
If the optical counterpart of the X-ray source is not detected because too faint, above the limiting magnitude of USNO B1.0 (V>21), then $\log(f_X/f_V) > (-0.27 \div -0.03)$, using an unabsorbed visual flux $f_V =$ $1.75 \times 10^{-12}$. This value does not exclude any source class as counterpart of the X-ray source, remember that a pulsar is characterized by $\log(f_X/f_{opt}) > 2$ [121].

| Parameter | Power-law model | Apec model | Black body model |
|---|---|---|---|
| $N_H{}^a$ | 0.08 (fixed) | 0.08 (fixed) | 0.08 (fixed) |
| $\Gamma$ | $1.69^{+0.51}_{-0.45}$ | - | - |
| kT (keV) | - | $6.78^{+38.45}_{-6.37}$ | $0.58^{+0.19}_{-0.16}$ |
| d.o.f. | 9 | 9 | 9 |
| $\chi^2_\nu$ | 0.717 | 0.744 | 1.12 |
| $\mathrm{Flux_{unabs}}{}^b$ | 2.31 | 2.31 | 1.34 |

$^a$in units of $10^{22}$ cm$^{-2}$

$^b$in units of $10^{-14}$ erg cm$^{-2}$ s$^{-1}$ and in the energy range of 0.3–10 keV

Table 5.15: Spectral parameters for the X-ray source #4 within the 2FGL 95% error ellipse of 2FGL J1744.1–7620. $Flux_{unabs}$ corresponds to the unabsorbed flux. We cannot assert if this X-ray source is a pulsar a star or an AGN because no spectral model can be rejected by the $\chi^2$ test.

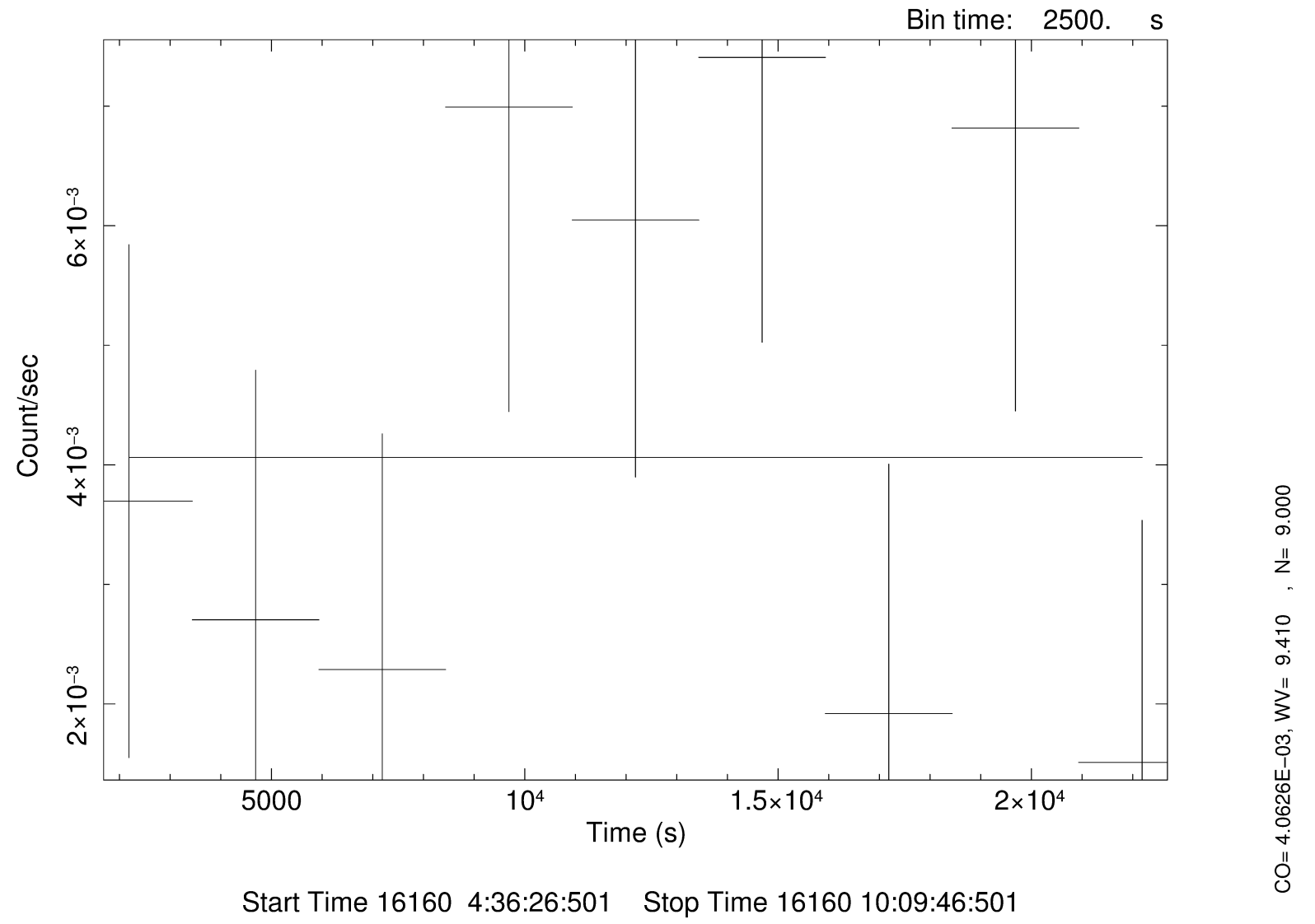

Figure 5.29: Light curve for the X-ray source #4 in the energy range between 0.3 keV and 10 keV within the $\sim$ 19 ks observation span. The black line depicts the best fit assuming a flat light curve. No significant X-ray variability within the observation span is detected on the basis of the $\chi^2$ test.

**Source #5:** The X-ray analysis yields 64 counts in the energy band 0.3–10 keV, with a source significance TS=75 in the PN and TS≃30 in the two

| Source #4 – optical/IR analysis | |
|---|---|
| Optical counterpart | USNO B1.0 |
| Detection year | 1990 |
| Distance | 2.2” |
| De-absorbed magnitudes (mag) | B=19.84 |
| | V=19.68[a] |
| | R=19.49 |
| Stellar spectral class | A |
| $f_{opt}$ (erg cm$^{-2}$ s$^{-1}$) | $7.4 \times 10^{-14}$ |
| $f_X/f_{opt}$ ($\log_{10}$) | -0.74÷-0.51 |
| Suggested class | star |

[a]value extrapolated as the average between absorbed B and R magnitudes

Table 5.16: Properties of the optical candidate counterpart of the X-ray source #4. For a description of each row see Table 5.5.

MOS; its count rate in the total energy band is $0.5 \times 10^{-2}$ cts s$^{-1}$. This object is situated within the 2FGL 95% confidence error ellipse at (RA, Dec)=(266°.00, -76°.32). In Figure 5.30 is shown the spectrum of this X-ray sources.

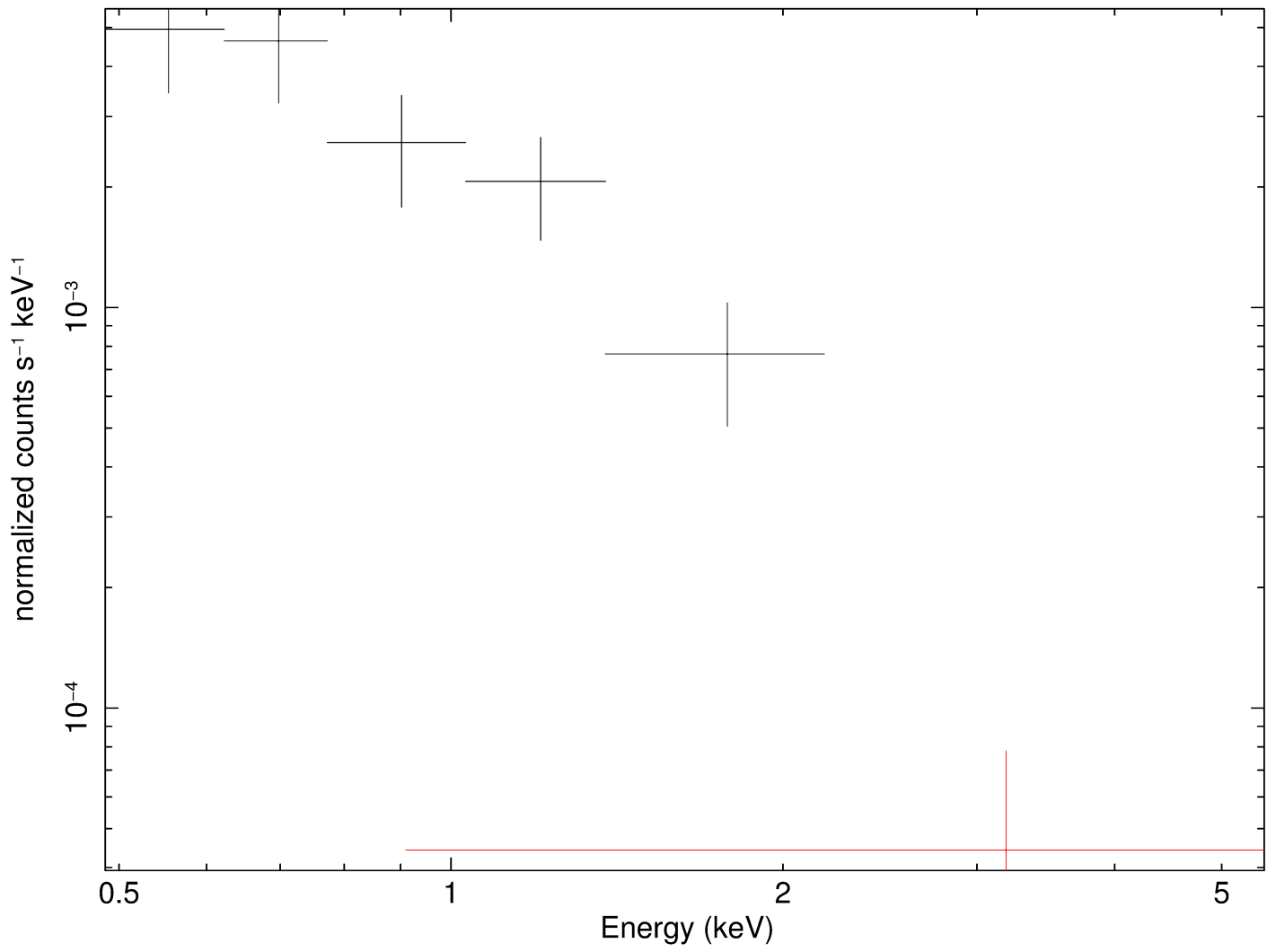

Figure 5.30: Spectrum of the X-ray source #5 binned with a minimum of 15 counts per bin. The spectrum shows that the X-ray object is rather absorbed at low energies and that the PN detected photons only until ≃2 keV.

The X-ray spectral fitting results are summarized in Table 5.17. We notice that only if we fix the interstellar medium absorption to $N_H = 0.08 \times 10^{22}$ cm$^{-2}$ [73] we find physical plausible best-fit spectral parameter values.

| Parameter | Power-law model | Apec model | Black body model |
|---|---|---|---|
| $N_H{}^a$ | 0.08 (fixed) | 0.08 (fixed) | 0.08 (fixed) |
| $\Gamma$ | $2.71^{+0.69}_{-0.64}$ | - | - |
| kT (keV) | - | $3.61^{+14.24}_{-1.73}$ | $0.23^{+0.08}_{-0.05}$ |
| d.o.f. | 2 | 2 | 2 |
| $\chi^2_\nu$ | 0.717 | 0.744 | 1.12 |
| $\text{Flux}_{\text{unabs}}{}^b$ | 1.92 | 1.63 | 1.01 |

$^a$in units of $10^{22}$ cm$^{-2}$

$^b$in units of $10^{-14}$ erg cm$^{-2}$ s$^{-1}$ and in the energy range of 0.3–10 keV

Table 5.17: Spectral parameters for the X-ray source #5 within the 2FGL 95% error ellipse of 2FGL J1744.1–7620. $Flux_{unabs}$ corresponds to the unabsorbed flux. We cannot assert if this X-ray source is a pulsar a star or an AGN because no spectral model can be rejected by the $\chi^2$ test especially because of the low statistics available (only 2 d.o.f). The $\gamma$-ray (E>100 MeV) to soft X-ray (0.3–10 keV) flux ratio is $F_\gamma/F_X \simeq (1000 \div 1900)$, which is compatible with the value of a typical $\gamma$-ray MSP and radio-loud pulsar [135].

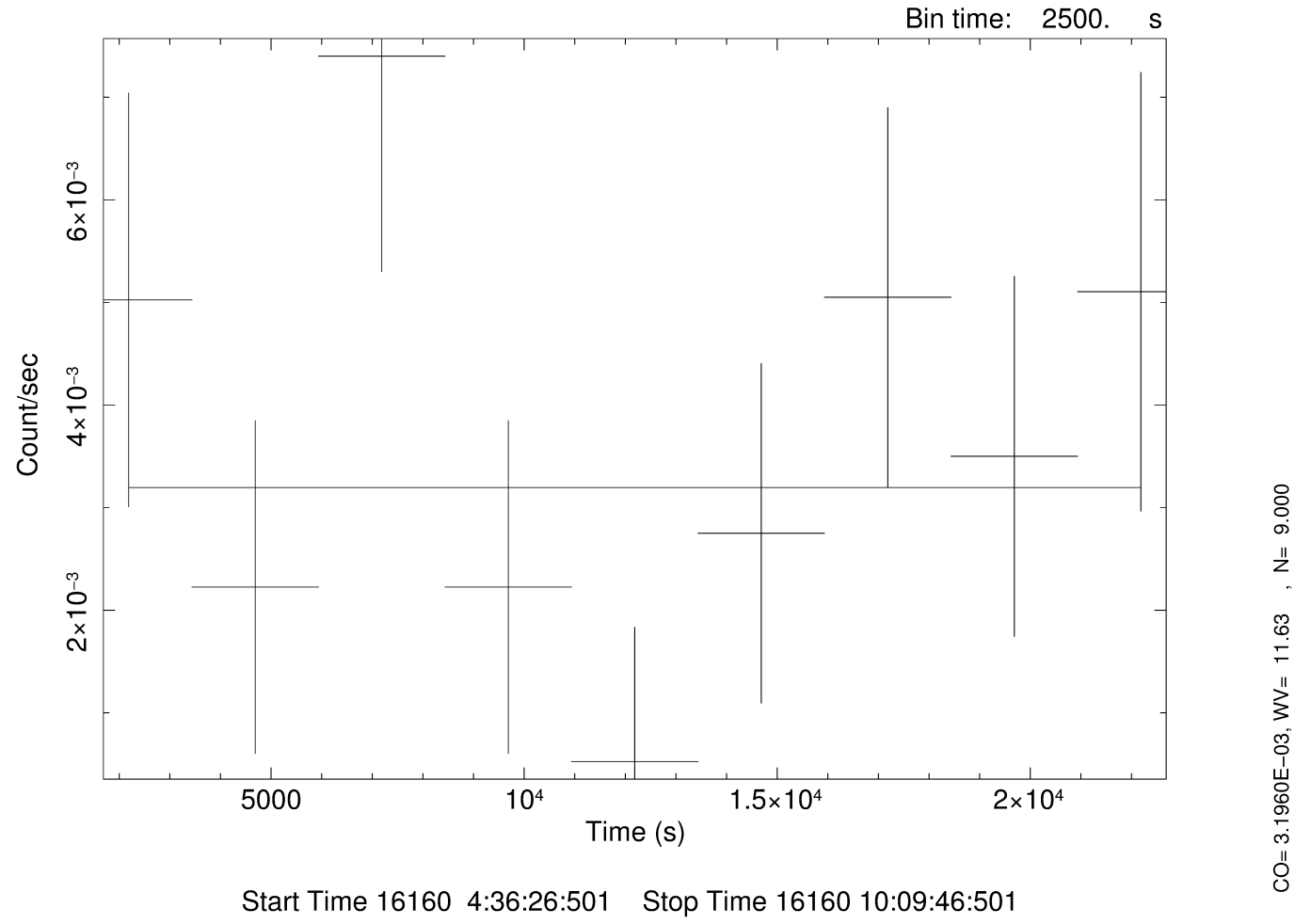

Figure 5.31: Light curve for the X-ray source #5 in the energy range between 0.3 keV and 10 keV within the $\sim$ 19 ks observation span. The black line depicts the best fit assuming a flat light curve. No significant X-ray variability within the observation span is detected on the basis of the $\chi^2$ test.

In Figure 5.31 is shown the light curved with a bin time of 2500 s.

Within 5″ radius X-ray error circle there is no optical/IR candidate counterpart of the source #10. If the optical counterpart of the X-ray source is not detected because too faint, above the limiting magnitude of USNO B1.0 (V>21), then $\log(f_X/f_V) > (-0.4 \div -0.12)$, using an unabsorbed visual flux $f_V = 1.75 \times 10^{-12}$. This value does not exclude any source class as counterpart of the X-ray source.

**Source hardness ratios distribution**

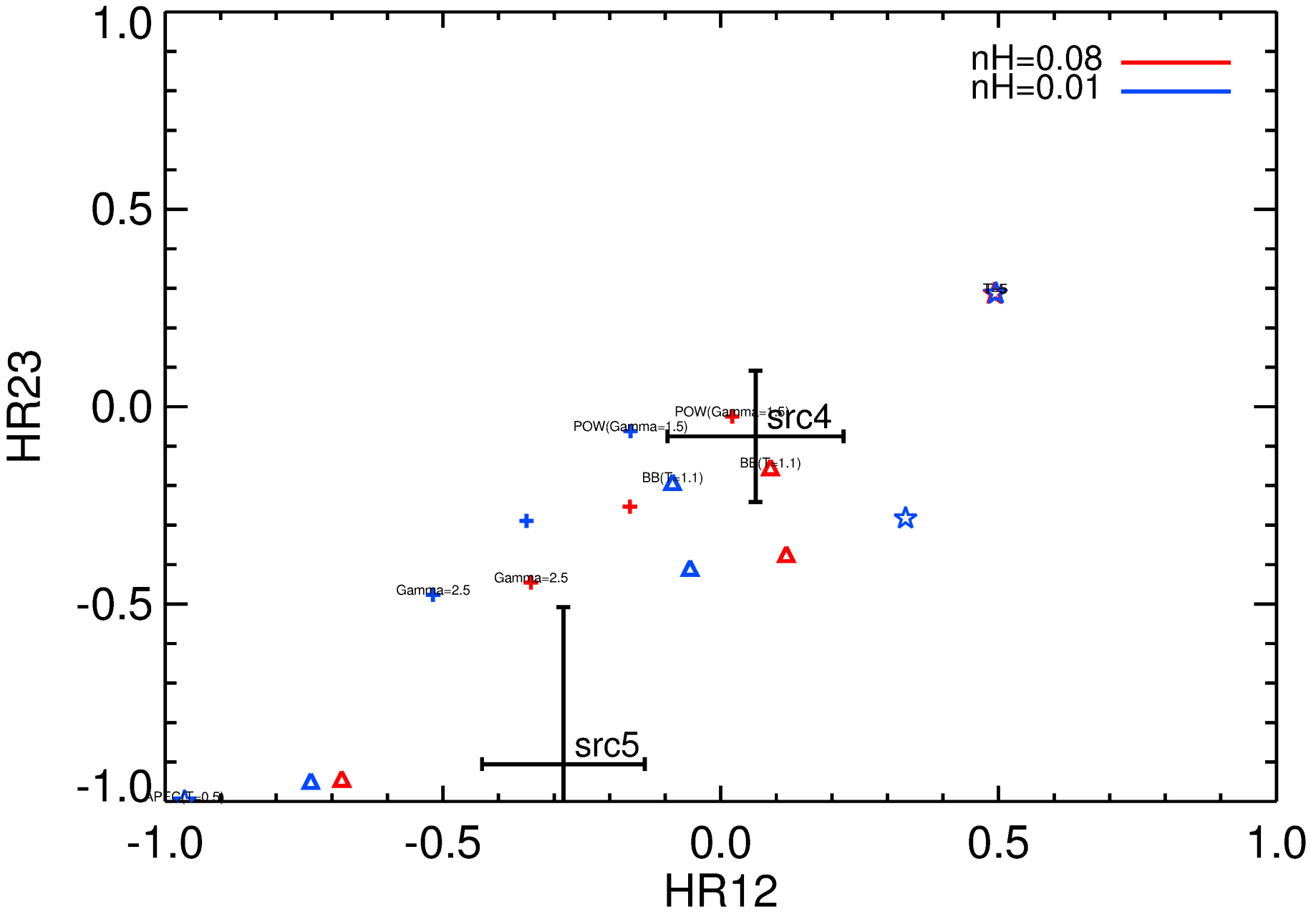

Figure 5.32: Distribution of HR12 vs. HR23 of the X-ray sources detected by the *XMM-Newton* within the 99% error circle of the *Fermi*-LAT source. Error bars are reported at $1\sigma$. Crosses indicate the expected HR12 vs. HR23 computed for power law spectra with $\Gamma$ from 1.5 and 2.5. Stars indicate the expected HR12 vs. HR23 computed for apec spectra with kT from 0.5 to 5.5 keV. Triangles indicate the expected HR12 vs. HR23 computed for black body spectra with kT from 0.1 to 1.1 keV. Each spectral model is computed using the interstellar medium absorption given by [73] (red) and one third of this value (blue)

In Figure 5.32 is shown the distribution of the HRs of the X-ray sources detected by the *XMM-Newton* within the 99% error circle of the *Fermi*-LAT source. Sources #4 is little absorbed and is characterized by a not very hard spectrum ($HR12 \simeq 0$ and $HR23 \simeq 0$). Source #5 is little absorbed and characterized by a very soft spectrum ($HR12 \simeq -0.3$ and $HR23 \simeq -1$), probably it is a nearby star. Source #4 is an interesting candidate counterpart of the 2FGL unidentified source.

The values of the expected HRs are overplotted in Figure 5.32. As can be seen, the distributions are compatible with a rather wide range of temperatures and photon indexes.

#### Discussion

In the *XMM-Newton* FoV of 2FGL J1744.1–7620 we have detected 31 X-ray sources between 0.3 and 10 keV, only 2 of them are situated within the 99% error circle of the *Fermi*-LAT source and have a source significance greater than 25, we can consider them as plausible counterparts of the putative MSPs. For each plausible X-ray counterpart we have not detected any significant X-ray variability and by an analysis of $f_\gamma/f_X$ each one seems to be a plausible counterpart of the 2FGL unidentified source. No optical or IR candidate counterpart was found for the source #5 but analyzing its X-ray spectrum it seems a nearby star. Source #4 is very interesting, it has an optical counterpart that may be identify as a star but analyzing the HRs it is evident that its spectrum is too hard in order to be associated with a star. This means that source #4 may be part of a binary system where the optical object is its companion. This means that 2FGL unidentified source may be an HMB or a "black-widow-type" millisecond pulsar as e.g. 2FGL J1311.7-3429 [178]. All these considerations make source #4 the most interesting plausible counterpart of the putative MSP.

If we hypothesize that the "real" counterpart of the unidentified source has not been detected above $3\sigma$ inside the 2FGL 99% error ellipse, we can derive the upper limit assuming an absorbed power law spectral model for the putative counterpart with $N_H = 0.08 \times 10^2 2$ cm$^2$ and $\Gamma = 2$. This means the putative counterpart should have an unabsorbed flux of $f_X \sim 3.6 \times 10^{-15}$ erg cm$^{-2}$ s$^{-1}$ in the energy range between 0.3 and 10 keV to be detected above $3\sigma$ in our observation. In this way we can obtain the inferior limit of the $\gamma$-ray flux to X-ray flux ratio as $f_\gamma/f_X \gtrsim 5500$.

## 5.2 PSR J0357+3205: a fast-moving pulsar with a very unusual X-ray trail

The Large Area Telescope onboard the *Fermi* satellite [22] has opened a new era for pulsar astronomy, by detecting $\gamma$-ray pulsations (at E > 100 MeV) from more than 120 pulsars[6], about 30% of which are not detected at radio wavelengths. The middle-aged PSR J0357+3205 (characteristic age $\tau_C \sim 0.54$ Myr) is one of the most interesting radio-quiet pulsars discovered in blind periodicity searches in *Fermi*-LAT data [6]. Its high $\gamma$-ray flux (it is included in the *Fermi*-LAT bright source list [7]), low spin-down luminosity ($\dot{E}_{rot} = 5\times10^{33}$ erg s$^{-1}$) and off-plane position (Galactic latitude $b \sim -16°$) point to a small distance of about 500 pc. The investigation of the field of PSR J0357+3205 with a joint X-ray and optical program with *Chandra* and the NOAO Mayall 4m telescope at Kitt Peak allowed to identify the soft X-ray counterpart of the pulsar as an unremarkable source, looking fainter (and colder) than other well known middle-aged pulsars [69]. More interestingly, the deep *Chandra* observation unveiled the existence of a very large, elongated feature of diffuse X-ray emission, apparently originating at the pulsar position and extending for more than $9'$ (corresponding to $\sim 1.3$ pc at the distance of 500 pc, assuming no inclination with respect to the plane of the sky), with a hard spectrum consistent both with a power law and with a hot thermal bremsstrahlung [69].

Elongated "trails" of diffuse emission have been associated with several rotation-powered pulsars [112] and explained as bow-shock pulsar wind nebulae (see [81] for a review), where their elongated morphology is "velocity-driven". Indeed, if the pulsar moves supersonically through the interstellar medium, the termination shock of the pulsar wind assumes a "bullet" morphology, due to ram pressure. Particles accelerated at the shock emit synchrotron radiation and cool down, confined by ram pressure in an elongated region aligned with the pulsar space velocity. However, explaining the nature of the nebula associated with PSR J0357+3205 turned out to be very challenging. As discussed in [69], the standard picture cannot apply here since the morphology is very different from the "cometary" shape which characterizes all other X-ray bow-shock nebulae. There is no emission in the surroundings of the pulsar, where the brightest portion (the termination shock) should be – indeed, the surface brightness grows as a function of the distance from PSR J0357+3205. Moreover, there are no evidences for spectral evolution as a function of the position, at odds with expectations for a population of particles injected at the shock and cooling via synchrotron radiation.

Other pictures could be explored. For instance, PSR B2224+65, the fast moving pulsar powering the well known "Guitar nebula" seen in $H_\alpha$ [59], displays an elongated X-ray feature which is reminiscent of the one of our target and cannot be a bow-shock nebula because it is misaligned by $\sim 118°$ with respect to the direction of the proper motion [104]. Thus, the possibility of a ballistic jet (similar to Active Galactic Nuclei), or the hypothesis of a nebula confined by a pre-existing, large scale magnetic field in the interstellar medium have been proposed [26, 108, 105].

[6]See `https://confluence.slac.stanford.edu/display/GLAMCOG/Public+List+of+LAT-Detected+Gamma-Ray+Pulsars/`

Indeed, a crucial piece of information in order to understand the physics of the huge elongated feature associated with PSR J0357+3205 is the direction of the pulsar proper motion. Detecting a pulsar angular displacement aligned with the nebula's main axis would link the morphology of the diffuse structure to the pulsar velocity. Conversely, if it were misaligned, the case of PSR J0357+3205 would become very similar to the one of PSR B2224+65 and would require a different explanation for the nature of the nebula.

Usually, a pulsar proper motion is measured in the radio band, or, more rarely, in the optical domani. Unfortunately, our target is radio-quiet and has no optical counterpart; moreover, timing analysis of $\gamma$-ray photons is not particularly sensitive to the proper motion (positional accuracy based on 5 yr of *Fermi*-LAT timing is estimated to be $\sim 2''$ [174]). The only way to search for a possible proper motion rests on the comparison of multi-epoch, high-resolution X-ray images. To this aim, we have obtained a multi-cycle observing campaign with *Chandra*, consisting of two observations to be performed at the end of 2011 and at the end of 2013. We will report here on the first observation of our program, as well as on a very recent observation of the field in the $H_\alpha$ band performed with GMOS instrument at the Gemini North telescope. Indeed, pulsars moving supersonically into warm interstellar gas can generate optical emission in the $H_\alpha$ line, due to collisional excitation of neutral hydrogen and charge exchange occurring at (and behind) the forward shock (see e.g. [59]), yielding a limb-brightened, arc-shaped bow-shock nebula, located at the apex of the forward shock, in the direction of the proper motion.

Moreover, the mere existence of the trail is problematic for a pulsar with such a low $\dot{E}_{rot}$ as J0357+3205 [65]. In order to assess the nature of the trail and its relationship with the pulsar, as well as to better constrain the pulsar emission properties and physics, we also obtained a deep observation of the pulsar and its peculiar trail with *XMM-Newton*. We studied the properties of the trail, in order to characterize the mechanisms responsible for its emission. Regarding the pulsar, *XMM-Newton*'s high spectral resolution allowed us to disentangle the components of the pulsar spectrum and to constrain its distance based on the absorption column density. Finally, a complete study of the timing behavior of PSR J0357+3205 in X-rays was carried out as well[7].

[7]In this thesis we will not discuss our studies on pulsar, the results are reported in the paper we have published on 2013 [137]

### 5.2.1 Measurement of the pulsar proper motion

The superb angular resolution of the *Chandra* optics makes it possible to measure tiny angular displacements of an X-ray source by performing relative astrometry on multi-epoch images. Indeed, such an approach has already been used to measure the proper motion of a few isolated neutron stars. See e.g. the work by [68], [154], [155], [110], [27], [208].

#### Chandra observations and data reduction

Our new observation of PSR J0357+3205 with *Chandra* was performed on 2011, December 24 (Obs.Id. 14007, 29.4 ks exposure time – hereafter tagged as "2011"). Previous observations were performed on 2009 October 25 (obs. id. 12008, 29.5 ks – hereafter "2009a") and on 2009 October 26 (obs. id. 11239, 47.1 ks – hereafter "2009b"). All data were collected using the Advanced CCD Imaging Spectrometer (ACIS) instrument in Timed Exposure mode with the VFAINT telemetry mode. We retrieve "Level 1" data from the *Chandra* Science Archive and reprocess them with the `chandra_repro`[8] script of the *Chandra* Interactive Analysis of Observation Software (CIAO v4.4)[9].

For each observation, we generate an image in the 0.3–8 keV energy range using the original ACIS pixel size ($0''.492$ pixel$^{-1}$). We perform a source detection using the `wavdetect`[10] task with wavelet scales ranging from 1 to 16 pixels with a $\sqrt{2}$ step size, setting a detection threshold of $10^{-6}$. In all observations, the target was imaged close to the aimpoint, on the backside-illuminated chip S3 of the ACIS-S array.

We cross-correlate the resulting source lists using a correlation radius of $3''$ and we extract a catalogue of common sources for each pair of observations. In view of the density of sources in each image, the possibility of a chance alignment of two false detections is $< 10^{-5}$. As a further step, we select sources within $4'$ of the aimpoint since the telescope point spread function deteriorates as a function of offaxis angle, hampering source localization accuracy (see discussion in [68]). Such an exercise yield 12 common sources (2011 vs. 2009a), 10 sources (2011 vs. 2009b) and 16 sources (2009a vs. 2009b), in addition to the target pulsar. The uncertainty on the source localization on each image depends on the signal to noise as well as on the offaxis angle and ranges from $\sim 0.08$ to $\sim 0.7$ pixel per coordinate.

[8] `http://cxc.harvard.edu/ciao/ahelp/chandra_repro.html`
[9] `http://cxc.harvard.edu/ciao/index.html`
[10] `http://cxc.harvard.edu/ciao/threads/wavdetect/`

| | 2011 vs. 2009a | 2011 vs. 2009b | 2009a vs. 2009b |
|---|---|---|---|
| Time baseline | 2.16 yr | 2.16 yr | 1 day |
| Number of ref.srcs | 11[a] | 10 | 16 |
| uncertainty on $X_{shift}$ (pixels) | 0.09 | 0.08 | 0.06 |
| $\chi^2$ (dof) | 13.6 (10) | 15.6 (9) | 7.8 (15) |
| uncertainty on $Y_{shift}$ (pixels) | 0.08 | 0.07 | 0.06 |
| $\chi^2$ (d.o.f.) | 8.1 (10) | 13.0 (9) | 13.8 (15) |
| PSR X displacement (pixels) | $0.53 \pm 0.12$ | $0.50 \pm 0.11$ | $0.10 \pm 0.10$ |
| PSR Y displacement (pixels) | $0.54 \pm 0.11$ | $0.50 \pm 0.10$ | $0.04 \pm 0.10$ |

[a] A reference source yielding high residuals was rejected.

Table 5.18: Results of X-ray relative astrometry.

### Relative astrometry

The positions of the selected, common sources (excluding the pulsar counterpart) are adopted as a reference grid to perform relative astrometry. We use the ACIS SKY reference system[11] (pixel coordinates with axes aligned along Right Ascension and Declination). Taking into account the corresponding uncertainties, we compute the best geometric transformation needed to superimpose the reference frames of two images collected at different epochs.

We superimpose the most recent 2011 data to first-epoch 2009a and (separately) 2009b data in order to measure the possible pulsar displacement over a baseline of $\sim$ 2.2 yr. We also superimpose 2009a data to 2009b data (no displacement is expected on a 1-day baseline) in order to check for systematics affecting our analysis. A simple translation yields a good superposition in all cases (see Table 5.18). The uncertainty on the frame registration turns out to be smaller than 50 mas per coordinate. Adding a further free parameter to the transformation (a rotation angle) does not result in a statistically compelling improvement of the fit.

We apply the best-fit transformation to the coordinates of the pulsar counterpart and we compute its displacement between different epochs. The error budget for the overall pulsar displacement includes the uncertainty on the pulsar position in each image as well as the uncertainty on the multi-epoch frame registration. Results are shown in Table 5.18. Displacement of the pulsar as well as residuals on the positions of the reference sources after frame registration are also shown in Figure 5.33. A significant and consistent displacement of the pulsar is apparent both in the 2011 vs. 2009a and in the 2011 vs. 2009b comparison, while no displacement is seen for the pulsar in the 2009a vs. 2009b one.

[11] `http://cxc.harvard.edu/ciao/ahelp/coords.html`

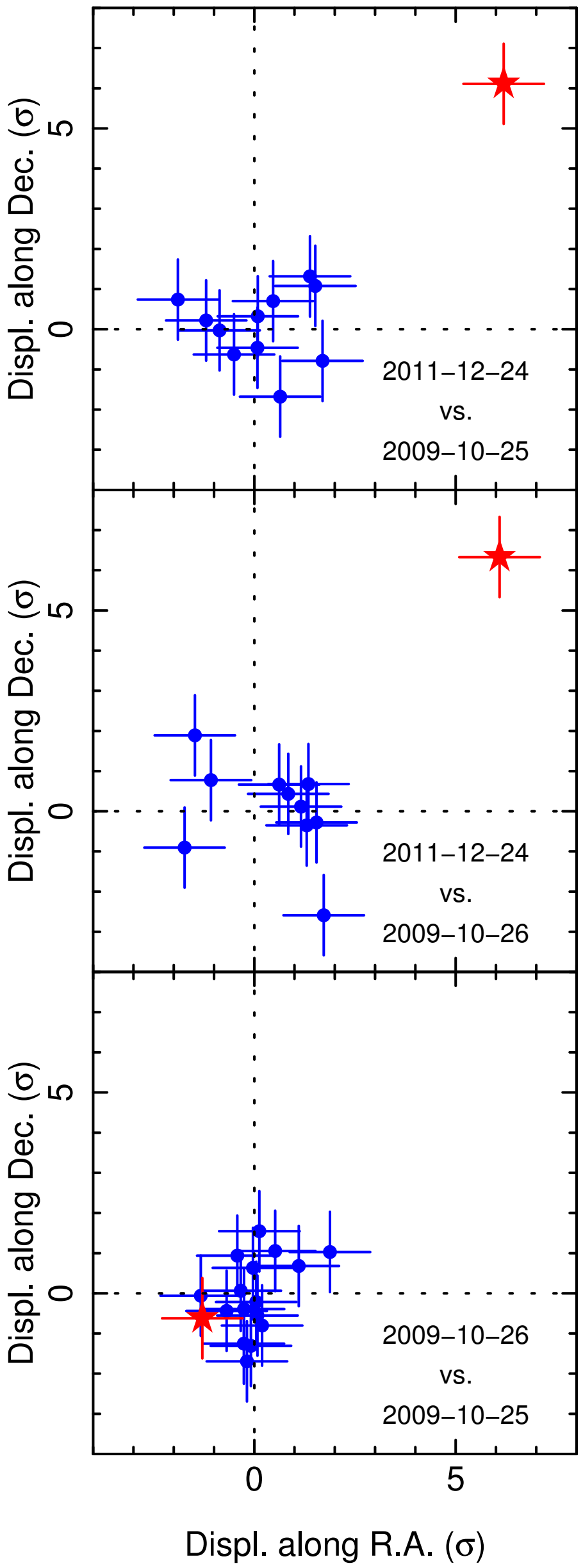

Figure 5.33: Results of relative astrometry. Two independent, first-epoch images (2009 October 25 and 2009 October 26) are compared to a second-epoch image (2011 December 24) in the upper and middle panels, respectively. Coordinate residuals after image superposition are shown for reference sources (blue dots) as well as for the pulsar counterpart (red star), in units of statistical errors. The displacement of the pulsar is apparent. The bottom box displays the comparison of the two first-epoch images showing, as expected, no displacements for any source.

Using the relative positions of the pulsar in the 2011 frame, we compute a best-fit proper motion $\mu_\alpha cos(\delta) = 117 \pm 20$ mas yr$^{-1}$ and $\mu_\delta = 115 \pm 20$ mas yr$^{-1}$. This corresponds to a total proper motion of 165 $\pm$ 30 mas yr$^{-1}$, translating to a (projected) space velocity of $\sim$ 390 d$_{500}$ km s$^{-1}$ (where d$_{500}$ is the distance to the pulsar in units of 500 pc). The position angle (P.A.) of the proper motion is $314° \pm 10°$ (north to east; see Figure 5.34).

**The position angle of the nebula**

In order to measure the sky orientation of the main axis of the nebula, we select nine contiguous rectangular images slices (see Figure 5.34, inset), aligned along R.A. and decl., having a width of $25''$–$40''$ along R.A. and a height of $8'.5$ along decl. (excluding the two bright sources located close to the southeastern end of the nebula). For each slice, we extract the image brightness profile in the north-south direction and, fitting a Gaussian+constant function to such profile, we evaluate the centroid of the nebular emission. A linear function describes very well ($\chi^2 = 7.5$, 7 dof) the nine resulting positions in R.A.–decl. plane, yielding a P.A. of $315°.5 \pm 1°.5$ (see Figure 5.34). Repeating the exercise using nine image slices oriented from east to west yields fully consistent results. Thus, the proper motion direction (P.A. = $314° \pm 8°$) is very well aligned with the main axis of the nebula, which is indeed an "X-ray trail".

## 5.2.2 Gemini H$_\alpha$ observations.

We observed the field of PSR J0357+3205 in the H$_\alpha$ band on September 23, 2012 with the Gemini North telescope on the Mauna Kea Observatory. We used the Gemini Multi-Object Spectrograph (GMOS) in its imaging mode, equipped with the interim upgrade e2v deep depletion (DD) detector, with enhanced response in the blue and red arm of the spectrum with respect to the original EEV detector. The e2v DD detector is an array of three chips, with an unbinned pixel size of $0''.072$ and covers an unvignetted Field–of–View of $5'.5 \times 5'.5$. We used the Ha_G0310 filter ($\lambda = 656$ nm; $\Delta\lambda = 7$ nm), centered on the H$_\alpha$ rest wavelength. The e2v DD detector was set in the $2 \times 2$ binning mode and read through the standard slow read-out/low gain mode with the six amplifiers. Observations were performed with an average airmass of 1.24, image quality of $\sim 0'.8$, and grey time. Two sets of six 500 s exposures were obtained, for a total integration time of 6000 s. Exposures were dithered in steps of $\pm$ 5 pixels to achieve a better Signal–to–Noise (S/N).

We process the GMOS images using the dedicated GMOS image reduction package available in IRAF. After downloading the closest bias and sky flat-field frames from the *Gemini* science archive[12], we use the tasks `gbias` and `giflat` to process and combine the bias and flat-field frames, respectively. We then reduce the single science frames using the task `gireduce` for bias subtraction, overscan correction, image trimming and flat-field normalization. From the reduced science images, we produce a mosaic of the three GMOS chips using the task `gmosaic` and we use the task `imcoadd` to average-stack the reduced image mosaics and filter out cosmic ray hits.

[12] `http://cadcwww.dao.nrc.ca/gsa/`

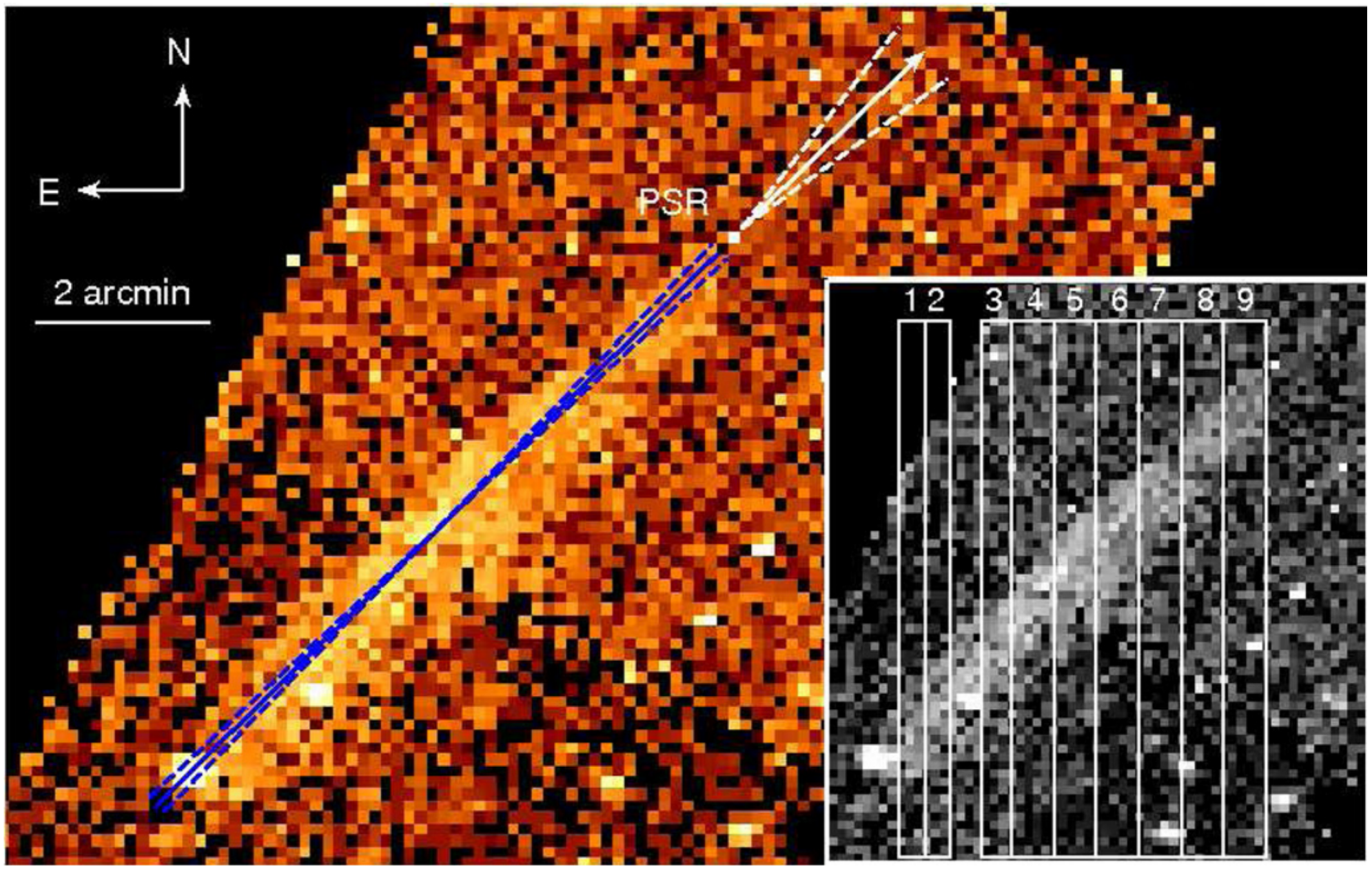

Figure 5.34: Field of PSR J0357+3205 as seen by Chandra in the 0.3–8 keV energy range. Data from the two observations performed in 2009 have been used (77 ks exposure time). The image has been rebinned to a pixel size of $8''$ in order to ease visibility of faint diffuse structures. The large, elongated nebula is apparent. The direction of the main axis of the nebula and its $1\sigma$ uncertainty (P.A. of $315^\circ.5 \pm 1^\circ.5$) are overplotted in blue. The pulsar counterpart is also marked, together with the direction of the proper motion (white arrow) and the $1\sigma$ errors (dashed lines), corresponding to a P.A. of $314^\circ \pm 8^\circ$. The inset shows the region used to evaluate the position angle of the nebula (see the text).

We compute the astrometry calibration using the *wcstools*[13] suite of programs, matching the sky coordinates of stars selected from the Two Micron All Sky Survey (2MASS) All-Sky Catalog of Point Sources [187] with their pixel coordinates computed by *Sextractor* [28]. After iterating the matching process applying a $\sigma$-clipping selection to filter out obvious mismatches, high-proper motion stars, and false detections, we obtained mean residuals of $\sim 0''.2$ in the radial direction, using up to 30 bright, but non-saturated, 2MASS stars. To this value we add in quadrature the uncertainty $\sigma_{tr} = 0''.08$ of the image registration on the 2MASS reference frame. This is given by $\sigma_{tr}=\sqrt{n/N_S}\sigma_{\rm S}$ (e.g. [124]), where $N_S$ is the number of stars used to compute the astrometric solution, $n$=5 is the number of free parameters in the sky–to–image transformation model, $\sigma_{\rm S} \sim 0''.2$ is the mean absolute position error of 2MASS [187]for stars in the magnitude range $15.5 \leq K \leq 13$. After accounting for the $0''.015$ uncertainty on the link of 2MASS to the International Celestial Reference Frame [187], we evaluate with the overall accuracy on our absolute astrometry to be of $\sim 0''.22$.

[13] `http://tdc-www.harvard.edu/wcstools/`

Unfortunately, no observations of spectro-photometric standards stars are available for the flux calibration of our $H_\alpha$ image. Thus, we cross-correlate instrumental magnitudes of more than 150 non-saturated stars in the Gemini image to their $R_F$ magnitudes listed in the Guide Star Catalogue v2.3.2 (GSC2, [123]), which yields a rather good fit with a r.m.s. of 0.16 mag. To assess the GSC2 flux zeropoint, we cross-correlate the GSC2 $R_F$ magnitude of more than 500 stars with their R magnitude as tabulated in the Stetson Standard photometric star archive[14]. We evaluate the transformation as $R_F$=0.97R–0.32 ($\chi^2 = 617$, 520 d.o.f.). Then, we compute the $H_\alpha$ fluxes of the 150 selected stars on our Gemini image using specific fluxes corresponding to their $R_F$ magnitudes and assuming a flat spectrum within the R filter bandpass. This yields a flux calibration of the $H_\alpha$ image with an uncertainty of ∼0.2 mag.

No structures of diffuse emission unambiguously related to the fast motion of PSR J0357+3205 can be discerned on our Gemini image – in particular, no arc-shaped "bow-shock" nebula is seen at the expected position in the direction of the pulsar proper motion (see Figure 5.35). We measure the background properties in the region surrounding the pulsar position. Considering a region of 10 square arcsec as a reference and taking into account the uncertainty in the flux calibration of our $H_\alpha$ image, we compute the $5\sigma$ upper limit to the surface brightness of an unseen bow-shock nebula to be of $5 \times 10^{-18}$ erg cm$^{-2}$ s$^{-1}$ arcsec$^{-2}$.

### 5.2.3 Trail characterization

Our result points to a direct link between the nebula morphology (and physics) and the space velocity of the pulsar – supersonic for any reasonable condition of the interstellar medium (typical values are of $\sim 1$, $\sim 10$ and $\sim 100$ km s$^{-1}$ for the cold, warm and hot components, respectively, see e.g. [120]). As already mentioned, "velocity-driven" pulsar wind nebulae are a well-known astrophysi-cal reality. Explaining the X-ray trail of PSR J0357+3205 as a bow-shock PWN has many problems. As discussed by [69], the lack of any evidence of the pulsar wind termination shock could be due to its confusion with the pulsar emission. Assuming a velocity of 390 $d_{500}$ km s$^{-1}$ and no inclination with respect to the plane of the sky, the expected angular separation between the pulsar and the forward and backward termination shock is $0''.3 n_{1,ISM}^{-1/2}$ and $2'' n_{1,ISM}^{-1/2}$, respectively [69], where $n_{1,ISM}$ is the density of the interstellar medium in units of 1 particle cm$^{-1}$. Thus, high inclination of the pulsar velocity and/or high density of the ISM, and/or a distance larger than expected should be assumed; any of these hypotheses could be possible (though rather unlikely). In any case, other peculiarities of the trail phenomenology cannot be accounted for in the standard scenario of synchrotron emission from shocked pulsar wind. We should explain why the maximum luminosity is generated at a distance as large as $\sim 3$ light years from the parent neutron star. This would require an increase of the density of the radiating particles on the same distance scale, or an increase of the magnetic field intensity, or a change of the angle between the magnetic field and the particle flow. Moreover, the lack of any spectral steepening across the trail [69] clashes with the expected, severe synchrotron energy losses of particles radiating in the keV range. Thus, either some very peculiar phenomenon is occurring in the magnetized pulsar particle outflow or the nature of the nebula is different (but linked to the pulsar velocity).

[14] http://www3.cadc-ccda.hia-iha.nrc-cnrc.gc.ca/community/STETSON/archive/

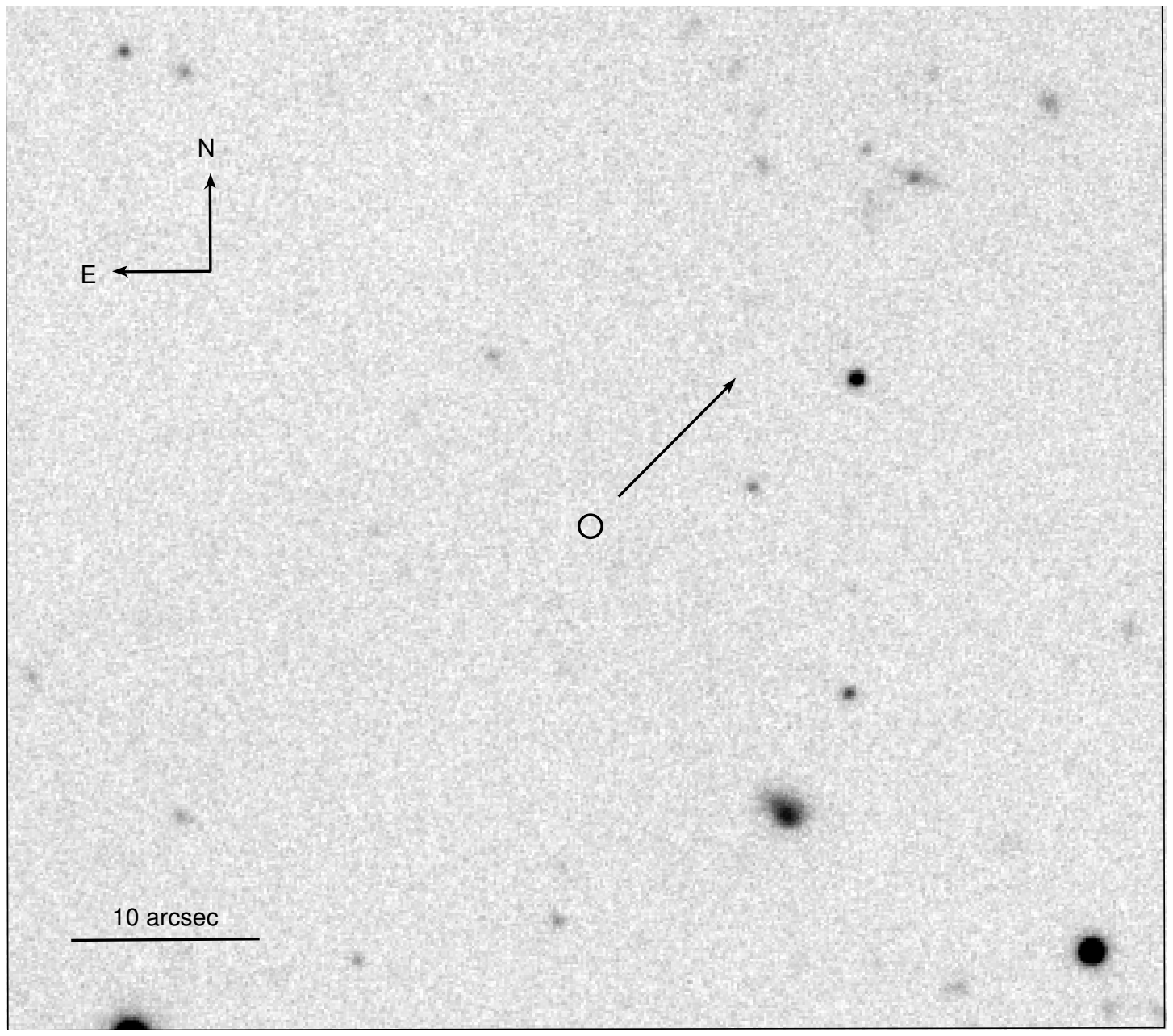

Figure 5.35: Inner portion of the field of PSR J0357+3205 as seen by the Gemini/GMOS instrument in the $H_\alpha$ band. The position of the pulsar is marked by a circle ($0''.6$ radius). No diffuse emission is seen, related to the proper motion direction of the pulsar, marked by an arrow. The upper limit to an undetected bow-shock nebula is $5 \times 10^{-18}$ erg cm$^{-2}$ s$^{-1}$ arcsec$^{-2}$.

In order to study and characterize in detail the mechanisms responsible for the X-ray trail of PSR J0357+32 emission we obtained a deep observation of the pulsar and its unusual trail with *XMM-Newton*.

### XMM-Newton observation and data reduction

Our deep *XMM-Newton* observation of PSR J0357+3205 started on 2011 September 15 at 02:37:18 UT and lasted 111.3 ks. The PN camera [194] of the EPIC instrument was operating in Large Window mode (time resolution of 47.7 ms over a $14' \times 27'$ Field–of–View (FoV)), while the MOS detectors [203] were set in full frame mode (2.6 s time resolution on a 15' radius FoV). The thin optical filter was used for the PN camera while we chose to use the medium filter for the MOS detectors due to the presence of moderately bright ($m_R \sim 9$) stars. We used the *XMM-Newton* Science Analysis Software (SAS) v11.0. After standard data processing, using the `epproc` and `emproc` tools, and screening for high particle background time intervals (following [70]), the good, dead-time corrected exposure time was 98.5 ks for the PN and 108.3 ks for the two MOS detectors.

### Spatial and spectral features

The extended feature of the bright nebula protruding from the pulsar and extending $\sim 9'$ in length can be easily studied in our *XMM-Newton* observation due to its lower particle background and higher spectral resolution, compared to the *Chandra* one. The worse spatial resolution of *XMM-Newton* prevented us from significantly constraining the properties of the nebula near the pulsar. We therefore concentrated on a higher-scale analysis of the extended feature.

Producing brightness profiles of the nebula along the pulsar motion and orthogonal to it as described in [137], we do not detect any spatial variations in the trail in the available observations. Along the direction of the pulsar motion, the trail emerges from the background within $50''$ of the pulsar and its flux increases gradually, reaching a flat maximum between $4'$ and $7'$. Its decrease is faster fading below the background level at about $9'$ from PSR J0357+3205. In the direction orthogonal to the main axis, the trail profile is clearly asymmetric, showing a fast increase in the northeast direction (rising to the maximum within $30''$), reaching a sharp maximum, in broad agreement with the neutron star proper motion direction. The further decrease is slower, fading at about $1'$–$1'.5$ in the nearest part of the nebula and $2'$ in the furthest.

Such a behavior recalls the projection of a cone, with the apex corresponding to the pulsar position (or somewhere in the vicinity of the pulsar).

After a detailed spectral analysis described in [137], we find that the spectrum of the nebula is well described ($\chi^2_\nu = 1.06$, 162 dof) by a power-law model with a photon index $\Gamma = 2.07 \pm 0.08$), absorbed by a column $N_H = (2.61 \pm 0.23) \times 10^{21}$ cm$^{-2}$ as well as a thermal bremsstrahlung model with a temperature $kT = 3.75^{+0.48}_{-0.40}$ keV and absorbed by a column density $N_H = (1.72 \pm 0.16) \times 10^{21}$ cm$^{-2}$. The thermal bremsstrahlung is the only model for which the value of $N_H$ obtained for the trail is not in conflict (at the $\sim 3\sigma$ level) with both the one obtained for the pulsar, as well as with the Galactic value. Finally, dividing the elliptical extraction region of the nebula in for semi-elliptical regions using the two axes, we do not detect any significant ($3\sigma$) variation of spectral parameters

## 5.2.4 Discussion

The morphology of the trail, its spatial extension – $\sim$ 1.3 pc at 500 pc assuming no inclination with respect to the plane of the sky ($i$) – and the lack of any other related Galactic source led [69] to associate unambiguously the nebular emission with the pulsar. Elongated X-ray trails coupled with pulsars are quite common (e.g. [111]) and are usually interpreted within the framework of bow-shock, ram-pressure-dominated pulsar wind nebulae (e.g. [81]). When a pulsar moves supersonically, the shocked pulsar wind is expected to flow in a region downstream of the termination shock (the cavity in the interstellar medium created by the moving Neutron Star and its wind), confined by ram pressure. X-rays are produced by synchrotron emission from the wind particles accelerated at the termination shock, which is typically seen (if it can be resolved) as the brightest part of the extended structure (see, e.g., [111]), according to the expectations from magnetohydrodynamic simulations [36, 207, 37]. We have detected and measured a very large proper motion for the pulsar: 165 $\pm$ 30 mas yr$^{-1}$ – 390 km s$^{-1}$ across the plane of the sky, at 500 pc – in the direction opposite to the trail, thus compatible with such a scenario.

The trail shows no other similarities with other, synchrotron-emitting nebulae (for a complete discussion about PSR J0357+3205's synchrotron-emitting model see [69]). The first problem arises from energetic requirements for the emitting particles. According to [88], the maximum potential drop between the pole and the light cylinder (in an aligned pulsar) is $\Delta\Phi = (3\,\dot{E}_{rot}/2c)^{1/2}$. In the PSR J0357+3205 case, the electrons are accelerated up to a maximum Lorentz factor $\gamma_{max} \sim 10^8$, which can be considered as an upper limit for the electrons injected in the pulsar wind nebula. The low $\dot{E}_{rot}$ of PSR J0357+3205 requires an ambient magnetic field as high as $\sim$50 $\mu$G – while the mean Galactic value is $\sim$1 $\mu$G ([107] and references therein). However, one must also take into ac-count the magnetic field carried by the particles themselves. An independent estimate of the magnetic field in the trail can be obtained from the measured synchrotron surface brightness [164]. Taking into account the uncertainty on the parameters we can conclude that a reasonable value of the magnetic field in the trail is in the range $20 - 100$ $\mu$G, consistent with the value of $\sim$50 $\mu$G obtained before.

The second problem is the lack of diffuse emission surrounding the pulsar, where the emission from the wind termination shock should be brightest, as observed in all the other known cases (e.g. [82, 143, 111]). Such a shock can be unresolved by *Chandra* only in the case of an extremely large ambient density (several hundred atoms per $cm^3$) and/or a pulsar speed higher than some thousands km $s^{-1}$. Taking into account the measure of the proper motion, the distance of the pulsar should be larger than 1.3 kpc or the inclination of the motion should be higher than 70°. Such conditions imply a trail at least 3.5 pc long, so that it becomes even more difficult to account for the energy of the particles. This picture implies also that a significant fraction of the point-like flux comes from a non-thermal termination shock, reducing dramatically the power-law component of the pulsar [137]. While it is acceptable for the non-thermal component of the pulsar to be a factor of some smaller than the value we obtained [136, 134], the lack of nebular emission around the pulsar, where the synchrotron emission should be stronger, remains unexplained.

The brightness profile along the main axis of the trail is consistent with no emission within $30''$–$50''$ of the pulsar, then it slowly increases to reach a broad peak at $\sim 4'$. This behavior is remarkably different from that observed in all the known synchrotron nebulae that show peaks close to the position of their parent pulsar (e.g. [111]), where the acceleration of the wind particles is the most important. It is very hard to explain in this scenario why the emission peaks at $\sim$1 pc from the pulsar.
The asymmetric brightness profile of the trail in the direction perpendicular to its main axis is characterized by a sharp northeastern edge and a slow decay toward the south-west direction. The synchrotron emission from the nebula is only marginally dependent on the ISM density, so that ad-hoc and large variations in the ISM are demanded, in order to explain this strange profile. Even if possible, such a case is hardly believable.
Furthermore, the inferred synchrotron cooling time implies a significant spatial-spectral evolution with respect to the distance from the pulsar, as observed in all synchrotron nebulae [111]. Thanks to the unique throughput of *XMM-Newton*, we could perform a deep spatial-spectral study of the trail, analyzing separately the furthest and the nearest parts of the nebula. No significant differences in the spectral index are detected (see the previous section). The lack of variation cannot be easily explained in terms of a synchrotron nebula. The lack of such a softening would require some particle re-acceleration mechanism within the trail itself.
The excellent throughput of the *XMM-Newton* instruments has revealed another problem of the synchrotron nebula scenario, connected to the value of the column density ($N_H$) obtained by fitting the non-thermal model of the nebula. While it is marginally compatible with the value obtained for the pulsar, the inferred $N_H$ appears to be larger than the Galactic one in that direction.

A thermal bremsstrahlung model fits equally well the spectrum of the trail and produces an estimate of $N_H$ in agreement with that of the pulsar and the Galactic one. In the bremsstrahlung scenario, the trail emission arises from the shocked ISM material heated up to X-ray temperatures. Under the assumption of a strong shock-wave and from the junction conditions at the shock front, we can estimate the speed of the pulsar directly from the temperature of the shocked ISM material at the head of the bow shock: $v_{psr} = \sqrt{16kT/(3\mu m_p)}$, where $\mu$ is the molecular weight of the ISM gas and we assumed an adiabatic index $\gamma_{ad} = 5/3$. The temperature obtained by fitting the trail spectrum with a thermal bremsstrahlung model, $kT \approx 3.75$ keV, implies a pulsar speed $v_{psr} \simeq 1376\mu^{-1/2}$ km s$^{-1}$. From the values of the abundances resulting from the spectral fit, we can assume that $\mu \simeq 0.54$, so the speed of the pulsar is about 1900 km s$^{-1}$, which would make it the largest ever observed in a pulsar. If we combine this result with the pulsar proper motion we can estimate the inclination angle $i$ of the pulsar motion (and thus of its trail), $i \approx \mathrm{acos}(0.2 \times d_{500})$. Moreover, through the value of the photon flux measured using a thermal bremsstrahlung model, it is possible to estimate the number density of the shocked ISM gas as $n \simeq 0.86 d_{500}^{-1/2}$ atoms cm$^{-3}$.

Thermal bremsstrahlung emission acts as a cooling mechanism. The time scale for cooling can be estimated as $t_{cool} \simeq (E_{thermal}/J)$, where $E_{thermal} \simeq nkT$ is the thermal energy per unit volume and $J$ is the volume emissivity. This approximation gives $t_{cool} \simeq 6 \times 10^3 T^{1/2}(n\bar{g}_B)^{-1}$ yr, where $\bar{g}_B$ is the frequency-averaged Gaunt factor that typically ranges between 1.1 and 1.5, depending on the plasma temperature [162]. From our estimates of the temperature and number density of the shocked ISM gas we expect $t_{cool} \simeq 10^7$ yr. Considering a pulsar proper motion of $165 \pm 30$ mas yr$^{-1}$ and an extension of the nebula of about $9'$, then the age of the trail ($\sim 3300$ yr) is much lower than its cooling time. This result can explain why we have not measured a significant evolution of the temperature of the trail as a function of the distance from the pulsar. On the other hand, the shorter than expected length of the trail can be explained assuming an ad-hoc decrease of a $\sim$3 factor in the number density $n$ beyond the end of the trail - the volume emissivity of the thermal bremsstrahlung emission is $J \propto n^2 T^{1/2}$. Such a strong dependence on the ISM density can explain the asymmetrical shape of the nebula in the direction orthogonal to its main axis: this could be a consequence of an inhomogeneity of the number density in the nebula, quite typical in the ISM.

From the conservation of mass, energy, and momentum on the two sides of the shock front and from the values measured in the trail (e.g. the pre-shock temperature $T_0$ and number density $n_0$), we can calculate the same physical quantities outside the nebula:

$$\frac{n}{n_0} = \frac{(\gamma+1)M^2}{(\gamma+1)+(\gamma-1)(M^2-1)} \tag{5.1}$$

$$\frac{T}{T_0} = \frac{[(\gamma+1)+2\gamma(M^2-1)]\,[(\gamma+1)+(\gamma-1)(M^2-1)]}{(\gamma+1)^2M^2} \tag{5.2}$$

where $M = 1/\sin(\alpha)$ is the Mach number and depends on the cone angle of the trail ($\alpha$) and an adiabatic index $\gamma_{ad} = 5/3$ is assumed for the ISM gas. In the strong shock-wave limit ($M \gg 1$) $n_0 = n/4 \simeq 0.22 d_{500}^{-1/2}$ atoms cm$^{-3}$ while $T_0$ depends on the Mach number and thus on the inclination angle $i$. From the geometry of the cone angle of the nebula, a plausible value of $T_0$ is in the range of $10^5$–$10^6$ K, for $i \gtrsim 60°$. The pre-shock temperature is consistent with that of the hot phase of the ISM, which fills a large fraction of the Galaxy (e.g. [31]). Such a scenario is also in full agreement with the requirement of a fully ionized ISM by the lack of detection of a H$\alpha$ nebula. In fact, using the scaling law proposed by [59] and [51], we estimated that for any reasonable value of the distance to PSR J0357+3205 (ranging from $\sim$ 200 pc, as suggested by the non negligible X-ray absorbing column, to $\sim$ 900 pc, where $\gamma$-ray luminosity would exceed the spin-down luminosity) and in a broad range of possible space velocities for the pulsar (assuming the inclination angle with respect to the plane of the sky in the $\pm 75°$ range), the lack of detection of a H$\alpha$ nebula can only be ascribed to a neutral fraction $X_{ISM}$ <0.01. We note that the density required by our model is slightly higher than expected for the hot ISM: this may result from the concerted action of massive stellar winds and supernova explosions. [218] report the presence of dense ($10^{-3}$–$10^{-1}$ atoms cm$^{-3}$), hot ($\sim 10^6$ K) envelopes of gas in our Galaxy. On the other hand, thermal models such as `mekal` (hot diffuse gas [147]), `nei` (collisional plasma, non-equilibrium, constant temperature [34]), and `pshock` (plane-parallel shocked plasma, constant temperature [34]) with redshift 0 fit equally well the spectrum of the trail but point to a very low metallicity of the ISM around the pulsar, making more difficult to explain this dense envelope as the result of a supernova explosion or stellar winds. Future multi-wavelength observations of the field and more detailed Galactic gas models could better constrain and explain the ISM composition around PSR J0357+3205.

A shock-wave scenario can easily explain the lack of diffuse emission surrounding the pulsar. Most of the energy in the pre-shock flow is carried by the

ions – the kinetic energy in the streaming of the electrons is less than 1/2000 of the total – but the electrons are generally responsible for the cooling of the plasma: this means that the electron temperature ($T_e$) is responsible for the dynamics of the shock and its emitted spectrum. The electrons must be heated by the ions before the emission becomes detectable and this process takes some time. Coulomb heating of the electrons behind the shock proceeds at a rate

$$\frac{dT_e}{dt} = \frac{T_i - T_e}{t_{eq}} \tag{5.3}$$

where $T_i$ is the ion temperature and $t_{eq} = 7.7(T_e^{3/2}/n)$ s ∼70 kyr is the equipartition time for a fully ionized plasma with a Coulomb logarithm of 30 (where $T_e$ is in units of K and $n$ in cm$^{-3}$). If all the post-shock energy is in the ions ($T_e \ll T_i$), then the increase in $\mathrm{T}_e$ due to collisional heating follows: $T_e(coll) = 1.7 \times 10^7 (n_0 t_3 v_{psr8}^2)^{2/5}$ K, where $t_3$ is the age of the trail in units of $10^3$ yr, $n_0$ is the pre-shock number density and $v_{psr8}$ is the pulsar speed in units of $10^8$ cm s$^{-1}$. Thus $T_e(coll) \lesssim 6.6 \times 10^7 d_{500}^{-1/5}$ K, not significantly different from the value observed in the trail. Moreover, $T_e(coll) \lesssim 5.5 \times 10^6$ K within $\sim 30''$ of the pulsar (corresponding to $\sim 200 - 300$ years, from the measure of the proper motion). This fully explains the lack of diffuse emission surrounding the pulsar.

So far, no evidence of other pulsars characterized by a thermal bremsstrahlung emitting trail has been reported. This could be explained by the peculiar conditions it requires, e.g. very fast pulsars and an hot and dense ISM. With its estimated velocity of 1900 km s$^{-1}$, J0357+3205 is the fastest moving pulsar known. Among all the pulsars listed in the ATNF Pulsar Catalogue [129], only 2 have a very high velocity ($\gtrsim$ 700 km/s), reliably measured through an estimate of the distance obtained by other methods than the dispersion measure. These two pulsars do not emit through thermal bremsstrahlung probably because of the ISM characteristics as described in detail in [137].

### 5.2.5 The first example of a thermally-emitting trail?

Relative astrometry on our multi-epoch *Chandra* images unveiled a significant proper motion of $165 \pm 30$ mas yr$^{-1}$ for PSR J0357+3205, corresponding to a velocity of $\sim 390\mathrm{d}_{500}$ km s$^{-1}$, along a direction almost coincident with the main axis of the elongated X-ray nebula.

The measure of the pulsar space velocity can be used, together with the upper limit to the surface brightness of any undetected diffuse structure in our Gemini $\mathrm{H}_\alpha$ image, to constrain the conditions of the medium in which PSR J0357+3205 is moving. The expected flux of the narrow-line component of a pulsar $\mathrm{H}_\alpha$ bow-shock nebula depends on the pulsar spin-down luminosity $\dot{E}_{rot}$, the pulsar space velocity $v_{psr}$, the distance to the source $D_{psr}$, and the neutral fraction of the medium $X_{ISM}$. Using the scaling law proposed by [59] and [51], we estimate that for any reasonable value of the distance to PSR J0357+3205 and in a broad range of possible space velocities for the pulsar the lack of detection of a $H_\alpha$ nebula can only be ascribed to a neutral fraction $X_{ISM}$ <0.01, i.e. the gas surrounding the pulsar is fully ionized. Thus, PSR J0357+3205 could be traveling across a region filled with ISM in the hot phase. Some contribution to the ionization of the medium from the pulsar itself is also possible, although its thermal emission is not particularly prominent [69].

Our *XMM-Newton* observation has allowed us to confirm the presence of a diffuse emission $\sim 9'$ long, the largest trail of X-ray emission associated with any rotation-powered pulsar. Such an extended emission cannot be explained in terms of the usual bow-shock ram-pressure-dominated pulsar wind nebula. In fact, the following problems arise:
- the existence of the trail is problematic for a pulsar with such a low $\dot{E}$ as PSR J0357+3205;
- no bow shock has been resolved around the pulsar;
- the lack of any nebular emission around the pulsar, where the particle acceleration is maximum, cannot be explained;
- the nebular spectrum lacks any spatial evolution, a necessary signature of the radiative cooling of the electrons accelerated at the wind termination shock;
- the very asymmetric brightness profile in the direction perpendicular to the main axis of the trail requires a large ad-hoc inhomogeneity of the ISM around the pulsar;
- fitting the *XMM-Newton* nebular spectra with a power law, the nebular $N_H$ has to be higher than the Galactic value and only marginally in agreement with the $N_H$ of the pulsar [137].

We propose a thermal bremsstrahlung model as an alternative explanation of the trail emission. In this scenario, the emission comes from the shocked ISM material heated up to X-ray temperatures. This model gives full account of the peculiar features of the trail:
- the lack of any detectable spatial evolution in the trail spectrum is due to the long bremsstrahlung cool-down time;
- the peculiar asymmetries of the brightness profile can be interpreted in terms of small changes in the ISM density that strongly affect the bremsstrahlung emissivity;
- most of the energy in the pre-shock flow is carried by the ions, while the electron temperature is responsible for the X-ray emission; the Coulomb heating time of the electrons behind the shock is fully in agreement with the lack of any detected nebular emission near the pulsar;
- the value of $N_H$ measured in the trail agrees both with the value obtained

for the pulsar and the Galactic value.

This scenario allows us to estimate some parameters of the pulsar and of the ISM around it. For a bremsstrahlung-emitting trail we estimate a pulsar velocity of $\sim$1900 km $s^{-1}$, in agreement with the pulsar proper motion for distances of some hundreds parsecs and a high inclination. The mean density of the ISM is required to be $\sim$0.2 atoms $cm^{-3}$ and the temperature of some $10^5$ K. This type of hot gas usually presents lower densities, but a denser phase (possibly detected by [218]) is predicted to be the result of the action of massive stellar winds and supernova explosions. However, the low metallicity we obtained from the spectral fit makes the explanation of this gas envelope more difficult.

For all these reasons, we believe PSR J0357+3205's nebula to be the first example of a new type of thermally-emitting trails. Until now we have no clear evidence of other pulsars characterized by a trail emitting via thermal bremsstrahlung, possibly for the requirements of a very fast pulsar and a hot, dense ISM. Moreover, energetic pulsars can also generate classic synchrotron nebulae, that may outshine a bremsstrahlung component.

## 5.3 Summary

In this Chapter a detailed description and results of multi-wavelength studies of three MSP radio-quiet candidates and of a middle-aged radio-quiet $\gamma$-ray pulsar are reported.

Blind search algorithm cannot be performed scanning the entire 2FGL 95% or 99% error circle because of the prohibitive request for computing power to search for a possible periodicity of some ms directly in $\gamma$-ray data. To increase the efficiency of the algorithm we have decided to perform a deep X-ray observation of the error box of three 2FGL unidentified sources which are very interesting MSP radio-quiet candidate. We have based the targets selection on their probability of being MSP by a hierarchical neural network analysis. Moreover, since MSPs are nearby objects, we have selected only sources situated far from the Galactic plane (b>5°). In the end they must have been the object of deep unsuccessful radio searches and they must be bright (TS>100) in order to run efficiently blind search. We have found that each 2FGL unidentified sources has some X-ray plausible counterparts within the 99% error circle. For these X-ray counterparts we have produced X-ray spectra, light curves and found optical counterparts. These analyses have brought us to identify and characterize some X-ray sources and to select only the best counterpart candidates of the putative MSPs. For each unidentified sources we have found at least one excellent X-ray counterpart candidates and we recommend to run blind search on much small sky area covered by the error regions of these detected X-ray sources.

Our multi-wavelength studies on PSR J0357+3205 have unveiled a significant proper motion of $165 \pm 30$ mas yr$^{-1}$, corresponding to a velocity of $\sim 390 d_{500}$ km s$^{-1}$, along a direction almost coincident with the main axis of the elongated X-ray nebula. The measure of the pulsar space velocity can be used, together with the upper limit to the surface brightness of any undetected diffuse structure in our Gemini $H_\alpha$ image, to constrain the conditions of the medium in which PSR J0357+3205 is moving. The study of the trail confirmed the lack of any extended emission near the pulsar itself. The trail shows a very asymmetric brightness profile and its spectrum does not vary as a function of the position. We have found such a nebular emission not to be consistent with a classical bow-shock, ram-pressure dominated pulsar wind nebulae. We proposed a thermal *bremsstrahlung* as an alternative model for the PSR J0357+3205's trail emission. In this scenario, the trail emission would come from the shocked interstellar medium (ISM) material heated up to X-ray temperature. This can fully explain the peculiar features of the trail in the case of a hot, moderately dense ISM around the pulsar. For a *bremsstrahlung*-emitting trail, we estimated the pulsar distance to be between 0.3 and 2 kpc and the trail length about 2 pc. A pulsar velocity of $\sim 1900$ km s$^{-1}$ is required, making PSR J0357+3205 the fastest pulsar known. Since the unique features of the pulsar's trail, we proposed pulsar's nebula to be the first example of a new class of thermally-emitting nebulae associated to fast pulsars.

# Conclusions and future perspectives

Gamma-ray sources are tracers of the most energetic processes in the Universe, they are very exotic objects, characterized by very intense magnetic fields and the presence of very high energy particles. For this reason, understanding the nature of the $\gamma$-ray unidentified sources is one of the most important open questions in high-energy astrophysics.

The purpose of my Ph.D. thesis has been to develop techniques complementary to those implemented by the LAT team during the construction of the $\gamma$-ray source catalogs in order to asses the likely association for each unidentified source, taking advantage of the large $\gamma$-ray source sample available and of the quality of the information about source locations, spectra, and timing variability. We have developed, applied, tested and compared two different classification algorithms, Logistic Regression (LR), which is characterized by a simple approach, and Artificial Neural Network (ANN), the natural extension of the LR approach.

As a first step we have implemented a LR analysis method. This technique uses identified objects to build up a classification analysis which provides the probability for an unidentified source to belong to a given class based on its $\gamma$-ray properties. We have tested this technique with the sources in the *First Fermi-LAT Source Catalog* (1FGL) providing classification probabilities for each unidentified source. Logistic Regression has been trained on known objects in order to predict the membership of a new object to a given class on the basis of its observables. We have trained the model using pulsars and AGNs identified in the 1FGL catalog because they are abundant and have different phenomenologies. Then, we have chosen the variables which distinguish more efficiently a pulsar from an AGN and the two classification thresholds to single out AGN and pulsar candidates in the 1FGL unassociated sample. The main input parameters were spectral and temporal information. The training sam-ple included 693 AGN and 63 pulsars. The spatial distribution has been used to confirm the accuracy of our classification. Applying the LR model to 1FGL unidentified sources we have been able to classify them as pulsar candidates or AGN candidates. We have tested the capability of the Logistic Regression algorithm for identifying the different source classes applying the LR analysis to the newly sources identified after the release of the 1FGL catalog. The LR efficiency at classifying new AGNs is very high ($\sim$80%) while it has a somewhat poorer success rate for finding new pulsars (efficiency $\sim$55%), this may be related to the bad characterization of some of new pulsars in the energy range of 0.3–10 GeV where $\sim$20% of them have only upper limits. Combining these results with those found by the LAT team using Classification Trees machine learning technique, the efficiency at classifying new sources significantly increases, $\sim$ 85% for new AGNs and $\sim$ 80% for new pulsars with a low rate of false negatives.

Encouraged by the results obtained by the application of the Logistic Regression, we have implemented a refined logistic regression aimed at discriminating pulsars and AGNs. In the newly released *Second Fermi Catalog* (2FGL), the larger sample, the richer statistics resulting from the use of two years of data, and the refinement of the $\gamma$-ray parameters of the sources in the 2FGL catalog have allowed us to significantly improve the performance of our classification technique in classifying 2FGL unidentified sources. The refined LR efficiency in classifying AGNs and pulsars identified after the release of the 2FGL is very high, especially for new bright young pulsars for which the efficiency is $\sim$80%. Performances of the upgraded LR method are poorer for low signal-to-noise sources.

Logistic regression is a generalized linear classifier and it classifies 2FGL sources through a linear distinction in the $\gamma$-ray variable space, this is a limit of this technique. Performances at classifying pulsar and AGN candidates in 2FGL source sample have been improved considerably using a neural network technique, which classifies 2FGL sources through a more complex non-linear separation with respect to the LR (a linear classifier). This has been the first time this technique has been applied for this purpose. The ANN efficiency in classifying new AGNs and pulsars is significantly higher and the false negative rate is lower with respect to those of the refined LR, this means that only a non-linear analysis is able to efficiently classify the 2FGL sources.

In view of the performances of the ANN algorithm and of its flexibility, we have implemented a more complex neural network architecture aimed at distinguishing the two pulsar subsamples, young pulsars and MSPs, and the two blazar subclasses, BL Lacs and FSRQs. For this purpose we have used a hierarchical neural network model. The efficiency of the hierarchical algorithm at classifying new MSP detections is very high (nearby 80%) and the absence of false negative is a very encouraging result. Conversely, efficiency at classifying new young pulsars is not extremely high ($\sim$60%), possibly owing to some peculiarities which make such pulsars somewhat different with respect to "standard" young pulsars. Although $\gamma$-ray properties of young pulsars seem very similar to MSPs, our model is able to efficiently discriminate these two source classes, this is a very important and exciting result because it gives us the opportunity to select targets for young pulsar and millisecond pulsar searches and because it tells us that $\gamma$-ray properties of these two pulsar subsamples are different (in particular, in spectral shape and flux distributions). This result needs further analysis because the results of the classification analysis can provide important guidance to plain follow-up observations. Regarding blazars, the efficiency of the hierarchical algorithm at classifying new BL Lacs and FSRQs is very high (nearby 80%), we have obtained that $\gamma$-ray observables which better discriminate the two blazar subclasses are those we expect from synchrotron-inverse Compton scenarios validating the accuracy of our approach. Our results give us the opportunity to improve the $\gamma$-ray emission models of blazar subclasses and to develop a theory aimed explaining the differences in the $\gamma$-ray emission from the two pulsar subclasses.

We have used the results of the hierarchical neural networks to select targets for radio-quiet millisecond (MSP) pulsar searches. Such sources are predicted by recent models but they have not yet been detected, their discovery will improve our knowledge of $\gamma$-ray emission mechanism and of $\gamma$-ray pulsar population. Multi-wavelength studies of the unidentified sources give us the opportunity to select the most probable counterparts of the $\gamma$-ray sources. This reduces by a factor $10^2$–$10^3$ the sky area to be considered for a blind periodicity search in $\gamma$-ray data, making such a search feasible . To this aim we have performed a deep X-ray observation of the error box of three radio-quiet MSP candidates: 2FGL J1036.1–6722, 2FGL J1539.2–3325 and 2FGL J1744.1–7620. For each 2FGL unidentified sources we have found some plausible X-ray counterparts within the $\gamma$-ray error circle. For each X-ray counterpart we have obtained an X-ray spectrum, assessed its variability and searched for optical/IR counterparts. These analyses have allowed us to select the best counterpart candidates of the putative radio-quiet MSPs. A blind search algorithm will be run as a future step.

The machine learning techniques we have developed, in particular ANNs, will be used as a standard analysis to be applied during the construction of future $\gamma$-ray source catalogs to provide the most probable association for each unidentified $\gamma$-ray on the basis of its properties. As a first application, the *Fermi*-LAT collaboration is developing the *Third Fermi-LAT Source Catalog* (3FGL) using five years of data and our statistical methods will be able to provide an additional information about the nature of the 3FGL unidentified sources. Our ANN tool will be upgraded and refined in future, for example adding other $\gamma$-ray parameters or multi-wavelength information from catalogs and follow-up programs, or incorporating the uncertainties in the input parameters. The efficiency at classifying unidentified sources will possibly increase combining the results obtained with different machine learning techniques.

In parallel to the work devoted to unidentified sources, I took part to the analysis and interpretation of multi-wavelength observations of isolated neutron stars. Multi-wavelength studies are very important to understand the physics of $\gamma$-ray pulsars characterizing and exploring their environment. In particular, the most important objects to constrain pulsar models are the "extreme" ones, accounting for the tails of the population distribution in energy, age and magnetic field. In this respect, we have studied the middle-aged PSR J0357+32051, a nearby, radio-quiet, bright $\gamma$-ray pulsar discovered by the *Fermi* mission. PSR J0357+32051 is one of the most interesting pulsars discovered by the LAT because it is one of the non-recycled $\gamma$-ray pulsars with the smallest rotational energy loss detected so far ($\dot{E}_{rot} = 5.9 \times 10^{33}$ erg s$^{-1}$). X-ray observations revealed a huge, very peculiar structure of diffuse X-ray emission originating at the pulsar position and extending for $> 9'$ on the plane of the sky. To better understand the nature of such a nebula, we have studied the proper motion of the parent pulsar. We performed relative astrometry on *Chandra* images of the field spanning a time baseline of 2.2 yr, unveiling a significant angular displacement of the pulsar counterpart, corresponding to a proper motion of $0''.165 \pm 0''.030$ yr$^{-1}$. The direction of the pulsar proper motion is aligned very well with the main axis of the X-ray nebula, pointing to a physical, yet elusive, link between the nebula and the pulsar space velocity. No optical emission in the $H_\alpha$ line is seen in a deep image collected at the Gemini telescope, which implies that the interstellar medium into which the pulsar is moving is fully ionized. Analyzing our *XMM-Newton* observation we have found pulsar's trail emission is consistent with a thermal *bremsstrahlung*. For a *bremsstrahlung*-emitting trail a pulsar velocity of $\sim 1900$ km s$^{-1}$ is required, making PSR J0357+32051 the fastest pulsar known. Because of the unique features of the pulsar's trail, we have proposed pulsar's nebula to be the first example of a new class of thermally-emitting nebulae associated to fast pulsars.

Appendix A

# The Fermi Gamma-ray Space Telescope

The *Fermi Gamma-ray Space Telescope* is an international and multi-agency observatory class mission that is exploring the Universe in the energy range from 10 keV up to more than 300 GeV, an energy band that has never been observed by a space telescope. *Fermi* is a product of a collaboration between NASA, the Department of Energy of the United Sates and other institutions in Italy, France, Germany, Sweden, Japan and United States. *Fermi*, whose original name before the launch was GLAST (*Gamma-ray Large Area Space Telescope*), was successfully launched on June 11, 2008 from the launch pad 17-B at Kennedy Space Flight Center (Florida, USA), into an initial orbit at about 565 km altitude with a 25.3° inclination and an eccentricity lower than 0.01. During its normal operations *Fermi* orbits around Earth with a period of 95 minutes and scans the sky with a rocking angle of about 35°. Figure A.1 shows a schematic view of the *Fermi* spacecraft with its two instruments: the *Large Area Telescope* (LAT) and the *Gamma-ray Burst Monitor* (GBM; formerly *GLAST Burst Monitor*).

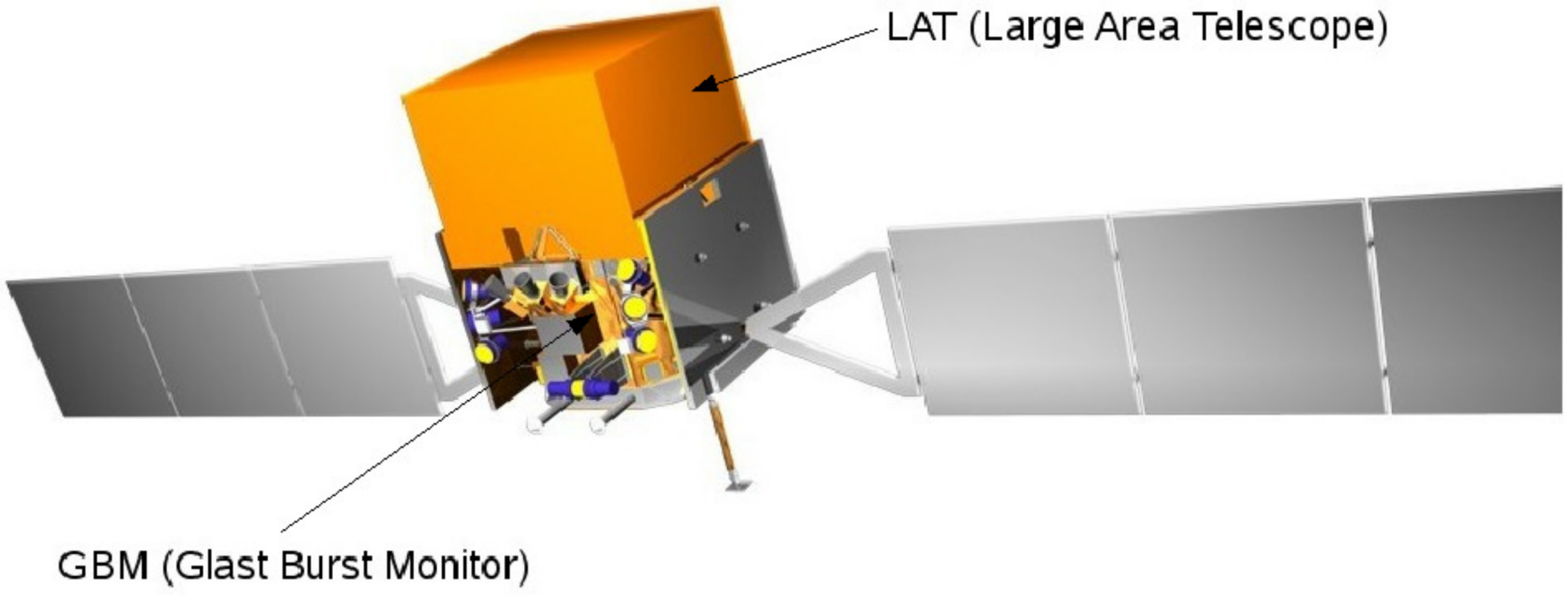

Figure A.1: Schematic view of the *Fermi* spacecraft and its two instruments.

The LAT, the main *Fermi* instrument, is a pair-conversion telescope based on high-precision detectors from High Energy Physics technology [22]. The LAT operates in an energy range from about 20 MeV to more than 300 GeV. The LAT is the successor of the EGRET telescope aboard the CGRO but its excellent performances in terms of effective area, angular resolution, energy resolution, Field-of-View (FoV) and dead-time provide a factor greater than 50 in sensitivity compared to EGRET [201].

The GBM is entirely devoted to the study of the transient $\gamma$-ray sources, such as Gamma-Ray Bursts and solar flares. It is made up by two kind of detectors based on scintilling materials, which together cover an energy range from 8 keV up to about 30 MeV. This energy range guarantee an energy overlap with the LAT.

*Fermi* represents the new generation of $\gamma$-ray telescopes, its performances are much better than those of the previous missions, such as EGRET onboard the CGRO. The *Fermi* mission is contributing in a decisive way in several topics of modern understanding of the $\gamma$-ray Universe, from the study of Galactic objects, such as pulsars, PWNe and SNRs, and extragalactic objects, such as blazars, to the detailed investigation on the nature of diffuse emission and transient sources, e.g. GRBs, answering the questions left open by the previous $\gamma$-ray missions.

## A.1 Overview of the Large Area Telescope

Pair production is the dominant mechanism of interaction between radiation and matter at the energies studied by the *Fermi Large Area Telescope* [22]. This process is the basis to measure the directions, the energies and the arrival times of the $\gamma$-ray photons entering the detector, while rejecting background from charged particles. For this reason a pair conversion telescope is made basically by a tracking system, a calorimeter and an anticoincidence system, as shown in Figure A.2. The EGRET experiment aboard CGRO and previous instruments aboard SAS-2 and COS-B missions shared the same detecting strategy. A $\gamma$ ray entering the LAT creates an electron-positron pair, whose energies and directions are reconstructed by the LAT subsystems. From these information it is possible to determine the energy, the arrival time and direction of the incoming photon using the conservation of four-momentum.

The tracks of the electrons and the positrons produced by a $\gamma$ ray are measured by the tracking system. In order to maximize the conversion probability in the detector, detecting planes are interleaved with *conversion foils* of particular thickness. Since the conversion probability increases with the atomic number ($Z$) as $Z^2$, the conversion foils are usually made by high Z material, EGRET used Tantalum (Ta) foils while LAT uses Tungsten (W) foils. In the calorimeter the electron-positron pair creates an electromagnetic shower, from the measurement of the shower the calorimeter determines the energy of the pair. The measurements gathered by the tracking system and by the calorimeter are then used to reconstruct the energy of the incoming $\gamma$ ray.

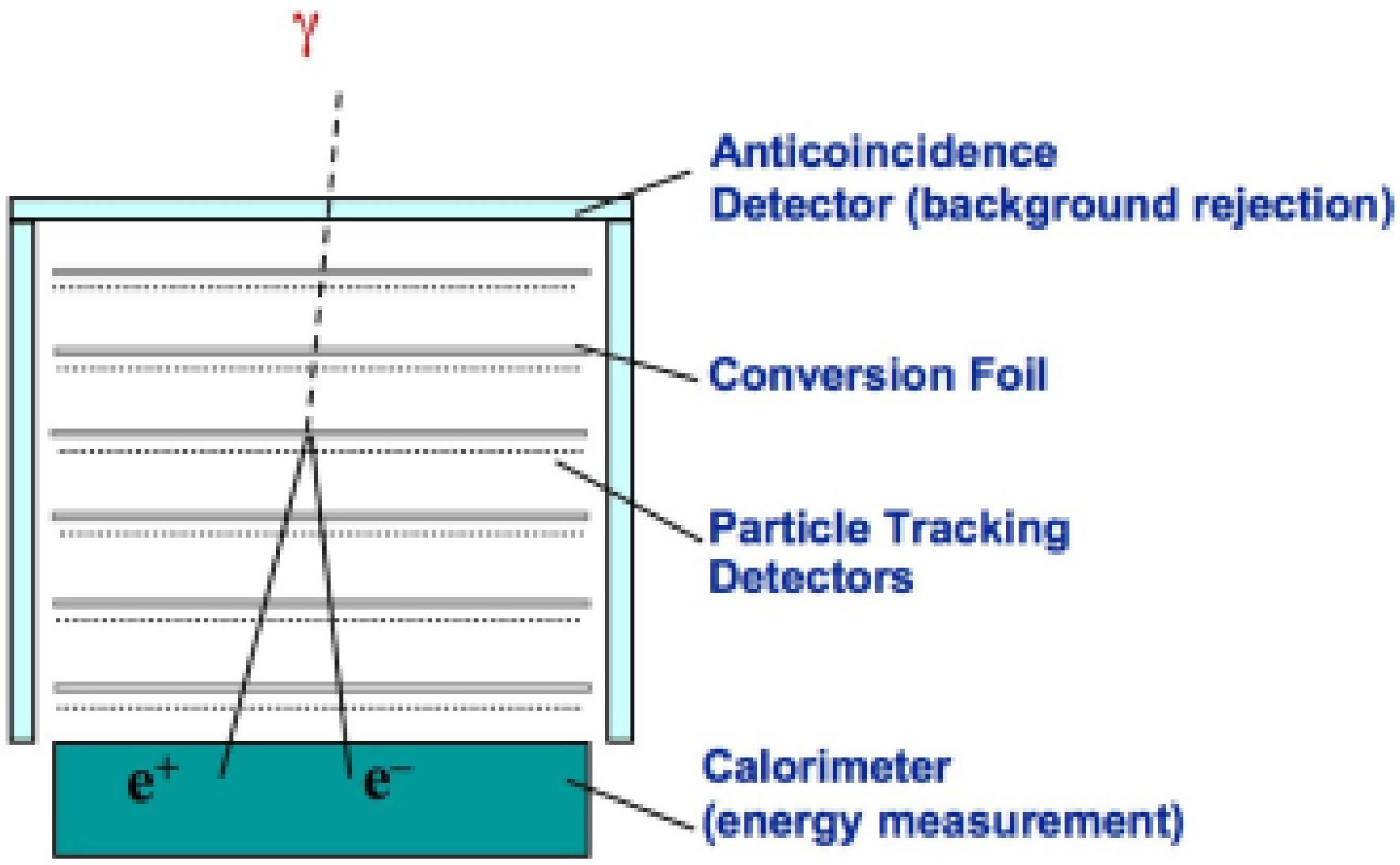

Figure A.2: Schematic view of a pair conversion telescope like the LAT. Credit: *Fermi*-LAT Team.

The orbit environment is extremely rich of charged particles that enter the detector with rates of the order of about $10^5$ times the rate of $\gamma$ rays. For this reason an anticoincidence detector is used to reject the charged particles background. The anticoincidence shield surrounds the detector and it is usually made by plastic scintillator. Since charged particles give a signal when crossing the scintillators while $\gamma$ rays do not, the anticoincidence shield reduce the charged particles background with high efficiency.

The *Fermi Large Area Telescope* (LAT) follows the same base principles but is based on a new generation $\gamma$-ray detectors developed for High Energy Physics. The main LAT subsystems are the *Tracker* (TKR), the *Calorimeter* (CAL), the *AntiCoincidence Detector* (ACD) and the *Data Acquisition System* (DAQ). The TKR is made up of alternated layers of converters, made up of Tungsten foils that allows the conversion of $\gamma$ rays, and trackers, made up of silicon microstrip detectors that allow the reconstruction of the electron positron tracks. The CAL is located below the TKR and measures the energy of the pair. The ACD covers the LAT in order to reduce the background due to charged particles discriminating their from $\gamma$-rays. The DAQ manages the main subsystems function, e.g. the reading procedure and the trigger control.

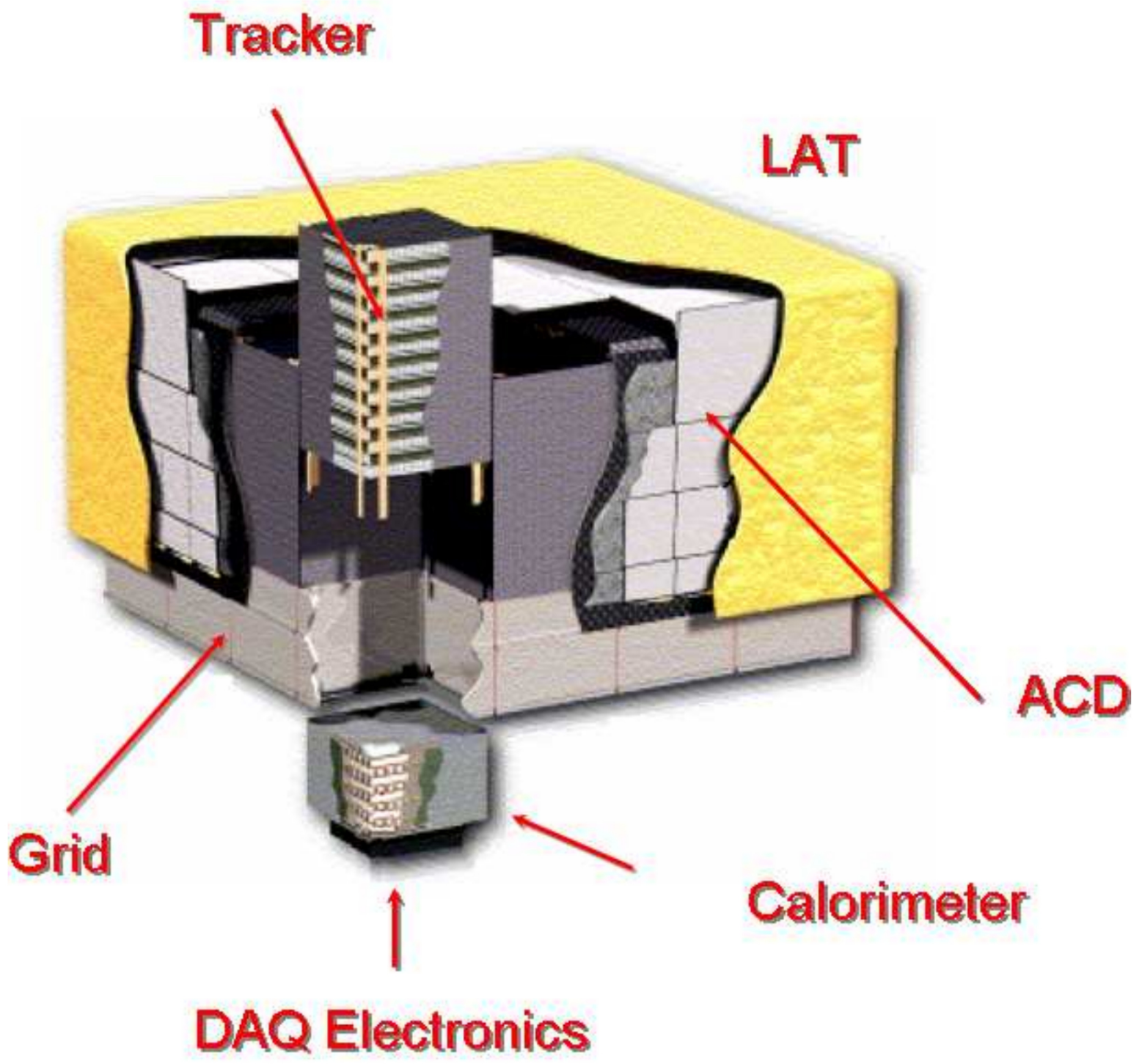

Figure A.3: Schematic diagram of the *Large Area Telescope*. The telescope's dimensions are 1.8 m × 1.8 m × 0.72 m. The power required and the mass are 650 W and 2,789 kg, respectively. Credit: *Fermi*-LAT Team.

The LAT is made of an array of 4 × 4 identical *towers* (see Figure A.3), each one made by a Tracker module, a Calorimeter module and a DAQ module; the 16 towers are surrounded by an outer segmented AntiCoincidence Detector. The main mechanical structure is a 16 cell aluminum grid that hold both the trackers towers and the calorimeter modules and all the electronic boxes are placed below the grid. The outside dimensions of the LAT are approximately 1.8 m × 1.8 m × 1 m and its mass is ∼3000 kg. The overall aspect ratio of the LAT tracker (height/width) is 0.4, allowing a large Field-of-View and ensuring that nearly all pair-conversion events initiated in the tracker pass into the calorimeter for energy measurement.

The LAT offer higher performances with respect to its predecessor EGRET

thanks to the new design strategy and the new detecting technologies.

The main innovation of the LAT is the introduction of the TKR based on solid-state detectors instead of spark chambers used for the EGRET tracker. These detectors have many advantages. First of all they provide a spatial resolution about 10 times better than spark chambers without many complications during fabrication. Additionally they offer a lower dead time of about 20 $\mu$s, with respect to the dead time of 100 ms of the EGRET spark chambers. Moreover, the silicon detectors used for the LAT Tracker are radiation hard and does not contain consumables, while EGRET used spark chambers for the tracking system and the gas deteriorated with time.

The LAT CAL is made by scintillation bars, in order to better reconstruct the electromagnetic shower development, while the EGRET calorimeter was based on a monolithic scintillating detector.

The segmented ACD detector is also another big LAT innovation, since the EGRET anticoincidence detector was made by a single scintillator panel. This segmentation will provide an higher detecting efficiency at energies greater than 10 GeV. At these energies the *self-veto* problem becomes important, because a particle from the electromagnetic shower can backscatter in the ACD producing a spurious signal. For this reason in the LAT the ACD is segmented, making possible to know roughly which ACD panel gave a signal, in order to determine if the panel has undergone a backsplash or not. In this way it is possible to avoid efficiency loss at high energies as was for the EGRET telescope.

In order to achieve its scientific goals the LAT must reject most of the background due to various contributions. The main contribution is due to cosmic rays that enter the detectors producing spurious signals. Another contribution comes from the albedo $\gamma$ rays from the Earth, that can be removed mainly by considering the position of the spacecraft with respect to our planet.

A comparison between the LAT performances and those of EGRET is shown in Table A.1.

### A.1.1 The Tracker

The TKR subsystem is the central detector of the LAT and serves to convert $\gamma$ rays into electron-positron pairs and to track the pair in order to measure the direction of the incoming $\gamma$ ray. The TKR consists of 16 modules, each one composed by planes of high $Z$ material (Tungsten) in which $\gamma$ rays incident on the LAT can convert into an electron-positron pair, interleaved with position-sensitive detectors (silicon strip detectors) that record the passage of charged particles measuring the tracks of the particles resulting from pair conversion.

The LAT tracker uses conversion foils of Tungsten ($Z$=74) because the pair production cross section is proportional to $Z^2$, then using an high $Z$ conversion foils the conversion probability is maximized. The technology employed in past years in High Energy Physics has been fundamental to choose the tracking detectors, since the alternatives were gas-filled trackers and scintillating fibers detectors. The silicon-based detectors were chosen for the LAT because of their higher sensitivity and angular resolution. The total silicon surface of the LAT tracker is of about 82 m$^2$. A scheme of the LAT TKR is displayed in Figure A.4. As for the other part of the LAT, carefully studies have produced the parameters for the Tracker in order to satisfy all the requirements and maintain the basic constraints as low consumed power, low detector noise and low computation power required.

| Quantity | LAT | EGRET |
|---|---|---|
| Energy Range | 20 MeV - 300 GeV | 10 MeV - 30 GeV |
| Peak Effective Area | $\sim 8000$ cm$^2$ | 1500 cm$^2$ |
| Field Of View | $\sim 2.4$ sr | 0.5 sr |
| Angular Resolution | $< 3^\circ.5$ (100 MeV) | $5^\circ.8$ (100 MeV) |
| (single photon) | $\sim 0^\circ.6$ (1 GeV) | $\sim 1^\circ.7$ (1 GeV) |
| (68% containment) | $< 0^\circ.15$ ($>$ 10 GeV) | |
| Energy Resolution | $< 10\%$ | 10% |
| Deadtime per Event | $< 100$ $\mu$sec | 100 msec |
| Source Localization | $< 0'.5$ | $15'$ |
| Point Source Sensitivity | $< 6 \times 10^{-9}$ cm$^{-2}$ sec$^{-1}$ | $\sim 10^{-7}$ cm$^{-2}$ sec$^{-1}$ |

Table A.1: LAT specifications and performances compared with EGRET. Quoted sensitivity for the LAT is referred to sources out of the Galactic plane and E $>$ 100 MeV.

The basic unit of the TKR is a square *Silicon Strip Detector* (SSD) with the size of 8.95 cm $\times$ 8.95 cm, where are implanted 384 parallel microstrips spaced by 228 $\mu$m. Four SSDs, each of them is 400 $\mu$m thick, are assembled in a *ladder*. In a ladder the end of each microstrip of a SSD can be connected to the end of the correspondent microstrip on the adjacent SSD in order to form a single longer microstrip. At this point 4 ladders are assembled to form a sensitive silicon microstrip layer, which is then inserted in a *tray*. A tray is a composite structure with a mechanical structure in carbon fiber that bring at both faces a sensitive silicon plane. The main components of a tray are a detecting silicon layer on the top face, an aluminum core, a Tungsten foil for the conversion of $\gamma$ rays and another silicon layer on the bottom face of the tray. The two silicon layers are mounted in a tray with parallel orientation of the microstrips. Each tray is then connected to the reading electronics. Trays are then piled up with a separation of 2 mm and each tray is rotated of 90° with respect to the adjacent tray. In this way the resulting system is made by a conversion foil followed by a couple of silicon layers with perpendicular microstrips in order to have XY detection capability. The resulting module form *tower* and it is made by 19 trays with 18 XY detection layers.

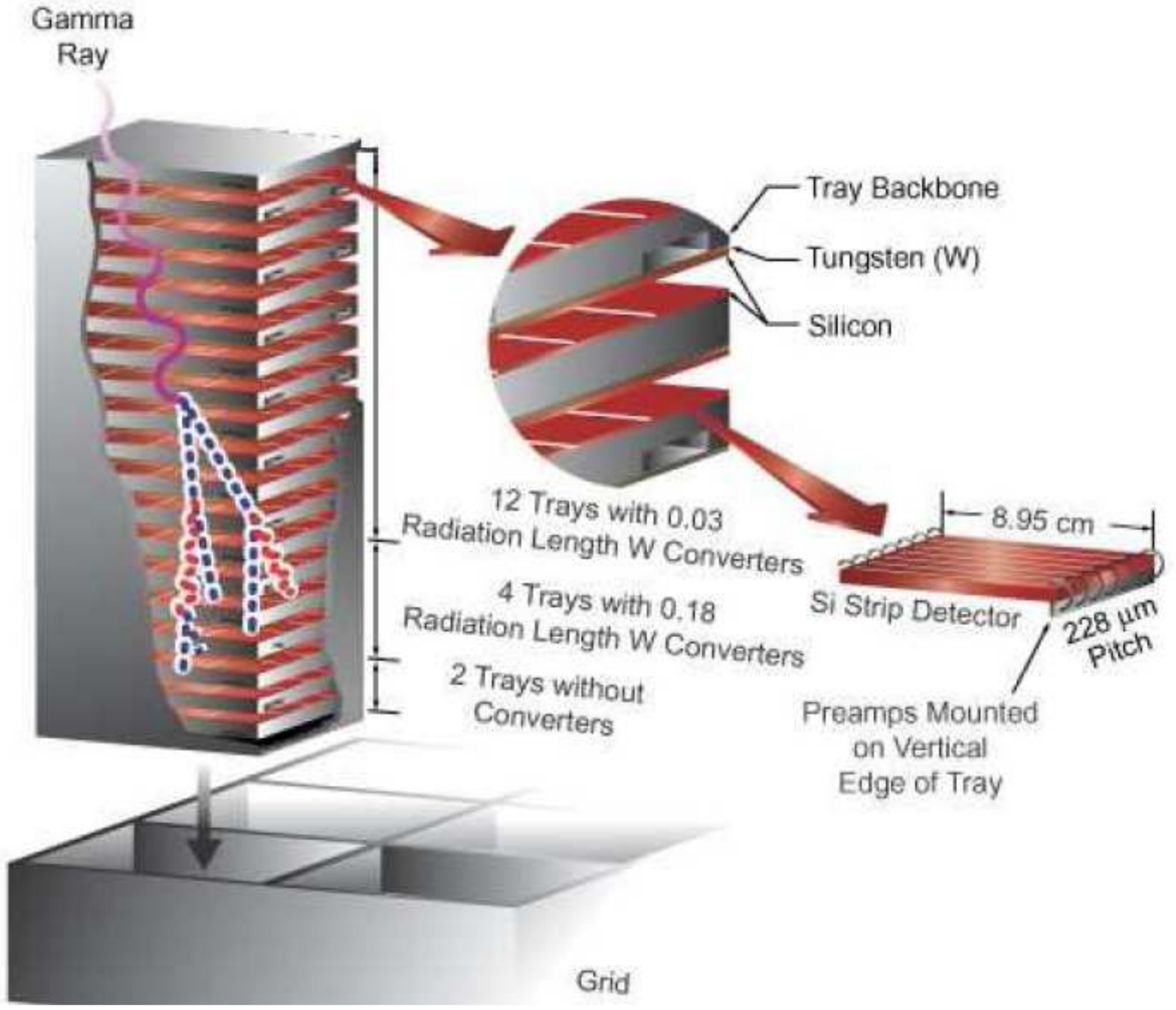

Figure A.4: Schematic view of the Large Area Telescope silicon tracker. Each tower includes 19 tray structures constituting the basic mechanical framework and housing both the silicon detection planes and the converter foils [22].

The layout of the converters (in terms of thickness) is organized into two different sections: a *Front* section, composed of 12 layers of 0.03 radiation length converter, and a *Back* section with 4 layers of 0.18 radiation length converters. The last two XY planes have no converter at all since the main *Fermi* trigger primitive requires three silicon layers in a row hit, so that a photon converted right above the last two planes would never trigger and the Tungsten would only introduce further useless multiple scattering. The two section provide measurements in a complementary manner: while the front has an excellent *Point Spread Function* (PSF), the back section greatly enhances the photon statistics, in such a way that a large effective area and a good angular resolution can be achieved, on average, at the same time.

### A.1.2 The Calorimeter

The Calorimeter [22] measures the energy of the electron-positron pair and gives information about the high-energy photons that have not converted in the Tracker. From the measure of the electron-positron energy it is possible to determine the energy of the primary photon using the conservations of the energy and momentum. A schematic view of the LAT Calorimeter is in Figure A.5.

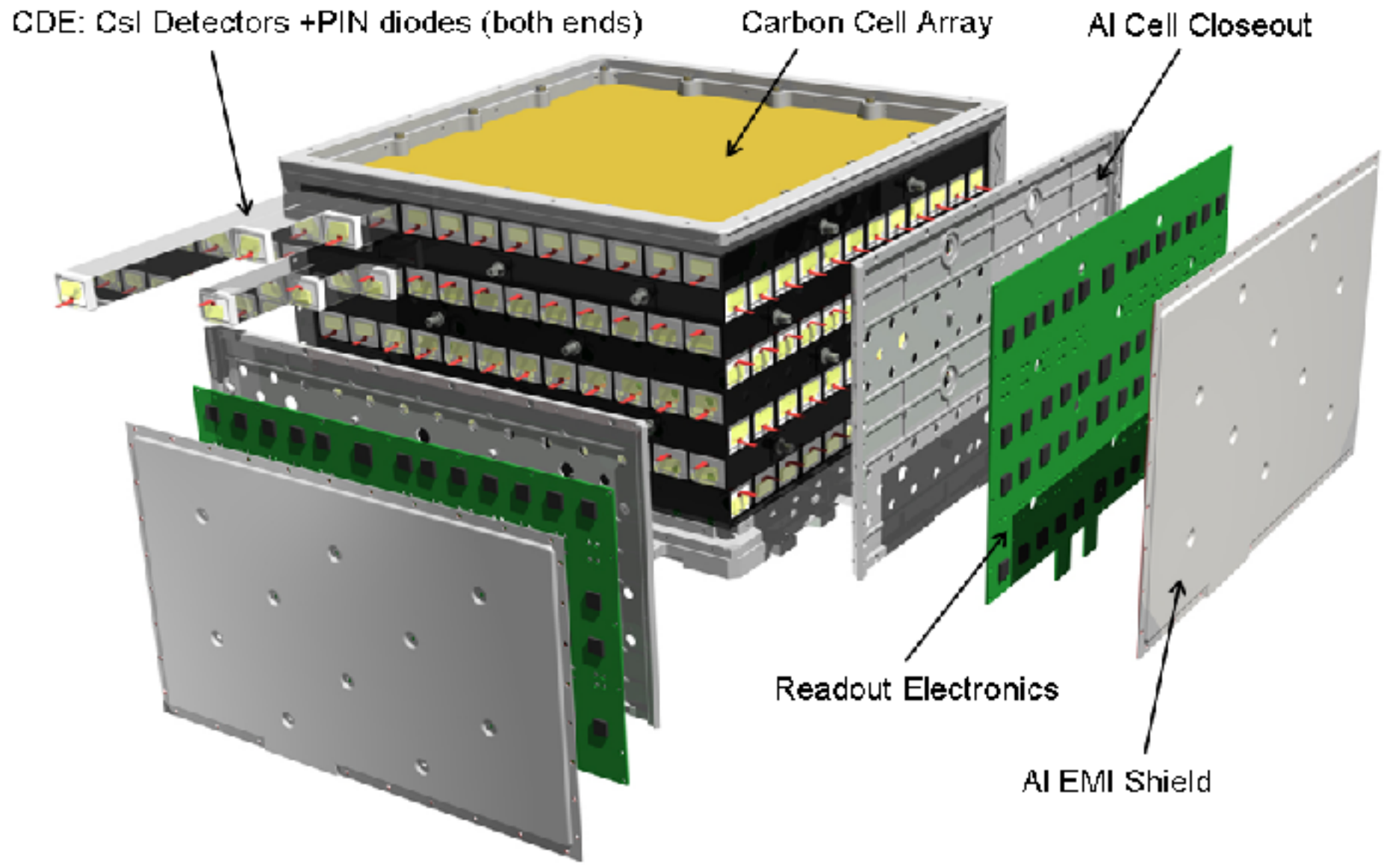

Figure A.5: LAT calorimeter module. The 96 CsI(Tl) scintillator crystal detector elements are arranged in 8 layers, with the orientation of the crystals in adjacent layers rotated by 90°. The total calorimeter depth (at normal incidence) is 8.6 radiation lengths [22].

The LAT calorimeter is composed of a set of CsI(Tl) crystals read by photodiodes. As the energy resolution strongly depends on depth, sampling and longitudinal segmentation, each CAL module is finely segmented both in depth and lateral directions. Each CAL tower will contain eight layers, each constituted by 12 crystals ($27 \times 20 \times 326$ mm$^3$), wrapped in reflective foils, for a total of 8.6 radiation lengths. As with the silicon detection planes in the tracker, each layer will be rotated by 90° with respect to the previous one (hodoscopic configuration), in order to achieve XY imaging capabilities. The lateral segmentation provides the necessary imaging capability to correlate the events in the tracker with the energy depositions in the calorimeter and derive loose (at the level of few degrees) directional information for those photons not converting in the tracker. On the other side the longitudinal segmentation allows to derive an estimate of the initial energy of the pair from the longitudinal shower profile by fitting the measurements to an analytical description of the energy-dependent mean longitudinal the showers and the subsequent leakage correction which highly enhance the response at high energy (up to several hundreds GeV) with respect to EGRET.

At both ends of each bar is placed a PIN photo diode used for reading, and the measurement of the relative intensity at both ends helps to determine the position where the energy deposition has taken place. The precision that can be obtained varies from some mm at low energies (about 10 MeV), up to less than a mm for energies above 1 GeV.

### A.1.3 The AntiCoincidence Detector

The purpose of the ACD is to detect incident charged cosmic ray particles that outnumber cosmic $\gamma$ rays by more than 5 orders of magnitude. When a $\gamma$-ray photon enter the LAT, it does not produce any signal in the ACD, but gives a signal in the TKR and in the CAL due to the produced pair. A charged particle behaves differently, since during the passage a signal also in the ACD is produced, then it is possible to recognize a $\gamma$ ray from a charged particle thanks to the different signature in the subsystems and in particular in the ACD. The events that give a signal in the TKR and in the CAL but not in the ACD can start the trigger, the other are refused as background events.

The LAT *AntiCoincidence Detector* (ACD) is made by a set of plastic scintillators coupled to Photo-Multipliers Tubes (PMTs) that use Wavelength-Shifting Fibers (WSFs) in order to increase the reading efficiency. With respect to the EGRET anticoincidence system, that was made by a single module, the LAT ACD is fine segmented. A scheme of the ACD assembly is displayed in Figure A.6.

The efficiency of background rejection, in particular for high-energy $\gamma$ rays, is increased thanks to the ACD segmentation. In EGRET it was necessary to reduce the triggers frequency in order to avoid gas consuming in the spark chambers. The EGRET ACD was implemented in the Level 1 trigger (see Section A.1.4). This reduced the working efficiency, mainly at GeV energies, where the *self-veto* becomes important. The self-veto happens when a member of the electromagnetic shower produced by the electron-positron pair is deflected and give a signal in the ACD (*backsplash*). The event has a signature of a cosmic ray and then it is rejected since it is confused with a background event. The segmentation helps to know exactly which scintillator has been hit, and it is then possible to compare the track direction with the position of the hit scintillator. In case of backsplash the position of the hit scintillator panel does not correspond exactly with the intersection of the track and the ACD, then a self-veto is recognized and avoided.

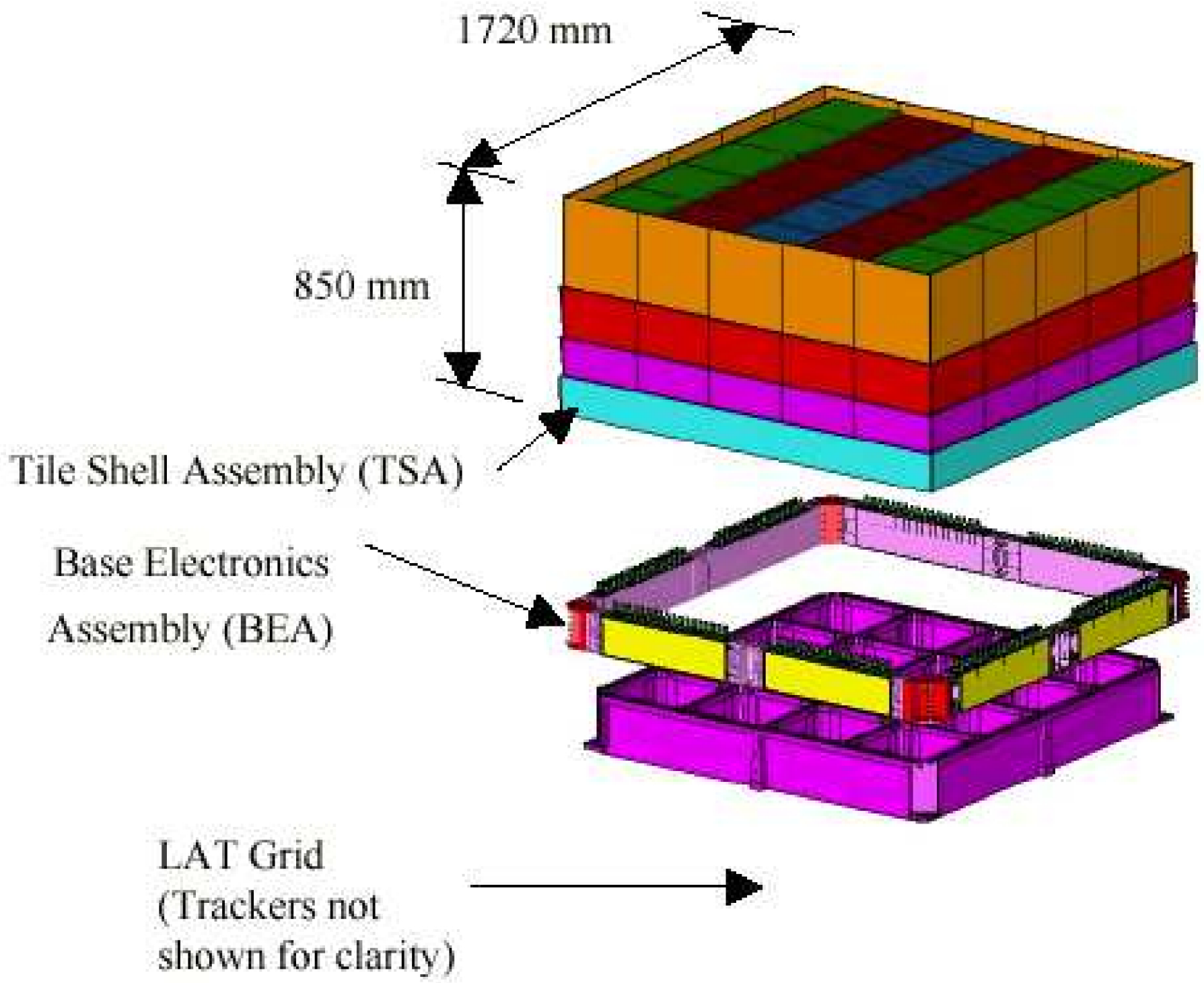

Figure A.6: Schematic view of the LAT ACD assembly [22].

Thanks to the lower dead time and to the absence of consumables in the TKR, the LAT can undergo a much higher Level 1 trigger frequency, then the ACD can be inserted in the Level 2 trigger. In this way each event can be analyzed with more care (as explained in the Section A.1.4) and the self-veto can be avoided in a very efficient way.

A total of 89 panels constitute the ACD, some of them are disposed in a 5 $\times$ 5 array on top of the TKR and the others are at the sides of the LAT. The assembling scheme of these panels has been designed with overlap in one dimension and scintillating fibers covering the gaps in the other dimension. Each scintillator is read out by an interleaved set of Wavelength Shifting (WS) fibers, with bundles connected to two phototubes, in order to guarantee redundancy.

The ACD is the first step in the background rejection scheme. The estimated Monte Carlo efficiency, confirmed by the beam tests, is greater then

0.9997.

### A.1.4 The Data Acquisition System and Trigger

The LAT Data Acquisition System (DAQ) has three main functions. It controls the trigger, it guides the event reading sequence and stores it in a temporary memory. The DAQ also manages the data elaboration and transfers to the ground. This system is also responsible for other functions, among others the control, monitoring and housekeeping of the instrument and the power management of the whole LAT.

The DAQ is made by 16 *Tower Electronic Modules* (TEM) located below each tower and two TEM specific for the ACD. Also two *Spacecraft Interface Unit* TEM are in this system and are located in the spacecraft below the LAT.

#### Trigger and background rejection

The LAT trigger must be very efficient on $\gamma$ rays and at the same time provide a high rejection power for the charged particles background. Because of the large uncertainties in cosmic ray fluxes the system flexibility is a particularly important feature.

The LAT trigger has a multi-level structure, in a similar way of the triggers employed in High Energy Physics experiments. The hardware trigger is based on special signals, called *primitives*, that originate from LAT subsystems. Primitives from Tracker, Calorimeter and Anticoincidence Detector are combined to decide if an event is recorded or not. The trigger of the LAT is very flexible in order to allow change of configuration to optimize trigger efficiency and versatile in order to accommodate various signatures of events.

The LAT trigger is organized in two levels. The first level, *Level 1 Trigger* (L1T), is a hardware trigger, based on special combinations of signals at the level of a single tower. The workhorse $\gamma$-ray trigger is the so called *three in a row*, consisting into 3 XY consecutive tracker planes sending a trigger request. There are also two different calorimeter based trigger primitives with different adjustable thresholds (nominally set at 100 MeV and 1 GeV of energy deposition in a crystal log). The ACD adds two other trigger signals: a veto signal and CNO signal. The latter has a threshold of several MIPs and is used to identify cosmic ions with $Z > 2$ for CAL on orbit calibration. In addition there are three other trigger sources: the *Periodic* trigger to sample detector noise and pedestals at regular interval, the *Solicited* trigger for special software trigger request and the External trigger for ground testing. An electronic module (the TEM) combines these signal in a 600 ns coincidence window and then "decides" if the event must be recorded and how to read out the detector: using or not the zero suppression, read all the four calorimeter ranges or just the "best" one, pre-scale this kind of events. The correspondence between trigger primitives coincidence and readout mode is configurable with a look-up table that allows up to 16 combinations (*engines*).

With exception of specific calibration event the typical read out time is about 26 $\mu$s allowing to trigger on almost all the particles that cross the LAT (whose rate is estimated of the order of few kHz). The fact that cosmic rays can be included in the trigger actually constitutes a sort of revolution with respect to the trigger scheme implemented in EGRET, in which that would not have been possible due to the high instrumental dead time in the spark chambers. In fact the only reason why further levels of data reduction are required onboard is the limited bandwidth of the telemetry.

The second trigger level, *Level 2 Trigger* (L2T), is software, two *Event Processing Unit* (EPUs) work in parallel to process LAT events. Multiple filters in succession are applied to each event, each filter optimized to select a different class of event (i.e. $\gamma$ rays or heavy ions for calibration). Within each filter events are accepted or rejected based on a sequence of test each one with tunable parameters. Together with the *Gamma Filter*, designed to accept $\gamma$ rays with high efficiency during normal operation, other filters are implemented to identify MIPs and heavy ions for instrument calibration.

## A.2 Overview of the Gamma-ray Burst Monitor

The *Gamma-ray Burst Monitor* (GBM), the successor of BATSE onboard the CGRO, is designed to detect transient objects, such as GRBs. The development of the GBM and the analysis of its observational data is a collaborative effort between the National Space Science and Technology Center in the U.S. and the Max Planck Institute for Extraterrestrial Physics (MPE) in Germany. The GBM consists of 12 detectors made of Sodium Iodide (NaI) for catching X rays and low energy $\gamma$ rays, and other two detectors made of Bismuth Germanate (BGO) for high-energy $\gamma$ rays placed as shown in Figure A.7. The two detectors together detect X rays and $\gamma$ rays in the energy range between 8 keV to 30MeV, overlapping with the lower energy limit of the LAT. Together the NaI and BGO detectors have similar characteristics to the combination of the BATSE large area and spectroscopy detectors but cover a wider energy range and have a smaller collection area.

The detectors do not block any part of the *Large Area Telescope* (LAT) Field-of-View nor interfere with the solar panels. They easily fit between the LAT and the shroud envelope on two sides of the spacecraft. The mounting arrangement is flexible with the two BGO detectors mounted on opposite sides of the spacecraft, and the NaI detectors mounted in 4 banks of 3 detectors in such a way that they sample a wide range of azimuth and elevation angles[1].

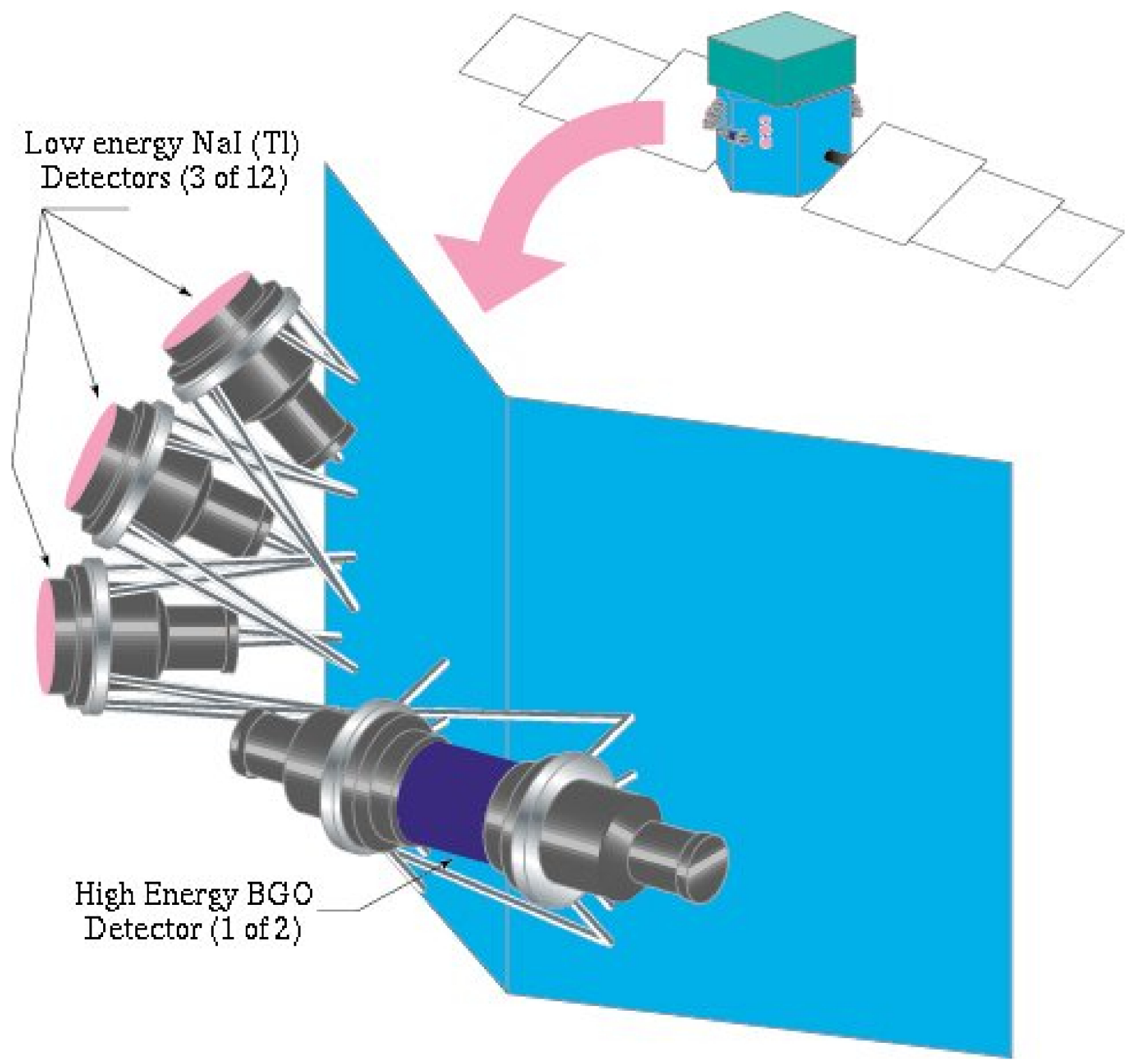

Figure A.7: Schematic view of the GBM in the *Fermi* spacecraft. Credit: *Fermi*-GBM Team

[1] A more specific description of the Gamma-ray Burst Monitor can be found in the following link: `http://fermi.gsfc.nasa.gov/science/instruments/gbm.html`

Appendix B

# Classification algorithms

In this appendix we describe in detail the theory of machine learning algorithms we have implemented and applied to classify $\gamma$-ray unidentified sources.

## B.1 Logistic Regression

Regression methods have become an integral component of any data analysis describing the relationship between a response variable and one or more explanatory variables. It is often the case that the outcome variable is discrete, taking two or more possible values. Over the last decades the logistic regression model has become, in many field, the standard method of analysis in this situation.

Logistic regression (LR) is part of a class of generalized linear models and is one of the simplest machine learning techniques. LR forms a multivariate relation between a dependent variable that can only take values from 0 to 1 and several independent variables. When the dependent variable has only two possible assignment categories, the simplicity of the LR method can be a benefit over other discriminant analyses.

Consider a collection of $p$ independent variables denoted by the vector $\mathbf{x} = [x_1, x_2, ...x_p]$. Let the conditional probability that the outcome is present be denoted by $P(Y = 1|\mathbf{x}) = \pi(\mathbf{x})$. The conditional probability can be defined by a multivariate logistic regression model:

$$\pi(\mathbf{x}) = \frac{e^{\beta_0+\beta_1 x_1+...+\beta_p x_p}}{1 + e^{\beta_0+\beta_1 x_1+...+\beta_p x_p}} \tag{B.1}$$

We define the *logit transformation* of $\pi(\mathbf{x})$ as:

$$z(\mathbf{x}) = \log\left[\frac{\pi(\mathbf{x})}{1 - \pi(\mathbf{x})}\right] = \beta_0 + \beta_1 x_1 + ... + \beta_p x_p \tag{B.2}$$

which shows that logistic regression is a standard linear regression model, once the dichotomous outcome is transformed by the logistic regression. The importance of this transformation is that $z(\mathbf{x})$ has many of the desirable properties of a linear regression model. The logit, $z(\mathbf{x})$, is linear in its parameters, may be continuous, and may range from $-\infty$ to $+\infty$, depending on the range of $\mathbf{x}$.

The above equations are for mean probabilities, and each data point will have an error term ($\epsilon$). We may express the value of the outcome variable given $\mathbf{x}$ as $y = \pi(\mathbf{x}) + \epsilon$. Since the dichotomous outcome variable, $\epsilon$ may assume one of two possible values. If $y = 1$ then $\epsilon = 1 - \pi(\mathbf{x})$ with probability $\pi(\mathbf{x})$, and if $y = 0$ then $\epsilon = -\pi(\mathbf{x})$ with probability $1 - \pi(\mathbf{x})$. Thus, $\epsilon$ has a distribution with mean zero and variance equal to $\pi(\mathbf{x})[1 - \pi(\mathbf{x})]$. That is, the conditional distribution of the outcome variable follows a binomial distribution with probability given by the conditional mean, $\pi(\mathbf{x})$.

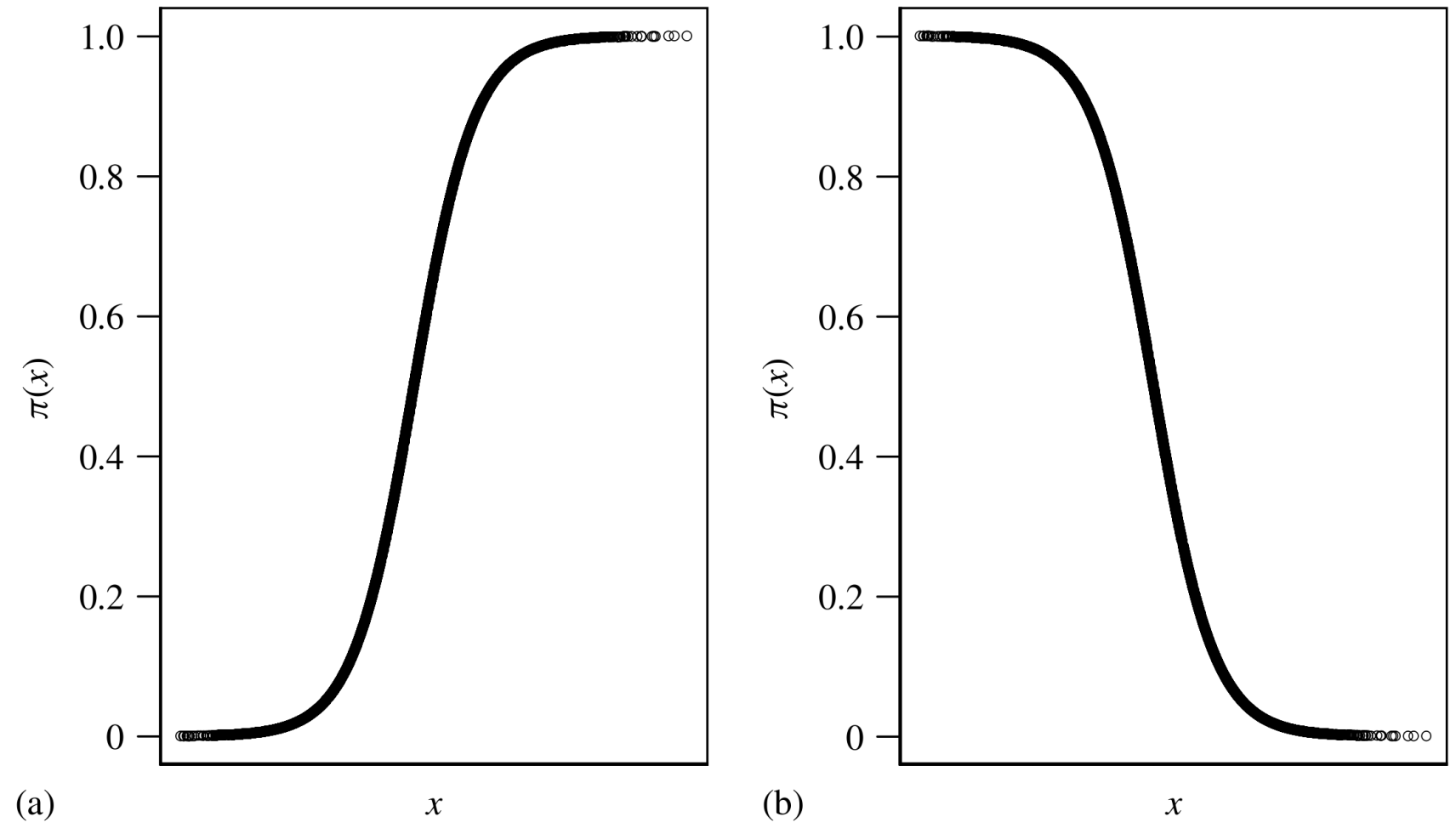

Figure B.1: Logistic regression function.

### B.1.1 Fitting the logistic regression model

Assume that we have a sample of $n$ independent observations $(\mathbf{x}_i, y_i)$, $i = 1, 2, ..., n$. where $y_i$ denotes the value of a dichotomous outcome variable and $x_i$ is the value of the independent variable for the $i^{th}$ observation. Furthermore, assume that the outcome variables has been coded as 0 or 1, representing the absence or presence of the characteristic, respectively. To fit the logistic regression model in Equation B.1 to a set of data requires that we estimate the values of $\boldsymbol{\beta} = (\beta_0, \beta_1, ..., \beta_p)$, the unknown parameters. The method of estimation used for the logistic regression model is the *maximum likelihood*, that yields values for the unknown parameters which maximize the probability of obtaining the observed set of data. In order to apply this method we must first construct the likelihood function, which expresses the probability of the observed data as a function of the unknown parameters. The maximum likelihood estimators of these parameters are chosen to be those values that maximize this function. Thus, the resulting estimators are those which agree most closely with the observed data. We now describe how to find these values from the logistic regression model.

It is easy to show that, if the observations are assumed to be independent, the likelihood function for the logistic regression model can be expressed as:

$$\mathcal{L}(\boldsymbol{\beta}) = \prod_{i=1}^{n} \pi_i(\mathbf{x}_i)^{y_i} \left[1 - \pi_i(\mathbf{x}_i)\right]^{1-y_i} \tag{B.3}$$

where the first term represents the probability that the outcome is equal to 1 for a given $\mathbf{x}$ and the second the probability that the outcome is equal to 0 for a given $\mathbf{x}$ [103].

The principle of maximum likelihood states that we use as our estimate of $\boldsymbol{\beta}$ the value which maximizes the expression in Equation B.3. However, it is easier mathematically to work with the log of Equation B.3. This expression, the *log likelihood*, is defined as:

$$\log[\mathcal{L}(\boldsymbol{\beta})] = \sum_{i=1}^{n} \left\{ y_i \log\left[\pi_i(\mathbf{x}_i)\right] + (1 - y_i) \log\left[1 - \pi_i(\mathbf{x}_i)\right] \right\} \tag{B.4}$$

To find the value of $\boldsymbol{\beta}$ that maximizes B.4 we must differentiate B.4 with respect $\beta_0, ..., \beta_p$ and set the resulting expressions equal to zero. These equations, called *likelihood equations*, are:

$$\sum_{i=1}^{n} \left[y_i - \pi(\mathbf{x}_i)\right] = 0 \sum_{i=1}^{n} x_{ij} \left[y_i - \pi(\mathbf{x}_i)\right] = 0 \tag{B.5}$$

for $j = 1, 2, ..., p$. The system of equation B.5 contains $p + 1$ likelihood equations non linear in $\beta_0, ..., \beta_p$ and thus requires special methods for their solution. These methods are iterative in nature and have been programmed into available logistic regression software. For example, we use the tool *glm*[1] (Generalized Linear Model), implemented in free software programming language **R**[2], a software environment for statistical computing and graphics, which fit generalized linear models as logistic regression. The value of $\boldsymbol{\beta}$ given by the solution to Equation B.5 is called the maximum likelihood estimate and will be denoted as $\hat{\boldsymbol{\beta}}$. In general, the use of the symbol $\hat{}$ denotes the maximum likelihood estimate of the respective quantity. For example, $\hat{\pi}(x_i)$ is the maximum likelihood estimate of $\pi(x_i)$.

[1] `http://stat.ethz.ch/R-manual/R-patched/library/stats/html/glm.html`
[2] http://www.r-project.org

The method of estimating the variances of the estimated coefficients follows from theory of maximum likelihood estimation [173]. This theory states that the estimators are obtained from the matrix of second partial derivate of the log likelihood function. These partial derivate have the following general form:

$$\frac{\partial^2 \log(\mathcal{L}(\beta))}{\partial \beta_j^2} = -\sum_{i=1}^{n} x_{ij}^2 \pi_i (1 - \pi_i) \tag{B.6}$$

and:

$$\frac{\partial^2 \log(\mathcal{L}(\beta))}{\partial \beta_j \partial \beta_l} = -\sum_{i=1}^{n} x_{ij} x_{jl} \pi_i (1 - \pi_i) \tag{B.7}$$

for $j$, $l = 0, 1, 2, ..., p$ where $\pi_i$ denotes $\pi(\mathbf{x}_i)$. Let the $(p+1) \times (p+1)$ matrix containing the negative of the terms given in equations B.6 and B.7 be denoted as $\mathbf{I}(\boldsymbol{\beta})$. The matrix is called the *observed information matrix.* The variances of the estimated coefficients are obtained from the inverse of this matrix which is denoted as $Var(\boldsymbol{\beta}) = \mathbf{I}^{-1}(\boldsymbol{\beta})$. The estimated standard errors of the estimated coefficients are denoted as:

$$\widehat{SE}\left(\hat{\beta}_j\right) = \left[\widehat{Var}\left(\hat{\beta}_j\right)\right]^{1/2} \tag{B.8}$$

for $j = 0, 1, 2, ..., p$.

### B.1.2 Testing for the significance of the model

Once we have fit a particular multivariable logistic regression model, the process of assessment of the significance of the variables in the model begins. This usually involves formulation and testing of a statistical hypothesis to determine whether the independent variables in the model are significantly related to the outcome variable.

One approach to testing for the significance of the coefficient of a variable in any model consists in comparing observed values of the response variable to predicted values obtained from models with and without the variable in question. In logistic regression, comparison of observed to predicted values is based on the log likelihood function defined in Equation B.4:

$$D = -2 \log \left[ \frac{\text{likelihood of the fitted model}}{\text{likelihood of the saturated model}} \right] \tag{B.9}$$

where the saturated model is one that contains as many parameters as there are data points. The quantity inside the large brackets in the Equation B.9 is called the *likelihood ratio*. Using minus twice its log is necessary to obtain a quantity whose di distribution is known and can therefore be used for hypothesis testing purpose. Such a test is called the *likelihood ratio test*. Using Equation B.4, Equation B.9 becomes:

$$D = -2\sum_{i=1}^{n}\left[y_i \log\left(\frac{\hat{\pi}_i}{y_i}\right) + (1-y_i)\log\left(\frac{1-\hat{\pi}_i}{1-y_i}\right)\right] \tag{B.10}$$

where $\hat{\pi}_i = \hat{\pi}(x_i)$ It easy to prove that in a setting of data where the values are either 0 or 1 the likelihood of the saturated model is 1.

For the purpose of assessing the significance of an independent variable we compare the the value of $D$ with and without the independent variable in the equation. The change in D due to the inclusion of the independent variable in the model is obtained as:

$$G = D(\text{model without the variable}) - D(\text{model with the variable}) \tag{B.11}$$

Because the likelihood of the saturated model is common on both values of $D$, it can be expressed as:

$$G = -2\log\left[\frac{\text{likelihood without the variable}}{\text{likelihood with variable}}\right] \tag{B.12}$$

Thus, it is easy to prove that [103]:

$$G = 2\left\{\sum_{i=1}^{n}[y_i\log(\hat{\pi}_i) + (1-y_i)\log(1-\hat{\pi}_i)] - [n_1\log(n_1) + n_0\log(n_0) - n\log(n)]\right\} \tag{B.13}$$

where $n_0 = \sum(1-y_i)$ and $n_1 = \sum y_i$. If under the null hypothesis that the $p$ coefficients in the model are equal to zero and that the size of the sample $n$ is sufficiently large, the distribution of G will be chi-square with $p$ degrees-of-freedom [216]. The p-value associated with this test is $P(\chi^2 > G)$.

## B.2 Artificial Neural Networks

An artificial neural network is a system composed of many simple processing elements operating in parallel whose function is determined by the network structure, connection strengths, and the processing performed at the computing elements or nodes.

An artificial neural network has a natural proclivity for storing experimental knowledge and making it available for use. The knowledge is acquired by the network through a learning process and the interneuron connection strengths – known as synaptic weights – are used to store the knowledge.

There are numerous types of artificial neural networks (ANNs) for addressing many different types of problems, such as modelling memory, performing pattern recognition, and predicting the evolution of dynamical systems. Most networks therefore perform some kind of data modelling.

The two main kinds of learning algorithms are: *supervised* and *unsupervised*. In the former the correct results (target values) are known and given to the ANN during the training so that the ANN can adjust its weights to try to match its outputs to the target values. In the latter, the ANN is not provided with the correct results during training. Unsupervised ANNs usually perform some kind of data compression, such as dimensionality reduction or clustering.

The two main kinds of network topology are *feed-forward* and *feed-back*. In feed-forward ANN, the connections between units do not form cycles and usually produce a relatively quick response to an input. Most feed-forward ANNs can be trained using a wide variety of efficient conventional numerical methods (e.g. conjugate gradients, Levenberg-Marquardt, etc.) in addition to algorithms invented by ANN researchers. In a feed-back or recurrent ANN, there are cycles in the connections. In some feed-back ANNs, each time an input is presented, the ANN must iterate for a potentially long time before producing a response.

### B.2.1 The multilayer perceptron

In the present work we have used one of the most important types of supervised neural networks, the *feed-forward multilayer perceptron* (MLP), in order to understand the nature of *Fermi*-LAT unidentified sources. The term *perceptron* is historical, and refers to the function performed by the nodes. An introduction on Artificial Neural Networks is provided by Sarle (1994) [181], and on multilayer Perceptron by Bailer-Jones et al. (2001) [25] and Sarle (1994) [180]. A comprehensive treatment of feed-forward neural networks is provided by Bishop (1995) [127].

In Figure B.2 the general architecture of a network is shown. The network is made up of nodes (analogous to human neurons) arranged in a series of layers. The nodes in a given layer are fully connected to the nodes in the next layer by links. The input layer consists of the input parameters, and the output layer consists of the classes. Any layer between the input and the output layers is called a “hidden layer”. The complexity (and non-linearity) of the ANN depends on the number of inputs, hidden nodes, layers, outputs and connections.

For each input pattern, the network produces an output pattern through the propagation rule, compares the actual output with the desired one and computes an error. The learning algorithm adjusts the weights of the connections by an appropriate quantity to reduce the error (sliding down the slope).

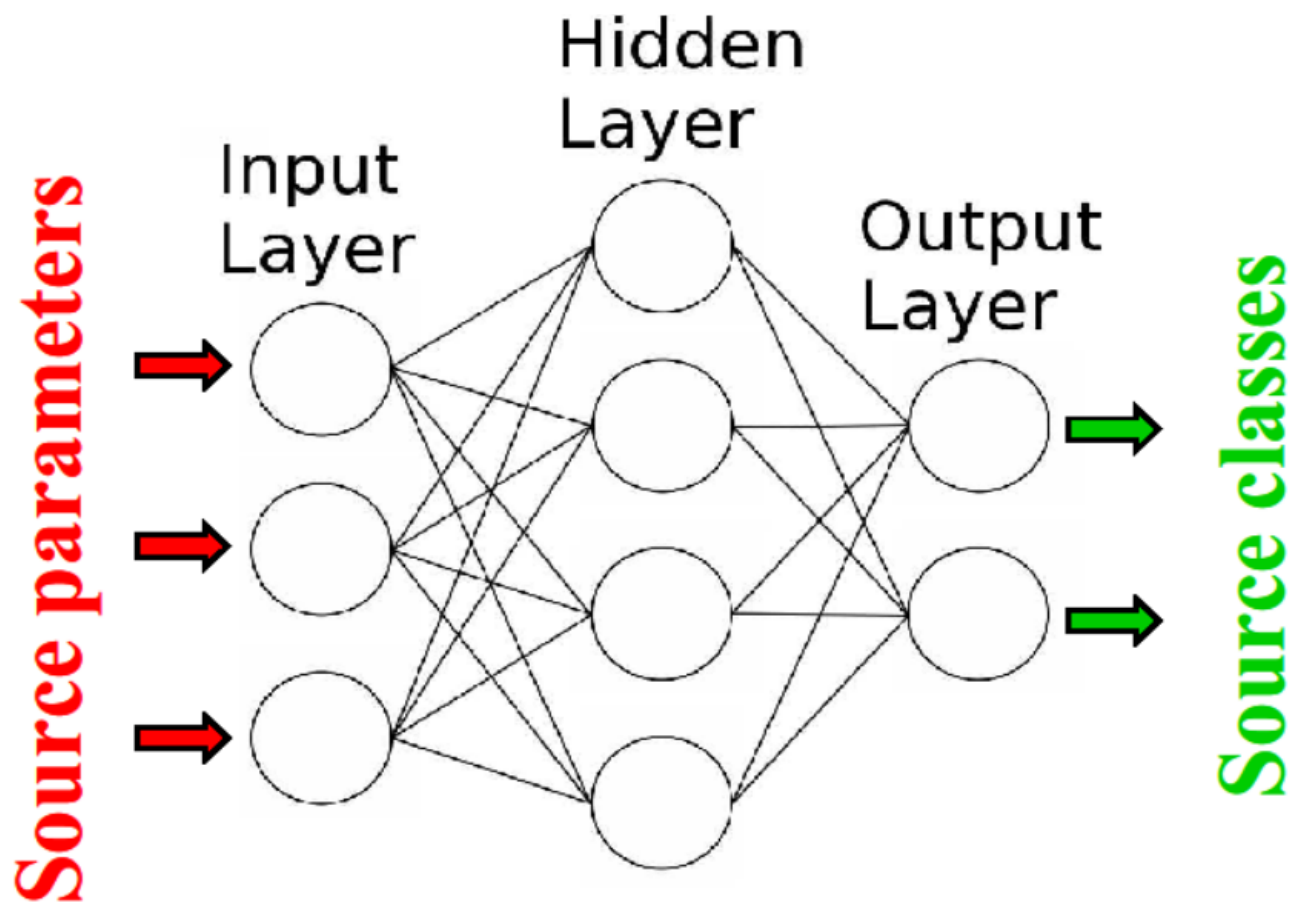

Figure B.2: Schematic view of a Two Layer Perceptron (2LP), the type of Artificial Neural Network we use. The data enter the 2LP through the Nodes in the Input Layer. The information travels from left to right across the Links and is processed in the Nodes. Each Node in the Output Layer returns the probability that a source belongs to a specific class.

This process continues until the error produced by the network is low, according to a given criterion.

### B.2.2 The propagation rule

The network operates as follows. Except for the nodes in the input layer, an input of a node at layer $s$ ($I_j^{(s)}$) is the combination of the output of the previous nodes ($o_i^{(s-1)}$) and the weights of the corresponding links ($w_{ij}^{(s)}$), the combination is linear: $I_j^{(s)} = \sum_i w_{ij}^{(s)} o_i^{(s-1)}$. Each node has a transform function (or activation function), which provides the output of the node as a function of the $I_j^{(s)}$. Nonlinear activation functions are needed to introduce nonlinearity into the network. We have used the *logistic* (or sigmoid) function: $out = 1/[1 + \exp(-I)]$ (in the interval [0, 1]) and the tanh function $out = tanh(I)$ (in the interval [−1, 1]), for all nodes. For the input nodes we decide to use a linear activation function. The propagation rule, from the input layer to the output layer, is a combination of activation functions.

No significant difference has been found in the training process between using the logistic and tanh functions, only that the training process is faster if we use the tanh function.

### B.2.3 Back-propagation of the error

The weights, $w$, are randomly initialized, they are the free parameters of the network and the goal is to minimize the total error function with respect to $w$ (maintaining a good generalization power, see below).

The error function in the weight space defines the multidimensional error surface and the objective is to find the global (or acceptable local) minima on this surface. The solution implemented in the present work is the *gradient descent*, within which the weights are adjusted (from small initial random values) in order to follow the steepest downhill slope. The error surface is not known in advance, so it is necessary to explore it in a suitable way.

The error function typically used is the sum-of-squares error, which for a single input vector, $n$, is:

$$E^{(n)} = \frac{1}{2} \sum_i \left( y_i^{(n)} - t_i^{(n)} \right)^2 \tag{B.14}$$

where $y_i$ is the output of the ANN and $t_i$ is the target output value for the $i$th output node and $n$ runs form 1 to the total number of examples in the training set. In the present work $i = 2$, two output nodes are used to understand the nature of 2FGL unidentified sources. In the gradient descent process the weight vector is adjusted in the negative direction of the gradient vector *backwards* from the output layer to one ore more hidden layers by a small change in each time-step:

$$\Delta \mathbf{w} = -\eta \frac{\partial E}{\partial \mathbf{w}} \tag{B.15}$$

and the new generic weight is:

$$w_{new} = w_{old} + \Delta w \tag{B.16}$$

The amplitude of the step on the error surface is set by the $\eta$-learning parameter: large values of $\eta$ mean large steps. Typically $\eta$ belongs to the interval [0, 1] (where the opening bracket means that the lower value is excluded). In our application a small value has been used (0.2). If $\eta$ is too small the training time becomes very long, while a large value can produce oscillations around a minimum or even lead to miss the optimal minimum in the error surface. The algorithm is stopped when the value of the error function has become sufficiently small.

The learning algorithm used in the present work is the standard *back-propagation*. It refers to the method for computing the gradient of the case-wise error function with respect to the weights for a feed-forward network. "*Standard backprop*" is a definition of the *generalized delta rule*, the training algorithm that remains one of the most widely used supervised training methods for neural network.

This learning algorithm implies that the error function is continuous and derivable, so that it is possible to calculate the gradient. For this reason the activation functions (and their final combination through the propagation rule) must be continuous and derivable. From the computational point of view, the derivative of the activation functions adopted in the present work is easily related to the value of the function out = F(net) itself (see Section B.2.2: F' $\propto$ out(1-out) in the case F = sigmoid or F' $\propto$ (1-out$^2$) if F = tanh).

When the network weights approach a minimum solution, the gradient becomes small and the step size diminishes too, giving origin to a very slow convergence. Adding a momentum (a residual of the previous weight variation) to the equations of the weight update, the minimization improves [127]:

$$w_{new} = w_{old} + \Delta w + \alpha \Delta w_{old} \tag{B.17}$$

where $\alpha$ is the momentum factor (set to 0.9 in our applications). This can reduce the decay in learning updates and cause the learning to proceed through the weight space in a fairly constant direction. Besides a faster convergence to the minimum, this method makes it possible to escape from a local minimum if there is enough momentum to travel through it and over the following hill (see Figure B.3). The generalized delta rule including the *momentum* is called the "*heavy ball method*" in the numerical analysis literature [29].

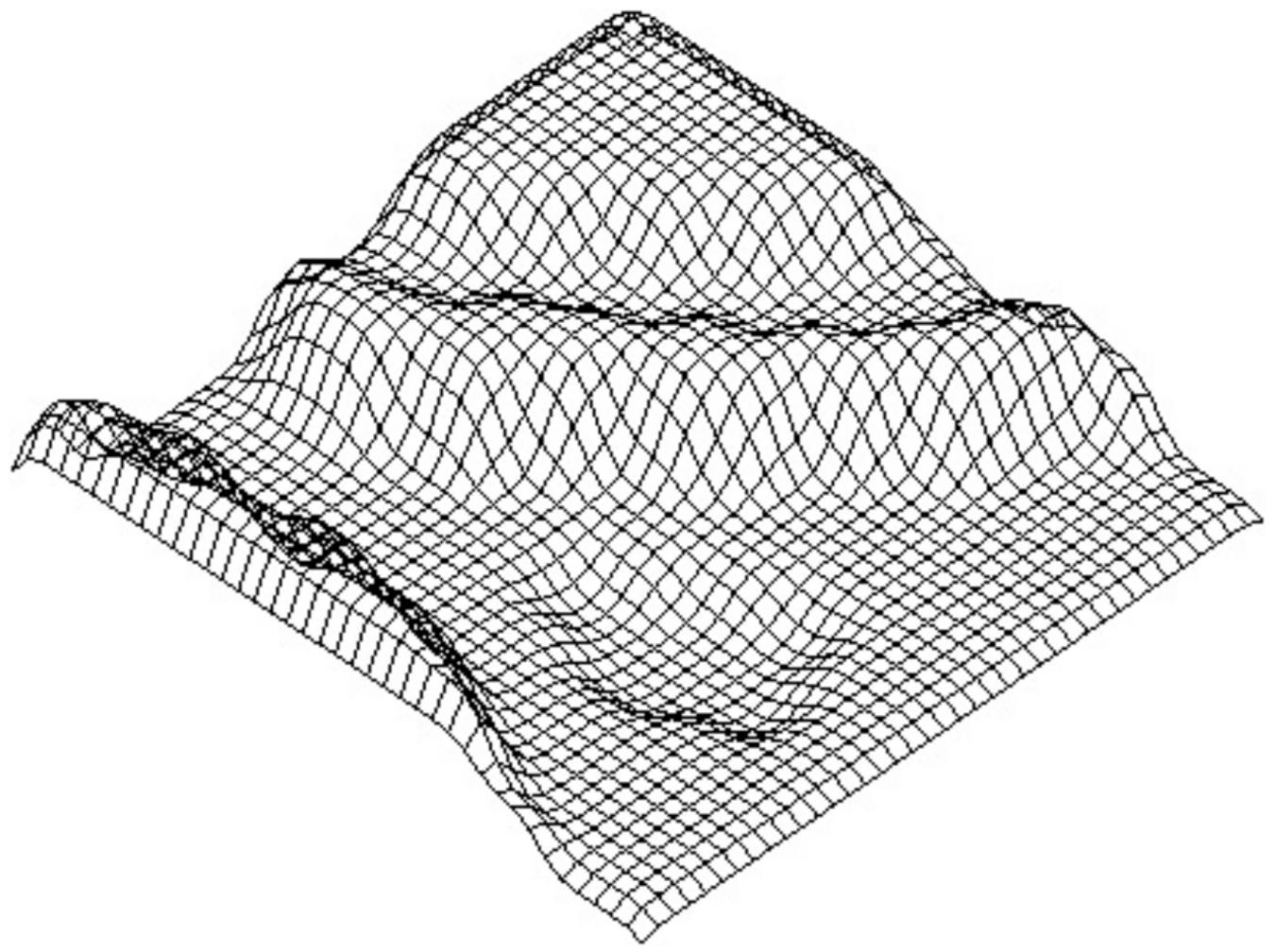

Figure B.3: A simplified representation of the error surface: the behavior of the error as a function of 2 weights.

The learning algorithm has been used in the so called online (or *incremental*) version, in which the weights of the connections are updated after each example is processed by the network. One epoch corresponds to the processing of all examples one time. The other possibility is to compute the training in the so called *batch learning* (or epoch learning), in which the weights are updated only at the end of each epoch (not used in the present application).

### B.2.4 The training technique

During the learning process, the output of a supervised neural network comes to approximate the target values given the inputs in the training set. This ability may be useful in itself, but more often the purpose of using a neural network is to generalize, i.e. to get some output from inputs that are *not* in the training set (*generalization*). ANNs, like other flexible nonlinear estimation methods such as kernel regression and smoothing splines, can suffer from either under fitting or over fitting. A network that is not sufficiently complex[3] can fail to fully detect the signal in a complicated data set, leading to under fitting: an *inflexible* model will have a large *bias*. On the other hand a network that is too complex may fit the noise, not just the signal, leading to over-fitting: a model that is too flexible in relation to the particular data set will produce a large *variance* [182]. The best generalization is obtained when the best compromise between these two conflicting quantities (bias and variance) is reached. There are several approaches to avoid under- and overfitting, and obtain a good generalization. Part of them aim to *regularize* the complexity of the network during the training phase, such as the *Early Stopping* and *weight-decay* methods (the size of the weights are tuned in order to produce a mapping function with small curvature, the large weights are penalized. Reducing the size of the weights reduces also the "effective" number of weights [150]).

#### Generalize error

The most commonly used method for estimating the generalization error in neural networks is to reserve part of the data as a *testing set*, which must not be used in *any* way during the training. After the training, the network is applied to the testing set, and the error on the testing set provides an unbiased estimate of the generalization error, provided that the testing set was chosen in a random way.

In order to avoid (possible) over-fitting during the training, another part of the data can be reserved as a *validation set* (independent both of the training and testing sets, not used for updating the weights), and used during the training to monitor the generalization error. The best epoch corresponds to the lowest validation error, and the training is stopped when the validation error rate "starts to go up" (*early stopping* method). The disadvantage of this technique is that it reduces the amount of data available for both training and validation, which is particularly undesirable if the available data set is small. Moreover, neither the training nor the validation make use of the entire sample.

[3]The complexity of a network is related to both the number of weights and the amplitude of the weights (the mapping produced by a ANN is an interpolation of the training data, a high order fit to data is characterized by large curvature of the mapping function, which in turn corresponds to large weights).

Appendix C

# Hierarchical neural network predictions for 2FGL unidentified sources

In this appendix we show the results obtained by the hierarchical ANN discussed in the Chapter 4 to classify each 2FGL unidentified source as a pulsar or AGN subclass on the basis of its $\gamma$-ray observables. Our hierarchical neural network model is composed by three simple neural networks. The first one (defined as *ANN1* and described in Section 4.3) is constructed to distinguish pulsars from AGNs, the second one (defined as *ANN2*) to distinguish young pulsars from MSPs and the third one (defined as *ANN3*) to distinguish BL Lacertae from FSRQs. In this way each 2FGL unidentified source "enters" the *ANN1*, if the *ANN1* predictor ($P_{ANN1}$) is greater than the threshold $C_{PSR} = 0.9962$ then the unidentified source is classified as a pulsar candidate and "enter" the *ANN2*, which decides if such object is a young pulsar candidate ($P_{ANN2} > 0.984$), a MSP candidate ($P_{ANN2} < 0.085$) or a $\gamma$-ray pulsar of uncertain type ($0.085 < P_{ANN2} < 0.984$). If $P_{ANN1}$ is less than the threshold $C_{AGN} = 3.4 \times 10^{-7}$ then the unidentified source is classified as an AGN candidate and "enter" the *ANN3*, which decides if such object is a BL Lac candidate ($P_{ANN3} > 0.758$), a FSRQ candidate ($P_{ANN3} < 0.429$) or an AGN of uncertain type ($0.429 < P_{ANN3} < 0.758$). If $3.4 \times 10^{-7} < P_{ANN1} < 0.9962$ the source remains unclassified. In Figure C.1 a schematic view of our optimized hierarchical ANN is shown.

Applying the optimized hierarchical neural network described in the Chapter 4 to the 576 2FGL unidentified sources we find that 75 (13%) are classified as young pulsar candidates, 34 (6%) as MSP candidates, 26 (4.5%) as pulsars of uncertain type, 84 (14.5%) as BL Lac candidates, 66 (11.5%) as FSRQ can-didates and 26 (4.5%) as blazars of uncertain type, while 269 (46%) remain unclassified. The number of sources unclassified and of uncertain type depends on the definition of the classification thresholds for each ANN.

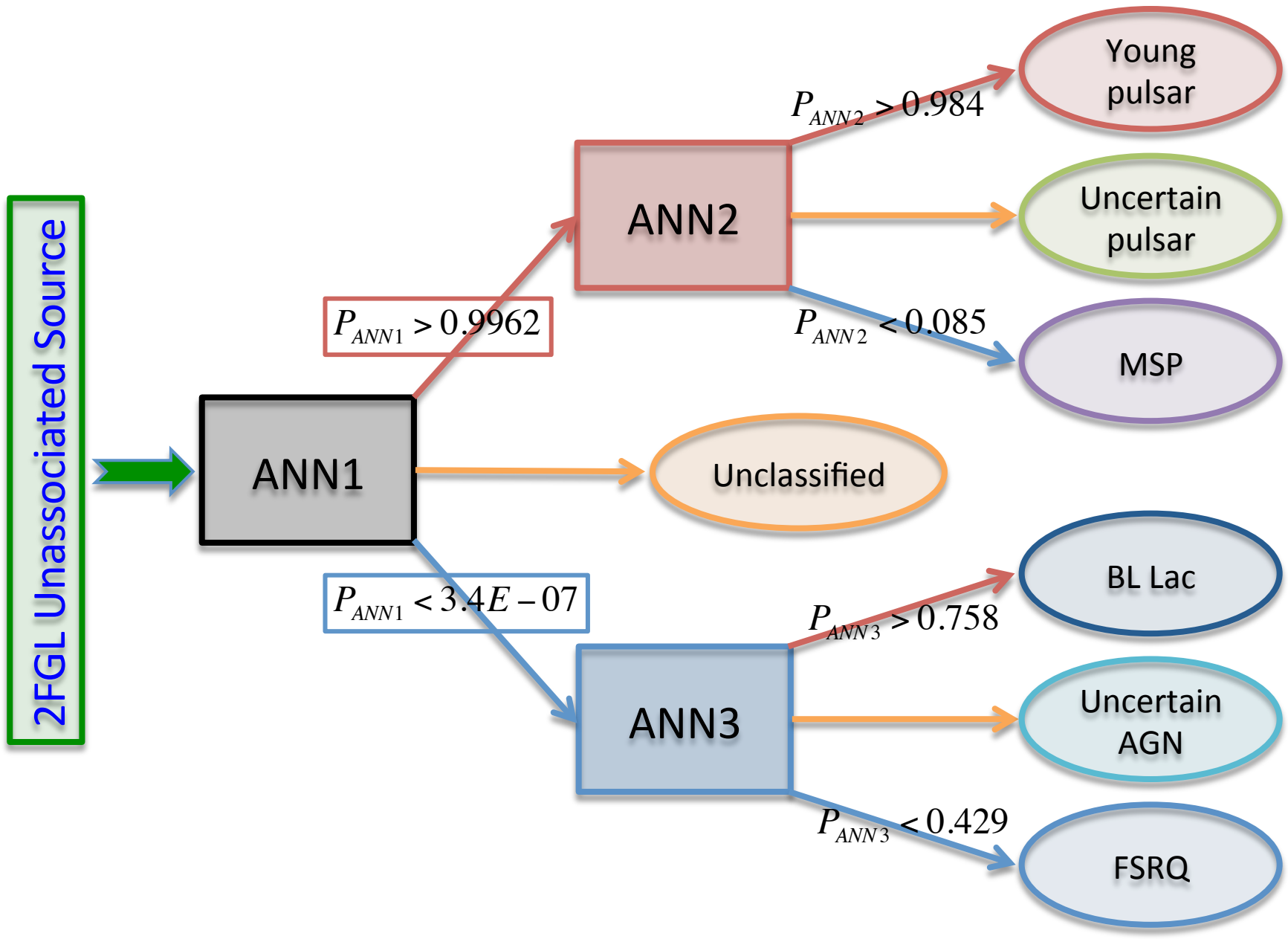

Figure C.1: Schematic view of our hierarchical ANN optimized to classify each 2FGL unidentified source as a specific pulsar or AGN subclass as described in the text.

In the Table C.1 we show the complete list of predictions and voting percentages for unidentified *Fermi*-LAT sources in the 2FGL catalog obtained using the hierarchical ANN technique described in the Chapter 4. Column (1): 2FGL unidentified source name. Col. (2) and (3): Galactic latitude and longitude. Col. (4): source significance in $\sigma$ units. Col. (5): energy flux at energies above 100 MeV. Col. (6): results of the ANN1 network described in the Section 4.3 and optimized to distinguish pulsars from AGNs. Col. (7): results of the ANN2 network described in the Section 4.5 and optimized to distinguish young pulsars from MSPs. Col. (8): results of the ANN3 network described in the Section 4.6 and optimized to distinguish BL Lacertae from FSRQs. Col. (9): hierarchical ANN predictions for 2FGL unidentified sources.

| Unidentified source | b (°) | l (°) | Source sign. ($\sigma$) | Energy flux (erg cm$^{-2}$ s$^{-1}$) | P$_{ANN1}$ | P$_{ANN2}$ | P$_{ANN3}$ | Prediction |
|---|---|---|---|---|---|---|---|---|
| 2FGL J0002.7+6220 | 0.00 | 117.31 | 13.7 | 2.517e-11 | 1 | 0.02347 | – | MSP |
| 2FGL J0004.2+2208 | -39.43 | 108.73 | 5.4 | 6.337e-12 | 1.388e-06 | – | – | Unclassified |
| 2FGL J0007.7+6825c | 5.89 | 118.91 | 6.2 | 1.749e-11 | 2.475e-07 | – | 0.2273 | FSRQ |
| 2FGL J0014.3–0509 | -66.31 | 99.37 | 4.2 | 7.103e-12 | 3.669e-11 | – | 0.3145 | FSRQ |
| 2FGL J0031.0+0724 | -55.11 | 114.10 | 4.4 | 4.035e-12 | 4.804e-11 | – | 0.9546 | BL Lac |
| 2FGL J0032.7–5521 | -61.57 | 308.52 | 17.6 | 1.816e-11 | 5.74e-14 | – | 0.1409 | FSRQ |
| 2FGL J0038.8+6259 | 0.16 | 121.51 | 5.6 | 1.435e-11 | 0.1789 | – | – | Unclassified |
| 2FGL J0039.1+4331 | -19.29 | 120.57 | 4.8 | 4.707e-12 | 4.261e-05 | – | – | Unclassified |
| 2FGL J0048.8–6347 | -53.33 | 303.41 | 4.1 | 3.905e-12 | 3.667e-08 | – | 0.8694 | BL Lac |
| 2FGL J0102.2+0943 | -53.05 | 127.35 | 5.5 | 6.043e-12 | 3.105e-09 | – | 0.9598 | BL Lac |
| 2FGL J0103.8+1324 | -49.36 | 127.56 | 4.9 | 4.865e-12 | 1.331e-05 | – | – | Unclassified |
| 2FGL J0106.5+4854 | -13.88 | 125.49 | 17.3 | 2.063e-11 | 0.999 | 0.008971 | – | MSP |
| 2FGL J0116.6–6153 | -54.99 | 297.75 | 5.5 | 4.522e-12 | 1.079e-10 | – | 0.982 | BL Lac |
| 2FGL J0124.6–2322 | -81.61 | 188.14 | 7.5 | 8.057e-12 | 1.661e-11 | – | 0.7824 | BL Lac |
| 2FGL J0129.4+2618 | -35.78 | 133.45 | 4.9 | 7.465e-12 | 0.01346 | – | – | Unclassified |
| 2FGL J0133.4–4408 | -71.00 | 279.23 | 5.6 | 4.97e-12 | 3.881e-09 | – | 0.8425 | BL Lac |
| 2FGL J0143.6–5844 | -57.10 | 290.47 | 14.2 | 1.613e-11 | 6.294e-10 | – | 0.9832 | BL Lac |
| 2FGL J0158.4+0107 | -57.47 | 155.33 | 4.8 | 5.795e-12 | 4.037e-12 | – | 0.4465 | Uncertain blazar |
| 2FGL J0158.6+8558 | 23.26 | 124.20 | 4.2 | 5.528e-12 | 4.217e-10 | – | 0.09168 | FSRQ |
| 2FGL J0200.4–4105 | -70.10 | 261.91 | 4.4 | 4.061e-12 | 9.257e-11 | – | 0.978 | BL Lac |
| 2FGL J0212.1+5318 | -7.67 | 134.94 | 15.1 | 1.495e-11 | 0.9999 | 0.001033 | – | MSP |
| 2FGL J0221.2+2516 | -33.31 | 147.35 | 4.4 | 5.118e-12 | 4.76e-05 | – | – | Unclassified |
| 2FGL J0221.3+6025c | -0.53 | 133.81 | 4.6 | 1.726e-11 | 0.0009095 | – | – | Unclassified |
| 2FGL J0222.7+6820 | 6.96 | 131.23 | 5.1 | 8.449e-12 | 3.772e-07 | – | – | Unclassified |
| 2FGL J0224.0+6204 | 1.13 | 133.55 | 14.4 | 5.923e-11 | 1 | 0.984 | – | Young pulsar |
| 2FGL J0225.9+6154c | 1.05 | 133.82 | 4.9 | 2.106e-11 | 8.163e-06 | – | – | Unclassified |
| 2FGL J0226.1+0943 | -46.58 | 158.10 | 6.5 | 7.617e-12 | 1.996e-11 | – | 0.7935 | BL Lac |
| 2FGL J0227.2+6029c | -0.22 | 134.47 | 6.1 | 2.078e-11 | 2.081e-05 | – | – | Unclassified |
| 2FGL J0227.7+2249 | -34.89 | 150.23 | 6.5 | 8.23e-12 | 4.976e-07 | – | – | Unclassified |
| 2FGL J0233.9+6238c | 2.07 | 134.40 | 8.9 | 2.794e-11 | 7.377e-12 | – | 0.2426 | FSRQ |
| 2FGL J0237.9+5238 | -6.91 | 138.84 | 7.5 | 1.246e-11 | 8.239e-05 | – | – | Unclassified |
| 2FGL J0239.5+1324 | -41.71 | 159.24 | 6.1 | 8.061e-12 | 3.797e-10 | – | 0.9427 | BL Lac |
| 2FGL J0248.5+5131 | -7.24 | 140.79 | 6.5 | 1.017e-11 | 6.313e-05 | – | – | Unclassified |
| 2FGL J0251.0+2557 | -29.60 | 153.96 | 4.5 | 6.476e-12 | 1.172e-06 | – | – | Unclassified |
| 2FGL J0253.9+5908 | -0.09 | 138.08 | 5.0 | 1.485e-11 | 7.6e-07 | – | – | Unclassified |
| 2FGL J0305.0–1602 | -57.15 | 200.15 | 5.3 | 5.579e-12 | 5.438e-07 | – | – | Unclassified |
| 2FGL J0307.4+4915 | -7.82 | 144.55 | 9.4 | 1.583e-11 | 2.137e-10 | – | 0.9653 | BL Lac |
| 2FGL J0308.7+5954 | 1.50 | 139.36 | 4.4 | 1.196e-11 | 0.003014 | – | – | Unclassified |
| 2FGL J0312.5–0914 | -52.22 | 191.45 | 5.8 | 7.568e-12 | 1.358e-06 | – | – | Unclassified |
| 2FGL J0312.8+2013 | -31.57 | 162.51 | 4.4 | 5.632e-12 | 1.11e-10 | – | 0.982 | BL Lac |
| 2FGL J0316.1–6434 | -46.08 | 281.50 | 7.9 | 8.129e-12 | 1.793e-09 | – | 0.9716 | BL Lac |
| 2FGL J0318.0+0255 | -43.62 | 178.36 | 6.8 | 9.216e-12 | 0.04173 | – | – | Unclassified |
| 2FGL J0330.3+5816c | 1.60 | 142.58 | 6.7 | 1.568e-11 | 0.9999 | 0.8569 | – | Uncertain PSR |
| 2FGL J0332.1+6309 | 5.73 | 139.96 | 5.5 | 1.041e-11 | 3.411e-09 | – | 0.9403 | BL Lac |
| 2FGL J0336.0+7504 | 15.55 | 133.06 | 9.2 | 1.202e-11 | 3.083e-05 | – | – | Unclassified |
| 2FGL J0338.2+1306 | -32.93 | 173.47 | 5.8 | 9.635e-12 | 1.348e-06 | – | – | Unclassified |
| 2FGL J0340.5+5307 | -1.73 | 146.75 | 16.3 | 4.006e-11 | 1 | 1 | – | Young pulsar |
| 2FGL J0340.7–2421 | -51.95 | 218.51 | 4.2 | 4.692e-12 | 2.393e-10 | – | 0.9044 | BL Lac |
| 2FGL J0341.8+3148c | -18.44 | 160.27 | 7.6 | 1.955e-11 | 4.219e-07 | – | – | Unclassified |
| 2FGL J0345.2–2356 | -50.85 | 218.27 | 9.6 | 1.143e-11 | 2.542e-08 | – | 0.07312 | FSRQ |
| 2FGL J0353.2+5653 | 2.35 | 145.87 | 4.6 | 9.393e-12 | 3.265e-09 | – | 0.8879 | BL Lac |
| 2FGL J0359.5+5410 | 0.84 | 148.29 | 13.5 | 2.581e-11 | 1 | 0.07725 | – | MSP |
| 2FGL J0404.0+3843 | -10.25 | 159.14 | 4.5 | 1.067e-11 | 2.503e-12 | – | 0.2191 | FSRQ |
| 2FGL J0404.6+5822 | 4.46 | 146.07 | 4.5 | 1.064e-11 | 6.167e-08 | – | 0.4154 | FSRQ |
| 2FGL J0409.5+0509 | -32.30 | 186.61 | 5.4 | 9.455e-12 | 1.505e-10 | – | 0.3329 | FSRQ |
| 2FGL J0409.8–0357 | -37.37 | 195.87 | 7.6 | 8.89e-12 | 1.978e-12 | – | 0.8433 | BL Lac |
| 2FGL J0414.9–0855 | -38.72 | 202.13 | 4.1 | 5.807e-12 | 5.33e-06 | – | – | Unclassified |
| 2FGL J0415.2+5518 | 3.19 | 149.21 | 5.4 | 1.317e-11 | 5.171e-08 | – | 0.7426 | Uncertain blazar |
| 2FGL J0416.0–4355 | -45.92 | 249.20 | 5.4 | 5.59e-12 | 4.285e-08 | – | 0.1074 | FSRQ |
| 2FGL J0418.9+6636 | 11.56 | 141.53 | 9.1 | 1.568e-11 | 3.388e-06 | – | – | Unclassified |
| 2FGL J0420.9–3743 | -45.14 | 240.30 | 5.8 | 7.669e-12 | 1.383e-12 | – | 0.5557 | Uncertain blazar |
| 2FGL J0423.4+5612 | 4.65 | 149.40 | 4.5 | 8.376e-12 | 1.24e-07 | – | 0.9827 | BL Lac |
| 2FGL J0426.7+5434 | 3.85 | 150.90 | 17.0 | 3.465e-11 | 1 | 0.9896 | – | Young pulsar |
| 2FGL J0427.2–6705 | -38.59 | 279.19 | 6.2 | 6.829e-12 | 1.092e-07 | – | 0.09804 | FSRQ |
| 2FGL J0428.0–3845 | -43.78 | 241.84 | 6.6 | 2.153e-11 | 2.643e-08 | – | 0.03631 | FSRQ |
| 2FGL J0430.2+3508c | -9.09 | 165.45 | 7.0 | 1.676e-11 | 0.2968 | – | – | Unclassified |
| 2FGL J0431.5+3622 | -8.06 | 164.72 | 4.5 | 9.727e-12 | 2.187e-07 | – | 0.8464 | BL Lac |
| 2FGL J0438.0–7331 | -35.17 | 286.09 | 6.1 | 7.417e-12 | 1.116e-07 | – | 0.988 | BL Lac |
| 2FGL J0439.8–1858 | -37.25 | 216.92 | 8.3 | 8.012e-12 | 6.121e-09 | – | 0.9832 | BL Lac |
| 2FGL J0440.5+2554c | -13.50 | 174.04 | 4.9 | 1.36e-11 | 0.9725 | – | – | Unclassified |
| 2FGL J0458.4+0654 | -21.31 | 192.76 | 5.5 | 8.393e-12 | 5.269e-13 | – | 0.1577 | FSRQ |
| 2FGL J0515.0–4411 | -35.34 | 249.59 | 5.4 | 7.132e-12 | 5.352e-08 | – | 0.03454 | FSRQ |
| 2FGL J0516.7+2634 | -6.63 | 178.49 | 8.1 | 2.019e-11 | 3.478e-07 | – | – | Unclassified |
| 2FGL J0523.3–2530 | -29.84 | 228.23 | 18.5 | 2.181e-11 | 0.4524 | – | – | Unclassified |
| 2FGL J0524.1+2843 | -4.08 | 177.64 | 6.1 | 1.216e-11 | 1.125e-08 | – | 0.8677 | BL Lac |
| 2FGL J0526.6+2248 | -6.89 | 182.89 | 7.1 | 2.625e-11 | 9.827e-08 | – | 0.08979 | FSRQ |
| 2FGL J0529.3+3821 | 2.20 | 170.24 | 4.6 | 1.117e-11 | 0.0001463 | – | – | Unclassified |
| 2FGL J0533.9+6759 | 18.19 | 144.77 | 11.1 | 8.442e-12 | 0.06242 | – | – | Unclassified |
| 2FGL J0534.8–0548c | -19.66 | 209.36 | 6.0 | 1.877e-11 | 0.005902 | – | – | Unclassified |
| 2FGL J0534.9–0450c | -19.22 | 208.45 | 4.9 | 1.287e-11 | 0.8498 | – | – | Unclassified |
| 2FGL J0538.5–0534c | -18.73 | 209.58 | 9.3 | 1.984e-11 | 1 | 0.9999 | – | Young pulsar |
| 2FGL J0539.3–0323 | -17.58 | 207.63 | 5.0 | 1.07e-11 | 0.002004 | – | – | Unclassified |
| 2FGL J0540.1–7554 | -30.57 | 287.31 | 4.1 | 5.164e-12 | 0.0006104 | – | – | Unclassified |
| 2FGL J0540.3+3549c | 2.67 | 173.56 | 11.4 | 2.469e-11 | 0.9999 | 0.1958 | – | Uncertain PSR |
| 2FGL J0541.8–0203c | -16.41 | 206.69 | 9.5 | 1.9e-11 | 0.2964 | – | – | Unclassified |
| 2FGL J0543.2–0120c | -15.75 | 206.20 | 5.5 | 1.553e-11 | 0.0002222 | – | – | Unclassified |
| 2FGL J0545.6+6018 | 15.74 | 152.50 | 7.9 | 9.807e-12 | 2.576e-05 | – | – | Unclassified |
| 2FGL J0547.1+0020c | -14.10 | 205.15 | 11.3 | 2.295e-11 | 1 | 0.8335 | – | Uncertain PSR |
| 2FGL J0547.4+3722 | 4.71 | 172.97 | 4.4 | 1.01e-11 | 0.004074 | – | – | Unclassified |
| 2FGL J0547.5–0141c | -14.97 | 207.03 | 6.1 | 1.311e-11 | 0.7088 | – | – | Unclassified |
| 2FGL J0555.9–4348 | -28.02 | 250.41 | 5.0 | 6.338e-12 | 2.467e-10 | – | 0.9205 | BL Lac |
| 2FGL J0600.8–1949 | -19.75 | 225.70 | 5.1 | 6.986e-12 | 6.004e-07 | – | – | Unclassified |
| 2FGL J0600.9+3839 | 7.64 | 173.19 | 5.1 | 7.354e-12 | 1.44e-09 | – | 0.9549 | BL Lac |
| 2FGL J0602.7–4011 | -25.96 | 246.78 | 9.6 | 1.242e-11 | 2.529e-07 | – | 0.9345 | BL Lac |
| 2FGL J0605.3+3758 | 8.08 | 174.20 | 7.6 | 7.14e-12 | 0.5748 | – | – | Unclassified |
| 2FGL J0607.5–0618c | -12.62 | 213.60 | 7.6 | 1.294e-11 | 3.156e-08 | – | 0.7015 | Uncertain blazar |
| 2FGL J0608.3+2037 | 0.28 | 189.77 | 12.5 | 2.89e-11 | 1 | 0.848 | – | Uncertain PSR |
| 2FGL J0616.6+2425 | 3.77 | 187.35 | 4.1 | 1.091e-11 | 0.0001144 | – | – | Unclassified |
| 2FGL J0620.8–2556 | -17.81 | 233.52 | 6.4 | 8.279e-12 | 4.644e-10 | – | 0.8141 | BL Lac |
| 2FGL J0621.2+2508 | 5.04 | 187.21 | 5.4 | 1.113e-11 | 0.6038 | – | – | Unclassified |
| 2FGL J0631.7+0428 | -2.32 | 206.70 | 10.2 | 2.69e-11 | 1 | 0.3635 | – | Uncertain PSR |
| 2FGL J0634.3+0356c | -1.99 | 207.47 | 5.4 | 2.241e-11 | 0.2787 | – | – | Unclassified |
| 2FGL J0637.0+0416c | -1.24 | 207.49 | 7.8 | 2.027e-11 | 0.9996 | 0.169 | – | Uncertain PSR |
| 2FGL J0641.1+1006c | 2.34 | 202.77 | 9.9 | 2.377e-11 | 0.6694 | – | – | Unclassified |
| 2FGL J0641.2+7315 | 25.17 | 141.48 | 7.8 | 9.077e-12 | 2.672e-12 | – | 0.4409 | Uncertain blazar |
| 2FGL J0642.9+0319 | -0.37 | 209.00 | 7.7 | 1.777e-11 | 0.6421 | – | – | Unclassified |
| 2FGL J0644.6+6034 | 22.60 | 155.07 | 9.3 | 9.962e-12 | 5.602e-12 | – | 0.9131 | BL Lac |
| 2FGL J0647.7+0032 | -0.58 | 212.03 | 9.8 | 1.923e-11 | 0.03055 | – | – | Unclassified |
| 2FGL J0658.4+0633 | 4.54 | 207.88 | 4.4 | 8.038e-12 | 5.462e-05 | – | – | Unclassified |
| 2FGL J0705.3–1043c | -1.81 | 224.05 | 4.1 | 1.563e-11 | 0.01402 | – | – | Unclassified |
| 2FGL J0708.5–1020c | -0.96 | 224.08 | 6.2 | 1.555e-11 | 0.1043 | – | – | Unclassified |
| 2FGL J0713.5–0952 | 0.36 | 224.23 | 7.1 | 1.774e-11 | 0.1556 | – | – | Unclassified |
| 2FGL J0719.2–5000 | -16.27 | 261.29 | 6.3 | 8.629e-12 | 1.295e-07 | – | 0.476 | Uncertain blazar |
| 2FGL J0723.9+2901 | 19.36 | 189.48 | 8.0 | 9.57e-12 | 1.56e-13 | – | 0.152 | FSRQ |
| 2FGL J0725.8–0549 | 4.93 | 222.06 | 4.1 | 5.299e-12 | 9.029e-10 | – | 0.9792 | BL Lac |
| 2FGL J0737.1–3235 | -5.56 | 246.86 | 6.9 | 1.662e-11 | 1.379e-06 | – | – | Unclassified |
| 2FGL J0737.5–8246 | -25.47 | 295.09 | 4.4 | 5.958e-12 | 1.597e-05 | – | – | Unclassified |
| 2FGL J0742.7–3113 | -3.87 | 246.24 | 5.2 | 1.272e-11 | 3.426e-06 | – | – | Unclassified |
| 2FGL J0744.1–2523 | -0.70 | 241.34 | 12.4 | 2.204e-11 | 1 | 0.01033 | – | MSP |
| 2FGL J0745.5+7910 | 29.14 | 135.00 | 5.5 | 4.743e-12 | 1.292e-08 | – | 0.4661 | Uncertain blazar |
| 2FGL J0746.0–0222 | 11.02 | 221.39 | 7.3 | 9.682e-12 | 9.861e-07 | – | – | Unclassified |
| 2FGL J0748.5–2204 | 1.84 | 238.97 | 5.2 | 1.304e-11 | 6.261e-07 | – | – | Unclassified |
| 2FGL J0753.2+1937 | 22.12 | 201.45 | 4.5 | 6.965e-12 | 1.784e-13 | – | 0.0389 | FSRQ |
| 2FGL J0753.2–1634 | 5.58 | 234.78 | 4.3 | 6.456e-12 | 5.36e-07 | – | – | Unclassified |
| 2FGL J0756.3–6433 | -17.76 | 277.27 | 4.5 | 4.359e-12 | 2.697e-08 | – | 0.9426 | BL Lac |
| 2FGL J0758.0–2615c | 1.54 | 243.66 | 5.1 | 1.146e-11 | 0.3265 | – | – | Unclassified |
| 2FGL J0758.8–1448 | 7.65 | 233.94 | 5.8 | 8.271e-12 | 6.598e-07 | – | – | Unclassified |
| 2FGL J0802.6–0940 | 11.05 | 229.93 | 5.2 | 5.895e-12 | 1.839e-09 | – | 0.9565 | BL Lac |
| 2FGL J0802.7–5615 | -13.17 | 270.04 | 5.3 | 8.776e-12 | 0.9994 | 0.0002362 | – | MSP |
| 2FGL J0803.2–0339 | 14.16 | 224.65 | 11.5 | 1.382e-11 | 3.527e-11 | – | 0.9543 | BL Lac |
| 2FGL J0807.0–6511 | -17.03 | 278.40 | 4.4 | 5.779e-12 | 2.177e-06 | – | – | Unclassified |
| 2FGL J0838.8–2828 | 7.82 | 250.60 | 8.2 | 1.216e-11 | 0.002041 | – | – | Unclassified |
| 2FGL J0841.3–3556 | 3.72 | 256.89 | 12.2 | 2.362e-11 | 0.002826 | – | – | Unclassified |
| 2FGL J0843.6+6715 | 35.61 | 147.75 | 7.4 | 6.612e-12 | 0.06308 | – | – | Unclassified |
| 2FGL J0844.9+6214 | 36.90 | 153.75 | 4.1 | 4.837e-12 | 6.466e-10 | – | 0.3647 | FSRQ |
| 2FGL J0846.0+2820 | 36.30 | 196.34 | 4.1 | 5.073e-12 | 5.1e-13 | – | 0.3594 | FSRQ |
| 2FGL J0846.7–4053 | 1.49 | 261.45 | 5.3 | 1.994e-11 | 2.322e-08 | – | 0.2662 | FSRQ |
| 2FGL J0848.7–4324 | 0.19 | 263.63 | 15.5 | 5.437e-11 | 1 | 0.9999 | – | Young pulsar |
| 2FGL J0850.1–4846 | -3.01 | 267.95 | 7.6 | 2.445e-11 | 0.2477 | – | – | Unclassified |
| 2FGL J0854.7–4501 | -0.01 | 265.57 | 13.6 | 4.132e-11 | 0.02644 | – | – | Unclassified |
| 2FGL J0858.0–4815 | -1.68 | 268.39 | 14.1 | 4.866e-11 | 1 | 1 | – | Young pulsar |
| 2FGL J0858.3–4333 | 1.43 | 264.87 | 14.1 | 4.422e-11 | 1 | 0.9999 | – | Young pulsar |
| 2FGL J0859.4–2532 | 13.25 | 251.12 | 4.6 | 6.541e-12 | 3.345e-12 | – | 0.8981 | BL Lac |
| 2FGL J0900.5–4441c | 0.99 | 265.98 | 5.7 | 2.527e-11 | 0.06443 | – | – | Unclassified |
| 2FGL J0900.9+6736 | 37.10 | 146.71 | 6.3 | 6.877e-12 | 3.041e-10 | – | 0.09606 | FSRQ |
| 2FGL J0901.7–4655 | -0.33 | 267.79 | 12.5 | 4.013e-11 | 0.9895 | – | – | Unclassified |
| 2FGL J0902.8–4741c | -0.70 | 268.50 | 7.0 | 1.369e-11 | 0.3336 | – | – | Unclassified |
| 2FGL J0904.0–4823c | -1.02 | 269.15 | 4.1 | 2.249e-11 | 1 | 1 | – | Young pulsar |
| 2FGL J0907.9–4716c | 0.22 | 268.76 | 7.5 | 3.094e-11 | 0.02604 | – | – | Unclassified |
| 2FGL J0914.1–4756 | 0.53 | 269.96 | 11.2 | 3.676e-11 | 1 | 0.9936 | – | Young pulsar |
| 2FGL J0922.2–5214c | -1.56 | 273.94 | 9.3 | 2.245e-11 | 0.9995 | 0.9998 | – | Young pulsar |
| 2FGL J0923.5+1508 | 40.42 | 215.85 | 4.2 | 4.683e-12 | 5.641e-10 | – | 0.1551 | FSRQ |
| 2FGL J0928.8–3530 | 11.23 | 263.02 | 4.4 | 6.301e-12 | 0.8516 | – | – | Unclassified |
| 2FGL J0934.0–6231 | -7.88 | 282.24 | 13.0 | 1.254e-11 | 0.9693 | – | – | Unclassified |
| 2FGL J0937.9–1434 | 27.21 | 248.49 | 4.1 | 4.642e-12 | 1.19e-07 | – | 0.939 | BL Lac |
| 2FGL J0952.7–3717 | 13.13 | 267.88 | 4.9 | 8.301e-12 | 0.002759 | – | – | Unclassified |
| 2FGL J0953.6–1504 | 29.66 | 251.84 | 6.7 | 8.978e-12 | 4.159e-05 | – | – | Unclassified |
| 2FGL J0955.0–3949 | 11.47 | 269.88 | 11.4 | 1.889e-11 | 0.2295 | – | – | Unclassified |
| 2FGL J1010.7–5643c | -0.46 | 282.21 | 8.7 | 3.347e-11 | 1 | 1 | – | Young pulsar |
| 2FGL J1012.1–4236 | 11.22 | 274.22 | 4.2 | 6.794e-12 | 0.9803 | – | – | Unclassified |
| 2FGL J1013.6+3434 | 55.54 | 190.58 | 4.1 | 3.872e-12 | 2.355e-07 | – | 0.8969 | BL Lac |
| 2FGL J1016.1+5600 | 49.97 | 155.96 | 4.6 | 4.988e-12 | 5.666e-09 | – | 0.02016 | FSRQ |
| 2FGL J1016.4–4244 | 11.56 | 274.96 | 4.3 | 6.804e-12 | 4.146e-07 | – | – | Unclassified |
| 2FGL J1020.0–6029 | -2.91 | 285.32 | 5.7 | 1.554e-11 | 0.5724 | – | – | Unclassified |
| 2FGL J1023.5–5749c | -0.43 | 284.25 | 7.4 | 7.505e-11 | 0.9959 | – | – | Unclassified |
| 2FGL J1027.4–5730c | 0.12 | 284.53 | 12.5 | 6.375e-11 | 1 | 1 | – | Young pulsar |
| 2FGL J1029.5–2022 | 31.38 | 263.34 | 4.3 | 5.101e-12 | 4.551e-10 | – | 0.9601 | BL Lac |
| 2FGL J1030.4–6015 | -2.02 | 286.28 | 6.1 | 2.146e-11 | 0.0005339 | – | – | Unclassified |
| 2FGL J1032.9–8401 | -22.16 | 299.27 | 4.2 | 6.328e-12 | 1.615e-08 | – | 0.7506 | Uncertain blazar |
| 2FGL J1033.5–5032 | 6.55 | 281.70 | 5.8 | 8.392e-12 | 2.335e-05 | – | – | Unclassified |
| 2FGL J1036.1–6722 | -7.84 | 290.45 | 17.8 | 1.985e-11 | 0.997 | 0.0005868 | – | MSP |
| 2FGL J1036.4–5828c | -0.11 | 286.05 | 4.5 | 2.649e-11 | 0.8212 | – | – | Unclassified |
| 2FGL J1038.2–2423 | 29.29 | 267.95 | 5.9 | 7.269e-12 | 8.47e-12 | – | 0.6698 | Uncertain blazar |
| 2FGL J1038.6–5850c | -0.28 | 286.47 | 5.4 | 2.701e-11 | 4.56e-05 | – | – | Unclassified |
| 2FGL J1045.0–5941 | -0.64 | 287.59 | 36.1 | 1.887e-10 | 1 | 1 | – | Young pulsar |
| 2FGL J1046.8–6005c | -0.88 | 287.98 | 4.9 | 2.604e-11 | 0.6008 | – | – | Unclassified |
| 2FGL J1050.3–5922c | -0.04 | 288.05 | 7.3 | 2.702e-11 | 0.9996 | 0.0001114 | – | MSP |
| 2FGL J1056.0–5853 | 0.71 | 288.49 | 12.0 | 3.702e-11 | 1 | 1 | – | Young pulsar |
| 2FGL J1056.2–6021 | -0.61 | 289.15 | 12.5 | 4.686e-11 | 0.9999 | 0.9849 | – | Young pulsar |
| 2FGL J1058.7–6621 | -5.93 | 291.96 | 5.6 | 1.11e-11 | 6.503e-07 | – | – | Unclassified |
| 2FGL J1059.3–6118c | -1.31 | 289.89 | 4.4 | 2.476e-11 | 0.003253 | – | – | Unclassified |
| 2FGL J1059.9–2051 | 34.94 | 270.73 | 4.2 | 4.956e-12 | 4.316e-12 | – | 0.8874 | BL Lac |
| 2FGL J1102.1–6308c | -2.85 | 290.96 | 5.1 | 2.141e-11 | 0.2415 | – | – | Unclassified |
| 2FGL J1104.7–6036 | -0.40 | 290.21 | 16.9 | 6.399e-11 | 1 | 0.9898 | – | Young pulsar |
| 2FGL J1105.4–7622 | -14.80 | 296.69 | 5.1 | 9.728e-12 | 1.04e-06 | – | – | Unclassified |
| 2FGL J1105.6–6114 | -0.93 | 290.55 | 10.0 | 4.378e-11 | 1 | 0.9997 | – | Young pulsar |
| 2FGL J1115.0–0701 | 48.64 | 265.11 | 4.6 | 4.668e-12 | 5.762e-07 | – | – | Unclassified |
| 2FGL J1117.2–5341 | 6.64 | 289.16 | 4.8 | 7.233e-12 | 3.215e-06 | – | – | Unclassified |
| 2FGL J1118.9–6027c | 0.39 | 291.77 | 8.6 | 2.771e-11 | 1 | 1 | – | Young pulsar |
| 2FGL J1120.0–2204 | 36.05 | 276.50 | 20.5 | 1.798e-11 | 0.9998 | 0.0002464 | – | MSP |
| 2FGL J1123.3–2527 | 33.29 | 279.04 | 4.5 | 5.523e-12 | 4.27e-05 | – | – | Unclassified |
| 2FGL J1129.0–0532 | 51.74 | 268.51 | 8.1 | 1.242e-11 | 2.307e-12 | – | 0.03232 | FSRQ |
| 2FGL J1129.5+3758 | 69.69 | 175.54 | 11.1 | 1.187e-11 | 1.029e-08 | – | 0.5159 | Uncertain blazar |
| 2FGL J1138.8–6233c | -0.86 | 294.67 | 4.9 | 2.269e-11 | 0.9839 | – | – | Unclassified |
| 2FGL J1156.7–0751 | 52.60 | 280.26 | 4.3 | 6.628e-12 | 9.387e-12 | – | 0.395 | FSRQ |
| 2FGL J1203.6–6243c | -0.35 | 297.49 | 9.8 | 4.196e-11 | 1 | 0.9998 | – | Young pulsar |
| 2FGL J1207.3–5055 | 11.34 | 295.87 | 5.8 | 9.498e-12 | 8.161e-07 | – | – | Unclassified |
| 2FGL J1208.5–6240 | -0.21 | 298.03 | 11.7 | 5.267e-11 | 0.9963 | 0.9982 | – | Young pulsar |
| 2FGL J1208.6–2257 | 38.86 | 290.25 | 4.4 | 7.106e-12 | 1.979e-12 | – | 0.1679 | FSRQ |
| 2FGL J1214.1–4410 | 18.19 | 295.90 | 6.0 | 9.083e-12 | 7.846e-05 | – | – | Unclassified |
| 2FGL J1221.4–0633 | 55.54 | 289.68 | 8.3 | 1.059e-11 | 9.381e-09 | – | 0.2796 | FSRQ |
| 2FGL J1223.3+7954 | 37.13 | 124.47 | 4.2 | 3.335e-12 | 8.989e-11 | – | 0.9888 | BL Lac |
| 2FGL J1226.0+2953 | 83.78 | 185.02 | 12.9 | 9.125e-12 | 0.1145 | – | – | Unclassified |
| 2FGL J1227.7–4853 | 13.80 | 298.92 | 24.3 | 3.338e-11 | 1 | 0.9114 | – | Uncertain PSR |
| 2FGL J1231.1–6512 | -2.42 | 300.81 | 8.1 | 1.853e-11 | 0.9999 | 0.902 | – | Uncertain PSR |
| 2FGL J1231.3–5112 | 11.54 | 299.72 | 10.1 | 1.354e-11 | 0.0003459 | – | – | Unclassified |
| 2FGL J1236.1–6155 | 0.90 | 301.14 | 5.1 | 2.389e-11 | 3.112e-06 | – | – | Unclassified |
| 2FGL J1240.6–7151 | -9.01 | 302.08 | 8.2 | 1.317e-11 | 3.703e-08 | – | 0.9825 | BL Lac |
| 2FGL J1243.9–6232 | 0.32 | 302.07 | 5.3 | 2.659e-11 | 0.4405 | – | – | Unclassified |
| 2FGL J1248.6–5510 | 7.70 | 302.53 | 5.2 | 1.135e-11 | 2.063e-10 | – | 0.4111 | FSRQ |
| 2FGL J1249.5–2811 | 34.69 | 302.42 | 5.4 | 5.949e-12 | 4.791e-08 | – | 0.9574 | BL Lac |
| 2FGL J1254.2–2203 | 40.81 | 303.80 | 5.4 | 8.326e-12 | 7.616e-11 | – | 0.788 | BL Lac |
| 2FGL J1255.8–5828 | 4.40 | 303.51 | 4.5 | 1.236e-11 | 0.9886 | – | – | Unclassified |
| 2FGL J1259.8–3749 | 25.01 | 304.77 | 5.5 | 7.357e-12 | 2.375e-09 | – | 0.9822 | BL Lac |
| 2FGL J1303.7–6316c | -0.43 | 304.32 | 5.8 | 2.736e-11 | 1.212e-07 | – | 0.9527 | BL Lac |
| 2FGL J1305.0+1152 | 74.41 | 315.45 | 5.0 | 6.253e-12 | 1.274e-05 | – | – | Unclassified |
| 2FGL J1306.2–6044 | 2.08 | 304.74 | 15.5 | 3.834e-11 | 0.9914 | – | – | Unclassified |
| 2FGL J1309.6–6230c | 0.28 | 305.03 | 6.1 | 3.555e-11 | 0.9893 | – | – | Unclassified |
| 2FGL J1311.7–3429 | 28.20 | 307.69 | 43.1 | 6.168e-11 | 1 | 0.881 | – | Uncertain PSR |
| 2FGL J1312.0–6458 | -2.19 | 305.11 | 4.5 | 1.716e-11 | 0.004969 | – | – | Unclassified |
| 2FGL J1312.9–2351 | 38.75 | 309.25 | 5.7 | 8.723e-12 | 1.393e-08 | – | 0.9831 | BL Lac |
| 2FGL J1315.6–0730 | 54.87 | 313.40 | 6.1 | 7.188e-12 | 1.264e-11 | – | 0.9697 | BL Lac |
| 2FGL J1317.2–6304 | -0.35 | 305.85 | 11.6 | 4.533e-11 | 1 | 0.9951 | – | Young pulsar |
| 2FGL J1320.1–5756 | 4.72 | 306.75 | 5.6 | 1.449e-11 | 0.0001307 | – | – | Unclassified |
| 2FGL J1324.4–5411 | 8.36 | 307.81 | 4.9 | 1.219e-11 | 1.641e-10 | – | 0.4093 | FSRQ |
| 2FGL J1328.5–4728 | 14.93 | 309.42 | 9.4 | 1.566e-11 | 2.527e-08 | – | 0.9885 | BL Lac |
| 2FGL J1329.5–3448 | 27.42 | 311.75 | 4.2 | 5.951e-12 | 8.071e-11 | – | 0.7193 | Uncertain blazar |
| 2FGL J1329.7–6108 | 1.39 | 307.54 | 5.7 | 2.066e-11 | 0.9997 | 0.9364 | – | Uncertain PSR |
| 2FGL J1335.3–4058 | 21.13 | 311.80 | 4.6 | 7.095e-12 | 1.89e-06 | – | – | Unclassified |
| 2FGL J1335.4–5658 | 5.39 | 308.94 | 6.0 | 1.671e-11 | 1.524e-07 | – | 0.262 | FSRQ |
| 2FGL J1339.2–2348 | 37.77 | 316.79 | 5.6 | 7.339e-12 | 8.822e-11 | – | 0.804 | BL Lac |
| 2FGL J1340.5–0412 | 56.50 | 325.51 | 4.3 | 4.785e-12 | 2.409e-09 | – | 0.9392 | BL Lac |
| 2FGL J1345.8–3356 | 27.58 | 315.65 | 5.4 | 7.181e-12 | 3.604e-09 | – | 0.8884 | BL Lac |
| 2FGL J1346.0–2605 | 35.18 | 317.96 | 5.0 | 6.653e-12 | 0.0001507 | – | – | Unclassified |
| 2FGL J1347.0–2956 | 31.40 | 317.05 | 5.0 | 7.498e-12 | 1.026e-07 | – | 0.9862 | BL Lac |
| 2FGL J1349.9–6222 | -0.27 | 309.65 | 11.4 | 5.451e-11 | 1 | 0.9812 | – | Uncertain PSR |
| 2FGL J1351.1–2749 | 33.20 | 318.73 | 4.4 | 6.373e-12 | 3.768e-12 | – | 0.2284 | FSRQ |
| 2FGL J1353.5–6640 | -4.55 | 309.03 | 7.0 | 1.174e-11 | 1.329e-05 | – | – | Unclassified |
| 2FGL J1358.8–6027c | 1.33 | 311.15 | 5.5 | 2.756e-11 | 0.8888 | – | – | Unclassified |
| 2FGL J1400.2–2412 | 36.04 | 322.43 | 4.3 | 5.535e-12 | 1.605e-06 | – | – | Unclassified |
| 2FGL J1400.7–1438 | 44.96 | 326.96 | 5.6 | 8.28e-12 | 0.6083 | – | – | Unclassified |
| 2FGL J1404.0–5244 | 8.57 | 313.93 | 4.3 | 1.006e-11 | 0.9058 | – | – | Unclassified |
| 2FGL J1405.5–6121 | 0.24 | 311.69 | 15.8 | 1.059e-10 | 1 | 1 | – | Young pulsar |
| 2FGL J1407.4–2948 | 30.22 | 322.01 | 5.3 | 7.835e-12 | 0.00511 | – | – | Unclassified |
| 2FGL J1407.6–5937c | 1.83 | 312.44 | 4.8 | 2.066e-11 | 1.687e-06 | – | – | Unclassified |
| 2FGL J1410.4+7411 | 41.83 | 115.84 | 9.1 | 7.379e-12 | 9.986e-09 | – | 0.9804 | BL Lac |
| 2FGL J1414.1–5450 | 6.13 | 314.74 | 5.3 | 1.412e-11 | 0.05314 | – | – | Unclassified |
| 2FGL J1415.7–6520 | -3.89 | 311.58 | 4.5 | 1.195e-11 | 5.677e-10 | – | 0.8158 | BL Lac |
| 2FGL J1417.5–4404 | 16.12 | 318.85 | 6.6 | 1.093e-11 | 4.185e-07 | – | – | Unclassified |
| 2FGL J1417.7–5028 | 10.08 | 316.68 | 5.0 | 1.01e-11 | 3.838e-10 | – | 0.8692 | BL Lac |
| 2FGL J1422.3–6841 | -7.27 | 311.06 | 7.0 | 1.5e-11 | 1.235e-10 | – | 0.3477 | FSRQ |
| 2FGL J1422.5–6137c | -0.64 | 313.54 | 12.2 | 3.037e-11 | 0.9996 | 3.869e-08 | – | MSP |
| 2FGL J1423.9–7842 | -16.68 | 307.53 | 4.9 | 7.02e-12 | 3.982e-08 | – | 0.8255 | BL Lac |
| 2FGL J1424.2–1752 | 39.68 | 332.08 | 6.5 | 8.07e-12 | 3.218e-07 | – | 0.964 | BL Lac |
| 2FGL J1427.6–6048c | -0.09 | 314.40 | 5.9 | 3.253e-11 | 0.03313 | – | – | Unclassified |
| 2FGL J1437.2–5211 | 7.37 | 318.91 | 4.9 | 1.213e-11 | 3.431e-05 | – | – | Unclassified |
| 2FGL J1446.6–5753 | 1.62 | 317.78 | 10.8 | 3.205e-11 | 0.9999 | 0.9984 | – | Young pulsar |
| 2FGL J1446.8–4701 | 11.40 | 322.53 | 7.2 | 1.291e-11 | 0.9999 | 8.155e-06 | – | MSP |
| 2FGL J1456.7–6247c | -3.32 | 316.72 | 4.4 | 1.491e-11 | 1.951e-08 | – | 0.8088 | BL Lac |
| 2FGL J1458.5–2121 | 32.58 | 338.53 | 4.1 | 6.472e-12 | 3.016e-09 | – | 0.667 | Uncertain blazar |
| 2FGL J1502.1+5548 | 52.95 | 92.70 | 6.8 | 7.726e-12 | 1.543e-09 | – | 0.05264 | FSRQ |
| 2FGL J1502.4+4804 | 57.07 | 81.25 | 9.5 | 1.015e-11 | 7.233e-14 | – | 0.05417 | FSRQ |
| 2FGL J1503.9–5800c | 0.46 | 319.77 | 6.2 | 3.587e-11 | 0.4299 | – | – | Unclassified |
| 2FGL J1506.9+1052 | 54.24 | 12.51 | 4.1 | 1.775e-11 | 1.03e-10 | – | 0.07304 | FSRQ |
| 2FGL J1507.0–6223 | -3.54 | 317.94 | 4.6 | 1.216e-11 | 5.271e-07 | – | – | Unclassified |
| 2FGL J1510.9–5808c | -0.11 | 320.50 | 4.4 | 3.378e-11 | 0.02675 | – | – | Unclassified |
| 2FGL J1511.8–0513 | 43.12 | 354.62 | 7.8 | 1.42e-11 | 3.405e-09 | – | 0.8709 | BL Lac |
| 2FGL J1512.5–6247c | -4.20 | 318.29 | 7.4 | 2.253e-11 | 0.9953 | – | – | Unclassified |
| 2FGL J1513.5–2546 | 27.01 | 338.90 | 4.4 | 7.346e-12 | 8.702e-06 | – | – | Unclassified |
| 2FGL J1513.9–2256 | 29.24 | 340.86 | 4.3 | 6.601e-12 | 4.663e-10 | – | 0.9689 | BL Lac |
| 2FGL J1517.2+3645 | 57.82 | 59.58 | 4.1 | 3.856e-12 | 9.431e-09 | – | 0.2059 | FSRQ |
| 2FGL J1518.4–5233 | 4.09 | 324.32 | 5.6 | 1.421e-11 | 0.004286 | – | – | Unclassified |
| 2FGL J1528.0–5841 | -1.77 | 322.09 | 7.3 | 3.176e-11 | 0.9372 | – | – | Unclassified |
| 2FGL J1536.4–4949 | 4.77 | 328.20 | 40.3 | 9.463e-11 | 0.9971 | 0.5259 | – | Uncertain PSR |
| 2FGL J1539.2–3325 | 17.53 | 338.74 | 10.8 | 1.056e-11 | 1 | 2.466e-06 | – | MSP |
| 2FGL J1539.3–4636 | 7.07 | 330.51 | 6.9 | 1.788e-11 | 2.606e-08 | – | 0.2697 | FSRQ |
| 2FGL J1543.7–0241 | 38.90 | 4.17 | 4.3 | 8.087e-12 | 3.707e-05 | – | – | Unclassified |
| 2FGL J1544.1–2554 | 22.61 | 344.75 | 8.5 | 1.601e-11 | 0.003358 | – | – | Unclassified |
| 2FGL J1544.5–1126 | 32.99 | 356.17 | 5.8 | 1.004e-11 | 1.053e-06 | – | – | Unclassified |
| 2FGL J1548.3+1453 | 47.18 | 25.59 | 10.4 | 1.309e-11 | 4.38e-11 | – | 0.9017 | BL Lac |
| 2FGL J1551.0–4636 | 5.83 | 332.11 | 7.1 | 1.662e-11 | 6.681e-08 | – | 0.6198 | Uncertain blazar |
| 2FGL J1551.3–5333c | 0.41 | 327.76 | 5.7 | 3.366e-11 | 1 | 0.1414 | – | Uncertain PSR |
| 2FGL J1552.8–4824 | 4.25 | 331.20 | 6.6 | 1.934e-11 | 1 | 0.999 | – | Young pulsar |
| 2FGL J1553.5–0324 | 36.53 | 5.38 | 5.2 | 1.044e-11 | 1.355e-08 | – | 0.9351 | BL Lac |
| 2FGL J1554.4–5317c | 0.32 | 328.29 | 7.9 | 5.616e-11 | 0.9999 | 0.9998 | – | Young pulsar |
| 2FGL J1601.1–4220 | 7.94 | 336.27 | 7.3 | 2.053e-11 | 1.609e-06 | – | – | Unclassified |
| 2FGL J1602.4+2308 | 46.83 | 38.63 | 5.2 | 6.293e-12 | 2.485e-06 | – | – | Unclassified |
| 2FGL J1610.1–4808 | 2.55 | 333.54 | 9.5 | 4.046e-11 | 0.9554 | – | – | Unclassified |
| 2FGL J1612.0+1403 | 41.61 | 27.77 | 4.7 | 6.372e-12 | 0.001667 | – | – | Unclassified |
| 2FGL J1614.8+4703 | 45.77 | 73.69 | 4.6 | 4.75e-12 | 1.586e-11 | – | 0.3955 | FSRQ |
| 2FGL J1614.9–5212 | -0.93 | 331.33 | 4.4 | 2.527e-11 | 0.0001741 | – | – | Unclassified |
| 2FGL J1615.2–5138 | -0.54 | 331.75 | 14.7 | 9.142e-11 | 0.9985 | 0.9993 | – | Young pulsar |
| 2FGL J1617.3–5336 | -2.19 | 330.61 | 4.8 | 2.169e-11 | 0.9997 | 3.371e-05 | – | MSP |
| 2FGL J1617.5–2657 | 16.66 | 349.69 | 4.2 | 1.102e-11 | 0.001445 | – | – | Unclassified |
| 2FGL J1617.6–4219 | 5.85 | 338.53 | 5.8 | 1.928e-11 | 7.609e-08 | – | 0.07733 | FSRQ |
| 2FGL J1618.0–5825 | -5.70 | 327.30 | 5.8 | 1.431e-11 | 6.359e-11 | – | 0.3751 | FSRQ |
| 2FGL J1619.0–4650 | 2.46 | 335.53 | 10.6 | 3.858e-11 | 1 | 1 | – | Young pulsar |
| 2FGL J1619.6–4509 | 3.59 | 336.80 | 5.5 | 1.964e-11 | 2.915e-10 | – | 0.9063 | BL Lac |
| 2FGL J1619.7–5040c | -0.35 | 332.92 | 6.4 | 6.349e-11 | 0.4945 | – | – | Unclassified |
| 2FGL J1620.5–2320c | 18.62 | 352.95 | 4.6 | 1.272e-11 | 0.9773 | – | – | Unclassified |
| 2FGL J1620.6–5111c | -0.82 | 332.66 | 9.9 | 6.844e-11 | 0.9998 | 0.9953 | – | Young pulsar |
| 2FGL J1620.8–4928 | 0.38 | 333.89 | 28.0 | 1.807e-10 | 1 | 1 | – | Young pulsar |
| 2FGL J1622.8–0314 | 30.71 | 10.71 | 6.0 | 9.953e-12 | 2.315e-07 | – | 0.7134 | Uncertain blazar |
| 2FGL J1622.8–5006 | -0.30 | 333.68 | 11.2 | 5.486e-11 | 1 | 0.9977 | – | Young pulsar |
| 2FGL J1623.2+4328 | 44.65 | 68.41 | 4.7 | 5.064e-12 | 2.021e-10 | – | 0.3179 | FSRQ |
| 2FGL J1624.0–4941c | -0.14 | 334.11 | 5.5 | 2.861e-11 | 1 | 0.0002919 | – | MSP |
| 2FGL J1624.1–4040 | 6.16 | 340.57 | 16.8 | 3.788e-11 | 1 | 0.328 | – | Uncertain PSR |
| 2FGL J1624.2–2124 | 19.25 | 355.10 | 8.2 | 1.536e-11 | 0.9995 | 1 | – | Young pulsar |
| 2FGL J1625.2–0020 | 31.83 | 13.92 | 20.5 | 1.615e-11 | 1 | 1.819e-06 | – | MSP |
| 2FGL J1626.4–4408 | 3.44 | 338.37 | 9.8 | 3.921e-11 | 0.9992 | 0.962 | – | Uncertain PSR |
| 2FGL J1627.8+3219 | 43.21 | 52.99 | 7.7 | 8.213e-12 | 3.275e-06 | – | – | Unclassified |
| 2FGL J1630.1–4615 | 1.51 | 337.29 | 11.3 | 4.913e-11 | 1 | 0.9999 | – | Young pulsar |
| 2FGL J1630.2–4752 | 0.39 | 336.13 | 5.1 | 3.578e-11 | 0.007554 | – | – | Unclassified |
| 2FGL J1630.3+3732 | 43.26 | 60.20 | 10.4 | 5.23e-12 | 0.9999 | 2.56e-06 | – | MSP |
| 2FGL J1631.0–1050 | 24.64 | 5.06 | 4.4 | 1.003e-11 | 0.982 | – | – | Unclassified |
| 2FGL J1631.6–2819 | 13.45 | 350.82 | 4.6 | 1.286e-11 | 0.476 | – | – | Unclassified |
| 2FGL J1632.4–4753c | 0.10 | 336.37 | 8.8 | 9.085e-11 | 0.7708 | – | – | Unclassified |
| 2FGL J1632.6–2328c | 16.45 | 354.77 | 6.2 | 1.759e-11 | 0.0009959 | – | – | Unclassified |
| 2FGL J1634.4–4743c | -0.03 | 336.71 | 8.1 | 5.271e-11 | 0.9992 | 1.325e-06 | – | MSP |
| 2FGL J1636.3–4740c | -0.24 | 336.96 | 13.3 | 1.138e-10 | 1 | 1 | – | Young pulsar |
| 2FGL J1636.6–0841 | 24.79 | 7.87 | 4.8 | 1.024e-11 | 1.105e-09 | – | 0.7955 | BL Lac |
| 2FGL J1638.0–4703c | -0.03 | 337.62 | 13.6 | 1.128e-10 | 1 | 0.995 | – | Young pulsar |
| 2FGL J1639.7–5504 | -5.56 | 331.81 | 5.9 | 1.748e-11 | 1.158e-07 | – | 0.1679 | FSRQ |
| 2FGL J1639.8–4921c | -1.79 | 336.10 | 4.4 | 2.507e-11 | 6.553e-05 | – | – | Unclassified |
| 2FGL J1639.8–5145 | -3.38 | 334.30 | 12.8 | 4.542e-11 | 9.466e-05 | – | – | Unclassified |
| 2FGL J1641.8–5319 | -4.64 | 333.33 | 6.2 | 1.961e-11 | 1 | 0.7592 | – | Uncertain PSR |
| 2FGL J1643.3–4928 | -2.29 | 336.40 | 5.9 | 3.341e-11 | 0.9719 | – | – | Unclassified |
| 2FGL J1645.7–2148c | 15.15 | 358.09 | 7.3 | 1.259e-11 | 0.998 | 0.6868 | – | Uncertain PSR |
| 2FGL J1646.7–1333 | 19.94 | 5.16 | 4.3 | 6.912e-12 | 7.799e-08 | – | 0.9817 | BL Lac |
| 2FGL J1647.0+4351 | 40.34 | 68.82 | 4.4 | 4.514e-12 | 3.612e-09 | – | 0.222 | FSRQ |
| 2FGL J1648.1–4930 | -2.91 | 336.88 | 4.6 | 1.614e-11 | 3.208e-06 | – | – | Unclassified |
| 2FGL J1649.2–3004 | 9.38 | 351.99 | 5.2 | 1.247e-11 | 0.0001902 | – | – | Unclassified |
| 2FGL J1650.6–4603c | -1.01 | 339.79 | 11.8 | 7.403e-11 | 1 | 0.9979 | – | Young pulsar |
| 2FGL J1651.8–4439c | -0.28 | 341.00 | 6.5 | 4.284e-11 | 0.9999 | 0.6851 | – | Uncertain PSR |
| 2FGL J1652.5–4351c | 0.13 | 341.70 | 7.8 | 3.695e-11 | 0.9999 | 0.6302 | – | Uncertain PSR |
| 2FGL J1653.6–0159 | 24.93 | 16.59 | 22.5 | 3.43e-11 | 1 | 0.325 | – | Uncertain PSR |
| 2FGL J1653.9–4627c | -1.71 | 339.84 | 6.5 | 3.527e-11 | 0.9604 | – | – | Unclassified |
| 2FGL J1656.4–0738 | 21.31 | 11.81 | 4.7 | 1.079e-11 | 3.897e-08 | – | 0.2716 | FSRQ |
| 2FGL J1657.5–4652 | -2.45 | 339.91 | 7.3 | 3.552e-11 | 0.0001716 | – | – | Unclassified |
| 2FGL J1658.1–4743 | -3.07 | 339.30 | 6.1 | 2.861e-11 | 0.007112 | – | – | Unclassified |
| 2FGL J1659.2–0142 | 23.86 | 17.66 | 5.7 | 1.07e-11 | 2.208e-08 | – | 0.9413 | BL Lac |
| 2FGL J1700.8–4912 | -4.33 | 338.41 | 4.3 | 1.611e-11 | 1.27e-06 | – | – | Unclassified |
| 2FGL J1702.5–5654 | -9.20 | 332.39 | 21.0 | 3.477e-11 | 1 | 0.7467 | – | Uncertain PSR |
| 2FGL J1704.3+1235 | 29.39 | 32.51 | 5.3 | 6.686e-12 | 9.496e-08 | – | 0.9074 | BL Lac |
| 2FGL J1704.6–0529 | 20.75 | 14.92 | 7.9 | 1.713e-11 | 4.792e-06 | – | – | Unclassified |
| 2FGL J1704.9–4618 | -3.11 | 341.14 | 9.3 | 3.33e-11 | 1 | 0.9997 | – | Young pulsar |
| 2FGL J1708.4+1003c | 27.42 | 30.37 | 4.7 | 7.994e-12 | 4.436e-10 | – | 0.1897 | FSRQ |
| 2FGL J1709.0–0821 | 18.32 | 12.96 | 5.4 | 1.421e-11 | 0.1753 | – | – | Unclassified |
| 2FGL J1710.0–0323 | 20.68 | 17.59 | 6.9 | 1.591e-11 | 0.9704 | – | – | Unclassified |
| 2FGL J1710.5–5020 | -6.26 | 338.44 | 8.3 | 1.683e-11 | 0.00417 | – | – | Unclassified |
| 2FGL J1716.6–0526c | 18.22 | 16.62 | 6.8 | 1.312e-11 | 0.155 | – | – | Unclassified |
| 2FGL J1717.3–2809 | 5.62 | 357.25 | 5.1 | 1.735e-11 | 3.686e-07 | – | – | Unclassified |
| 2FGL J1721.0+0711 | 23.36 | 29.03 | 7.3 | 1.311e-11 | 2.434e-05 | – | – | Unclassified |
| 2FGL J1721.5–0718c | 16.23 | 15.60 | 6.0 | 1.061e-11 | 0.1542 | – | – | Unclassified |
| 2FGL J1722.5–0420 | 17.52 | 18.40 | 9.0 | 1.681e-11 | 1 | 0.02525 | – | MSP |
| 2FGL J1724.9–0508c | 16.60 | 17.99 | 4.4 | 1.01e-11 | 1.641e-07 | – | 0.8912 | BL Lac |
| 2FGL J1726.6–3545 | -0.31 | 352.10 | 8.4 | 4.223e-11 | 0.9998 | 0.999 | – | Young pulsar |
| 2FGL J1727.6+0647 | 21.70 | 29.44 | 4.8 | 9.243e-12 | 2.26e-10 | – | 0.1661 | FSRQ |
| 2FGL J1727.8–2308 | 6.47 | 2.76 | 5.8 | 1.842e-11 | 3.898e-07 | – | – | Unclassified |
| 2FGL J1728.0–2737c | 3.97 | 359.02 | 5.4 | 2.627e-11 | 0.005793 | – | – | Unclassified |
| 2FGL J1729.5–0854 | 13.71 | 15.23 | 10.2 | 1.933e-11 | 1 | 0.9984 | – | Young pulsar |
| 2FGL J1730.6–0353 | 16.00 | 19.85 | 7.1 | 8.272e-12 | 0.8272 | – | – | Unclassified |
| 2FGL J1730.6–2409 | 5.38 | 2.25 | 8.7 | 2.083e-11 | 0.9999 | 0.289 | – | Uncertain PSR |
| 2FGL J1730.8+5427 | 33.31 | 82.17 | 4.6 | 5.448e-12 | 1.052e-05 | – | – | Unclassified |
| 2FGL J1731.9–2703c | 3.55 | 359.98 | 7.0 | 3.208e-11 | 0.001852 | – | – | Unclassified |
| 2FGL J1733.2–2913c | 2.14 | 358.31 | 5.6 | 2.942e-11 | 3.221e-05 | – | – | Unclassified |
| 2FGL J1733.4–2812c | 2.65 | 359.20 | 6.2 | 3.059e-11 | 0.0001025 | – | – | Unclassified |
| 2FGL J1734.7–2533 | 3.84 | 1.58 | 10.4 | 3.292e-11 | 1 | 0.9998 | – | Young pulsar |
| 2FGL J1738.9+8716 | 27.95 | 120.00 | 12.3 | 1.364e-11 | 1.14e-13 | – | 0.5108 | Uncertain blazar |
| 2FGL J1739.6–2726 | 1.91 | 0.59 | 15.2 | 5.66e-11 | 1 | 0.9401 | – | Uncertain PSR |
| 2FGL J1741.0+1347 | 21.73 | 37.76 | 4.1 | 5.728e-12 | 2.765e-07 | – | 0.9746 | BL Lac |
| 2FGL J1741.1–6750 | -18.72 | 325.24 | 5.0 | 7.549e-12 | 1.14e-07 | – | 0.09241 | FSRQ |
| 2FGL J1742.0–2540c | 2.39 | 2.36 | 6.1 | 2.66e-11 | 0.0001605 | – | – | Unclassified |
| 2FGL J1742.5–3323 | -1.78 | 355.86 | 10.5 | 3.572e-11 | 1 | 0.9933 | – | Young pulsar |
| 2FGL J1743.2–2304 | 3.53 | 4.72 | 7.2 | 2.722e-11 | 0.003215 | – | – | Unclassified |
| 2FGL J1743.9–3039c | -0.59 | 358.35 | 4.8 | 4.16e-11 | 0.06542 | – | – | Unclassified |
| 2FGL J1744.1–7620 | -22.48 | 317.09 | 20.8 | 2.001e-11 | 1 | 1.567e-05 | – | MSP |
| 2FGL J1745.6+0203 | 15.56 | 27.16 | 4.2 | 9.828e-12 | 0.002126 | – | – | Unclassified |
| 2FGL J1746.5–3238 | -2.10 | 356.94 | 20.4 | 8.12e-11 | 1 | 0.9995 | – | Young pulsar |
| 2FGL J1746.6–2851c | -0.16 | 0.19 | 5.9 | 7.326e-11 | 0.9971 | 0.992 | – | Young pulsar |
| 2FGL J1747.2–3507 | -3.51 | 354.89 | 5.9 | 2.338e-11 | 1.773e-05 | – | – | Unclassified |
| 2FGL J1747.3–2825c | -0.07 | 0.63 | 10.6 | 6.926e-11 | 1 | 0.982 | – | Uncertain PSR |
| 2FGL J1747.6+0324 | 15.72 | 28.63 | 5.8 | 1.2e-11 | 1.078e-08 | – | 0.8111 | BL Lac |
| 2FGL J1748.6–2913 | -0.72 | 0.09 | 14.4 | 1.097e-10 | 1 | 1 | – | Young pulsar |
| 2FGL J1748.8+3418 | 27.08 | 59.63 | 6.7 | 7.751e-12 | 1.217e-12 | – | 0.4239 | FSRQ |
| 2FGL J1748.9–3923 | -5.98 | 351.38 | 4.3 | 1.189e-11 | 1.404e-07 | – | 0.941 | BL Lac |
| 2FGL J1749.1+0515 | 16.24 | 30.53 | 4.6 | 9.909e-12 | 5.108e-09 | – | 0.4034 | FSRQ |
| 2FGL J1749.7–3134c | -2.13 | 358.20 | 4.7 | 2.772e-11 | 0.7204 | – | – | Unclassified |
| 2FGL J1753.8–4446 | -9.45 | 347.12 | 5.8 | 1.174e-11 | 4.636e-07 | – | – | Unclassified |
| 2FGL J1754.1–2930 | -1.89 | 0.46 | 9.1 | 3.833e-11 | 1 | 0.9997 | – | Young pulsar |
| 2FGL J1754.4–2538c | 0.00 | 3.83 | 4.3 | 1.695e-11 | 0.9931 | – | – | Unclassified |
| 2FGL J1757.5–6028 | -17.13 | 333.03 | 5.8 | 7.243e-12 | 0.003633 | – | – | Unclassified |
| 2FGL J1758.8–2402c | -0.06 | 5.72 | 8.5 | 5.305e-11 | 0.5663 | – | – | Unclassified |
| 2FGL J1759.2–3853 | -7.47 | 352.81 | 4.3 | 1.193e-11 | 0.0005122 | – | – | Unclassified |
| 2FGL J1759.4–2954 | -3.09 | 0.69 | 5.3 | 2.291e-11 | 8.331e-08 | – | 0.5954 | Uncertain blazar |
| 2FGL J1759.5–0521 | 8.99 | 22.14 | 4.7 | 1.628e-11 | 1.408e-06 | – | – | Unclassified |
| 2FGL J1800.8–2400 | -0.43 | 5.97 | 8.8 | 7.134e-11 | 0.8608 | – | – | Unclassified |
| 2FGL J1802.8–6706 | -20.32 | 326.85 | 4.3 | 5.879e-12 | 3.867e-06 | – | – | Unclassified |
| 2FGL J1803.3–2148 | 0.16 | 8.17 | 15.5 | 1.087e-10 | 1 | 0.9982 | – | Young pulsar |
| 2FGL J1805.0–0845 | 6.16 | 19.80 | 5.9 | 1.859e-11 | 1.172e-06 | – | – | Unclassified |
| 2FGL J1805.8+0612 | 12.95 | 33.33 | 7.9 | 1.449e-11 | 0.1508 | – | – | Unclassified |
| 2FGL J1807.7–0419 | 7.68 | 24.03 | 6.7 | 1.959e-11 | 0.0002352 | – | – | Unclassified |
| 2FGL J1808.3–3356 | -6.70 | 358.08 | 6.8 | 1.715e-11 | 0.9912 | – | – | Unclassified |
| 2FGL J1808.5–2037c | -0.32 | 9.80 | 8.8 | 7.674e-11 | 1 | 0.9456 | – | Uncertain PSR |
| 2FGL J1811.3–2421 | -2.69 | 6.83 | 6.4 | 2.824e-11 | 1 | 0.9963 | – | Young pulsar |
| 2FGL J1813.6–2821 | -5.04 | 3.56 | 4.4 | 1.316e-11 | 0.007354 | – | – | Unclassified |
| 2FGL J1813.7–1139c | 2.90 | 18.26 | 9.5 | 3.785e-11 | 1 | 0.9973 | – | Young pulsar |
| 2FGL J1814.1–1735c | -0.02 | 13.09 | 13.1 | 8.423e-11 | 1 | 0.9995 | – | Young pulsar |
| 2FGL J1816.5+4511 | 24.74 | 72.85 | 13.0 | 1.531e-11 | 1.705e-05 | – | – | Unclassified |
| 2FGL J1817.6–1651c | -0.41 | 14.13 | 6.2 | 4.363e-11 | 0.9993 | 0.9017 | – | Uncertain PSR |
| 2FGL J1819.3–1523 | -0.07 | 15.62 | 19.3 | 1.11e-10 | 1 | 0.9999 | – | Young pulsar |
| 2FGL J1820.6–3219 | -8.23 | 0.73 | 4.5 | 1.147e-11 | 0.000336 | – | – | Unclassified |
| 2FGL J1821.8+0830 | 10.42 | 37.22 | 4.6 | 1.122e-11 | 8.675e-07 | – | – | Unclassified |
| 2FGL J1823.1–1338c | -0.06 | 17.60 | 10.4 | 9.157e-11 | 0.9999 | 0.9983 | – | Young pulsar |
| 2FGL J1824.5+1013 | 10.56 | 39.09 | 4.6 | 7.711e-12 | 0.09768 | – | – | Unclassified |
| 2FGL J1827.4–0846 | 1.28 | 22.39 | 6.3 | 2.81e-11 | 0.6787 | – | – | Unclassified |
| 2FGL J1827.4–1445c | -1.51 | 17.10 | 4.6 | 3.143e-11 | 0.08663 | – | – | Unclassified |
| 2FGL J1827.6+1149 | 10.58 | 40.89 | 5.1 | 1.06e-11 | 0.0007372 | – | – | Unclassified |
| 2FGL J1828.7+3231 | 18.63 | 60.63 | 5.5 | 7.691e-12 | 0.02674 | – | – | Unclassified |
| 2FGL J1829.1–0340c | 3.26 | 27.12 | 4.2 | 2.165e-11 | 0.962 | – | – | Unclassified |
| 2FGL J1829.8–0204c | 3.86 | 28.61 | 6.6 | 2.03e-11 | 0.9929 | – | – | Unclassified |
| 2FGL J1830.4–1634 | -2.98 | 15.82 | 8.3 | 3.811e-11 | 3.197e-07 | – | 0.2427 | FSRQ |
| 2FGL J1830.9–3132 | -9.84 | 2.42 | 4.1 | 8.359e-12 | 0.612 | – | – | Unclassified |
| 2FGL J1831.2–1518 | -2.57 | 17.04 | 4.9 | 3.313e-11 | 2.272e-05 | – | – | Unclassified |
| 2FGL J1832.0–0200 | 3.40 | 28.92 | 9.2 | 4.557e-11 | 0.8027 | – | – | Unclassified |
| 2FGL J1832.2–6502 | -22.46 | 330.00 | 5.2 | 6.925e-12 | 0.9458 | – | – | Unclassified |
| 2FGL J1833.1–0437c | 1.93 | 26.73 | 7.9 | 3.951e-11 | 0.9982 | 0.001452 | – | MSP |
| 2FGL J1835.4+1036 | 8.34 | 40.63 | 4.2 | 9.86e-12 | 6.79e-10 | – | 0.3721 | FSRQ |
| 2FGL J1835.4+1349 | 9.74 | 43.56 | 5.3 | 1.22e-11 | 6.596e-09 | – | 0.6459 | Uncertain blazar |
| 2FGL J1835.5–0649 | 0.39 | 25.05 | 5.2 | 4.599e-11 | 0.0008108 | – | – | Unclassified |
| 2FGL J1837.3–0700c | -0.09 | 25.10 | 8.2 | 6.089e-11 | 0.9986 | 0.9993 | – | Young pulsar |
| 2FGL J1837.9+3821 | 18.91 | 67.09 | 4.8 | 6.704e-12 | 7.543e-08 | – | 0.07341 | FSRQ |
| 2FGL J1839.0–0102 | 2.28 | 30.60 | 9.6 | 3.669e-11 | 0.9998 | 0.4378 | – | Uncertain PSR |
| 2FGL J1839.0–0539 | 0.16 | 26.49 | 23.6 | 2.353e-10 | 1 | 1 | – | Young pulsar |
| 2FGL J1839.3–0558c | -0.04 | 26.23 | 9.4 | 9.924e-11 | 1 | 0.9999 | – | Young pulsar |
| 2FGL J1842.3+2740 | 14.07 | 57.11 | 4.9 | 7.617e-12 | 0.008396 | – | – | Unclassified |
| 2FGL J1842.3–5839 | -21.78 | 336.97 | 6.0 | 7.583e-12 | 5.419e-07 | – | – | Unclassified |
| 2FGL J1842.8–0359c | 0.08 | 28.41 | 6.0 | 3.545e-11 | 1 | 0.9981 | – | Young pulsar |
| 2FGL J1843.7–0312c | 0.25 | 29.19 | 8.3 | 5.817e-11 | 1 | 1 | – | Young pulsar |
| 2FGL J1844.3+1548 | 8.68 | 46.32 | 12.5 | 2.621e-11 | 2.859e-08 | – | 0.7323 | Uncertain blazar |
| 2FGL J1844.3–0343c | -0.14 | 28.81 | 8.6 | 5.685e-11 | 1 | 0.9936 | – | Young pulsar |
| 2FGL J1844.9–1116 | -3.70 | 22.15 | 8.4 | 2.321e-11 | 0.04539 | – | – | Unclassified |
| 2FGL J1846.6–2519 | -10.26 | 9.63 | 7.2 | 1.663e-11 | 9.551e-05 | – | – | Unclassified |
| 2FGL J1847.2–0236 | -0.26 | 30.13 | 13.8 | 9.168e-11 | 1 | 0.9999 | – | Young pulsar |
| 2FGL J1848.2–0139c | -0.05 | 31.10 | 11.4 | 1.206e-10 | 0.9999 | 0.9999 | – | Young pulsar |
| 2FGL J1849.9–0125c | -0.32 | 31.50 | 7.5 | 6.028e-11 | 1 | 1 | – | Young pulsar |
| 2FGL J1852.8+0156c | 0.57 | 34.81 | 7.3 | 3.811e-11 | 1 | 1 | – | Young pulsar |
| 2FGL J1856.2+0450c | 1.14 | 37.79 | 12.3 | 6.532e-11 | 1 | 1 | – | Young pulsar |
| 2FGL J1857.2+0055c | -0.87 | 34.42 | 12.2 | 6.005e-11 | 0.9999 | 0.9979 | – | Young pulsar |
| 2FGL J1857.6+0211 | -0.39 | 35.59 | 17.8 | 9.869e-11 | 1 | 0.9999 | – | Young pulsar |
| 2FGL J1857.8+0355c | 0.36 | 37.16 | 11.6 | 5.321e-11 | 1 | 0.9353 | – | Uncertain PSR |
| 2FGL J1858.3–2218 | -11.42 | 13.56 | 6.2 | 1.157e-11 | 0.03534 | – | – | Unclassified |
| 2FGL J1858.5+0129c | -0.89 | 35.07 | 7.0 | 3.621e-11 | 1 | 0.9998 | – | Young pulsar |
| 2FGL J1859.3+0312c | -0.30 | 36.69 | 6.3 | 3.813e-11 | 0.0006497 | – | – | Unclassified |
| 2FGL J1901.1+0427 | -0.13 | 38.01 | 16.3 | 9.058e-11 | 1 | 0.9988 | – | Young pulsar |
| 2FGL J1902.3–1106 | -7.43 | 24.23 | 5.6 | 1.549e-11 | 1.789e-08 | – | 0.2153 | FSRQ |
| 2FGL J1902.7–7053 | -26.44 | 324.34 | 9.9 | 1.288e-11 | 1.218e-07 | – | 0.848 | BL Lac |
| 2FGL J1904.8–0705 | -6.21 | 28.12 | 10.3 | 2.763e-11 | 0.09346 | – | – | Unclassified |
| 2FGL J1904.9–3720c | -18.55 | 359.78 | 8.4 | 1.015e-11 | 0.7889 | – | – | Unclassified |
| 2FGL J1906.5+0720 | -0.00 | 41.19 | 24.0 | 1.296e-10 | 1 | 0.9999 | – | Young pulsar |
| 2FGL J1908.8–0132 | -4.57 | 33.53 | 7.4 | 2.294e-11 | 6.126e-06 | – | – | Unclassified |
| 2FGL J1913.8–1237 | -10.62 | 24.08 | 4.2 | 1.057e-11 | 0.0002155 | – | – | Unclassified |
| 2FGL J1914.0+1436 | 1.72 | 48.48 | 7.4 | 2.812e-11 | 0.8387 | – | – | Unclassified |
| 2FGL J1914.4+0951c | -0.57 | 44.31 | 5.5 | 2.956e-11 | 1 | 0.1669 | – | Uncertain PSR |
| 2FGL J1917.0–3027 | -18.47 | 7.51 | 6.1 | 9.776e-12 | 0.0003509 | – | – | Unclassified |
| 2FGL J1919.5–7324 | -27.91 | 321.64 | 4.4 | 7.606e-12 | 7.819e-09 | – | 0.02879 | FSRQ |
| 2FGL J1921.1+1436c | 0.20 | 49.28 | 8.2 | 5.031e-11 | 1 | 0.9999 | – | Young pulsar |
| 2FGL J1921.3+0131 | -5.95 | 37.71 | 7.6 | 2.059e-11 | 5.344e-05 | – | – | Unclassified |
| 2FGL J1923.4+2013 | 2.37 | 54.50 | 5.7 | 2.347e-11 | 0.1445 | – | – | Unclassified |
| 2FGL J1924.8+1724c | 0.75 | 52.18 | 7.5 | 2.639e-11 | 0.9998 | 0.9935 | – | Young pulsar |
| 2FGL J1924.9–1036 | -12.20 | 27.14 | 8.8 | 1.759e-11 | 5.093e-05 | – | – | Unclassified |
| 2FGL J1925.7–7836c | -28.25 | 315.75 | 6.0 | 9.476e-12 | 6.082e-09 | – | 0.6409 | Uncertain blazar |
| 2FGL J1931.8+1325 | -2.63 | 49.48 | 5.8 | 1.814e-11 | 0.9993 | 0.2261 | – | Uncertain PSR |
| 2FGL J1936.5–0855 | -14.06 | 29.97 | 4.8 | 1.015e-11 | 0.0015 | – | – | Unclassified |
| 2FGL J1942.7–8049c | -28.79 | 313.16 | 7.3 | 1.337e-11 | 1.91e-05 | – | – | Unclassified |
| 2FGL J1942.9–3528 | -25.23 | 4.28 | 4.8 | 8.793e-12 | 2.833e-08 | – | 0.9233 | BL Lac |
| 2FGL J1944.3+7325 | 22.27 | 105.47 | 4.5 | 6.902e-12 | 3.868e-07 | – | – | Unclassified |
| 2FGL J1946.4–5402 | -29.55 | 343.90 | 12.2 | 9.851e-12 | 0.9898 | – | – | Unclassified |
| 2FGL J1946.7–1118 | -17.33 | 28.86 | 4.8 | 9.589e-12 | 0.02794 | – | – | Unclassified |
| 2FGL J1947.8–0739 | -16.01 | 32.43 | 6.5 | 1.264e-11 | 0.002925 | – | – | Unclassified |
| 2FGL J1949.4–1457 | -19.46 | 25.66 | 5.2 | 9.286e-12 | 2.462e-05 | – | – | Unclassified |
| 2FGL J1949.7+2405 | -1.06 | 60.84 | 6.9 | 1.859e-11 | 0.006452 | – | – | Unclassified |
| 2FGL J1950.3+1223 | -7.06 | 50.78 | 9.5 | 2.167e-11 | 2.313e-05 | – | – | Unclassified |
| 2FGL J1952.6–3252 | -26.40 | 7.69 | 4.4 | 8.152e-12 | 1.396e-06 | – | – | Unclassified |
| 2FGL J1955.9–0241 | -15.57 | 37.97 | 5.7 | 1.166e-11 | 4.687e-06 | – | – | Unclassified |
| 2FGL J1958.6+4020 | 5.70 | 75.77 | 5.5 | 1.197e-11 | 2.684e-09 | – | 0.7361 | Uncertain blazar |
| 2FGL J1959.9+3336c | 1.95 | 70.14 | 7.5 | 2.099e-11 | 0.001675 | – | – | Unclassified |
| 2FGL J2002.8–2150 | -25.02 | 20.09 | 4.1 | 6.044e-12 | 3.079e-07 | – | 0.8496 | BL Lac |
| 2FGL J2004.4+3339c | 1.18 | 70.69 | 9.1 | 3.217e-11 | 0.0692 | – | – | Unclassified |
| 2FGL J2004.6+7004 | 19.47 | 102.86 | 9.5 | 1.313e-11 | 5.845e-09 | – | 0.9662 | BL Lac |
| 2FGL J2006.2–0929 | -20.90 | 32.78 | 4.8 | 8.651e-12 | 7.617e-05 | – | – | Unclassified |
| 2FGL J2006.5–2256 | -26.21 | 19.27 | 6.4 | 1.118e-11 | 2.974e-11 | – | 0.04227 | FSRQ |
| 2FGL J2006.9–1734 | -24.34 | 24.84 | 4.7 | 9.571e-12 | 0.04139 | – | – | Unclassified |
| 2FGL J2009.1–0339 | -18.93 | 38.67 | 5.9 | 1.082e-11 | 0.01449 | – | – | Unclassified |
| 2FGL J2009.2–1505 | -23.89 | 27.60 | 6.5 | 8.951e-12 | 2.828e-11 | – | 0.2019 | FSRQ |
| 2FGL J2009.8+2747 | -2.95 | 66.37 | 4.6 | 1.171e-11 | 0.9997 | 0.9998 | – | Young pulsar |
| 2FGL J2012.4+3955c | 3.23 | 76.83 | 8.6 | 2.334e-11 | 1 | 0.9981 | – | Young pulsar |
| 2FGL J2013.8+4115c | 3.75 | 78.09 | 8.6 | 2.639e-11 | 0.9996 | 0.9994 | – | Young pulsar |
| 2FGL J2017.4–3215 | -31.27 | 9.94 | 5.4 | 9.349e-12 | 5.352e-12 | – | 0.1918 | FSRQ |
| 2FGL J2017.5–1618 | -26.21 | 27.22 | 9.2 | 1.285e-11 | 0.8296 | – | – | Unclassified |
| 2FGL J2018.0+3626 | 0.39 | 74.54 | 18.5 | 7.501e-11 | 1 | 0.9998 | – | Young pulsar |
| 2FGL J2020.0+4159 | 3.19 | 79.35 | 9.4 | 3.996e-11 | 0.7151 | – | – | Unclassified |
| 2FGL J2021.5+0632 | -16.67 | 49.61 | 5.2 | 8.307e-12 | 0.02509 | – | – | Unclassified |
| 2FGL J2028.3+3332 | -3.01 | 73.37 | 25.8 | 6.084e-11 | 1 | 0.07751 | – | MSP |
| 2FGL J2030.7+4417 | 2.92 | 82.34 | 18.6 | 5.62e-11 | 1 | 0.9881 | – | Young pulsar |
| 2FGL J2031.4–1842 | -30.17 | 26.10 | 4.1 | 6.643e-12 | 2.905e-12 | – | 0.04623 | FSRQ |
| 2FGL J2033.6+3927 | -0.38 | 78.77 | 13.0 | 6.114e-11 | 0.5593 | – | – | Unclassified |
| 2FGL J2034.7–4201 | -36.39 | 359.09 | 4.7 | 5.33e-12 | 3.192e-09 | – | 0.9485 | BL Lac |
| 2FGL J2034.9+3632 | -2.33 | 76.59 | 5.5 | 1.699e-11 | 0.9669 | – | – | Unclassified |
| 2FGL J2036.0+4224c | 1.02 | 81.42 | 6.1 | 3.985e-11 | 0.02505 | – | – | Unclassified |
| 2FGL J2038.0+4145c | 0.33 | 81.12 | 8.8 | 5.421e-11 | 1 | 1 | – | Young pulsar |
| 2FGL J2039.8–5620 | -37.18 | 341.19 | 17.6 | 2.487e-11 | 1 | 0.003664 | – | MSP |
| 2FGL J2040.1+4105c | -0.39 | 80.83 | 4.5 | 3.011e-11 | 0.001657 | – | – | Unclassified |
| 2FGL J2041.2+4735 | 3.43 | 86.10 | 11.7 | 2.884e-11 | 1 | 0.8175 | – | Uncertain PSR |
| 2FGL J2042.0+4252c | 0.43 | 82.45 | 9.9 | 4.705e-11 | 1 | 1 | – | Young pulsar |
| 2FGL J2042.8–7317 | -33.80 | 320.73 | 5.1 | 5.693e-12 | 5.311e-08 | – | 0.9169 | BL Lac |
| 2FGL J2044.4–4757 | -38.38 | 351.73 | 7.3 | 9.567e-12 | 0.7926 | – | – | Unclassified |
| 2FGL J2046.2–4259 | -38.58 | 358.09 | 4.1 | 4.907e-12 | 4.663e-07 | – | – | Unclassified |
| 2FGL J2047.9+4536c | 1.30 | 85.26 | 6.7 | 2.17e-11 | 0.9757 | – | – | Unclassified |
| 2FGL J2051.8+5054 | 4.16 | 89.77 | 6.4 | 2.113e-11 | 0.9952 | – | – | Unclassified |
| 2FGL J2053.2+1212c | -20.15 | 59.11 | 5.2 | 7.25e-12 | 3.446e-11 | – | 0.4958 | Uncertain blazar |
| 2FGL J2055.8+4754 | 1.73 | 87.89 | 4.4 | 1.197e-11 | 3.973e-06 | – | – | Unclassified |
| 2FGL J2103.3+4357c | -1.84 | 85.76 | 8.0 | 2.522e-11 | 0.4246 | – | – | Unclassified |
| 2FGL J2103.5–1112 | -34.37 | 37.86 | 5.0 | 6.681e-12 | 8.384e-08 | – | 0.8889 | BL Lac |
| 2FGL J2104.9+3555 | -7.41 | 79.94 | 5.4 | 9.487e-12 | 5.39e-10 | – | 0.4998 | Uncertain blazar |
| 2FGL J2107.6+2506 | -14.94 | 71.99 | 4.2 | 7.119e-12 | 3.082e-06 | – | – | Unclassified |
| 2FGL J2107.8+3652 | -7.21 | 81.05 | 5.0 | 6.55e-12 | 5.636e-09 | – | 0.9779 | BL Lac |
| 2FGL J2107.9+5207c | 3.06 | 92.33 | 12.9 | 3.595e-11 | 1 | 0.2827 | – | Uncertain PSR |
| 2FGL J2110.3+3822 | -6.57 | 82.50 | 4.6 | 7.364e-12 | 4.367e-09 | – | 0.9522 | BL Lac |
| 2FGL J2111.3+4605 | -1.45 | 88.29 | 22.2 | 3.882e-11 | 1 | 0.0001281 | – | MSP |
| 2FGL J2112.3–4832 | -42.97 | 350.59 | 5.6 | 7.667e-12 | 3.517e-12 | – | 0.134 | FSRQ |
| 2FGL J2112.5–3042 | -42.45 | 14.93 | 19.7 | 1.89e-11 | 1 | 2.587e-05 | – | MSP |
| 2FGL J2114.1+5440 | 4.12 | 94.84 | 4.3 | 1.356e-11 | 1.238e-06 | – | – | Unclassified |
| 2FGL J2115.4+1213 | -24.50 | 62.60 | 5.1 | 6.751e-12 | 7.655e-09 | – | 0.6152 | Uncertain blazar |
| 2FGL J2117.5+3730 | -8.19 | 82.84 | 10.5 | 1.081e-11 | 0.02025 | – | – | Unclassified |
| 2FGL J2124.0–1513 | -40.57 | 35.81 | 4.2 | 6.43e-12 | 1.508e-13 | – | 0.05749 | FSRQ |
| 2FGL J2125.0–4632 | -45.34 | 352.99 | 6.9 | 9.022e-12 | 3.515e-10 | – | 0.07837 | FSRQ |
| 2FGL J2128.7+5824 | 5.33 | 98.91 | 8.4 | 1.73e-11 | 0.9996 | 0.9686 | – | Uncertain PSR |
| 2FGL J2131.0–5417 | -44.71 | 341.98 | 9.0 | 1.171e-11 | 1.231e-13 | – | 0.03078 | FSRQ |
| 2FGL J2132.5+2605 | -18.44 | 76.69 | 5.8 | 7.652e-12 | 2.184e-10 | – | 0.8589 | BL Lac |
| 2FGL J2133.5–6431 | -41.32 | 328.73 | 6.2 | 8.578e-12 | 0.001868 | – | – | Unclassified |
| 2FGL J2133.9+6645 | 10.96 | 105.17 | 6.7 | 1.123e-11 | 1.249e-07 | – | 0.9807 | BL Lac |
| 2FGL J2134.6–2130 | -45.08 | 28.95 | 7.2 | 8.508e-12 | 2.033e-09 | – | 0.8747 | BL Lac |
| 2FGL J2139.8+4714 | -4.03 | 92.60 | 15.5 | 2.224e-11 | 1 | 0.004044 | – | MSP |
| 2FGL J2200.1–6931 | -40.96 | 321.28 | 7.6 | 8.897e-12 | 4.337e-11 | – | 0.03542 | FSRQ |
| 2FGL J2201.2+5926 | 3.40 | 102.79 | 4.6 | 1.415e-11 | 2.371e-05 | – | – | Unclassified |
| 2FGL J2210.1+5913 | 2.56 | 103.59 | 5.8 | 1.878e-11 | 0.000197 | – | – | Unclassified |
| 2FGL J2212.6+0702 | -38.58 | 68.74 | 9.1 | 1.167e-11 | 0.0006597 | – | – | Unclassified |
| 2FGL J2213.7–4754 | -53.05 | 348.01 | 4.9 | 4.905e-12 | 4.623e-09 | – | 0.9188 | BL Lac |
| 2FGL J2219.6+5850 | 1.55 | 104.37 | 6.6 | 2.15e-11 | 0.06338 | – | – | Unclassified |
| 2FGL J2221.0+6307 | 5.05 | 106.86 | 11.3 | 2.334e-11 | 1 | 0.5141 | – | Uncertain PSR |
| 2FGL J2227.8+0051 | -45.56 | 66.07 | 4.2 | 5.537e-12 | 5.088e-07 | – | – | Unclassified |
| 2FGL J2228.6–1633 | -55.36 | 43.22 | 6.2 | 6.281e-12 | 1.327e-11 | – | 0.9178 | BL Lac |
| 2FGL J2231.0+6512 | 6.23 | 108.90 | 8.3 | 2.029e-11 | 3.263e-06 | – | – | Unclassified |
| 2FGL J2237.2+6316 | 4.22 | 108.50 | 6.3 | 1.777e-11 | 5.112e-05 | – | – | Unclassified |
| 2FGL J2246.3+1549 | -37.41 | 83.97 | 8.2 | 1.838e-11 | 5.395e-11 | – | 0.1695 | FSRQ |
| 2FGL J2249.1+5758 | -1.13 | 107.29 | 5.3 | 1.226e-11 | 0.9861 | – | – | Unclassified |
| 2FGL J2250.7+6305c | 3.35 | 109.76 | 5.7 | 1.678e-11 | 5.623e-05 | – | – | Unclassified |
| 2FGL J2251.1–4927 | -57.90 | 340.73 | 5.9 | 6.032e-12 | 7.725e-07 | – | – | Unclassified |
| 2FGL J2251.8+4211 | -15.39 | 100.40 | 4.6 | 8.038e-12 | 1.226e-11 | – | 0.1444 | FSRQ |
| 2FGL J2257.5+6222c | 2.37 | 110.15 | 7.4 | 1.438e-11 | 1 | 0.01532 | – | MSP |
| 2FGL J2257.9–3646 | -64.19 | 3.90 | 5.3 | 4.914e-12 | 3.81e-09 | – | 0.9703 | BL Lac |
| 2FGL J2259.0–8254 | -33.33 | 306.92 | 5.5 | 8.114e-12 | 0.9899 | – | – | Unclassified |
| 2FGL J2300.0–3553 | -64.80 | 5.65 | 4.6 | 5.287e-12 | 4.757e-13 | – | 0.4575 | Uncertain blazar |
| 2FGL J2319.3–3830 | -67.64 | 356.51 | 4.5 | 4.798e-12 | 2.359e-06 | – | – | Unclassified |
| 2FGL J2332.5–5535 | -58.18 | 324.13 | 7.2 | 7.135e-12 | 0.0488 | – | – | Unclassified |
| 2FGL J2339.6–0532 | -62.47 | 81.36 | 32.2 | 3.047e-11 | 0.9996 | 2.556e-06 | – | MSP |
| 2FGL J2343.3–4752 | -65.37 | 331.05 | 6.8 | 6.819e-12 | 2.524e-12 | – | 0.5763 | Uncertain blazar |
| 2FGL J2347.2+0707 | -52.39 | 96.22 | 7.2 | 8.445e-12 | 1.949e-07 | – | 0.9566 | BL Lac |
| 2FGL J2351.6–7558 | -40.60 | 307.65 | 4.1 | 4.604e-12 | 1.229e-06 | – | – | Unclassified |
| 2FGL J2353.3+6643c | 4.49 | 117.23 | 5.4 | 1.511e-11 | 0.229 | – | – | Unclassified |
| 2FGL J2354.2–6615 | -49.87 | 311.80 | 4.9 | 4.986e-12 | 3.625e-08 | – | 0.7016 | Uncertain blazar |
| 2FGL J2356.0–5256 | -62.21 | 320.94 | 11.7 | 1.31e-11 | 3.912e-13 | – | 0.05585 | FSRQ |
| 2FGL J2358.4–1811 | -74.88 | 66.37 | 4.1 | 4.321e-12 | 0.005714 | – | – | Unclassified |
| 2FGL J2359.4+6751c | 5.49 | 118.05 | 7.7 | 1.744e-11 | 0.7372 | – | – | Unclassified |
| 2FGL J2359.6+6543c | 3.39 | 117.64 | 7.7 | 2.217e-11 | 0.08554 | – | – | Unclassified |

Table C.1: Hierarchical ANN predictions for 2FGL unidentified sources as discussed in the text.

# Appendix D

# Study of two peculiar X-ray galactic objects

As mentioned in this Ph.D. thesis, n parallel to the work devoted to unidentified sources, I took part to the analysis and interpretation of multi-wavelength observations of isolated neutron stars. This appendix contains two papers we have published on astrophysical journals about the analyses of the X-ray emission from two interesting isolated neutron stars, the *Central Compact Object* (CCO) RX J0822–4300 and the *Magnetar* SGR 0418+5729.

CCOs are supposed to be young, isolated, radio-quiet neutron stars located to the geometrical centers of non-plerionic supernova remnants (SNRs) with no counterparts at any other wavelength. RX J0822–4300 is the CCO in the Puppis A SNR. We performed a very deep (130-ks) observation with XMM–Newton, which allowed us to study in detail the phase-resolved properties of RXJ0822–4300. Our data confirm the existence of an emission line at 0.8 keV, best modelled as an emission line, only seen in the 'soft'-phase interval – when the cooler region is best aligned to the line of sight. Surprisingly, we detect an evident variation in the emission line component, which can be modelled as a decrease in the central energy from $\sim 0.80$ keV in 2001 to $\sim 0.73$ keV in 2009–10. The line could be generated via cyclotron scattering of thermal photons in an optically-thin layer of gas, or, alternatively, it could originate in low-rate accretion by a debris disc. In any case, a variation in energy, pointing to a variation of the magnetic field in the line-emitting region, cannot be easily accounted for.

A magnetar is a type of neutron star with a relatively slow spin rate and an extremely powerful magnetic field ($10^{12}$–$10^{15}$ Gauss), the decay of which generates occasional large blasts of X-rays. Our long-term X-ray monitoring of the outburst decay of the low magnetic field magnetar SGR 0418+5729 from the discovery of the source in 2009 June up to 2012 August allowed us to obtain the first measurement of the period derivative of SGR 0418+5729: $\dot{P} = 4(1) \times 10^{-15}$ s s$^{-1}$, significant at a $\sim 3.5\sigma$ confidence level. This leads to a surface dipolar magnetic field of $B_{dip} \simeq 6 \times 10^{12}$ Gauss. This measurement confirms SGR 0418+5729 as the lowest magnetic field magnetar. Following the flux and spectral evolution from the beginning of the outburst up to $\sim$ 1200 days, we observe a gradual cooling of the tiny hot spot responsible for the X-ray emission, from a temperature of $\sim$ 0.9 to 0.3 keV. Simultaneously, the X-ray flux decreased by about three orders of magnitude: from about $1.4 \times 10^{-11}$ to $1.2 \times 10^{-1}4$ erg s$^{1}$ cm$^{-2}$. By modeling the magneto-thermal secular evolution of SGR 0418+5729, we infer a realistic age of $\sim$ 550 kyr, and a dipolar magnetic field at birth of $\sim 10^{14}$ Gauss. The outburst characteristics suggest the presence of a thin twisted bundle with a small heated spot at its base. The bundle untwisted in the first few months following the outburst, while the hot spot decreases in temperature and size. We estimate the outburst rate of low magnetic field magnetars to be about one per year per galaxy, and we briefly discuss the consequences of such a result in several other astrophysical contexts.

# A time-variable, phase-dependent emission line in the X-ray spectrum of the isolated neutron star RX J0822−4300

A. De Luca,[1,2,3]⋆ D. Salvetti,[1,3,4] A. Sartori,[1,3,4] P. Esposito,[5] A. Tiengo,[1,2] S. Zane,[6] R. Turolla,[6,7] F. Pizzolato,[1] R. P. Mignani,[6,8] P. A. Caraveo,[1,3] S. Mereghetti[1] and G. F. Bignami[1,2]

[1]*INAF – Istituto di Astrofisica Spaziale e Fisica Cosmica - Milano, via E. Bassini 15, I-20133 Milano, Italy*
[2]*Istituto Universitario di Studi Superiori, Viale Lungo Ticino Sforza 56, I-27100 Pavia, Italy*
[3]*Istituto Nazionale di Fisica Nucleare, Sezione di Pavia, Via Bassi 6, I-27100 Pavia, Italy*
[4]*Dipartimento di Fisica Nucleare e Teorica, Università degli Studi di Pavia, Via Bassi 6, I-27100 Pavia, Italy*
[5]*INAF – Osservatorio Astronomico di Cagliari, località Poggio dei Pini, strada 54, I-09012 Capoterra, Italy*
[6]*Mullard Space Science Laboratory, University College London, Holmbury St. Mary, Dorking, Surrey RH5 6NT*
[7]*Dipartimento di Fisica e Astronomia, Università degli Studi di Padova, Via Marzolo 8, I-35131 Padova, Italy*
[8]*Kepler Institute of Astronomy, University of Zielona Góra, Lubuska 2, 65-265 Zielona Góra, Poland*

## ABSTRACT

RX J0822−4300 is the central compact object associated with the Puppis A supernova remnant. Previous X-ray observations suggested RX J0822−4300 to be a young neutron star with a weak dipole field and a peculiar surface temperature distribution dominated by two antipodal spots with different temperatures and sizes. An emission line at 0.8 keV was also detected. We performed a very deep (130-ks) observation with *XMM–Newton*, which allowed us to study in detail the phase-resolved properties of RX J0822−4300. Our new data confirm the existence of a narrow spectral feature, best modelled as an emission line, only seen in the 'soft'-phase interval – when the cooler region is best aligned to the line of sight. Surprisingly, comparison of our recent observations to the older ones yields evidence for a variation in the emission-line component, which can be modelled as a decrease in the central energy from ∼0.80 keV in 2001 to ∼0.73 keV in 2009–10. The line could be generated via cyclotron scattering of thermal photons in an optically-thin layer of gas, or, alternatively, it could originate in low-rate accretion by a debris disc. In any case, a variation in energy, pointing to a variation of the magnetic field in the line-emitting region, cannot be easily accounted for.

## 1 INTRODUCTION

Central compact objects (CCOs) in supernova remnants (SNRs) are a handful of about 10 point-like, thermally emitting X-ray sources located close to the geometrical centres of non-plerionic SNRs, with no counterparts at any other wavelength. CCOs are supposed to be young, isolated, radio-quiet neutron stars (see De Luca 2008, for a review).

While the first discovered CCO (the one in the RCW103 SNR) turned out to be a unique object (De Luca et al. 2006), results of X-ray timing on a subsample of sources with fast periodicities have recently set very useful constraints for a general picture of CCOs as a class. Analysis of multi-epoch *XMM–Newton* and *Chandra* observations of 1E 1207.4−5209 inside G296.5+10.0 and of CXOU J185238.6+004020 at the centre of Kes 79 ($P = 424$ and 105 ms, respectively) yielded unambiguous evidence for very small period derivatives (Halpern & Gotthelf 2010, 2011). This implies, under standard rotating dipole assumptions, characteristic ages exceeding the age of the host SNRs by more than three orders of magnitude, as well as very small ($\lesssim 10^{11}$ G) dipole magnetic fields. This points to an interpretation of such sources as 'anti-magnetars' – weakly magnetized isolated neutron stars – born with a spin period almost identical to the currently observed one.

Such a picture has been recently strengthened by the discovery of 112-ms pulsations from RX J0822−4300, the CCO in the Puppis A SNR (Gotthelf & Halpern 2009), in two archival *XMM–Newton* data sets collected in 2001, previously used by Hui & Becker (2006) for a comprehensive spectral analysis of the neutron star and of the surrounding SNR. Gotthelf, Perna & Halpern (2010), using a 2010

⋆E-mail: deluca@iasf-milano.inaf.it

*Chandra* observation, set a $2\sigma$ upper limit of $3.5 \times 10^{-16}$ s s$^{-1}$ to the period derivative, corresponding to a dipole magnetic field $B < 2 \times 10^{11}$ G, and to a characteristic age $\tau_c > 5$ Myr, much larger than the age of the host SNR (3.7 kyr, Winkler et al. 1988). Thus, RX J0822−4300 can be included in the antimagnetar family.

The X-ray emission properties of RX J0822−4300 are very peculiar. Gotthelf & Halpern (2009) report on a 180° shift in the phase of the pulse peaks occurring abruptly at ∼1.2 keV. This has been interpreted as due to the existence of two antipodal spots on the star surface, with two different temperatures (‘warm’ and ‘hot’) – indeed seen in the emission spectrum as two different blackbody components. The star rotation bringing into view or hiding such regions makes the observed spectrum to change from a soft phase (maximum alignment of the warm spot with the line of sight) to a hard phase (maximum alignment of the hot spot). Even more intriguing, Gotthelf & Halpern (2009) report on the evidence for an emission line at ∼0.8 keV, possibly associated with the warm spot.

Here we report on a very deep *XMM–Newton* observation of RX J0822−4300, performed in 2009 and 2010, which allows us to study in detail the phase-resolved emission properties of the neutron star and to check for any time-variability.

## 2 OBSERVATIONS AND DATA REDUCTION

Our study is based on a very deep (∼130 ks) observation with *XMM–Newton*, originally scheduled to fit into a single satellite orbit (revolution number 1836), started on 2009 December 17. However, the launch of the *Helios 2B* spacecraft required support from *XMM–Newton* ground stations, which resulted in an ∼50 ks data gap in the middle of the orbit. The observation was completed on 2010 April 5. We also included in our study the two archival, shorter data sets, obtained on 2001 April 15 (∼29 ks) and 2001 November 8 (∼24 ks), used by Gotthelf & Halpern (2009) in their previous investigations. A summary of the observations is reported in Table 1.

We focus on data obtained by the pn detector (Strüder et al. 2001) of the European Photon Imaging Camera (EPIC). The detector was operated in the small-window mode (time-resolution of 5.7 ms, field of view of $4.3 \times 4.3$ arcmin$^2$). The thin optical filter was used in all observations. EPIC/pn observation data files were processed with the *XMM–Newton* SAS v11 (Science Analysis Software) using standard pipelines.

## 3 TIMING ANALYSIS

For our timing analysis, we selected the pn source events from a circular region of 30 arcsec radius, including only 1- and 2-pixel events (PATTERN 0 to 4) with the default flag mask. Photon arrival times were converted to the Solar system barycentre TBD time using the *Chandra* position (Gotthelf & Halpern 2009).

Before the analysis presented in Gotthelf & Halpern (2009), the pulsations in RX J0822−4300 eluded detection for many years, because of a phase shift of about half a cycle between the nearly sinusoidal profiles in the soft (<1.2 keV) and hard bands. By computing $Z_1^2$ periodograms (Buccheri et al. 1983), we found that the energy bands that maximize the pulsed signal are 0.46–1.17 and 1.25–5.12 keV. The resulting soft and hard pulse profiles of the individual *XMM–Newton* observations (Table 1) were cross-correlated, shifted and summed to create two distinct pulse profile templates. Owing to the higher signal-to-noise ratio of the 1.25–5.12 keV profiles, we carried out the analysis in the hard band, but we checked that the results are fully consistent with those obtained in the soft band. The hard pulse profiles from temporal segments of the *XMM–Newton* observations were cross-correlated with the template to determine times of arrival at each epoch. By means of a phase-fitting analysis (e.g. Dall’Osso et al. 2003), we measured the periods given in Table 1 (we also confirm the measurements by Gotthelf & Halpern 2009 for observations A and B).

We attempted to obtain a full phase-connected timing solution (i.e. a solution that accounts for all the spin cycles of the pulsar) for the longest possible time. Unfortunately, it was not possible to univocally phase-connect the solution found for the two adjacent data sets C and D to other observations (the uncertainty on the phase – propagated along the large time-span to other observations – largely exceeds 1 cycle).

A linear fit of the periods in Table 1 gives for the period derivative $(9 \pm 12) \times 10^{-17}$ s s$^{-1}$, which translates into $3\sigma$ limits $-2.7 \times 10^{-16} < \dot{P} < 4.5 \times 10^{-16}$ s s$^{-1}$ (in agreement with the limits recently published in Gotthelf et al. 2010). The period derivative at the Solar system barycentre results from the *intrinsic* pulsar spin-up/spin-down plus the contributions due to any external gravitational field and the pulsar proper motion (e.g. Phinney 1992). In the case of RX J0822−4300, for an assumed distance of 2.2 kpc (Reynoso et al. 1995), the Galactic contribution is negative and negligible ($\sim -2 \times 10^{-20}$ s s$^{-1}$) and that produced by the proper motion ($165 \pm 25$ mas yr$^{-1}$; Winkler & Petre 2007) is $1.6^{+0.6}_{-0.4} \times 10^{-17}$ s s$^{-1}$. While the total (Galactic + proper motion) contribution does not impact significantly the current limits on the $\dot{P}$ of RX J0822−4300, we note that it is larger than the period derivative measured for the CCO in Kes 79 ($\sim 8.7 \times 10^{-18}$ s s$^{-1}$; Halpern & Gotthelf 2010).

The phase of the pulse peak is energy-dependent (see Fig. 1), with an offset of $0.44 \pm 0.02$ between the profiles seen at lower ($E < 1.17$ keV) and higher energy ($E > 1.25$ keV), the transition occurring quite abruptly at ∼1.2 keV, consistent with the findings of Gotthelf & Halpern (2009). As shown in Fig. 2, the pulsed fraction (PF) decreases from ∼15 per cent in the 0.4–0.6 keV energy range to <2 per cent in the 1.1–1.3 keV range, then grows again to ∼15 per cent at $E > 2$ keV, in good agreement with the model by Gotthelf et al. (2010).

**Table 1.** Journal of the *XMM–Newton* observations.

| Data set | Observation ID | Date[a] (MJD TBD) | Duration[b] (ks) | Pulse period[c] (ms) |
|---|---|---|---|---|
| A | 0113020101 | 52014.471 | 25.1 | 112.799 42(5)[d] |
| B | 0113020301 | 52221.898 | 23.0 | 112.799 44(4)[d] |
| C | 0606280101[e] | 55183.064 | 41.9 | 112.799 453(2) |
| D | | 55184.065 | 43.0 | |
| E | 0606280201 | 55291.622 | 42.2 | 112.799 45(1) |

[a]Mid-point of the observation.
[b]Time between the first and last events.
[c]$1\sigma$ errors in the last digit are quoted in parentheses.
[d]Gotthelf & Halpern (2009).
[e]This observation was broken into two segments (see text).

## 4 PHASE-RESOLVED SPECTROSCOPY

Thorough phase-integrated spectroscopy of RX J0822−4300 has been published by Hui & Becker (2006). We will focus here on phase-resolved spectroscopy.

RX J0822−4300 lies in a very complex environment, which makes background subtraction a critical task. Using phase-integrated data, we evaluated an optimal selection of source events by maximizing the signal-to-noise ratio in the 0.3–10 keV range as a function of source extraction radius. The best choice turned out to be a circle of radius 17.5 arcsec centred at the *Chandra* position. We extracted the background spectrum from an annular region with radii of 28 and 35 arcsec centred on the source. Different sources and/or background regions yield consistent results within $\sim 2\sigma$.

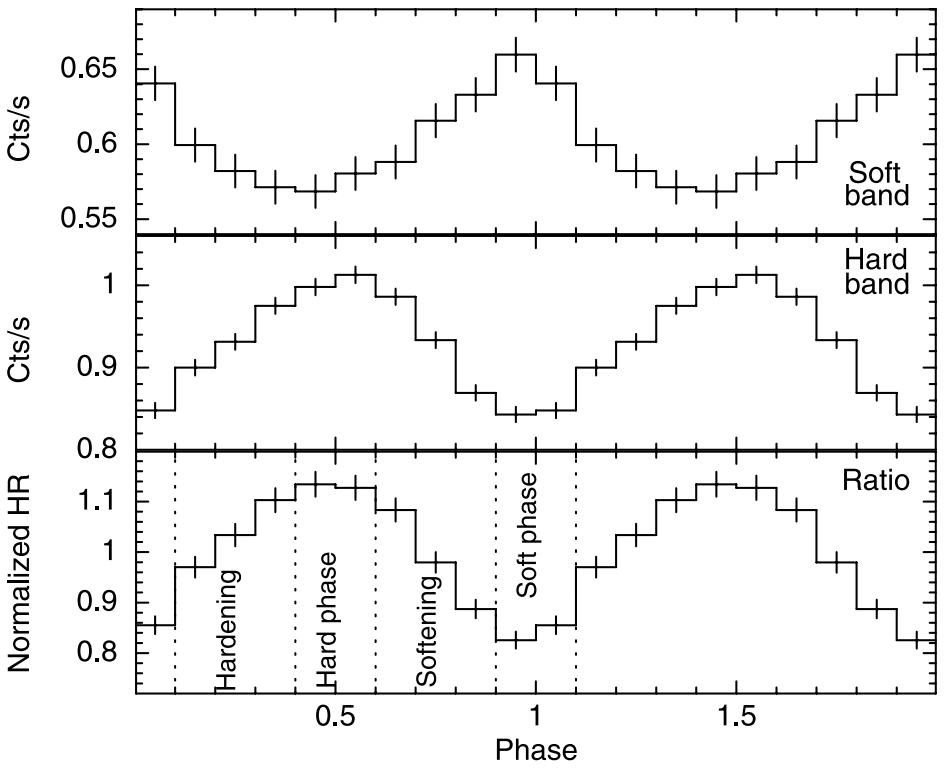

**Figure 1.** Background-subtracted folded light curves for RX J0822–4300 in the soft energy range (0.46–1.17 keV, upper panel) and in the hard energy range (1.25–5.12 keV, middle panel). The lower panel shows the hardness ratio (hard to soft), normalized to its average value. Phase intervals used for phase-resolved spectroscopy are also marked.

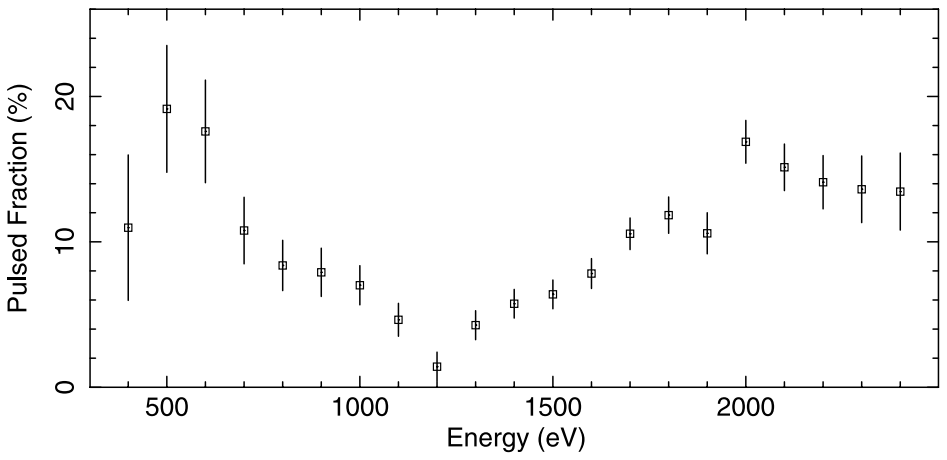

**Figure 2.** Energy dependence of PF. The PF was evaluated in overlapping 0.2 keV energy bins, incremented in 0.1 keV steps, as the ratio between the number of counts above the minimum and the total number of counts. Background has been subtracted. A clear trend is apparent, with a minimum in the 1.1–1.3 keV energy range.

We aligned the photon phases for each observation by cross-correlating the pulse profiles (see Section 3). We generated a combined event list for first-epoch observations (i.e. data collected in 2001 and used by Gotthelf & Halpern 2009) and one for second-epoch observations (i.e. our new data, collected in 2009–10), in view of the large intercurring time-span. We extracted four phase-resolved spectra from both first-epoch and second-epoch data, based on the intervals marked as 'Hard phase', 'Soft phase', 'Hardening' and 'Softening' in Fig. 1. Phase-resolved spectra were rebinned with at least 100 counts per bin and so that the energy resolution was not oversampled by more than a factor of 3. Response matrices and effective area files for each epoch were generated by combining (weighted by exposure time) the corresponding files generated using the SAS tools RMFGEN and ARFGEN. Spectral modelling was performed with the XSPEC v12.6.0 package in the 0.3–4.0 keV energy range. To account for the effects of interstellar absorption, we used the tbabs model in XSPEC, with the abundances by Wilms, Allen & McCray (2000). We quote errors at 90 per cent confidence level for a single interesting parameter, unless otherwise specified.

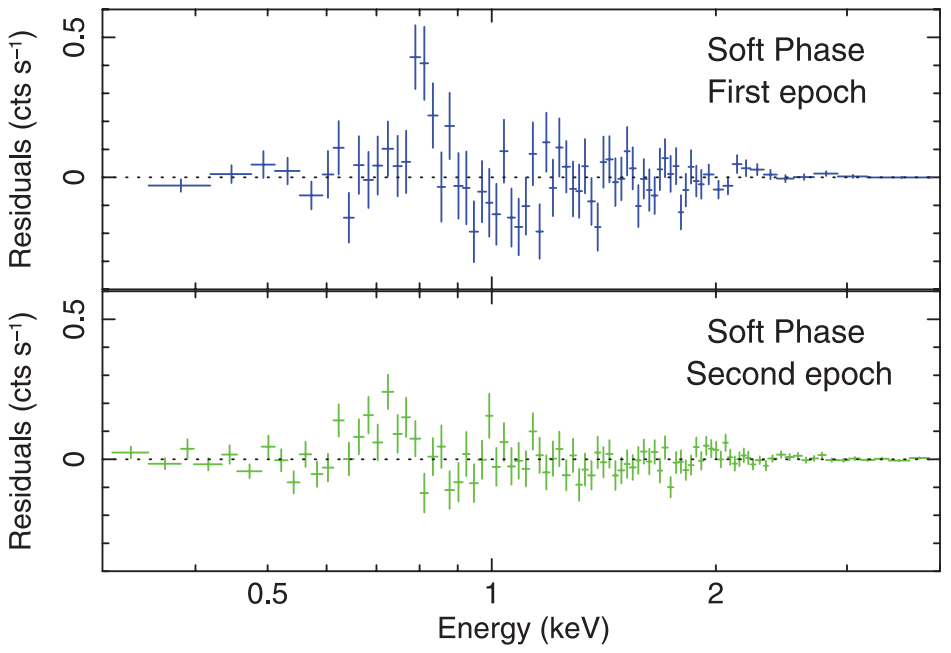

**Figure 3.** Residuals (in counts $s^{-1}$) of two-blackbody fits to the soft-phase spectra (see Fig. 1). A structure in the 0.6–0.9 keV range is seen in both epochs, although with different shape and intensity.

First, we repeated the exercise performed by Gotthelf & Halpern (2009), fitting a double-blackbody model to our data. A simultaneous fit to the four phase-resolved spectra was performed for each epoch. The blackbody normalizations were allowed to vary as a function of phase, while the temperatures and $N_{\rm H}$ were held fixed in order to constrain a single value for all phase ranges. This model yields a reduced $\chi^2$ of 1.20 for 298 degrees of freedom (d.o.f.) and of 1.23 for 389 d.o.f for the first- and second-epoch data, respectively. Although modulation of the emitting radii accounts for the bulk of the spectral variation, structured residuals in the 0.6–0.9 keV range are apparent in the soft phase both in the 2001 data set (as already reported by Gotthelf & Halpern 2009) and in our deeper 2009–10 data set, which confirms the existence of a phase-dependent spectral feature.

Very interestingly, while the phase-resolved best-fitting parameters do not change as a function of epoch – they can be linked in a simultaneous two-epoch fit (more details below) – the deviation from the continuum in the soft phase has a somewhat different shape in 2001 with respect to 2009–10 (see Fig. 3), suggesting a possible time-variability of the spectral feature.

Indeed, such variation is fully apparent when plotting together the two soft-phase spectra (see Fig. 4). To quantify the significance of the spectral change in a model-independent way, we compared the distributions of the source events' energies observed in the two epochs using the Kolmogorov–Smirnov test. The probability of a statistical fluctuation producing the apparent difference in the energy range where the feature is seen ($\sim$0.6–0.9 keV) turned out to be $3 \times 10^{-6}$.

As a second step, in order to model the feature, we focused on the two soft-phase spectra. Following Gotthelf & Halpern (2009), we added a Gaussian emission line to the two-blackbody model. Indeed, this yields a much better fit with no structured residuals in the 0.6–0.9 keV range. As expected, the line component varies as a function of time, its central energy being higher in the first epoch ($\sim$0.80 keV) than in the second epoch ($\sim$0.73 keV). The significance of such line components was studied by calibrating the F-statistics using simulations of the null model (the double-blackbody model) as suggested by Protassov et al. (2002). Each epoch was studied individually. Running $10^4$ Monte Carlo simulations, we estimated that the significance of the line is greater than 99.99 per cent in the second-epoch spectrum, and greater than 99.97 per cent in the first-epoch spectrum. Performing a simultaneous fit to the two spectra, the two-blackbody plus emission-line model yields a reduced $\chi^2$ of 1.17 (155 d.o.f.) when all parameters (both continuum and line) are linked, while by allowing the line central energy to vary between the first and second epochs, the fit improves to a reduced $\chi^2$ of 1.02 (154 d.o.f.) (see Fig. 4). The chance occurrence probability of such an improvement is $\sim 3 \times 10^{-6}$, as evaluated by an F-test. Such a result is confirmed by Monte Carlo simulations: assuming the best-fitting model of second-epoch data, a line centroid as observed in first-epoch data could be obtained by chance with a probability as low as $\sim 5 \times 10^{-6}$. The fit does not improve significantly by allowing further line parameters to vary. There is no evidence for any variation in the continuum parameters between the two epochs.

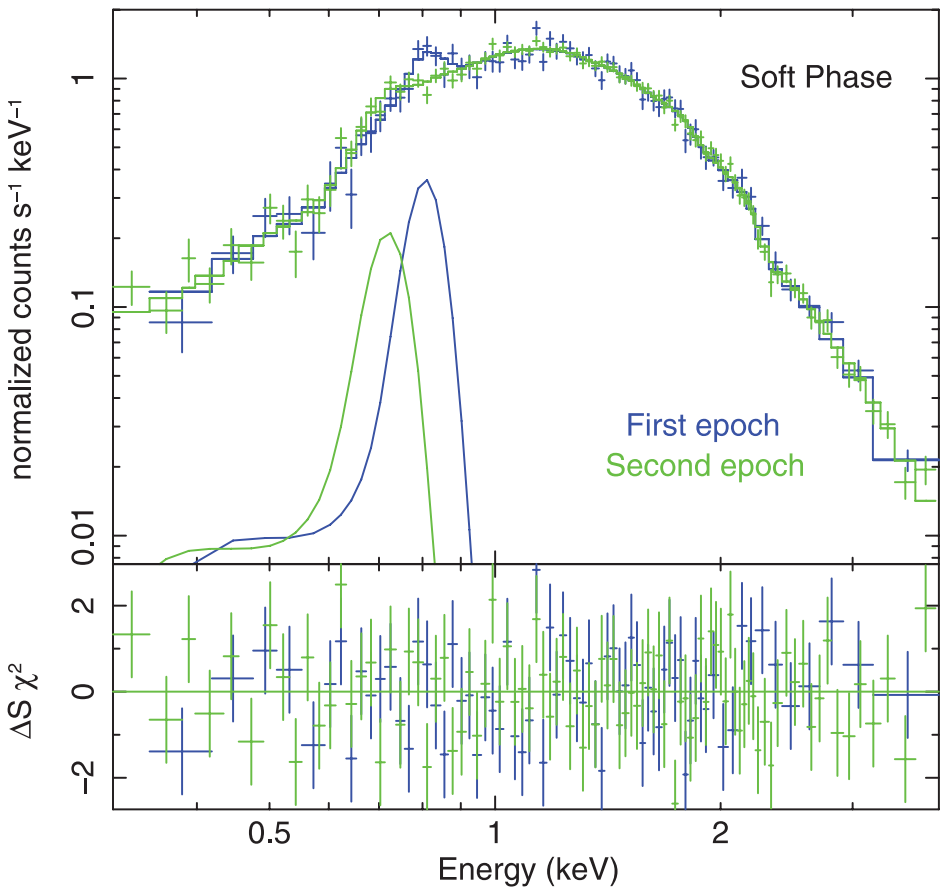

**Figure 4.** Upper panel: soft-phase spectra for the two epochs. First epoch: blue. Second epoch: green. The variation in the 0.6–0.9 keV range is apparent. The best-fitting model, two blackbody components and a variable Gaussian emission line, is superimposed. The line component at both epochs is also shown. The continuum components do not change as a function of time. Lower panel: residuals to the best-fitting model. The model yields a very good description of the two spectra ($\chi^2_\nu = 1.02$, 154 d.o.f.).

We also tried to model the two soft-phase spectra using variable *absorption* features. Adding a single Gaussian absorption line (gabs in XSPEC) at a higher energy than the one required by the emission-line model (∼1.00 and ∼0.89 keV in the first and second epochs, respectively) yields a much worse fit than the emission-line model (reduced $\chi^2$ of 1.20 for 154 d.o.f.). Adding *two* lines at ∼0.97 and ∼0.63 keV in the first epoch and at ∼0.88 and ∼0.55 keV in the second epoch still yields a remarkably worse description of the data (reduced $\chi^2 = 1.15$ for 150 d.o.f.). As a further test, we tried low-*B* neutron star atmosphere models for the continuum (nsa, Zavlin, Pavlov & Shibanov 1996; nsatmos, Heinke et al. 2006). As for the blackbody case, two components with different temperatures are needed. Addition of a variable Gaussian emission line is still favoured ($\chi^2 = 1.05$ for 154 d.o.f.) with respect to a single variable absorption line, as well as to two variable absorption lines ($\chi^2 = 1.23$ for 150 d.o.f. for the latter model).

Then, the whole analysis was repeated for the other phase intervals (hard, softening and hardening). No significant improvement in the fit was obtained by adding an emission line to the two-blackbody model (the same is true using absorption features). We assessed that, in each phase interval, the continuum did not change as a function of epoch and that there are no systematic variations between the first and second epochs in the 0.5–1 keV energy range (such results indicate that the long-term variability of the feature cannot have an instrumental origin).

As a final step, we performed a simultaneous fit to all the spectra based on the results described above. We used the two-blackbody plus emission-line model. The blackbody normalizations are phase-dependent, but not epoch-dependent; the line normalization is phase-dependent and its central energy is epoch-dependent. This yields a reduced $\chi^2 = 1.13$ for 691 d.o.f. In such models, $N_{\rm H}$ is $(5.0 \pm 0.1) \times 10^{21}$ cm$^{-2}$, the warm blackbody has a temperature $kT_{\rm W} = 265 \pm 15$ eV and a radius ranging from 2.27 km (soft phase) to 2.04 km (hard phase), and the hot blackbody has a temperature $kT_{\rm H} = 455 \pm 20$ eV and a radius ranging from 0.53 km (soft phase) to 0.65 km (hard phase). The line component is narrow ($\sigma < 40$ eV). In 2001, the line energy is $0.80 \pm 0.01$ keV and the equivalent width (EW) ranges from ∼53 eV in the soft phase to <10–15 eV in other intervals (where the line is not significant). In 2009–10, the line energy is $0.73^{+0.01}_{-0.02}$ keV and the EW ranges from ∼45 eV (soft phase) to <12–18 eV in other intervals. The unabsorbed flux of the line between 0.3 and 10 keV is ∼3 per cent of the flux of the continuum in the same energy range.

## 5 DISCUSSION

Our multi-epoch *XMM–Newton* analysis shows a phase-dependent, *time-variable* spectral feature, best modelled as an emission line with a variable central energy, in the X-ray spectrum of the 'anti-magnetar' candidate RX J0822−4300.

To put such a peculiar result in context, we first note that our observations confirm the picture of RX J0822−4300 as a weakly magnetized neutron star. Indeed, we improved the upper limit on $\dot{P}$, bringing the dipole component of the magnetic field down to $<2.3 \times 10^{11}$ G at $3\sigma$ level. Based on larger statistics, we also confirm the geometric model by Gotthelf et al. (2010), explaining the phase-resolved thermal emission with two antipodal spots of different temperature (compare e.g. our Fig. 2 to their fig. 6). Lack of any measurable time-variation in the continuum properties suggests that the warm and hot regions are intrinsic, persistent features in the thermal map of the star. To explain such a large surface anisotropy for RX J0822−4300 (and for CCOs in general), we may consider that a large difference in intensity could exist between the internal and the external magnetic fields of the neutron star, as proposed by Turolla et al. (2011) to explain the properties of the low magnetic field soft gamma repeater SGR 0418+5729 (Rea et al. 2010). In our case, an internal (toroidal + poloidal) field of a few $10^{13}$ G would be large enough to effectively channel the heat flux from the core (Geppert, Küker & Page 2004, 2006), but would not induce crustal fractures with consequent magnetar-like bursting activity.

The most natural interpretation of the variable emission line is that of a cyclotron feature produced by electrons. If its central energy is associated with the fundamental $e^-$ cyclotron frequency, the magnetic field in the line-emitting region would be (6–7) × $10^{10}(1 + z)$ G (where $z \sim 0.25$ is the gravitational redshift). This is quite compatible with the upper limit from the spin-down rate.

The line is very narrow ($\sigma \leq 40$ eV). If it is a cyclotron line, then $\Delta E/E = \Delta B/B$ and the relative variation of $B$ over the emitting region (conservatively) needs to be $\leq$10 per cent. Thus, the line should be produced in a very compact region. A variation of the central energy of the feature would require a change either in the position of the emitting plasma within a non-variable magnetic field, or in the intensity of the magnetic field itself. Lack of changes in the phase-resolved continuum emission rules out simpler, purely geometric explanations such as precession of the neutron star.

To explain the generation of an emission line in the spectrum of an isolated neutron star, the possibility of cyclotron scattering of surface thermal photons by a geometrically-thin, optically-thin layer of plasma could be considered. However, under simple assumptions (emission from the entire star surface; plane-parallel geometry; pure, conservative scattering), a scattering layer would produce an absorption line. One might invoke a spatially limited scattering medium, possibly some distance away from the star surface. Photons coming from the part of the surface not covered by the layer could be scattered along the line of sight giving rise to an emission feature. The value of $B$ derived from the line energy is somehow smaller than the upper limit on the surface field, so the line could indeed form at some height in the magnetosphere. A confined medium seems also to be required by the results of phase-resolved spectroscopy which shows that the emission line is seen mostly when the cooler spot is into view. Still, such a picture seems rather contrived, since the nature of the layer and the mechanism keeping the plasma suspended and confined in a compact blob remains to be understood.

To ease the problem, an energy source unrelated to the surface thermal emission should be invoked to excite $e^-$ to higher Landau levels in the line-emitting region. Indeed, Nelson, Salpeter & Wasserman (1993) predicted that for neutron stars accreting at a low rate ($L_{accr} < 10^{34}$ erg s$^{-1}$), and endowed with magnetic fields of $10^{11}$–$10^{13}$ G, accreting ions may lose energy to atmospheric electrons via magnetic Coulomb collisions. Electrons, excited to high Landau levels, radiatively decay and part of the cyclotron photons are expected to escape producing an emission line. According to Nelson et al. (1993), at $B < 10^{12}$ G, the fraction of the accretion-powered flux escaping in the line is expected to be very small ($\ll$5 per cent), the largest part being reprocessed and emitted in a thermal continuum. Thus, one should postulate that the bulk of the X-ray luminosity of RX J0822−4300 is accretion powered, at variance with observations. It would require an accretion rate of $\sim 2 \times 10^{13}$ g s$^{-1}$ implying – under standard relations for propeller spin-down (Menou et al. 1999) – a $\dot{P}$ value more than 10 times larger than our upper limit. Although the model by Nelson et al. (1993) does not fit to our case, low-level accretion of supernova fallback material (which cannot be ruled out, based on X-ray timing, as well as on the optical upper limits set by Mignani et al. 2009) could play some role in generating an emission line in a low-$B$ atmosphere. A detailed investigation of such possibility is beyond the scope of this Letter.

We will not go into further speculations about the line-emitting mechanism. We stress that the evidence for time-evolution of the spectral feature is model-independent and represents to date the first evidence for variability in an 'antimagnetar' candidate. Likely, such 'activity' is related to a variation in the magnetic field of the star. An $\sim$10 per cent decrease in $\sim$8 yr seems vastly too steep to be attributed to the large-scale dipole field. Possibly, we are witnessing evolution of a localized multipole component, dominating close to the star surface. This would hint at the presence of a large internal field, as proposed to explain the anisotropic thermal map of the star. Precise X-ray timing, assessing the $\dot{P}$ and measuring the star dipole field will add a crucial piece of information. Coupled with further sensitive phase-resolved spectroscopy to monitor spectral variability, this could help to solve the puzzles set by our results on RX J0822−4300 which would have important implications for the understanding of the nature of CCOs and of their relations with other families of neutron stars.

## ACKNOWLEDGMENTS

This study was based on observations obtained with *XMM–Newton*, an ESA science mission with instruments and contributions directly funded by ESA Member States and NASA. We thank F. Gastaldello for useful advice about statistics and A. Possenti for discussions. This work was partially supported by the ASI-INAF I/009/10/0 agreement. PE acknowledges financial support from the Autonomous Region of Sardinia through a research grant under the programme PO Sardegna FSE 2007–2013, L.R. 7/2007.

This paper has been typeset from a TEX/LATEX file prepared by the author.

# THE OUTBURST DECAY OF THE LOW MAGNETIC FIELD MAGNETAR SGR 0418+5729

N. Rea[1], G. L. Israel[2], J. A. Pons[3], R. Turolla[4,5], D. Viganò[3], S. Zane[5], P. Esposito[6], R. Perna[7], A. Papitto[1], G. Terreran[1,4], A. Tiengo[6,8,9], D. Salvetti[6,9,10], J. M. Girart[1], Aina Palau[1], A. Possenti[11], M. Burgay[11], E. Göğüş[12], G. A. Caliandro[1], C. Kouveliotou[13], D. Götz[14], R. P. Mignani[5,6,15], E. Ratti[16], and L. Stella[2]
[1] Institute of Space Sciences (CSIC–IEEC), Campus UAB, Faculty of Science, Torre C5-parell, E-08193 Barcelona, Spain
[2] INAF/Osservatorio Astronomico di Roma, via Frascati 33, I-00040 Monteporzio Catone, Italy
[3] Department de Fisica Aplicada, Universitat d'Alacant, Ap. Correus 99, E-03080 Alacant, Spain
[4] Dipartimento di Fisica e Astronomia, Università di Padova, via F. Marzolo 8, I-35131 Padova, Italy
[5] Mullard Space Science Laboratory, University College London, Holmbury St. Mary, Dorking, Surrey RH5 6NT, UK
[6] INAF/Istituto di Astrofisica Spaziale e Fisica Cosmica-Milano, via E. Bassini 15, I-20133 Milano, Italy
[7] JILA and Department of Astrophysical and Planetary Sciences, University of Colorado, Boulder, CO 80309, USA
[8] IUSS-Istituto Universitario di Studi Superiori, Piazza della Vittoria 15, I-27100 Pavia, Italy
[9] INFN Istituto Nazionale di Fisica Nucleare, Sezione di Pavia, Via Bassi 6, I-27100 Pavia, Italy
[10] Dipartimento di Fisica Nucleare e Teorica, Università degli Studi di Pavia, via Bassi 6, I-27100 Pavia, Italy
[11] INAF/Osservatorio Astronomico di Cagliari, località Poggio dei Pini, strada 54, I-09012 Capoterra, Italy
[12] Sabancı University, Orhanlı-Tuzla, 34956 İstanbul, Turkey
[13] NASA Marshall Space Flight Center, Huntsville, AL 35812, USA
[14] AIM (UMR 7158 CEA/DSM-CNRS-Université Paris Diderot) Irfu/Service d'Astrophysique, Saclay, F-91191 Gif-sur-Yvette Cedex, France
[15] Kepler Institute of Astronomy, University of Zielona Góra, Lubuska 2, 65-265 Zielona Góra, Poland
[16] SRON-Netherlands Institute for Space Research, Sorbonnelaan 2, 3584 CA Utrecht, The Netherlands

## ABSTRACT

We report on the long-term X-ray monitoring of the outburst decay of the low magnetic field magnetar SGR 0418+5729 using all the available X-ray data obtained with *RXTE*, *Swift*, *Chandra*, and *XMM-Newton* observations from the discovery of the source in 2009 June up to 2012 August. The timing analysis allowed us to obtain the first measurement of the period derivative of SGR 0418+5729: $\dot{P} = 4(1) \times 10^{-15}$ s s$^{-1}$, significant at a $\sim 3.5\sigma$ confidence level. This leads to a surface dipolar magnetic field of $B_{\rm dip} \simeq 6 \times 10^{12}$ G. This measurement confirms SGR 0418+5729 as the lowest magnetic field magnetar. Following the flux and spectral evolution from the beginning of the outburst up to $\sim$1200 days, we observe a gradual cooling of the tiny hot spot responsible for the X-ray emission, from a temperature of $\sim$0.9 to 0.3 keV. Simultaneously, the X-ray flux decreased by about three orders of magnitude: from about $1.4 \times 10^{-11}$ to $1.2 \times 10^{-14}$ erg s$^{-1}$ cm$^{-2}$. Deep radio, millimeter, optical, and gamma-ray observations did not detect the source counterpart, implying stringent limits on its multi-band emission, as well as constraints on the presence of a fossil disk. By modeling the magneto-thermal secular evolution of SGR 0418+5729, we infer a realistic age of $\sim$550 kyr, and a dipolar magnetic field at birth of $\sim 10^{14}$ G. The outburst characteristics suggest the presence of a thin twisted bundle with a small heated spot at its base. The bundle untwisted in the first few months following the outburst, while the hot spot decreases in temperature and size. We estimate the outburst rate of low magnetic field magnetars to be about one per year per galaxy, and we briefly discuss the consequences of such a result in several other astrophysical contexts.

## 1. INTRODUCTION

Neutron stars showing magnetar-like activity (comprising the anomalous X-ray pulsars, soft gamma repeaters, and a high magnetic field pulsar) are a small group of X-ray pulsars (about 20 objects) with spin periods between 0.3–12 s, whose strong persistent and/or flaring emission are hard to explain using the common scenarios for rotation-powered pulsars or accreting pulsars. In fact, the very strong X-ray emission of these objects ($L_{\rm X} \sim 10^{35}$ erg s$^{-1}$) is too high and/or variable to be fed by the rotational energy alone (as in the radio pulsars), and no evidence for a companion star has been found, hence ruling out accretion in a binary. Accretion from a fossil disk remnant of the supernova explosion might be responsible for part of the observational properties of these objects, but it fails to explain some of their characteristics, such as the flaring X-ray activity. Their inferred magnetic fields, under the typical assumption of magnetic dipolar losses alone, appear to be as high as $B_{\rm dip} \simeq 3.2 \times 10^{19}\sqrt{P\dot{P}} \sim 10^{14}$–$10^{15}$ G (see Mereghetti 2008 for a review). These strong fields are believed to either form via a dynamo action in a rapidly rotating protoneutron star (<3 ms; Thompson & Duncan 1995) or be fossil field remnants of a highly magnetic massive star ($\sim$1 kG; Ferrario & Wickramasinghe 2006). Because of these high $B$ fields, the emission of "magnetars" is thought to be powered by the decay and the instability of their strong fields (Duncan & Thompson 1992; Thompson & Duncan 1993; Thompson et al. 2002). Their powerful X-ray output is usually well modeled by thermal emission from the neutron star hot surface, reprocessed in a twisted magnetosphere through resonant cyclotron scattering (Thompson et al. 2002; Nobili et al. 2008; Rea et al. 2008; Zane et al. 2009), a process favored under these extreme magnetic conditions. On top of their persistent X-ray emission, magnetars emit very peculiar flares and outbursts on several timescales, from fractions of a second to years reaching very high, super-Eddington luminosities ($10^{38}$–$10^{46}$ erg s$^{-1}$). These flares are most probably caused by rearrangements of the twisted magnetic field lines, either accompanied or triggered by fractures of the neutron-star crust (Thompson & Duncan 1995; Perna & Pons 2011).

Transient events are a characteristic signature of magnetar emission, and are one of the main ways to discover new sources of this class and study its physics. From the discovery of the first transient less than a decade ago, we now count about a dozen outbursts, which increased the number of known magnetars by a third in six years (see Rea & Esposito 2011; Rea 2013 for recent reviews). Magnetar outbursts might involve their multi-band emission resulting in an increased activity from the radio to hard X-ray, usually with a soft X-ray flux increase of a factor of 10–1000 with respect to the quiescent level. An associated X-ray spectral evolution is often observed, with a spectral softening during the outburst decay (Rea et al. 2009). The flux decay timescale varies substantially from source to source, ranging from a few weeks to several years (Rea & Esposito 2011; Pons & Rea 2012).

The extensive follow-up of magnetars undergoing an outburst yielded the most unexpected discovery of recent years in the magnetar field. Prompted by the detection of typical magnetar-like bursts and a powerful outburst, a new transient magnetar with a spin period of ∼9 s was discovered in 2009, namely SGR 0418+5729 (van der Horst et al. 2010; Esposito et al. 2010). However, after more than two years of extensive monitoring, no period derivative was detected. This led to an upper limit on the source surface dipolar field of $B_{\rm dip} < 7.5 \times 10^{12}$ G (Rea et al. 2010). For the first time, we detected a magnetar with a low dipolar magnetic field, showing that a critical magnetic field is not necessary for a neutron star in order to display magnetar-like activity. In turn, this means that many seemingly normal pulsars could turn out to be magnetars at anytime (this was supported by the discovery of a second low-$B$ magnetar following soon after that of the first; Rea et al. 2012; Scholz et al. 2012). After the discovery of this low dipolar magnetic field soft gamma repeater, several models were put forward to explain its puzzling emission. They involve the possible presence of a fallback disk slowing down the pulsar up to the current spin period (Alpar et al. 2011), a tiny inclination angle between the magnetic and rotational axis resulting in a higher inferred magnetic field (Tong & Xu 2012), a pulsar with a strongly magnetized core (Soni 2012), an old quark nova (Ouyed et al. 2011), or a massive highly magnetized, rotating white dwarf (Malheiro et al. 2012). In Rea et al. (2010) and Turolla et al. (2011), we suggested that a non-dipolar component of the field, larger than the measured dipolar one, can be responsible for the behavior of this magnetar, if it has a relatively old age (≃1 Myr).

In this paper we present the complete study of the outburst of the low dipolar magnetic field magnetar SGR 0418+5729, from the first outburst phases until about three years after its onset. This long-term monitoring campaign using several X-ray satellites, allowed us to estimate SGR 0418+5729's period derivative, and follow the cooling of its surface temperature during the outburst decay up to the (probable) quiescent level. Furthermore, we inferred limits on its emission in the radio, millimeter, optical, and gamma-ray bands. We discuss our findings in terms of the magneto-thermal history of this magnetar, discuss the current limits on the presence of a fossil disk, and present some discussion on the broader consequences of the discovery of low magnetic field magnetars.

## 2. X-RAY OBSERVATIONS AND DATA REDUCTION

In this study, we used data obtained from several different satellites (see Table 1 and Figure 1 for a summary). We describe below the observations and data analysis. Parts of the data we used in this paper were already published by van der Horst et al. (2010), Esposito et al. (2010), and Rea et al. (2010).

### 2.1. *Swift* Data

The X-Ray Telescope (XRT; Burrows et al. 2005) on board *Swift* uses a front-illuminated CCD detector sensitive to photons between 0.2 and 10 keV. Two main readout modes are available: photon counting (PC) and windowed timing (WT). PC mode provides two-dimensional imaging information and a 2.5073 s time resolution; in WT mode only one-dimensional imaging is preserved, achieving a time resolution of 1.766 ms. The XRT data were uniformly processed with XRTPIPELINE (version 12, in the HEASOFT software package version 6.11), and filtered and screened with standard criteria, correcting for effective area, dead columns, etc. The source counts were extracted within a 20 pixel radius (one XRT pixel corresponds to about 2″.36). For the spectroscopy, we used the spectral redistribution matrices in CALDB (20091130; matrices version v013 and v014 for the PC and WT data, respectively), while the ancillary response files were generated with XRTMKARF, and they account for different extraction regions, vignetting, and point-spread function (PSF) corrections.

### 2.2. *RXTE* Data

The Proportional Counter Array (PCA; Jahoda et al. 1996) on board *RXTE* consists of five collimated xenon/methane multi-anode Proportional Counter Units (PCUs) operating in the 2–60 keV energy range. Raw data were reduced using the FTOOLS package (version 6.11). To study the timing properties of SGR 0418+5729, we restricted our analysis to the data in Good Xenon mode, with a time resolution of 1 $\mu$s and 256 energy bins. The event-mode data were extracted in the 2–10 keV energy range from all active detectors (in a given observation) and all layers, and binned into light curves of 0.1 s resolution. We use here 46 *RXTE*/PCA observations of SGR 0418+5729, spanning the first six months of the outburst, until the source flux decayed below the instrument detection level. The total 194.2 ks exposure time is divided in observations of 0.6 to 13.6 ks exposure each. See Esposito et al. (2010) for further details on the *Swift* and *RXTE* observations.

### 2.3. *Chandra* Data

The *Chandra X-Ray Observatory* monitored SGR 0418+5729 five times during the past three years. The first observation was with the High Resolution Imaging Camera (HRC-I; Zombeck et al. 1995) and the following four observations were with the Advanced CCD Imaging Spectrometer (ACIS-S; Garmire et al. 2003). Data were analyzed using standard cleaning procedures[17] and CIAO version 4.4. The HRC-I camera does not have a sufficient spectral resolution, and it was used only for the timing analysis; it has a timing resolution of ∼16 $\mu$s. All ACIS-S observations were performed in VERY FAINT mode, with only the S7 CCD on, resulting in a timing resolution of 0.44 s. Photons were extracted from a circular region with a radius of 3″ around the source position, including more than 90% of the source photons, and background was extracted from a similar region far from the source position.

[17] http://asc.harvard.edu/ciao/threads/index.html

**Table 1**
Journal of All the X-Ray Observations of SGR 0418+5729

| Instrument | ObsID | Starting Date | Exp. (ks) | Counts s$^{-1}$ | Flux[c] | $kT_{\rm BB}$[d] (keV) | BB Norm.[e] |
|---|---|---|---|---|---|---|---|
| *RXTE*/PCA[a] | 94048 | 2009 06-11/11-24 | 194.2 | | ... | ... | ... |
| *Swift*/XRT | 00031422001 | 2009-07-08 20:48:01 | 2.9 | $0.229 \pm 0.008$ | $13.4 \pm 1.0$ | $0.88 \pm 0.05$ | $2.2 \pm 0.2$ |
| *Swift*/XRT (PC) | 00031422002 | 2009-07-09 00:04:01 | 10.6 | $0.245 \pm 0.003$ | $13.8 \pm 0.7$ | $0.94 \pm 0.03$ | $1.7 \pm 0.1$ |
| *Swift*/XRT (PC) | 00031422003 | 2009-07-10 00:15:01 | 5.6 | $0.179 \pm 0.005$ | $11.0 \pm 0.7$ | $0.95 \pm 0.05$ | $1.33 \pm 0.13$ |
| *Swift*/XRT (WT) | 00031422004 | 2009-07-12 00:27:01 | 7.1 | $0.218 \pm 0.006$ | $11.5 \pm 0.7$ | $0.91 \pm 0.04$ | $1.68 \pm 0.14$ |
| *Chandra*/HRC-I[a] | 10168 | 2009-07-12 06:06:43 | 24.1 | $0.317 \pm 0.005$ | ... | ... | ... |
| *Swift*/XRT (WT) | 00031422006 | 2009-07-15 00:48:39 | 7.7 | $0.252 \pm 0.006$ | $12.7 \pm 1.0$ | $0.93 \pm 0.03$ | $1.69 \pm 0.12$ |
| *Swift*/XRT (WT) | 00031422007 | 2009-07-16 00:53:01 | 16.4 | $0.217 \pm 0.004$ | $11.7 \pm 0.8$ | $0.93 \pm 0.02$ | $1.57 \pm 0.08$ |
| *XMM-Newton*/EPIC* | 0610000601 | 2009-08-12 21:09:12 | 67.1 | $1.281 \pm 0.005$ | $6.75 \pm 0.07$ | $0.897 \pm 0.007$ | $1.038 \pm 0.018$ |
| *Swift*/XRT (PC) | 00031422008 | 2009-09-20 21:09:00 | 9.4 | $0.066 \pm 0.002$ | $3.66 \pm 0.30$ | $0.82 \pm 0.05$ | $0.79 \pm 0.09$ |
| *Swift*/XRT (PC) | 00031422009 | 2009-09-22 00:43:00 | 7.6 | $0.072 \pm 0.003$ | $3.58 \pm 0.40$ | $0.82 \pm 0.05$ | $0.79 \pm 0.11$ |
| *Swift*/XRT (PC) | 00031422010 | 2009-11-08 00:36:01 | 15.1 | $0.043 \pm 0.002$ | $2.14 \pm 0.20$ | $0.82 \pm 0.05$ | $0.47 \pm 0.05$ |
| *Swift*/XRT (PC) | 00031422011[b1] | 2010-01-14 08:06:01 | 3.6 | $0.019 \pm 0.001$[b] | $1.05 \pm 0.10$ | $0.75 \pm 0.07$ | $0.32 \pm 0.06$ |
| *Swift*/XRT (PC) | 00031422012[b1] | 2010-01-15 13:08:01 | 3.7 | $0.019 \pm 0.001$[b] | " | " | " |
| *Swift*/XRT (PC) | 00031422013[b1] | 2010-01-16 08:14:01 | 4.0 | $0.019 \pm 0.001$[b] | " | " | " |
| *Swift*/XRT (PC) | 00031422014[b1] | 2010-01-17 06:47:01 | 3.8 | $0.019 \pm 0.001$[b] | " | " | " |
| *Swift*/XRT (PC) | 00031422015[b2] | 2010-02-14 17:33:01 | 4.5 | $0.0172 \pm 0.0008$[c] | $0.76 \pm 0.11$ | $0.74 \pm 0.04$ | $0.25 \pm 0.03$ |
| *Swift*/XRT (PC) | 00031422016[b2] | 2010-02-15 17:37:01 | 4.5 | $0.0172 \pm 0.0008$[c] | " | " | " |
| *Swift*/XRT (PC) | 00031422017[b2] | 2010-02-16 01:38:01 | 4.6 | $0.0172 \pm 0.0008$[c] | " | " | " |
| *Swift*/XRT (PC) | 00031422018[b2] | 2010-02-17 09:49:01 | 4.6 | $0.0172 \pm 0.0008$[c] | " | " | " |
| *Swift*/XRT (PC) | 00031422019[b2] | 2010-02-18 16:14:01 | 3.9 | $0.0172 \pm 0.0008$[c] | " | " | " |
| *Swift*/XRT (PC) | 00031422020[b2] | 2010-02-19 00:23:01 | 3.2 | $0.0172 \pm 0.0008$[c] | " | " | " |
| *Swift*/XRT (PC) | 00031422021[b3] | 2010-07-09 06:50:01 | 3.6 | $0.0023 \pm 0.0003$[d] | $0.10 \pm 0.04$ | $0.47 \pm 0.13$ | $0.23 \pm 0.13$ |
| *Swift*/XRT (PC) | 00031422022[b3] | 2010-07-10 18:11:00 | 5.2 | $0.0023 \pm 0.0003$[d] | " | " | " |
| *Swift*/XRT (PC) | 00031422023[b3] | 2010-07-11 05:19:01 | 5.0 | $0.0023 \pm 0.0003$[d] | " | " | " |
| *Swift*/XRT (PC) | 00031422024[b3] | 2010-07-11 23:06:01 | 5.4 | $0.0023 \pm 0.0003$[d] | " | " | " |
| *Swift*/XRT (PC) | 00031422025[b3] | 2010-07-13 00:47:01 | 4.9 | $0.0023 \pm 0.0003$[d] | " | " | " |
| *Chandra*/ACIS-S | 12312 | 2010-07-23 15:04:09 | 30.0 | $0.0017 \pm 0.0008$ | $0.13 \pm 0.02$ | $0.68 \pm 0.04$ | $0.061 \pm 0.008$ |
| *XMM-Newton*/EPIC | 0605852201 | 2010-09-24 01:54:56 | 34.2 | $0.0370 \pm 0.0020$ | $0.16 \pm 0.02$ | $0.69 \pm 0.05$ | $0.07 \pm 0.01$ |
| *Chandra*/ACIS-S | 13148 | 2010-11-29 05:59:57 | 30.0 | $0.0038 \pm 0.0004$ | $0.021 \pm 0.002$ | $0.38 \pm 0.11$ | $0.12 \pm 0.06$ |
| *XMM-Newton*/EPIC | 0672670201 | 2011-03-10 03:15:53 | 35.0 | $0.0071 \pm 0.0007$ | $0.015 \pm 0.002$ | $0.32 \pm 0.05$ | $0.21 \pm 0.08$ |
| *Chandra*/ACIS-S | 13235 | 2011-07-20 02:26:12 | 77.0 | $0.0033 \pm 0.0002$ | $0.015 \pm 0.003$ | $0.37 \pm 0.04$ | $0.11 \pm 0.02$ |
| *XMM-Newton*/EPIC[b4] | 0672670401 | 2011-09-09 15:27:23 | 33.0 | $0.0071 \pm 0.0006$[e] | $0.016 \pm 0.002$ | $0.28 \pm 0.05$ | $0.34 \pm 0.13$ |
| *XMM-Newton*/EPIC[b4] | 0672670501 | 2011-09-11 21:47:41 | 48.5 | $0.0071 \pm 0.0006$[e] | " | " | " |
| *Chandra*/ACIS-S | 13236 | 2011-11-26 11:48:02 | 75.0 | $0.0026 \pm 0.0002$ | $0.015 \pm 0.002$ | $0.35 \pm 0.07$ | $0.13 \pm 0.05$ |
| *XMM-Newton*/EPIC* | 0693100101 | 2012-08-25 14:18:08 | 78.2 | $0.0058 \pm 0.0004$ | $0.012 \pm 0.001$ | $0.32 \pm 0.05$ | $0.16 \pm 0.05$ |

**Notes.**
[a] The *RXTE*-PCA and the *Chandra* HRC-I were used only for the timing analysis.
[b] These observations were merged in the timing and spectral analysis to improve statistics.
[c] Absorbed flux in the 0.5–10 keV energy range, and in units of $10^{-12}$ erg s$^{-1}$ cm$^{-2}$. Errors in the table are at 90% confidence level.
[d] Fitted model is `phabs*bbodyrad`; $N_{\rm H} = (1.15 \pm 0.06) \times 10^{21}$ cm$^{-2}$ and $\chi^2_\nu = 1.19$ (for 940 dof).
[e] The BB radius in km is the square root of this BB normalization, times the distance in units of 10 kpc.
* See Section 3.1 for details on the modeling of these observations.

### 2.4. *XMM-Newton* Data

SGR 0418+5729 was observed six times with *XMM-Newton* (Jansen et al. 2001). Data have been processed using SAS version 12, and we have employed the most up-to-date calibration files available at the time the reduction was performed (2012 August). Standard data screening criteria are applied in the extraction of scientific products. For our spectral analysis we used only the EPIC-pn camera (Turner et al. 2001) which provides the spectra with the best statistics, while the MOS cameras (Strüder et al. 2001) were added in the timing analysis. The EPIC-pn camera was set in Small Window (timing resolution of 6 ms) and Full Frame (73 ms) modes in the first two observations, respectively, and in Large Window mode for all the following ones (48 ms), with the source at the aim-point of the camera, and the MOS cameras in Small Window mode (0.3 s). We extracted the source photons from a circular region of 30″ radius, and a similar region was chosen for the background in the same CCD. We restricted our spectral analysis to photons having PATTERN $\leqslant 4$ and FLAG $= 0$ for the EPIC-pn data.

## 3. RESULTS OF THE X-RAY MONITORING

### 3.1. X-Ray Spectral Modeling

For the spectral analysis we used source and background photons from the *Swift*, *Chandra*, and *XMM-Newton* observations extracted as described in the previous section (we also checked our results using a larger extraction region for the background spectra). The response matrices were built using ad hoc bad-pixel files built for each observation. We used the `XSPEC` package (version 12.4) for all fittings, and the `phabs` absorption model with the Anders & Grevesse (1989) abundances, and the Balucinska-Church & McCammon (1992) photoelectric

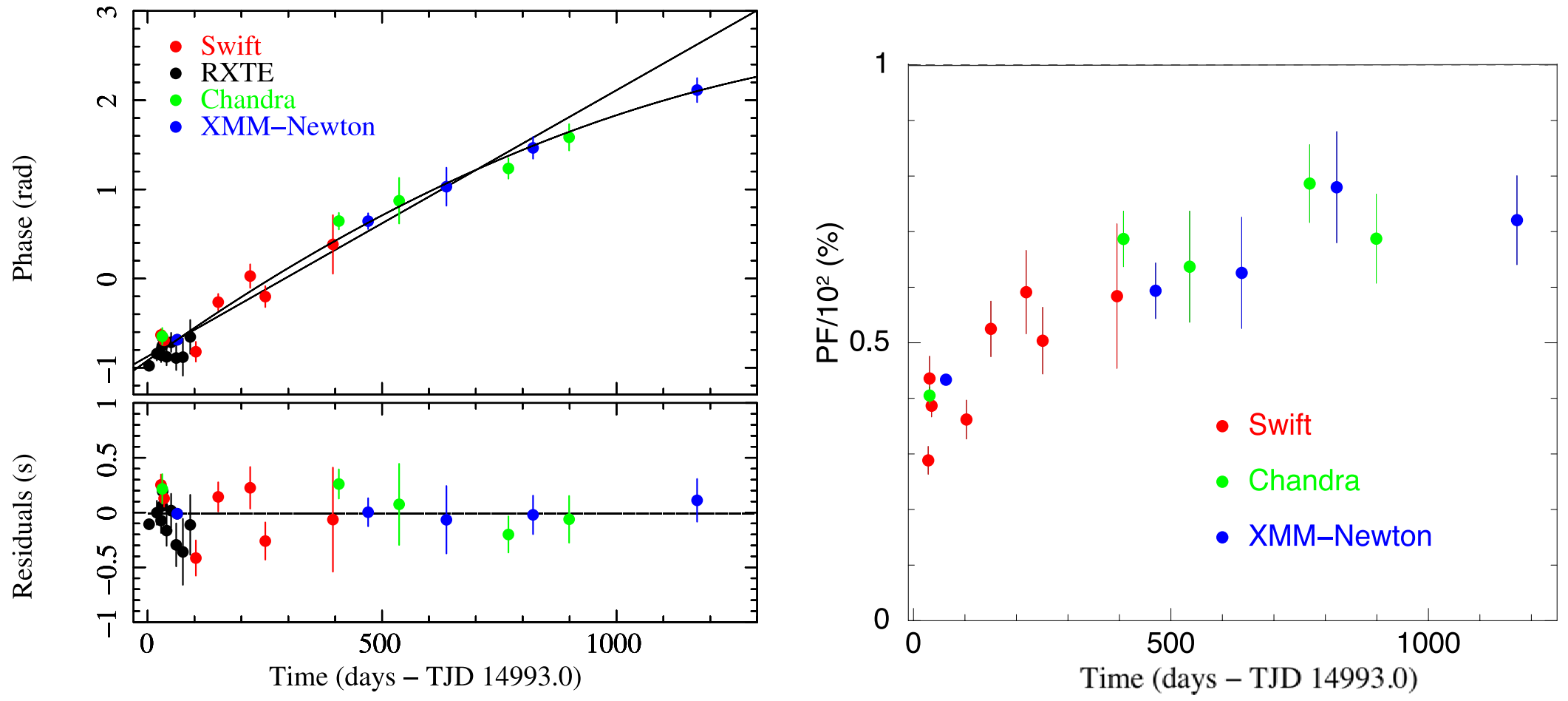

**Figure 1.** Left panel: evolution of the pulse phases with time (upper panel). The solid lines represent the timing solution without (linear) and with a $\dot{P}$ component (quadratic). The time residuals (lower panel) are relative to the quadratic fit. Right panel: pulsed fraction evolution in the 0.5–10 keV band for the *Swift* (red), *Chandra* (green), and *XMM-Newton* (blue) observations.

(A color version of this figure is available in the online journal.)

cross-sections. We restricted our spectral modeling to the 0.7–10 keV energy band, excluding bad channels when needed. The *Swift* spectra were binned in order to have at least 30 counts per spectral bin. *XMM-Newton* spectra were grouped such to have at least 100, 50, and 40 photons per bin in the first three observations, respectively, and a minimum of 30 counts in the subsequent observations. On the other hand, all *Chandra* spectra have at least 25 counts per bin.

We started the spectral analysis by fitting all the spectra together (see Table 1) with a single component model: an absorbed blackbody or a power-law model. While the former gave a good fit, a single power law could not reproduce all the spectra at the same time. Fixing the absorption value to be the same for all spectra, for a single blackbody model (`phabs*bbodyrad`) we find an acceptable fit with $N_{\rm H} = (1.15 \pm 0.06) \times 10^{21}$ cm$^{-2}$ and $\chi^2_\nu = 1.19$ (940 dof; errors on the spectral parameters are all reported at 90% confidence level). However, not unexpectedly, the best collected spectrum (the first *XMM-Newton* observation on 2009-08-12; see Table 1) gave bad residuals at lower and higher energies (see Figure 2, left panel). We tried to model this observation alone, and indeed a single absorbed blackbody or power-law components were not reproducing this spectrum properly ($\chi^2_\nu > 2$). We then used a composite model. Good fits were found using both an absorbed blackbody plus a power law (`phabs*(bbodyrad + power)`; $N_{\rm H} = (6.32 \pm 0.04) \times 10^{21}$ cm$^{-2}$, $kT = 0.91 \pm 0.07$ keV, $\Gamma = 2.82 \pm 0.16$, and $\chi^2_\nu = 0.97$ for 392 dof) and an absorbed resonant cyclotron scattering model (RCS: Rea et al. 2007, 2008, or NTZ: Zane et al. 2009). Two blackbodies were also producing acceptable reduced chi-square values ($\chi^2_\nu = 1.01$ for 392 dof) but with worse residuals at higher energies (this is compatible with what was found in Esposito et al. 2011 and Turolla et al. 2011). The parameters we found for the resonant cyclotron scattering models are $N_{\rm H} = (1.9 \pm 0.3) \times 10^{21}$ cm$^{-2}$, $\tau = 8.8 \pm 1.2$, $\beta = 0.21 \pm 0.08$, and $kT = 0.63 \pm 0.11$ keV ($\chi^2_\nu = 1.08$ for 392 dof) for the RCS model; and $N_{\rm H} = (1.8 \pm 0.2) \times 10^{21}$ cm$^{-2}$, $\Delta\phi = 1.9 \pm 1.0$, $\beta_{\rm bulk} = 0.13 \pm 0.05$, and $kT = 0.88 \pm 0.07$ keV ($\chi^2_\nu = 1.11$ for 392 dof) for the NTZ model.

We then continued our spectral modeling of all spectra together by adding a further component only for this observation (adding a further component for all spectra did not significantly change the goodness of the fit; $\chi^2_\nu = 1.13$ (938 dof); see Figure 2, middle and right panels). In Table 1 we report the values of the single absorbed blackbody model (see also Figure 2, left panel, and Figures 3 and 4), since when fitting all the spectra by using a composite model only for the first *XMM-Newton* observation, we find no change in the parameters of the other spectra with respect to the single blackbody fit. However, although a blackbody plus power-law model gives a good fit when fitting the first *XMM-Newton* observation alone, it is not so when fitting all data together. This is because the power-law component produces an unrealistic $N_{\rm H}$ increase, which does not match the value required by all the other observations modeled by a single blackbody. We then use one of the resonant cyclotron scattering models, the RCS model, for the joint fit as an empirical model for the first *XMM-Newton* observation.[18]

In addition to the joint spectral modeling, we also fitted all the spectra individually. Beside the first *XMM-Newton* observation discussed above, the last *XMM-Newton* observation, when fitted alone with a single blackbody model, did not give a good chi-square ($\chi^2_\nu = 2.2$ for 16 dof). Given the low number of counts in the spectrum of this observation (∼400 background-subtracted counts), this deviation from the blackbody model had only a marginal effect on the joint fit. A better fit was found adding a second blackbody ($\chi^2_\nu = 1.2$ for 14 dof) or (with a slightly worse chi-square) a power-law component ($\chi^2_\nu = 1.5$ for 14 dof). However, given the reduced number of counts collected in this

[18] Note that both the RCS and NTZ models are built for higher surface dipolar fields, hence the fact that they provide a very good fit to the data is probably just an indication of the presence of some magnetospheric distortion. However, no real physical information can be derived from the resulting magnetospheric parameters. For the purpose of this work, we are mainly interested in the surface thermal cooling of the source.

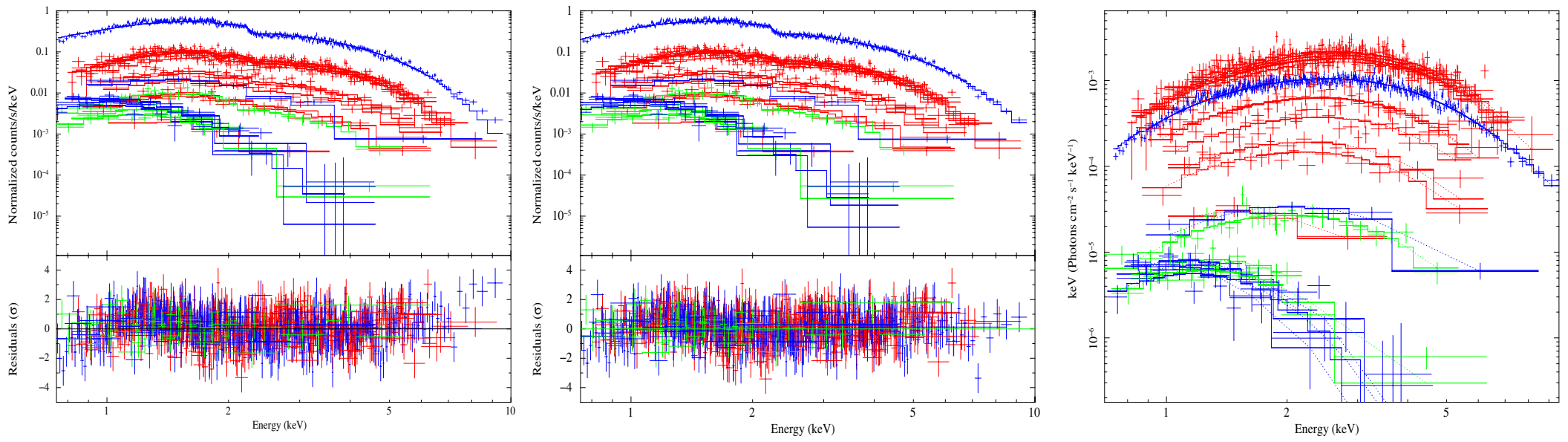

**Figure 2.** Spectral modeling of all observations listed in Table 1. Left and middle panels: spectra and residuals for all observations (*Swift* (red), *Chandra* (green), and *XMM-Newton* (blue)) fitted simultaneously with a single blackbody model (left) and using an RCS model only for the first *XMM-Newton* observation (center). Right panel: unfolded spectrum relative to the modeling shown in the central panel.

(A color version of this figure is available in the online journal.)

observation, a detailed modeling of the quiescent spectrum of SGR 0418+5729 will be possible only when more data is accumulated.

We also tried to (1) model all the spectra with two blackbodies leaving one of the blackbodies with a fixed area mimicking the whole surface emission, and (2) fix one blackbody to the value observed in the last observation (see Table 1) and leave the second blackbody free to vary. In neither of those two cases could we find any improvement in the modeling of the data. Note that although a joint two blackbody model can fit the first few observations (Turolla et al. 2011), this is no longer the case when modeling together all the data collected in the entire 1200 day time span.

### 3.2. X-Ray Timing Analysis

All the *Chandra* and *XMM-Newton* event files collected between 2010 November and 2012 August were used in order to extend the coherent timing solution we derived in Rea et al. (2010): $P = 9.07838827(4)$ s 90% c.l., and $3\sigma$ first period derivative upper limit of $|\dot{P}| < 6.0 \times 10^{-15}$ s s$^{-1}$ at epoch 54993.0 MJD. Photon arrival times were corrected to the barycenter of the solar system.[19] Timing analysis was carried out by means of a phase-fitting technique (details on this technique are given in Dall'Osso et al. 2003; see also Esposito et al. 2010 for further details on this source). Given the intrinsic variability of the pulse shape as a function of time (see Figure 6), we inferred the phase of the modulation by fitting the average pulse shape of each observation with a number of harmonics, the exact number of which is variable and determined by requesting that the inclusion of any higher harmonic is statistically significant (by means of an $F$ test). All data reported in Table 1 were folded using a reference period 9.07838880562798 s at epoch 54993 MJD, and fitted with one or more harmonics. In Figure 1 we plot the phases at which the fundamental sine function is equal to zero in its ascending part (positive derivative).

The fit of the resulting pulse phases with a linear component gives a reduced $\chi_r^2 \sim 3.2$ for 26 degrees of freedom (dof hereafter). The inclusion of a quadratic term, corresponding to a first period derivative component, was found to be significant at a confidence level of $3.5\sigma$ (by means of an $F$ test). The resulting best-fit solution corresponds to $P = 9.07838822(5)$ s ($1\sigma$ c.l., two parameters of interest; epoch 54993.0 MJD) and $\dot{P} = 4(1) \times 10^{-15}$ s s$^{-1}$ with a reduced $\chi_r^2 \sim 2.1$ (for 25 dof; see also Figure 1). The new timing solution implies an rms variability of only 0.2 s. As depicted above, the time evolution of the phase can be described by a relation of the form $\phi = \phi_0 + 2\pi(t - t_0)/P - \pi(t - t_0)2\dot{P}/P^2$.

To further assess the significance of the quadratic component, reflecting the period derivative, we performed detailed Monte Carlo (MC) simulations assuming as the null model a simple linear relation (see Protassov et al. 2002 for further details). By running $10^5$ MC simulations we verified that the quadratic component is significant at >99.96% confidence level, which is in very good agreement with what was estimated by means of the $F$ test. We notice that examining the simulated data with different sampling distributions does not significantly change the results. Furthermore, we performed the same MC simulations using the whole set of observations but without considering the last *XMM-Newton* observation. We find a chance probability for the addition of a quadratic component of 0.65% (<$3\sigma$). Given the strong influence of the last of our observations in the determination of the period derivative, we will perform further X-ray observations in the next few years in order to increase the significance of the current $\dot{P}$ measurement.

Using this $\dot{P}$ measurement, we infer a surface dipolar magnetic field strength of $B_{\rm dip} = (6 \pm 2) \times 10^{12}$ G, calculated at the neutron star equator. This value is fully consistent with the $3\sigma$ upper limit reported in Rea et al. (2010). We also estimate a characteristic age of $\tau_c \simeq P/2\dot{P} \sim 35$ Myr, and a rotational power of $\dot{E} \simeq 3.9 \times 10^{46}\dot{P}/P^3 \sim 2 \times 10^{29}$ erg s$^{-1}$.

Based on the above phase coherent timing solution, we also studied the pulse shape and pulsed fraction evolution. Figure 1 shows the pulsed fraction evolution as a function of time. There is an evident increase starting soon after the burst detection with a recovery toward an asymptotic quiescent value which appears to be at about the 70%–80% level.

In Figures 1, 5, and 6 we study in detail the shape of the pulse profile as it evolves in time and in energy. By looking at the profile shapes of all X-ray observations performed so far, the source appears to be switching among a triple/double/single peak shape during the early outburst phases, with no clear trend in time (Figure 1). The pulse profile stabilizes to a single peak about three months after the outburst onset. However, studying in detail the first and the last *XMM-Newton* observations, a few key pieces of information can be extracted: (1) at lower energies (<1 keV), the pulse profile is mainly single peaked,

[19] We have corrected the arrival times of the last *XMM-Newton* observation for the 2012 June 30 leap second (see http://xmm.esa.int/sas/current/watchout/12.0.0/leapsec_2012.shtml for further details).

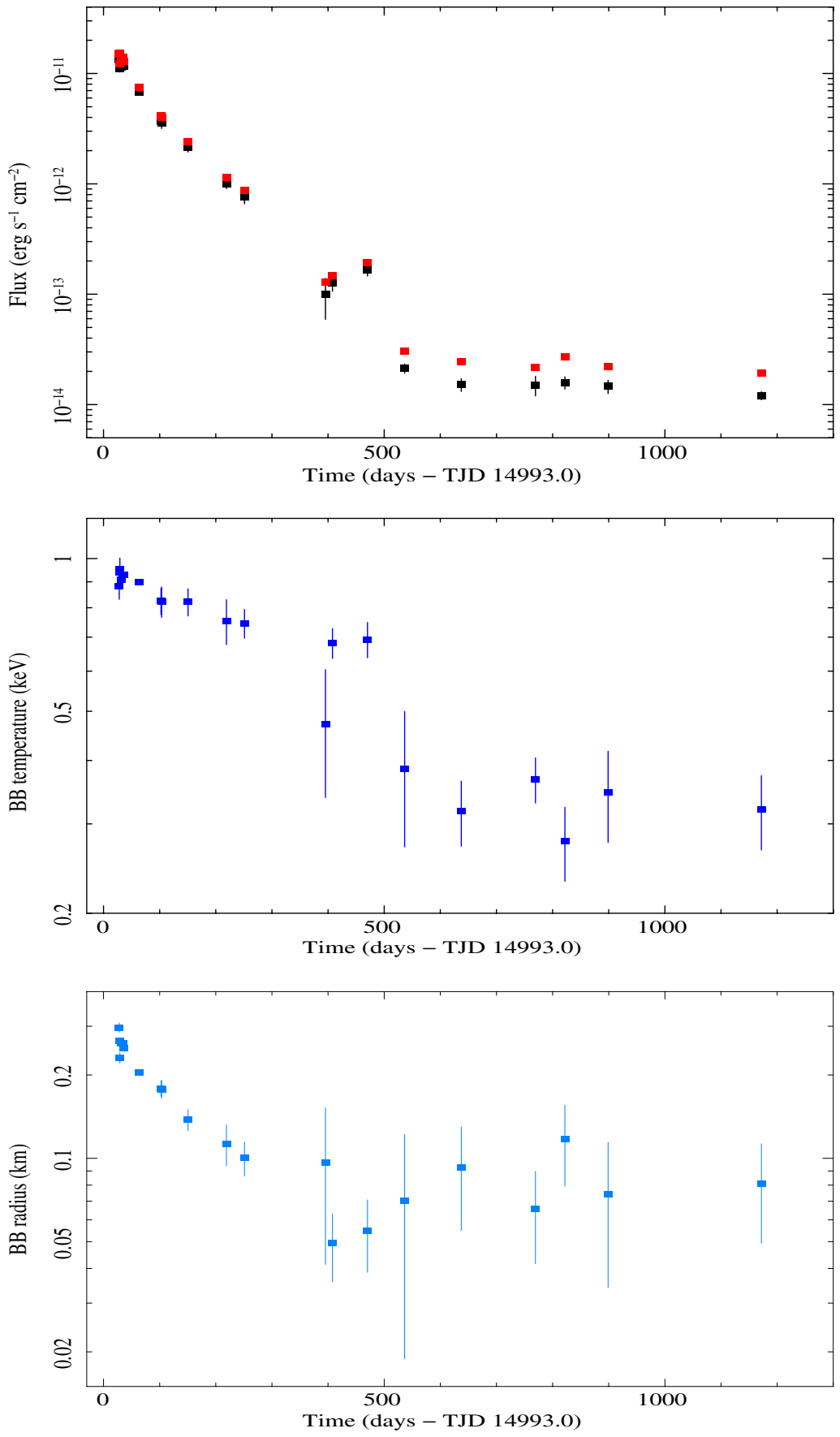

**Figure 3.** Spectral evolution with time. Top panel: flux evolution for the absorbed 0.5–10 keV flux (black), and for the bolometric unabsorbed flux (red). Middle and bottom panels: evolution of the blackbody temperature and radius, calculated at infinity (the latter assuming a 2 kpc distance).

(A color version of this figure is available in the online journal.)

while a second, and possibly also a third peak, appears at higher energies (see Figure 5); (2) the main component of the pulse profile continues to be at the same phase over the whole outburst decay (see Figure 7).

### 3.3. Pulse Phase Spectroscopy

We performed a pulse phase spectroscopy of the first *XMM-Newton* observation. A clear pulse phase dependence of the spectrum is already observed by simply looking at the pulse profile changes as a function of the energy (see Figure 5). In order to quantify the spectral variability as a function of the rotational phase, we performed a pulse phase spectroscopy extracting the spectra from phases 0–0.4, 0.4–0.6, and 0.6–1. These phase intervals were chosen by looking at Figure 5 in order to isolate the dip in the 1–4 keV pulse profiles at phase ∼0.55. In Table 2 we report the results of our modeling. The phase-averaged spectrum is not well fit by either a single blackbody nor a power law. We then used an absorbed blackbody

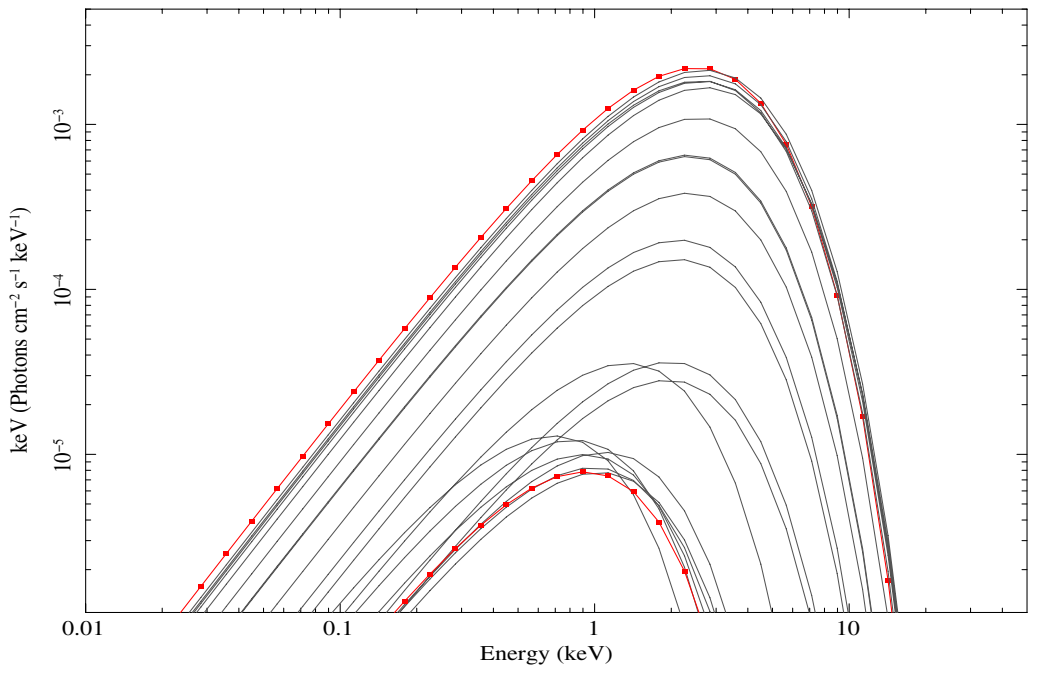

**Figure 4.** Fitted blackbody models (see Table 1), with the first and last observations labeled as red squares (see Figure 3 and text for details).

(A color version of this figure is available in the online journal.)

**Table 2**
Pulse Phase Spectroscopy of the First *XMM-Newton* Observation of SGR 0418+5729

| Phase | Counts $s^{-1}$ | Flux[a] | Photon Index[b] |
|---|---|---|---|
| 0.0–0.4 | $1.46 \pm 0.01$ | $7.5 \pm 0.1$ ($1.71 \pm 0.02$) | $2.86 \pm 0.15$ |
| 0.4–0.6 | $1.00 \pm 0.01$ | $5.5 \pm 0.1$ ($1.20 \pm 0.02$) | $4.08 \pm 0.44$ |
| 0.6–1.0 | $1.26 \pm 0.01$ | $6.6 \pm 0.1$ ($1.47 \pm 0.02$) | $3.11 \pm 0.21$ |

**Notes.**
[a] Absorbed flux in the 0.5–10 keV energy range, and in units of $10^{-12}$ erg $s^{-1}$ ($10^{-3}$ photons $cm^{-2}$ $s^{-1}$). See also Section 3.3.
[b] Fitted model is `phabs*(bbodyrad+power)`; $N_H = (6.9 \pm 0.5) \times 10^{21}$ $cm^{-2}$, $kT_{BB} = 0.91 \pm 0.01$ keV, BB norm= $0.82 \pm 0.05$, and $\chi^2_\nu = 0.96$ (for 680 dof).

plus power-law modeling (note that the RCS and NTZ models are not suited for phase resolved analysis since they are intrinsically phase-average), using for the photoelectric absorption model the same cross-section and abundances as for the phase-average spectrum (see Section 3.1). The blackbody temperature and radius were consistent in all three spectra, hence we fixed them to be the same for all spectra ($kT_{BB} = 0.91 \pm 0.01$ keV, BB norm $= 0.82 \pm 0.05$), while a variability $>3\sigma$ has been observed in the photon index (it changed from about 2.9 to 4.1 between the spectra of the first peak and the dip).

However, from Figure 7 it is clear that the main difference in the spectra is at lower energies. In particular, above 5 keV the three spectra are very similar, while the 0.4–0.6 phase-resolved spectrum seems to have less counts than the other below such energy.

## 4. GREEN BANK TELESCOPE RADIO OBSERVATIONS

We observed SGR 0418+5729 using the 101 m Green Bank Telescope (GBT) on 2012 October 4, during the return to quiescence. Data were acquired with the Green Bank Ultimate Pulsar Processing Instrument (GUPPI; DuPlain et al. 2008) at a central frequency of 2.0 GHz (with a bandwidth of 800 MHz, integration time of ∼5400 s and sampling time of 655 $\mu$s) and 820 MHz (with a bandwidth of 200 MHz, integration time of ∼5600 s, and sampling time of 655 $\mu$s). To minimize the dispersive effects of the interstellar medium, the bandwidths were split into 2048 and 512 channels, respectively. The working of the system was checked looking at the pulsar PSR B0450+55. A mask was first applied to the full resolution data for reducing the effects of impulsive RFI and of bad channels. Then the cleaned data were downsampled a factor two in sampling time, matching the frequency resolution in order to have a maximum dispersion smearing of order 1.3 ms in each channel for a pulsar with dispersion measure (DM) ∼100 pc $cm^{-3}$.

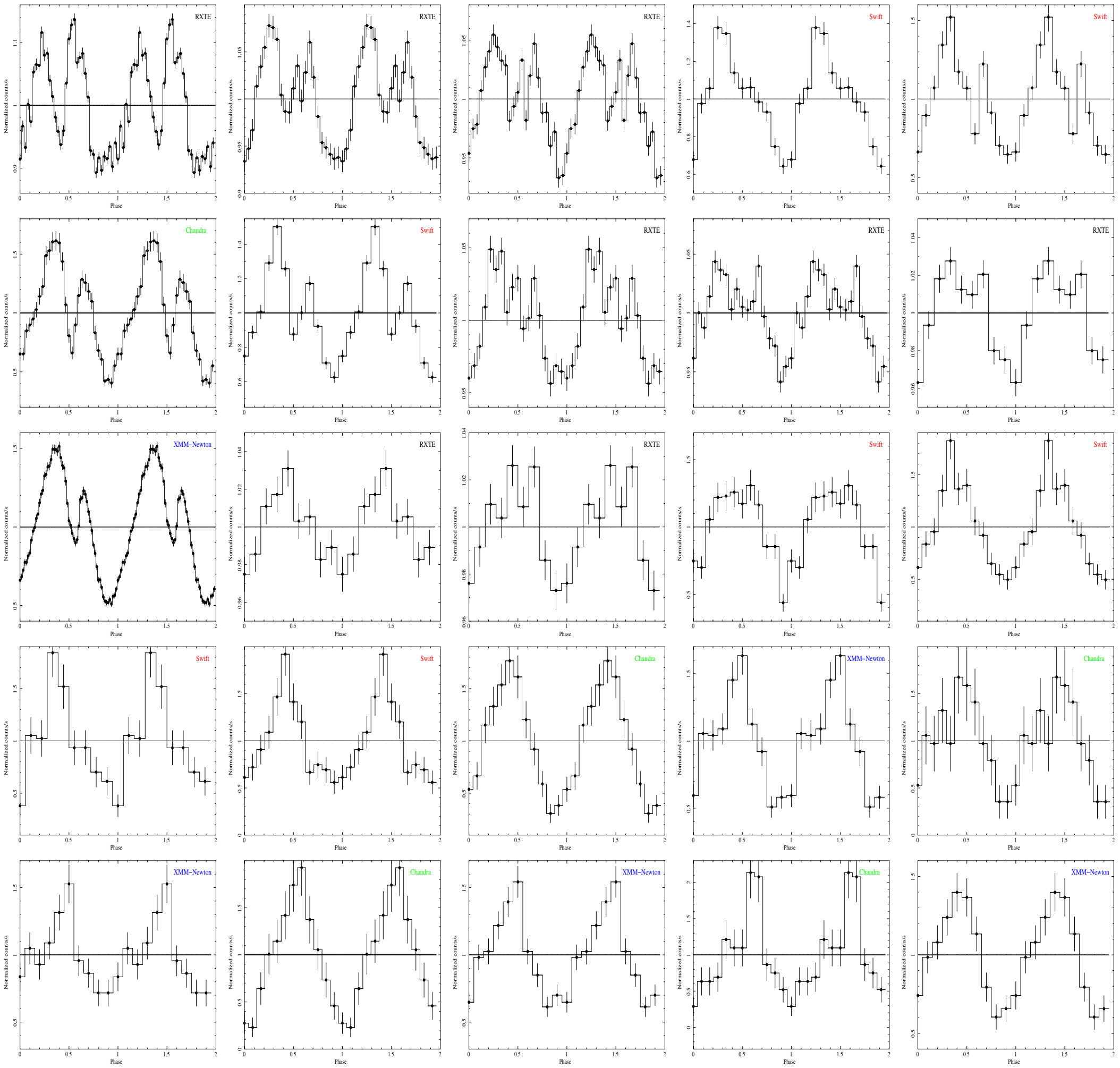

**Figure 5.** Pulse profile (normalized counts $s^{-1}$ vs. phase) as a function of energy, relative to the first *XMM-Newton* observation (see Table 1).
(A color version of this figure is available in the online journal.)

The ephemerides acquired from the X-ray observations (see Section 3.2), were used to fold the resulting data in ∼3 minutes long sub-integrations at the known magnetar period. We also folded the data at half, one-third, and a quarter of the nominal period in order to detect putative higher harmonics components of the intrinsic signal, in case the latter were deeply contaminated by interference (RFI). Folding was done using dspsr (van Straten & Bailes 2010). The sub-integrations and the frequency channels, cleaned from RFI, were then searched around the pulsar period $P$ and over a wide range of DM values (from 0 to 1000 pc $cm^{-3}$) to find the $P$–DM combination maximizing the signal-to-noise ratio. No dispersed signal was found in the data down to a signal-to-noise limit of 10 in both data sets. Given the parameters of the antenna and of the receivers,[20] and assuming a pulsar with a duty cycle of 10%, that translates to flux densities of ∼0.02 mJy and ∼0.05 mJy, for the 2 GHz and 820 MHz observations, respectively. Data were also blindly searched for a periodic signal in the Fourier domain, and for single de-dispersed pulses (within a DM range from 0 to 200 pc $cm^{-3}$). No signal was found in either the Fourier domain (down to a spectral signal-to-noise ratio of 4) or in the single pulse searches (down to a signal-to-noise ratio of 5 for the individual pulses).

No previous search for pulsed radio emission had been performed at 2 GHz, whereas the flux density value at 820 MHz improves by ~15% the limit of the observation at 820 MHz performed on 2009 July 19 (Lorimer et al. 2009, ATel 2096), when the source was in the phase of outburst. Assuming a typical pulsar spectral index of 1.7, a typical duty cycle ~10% and a distance of 2 kpc (van der Horst et al. 2010), the observations at 820 MHz sampled more than 97% of the luminosity distribution of the population of known ordinary pulsars with rotational period longer than 100 ms, as derived from the ATNF pulsar catalog.[21]

[20] http://www.gb.nrao.edu/gbtprops/man/GBTpg.pdf

## 5. PLATEAU DE BURE MM OBSERVATIONS

SGR 0418+5729 was observed with the Plateau de Bure Interferometer (PdBI) at 1.8 mm (166.50 GHz) in the D configuration between 2011 June and July (June 27, and July 9, 10, 15, and 16). This configuration provides baselines between 22.1 and 95.6 m. The phase center of the observations was 04:18:33.867, +57:32:22.910. The dominant track was July 15 (eight-hour track and excellent weather conditions). The system temperatures were typically in the 150 to 200 K range, and the averaged atmospheric precipitable water vapor was 2 mm. The gain calibration was performed observing the quasars B0552+398 and J0512+294. After calibration, the phase rms was 20°–60°. The bandpass calibrator used was B0851+202. The adopted flux density for the flux calibrator 3C273 was 16.57 Jy. Calibration and imaging were performed using the standard procedures in the CLIC and MAPPING packages of the GILDAS[22] software. The resulting final map, obtained combining all the data, yields a synthesized beam size of 3″.95 × 3″.16 with a position angle of P.A. = 0°.0. The rms noise achieved using the full 3.6 GHz provided by the WideX correlator is 60 $\mu$Jy beam$^{-1}$. The primary beam of the PdBI at 166.50 GHz is 30″.3.

We did not detect continuum emission within the PdBI primary beam toward SGR 0418+5729, and obtained an upper limit of 0.24 mJy beam$^{-1}$ at a 4$\sigma$ level (see Figure 8). The only detected source is at R.A. = 04:18:30.077, decl. = 57:32:52.00, which corresponds to an offset of (30″.5, 29″.1) with respect to the phase center, or a total offset of 42″.1. The flux density of this millimeter source is 0.34 ± 0.06 mJy (from a Gaussian fit in the *uv* plane and without correcting for the primary beam response). In addition, we looked for possible "pulses" of emission at 1.8 mm. In order to do that, we checked the calibrated amplitude versus time for the longest track (July 15). By averaging the visibilities in intervals of 1 minute, we found no hints of variable emission at an upper limit of roughly ~10 mJy.

We also searched the NRAO Very Large Array Sky Survey at 21 cm and found no source within 15′ of the millimeter source (Condon et al. 1998; limiting brightness: 2.0 mJy beam$^{-1}$).

## 6. WILLIAM HERSCHEL TELESCOPE OPTICAL OBSERVATIONS

We acquired four 300 s *r*-band images of the field containing SGR 0418+5729 on 2009 August 16, using the ACAM imager mounted at the 4.2 m William Herschel Telescope (WHT) on La Palma. The average seeing was 1″ and airmass 1.26. Observations of a nearby field containing Sloan Digital Sky Survey calibrated stars were obtained for the absolute photometric calibration, while astrometry was performed against Two Micron All Sky Survey sources, resulting in an accuracy of ~0″.1 on both R.A. and decl.

No source was detected within the 95% confidence down to a limit of $r$ = 24. PSF photometry reveals that the nearest object is detected at a magnitude $r = 22.7 \pm 0.1$ and center coordinates R.A. = 04:18:34.0, decl. = 57:32:23.5. This source is consistent with the near-infrared source reported by Wachter et al. (2009), but its distance from the SGR 0418+5729 position (~1″.4), makes the association with the magnetar rather unlikely.

## 7. *FERMI*-LAT GAMMA-RAY OBSERVATIONS

We used data from the Large Area Telescope (LAT) on board *Fermi* (Atwood et al. 2009) from 2008 August 4 until 2012 October 24. The *Fermi* SCIENCE TOOLS SC09-28-00 package was used to analyze the "`source`" event class data. We selected events within a circular region of interest (ROI) of 10° radius centered on the position of SGR 0418+5729, and in the energy range 100 MeV–100 GeV. The good time intervals are defined so that the ROI does not fall below the gamma-ray-bright Earth limb (defined at 100° from the zenith angle), and the source is always inside the LAT field of view, namely in a cone angle of 66°. The "`P7SOURCE_V6`" instrument response functions (IRFs) are applied in the analysis.

The likelihood analysis of SGR 0418+5729 was performed by means of the binned maximum-likelihood method (Mattox et al. 1996), using the official tool `gtlike` released by the *Fermi*-LAT collaboration. The spectral-spatial model created for the likelihood analysis includes the Galactic, and the isotropic diffuse emission models, as well as all the 2FGL sources within a radius of 15° from SGR 0418+5729. Since there is no 2FGL source that is positionally associated to the magnetar, we added in the spectra-spatial model a point-like source modeled with a simple power-law with the coordinates of SGR 0418+5729. The 2FGL sources within 3° of SGR 0418+5729 (3 sources) are modeled with the flux parameter allowed to vary, while the others 34 sources had all their parameters fixed to the value from the 2FGL catalog. Figure 8 shows the diffuse subtracted TS map of the 7° × 7° region centered on SGR 0418+5729. It was obtained associating to each pixel (of size 0°.1 × 0°.1) the TS value calculated assuming a point-like testing source in its center. Diffuse subtracted TS map means that the spectral-spatial model for the null hypothesis includes only the Galactic and isotropic emission models, so that the point-like sources should be visible in the map.

As no significant gamma-ray counterpart to SGR 0418+5729 is identified, 95% flux upper limit is derived using the Bayesian method developed by Abdo et al. (2010). The 95% flux upper limit for $E > 100$ MeV is $F < 1.3 \times 10^{-8}$ photons cm$^{-2}$ s$^{-1}$, including systematics.

The non-detection of SGR 0418+5729 at energies >100 MeV is not surprising, given a similar non-detection of all other known magnetars (Abdo et al. 2010).

## 8. DISCUSSION

We have presented here a detailed X-ray study of the outburst of the low magnetic field soft gamma repeater SGR 0418+5729. The long-term monitoring we performed over 1200 days allowed us to measure the period derivative of this pulsar ($\dot{P} = 4(1) \times 10^{-15}$ s s$^{-1}$) with a 3.5$\sigma$ significance (Figure 1). This yields an estimate of its dipolar magnetic field of $B_{\rm dip} \sim 6 \times 10^{12}$ G, and confirms this object as the magnetar with the lowest dipolar magnetic field ever discovered.

[21] http://www.atnf.csiro.au/research/pulsar/psrcat/
[22] The GILDAS data reduction package is available at http://www.iram.fr/IRAMFR/GILDAS.

Assuming that SGR 0418+5729 attained its quiescent state in the last few observations, the X-ray quiescent emission appears to be dominated by a very small spot at $kT \sim 0.3$ keV and of radius ∼0.16 km (assuming a 2 kpc distance), which corresponds to a cap of semi-aperture ∼1°–2°, similar to what observed in old radio pulsars. However, in the present case the rotation power ($\dot{E} \approx 10^{29}$ erg s$^{-1}$) is about two orders of magnitude smaller than the observed X-ray luminosity (≈$10^{31}$ erg s$^{-1}$), thus indicating a different origin for the emission, most likely magnetic. Actually, it is quite likely that most of the surface is at a much lower temperature and is therefore invisible at energies between 0.5–10 keV. This implies that the quiescent bolometric flux may be severely underestimated (see also below).

The study of the spectral evolution during the outburst shows the presence of a non-thermal component (probably magnetospherical) at the beginning of the outburst, which fades away after a few hundred days. On the other hand, the temperature of the small region responsible for the surface anisotropy fades from 0.9 to 0.3 keV within a timescale of a few years (see Table 1 and Figure 3).

The pulse profile evolution during the outburst decay shows some interesting features. As Figure 6 shows, there is an overall trend toward a simplification of the pulse, which starts with a complex, three-peaked shape and ends with a fairly sinusoidal pattern. Furthermore, the study of the pulse profiles as a function of energy (Figure 5) in the first outburst stages, shows a great variability too. A large dip in an otherwise rather sinusoidal profile is observed at energies between 1–4 keV.

The large pulsed fraction of 40%–70% (Figure 1), the evidence of nearly phase aligned spots responsible for the 0.9 keV thermal emission in the early outburst phases, as well as the 0.3 keV emission at late times, disfavor the presence of two spots at different temperatures, while favoring the presence of a single tiny spot cooling down (from 0.9 to 0.3 keV) and reducing its size (from 0.21 to 0.16 km) during the return to quiescence. This means that the multi-peaked pulse profile is probably due to anisotropies in the magnetospheric electrons distribution (on top of a non-isotropic surface thermal emission).

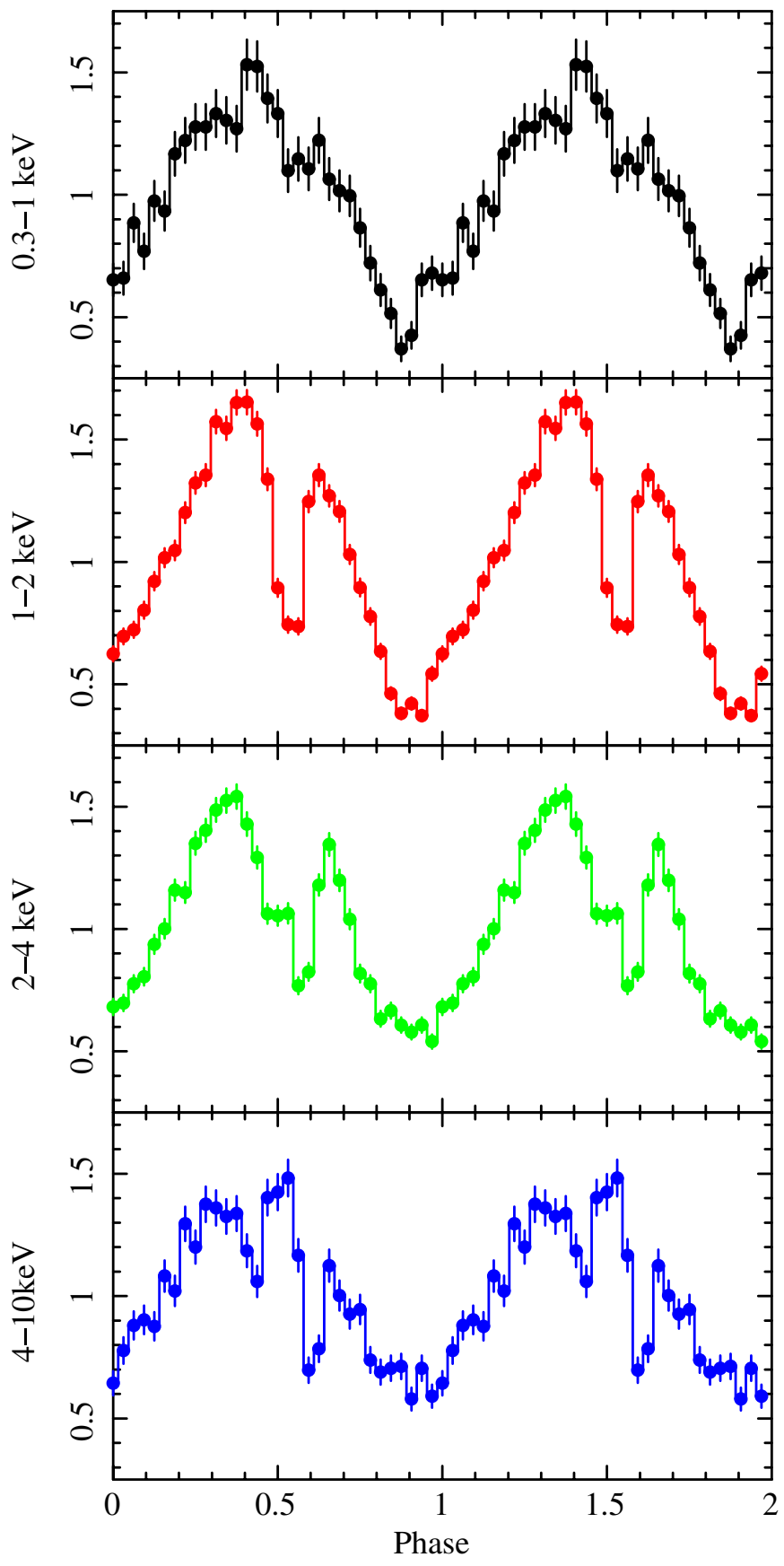

**Figure 6.** Pulse profiles evolution in the 2–10 keV (*RXTE*) and 0.5–10 keV (*Chandra* and *XMM-Newton*) energy ranges, for most of the observations reported in Table 1. Epoch increases from left to right, and top to bottom.
(A color version of this figure is available in the online journal.)

### *8.1. SGR 0418+5729 as an Evolved Magnetar*

In Turolla et al. (2011) it was shown that the rotational properties of SGR 0418+5729 can be reproduced if the source is an aged magnetar, which experienced substantial field decay but still retains a strong enough internal toroidal field. The most updated magneto-thermal evolutionary models discussed in Viganò et al. (2013; but see also Pons et al. 2009 and Aguilera et al. 2008), confirm this scenario. The evolution of an initial dipolar magnetic field of $B^0_{\rm dip} \sim 1.5 \times 10^{14}$ G (surface value at the pole) correctly provides the observed $P$ and $\dot{P}$ at an age of ∼550 kyr, which is probably the real age of this source.

Although different combinations of the initial components of the magnetic field are possible, in all the models the magnetic field must have been large in the past ($\gtrsim 10^{14}$ G) to explain at the same time the long spin period, the bright X-ray emission at this old age, and the flaring activity of the source. The characteristic age overestimates the real one by almost two orders of magnitude. In Figure 9, we show the evolution of period, period derivative, the source track in the $P$–$\dot{P}$ diagram, and the bolometric thermal luminosity. In this scenario we estimate that SGR 0418+5729's mean surface temperature should now be ∼0.05 keV, unfortunately undetectable by current X-ray observations (which are observing only a hot tiny region on the star surface).

For the evolution of the timing properties, we assume the magneto-dipole braking formula given by Spitkovsky (2006): $I\Omega\dot{\Omega} \approx (B_d^2 R^6 \Omega^4/4c^3)(1 + \sin\chi^2)$, where $R$ is the NS radius, $\chi$ is the angle between the rotational and the magnetic axis, $c$ is the speed of light, $\Omega = 2\pi/P$ is the angular velocity, and $I$ is the moment of inertia of the star.

An alternative possibility is that the neutron star was born with an external magnetic field close to the present one, but its large core poloidal field slowly diffuses out, i.e., by ambipolar diffusion (Soni 2012). Although from the timing properties alone it is hard to discriminate between a hidden strong crustal magnetic field, a hidden strong core field or an intrinsically low-$B$ neutron star, the magnitude of the X-ray luminosity, and the spectral properties and light curves may be used to distinguish the different scenarios.

In particular, the low magnetic field scenario cannot explain the high luminosity, large pulsed fraction, and the flaring activity of the source. As a matter of fact, if there is little field decay and

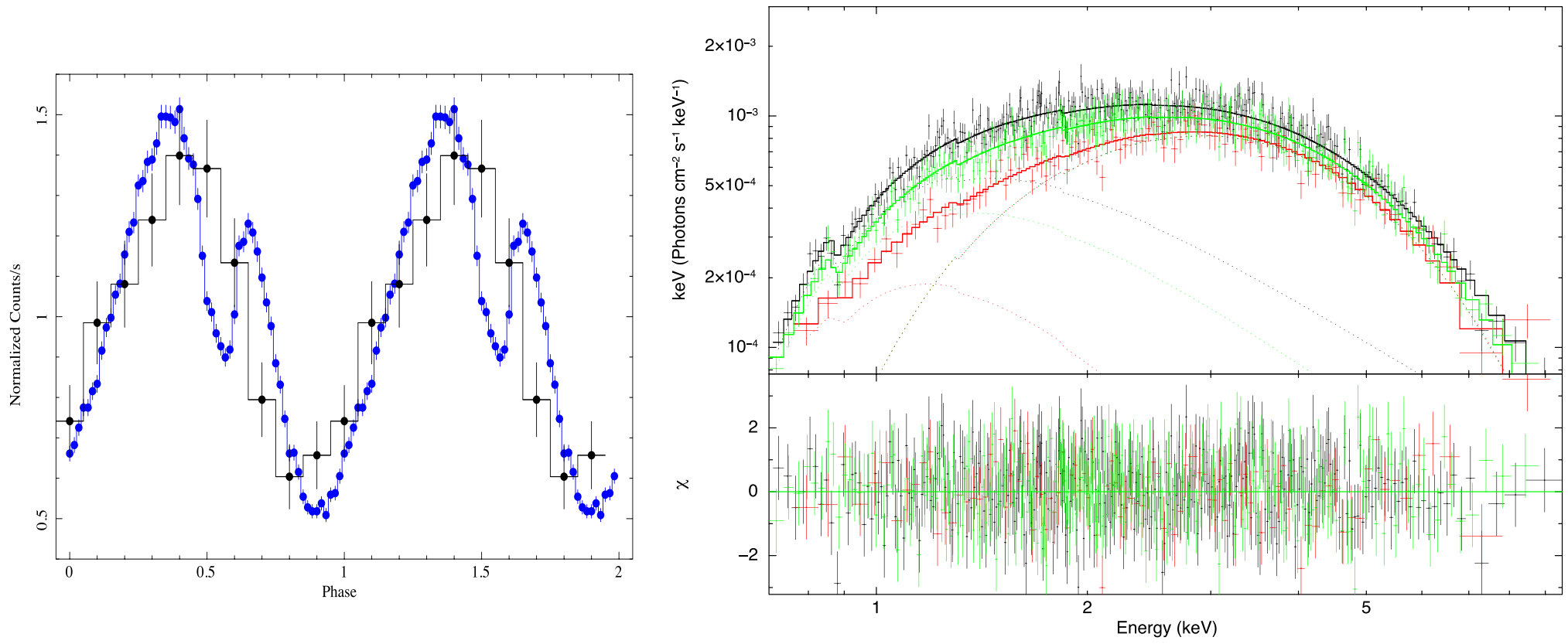

**Figure 7.** Left panel: pulse profile of the first (blue) and last (black) *XMM-Newton* observations in the 0.5–10 keV energy band. Right panel: phase-resolved unfolded spectra for the first *XMM-Newton* observation. The spectra are relative to phases 0.0–0.4 (black), 0.4–0.6 (red), and 0.6–1 (green). The phase ranges are relative to the blue pulse profile in the left panel.

(A color version of this figure is available in the online journal.)

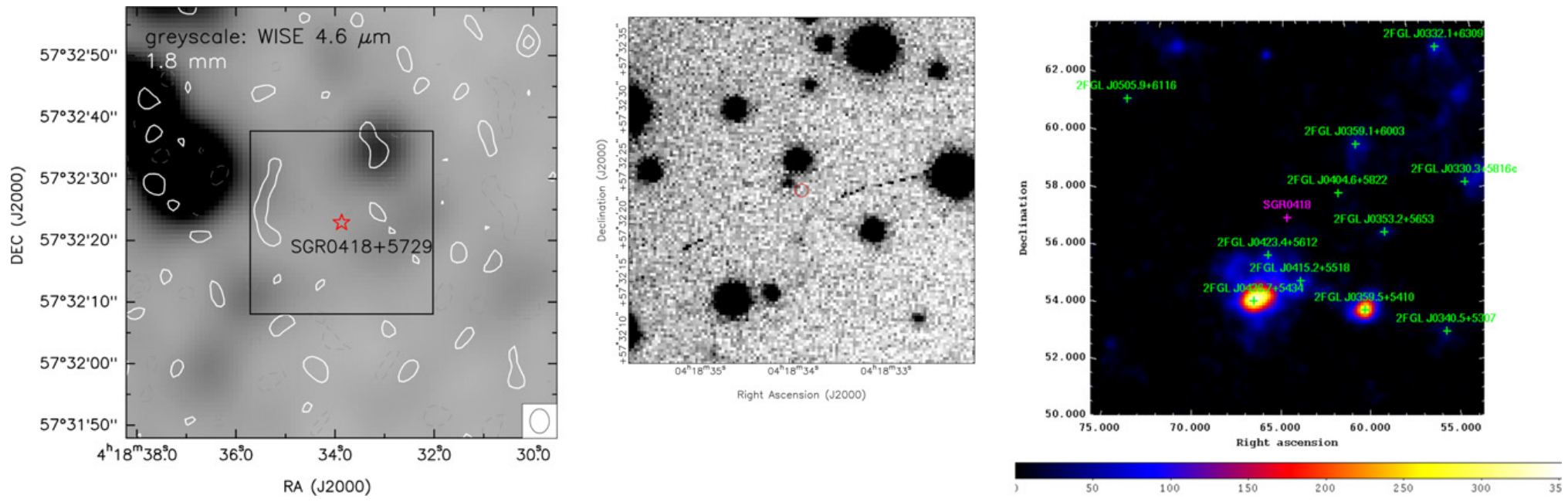

**Figure 8.** Left panel: contours of the 1.8 mm Plateau de Bure emission of the field of SGR 0418+5729. Contours are −4, −2 (dashed gray), 2, and 4 (white) times the rms noise of the map, 60 $\mu$Jy beam$^{-1}$, and they are over-plotted on the *Wide-field Infrared Survey Explorer* image at 4.6 $\mu$m. The star symbol indicates the position of SGR 0418+5729, and its size corresponds to its positional uncertainty (∼1″.2 in diameter). The synthesized beam, of 3″.95 × 3″.16, at P.A. = 0°, is shown in the bottom right corner. The square indicates the field of view of the *r*-map acquired with the William Herschel Telescope (central panel). Central panel: William Herschel Telescope *r*-band field of SGR 0418+5729. Right panel: *Fermi*-LAT (0.1–100 GeV). Diffuse subtracted TS map of the 7° × 7° sky region centered on the magnetar position. The map is calculated for $E > 300$ MeV. The 2FGL sources are labeled in green, while SGR 0418+5729 in magenta.

(A color version of this figure is available in the online journal.)

the real age corresponds to the characteristic age (which in this scenario would be needed to reach the present period of 9 s), no existing non-magnetic cooling model can account for an X-ray luminosity of $\approx 10^{31}$ erg s$^{-1}$ at a characteristic age of ≈35 Myr.

The scenario in which a large core field diffuses out has the same problem: if the real age of the star is similar to its old characteristic age, no cooling model in the literature predicts such high quiescent luminosity. In addition, while the timescales used in Soni (2012) are correct for normal, non-superfluid nuclear matter, recent work (Glampedakis et al. 2011) shows that in the presence of superfluidity in the neutron star core, the timescales for ambipolar diffusion are many orders of magnitude longer, and therefore ambipolar diffusion does not play any role during the active age of the star.

### 8.2. SGR 0418+5729 Outburst Rate

An important question is whether a relatively low dipolar field is consistent with the star-quake model in which the primary cause of the outburst is an internal deposition of energy following a crust fracture. It is often overlooked that the magnetic stress needed to break the crust is strongly dependent on density (it is much easier to break the outer crust than the inner crust) and that the crust thickness grows as the temperature drops with age. In Figure 10 we show an estimate of the minimum magnetic field variation required to induce a fracture. As assumed in Perna & Pons (2011), this estimate is obtained assuming that the crust moves through a series of equilibrium states in which its elastic stress balances the (time-dependent) magnetic stress. The deviation of the magnetic field with respect to the last unstressed configuration ($\delta B_c$) may be large enough to break the crust when

$$\delta B_c \approx \left(4\pi\sigma_b^{\max}\right)^{1/2} \quad (1)$$

where $\sigma_b^{\max}$ is the maximum stress that a neutron star crust can sustain. For young, relatively hot magnetars (crustal temperatures of $5 \times 10^8$ K), only the inner crust is solid, and strong

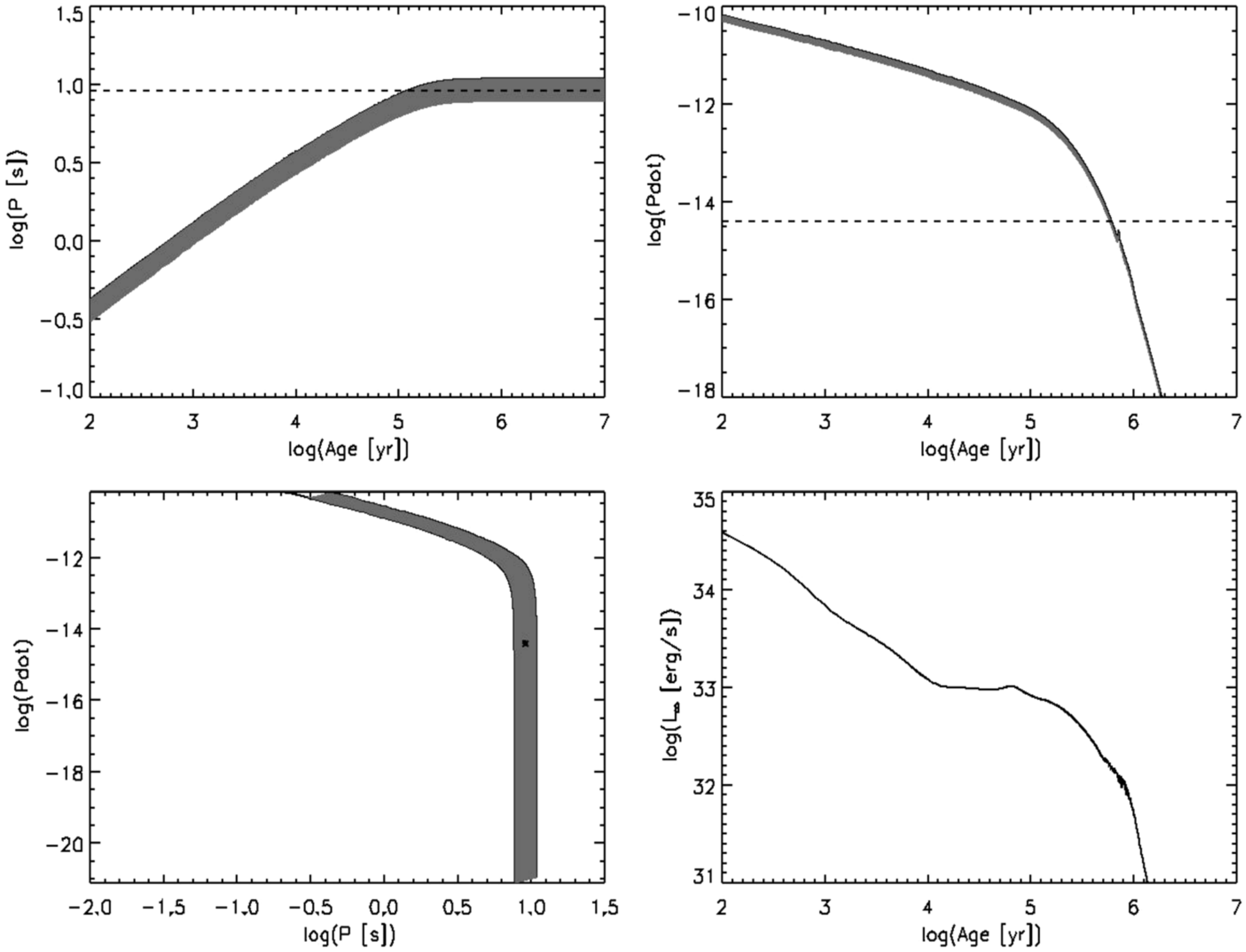

**Figure 9.** Magneto-thermal evolution of a neutron star with an initial poloidal field of $B_{\rm dip} = 1.5 \times 10^{14}$ G: period (left-top), period derivative (right-top), evolution in the $P - \dot{P}$ diagram (left-bottom), and bolometric thermal luminosity (right-bottom). The gray band corresponds to the uncertainty of the angle-dependent spin-down formula.

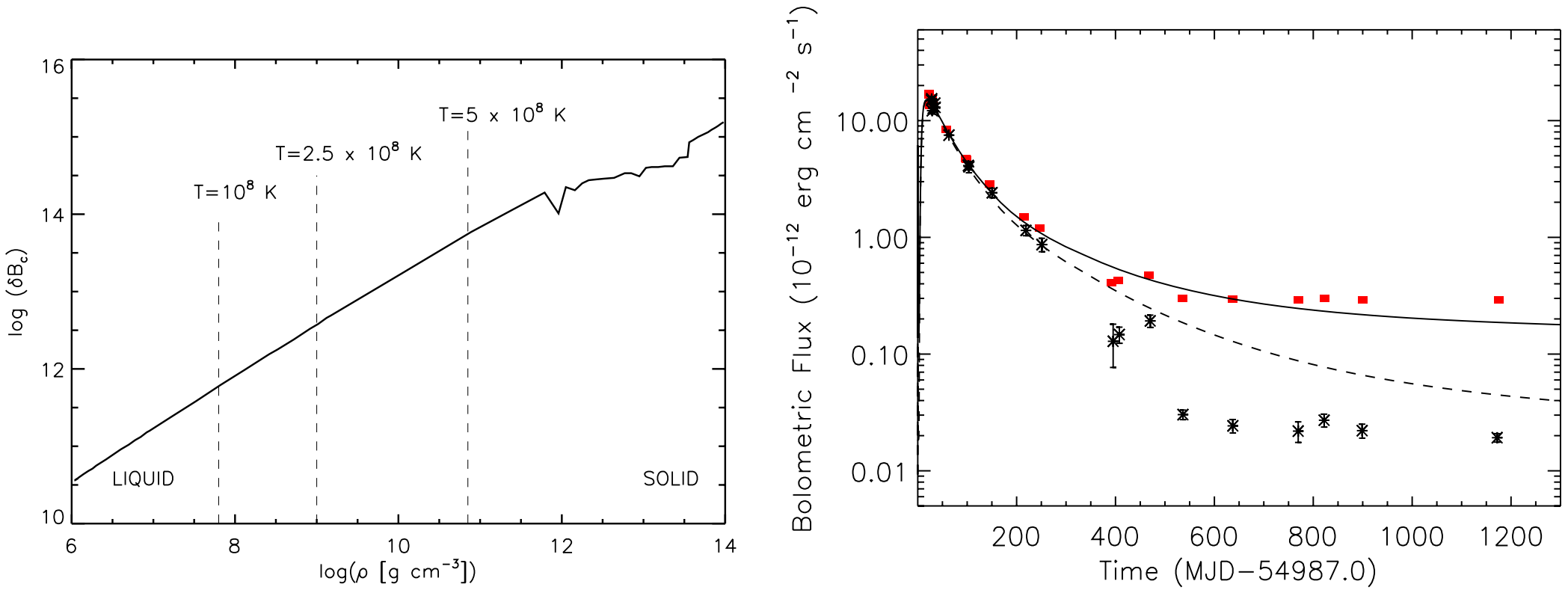

**Figure 10.** Left panel: minimum variation of the magnetic field required to break the crust by magnetic stresses as a function of density. The vertical dashed lines delimit the transition from solid to liquid for three different temperatures. Right panel: outburst modeling following Pons & Rea (2012). Black data are the 0.5–10 keV unabsorbed flux, while red squares are the bolometric unabsorbed flux with the addition of the flux of a thermal component at $kT = 0.05$ keV from the entire neutron star. Solid and dashed lines refer to the outburst model for the bolometric and 0.5–10 keV thermal flux, respectively (see text for details).

(A color version of this figure is available in the online journal.)

field variations $\delta B_c \gtrsim 10^{14}$ G are required to fracture the crust. However, for old, cold neutron stars (crustal temperatures of $10^8$ K), the solid crust extends down to $10^8$ g cm$^{-3}$, and is much easier to break, even with variations of the magnetic field of the order of $\delta B_c \gtrsim 10^{12}$ G. Note also that fractures close to the inner crust are much more energetic (because of both the higher available elastic energy and larger volume involved) than fractures in the low density region. In the first case, one can reach up to $10^{44}$ erg, while in the second case, events of $\approx 10^{41}$ erg are expected. A rough prediction of the expected outburst rate (Perna & Pons 2011; Pons & Perna 2011) for the solution model mentioned above gives $\lesssim 10^{-3}$ star-quakes yr$^{-1}$ for an object as SGR 0418+5729. Assuming that there are about $10^4$ neutron stars in the Galaxy with similar age, and that (approximately) 10% of them are born as magnetars, a naive extrapolation of this event rate to the whole neutron star population leads to an expected low-$B$ magnetar outburst rate of $\lesssim 1$ per year. Therefore, we expect that more and more objects of this class will be discovered in the upcoming years (as, e.g., Swift 1822.3–1606; Rea et al. 2012; Scholz et al. 2012).

### 8.3. SGR 0418+5729 Outburst Decay

By modeling the flux evolution in time we tested if the crustal cooling model presented in Pons & Rea (2012) can fit the flux decay of SGR 0418+5729, on the wave of what done for Swift J1822.3–1606 (Rea et al. 2012). We assume a dipolar field of $6 \times 10^{12}$ G (equatorial), an internal toroidal field of $10^{14}$ G (at maximum) as inferred in Section 8.1, and an average surface temperature of 0.05 keV, which is the temperature we expect for the surface of such an old magnetar (note that in the 0.5–10 keV band we are only seeing a tiny hot spot). The best modeling was found by injecting $2.5 \times 10^{26}$ erg cm$^{-3}$ in a thin layer in the outer crust between $4.5 \times 10^9$ and $10^{10}$ g cm$^{-3}$, and in the region contained within a cone with axis in the direction of the magnetic pole, and aperture $a \approx 0.4$ rad, for a total energy deposition of $2.5 \times 10^{41}$ erg (compatible with typical magnetar outbursts; Pons & Rea 2012). The evolution of the bolometric, and of the 0.5–10 keV flux is shown in Figure 10 (solid and dashed lines, respectively). We have shown with red squares how the observed flux decay would appear when adding the contribution of a blackbody component at 0.05 keV, mimicking the entire neutron star surface. It is clear that crustal cooling can easily explain the decay only if this further component is taken into account. In particular, the solid line is fitting the red points because the entire neutron star surface is taken into account, while it is not in the observed black data, which in fact cannot be reproduced by the dashed line. This is indicative of the difficulty of comparing theoretical cooling curves with data obtained in a certain energy band. We also note that no theoretical model predicts a surface temperature as high as 0.3 keV on a timescale of years, unless a continuous energy release is assumed (e.g., by long-lived internal currents). We finally mention that all the previous considerations are based on the (implicit) assumption that the blackbody temperature is a measure of the physical temperature of the emitting region. If this turned out not to be the case, e.g., because the spectrum is thermal but not Planckian so that a color correction is required, the physical surface temperature may be smaller than the measured blackbody temperature.

An alternative model to the crustal cooling scenario consists in the presence of currents flowing into the magnetosphere through a gradually shrinking magnetic bundle heating the neutron star surface from the top. In particular, the deepest available *XMM-Newton* observation of SGR 0418+5729, performed two months after the outburst onset, revealed that the 0.5–10 keV spectrum of SGR 0418+5729 is best reproduced by a blackbody component plus an additional non-thermal component, or by a resonant cyclotron scattering model. This suggests the presence of twisted magnetic field lines, at least in the first outburst stages. Furthermore, the limited spatial extent of the heated region (<1 km) is suggestive of a scenario in which the twist is confined within a small part of the magnetosphere, a thin current-carrying bundle, or $j$-bundle (Beloborodov 2009). As the $j$-bundle untwists during the outburst decay, the spectrum becomes more and more blackbody-like, as indeed observed.

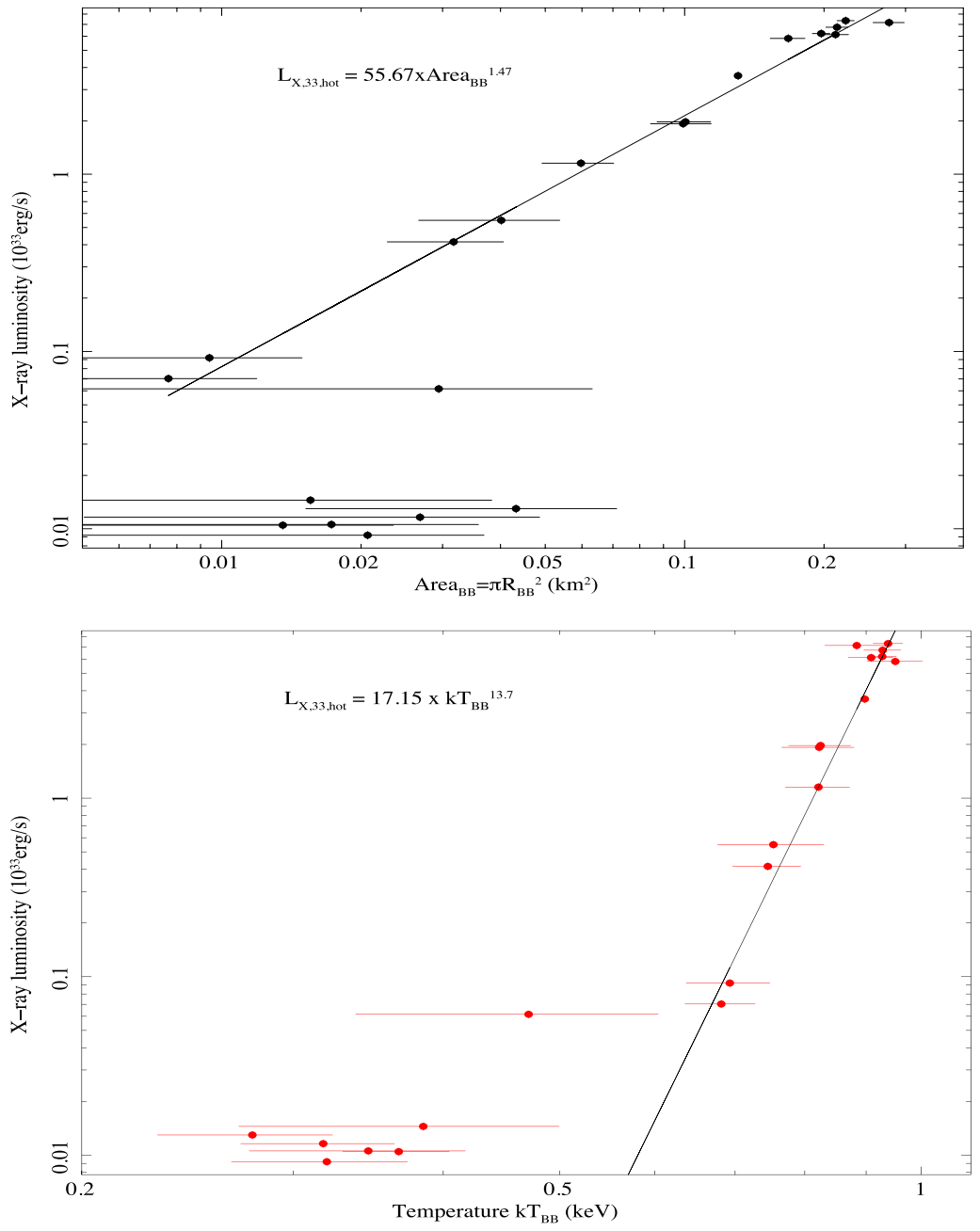

**Figure 11.** X-ray luminosity evolution as a function of the blackbody emitting area and temperature (see also Table 1; we assume a 2 kpc distance).
(A color version of this figure is available in the online journal.)

Resonant cyclotron scattering from a thin, decaying $j$-bundle appears also capable of explaining (qualitatively) the spectrum and its evolution during the first outburst stages, but whether it can explain the double-peaked pulse profiles is unclear. However, this scenario has some further difficulties: (1) the total luminosity produced by currents in the bundle is, for such a low-$B$ and a small thermal spot, well below the one observed at early times, at least if the spot is at the polar cap (see again Beloborodov 2009 and also Turolla et al. 2011), (2) the timescale for the twist decay is much shorter (<1 yr) than that implied by the long outburst of SGR 0418+5729, and (3) an approximate relation between the emitting area and the luminosity exists ($A \sim L^2$; Beloborodov 2009) if most of the luminosity is produced by current dissipation, but SGR 0418+5729 data show a somewhat flatter dependence when the first stages of the outburst are fitted (see Figure 11).

In summary, the crustal cooling model, when including also a possible hidden contribution from the entire neutron star surface, appears favorable in explaining the outburst decay of SGR 0418+5729. However, it is likely that a combination of crustal cooling and magnetospheric untwisting bundle can be operating at the first stages of the outburst.

### 8.4. Constraints on the Presence of a Fossil Disk Surrounding SGR 0418+5729

A fallback disk around SGR 0418+5729 was suggested by Alpar et al. (2011) as a way to aid the spin-down of the pulsar and explain the 9 s periodicity of this source. As an alternative, our results show (see Section 8.1) that both the thermal and the timing properties of this source can be reproduced for a pulsar age of ∼550 kyr by properly accounting for magnetic field evolution and dissipation, which also imply that the neutron star was born with a much higher dipolar field than the one measured today (see Section 8.1). Hence, in principle, the timing properties of this source would not necessarily require an additional spin down torque by a disk. However, given the suggestion that fallback disks around isolated neutron stars might be common (Michel 1988; Chevalier 1989; Lin et al. 1991), it is worthwhile to use the current multi-band upper limits to set constraints on the presence of a fallback disk around SGR 0418+5729.

Since the pulsar is currently spinning down, any disk-magnetosphere interaction must probably occur in the propeller regime, with the pulsar transferring angular momentum to the disk. For this condition to be satisfied, the inner boundary of the disk, located at about the magnetospheric radius $R_m = 2.5 \times 10^8[\dot{M}/(10^{16}\ \mathrm{g\,s^{-1}})]^{-2/7}(M_{\rm NS}/M_\odot)^{-1/7}[B/(10^{12}\ \mathrm{G})]^{4/7}$, must be equal to or larger than the corotation radius $R_{\rm co} = (GM_{\rm NS})^{1/3}\Omega^{2/3}$. The strongest constraint on the disk emission is obtained when the inner radius of the disk obtains its minimum value, i.e., $R_{\rm in} = R_m = R_{\rm co}$. For the outer radius, we assume $R_{\rm out} = 10^{14}$ cm. We found that larger values do not result in appreciably larger emission at the frequency of interest, and hence this value allows to set the tightest constraint on the disk emission.

With the inner and outer disk radii fixed as discussed above, the emission spectra from the fallback disk is computed using the model of Perna et al. (2000). The disk is assumed to be optically thick and geometrically thin, and the anisotropy in the X-ray luminosity from the source (which irradiates the disk) is neglected, since it is found to be of second order (Perna & Hernquist 2000). The disk is assumed to be still "active," i.e., viscously accreting (see Menou et al. 2001). The disk emission is the result of both viscous dissipation and re-radiation of the pulsar X-ray luminosity. In order for the magnetospheric radius not to exceed the corotation radius, the accretion rate must be limited to $\dot{M} \lesssim 10^{15}$ g s$^{-1}$. With this value, the disk luminosity in the millimeter band is dominated by reprocessing of the pulsar X-ray luminosity. At 166 GHz, the predicted flux is about 0.01 mJy for a face-on disk, below the measured limit of 0.24 mJy. Hence the presence of a fossil disk cannot be ruled out by the current millimeter measurements. Even adding the contribution from the whole surface of the star by a putative thermal component at 0.05 keV, would bring the predicted millimeter flux just around the measured flux limit for a face-on disk.

The field around SGR 0418+5729 was also observed with the Grantecan and *Hubble* telescopes (Esposito et al. 2010; Durant et al. 2011). In particular, the latter observations were performed in two wide filters, the optical, with a pivot wavelength of 5921 Å, and in the NIR, with pivot wavelength of 11534 Å. The source was not detected down to the flux of $f_O < 2.3 \times 10^{-31}$ erg s$^{-1}$ cm$^{-2}$ Hz$^{-1}$ and $f_{\rm NIR} < 4.4 \times 10^{-31}$ erg s$^{-1}$ cm$^{-2}$ Hz$^{-1}$, respectively. We found that this optical limit (nor the Grantecan or WHT limit) is not sufficiently constraining for a disk with the properties described above (the predicted emission for a face-on disk is about a factor of four below the limit). On the other hand, in the NIR, the observational limit is already able to rule out a face-on disk, which would yield an emission about twice larger than the measured flux limit. However, a disk inclined with respect to the observer by an angle $\cos\theta \lesssim 0.5$ would still be allowed by the observations (although falling short in explaining the X-ray bursts of this object).

### 8.5. Conclusions

At the time of this writing, in the ATNF pulsar catalogue (Manchester et al. 2005) 138 isolated radio pulsars have a dipolar magnetic field larger than that inferred for SGR 0418+5729. Our results imply that some of these objects might hide a strong toroidal component of the internal field, not measurable via the pulsar timing properties. A hint for such strong fields might be a high surface temperature, hotter than what would be predicted by standard cooling models at the pulsar age. However, only a few of those pulsars have had dedicated X-ray observations, and the shallow surveys do not suffice to detect such emission (expected to be as luminous as $L_X \sim 10^{31}$ erg s$^{-1}$). Furthermore, our calculation of the outburst rate of a low magnetic field magnetar also suggests that roughly once a year a quiet neutron star might turn on with magnetar-like activity.

On the other hand, if indeed a large number of neutron stars is hiding a strong magnetic field component, there would be important consequences also for other branches of astrophysics. In particular, it would imply that supernova explosions should generally produce strong magnetic fields, and that most massive stars are either producing fast rotating cores during the explosion to activate a dynamo, or are strongly magnetized themselves. Furthermore, in this scenario a non-negligible fraction of gamma-ray bursts might be due to the formation of magnetars, and the gravitational wave background produced by magnetar births should then be larger than predicted so far (important for future instruments as the Advanced-LIGO).

We are indebted to the *Chandra*, *Swift*, *XMM-Newton*, GBT, WHT, and PdBI support teams, with a special thank to Michael Bremer, for the extraordinary job in planning the PdB observations presented in this paper. We thank S. Mereghetti and H. Tong for their valuable comments on the manuscript. N.R. is supported by a Ramon y Cajal Research Fellowship, and by grants AYA2009-07391, AYA2012-39303, SGR2009-811, TW2010005, and iLINK 2011-0303. J.A.P. and D.V. acknowledge support from the grants AYA 2010-21097-C03-02 and Prometeo/2009/103. R.T. and S.M. are partially funded through an INAF 2011 PRIN grant. A.P. is supported by a JAE-Doc CSIC fellowship co-funded with the European Social Fund under the program "Junta para la Ampliación de Estudios," by the Spanish MICINN grant AYA2011-30228-C03-02 (co-funded with FEDER funds), and by the AGAUR grant 2009SGR1172 (Catalonia).

# Acknowledgements

Desidero ringraziare la Dott.ssa Patrizia A. Caraveo per la fiducia, l'aiuto concreto e per avermi inserito nella collaborazione *Fermi*-LAT, tutto questo ha permesso lo sviluppo ed il completamento di questo lavoro. Un grazie speciale ad Andrea De Luca che mi ha insegnato cosa vuol dire essere un ricercatore, come pormi di fronte ad un problema astrofisico e le innumerevoli tecniche per cercare di risolverlo, tutto questo sarà mio bagaglio per tutta la vita.

Un grazie speciale a Martino Marelli che mi ha insegnato tutto quello che c'è da sapere, e anche di più, sull'analisi dati degli strumenti X quali *XMM-Newton*, *Swift* e *Chandra*. Ringrazio inoltre Pablo Saz Parkinson per avermi ospitato alla University of California, Santa Cruz per diversi mesi, questa esperienza mi ha permesso di capire come viene fatta la ricerca all'estero. Un grazie sentito ad Andrea ("Marietto") Belfiore senza il quale non sarebbe stato possibile lo sviluppo del codice di reti neurali.

Tutta la collaborazione *Fermi*-LAT per la possibilità di sfruttare appieno i risultati ottenuti dallo strumento LAT e per i suggerimenti, più o meno utili, e per gli scambi di opinione.

Tra gli astrofisici dell'INAF-IASF di Milano ringrazio in particolar modo Nicola ("Nik") Sartore per i suggerimenti riguardanti la programmazione e i confronti a riguardo di diverse problematiche astrofisiche, Fabio Gastaldello ("Gasta") per gli innumerevoli consigli e lo scambio di opinioni riguardo le sterminate tecniche statistiche, Nicola La Palombara per gli ottimi suggerimenti riguardanti l'analisi a multilunghezza d'onda e Paolo Esposito per avermi dato la possibilità di ampliare la mia conoscenza di oggetti astrofisici estremamente energetici.

Un grazie al Prof. Matteo Matteucci, al Prof. Giuseppe De Nicolao, al Dr. Pablo Genova e al Dr. Daniele Loiacono per lo scambio di opinioni a riguaro dell'applicazione delle reti neurali.

Ringrazio inoltre la famiglia, tutti gli amici e, di cuore, Chiara, senza la quale non avrei mai concluso questo lavoro in tempo.